\documentclass{aastex62}

\usepackage{epstopdf}

\def\g{\gamma}
\newcommand{\upp}

\newcommand{\swift}{{\it Swift }}

\begin{document}
\title{The shallow decay segment of GRB X-ray afterglow revisited\\
 \vspace{0.5 cm} Submitted to APJ 2019 March 9; Accepted 2019 June 27}

\correspondingauthor{He Gao}
\email{gaohe@bnu.edu.cn}

\author[0000-0002-0786-7307]{Litao Zhao}
\affil{Department of Astronomy ,Beijing Normal University, Beijing, China;}

\author{Binbin Zhang}
\affiliation{School of Astronomy and Space Science, Nanjing University, Nanjing 210093, China;}

\author{He Gao}
\affiliation{Department of Astronomy ,
Beijing Normal University, Beijing, China;}

\author{Lin Lan}
\affiliation{Guangxi Key Laboratory for Relativistic Astrophysics, Department of Physics, Guangxi University, Nanning 530004, China;}

\author{Houjun L\"{u}}
\affiliation{Guangxi Key Laboratory for Relativistic Astrophysics, Department of Physics, Guangxi University, Nanning 530004, China;}

\author{Bing Zhang}
\affiliation{Department of Physics and Astronomy, University of Nevada, Las Vegas, NV 89154, USA}

\begin{abstract}
Based on the early-year observations from Neil Gehrels \emph{Swift} Observatory, \cite{liang07} performed a systematic analysis for the shallow decay component of gamma-ray bursts (GRBs) X-ray afterglow, in order to explore its physical origin. Here we revisit the analysis with an updated sample (with Swift/XRT GRBs between February 2004 and July 2017). We find that with a larger sample, 1) the distributions of the characteristic properties of the shallow decay phase (e.g. $t_{b}$ , $S_{X}$,  $\Gamma_{X,1}$, and $\alpha_{X,1}$) still accords with normal or lognormal distribution; 2) $\Gamma_{X,1}$ and $\Gamma_{\gamma}$ still show no correlation, but the tentative correlations of durations, energy fluences, and isotropic energies between the gamma-ray and X-ray phases still exist; 3) for most GRBs, there is no significant spectral evolution between the shallow decay segment and its follow-up segment, and the latter is usually consistent with the external-shock models; 4) assuming that the central engine has a power-law luminosity release history as $L(t)=L_{0}(\frac{t}{t_{0}})^{-q}$, we find that the value $q$ is mainly distributed between -0.5 and 0.5, with an average value of 0.16$\pm$ 0.12; 5) the tentative correlation between $E_{\rm{iso},X}$ and $t'_{b}$ disappears, so that the global 3-parameter correlation ($E_{\rm{iso},X}-E'_{p}-t'_{b}$) becomes less significant; 6) the anti-correlation between $L_{X}$ and $t'_{b}$ and the three-parameter correlation ($E_{\rm{iso},\gamma}-L_{X}-t_{b}$) indeed exist with a high confidence level. Overall, our results are generally consistent with \cite{liang07}, confirming their suggestion that the shallow decay segment in most bursts is consistent with an external forward shock origin, probably due to a continuous energy injection from a long-lived central engine.
\end{abstract}

\section{Introduction} \label{sec:intro}

Gamma-ray bursts (GRBs) are considered as the most extreme explosive events in the universe, which contains two phenomenological emission phases: prompt phase (with an initial prompt $\g$-ray emission) and afterglow phase (with a longer-lived broadband emission) \citep{zhang19}. Although there are many uncertainties in the detailed physics of the prompt emission, mainly due to our poorly understanding the degree of magnetization of the GRB jet \citep{zhang14prompt,kumarzhang15}, a generic synchrotron external shock model has been constructed for interpreting the broadband afterglow data \citep{rees92,rees94,meszaros93,meszaros97,gao13,wang15}.

In the pre-\swift era, the simple external shock signal was found to successfully explain a bunch of late-time afterglow data \citep{wijers97,waxman97a,wijers99,huang99,huang00,panaitescu01,panaitescu02,yost03}. With the successful launch of the Neil Gehrels \emph{Swift} Observatory, unprecedented new information about early-time afterglows was revealed \citep{tagliaferri05,burrows05a,zhang06,nousek06,obrien06,evans09}, especially in the X-ray band, thanks mainly to the rapid slewing and precise localization capability of its on-board X-Ray Telescope (XRT) \citep{burrows05b}. After 6 months data accumulating, a five-segment canonical X-ray afterglow light curve was proposed \citep{zhang06}, including a distinct rapidly decaying component, a shallow decay component, a normal decay component, a post jet break component and X-ray flares. With 2 years data collecting (with $\sim100$ GRBs), a series of systematic analysis of the Swift XRT data was performed, in order to explore the physical origin for each segment of the canonical light curve \citep{zhangbb07,liang07,willingale07,liang08,evans09}.

After fourteen years of operation, thousands of \swift GRBs were detected by XRT, it is of great interest to revisit the early analysis and see whether the previous results still consistent with the larger sample. In this work, we focus on the analysis of shallow decay component from \cite{liang07} (hereafter L07). L07 made a comprehensive analysis of the properties of the shallow decay segment with a sample of 53 long Swift GRBs detected before 2007 February. Their statistic results are summarized as follows:
(1) the distributions of shallow decay segment parameters are normal and lognormal, with $\rm{1og}_{10}t_{b}(s)=4.09\pm0.61$, $\rm{1og}_{10} S_{X}(\rm{ergs}\ \rm{cm}^{-2} ) = -6.52\pm0.69$, $\Gamma_{X,1}=2.09\pm 0.21$ and $\alpha_{1}=0.35\pm0.35$; (2) The spectrum of the shallow decay phase is softer than that of the prompt gamma-ray phase, while the X-ray fluence and isotropic energy are almost linearly correlated with gamma-ray fluence and gamma-ray energy, respectively; (3) there is no significant spectral evolution between the shallow decay segment and its follow-up segment, and the follow-up segment in most bursts is consistent with the closure relations of external-shock models; (4) within the scenario of refreshed external shocks, the average energy injection index q $\sim$ 0, suggesting a roughly constant injection luminosity from the central engine; (5) there is an empirical multivariate relation between parameters $E_{\rm{iso},X}-E'_{p}-t'_{b}$ (henceforth the prime marks properties in the burst rest frame). Based on these results, L07 proposed that the shallow decay segment in most bursts is consistent with an external forward shock origin with continuous energy injection from a long-lived central engine. For a small fraction of bursts, whose post-break phase significantly deviate from the external-shock models, the shallow decay phase might be of  internal origin and demand a long-lived emission component directly from the central engine.

In this work, we revisit the results of L07 with \swift observed GRBs between February 2004 and July 2017. In addition, it has been discovered that there exists a rough anti-correlation between the rest frame X-ray plateau end time ($t'_{b}$) and X-ray luminosity $L_{X}$ \citep{dainotti08,dainotti10,dainotti11a}. The slope is roughly $-1$(\cite{dainotti13a}). This suggests that the total plateau energy has relatively small scatter: a longer plateau tends to have a lower luminosity and vice versa. \cite{dainotti11b} analysed and suggested correlations between $L_{X}$ with several prompt emission parameters, including the isotropic energy $E_{\rm{iso},\gamma}$. \cite{xuhuang12} claimed that a three-parameters collection, expressed as $L_{X} \varpropto t_{b}^{0.87}E_{\rm{iso},\gamma}^{0.88}$, becomes tighter. On the other hand, \cite{dainotti16, dainotti17} extend the $L_{X}-t'_{b}$ correlation by adding a third parameter (the peak luminosity in the prompt emission $L_{\rm peak}$), and find that the new $L_{X}-t'_{b}-L_{\rm peak}$ correlation becomes much tighter, which is 37\% tighter than the \cite{xuhuang12} correlation. In this work, we also test these relations based on our new sample.

\section{Data Reduction And Sample Selection}

The XRT light curves data were downloaded from the Swift/XRT team website\footnote{\url{http://www.swift.ac.uk/xrt\_ curves/}}\citep{belokurov09}, and processed with HEASOFT v6.12. 1291 Swift GRBs were detected by Swift/XRT between February 2004 and July 2017, with 625 GRBs having well-sampled XRT light curves, which contains at least 6 data points, excluding upper limits. We then fit the XRT light curves (in logarithmic scale) with multi-segment broken power law function. Here the multivariate adaptive regression spline (MARS) technique (e.g. \cite{friedman91}) is adopted, which can automatically determine both variable selection and functional form, resulting in an explanatory predictive model. It has been proven that MARS can automatically fit the XRT light curve with multi-segment broken power-law function (results consistent with fitting results provided by the XRT GRB online catalog), detect and optimize all breaks, and record all break times and power-law indices for each segment (see \cite{zhang14} and \cite{gao17} for details). We find that the distribution of the index difference between adjacent segments for all GRBs could be fitted with a Gaussian function (with mean value as 0.6 and standard deviation as 0.3). To avoid the potential over-fitting problem from MARS technique, we treat the adjacent segments with index difference smaller than 0.3 as one component when calculating the segment time span.

We define segments with decay slope shallower than 0.8 and time span in log scale larger than 0.4 dex\footnote{We have tested that reasonably adjusting the criteria values of 0.8 and 0.4 might slightly change the size of the final sample but would not affect the main conclusions.} as the shallow decay component candidates. 284 GRBs are excluded due to the lack of any shallow decay candidates. Among the rest 341 GRBs, 67 bursts are further excluded since there are too few data points or there seem seems to be weak flare signatures in the shallow decay segments. Moreover, we excluded 54 GRBs whose shallow decay candidate.segments  lacks of a follow-up segment, namely the shallow decay segment is the only segment or the last segment in the light curve. 18 GRBs are excluded since they consist more than one nonadjacent shallow decay candidates, which would be discussed in a separate work. GRB 060218 is excluded since its early shallow decay behavior in the X-ray light curve should be mainly determined by the shock break-out emission instead of external shock emission \citep{campana06}.

For the remaining 201 GRBs, we record the beginning time ($t_{1}$) of the shallow decay segment and the end time ($t_{2}$) of its follow-up segment. We then fit the light curve in the time interval [$t_{1}$, $t_{2}$] with a smoothed broken power law (BPL) function
\begin{eqnarray}
\label{Eq1}
f(t)=f_{0}[(\frac{t}{t_{b}}) ^{w\alpha_{1}} +   (\frac{t}{t_{b}}) ^{w\alpha_{2}}]^{-1/w}
\end{eqnarray}
with $\alpha_{1}$ and $\alpha_{2}$ representing the decay scopes of pre-break segment and post-break segment, and $w$ determining the sharpness of the break. Here we adopt $w=3$ as suggested in L07, in order to make better comparison with their results. All fitting light curves are shown in Fig.\ref{fig-1-1}, and the best fitting parameters are collected in Table.\ref{table-1} 
. The X-ray fluence ($S_{X}$) of the shallow decay segment is obtained by integrating the fitted light curve from $t_{1}$ to $t_{b}$.  Note that here we further exclude three bursts in our sample since the difference between their $\alpha_{1}$ and $\alpha_{2}$ are smaller than 0.3.

 There is no spectral evolution during the shallow decay and followed normal decay phases\citep{liang07}, so that we extract the time-integrated spectra in the time interval of [$t_{1}$, $t_{b}$] and [$t_{b}$, $t_{2}$]. We fit the X-ray spectrum with the Xspec package, and an absorbed power-law model is invoked to fit the spectrum, i.e., abs$\times$zabs$\times$zdust$\times$power law, where zdust is the dust extinction of the GRB host galaxy, abs and zabs are the absorption models for the Milky Way and the GRB host galaxy, respectively. Then, we derived the photon indices $\Gamma_{X,1}$ and $\Gamma_{X,2}$ for shallow decay and followed normal decay phases.

\begin{figure*}
\begin{center}
\setlength{\abovecaptionskip}{0.cm}
\setlength{\belowcaptionskip}{-0.cm}
\hspace{0cm}
\graphicspath{{oneparametre/}}
\figurenum{2}
\includegraphics[width=6.7cm,height=5cm]{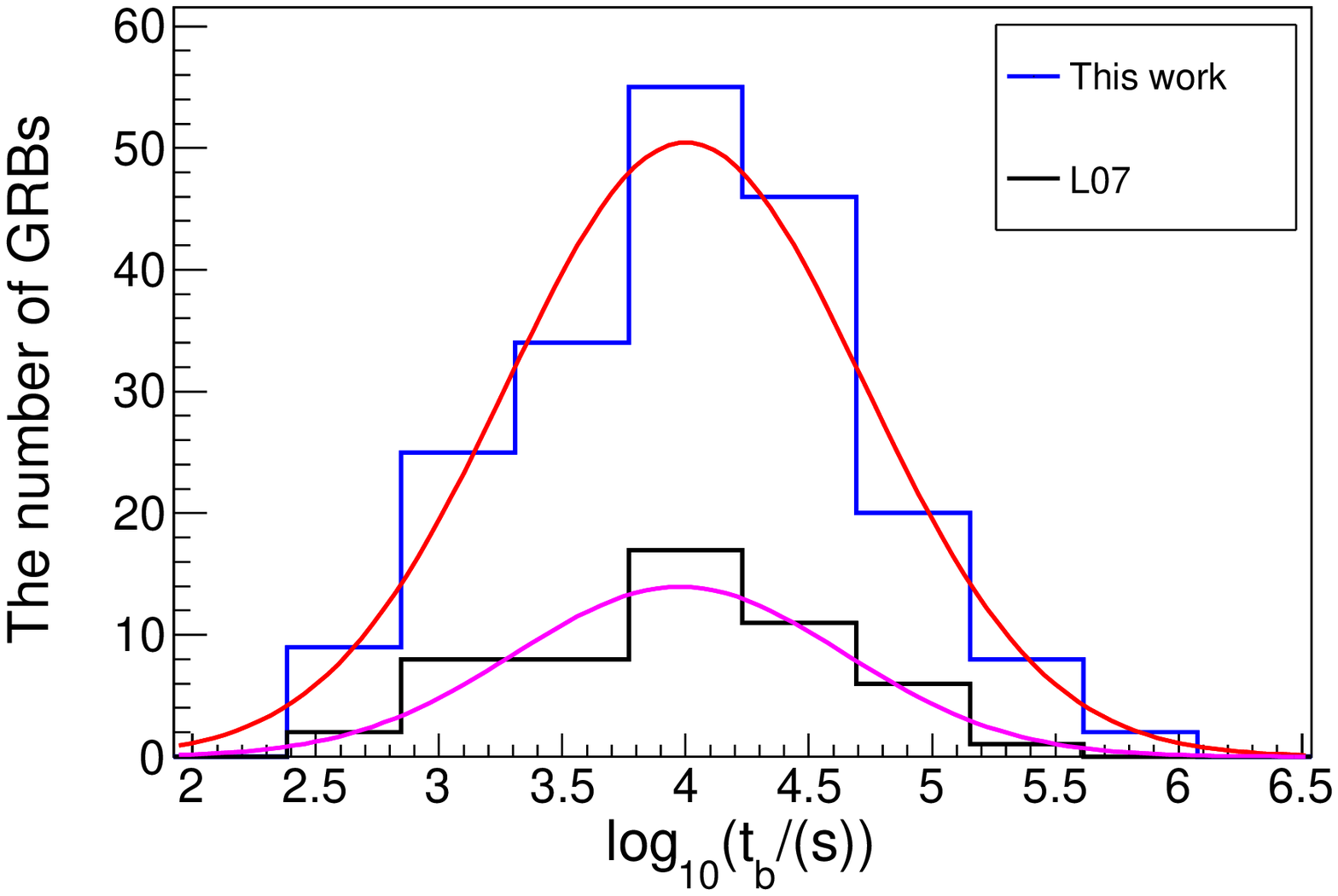}
\includegraphics[width=6.7cm,height=5cm]{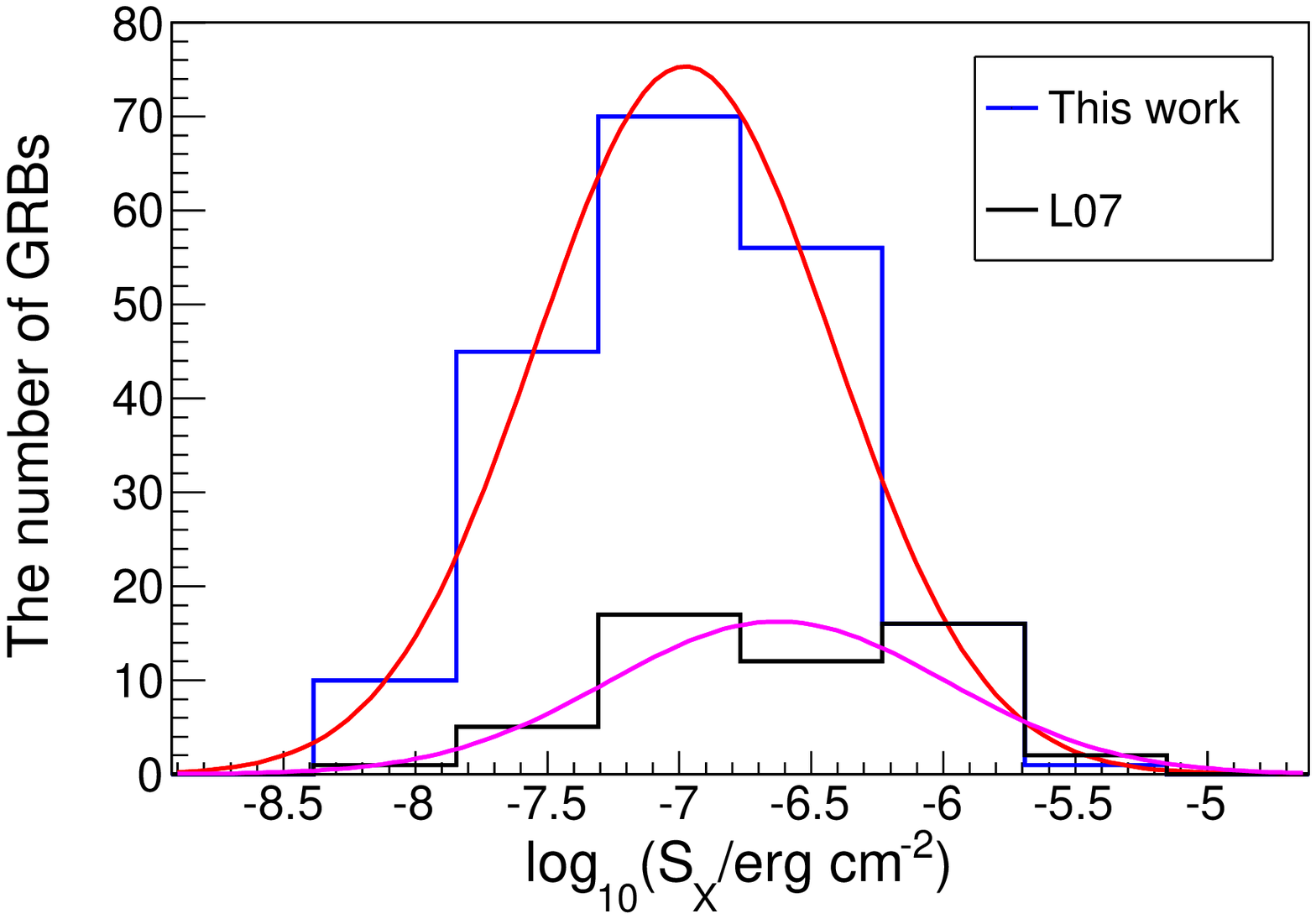}
\includegraphics[width=6.7cm,height=5cm]{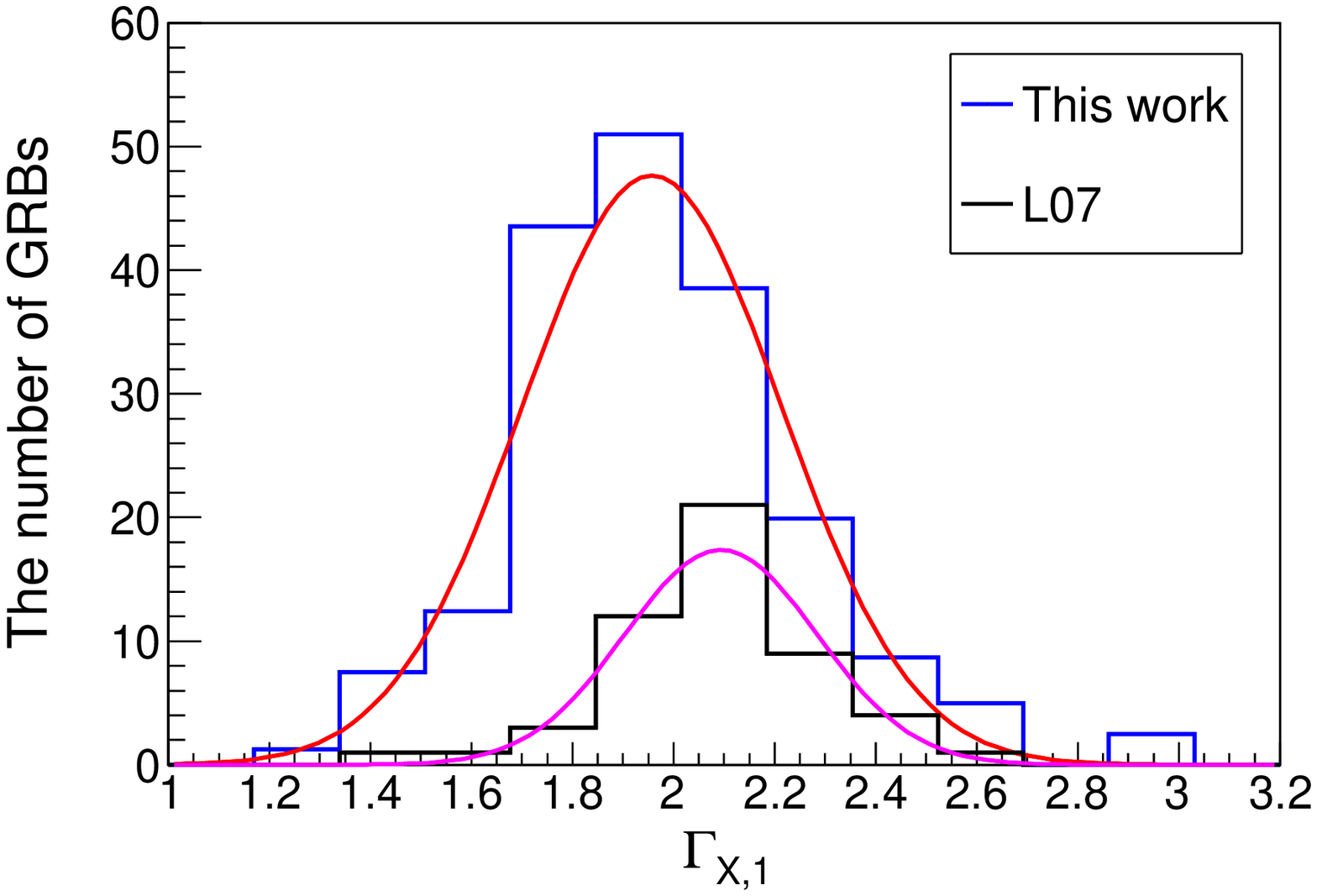}
\includegraphics[width=6.7cm,height=5cm]{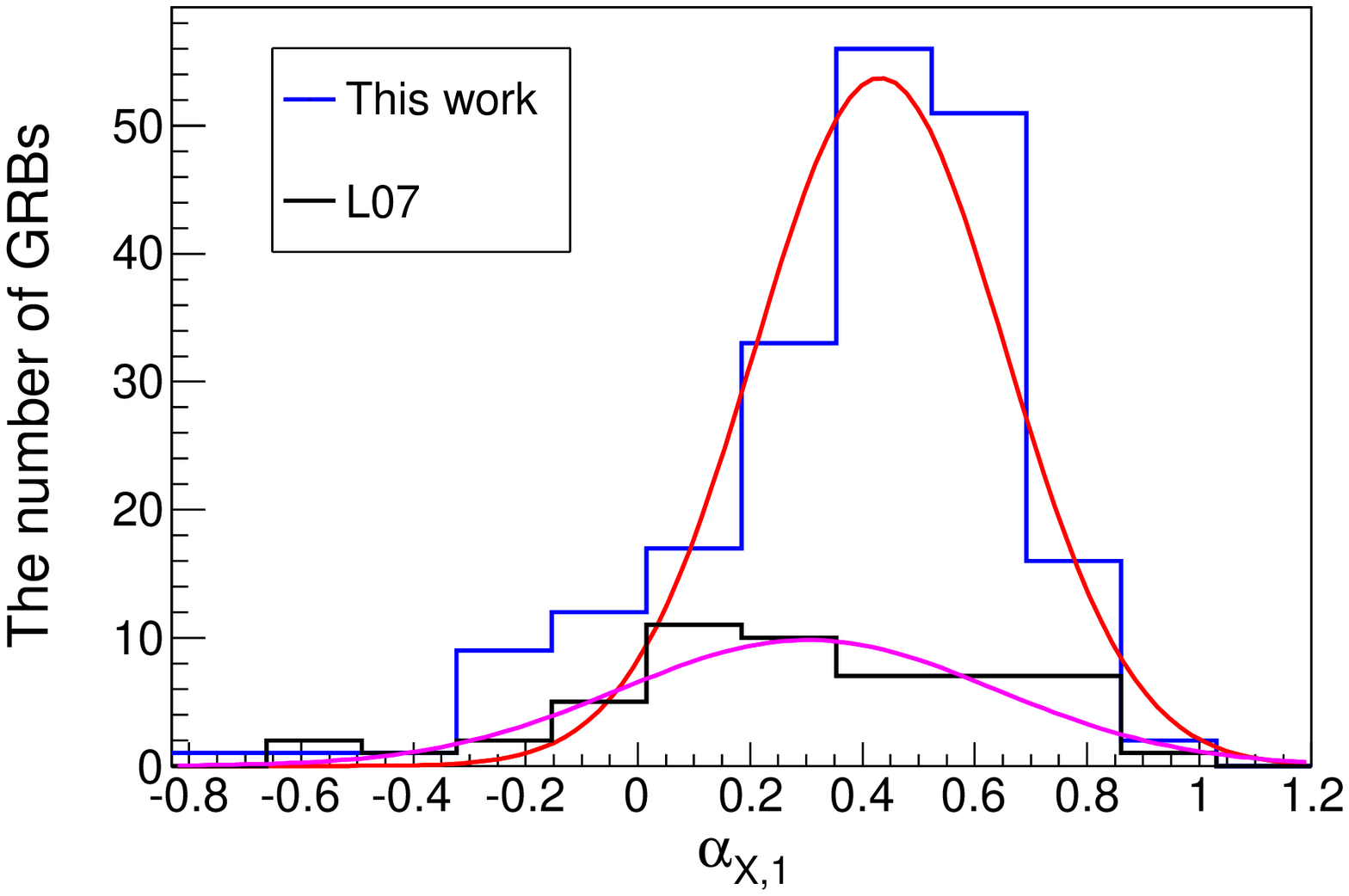}
\caption{Distributions of the characteristic properties of the shallow decay phase for GRBs in our sample, comparing with L07's results. The solid red and pink lines are the results of Gaussian fits.}
\label{fig-2}
\end{center}
\end{figure*}

\section{L07 revisited}

\subsection{Distributions of the characteristic properties of the shallow decay phase}

For our sample, we find that the distributions of the characteristic properties of the shallow decay phase (e.g. $t_{b}$ , $S_{X}$,  $\Gamma_{X,1}$, and $\alpha_{X,1}$) accords with normal or lognormal distribution\footnote{In our analysis, we have checked several different distribution functions, including Weibull distribution, Beta distribution, Gamma distribution, normal distribution, lognormal distribution, Exponential distribution and so on. We compare these distribution functions by comparing their fitting square error (sum of squared discrepancies between histogram frequencies and fitted-distribution frequencies), and find that normal or lognormal functions are always the best or very close to the best to fit our interested distributions. In order to better compare our results with L07 findings, and considering that normal or lognormal functions have stronger physical meaning than other functions, here we only use normal and lognormal functions to fit relevant distributions.}, with $\rm{log}_{10}(t_{b}) =4\pm0.72$, $\rm{log}_{10}(S_{X}(\rm{ergs \ cm^{-2})})= -6.97\pm0.56$, $\Gamma_{X,1} =1.96\pm0.26$ and $\alpha_{X,1}=0.43\pm0.22$. For comparison, we plot our result together with L07's result in Fig.\ref{fig-2}. It is found that when larger sample being involved, the distributions for all the parameters become slightly broader, but the peaks of each distribution barely change, for instance, the mean value for distributions of $\rm{log}_{10}(t_{b})$, $\rm {log}_{10}(S_{X})$, $\Gamma_{X,1}$, and $\alpha_{X,1}$ differs by $3\%$, $7\%$, $6\%$, and $23\%$ respectively, between L07 results and our results.

\subsection{Relationship of the shallow decay phase to the prompt gamma-ray phase} \label{sec:style}

\begin{figure*}
\begin{center}
\setlength{\abovecaptionskip}{0.cm}
\setlength{\belowcaptionskip}{-0.cm}
\hspace{0cm}
\graphicspath{{twoparametre/}}
\figurenum{3}
\includegraphics[width=6.7cm,height=6.7cm]{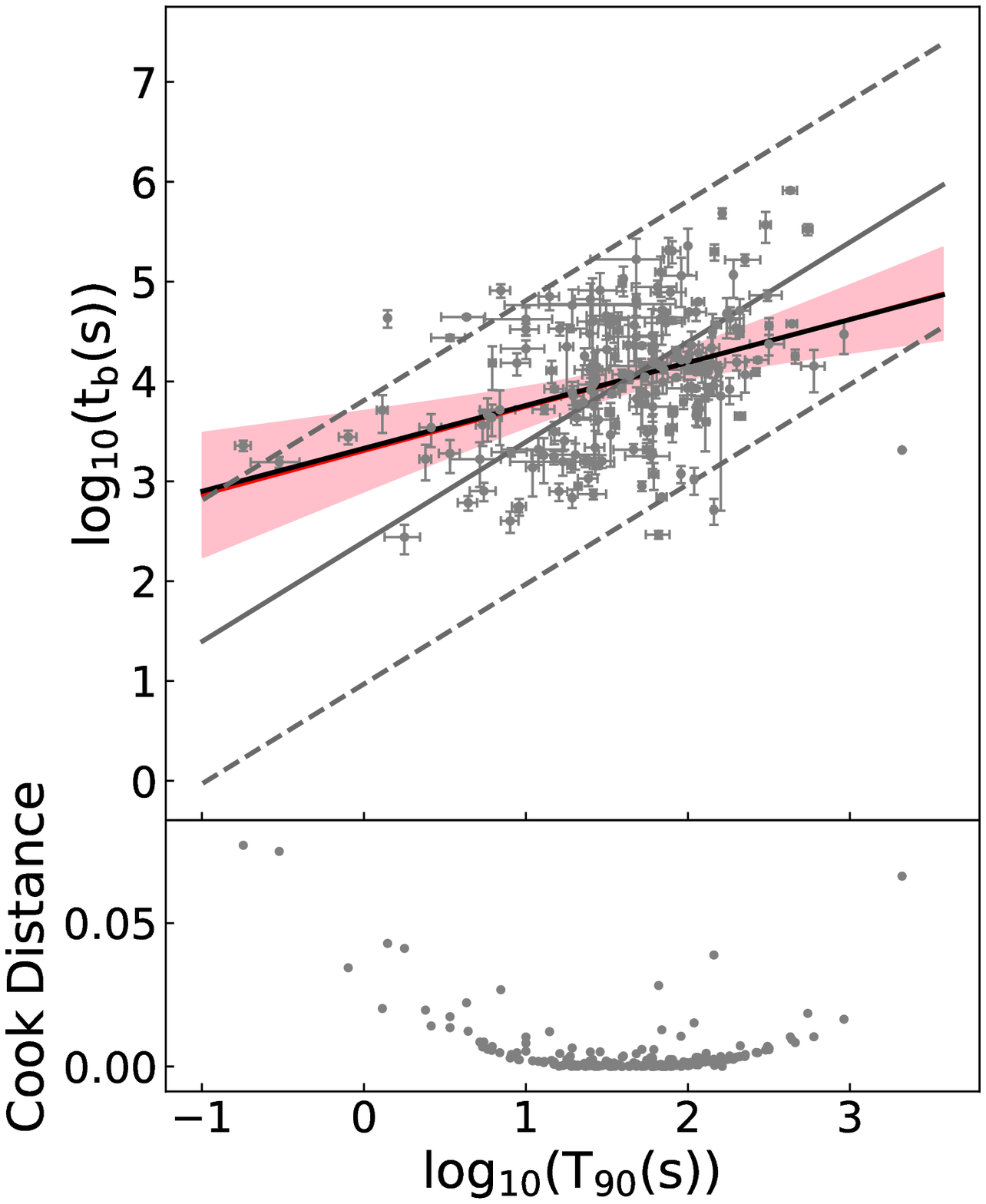}
\includegraphics[width=6.7cm,height=6.7cm]{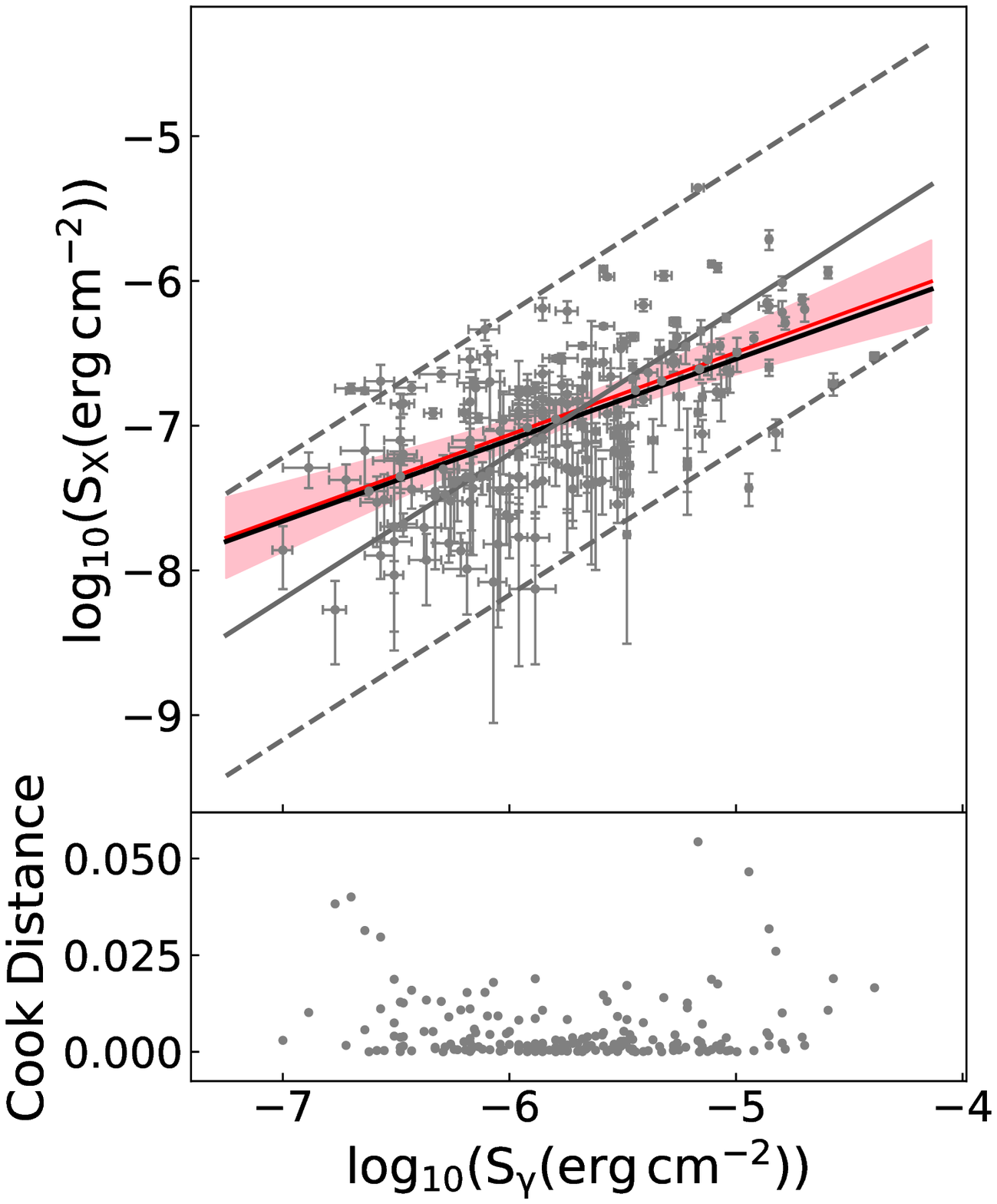}
\includegraphics[width=6.7cm,height=6.7cm]{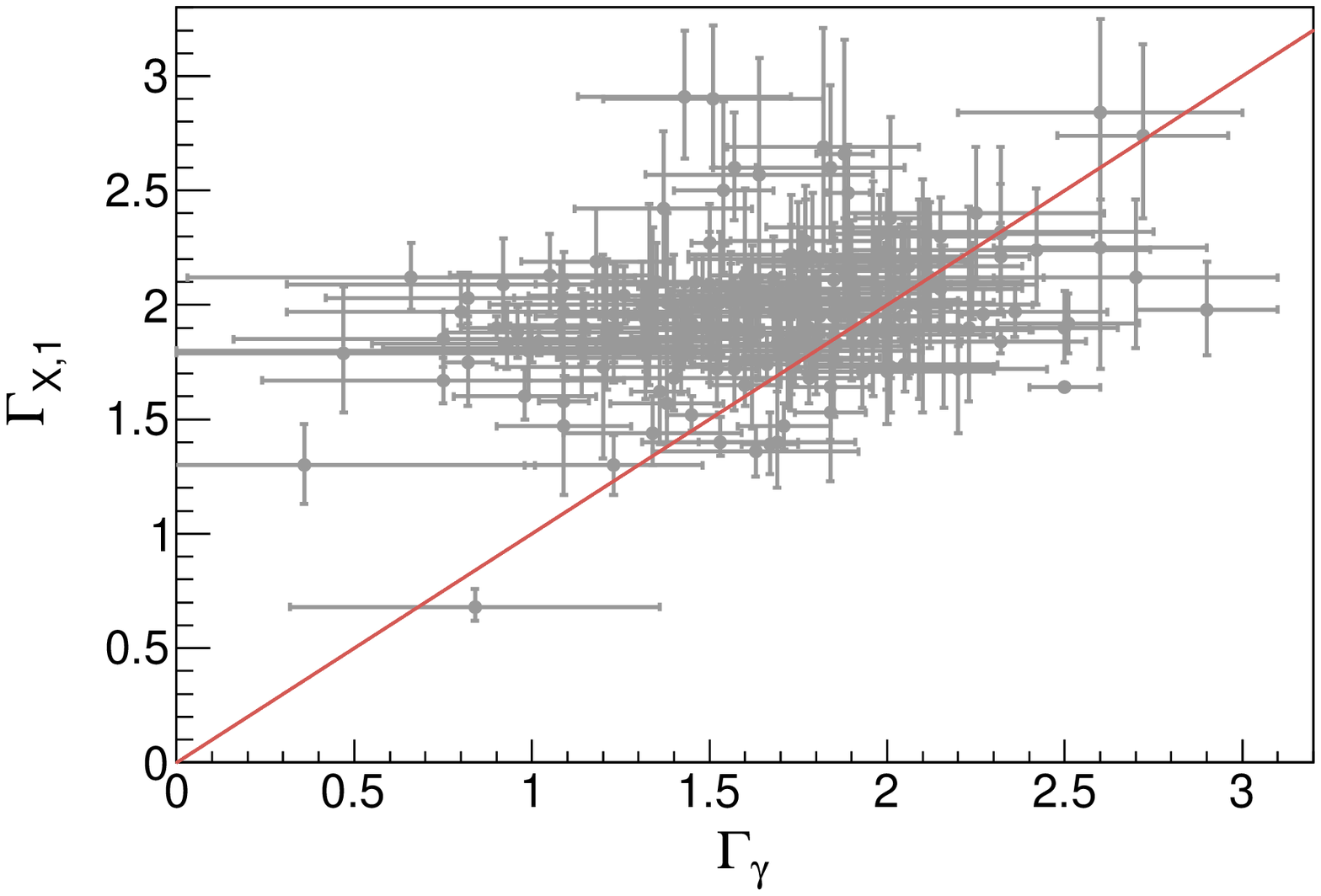}
\includegraphics[width=6.7cm,height=6.7cm]{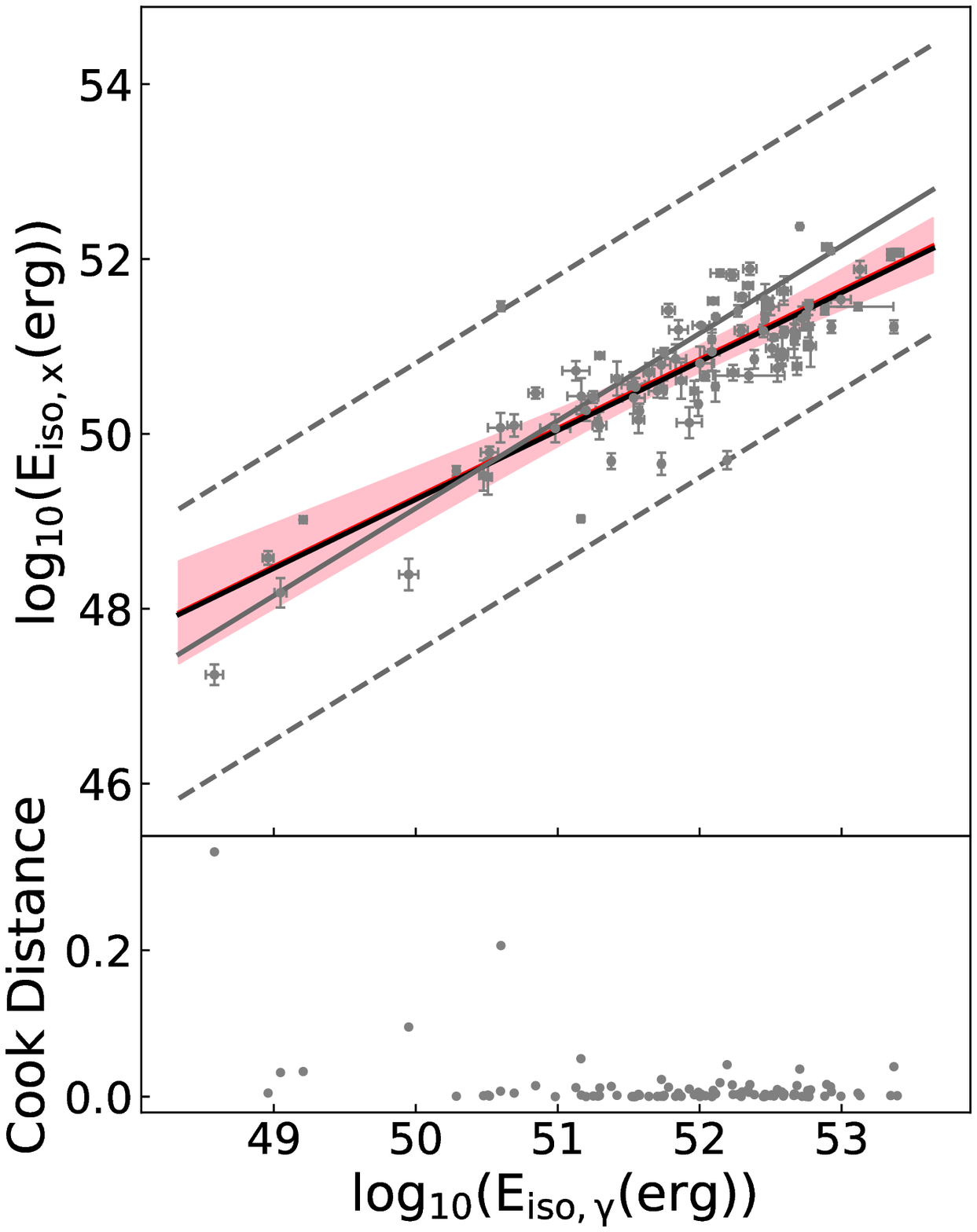}
\caption{Parameters relationship between shallow decay phase and prompt emission phase, including  $t_b$- $T_{\rm{90}}$, $S_{X}$- $S_{\gamma}$, $\Gamma_{X,1}-\Gamma_{\gamma}$ and $E_{\rm{iso},X}$-$E_{\rm{iso},\gamma}$. The black solid line represents the best fit with least square regression method. The red solid line present the best fit with the bivariate linear regression method and the pink shadowed region show the intrinsic scatter to the population $3\sigma_{\rm int}$. The grey lines mark the best fitting results (solid) and its 2$\sigma$ linear correlation region (dotted), which is defined with $y=x+A\pm 2\delta_{A}$, where y and x are the two quantities in question and A and $\sigma_{A}$ are the mean and its $1\sigma$ standard error for the y-x correlation. The orange solid lines show $y=x$.}
\label{fig-3}
\end{center}
\end{figure*}

In order to investigate the relationship of the shallow decay phase to the prompt gamma-ray phase, we first collect relevant parameters of prompt gamma-ray emission for our sample, including the duration ($T_{90}$), fluence ($S_{\gamma}$), photon indices ($\Gamma_{\gamma}$), and the peak energy of the $\nu F_{\nu}$ spectrum ($E_{p}$) through published articles and GCN Circulars (the results are presented in Table \ref{table-2}). For BAT spectrum that can be fitted by a single power-law, we estimate $E_{p}$ through the correlation between the BAT photon index $\Gamma$ and $E_{p}$ (\citep{zhangb07,sakamoto09c,virgili12}), namely
\begin{eqnarray}
\label{Eq7}
\rm{log}_{10}(E_{p})=(4.34\pm0.475)-(1.32\pm0.219)\rm{log}_{10}(\Gamma)
\end{eqnarray}

Within our sample, 97 GRBs have redshift measurements, whose isotropic-equivalent radiation energies ($E_{\rm{iso},\gamma}$, $E_{\rm{iso},X}$) in the prompt phase and in the shallow decay phase could be calculated as
 \begin{eqnarray}
\label{Eq3}
E_{\rm{iso},(\gamma,X)}=\frac{4\pi D_{L}^{2}S_{(\gamma,X)}}{1+z}
\end{eqnarray}
where $D_{L}$ is the luminosity distance. Here $H_{0}$ =71km $s^{-1} Mpc^{-1}, \Omega_{M}=0.3,   \Omega_{A}=0.7$ are adopted for calculating $D_{L}$. We then calculate the bolometric energy $E_{\rm{iso},(\gamma,X)}^{b}$ in the 1$\sim10^{4}$ keV band with the k-correction method proposed by \cite{bloom01},
\begin{eqnarray}
\label{Eq6}
E_{\rm{iso},(\gamma,X)}^{b}=kE_{\rm{iso},(\gamma,X)},
\end{eqnarray}
where
\begin{eqnarray}
\label{Eq5}
k=\frac{\int_{1/(1+z)}^{10^{4}/(1+z)}EN(E)dE}{\int_{15}^{150}EN(E)dE}.
\end{eqnarray}
assuming the gamma-ray spectrum for all GRBs following Band function with photon indices $-1$ and $-2.3$ before and after $E_p$.

Fig.\ref{fig-3} shows parameter relationships of the shallow decay phase to the prompt gamma-ray phase, including $t_b$- $T_{90}$, $S_{X}$- $S_{\gamma}$, $\Gamma_{X,1}-\Gamma_{\gamma}$ and $E_{\rm iso,X}$-$E_{\rm iso,\gamma}$ relationships. We find that with larger sample, $\Gamma_{X,1}$ and $\Gamma_{\gamma}$ still show no correlation, with $\Gamma_{X,1}$ systemically larger than $\Gamma_{\gamma}$. The result is consistent with the relevant findings in \cite{dainotti15}. The tentative correlations of duration, energy fluences, and isotropic energies between the gamma-ray and X-ray phases still exist, with the best fit as $\rm{log}_{10}(t_{b})=(0.43\pm0.08)log_{10}(T_{90})+(3.33\pm0.15)$ (Spearman correlation coefficient $r=0.39$, significance level $p<10^{-4}$ for $N=198$, fraction of the variance $R^2=14.1\%$, and significance of Anderson-Darling test $p_{\rm{AD}}=0.647$), $\rm{log}_{10}(S_{X})=(0.56\pm0.06)\rm{log}_{10}(S_{\gamma})+(-3.74\pm0.37)$ ($r=0.55$, $p<10^{-4}$ for $N=198$, $R^2=35.3\%$ and $p_{\rm{AD}}=0.44$) and $\rm{log}_{10}(E_{\rm{iso},X})=(0.788\pm0.05)log_{10}(E_{\rm{iso},\gamma})+(9.85\pm2.58)$ ($r=0.77$, $p<10^{-4}$ for $N=198$, $R^2=72.4\%$ and $p_{\rm{AD}}=0.566$). With the exception of $E_{\rm iso,X}$-$E_{\rm iso,\gamma}$, the correlations of $t_b$- $T_{90}$  and $E_{\rm iso,X}$-$E_{\rm iso,\gamma}$ become weaker with lager sample, and the correlation coefficients were reduced by approximately 18\% (for duration) and 21\% (for energy fluence). The fitting results and Cook distance for the fitting are shown in figure \ref{fig-3}. Here we also apply the bivariate linear regression procedure to fit the data \cite[][for details]{kelly07}. We find that the best fit results for different regression methods are consistent. The intrinsic scatter to the population $\sigma_{\rm int}$ is shown in Fig.\ref{fig-3}. In order to compare with L07, we also check the possible linear correlations for the quantities in the two phases by defining $2\sigma$ linear correlation regions\footnote{$2\sigma$ linear correlation regions is defined with $y=x+A\pm 2\delta_{A}$, where y and x are the two quantities in question and A and $\sigma_{A}$ are the mean and its $1\sigma$ standard error for the y-x correlation} (see Figure.\ref{fig-3} for details), and we find that most of our GRBs fall in this region, suggesting that the radiation during the shallow decay phase might indeed be correlated with that in the prompt gamma-ray phase.

\subsection{Test physical origin of the shallow decay segment} \label{sec:style}

\begin{figure*}
\begin{center}
\setlength{\abovecaptionskip}{0.cm}
\setlength{\belowcaptionskip}{-0.cm}
\hspace{0cm}
\figurenum{4}
\graphicspath{{spectralevolution/}}
\includegraphics[width=5.5cm,height=5cm]{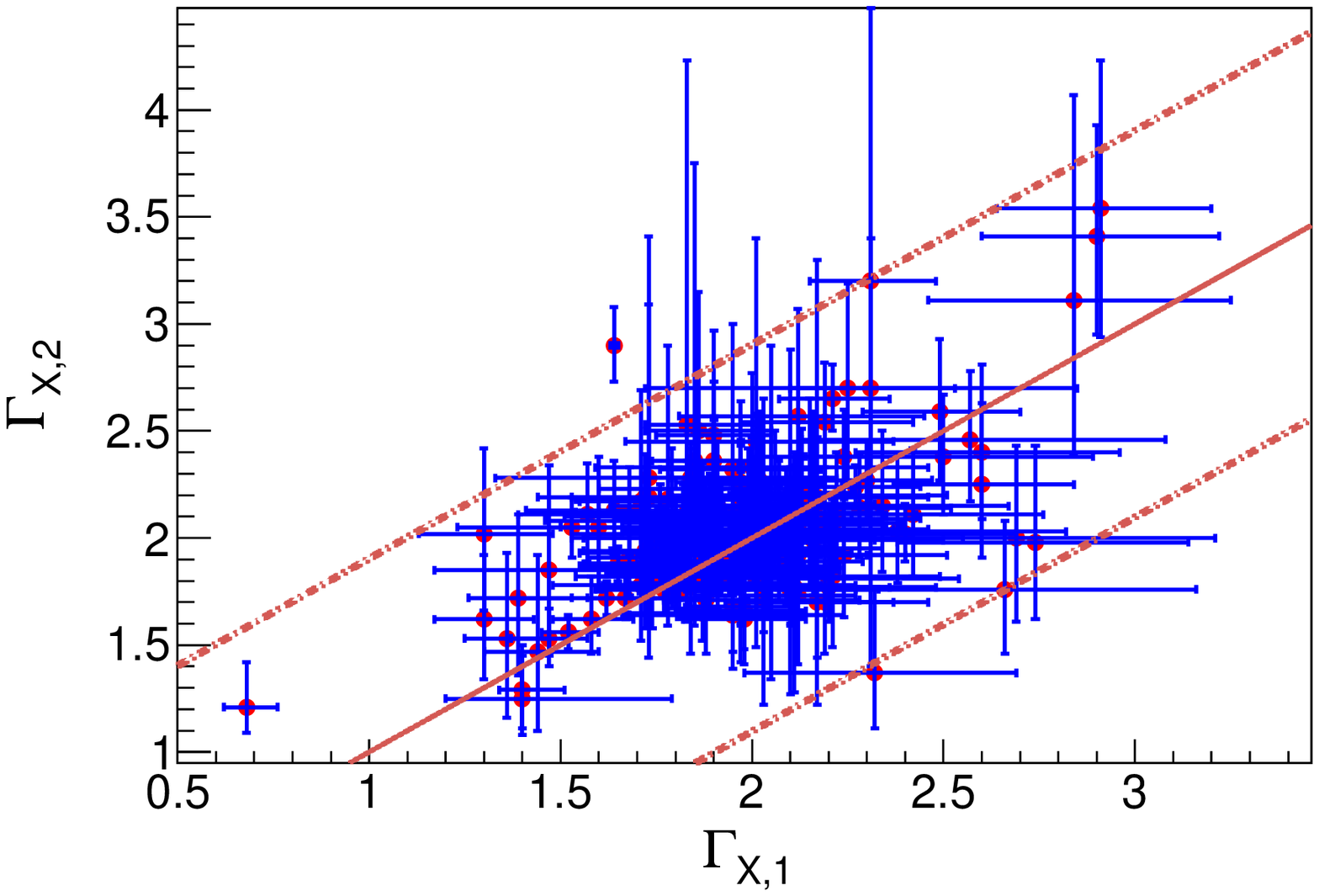}
\includegraphics[width=5.5cm,height=5cm]{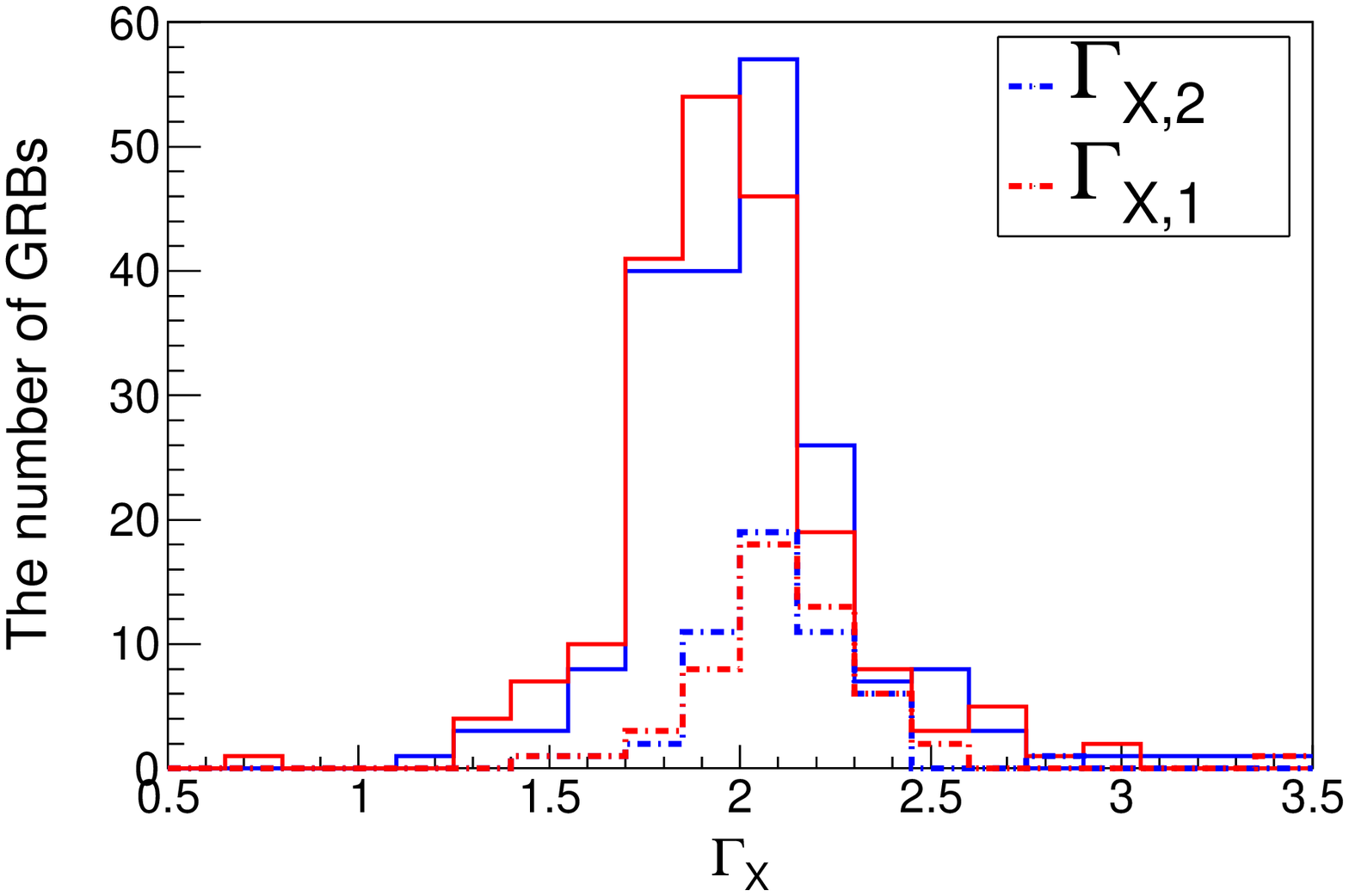}
\includegraphics[width=5.5cm,height=5cm]{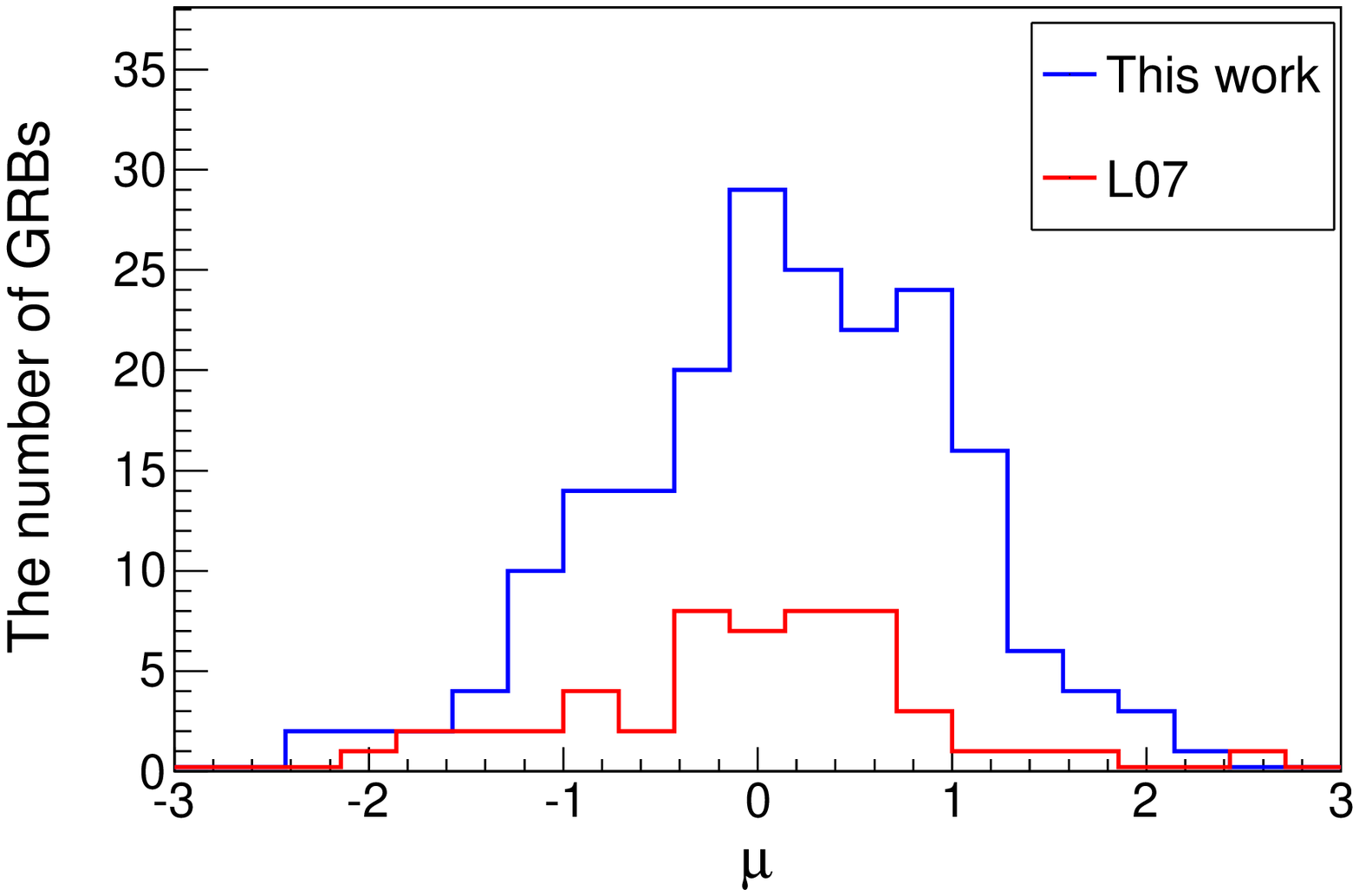}
\caption{Left panel: Comparison between $\Gamma_{X,1}$ and $\Gamma_{X,2}$. The  orange lines represent $\Gamma_{X,1}=\Gamma_{X,2}$ (solid)and its 2 $\sigma$ region(dotted). Middle panel:  The distribution of $\Gamma_{X,1}$ and $\Gamma_{X,2}$, with solid lines showing our results and dotted lines showing L07's results. Right panel: The distribution of $\mu$.}
\label{fig-4}
\end{center}
\end{figure*}

We first check whether there is  any spectral evolution between shallow decay segment and its follow-up segment. The left panel of  Fig.\ref{fig-4} plots $\Gamma_{X,2}$ as a function of $\Gamma_{X,1}$. We find that all GRBs in our sample fall in the $2\sigma$ linear correlation regions. The middle of Fig.\ref{fig-4} shows the comparison between the distributions of $\Gamma_{X,1}$ and $\Gamma_{X,2}$. For L07's sample, the Kolmogorov-Smirnov test suggests that these two distributions being consistent with the significance level of 0.96. With the larger sample, we find that the two distributions are still consistent, but the
significance level of this consistency decreases to 0.40. As suggested by L07, here we define a new parameter as
\begin{eqnarray}
\label{Eq8}
\mu=\frac{\Gamma_{X,2}-\Gamma_{X,1}}{\sqrt{ \delta\Gamma_{X,1}^{2}+\delta\Gamma_{X,2}^{2} }},
\end{eqnarray}
to verify this consistency within the observational uncertainty for individual bursts. The right panel of Fig.\ref{fig-4} shows $\mu$ distribution. We find that 84\% GRBs in our sample have $\mu\leq1$, and only 2 GRBs (GRB 080903 and GRB 150201A) show significant spectral evolution ($\mu>3$). Considering all these evidence, we confirm L07's conclusion that there is no significant spectral evolution between the shallow decay segment and its follow-up segment, indicating that the shallow decay phase is a refreshed forward shock\citep{rees98,dailu98,zhangmeszaros01,zhang06,nousek06}.

In this scenario, the shallow decay's follow-up segment should be consistent with the predictions of the forward shock models, namely the observed spectral index $\beta_{X,2}=\Gamma_{X,2}-1$ and temporal decay index $\alpha_2$ should follow the so called ``closure relations" \cite[][for a review]{gao13}. Although the closure correlations vary for different spectral regimes, different cooling schemes or different properties of the ambient medium, as illustrated in L07, three regimes are relevant here:
\begin{itemize}
\item regime 1: $\nu_x>{\rm max}(\nu_{m},\nu_{c})$ (for either ISM or wind medium), where $\alpha_2=(3\beta_2-1)/2$;
\item regime 2: $\nu_{m}<\nu_x<\nu_{c}$ (for ISM medium), where $\alpha_2=3\beta/2$;
\item regime 3: $\nu_{m}<\nu_x<\nu_{c}$ (for wind medium), where $\alpha_2=(3\beta_2+1)/2$.
\end{itemize}
We define a new parameter $\phi_{i}~(i=1,2,3)$ to test how well individual bursts being consistent with the closure relations under regime $(i=1,2,3)$:
\begin{eqnarray}
\label{eq1}
\phi_{i}=  \frac{{|\alpha}^{\rm{obs}}-\alpha_{i}(\beta^{\rm{obs}})|}{\sqrt { (\delta \alpha^{\rm{obs}})^{2} +[ \delta \alpha_{i}(\beta ^{\rm{obs}}) ]^{2} } }
\end{eqnarray}
where $\alpha^{obs}$ and $\alpha_{i}(\beta^{obs})$ are the temporal decay slope from the observations and from the closure relations, and $\delta \alpha^{obs}$ and $\delta(\alpha_{i}({ \beta^{obs}} ))$ represent their uncertainties. For an individual burst, if there existing $\phi_{i}<3$, we conclude that this burst belongs to regime i model within $3-\sigma$ significance. For those bursts that have more than one regimes satisfying $\phi_{i}<3$, we choose the regime with smallest $\phi$ value.

\begin{figure}[!htb]
\centering
\figurenum{5}
\includegraphics[width=7cm,height=6cm]{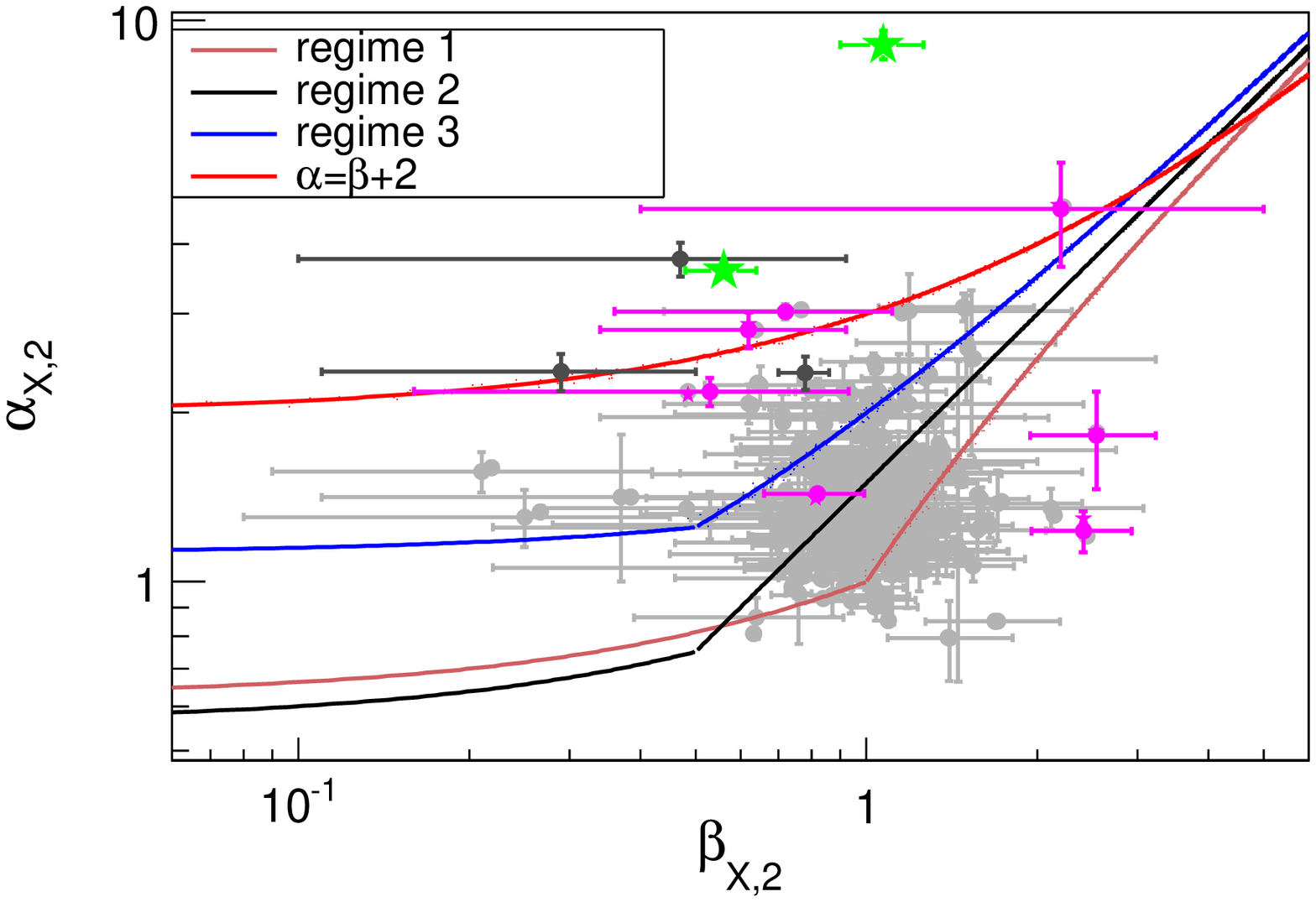}
\caption{Temporal decay index $\alpha_{X,2}$ as a function of the spectral index $\beta_{X,2}$ for shallow decay's follow-up segment, compared with the closure correlations of various spectral regimes. The gray dots show GRBs  belonging to regime 1-3($case^{1}$).
Pink dots show GRBs whose shallow decay segment and follow-up segment belongs to different spectrum regimes.($case^{2}$). Black dots show GRBs not belonging to regime 1-3 but cannot be excluded within $3\sigma$ significance by curvature effect regime($case^{3}$). Green dots show GRBs that could be excluded within $3\sigma$ significance by curvature effect regime($case^{4}$). }
\label{fig-5}
\end{figure}

Overall, we find that 66/198 bursts belong to regime 1, 82/198 bursts belong to regime 2, and 45/198 bursts belong to regime 3. For the other 5 bursts, we test their consistency with the curvature effect curve $\alpha=\beta+2$ \citep{Kumar00,Panaitescu06} with the $\phi$ parameter. We find that three bursts cannot be excluded within $3\sigma$ significance by curvature effect regime. These bursts could still be interpreted with external shock model once the jet break effect is invoked. We name regime 4 for these bursts. The other two bursts that could be excluded within $3\sigma$ significance by curvature effect regime are difficult to be interpreted under external shock framework. As suggested by many previous works, the early X-ray plateau for these cases should be of internal origin and is directly connected to a long-lasting central engine \citep{troja07,rowlinson10,rowlinson13,rowlinson14, dallosso11,lv14,lv15,gibson17,gibson18,gompertz13,gompertz14,gompertz15,rea15,gompertzfruvhter17,stratta18}. We did not assign these bursts into either regime.  Above all, we divided the GRBs of our sample  into four groups, $case^{1}$, $case^{2}$ ,$case^{3}$ and $ case ^{4}$. Which case of each GRB in our sample belongs to and which regime  is consistent with are listed in Table \ref{table-2}. The relationship of $\alpha_{X,2}$ and $\beta_{X,2}$ in difference cases are shown in Fig.\ref{fig-5}.

The normal decay phases for 196/198 bursts in our sample are consistent with the external-shock models, inferring that their shallow decay phase might also be of external origin and may be related to continuous energy injection from the central engine. Here we assume that the central engine has a power-law luminosity release history as $L(t)=L_{0}(\frac{t}{t_{0}})^{-q}$, so that the injected energy would be $E_{\rm{inj}}=\frac{L_{0}t_{0}^{q}}{1-q}t^{1-q}$(\cite{zhangmeszaros01}). When $E_{\rm{inj}}$ is larger than the impulsively injected energy during the prompt emission phase, the dynamics of the external shock wave would be related to $q$ as $\Gamma \propto t^{-\frac{q+2}{8}}$ ($\Gamma \propto t^{-\frac{q}{4}}$) and $R \propto t^{\frac{2-q}{4}}$ ($R \propto t^{\frac{2-q}{2}}$) for ISM (wind) case. Consequently, we have $\nu_{m}\propto t^{-1-q/2}$($t^{-1-q/2}$), $\nu_{c}\propto t^{-1+q/2}$($t^{1-q/2}$) and $F_{\nu,\rm{max}}\propto t^{1-q}$($t^{-q/2}$) for the ISM (wind) models for p$>$2, and $\nu_{m}\propto t^{-\frac{(q+2)(p+2)}{8(p-1)}}$($t^{\frac{4+pq}{4(1-q)}}$), $\nu_{c}\propto t^{-1+q/2}$($t^{1-q/2}$) and $F_{\nu,\rm{max}}\propto t^{1-q}$($t^{-q/2}$) for the ISM (wind) models for $1<p<2$, where p is the electronic power law index, and $F_{\nu,\rm{max}}$ is the observed peak flux. The relationship between decay slope $\alpha$ and energy injection index $q$ for different spectral regimes could thus be derived (see results in Tables 13-16 of \cite{gao13}).

\begin{figure*}
\begin{center}
\setlength{\abovecaptionskip}{0.cm}
\setlength{\belowcaptionskip}{-0.cm}
\hspace{0cm}
\graphicspath{{inject/}}
\figurenum{6}
\includegraphics[width=6.7cm,height=5cm]{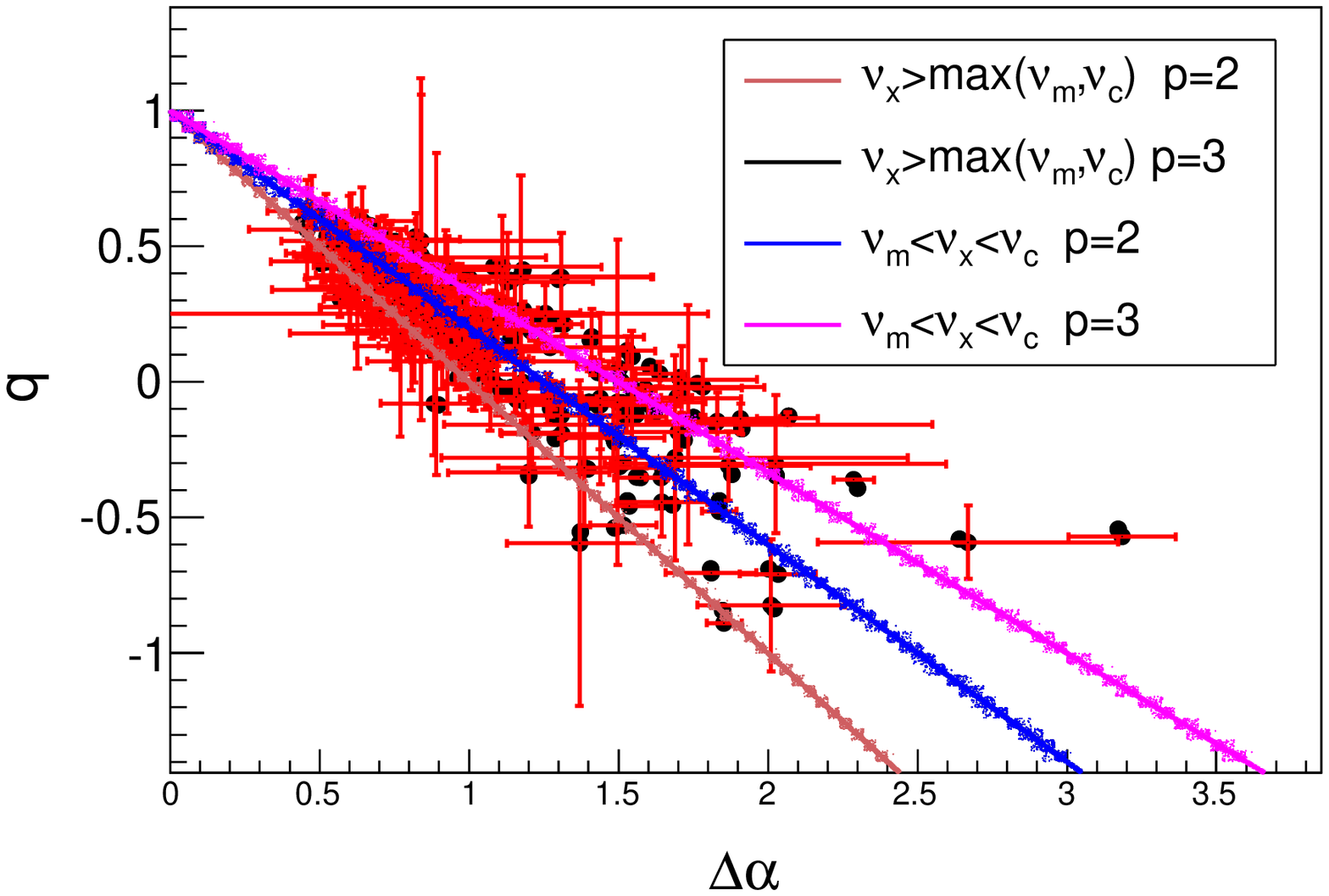}
\includegraphics[width=6.7cm,height=5cm]{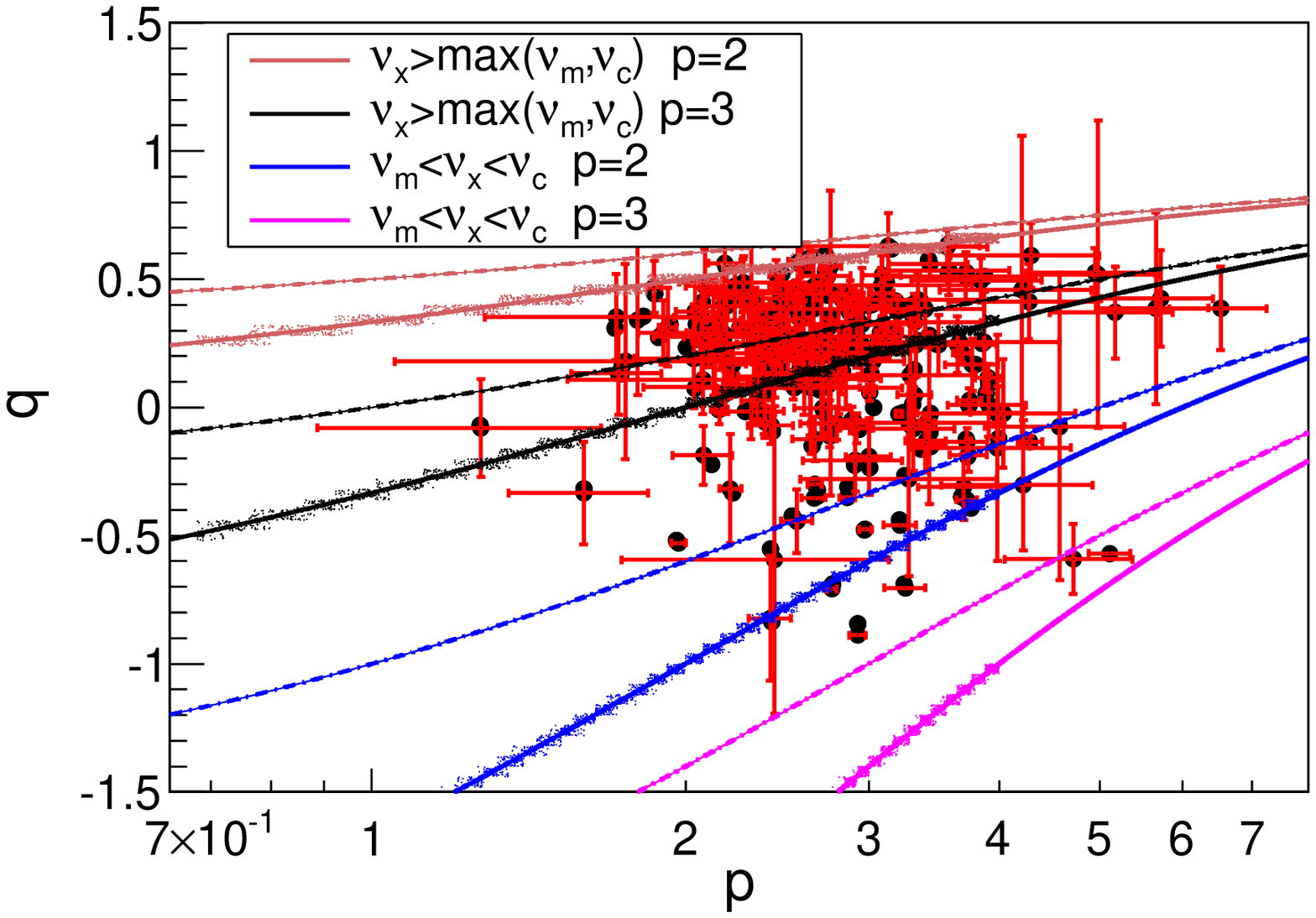}
\includegraphics[width=6.7cm,height=5cm]{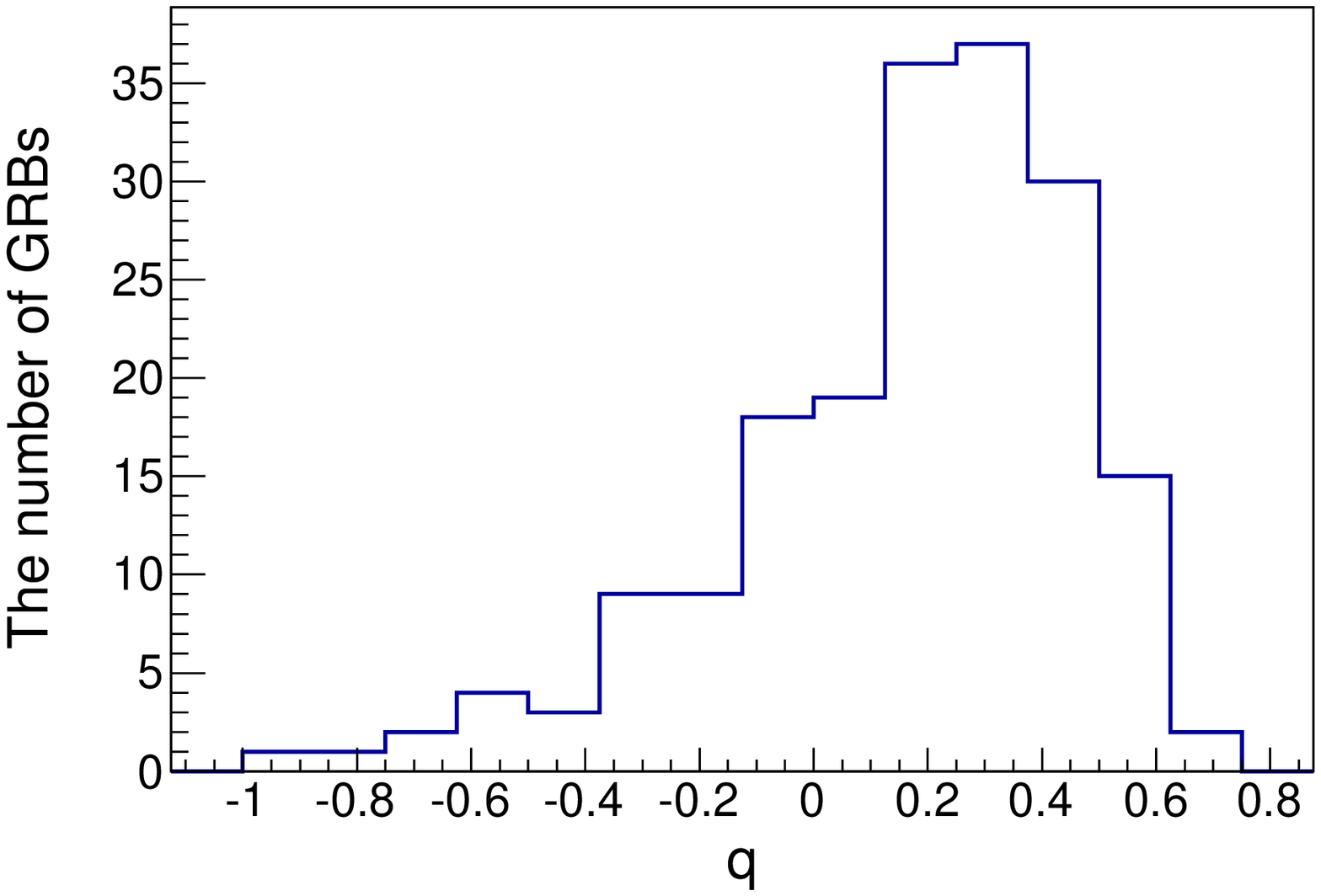}
\includegraphics[width=6.7cm,height=5cm]{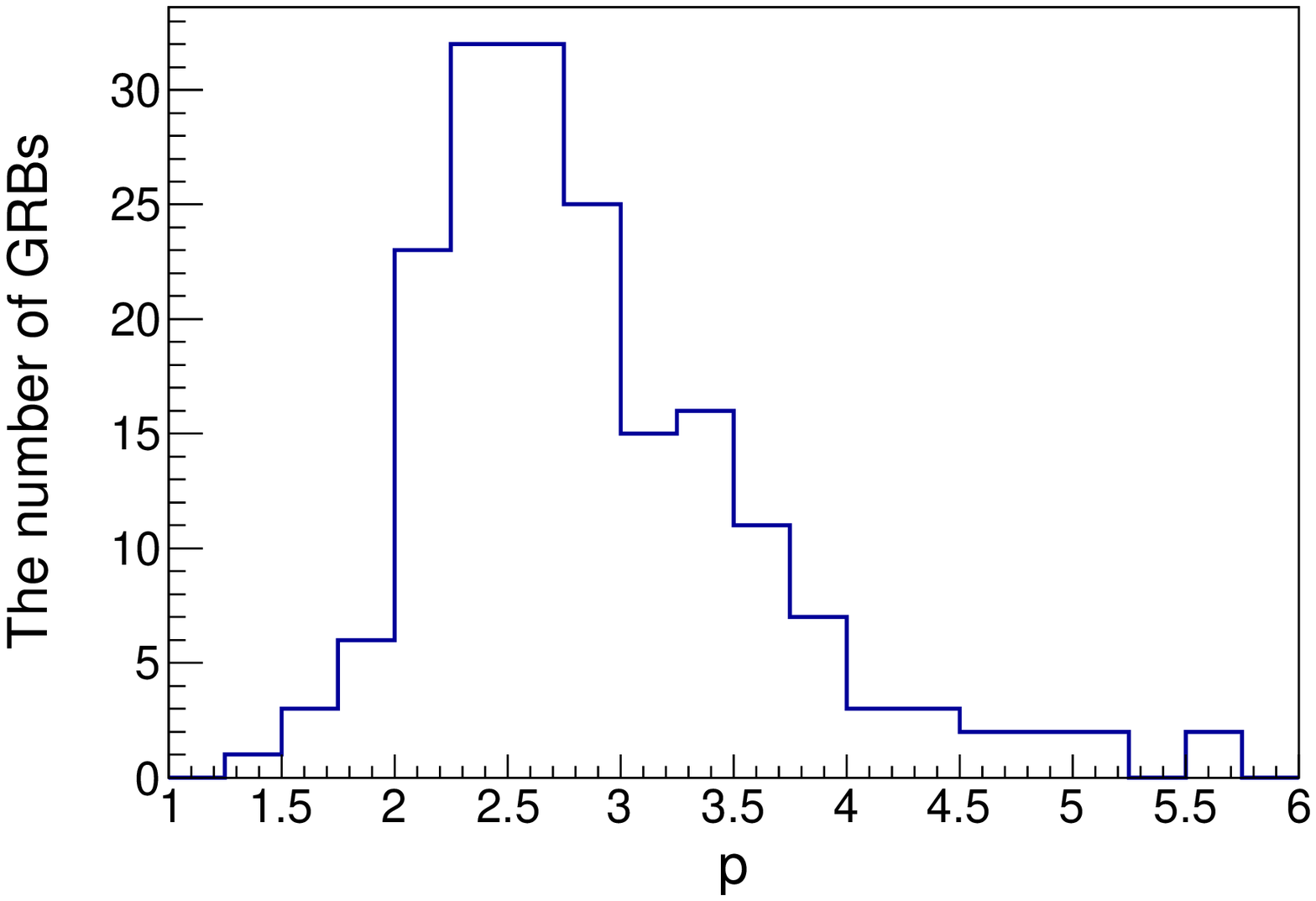}
\caption{Upper panels: distributions of bursts in regimes 1-3 in the two-dimensional q-$\Delta \alpha$ and q-p planes, along with the model prediction for $ \upsilon>max(\upsilon_{m},\upsilon_{c})$ (thick  solid lines)  and $\upsilon_{m} <\upsilon< \upsilon_{c}$ (thin  solid lines).  Lower panels:  distributions of p values and q values.}
\label{fig-6}
\end{center}
\end{figure*}

In this case, the decay indices difference between shallow decay segment and its follow-up segment in different regimes would be \citep{gao13}.
\begin{eqnarray}
\label{Eq9}
\Delta\alpha=\frac{1}{16}(p+14)(1-q), {\rm regime \  1}, 1<p<2 \nonumber \\
\Delta\alpha= \frac{1}{16}(p+18)(1-q), {\rm regime  \ 2}, 1<p<2 \nonumber \\
\Delta\alpha= \frac{1}{8}(p+4)(1-q), {\rm regime  \ 3}, 1<p<2 \nonumber \\
\Delta\alpha= \frac{1}{4}(p+2)(1-q) , {\rm regime \  1}, p>2 \nonumber \\
\Delta\alpha= \frac{1}{4}(p+3)(1-q) , {\rm regime  \ 2}, p>2 \nonumber \\
\Delta\alpha= \frac{1}{4}(p+1)(1-q) , {\rm regime  \ 3}, p>2
\end{eqnarray}
where p could be derived from the observed spectral index, depending on the observed spectral regime. Given the regime models for each burst (here we only test bursts in regimes 1-3), its energy injection parameter q could be thus derived based on $\alpha_{X,1}$ and $\alpha_{X,2}$. The upper panels of  Fig.\ref{fig-6} shows the distributions of these GRBs in the two-dimensional q-$\Delta \alpha$ and q-p planes along with the contours of constant p and $\Delta \alpha$. The distribution of derived p and q values are plotted in the lower panels of  Fig.\ref{fig-6}. We find that q values are mainly distributed between -0.5 and 0.5, with an average value of 0.16$\pm$ 0.12, which is well consistent with the model prediction (q-value around zero) where a spinning-down pulsar as the energy injection central engine \citep{dailu98,zhangmeszaros01}. With the derived q value, we can further test whether the $\alpha_{X,1}$ and $\beta_{X,1}$ satisfy the corresponding closure relations with energy injection (see Table 14 and 16 in \cite{gao13}) with $\phi$ parameter. It turns out the shallow decay phases for all 194 GRBs satisfy the closure relations with energy injection, and their corresponding spectral regime for 189/196 (96.4\%) bursts are consistent with their follow-up phases, which is expected since there are no systematic spectral evolution between these two segments\footnote{There are 7 GRBs (050401,  060413, 071118, 100508A, 100902A, 110411A and 161202A) whose shallow decay phase and normal decay phase belong to different spectral regimes. For GRBs 050401, 071118, 110411A, and 161202A, too few data points around $t_b$ might cause bigger uncertainty for $\alpha_{X,2}$ and/or $\beta_{X,2}$.  GRBs 060413, 100508A and 100902A    marginally belongs to regime 3 but with a large decay slope for the normal decay phase. }. The results reinforce the conclusion that the shallow decay segment in most bursts is consistent with an external forward shock origin, probably due to a continuous energy injection from a long-lived central engine.

\subsection{Empirical relations revisited} \label{sec:style}

\subsubsection{Empirical relation among $E_{\MakeLowercase{\rm{iso}}},  E'_{\MakeLowercase{p}}$  and $ \MakeLowercase{t'}_{\MakeLowercase{b}}$ } \label{sec:style}

 \begin{figure*}
\begin{center}
\setlength{\abovecaptionskip}{0.cm}
\setlength{\belowcaptionskip}{-0.cm}
\hspace{0cm}
\figurenum{7}
\graphicspath{{liangzhang/}}
\includegraphics[width=6.7cm,height=6.7cm]{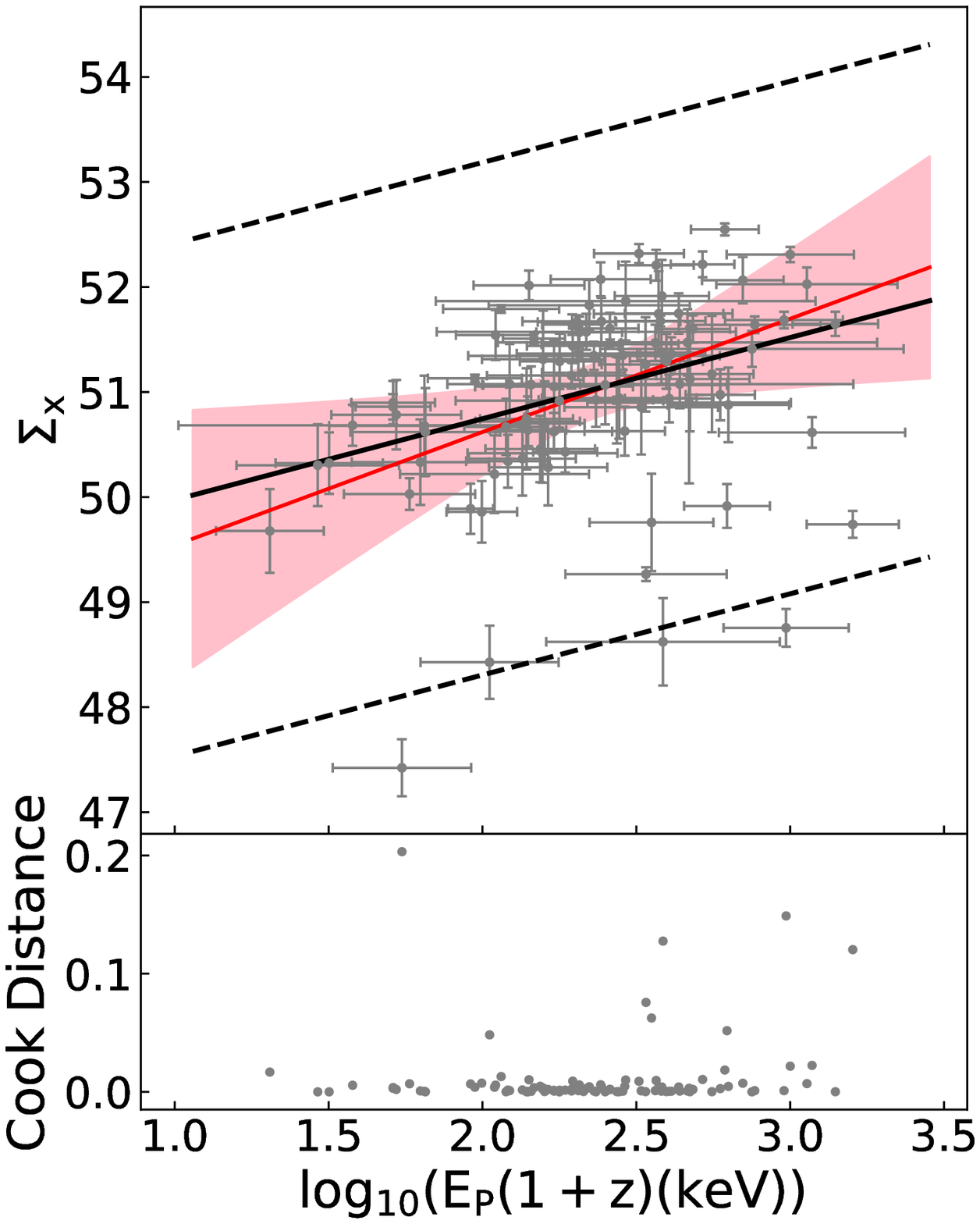}
\includegraphics[width=6.7cm,height=6.7cm]{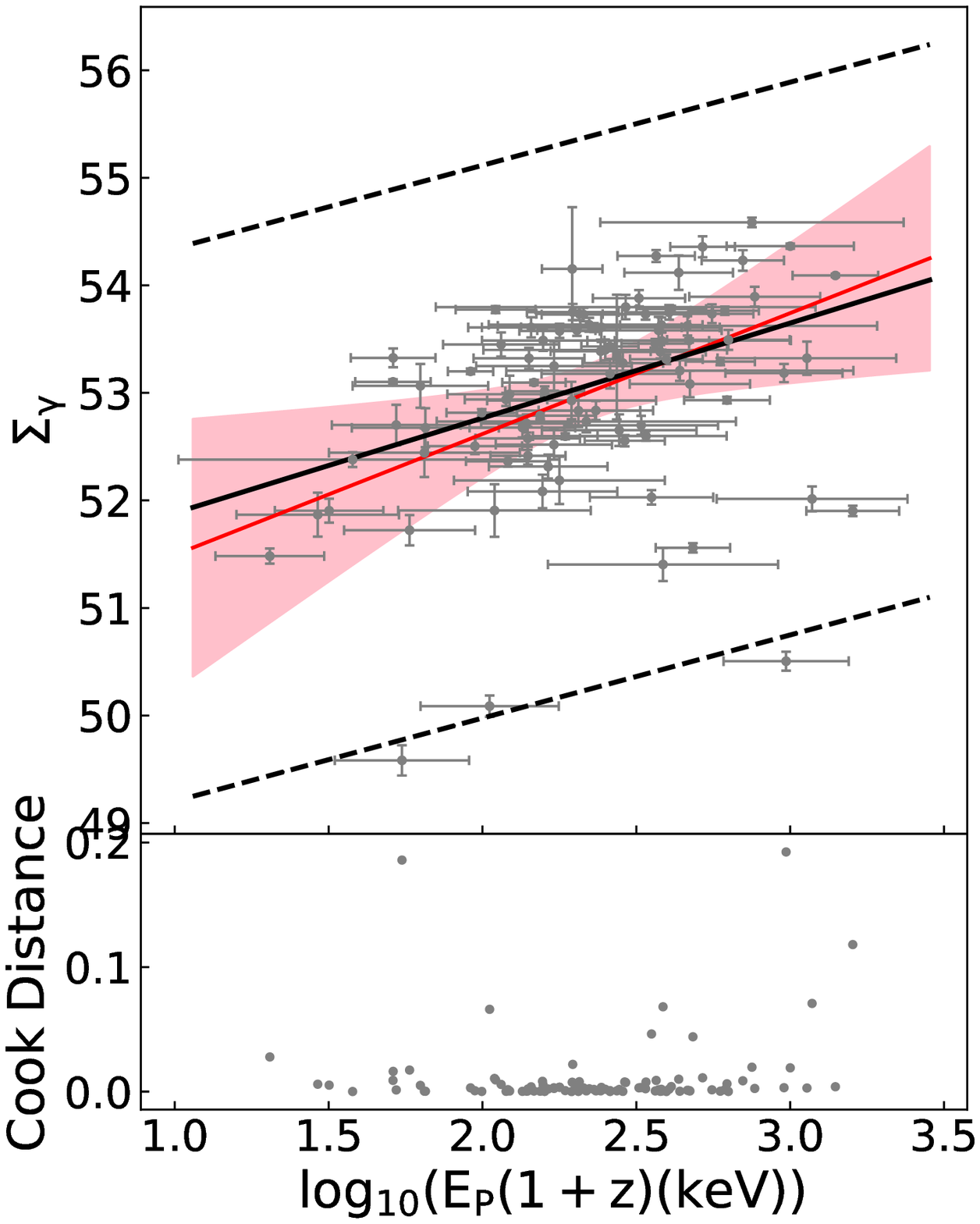}

\caption{Left panel: the correlation between $\Sigma_{X}=\rm{log}_{10}(E_{\rm{iso},X})-\kappa_{1} \rm{log}_{10}(t'_{b})$ and $E'_{p}$. Right panel: the correlation between $\Sigma_{\gamma}= \rm{log}_{10}(E_{\rm{iso},\gamma}^{b})-\kappa_{2}\rm{log}_{10}(t'_{b})$ and $E'_{p}$. The black lines represent the best fitting results (solid) and its 3$\sigma$ uncertainty region (dotted) with least square regression method. The red line presents the best fitting results with the bivariate linear regression method and the pink shadowed region show the intrinsic scatter to the population $3\sigma_{\rm int}$.}
\label{fig-7}
\end{center}
\end{figure*}

Based on their sample, L07 has investigated the relations among the energies $E_{\rm{iso},X}$ and $E_{\rm{iso},\gamma}$ and other parameters $E'_{p}$ and $t'_{b}$, with a regression model as
\begin{eqnarray}
\label{Eq12}
\rm{log}_{10}(E_{\rm{iso},(X,\gamma)})= \kappa_{0}+\kappa_{1}\rm{log}_{10}(E'_{p})+\kappa_{2}\rm{log}_{10}(t'_{b})
\end{eqnarray}
where $\kappa_{0}$, $\kappa_{1}$ and $\kappa_{2}$ are multiple regression coefficients. The probability from a t-test ($p_{t}$) was invoked to measure the significance of the dependence of each variable on the model and the probability from a F-test ($p_{F}$) was invoked to justify the global regression of the three parameter relation. For $E_{\rm{iso},X}-E'_{p}-t'_{b}$ relation, L07 yields $\kappa_{0}= 44.0\pm1.1$ ($p_{t}<10^{-4}$) and $\kappa_{1} =1.82\pm0.33~(p_{t}<10^{-4})$, $\kappa_{2} =0.61\pm0.18~ (p_{t}<3\times10^{-3})$, and the $p_{F}$ is less than $10^{-4}$. Their results suggest a strong correlation between $E_{\rm{iso},X}$ and $E'_{p}$ but a tentative correlation between $E_{\rm{iso},X}$ and $t'_{b}$, and thus a tentative global 3-parameter correlation. We apply the regression model to our sample, we yield $\kappa_{0}= 49.2\pm0.77$ ($p_{t}<10^{-4}$), $\kappa_{1} =0.77\pm0.22~(p_{t}=0.001)$ and $\kappa_{2} =-0.051\pm0.12~(p_{t}=0.683)$, with $p_{F}=0.002$, $R^2=12.5\%$ and $p_{\rm AD}=0.555$. We find that with a larger sample, a strong correlation between $E_{\rm{iso},X}$ and $E'_{p}$ still exists but the tentative correlation between $E_{\rm{iso},X}$ and $t'_{b}$ is completely disappeared, so that the global 3-parameter correlation becomes less significant. Note that already in \cite{dainotti11b} it was clear that $E_{\rm{iso},X}-t'_{b}$ was a weak correlation, while $L_{\rm{peak}}-t'_{b}$ was a stronger correlation. \cite{dainotti16} thus proposed the $L_{X}-L_{\rm peak}-t'_{b}$ correlation given the already established two-parameter $L_{\rm peak}-t'_{b}$ \cite[e.g.][]{dainotti08} and $L_{\rm{peak}}-L_{X}$ correlations \citep{dainotti11b,dainotti15}. With an extended sample (updated to July 2016) of long GRBs with X-ray plateau, \cite{dainotti17} yielded ${\rm log}_{10}L_X=(17.67\pm5.7)-(0.83\pm0.10){\rm log}_{10}t'_{b}+(0.64\pm0.11){\rm log}_{10}L_{\rm peak}$, with the Pearson correlation coefficient $r=0.90$ with a probability of the same sample occurring by chance, $P=1.75\times10^{-17}$.

For $E_{\rm{iso},\gamma}-E'_{p}-t'_{b}$ relation, L07 yields $\kappa_{0}= 48.3\pm0.8$ ($p_{t}<10^{-4}$) and $\kappa_{1} =1.7\pm0.25~(p_{t}<10^{-4})$, $\kappa_{2} =0.07\pm0.13~(p_{t}=0.486)$, and the $p_{F}$ is less than $10^{-4}$. Their results suggest a strong correlation between $E_{\rm{iso},\gamma}$ and $E'_{p}$ but no correlation between $E_{\rm{iso},\gamma}$ and $t'_{b}$. A tentative global 3-parameter correlation might exist.  With a larger sample, we yield $\kappa_{0}= 51\pm0.76$ ($p_{t}<10^{-4}$), $\kappa_{1} =0.883\pm0.22~(p_{t}<10^{-4})$, and $\kappa_{2} =-0.19\pm0.12~(p_{t}=0.11)$, with $p_{F}<10^{-4}$, $R^2=19.5\%$ and $p_{\rm AD}=0.403$. We confirm that a strong correlation between $E_{\rm{iso},\gamma}$ and $E'_{p}$ and a tentative global 3-parameter correlation exist and there is no correlation between $E_{\rm{iso},\gamma}$ and $t'_{b}$. The data and regression modeling results are shown in Fig.\ref{fig-7}.It is worth noticing that \cite{dainotti11b} also investigated the $E^{b}_{\rm{iso},\gamma}$ -$t'_{b}$ relation with 62 long GRBs with known redshift, and the correlation coefficient and the random occurrence probability p for their sample are $r=-0.19$ and $p=0.1$, also inferring that the correlation between $E_{\rm{iso},\gamma}$ and $t'_{b}$ was very weak.

\begin{figure*}
\begin{center}
\setlength{\abovecaptionskip}{0.cm}
\setlength{\belowcaptionskip}{-0.cm}
\hspace{0cm}
\figurenum{8}
\graphicspath{{zhang/}}
\includegraphics[width=5.5cm,height=6.7cm]{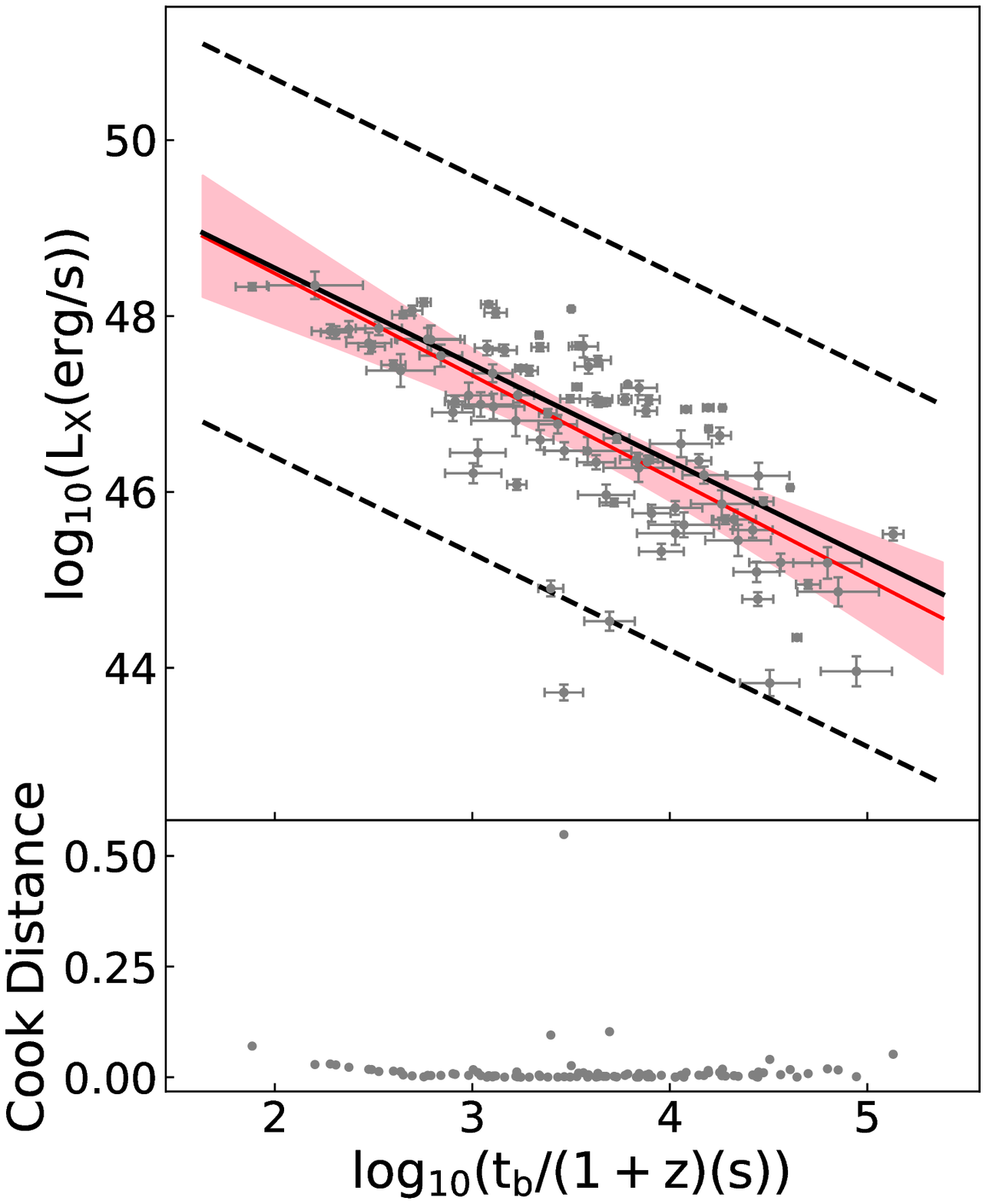}
\includegraphics[width=5.5cm,height=6.7cm]{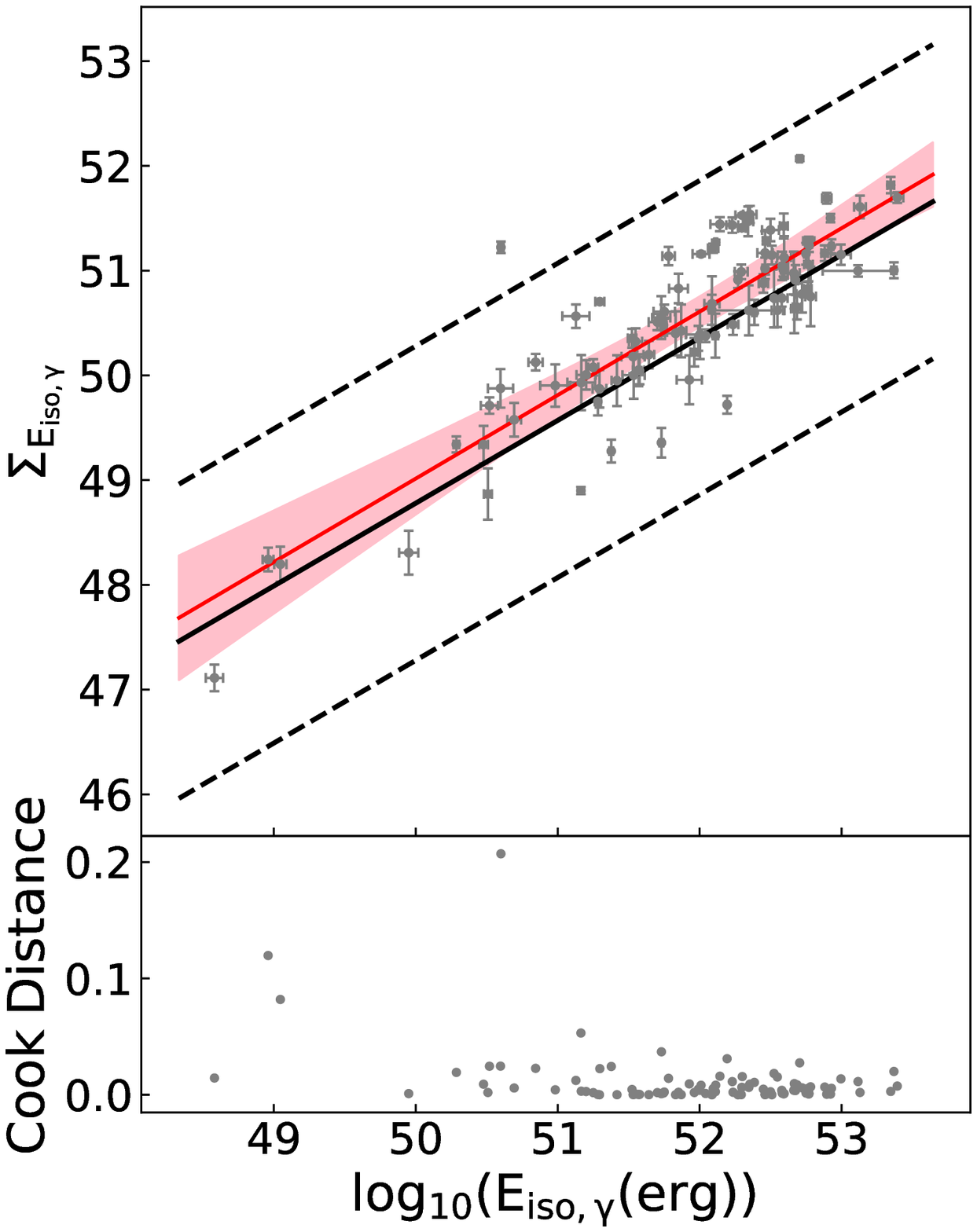}
\includegraphics[width=5.5cm,height=6.7cm]{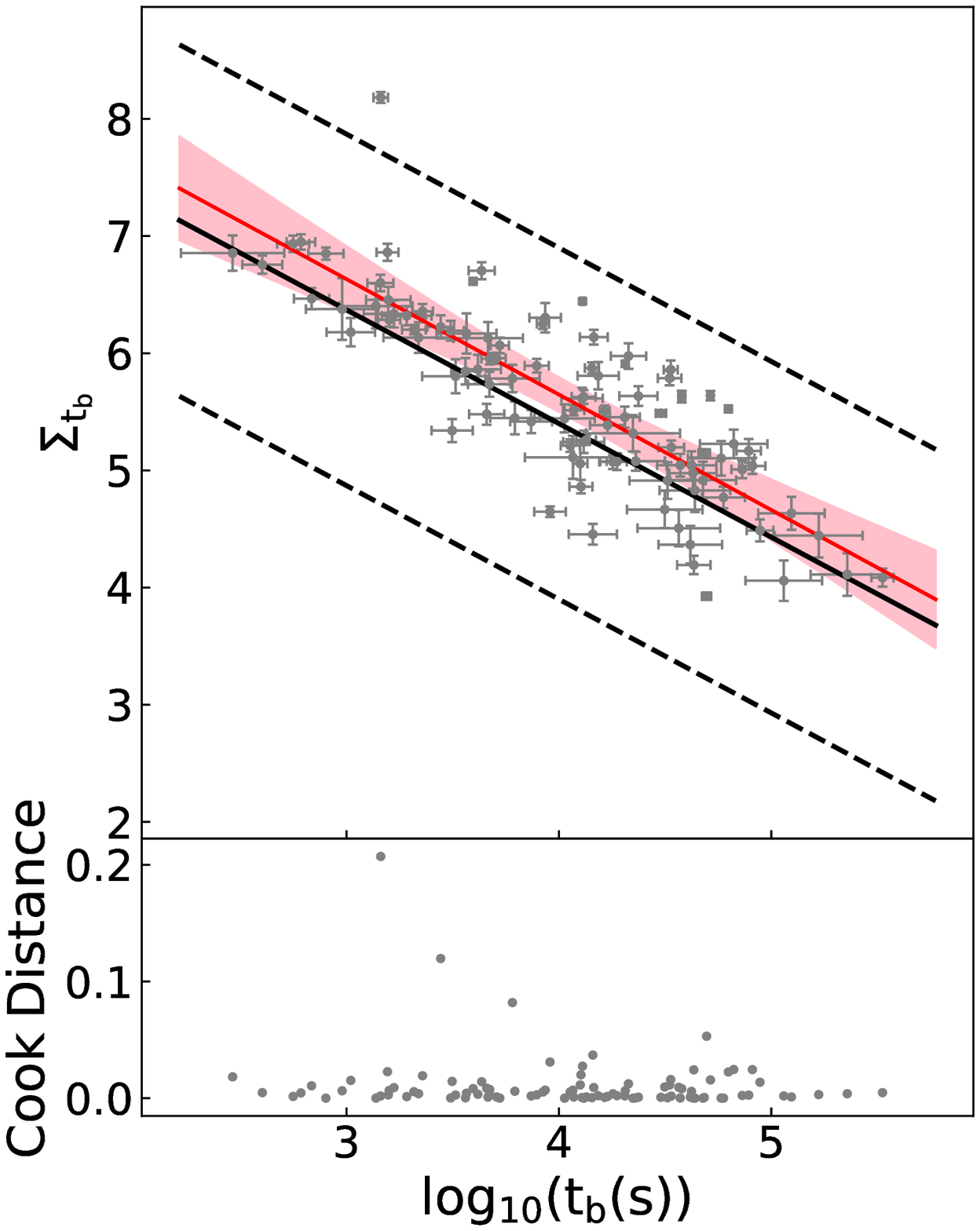}
\caption{Left panel:  the correlation between $L_{X}$ and $t'_{b}$. Middle panel: the correlation between $\Sigma_{E_{ \rm{iso},\gamma}}=\rm{log}_{10}(L_{X})-\kappa_{1} \rm{log}_{10}(t_{b})$ and $E_{\rm{iso},\gamma}$. Right panel:  the correlation between $\Sigma_{t_{b}}= \rm{log}_{10}(L_{X})-\kappa_{2}\rm{log}_{10}(E_{iso,\gamma})$ and $t_{b}$. The black lines represent the best fitting results (solid) and its 3$\sigma$ uncertainty region (dotted) with least square regression method. The red line presents the best fitting results with the bivariate linear regression method and the pink shadowed region show the intrinsic scatter to the population $3\sigma_{\rm int}$. }
\label{fig-8}
\end{center}
\end{figure*}

\subsubsection{Revisit correlations between $t'_{b}$ ($t_{b}$), $L_{X}$ and $E_{\rm{iso}\gamma}$} \label{sec:style}

\cite{dainotti08} discovered a formal anti-correlation between X-ray luminosity at the end of plateau $L_{X}$ and rest frame plateau end time $t'_{b}$,  described as $\rm{log}_{10}(L_{X})=\kappa_{0}+\kappa_{1}\rm{log}_{10}(t'_{b})$.  \cite{dainotti10} fitted the correlation between $L_{X} $ and $t'_{b}$ and achieved the results $\kappa_{0}=51.06\pm1.02$, $\kappa_{1}=-1.06^{+0.27}_{-0.28}$.
It is of interest to justify this empirical relationship with our updated sample. With multivariate linear regression method, we find that the anti-correlation indeed exist between $L_{X}$ and $t'_{b}$, with $\kappa_{0}= 50.7\pm0.39$ ($p_{t}<10^{-4}$) and $\kappa_{1} =-1.15\pm 0.1 (p_{t}<10^{-4})$, where $p_{F}<10^{-4}$, $R^2=54.9\%$ and $p_{\rm AD}=0.262$.  In addition, the $L_{X}$ for individual GRB in our sample are listed in Table \ref{table-3}. Basically speaking, our results in general agree well with \cite{dainotti13b}, indicating that an intrinsic relationship between $L_{X}$ and $(t'_{b})$ indeed exist. This correlation has been tested against selection bias robustly \citep{dainotti13b}. It is worth noticing that $L_{X} $ is roughly inversely proportional to the timescale of the energy injection, inferring that the energy reservoir should be roughly a constant. This is consistent with the energy injection model, where the central engine is a newborn magnetars \citep{dai04}.

\cite{xuhuang12} introduced a third parameter parameter $E_{\rm{iso},\gamma}$ to the above relation (and replacing $t'_{b}$ by $t_{b}$) and claimed that the three-parameter correlation would be tighter. The relation could be expressed as $L_{X} =\kappa_{0} +\kappa_{1}\rm{log}_{10}(t_{b})+\kappa_{2}\rm{log}_{10}(E_{\rm{iso},\gamma})$, with $\kappa_{0}=4.14$, $ \kappa_{1} =-0.87$ and $\kappa_{2}=0.88$. We have also justified this empirical relationship with our updated sample. With multiple regression analysis methods, we find that the three-parameter correlation indeed exist, with $\kappa_{0}=9.26\pm2.59~(p_{t}<10^{-4})$ , $\kappa_{1}=-0.97\pm0.07~(p_{t}<10^{-4})$ and $\kappa_{2}=0.79\pm0.05$ ($p_{t}<10^{-4}$), where $p_{F}<10^{-4}$, $R^2=73.3\%$ and $p_{\rm AD}=0.372$. The data and fitting results regarding this relationship are shown in Fig.\ref{fig-8}. It is interesting to note that besides the \cite{xuhuang12} correlation, more recently \cite{dainotti16, dainotti17} showed that the $L_{X}-L_{\rm{peak}}-t'_{b}$ correlation is 37\% tighter than the \cite{xuhuang12} correlation, making the $L_{X}-L_{\rm{peak}}-t'_{b}$ correlation the tighest correlation in the literature so far involving the plateau emission.

\section{conclusions and discussion}

With Swift/XRT light curves of GRBs between February 2004 and July 2017, we revisit the analysis of shallow decay component from Liang et al. (2007). Our results and the comparison with L07's results are summarized as follows:

\begin{itemize}
\item We find that with a larger sample, the distributions of the characteristic properties of the shallow decay phase (e.g. $t_{b}$ , $S_{X}$,  $\Gamma_{X,1}$, and $\alpha_{X,1}$) still accords with normal or lognormal distribution with  $\rm{log}_{10}(t_{b}) =4\pm0.72$, $\rm{log}_{10}(S_{X}(\rm{ergs \ cm^{-2})})= -6.97\pm0.56$, $\Gamma_{X,1} =1.96\pm0.26$ and $\alpha_{X,1}=0.43\pm0.22$.  Comparing with L07's results, the distributions for all the parameters become slightly broader but the peaks of each distribution barely change.

\item We find that with a larger sample, $\Gamma_{X,1}$ and $\Gamma_{\gamma}$ still show no correlation, with $\Gamma_{X,1}$ systemically larger than $\Gamma_{\gamma}$. The tentative correlations of durations, energy fluences, and isotropic energies between the gamma-ray and X-ray phases still exist, but all correlations become significantly weaker than L07's results.

\item We find that for most GRBs, there is no significant spectral evolution between the shallow decay segment and its follow-up segment, and the follow-up segment for most bursts in our sample are consistent with the external-shock models. These two findings are consistent with L07's results, which infer that the shallow decay phase for most GRBs should be of external origin and may be related to continuous energy injection from the central engine.

\item Assuming that the central engine has a power-law luminosity release history as $L(t)=L_{0}(\frac{t}{t_{0}})^{-q}$, we find that q values are mainly distributed between -0.5 and 0.5, with an average value of 0.16$\pm$ 0.12, which is consistent with L07's results and is well consistent with the model prediction (q-value around zero) where a spinning-down pulsar as the energy injection central engine \citep{dailu98,zhangmeszaros01}. With the derived q value, we find that the shallow decay phases for 196/198 GRBs satisfy the closure relations with energy injection, and their corresponding spectral regime for most bursts (96.4\%) is consistent with their follow-up phases.

\item We find that with a larger sample, L07 suggested correlation between $E_{\rm{iso},X}$ and $E'_{p}$ still exists but the tentative correlation between $E_{\rm{iso},X}$ and $t'_{b}$ is completely disappeared, so that the global 3-parameter correlation becomes less significant. On the other hand, we confirm that a strong correlation between $E_{\rm{iso},\gamma}$ and $E'_{p}$ and a tentative global 3-parameter correlation ($E_{\rm{iso},\gamma}-E'_{p}-t'_{b}$) exist but there is no correlation between $E_{\rm{iso},\gamma}$ and $t'_{b}$.

\item We find that the anti-correlation indeed exist between $L_{X}$ and $(t'_{b})$ (as suggested by \cite{dainotti08} ), with $\kappa_{0}= 50.7\pm0.39$ ($p_{t}<10^{-4}$) and $\kappa_{1} =-1.15\pm0.1~(p_{t}<10^{-4})$, where the $p_{F}$ is less than $10^{-4}$.  \cite{xuhuang12} introduced three-parameter correlation by involving $E_{\rm{iso},\gamma}$ into the above relation (and replacing $t'_{b}$ by $t_{b}$) and we find that with our updated sample, the three-parameter correlation indeed exist, with $\kappa_{0}=9.26\pm2.59~(p_{t}<10^{-4})$ , $\kappa_{1}=-0.97\pm0.07~(p_{t}<10^{-4})$ and $\kappa_{2}=0.79\pm0.05$ ($p_{t}<10^{-4}$), where the $p_{F}$ is $p_{F} <10^{-4}$. More recently \cite{dainotti16, dainotti17} showed that the $L_{X}-L_{\rm{peak}}-t'_{b}$  correlation is 37\% tighter than the \cite{xuhuang12} correlation, making the $L_{X}-L_{\rm{peak}}-t'_{b}$  correlation the tighest correlation in the literature involving plateau emission. In addition, we note that this correlation is the results of two correlation which have been tested for selection bias.
\end{itemize}

In conclusion, with an updated sample, our results are consistent with most of L07's results and confirm their suggestion that the shallow decay segment in most bursts is consistent with an external forward shock origin, probably due to a continuous energy injection from a long-lived central engine.

\acknowledgments
This work was supported by the National Natural Science Foundation of China (under Grant No. 11722324, 11603003, 11633001, 11690024, 11603006), and the Strategic Priority Research Program of the Chinese Academy of Sciences (Grant No. XDB23040100). L. H. J. acknowledges support by the GuangXi Science Foundation (grant Nos. 2017GXNSFFA198008 and 2016GXNSFCB380005),  the One-Hundred-Talents Program of GuangXi colleges, and Bagui Young Scholars Program of GuangXi. BBZ acknowledges support from National Thousand Young Talents program of China and National Key Research and Development Program of China (2018YFA0404204) and The National Natural Science Foundation of China (Grant No. 11833003).

\software{XSPEC(\cite{arnaud96}), HEAsoft(v6.12;\cite{heasarc14}), root(v5.34;\cite{BrunRademakers97}), Linmix\_err(\cite{kelly07})}

\begin{figure*}
\begin{center}
\setlength{\abovecaptionskip}{0.cm}
\setlength{\belowcaptionskip}{-0.cm}
\hspace{0cm}
\graphicspath{{lightcurve/}}
\figurenum{1}
\includegraphics[width=5.5cm,height=5cm]{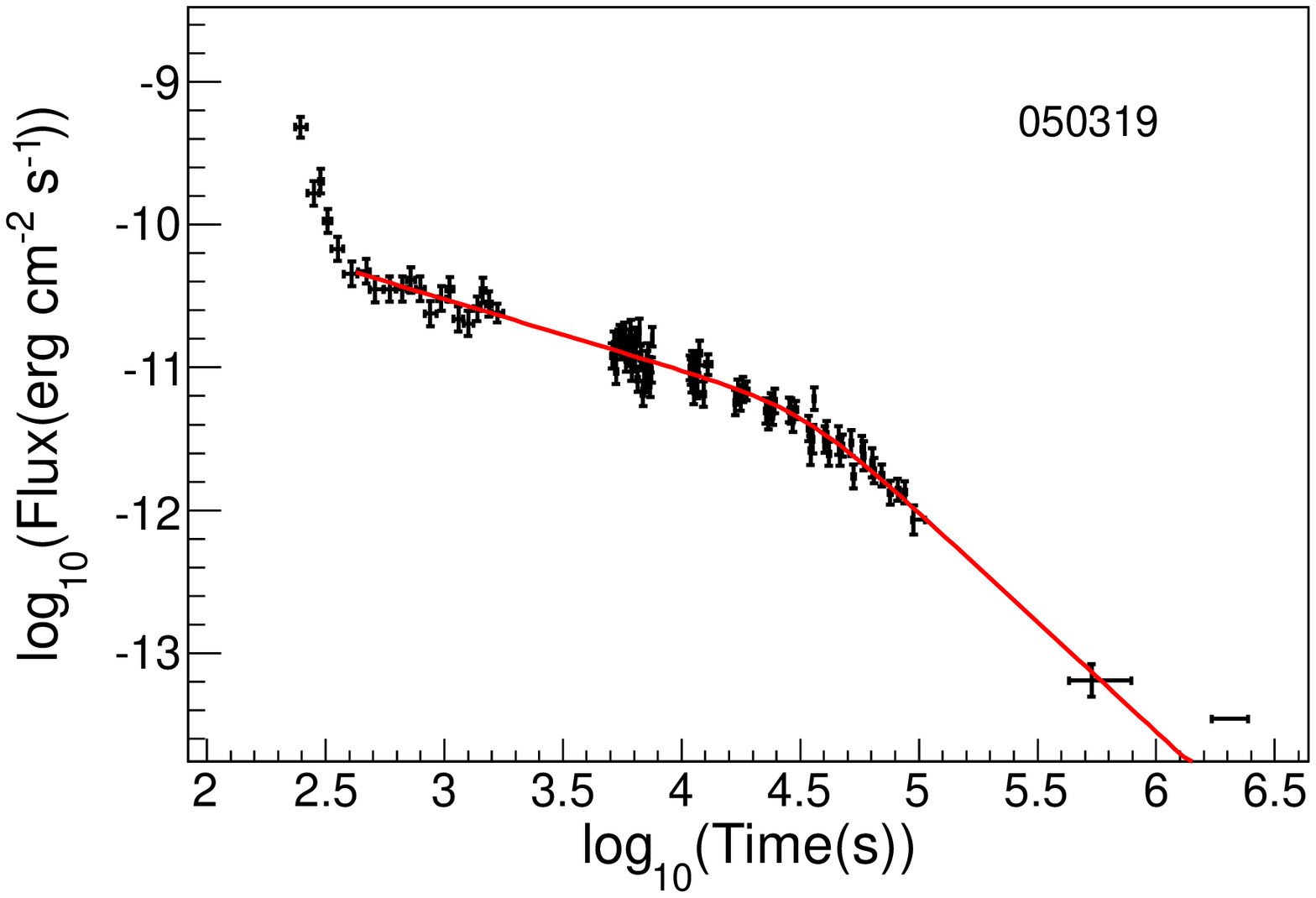}
\includegraphics[width=5.5cm,height=5cm]{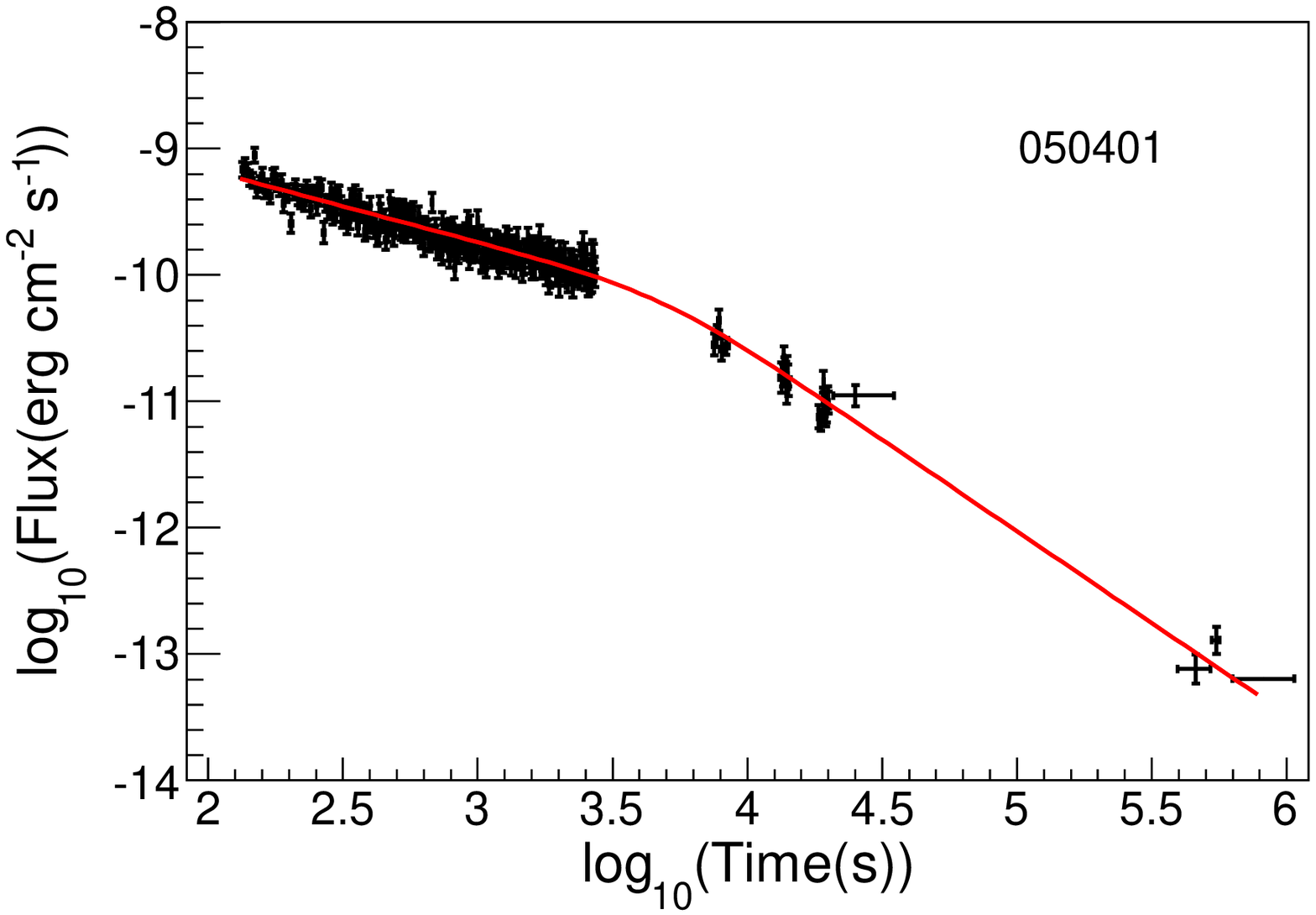}
\includegraphics[width=5.5cm,height=5cm]{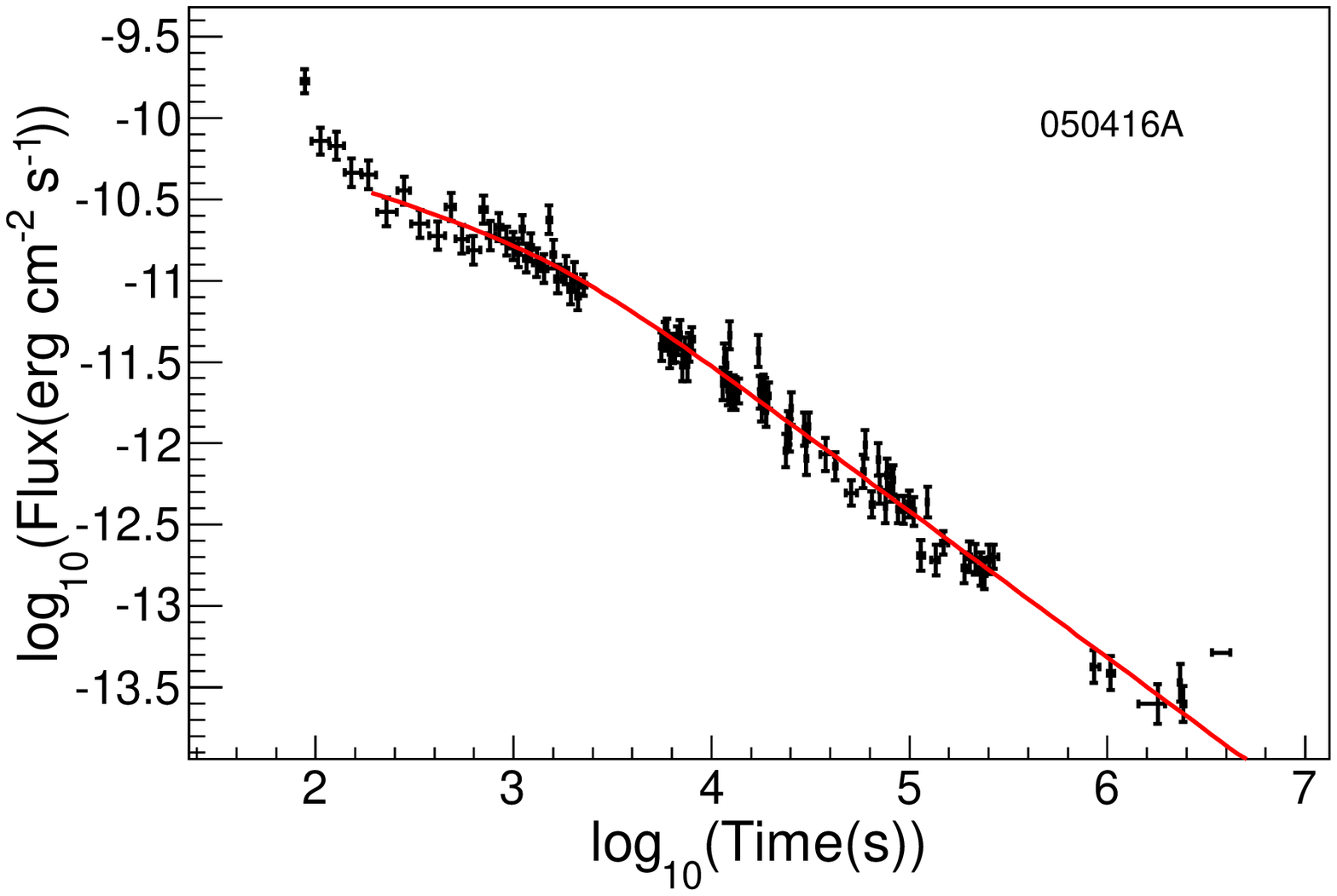}
\includegraphics[width=5.5cm,height=5cm]{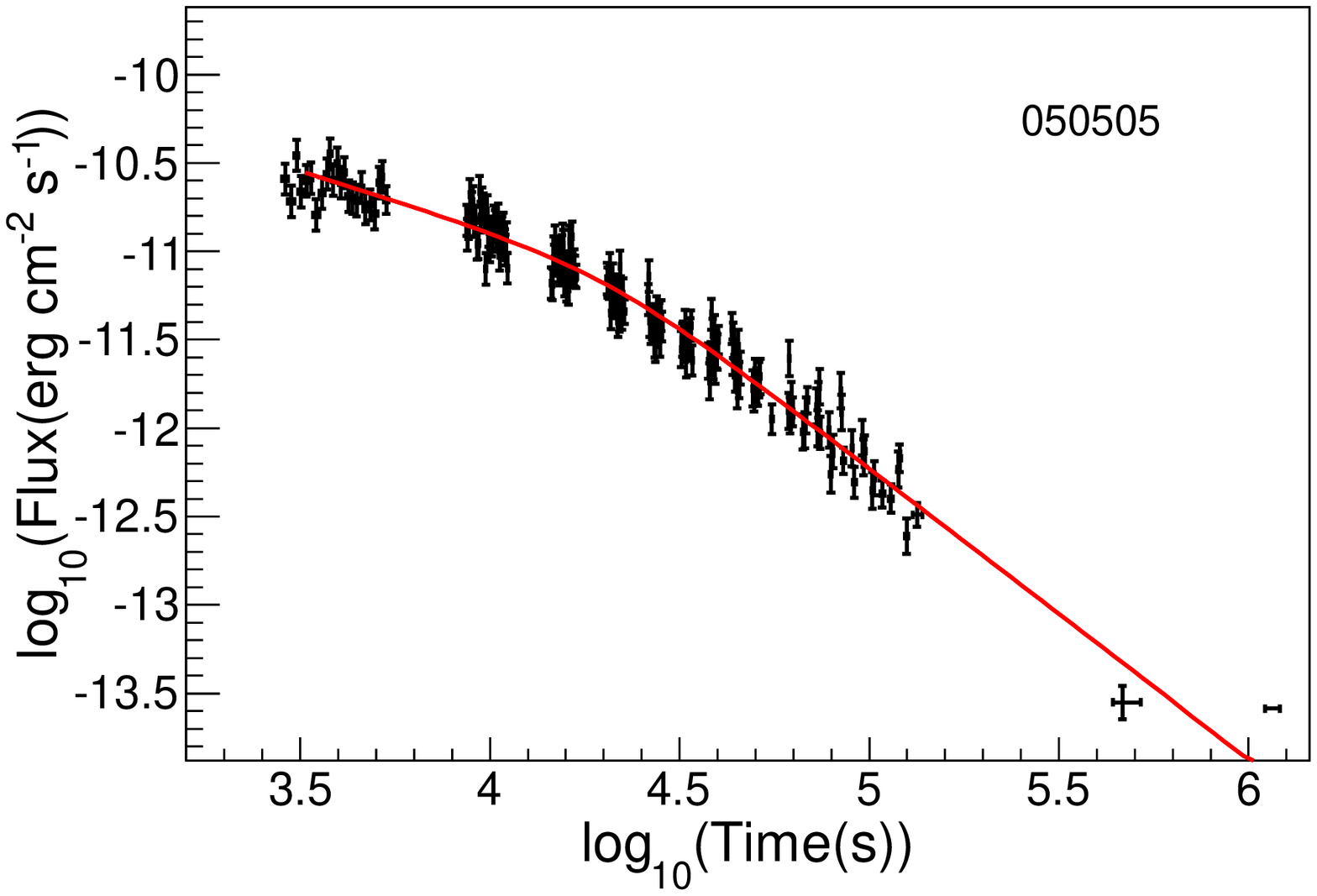}
\includegraphics[width=5.5cm,height=5cm]{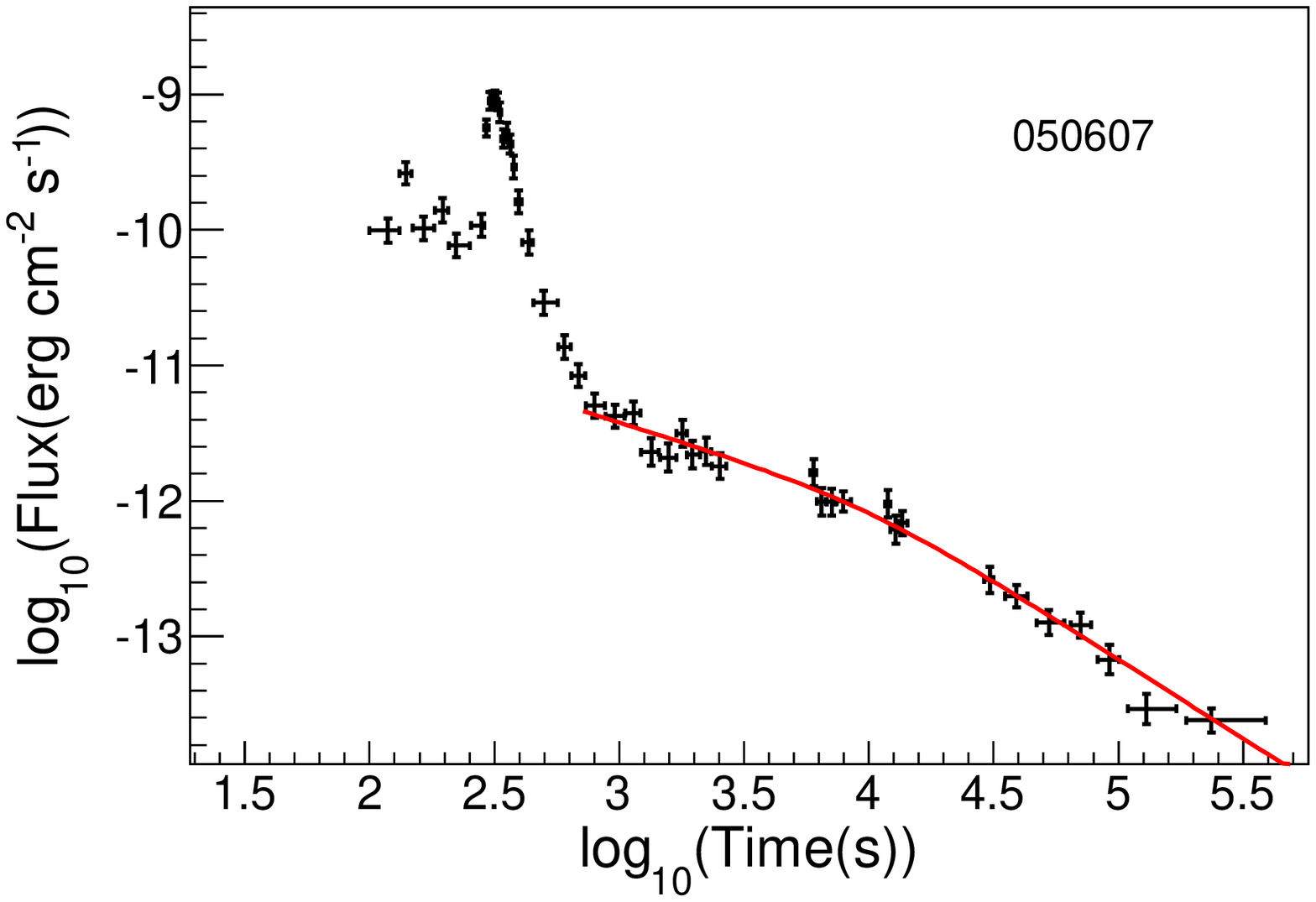}
\includegraphics[width=5.5cm,height=5cm]{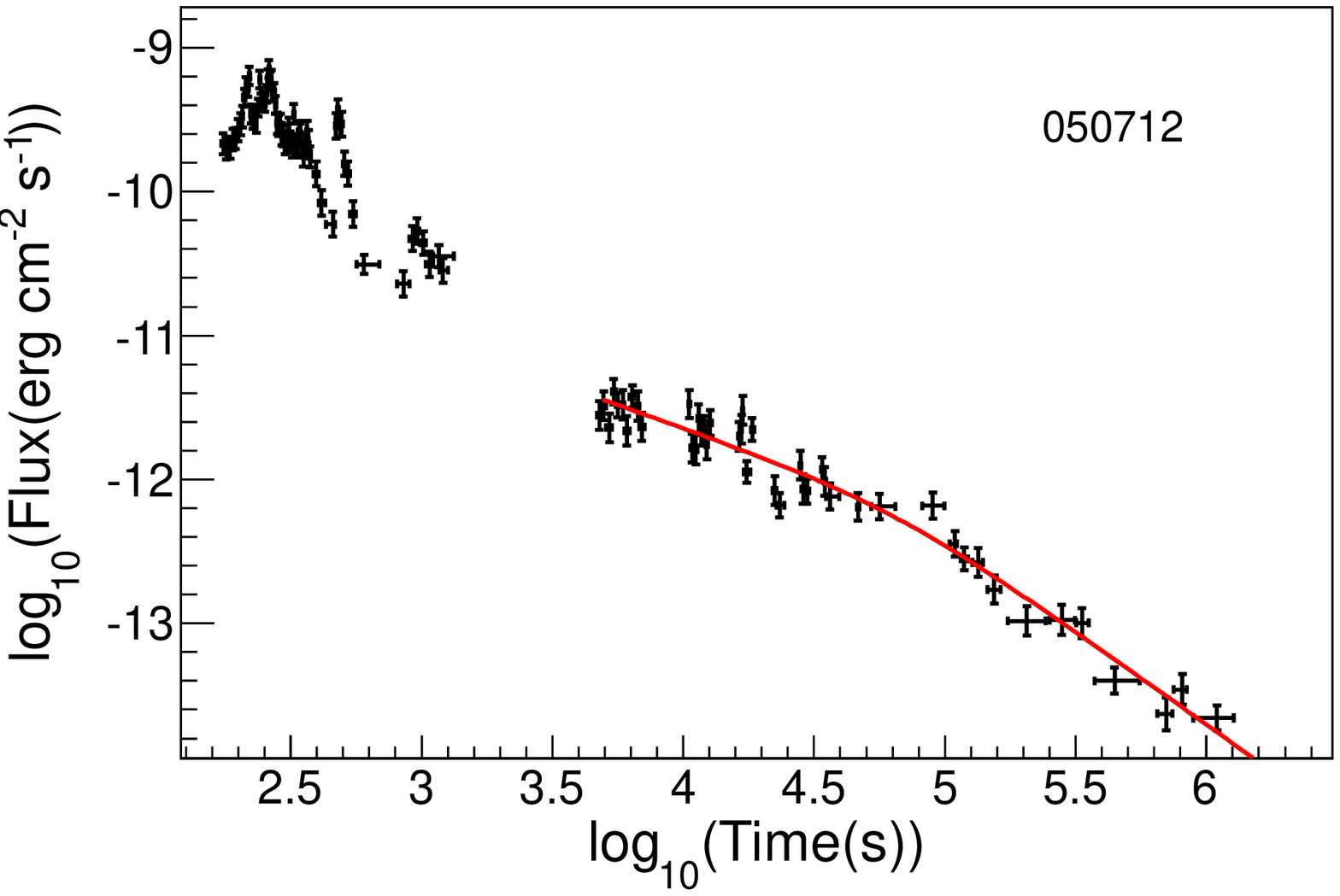}
\includegraphics[width=5.5cm,height=5cm]{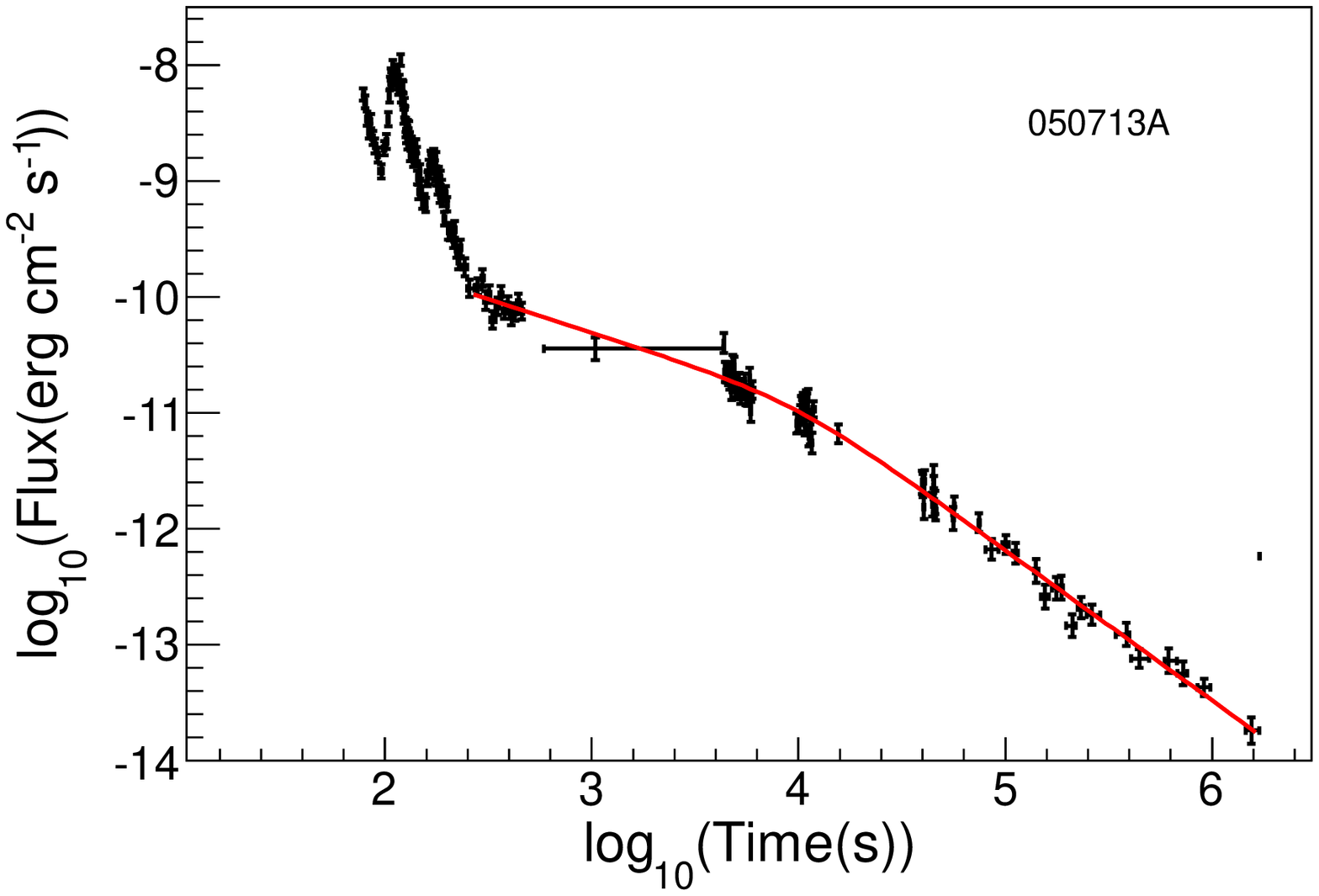}
\includegraphics[width=5.5cm,height=5cm]{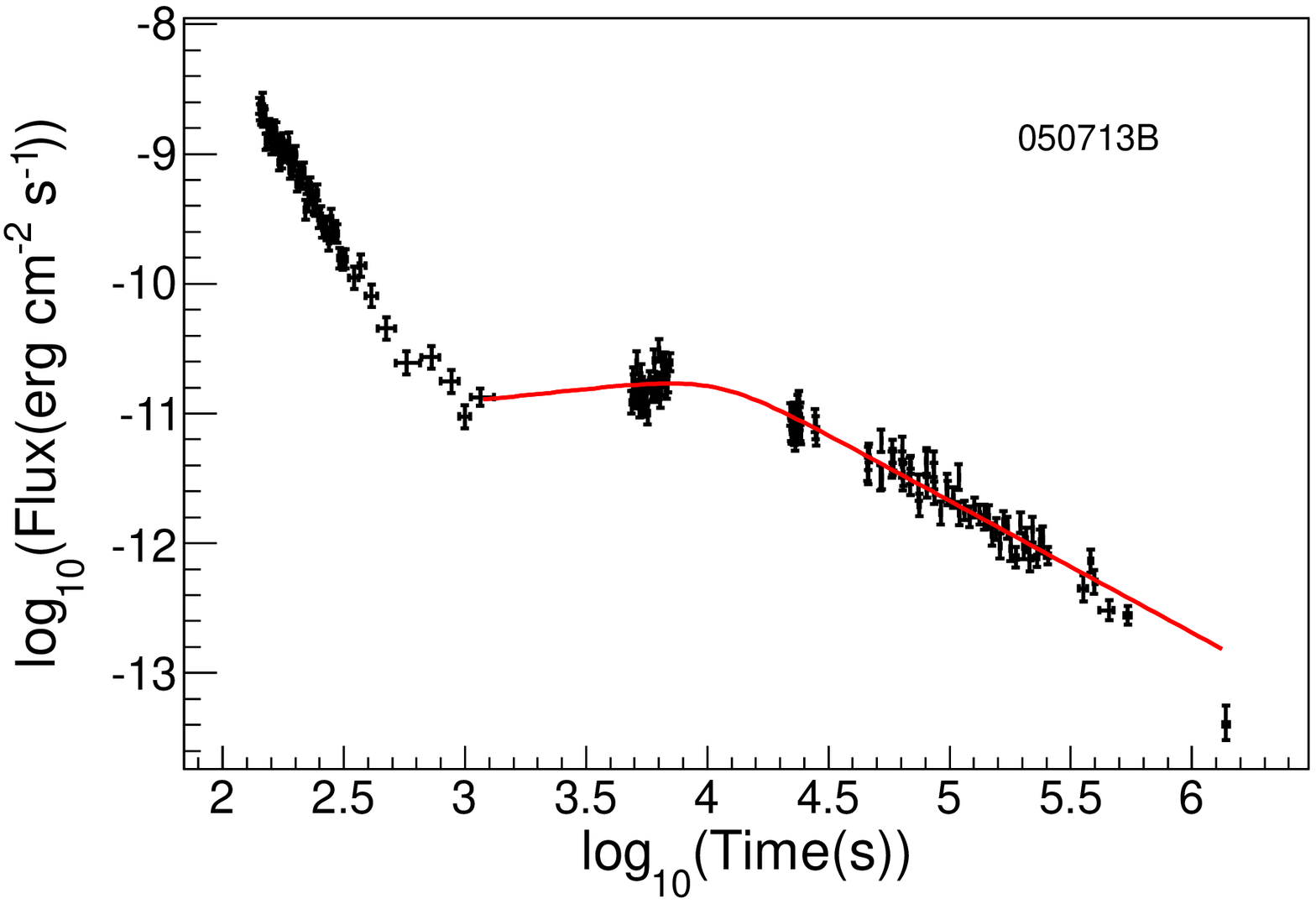}
\includegraphics[width=5.5cm,height=5cm]{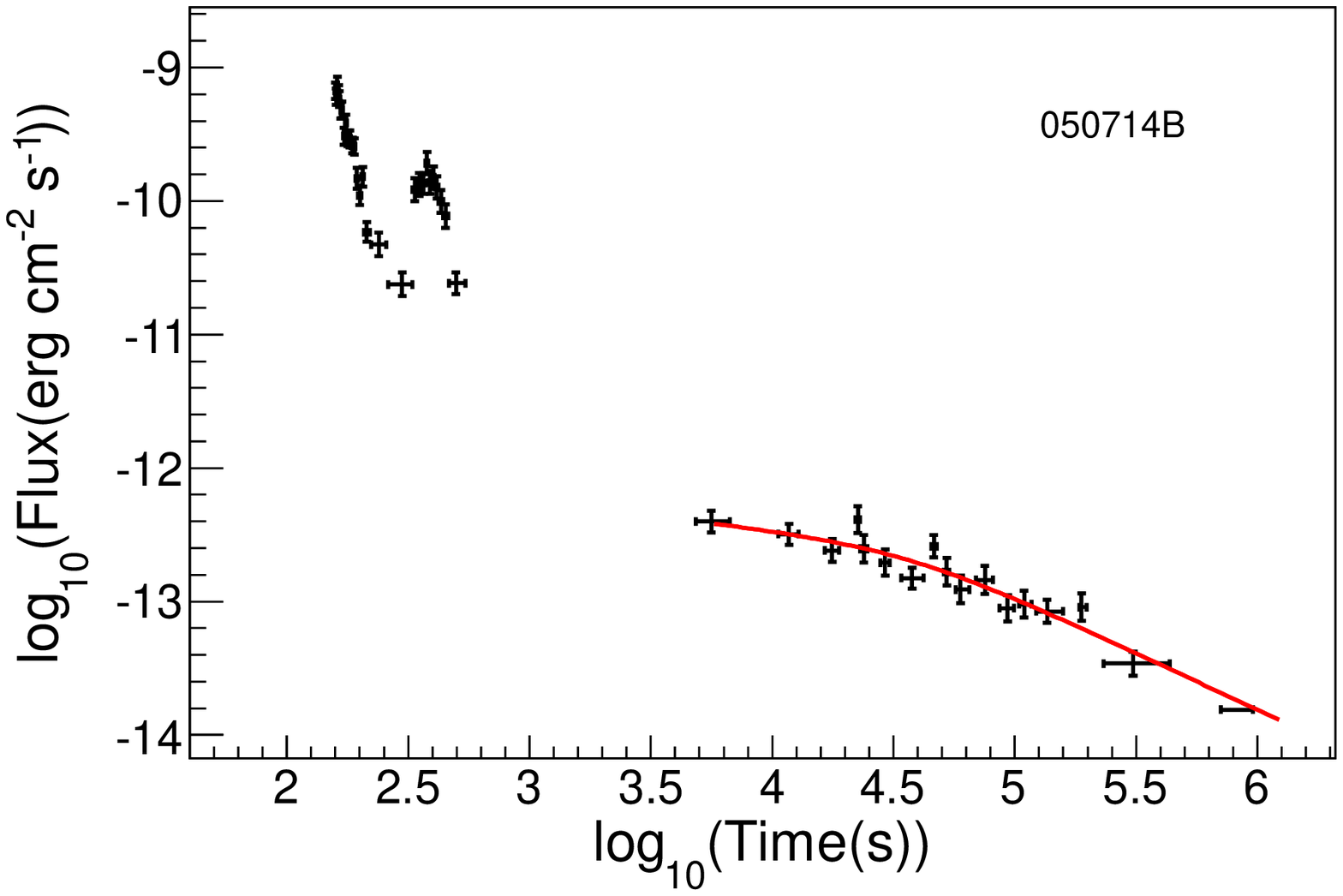}
\includegraphics[width=5.5cm,height=5cm]{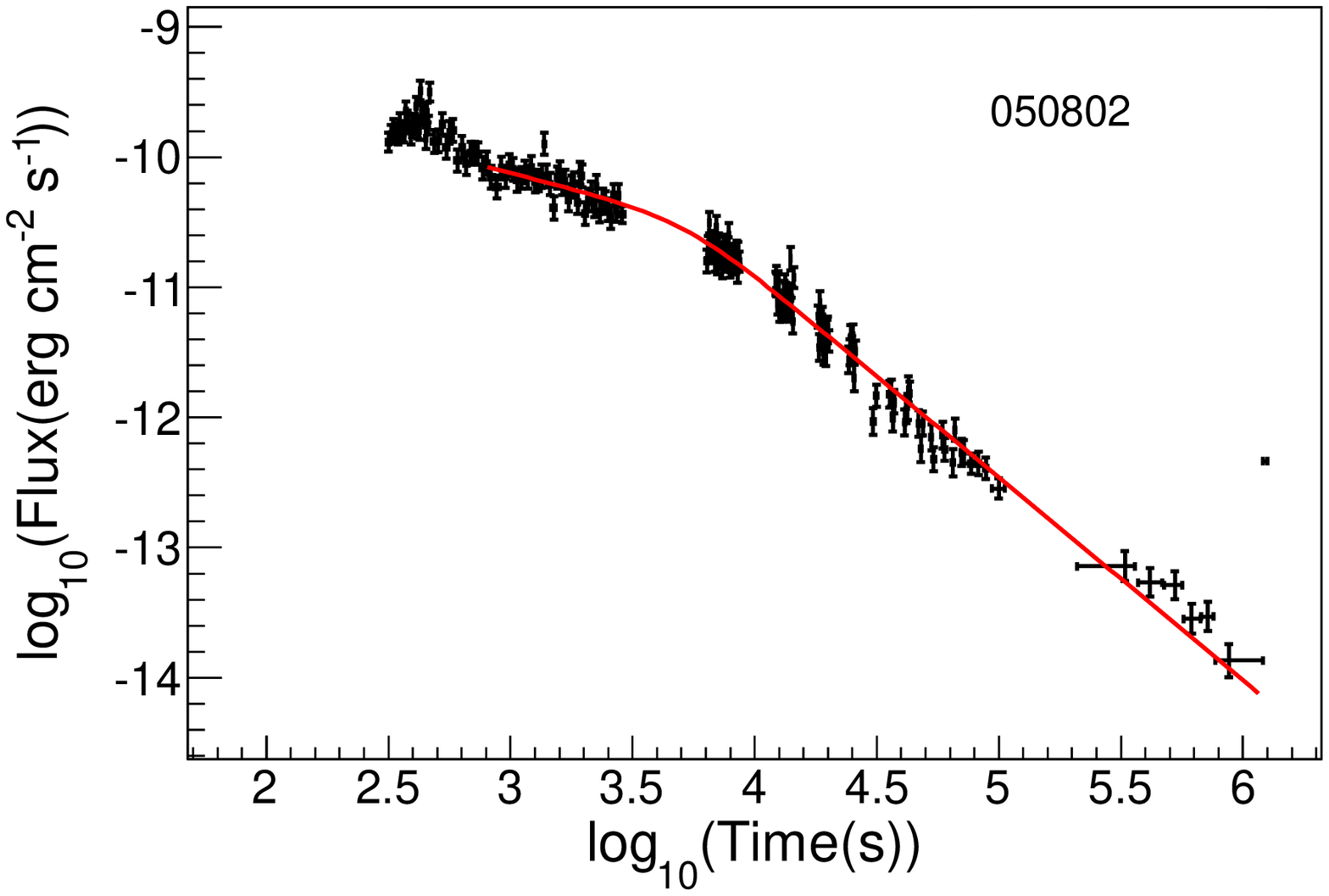}
\includegraphics[width=5.5cm,height=5cm]{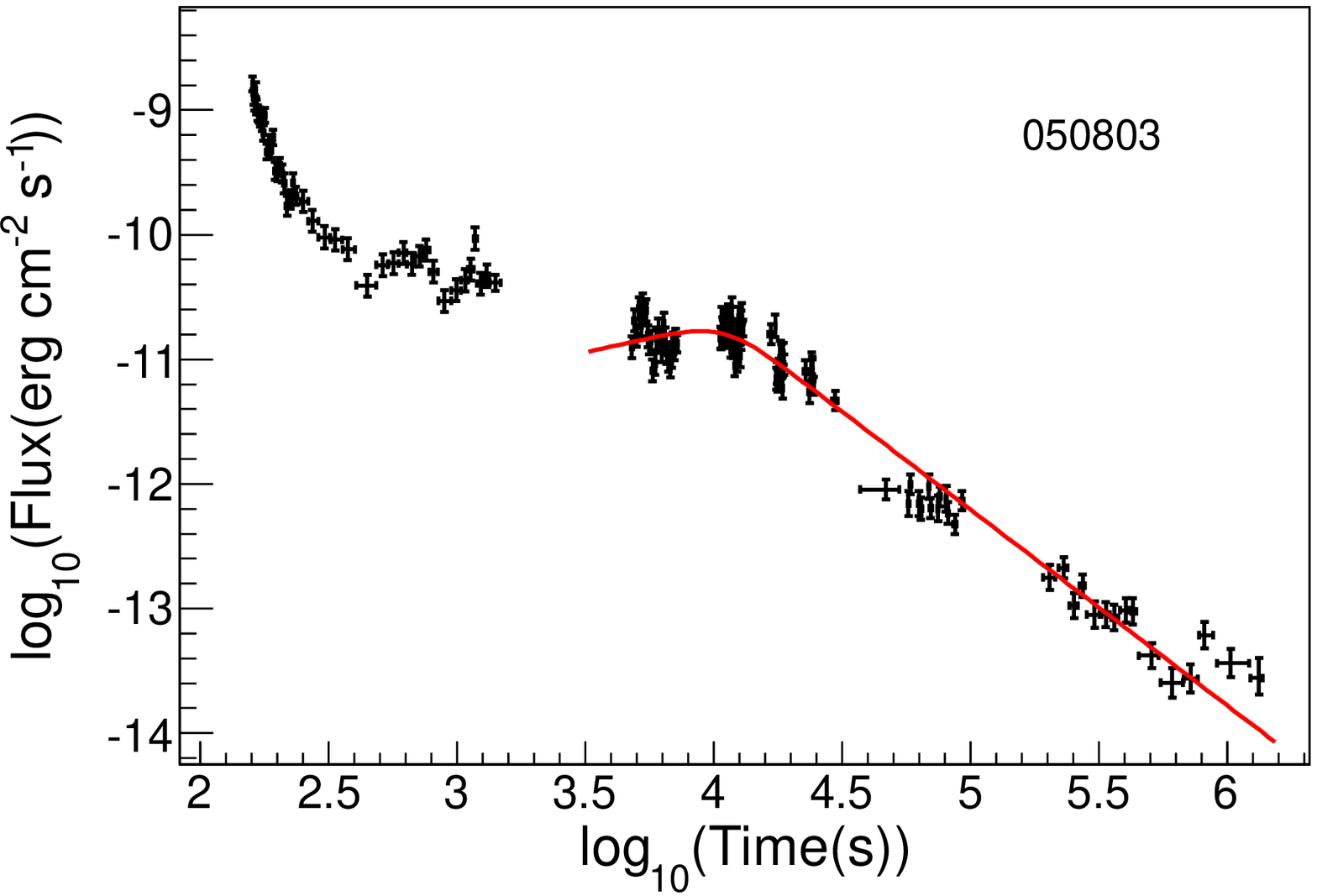}
\includegraphics[width=5.5cm,height=5cm]{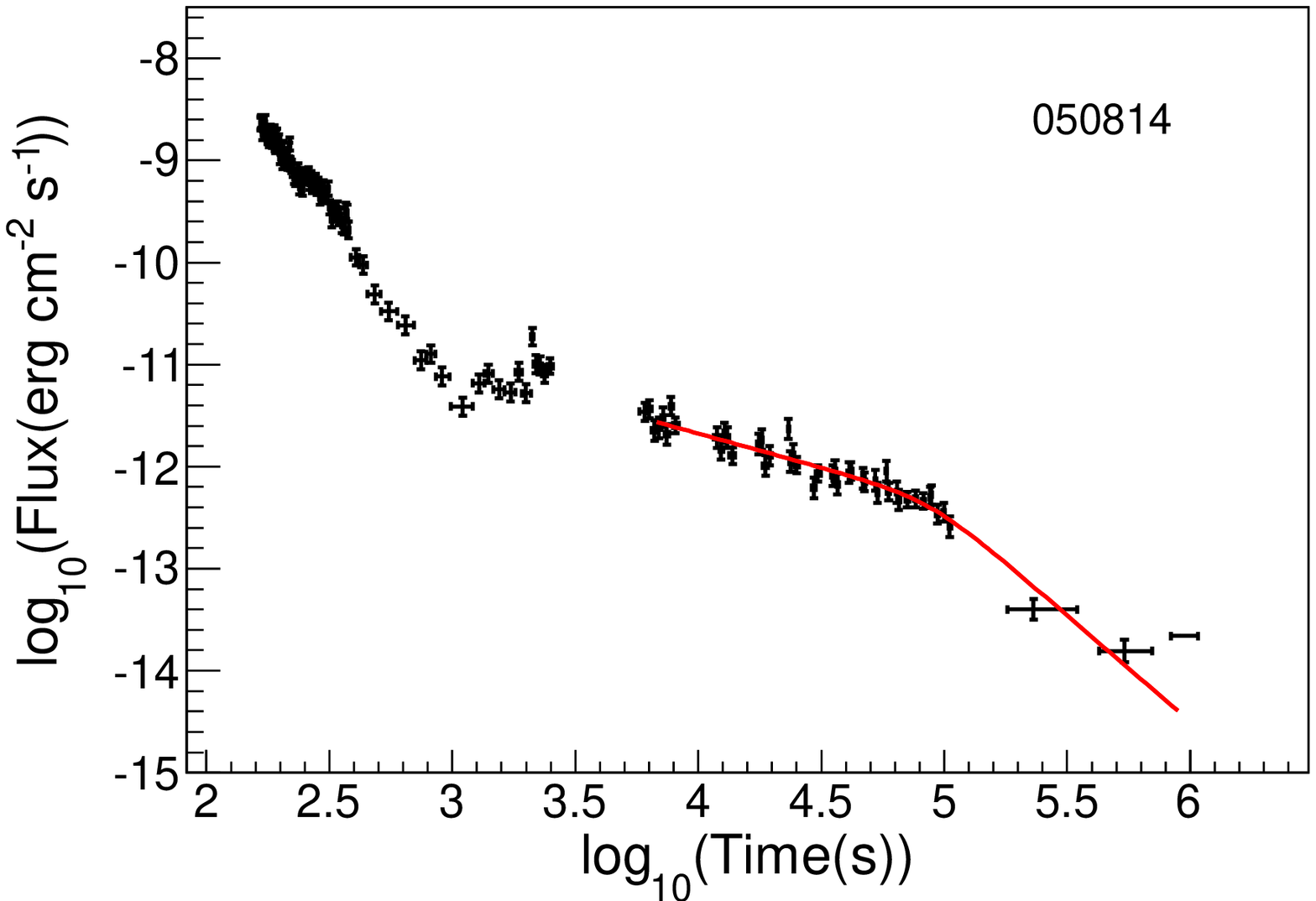}
\caption{XRT light curves for the bursts in our sample. The solid red lines are the best fits with a smooth broken power law for the shallow decay phase and its follow-up decay phase.}
\label{fig-1-1}
\end{center}
\end{figure*}

\begin{figure*}
\begin{center}
\setlength{\abovecaptionskip}{0.cm}
\setlength{\belowcaptionskip}{-0.cm}
\hspace{0cm}
\figurenum{1}
\graphicspath{{lightcurve/}}
\includegraphics[width=5.5cm,height=5cm]{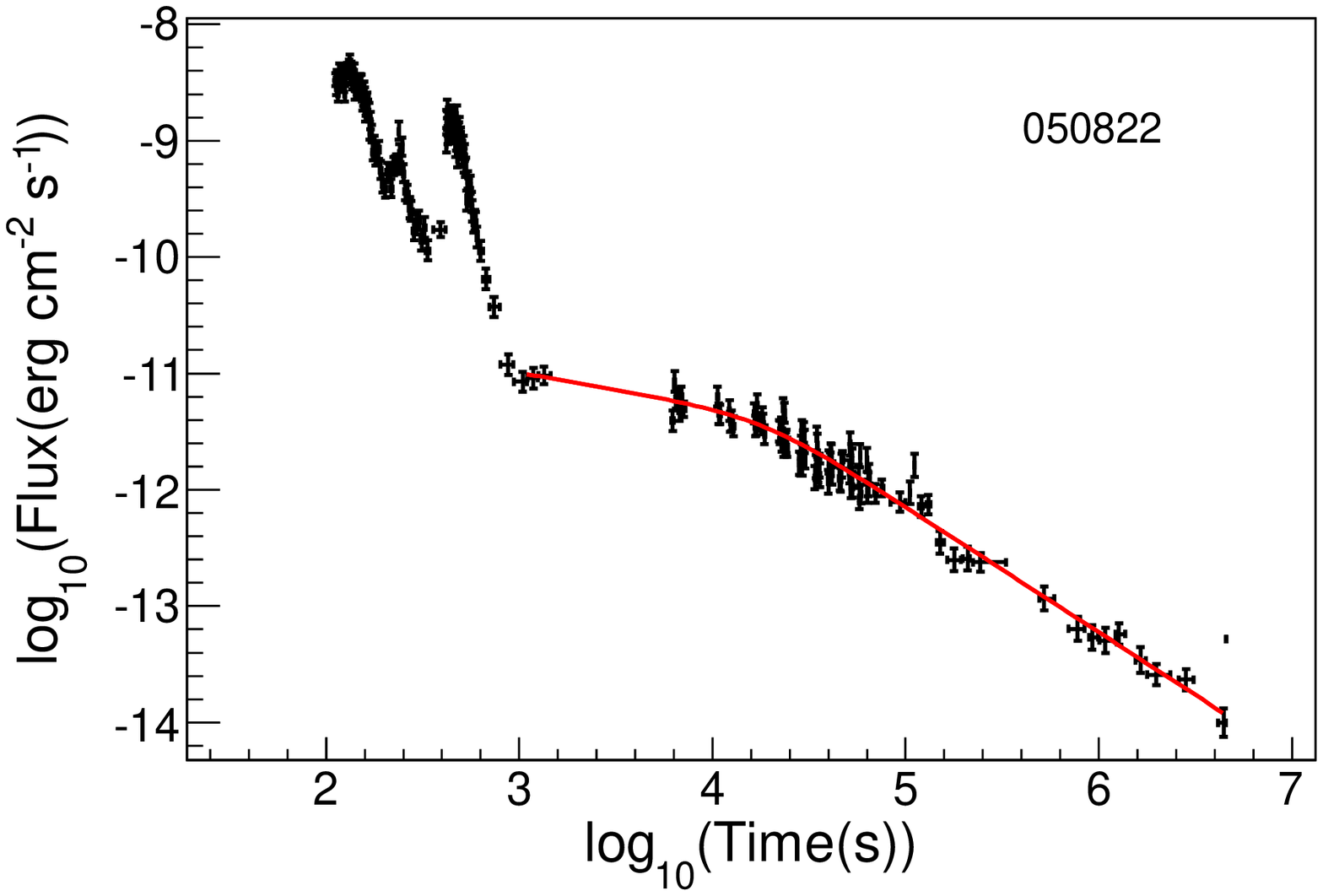}
\includegraphics[width=5.5cm,height=5cm]{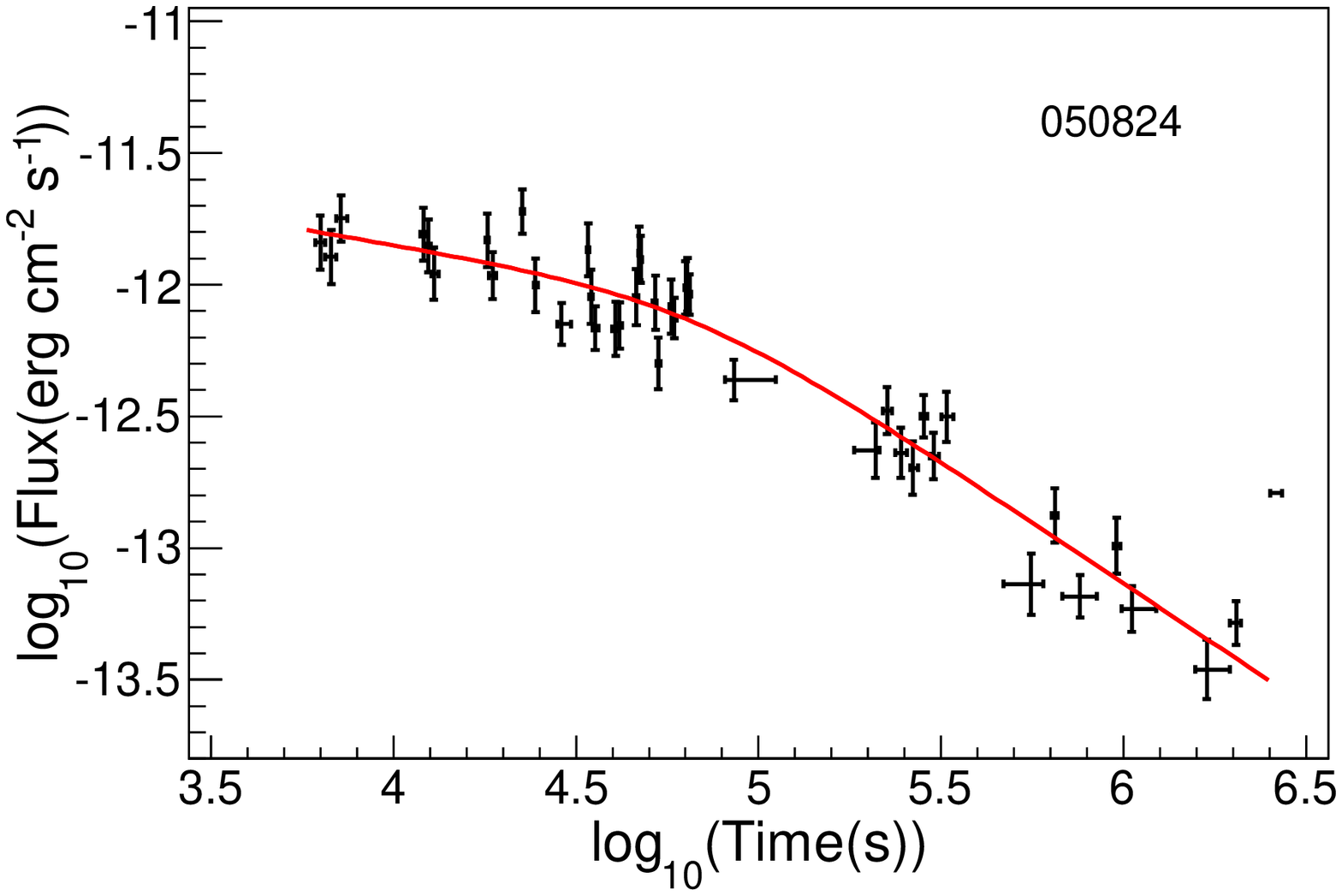}
\includegraphics[width=5.5cm,height=5cm]{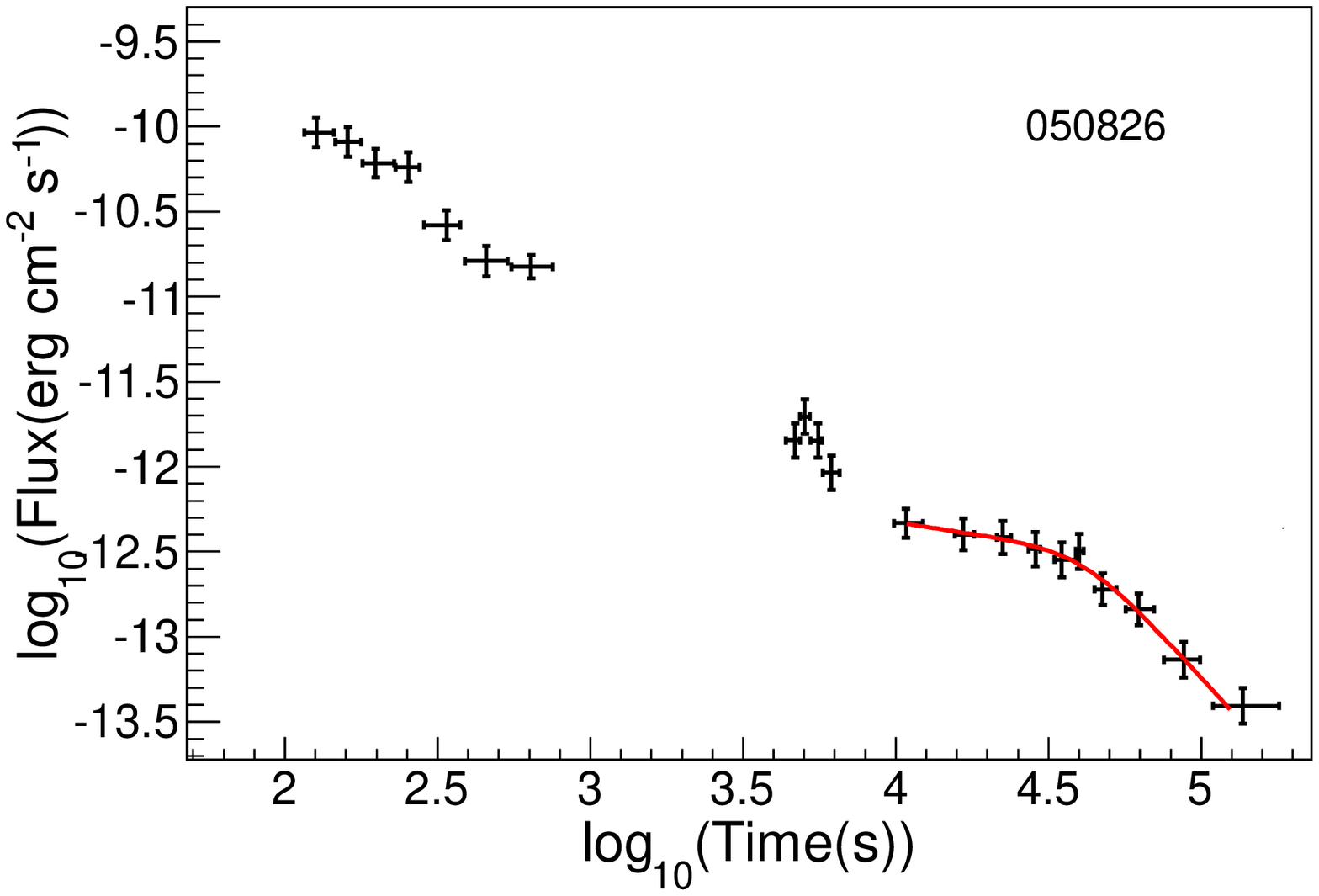}
\includegraphics[width=5.5cm,height=5cm]{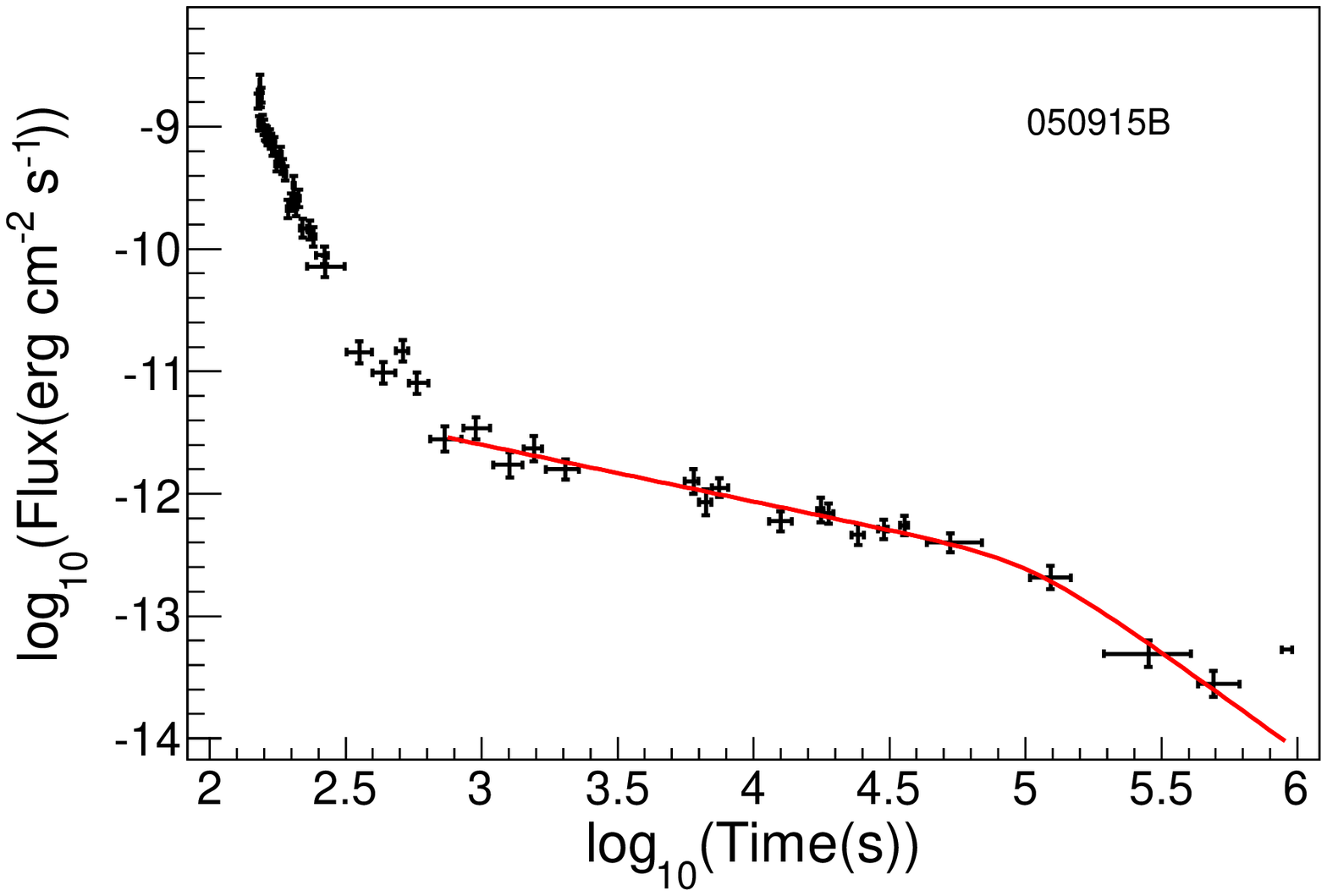}
\includegraphics[width=5.5cm,height=5cm]{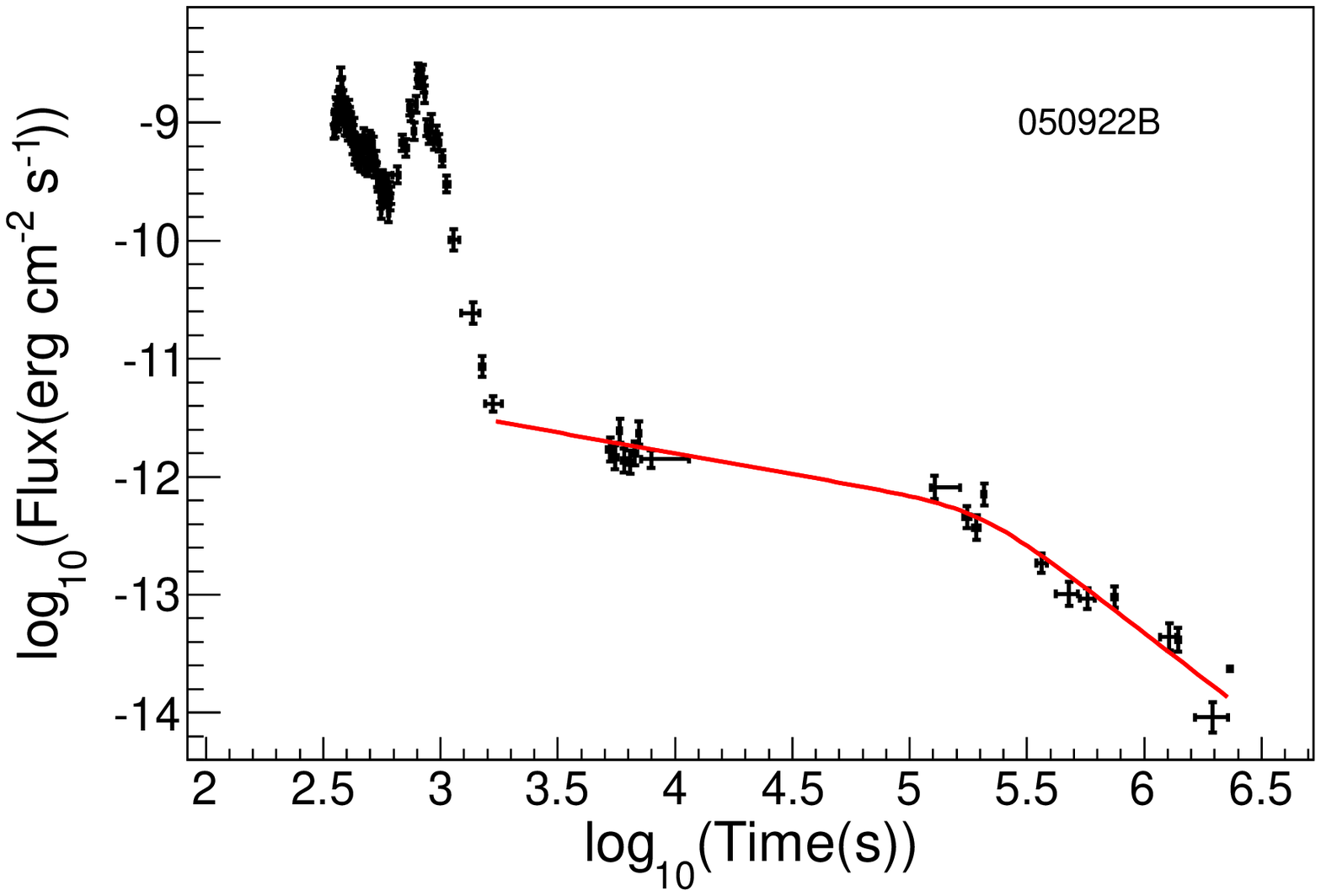}
\includegraphics[width=5.5cm,height=5cm]{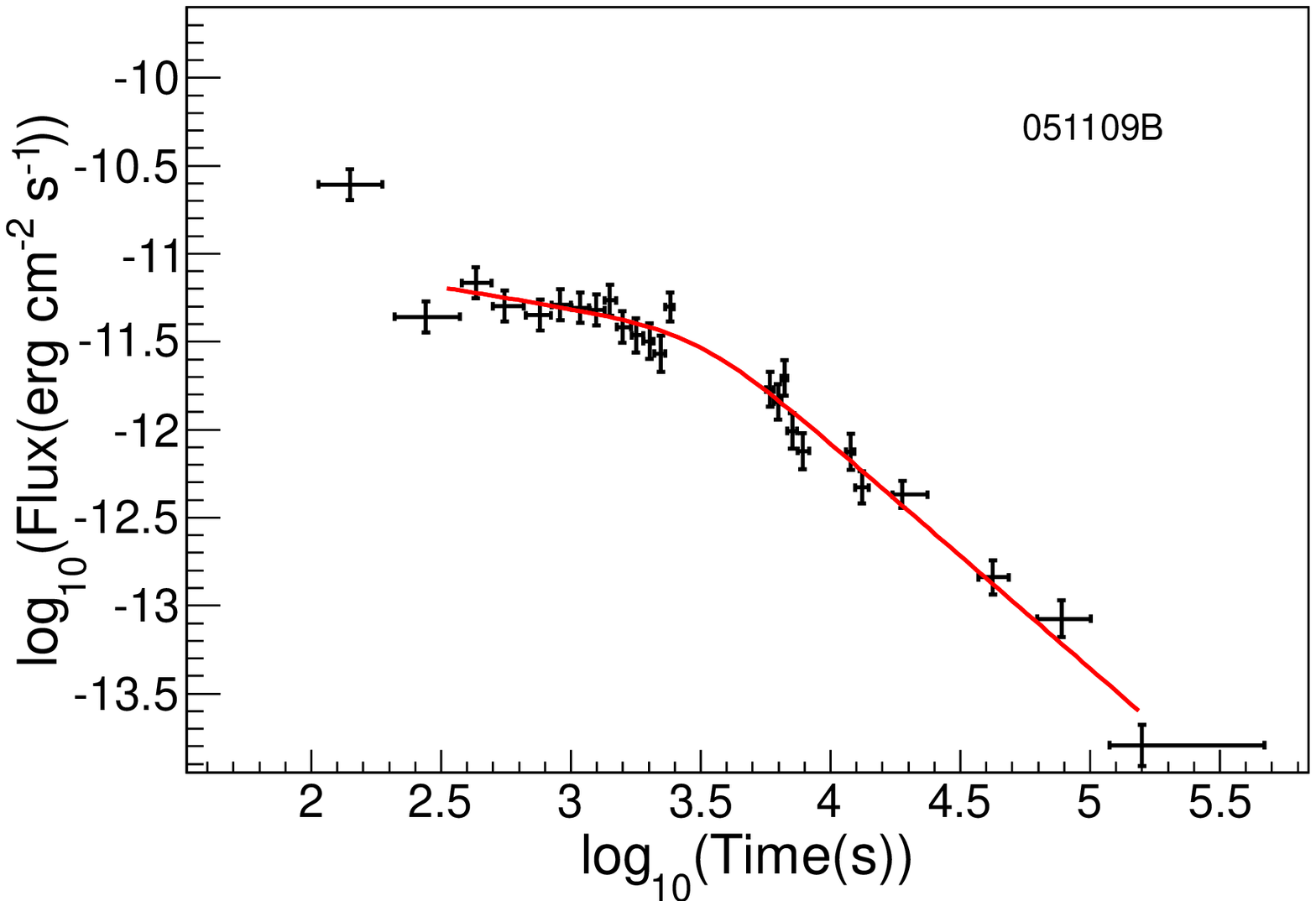}
\includegraphics[width=5.5cm,height=5cm]{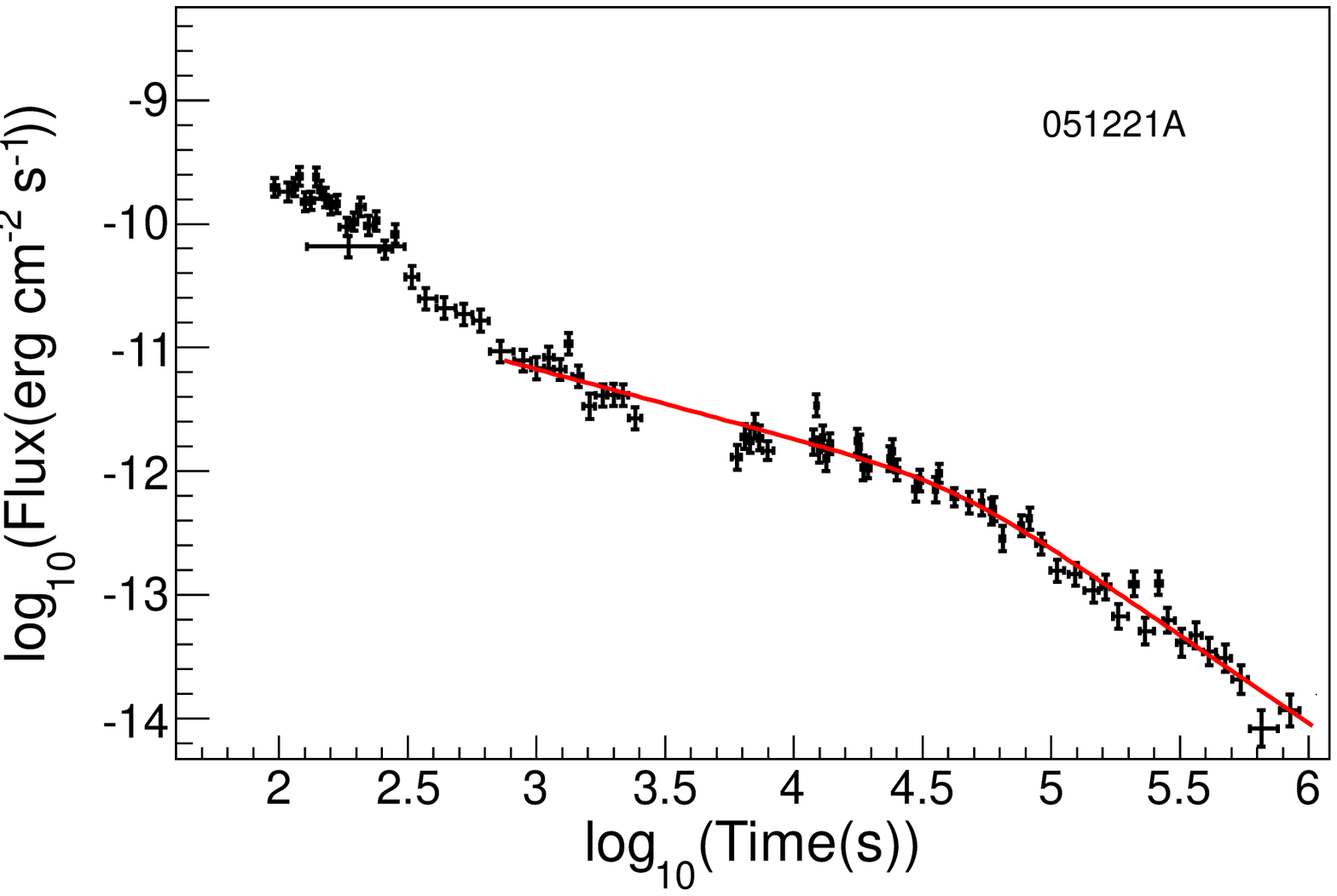}
\includegraphics[width=5.5cm,height=5cm]{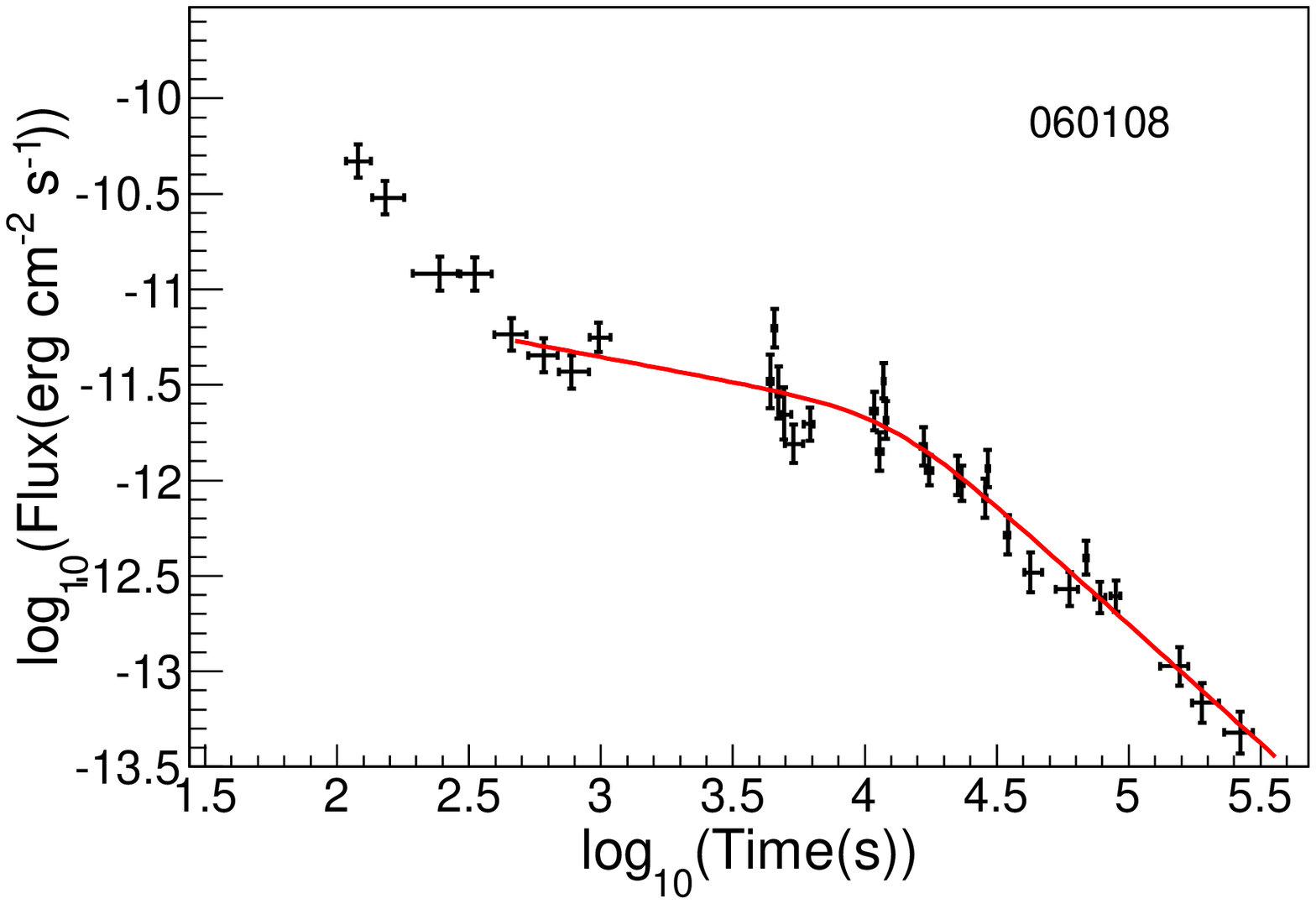}
\includegraphics[width=5.5cm,height=5cm]{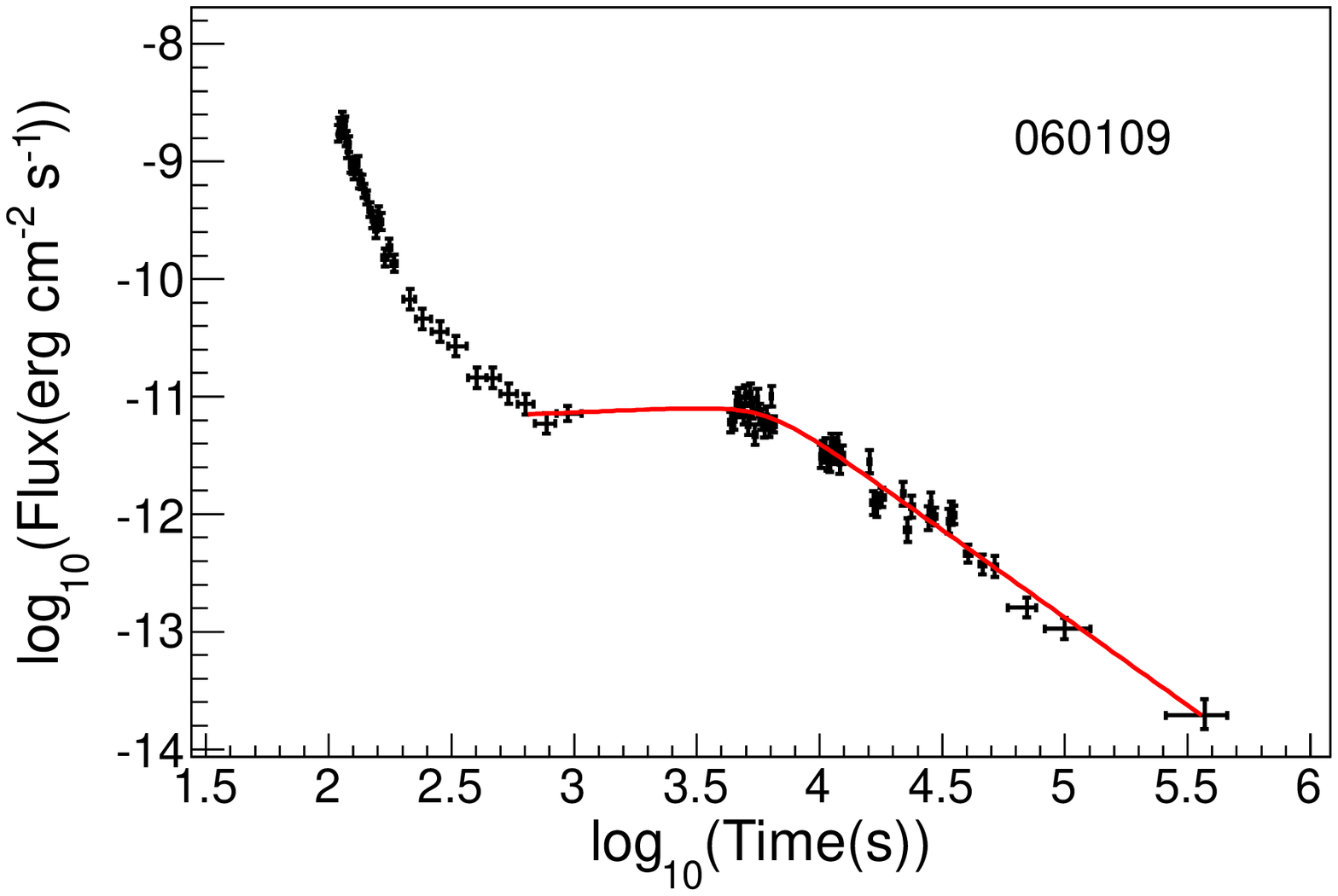}
\includegraphics[width=5.5cm,height=5cm]{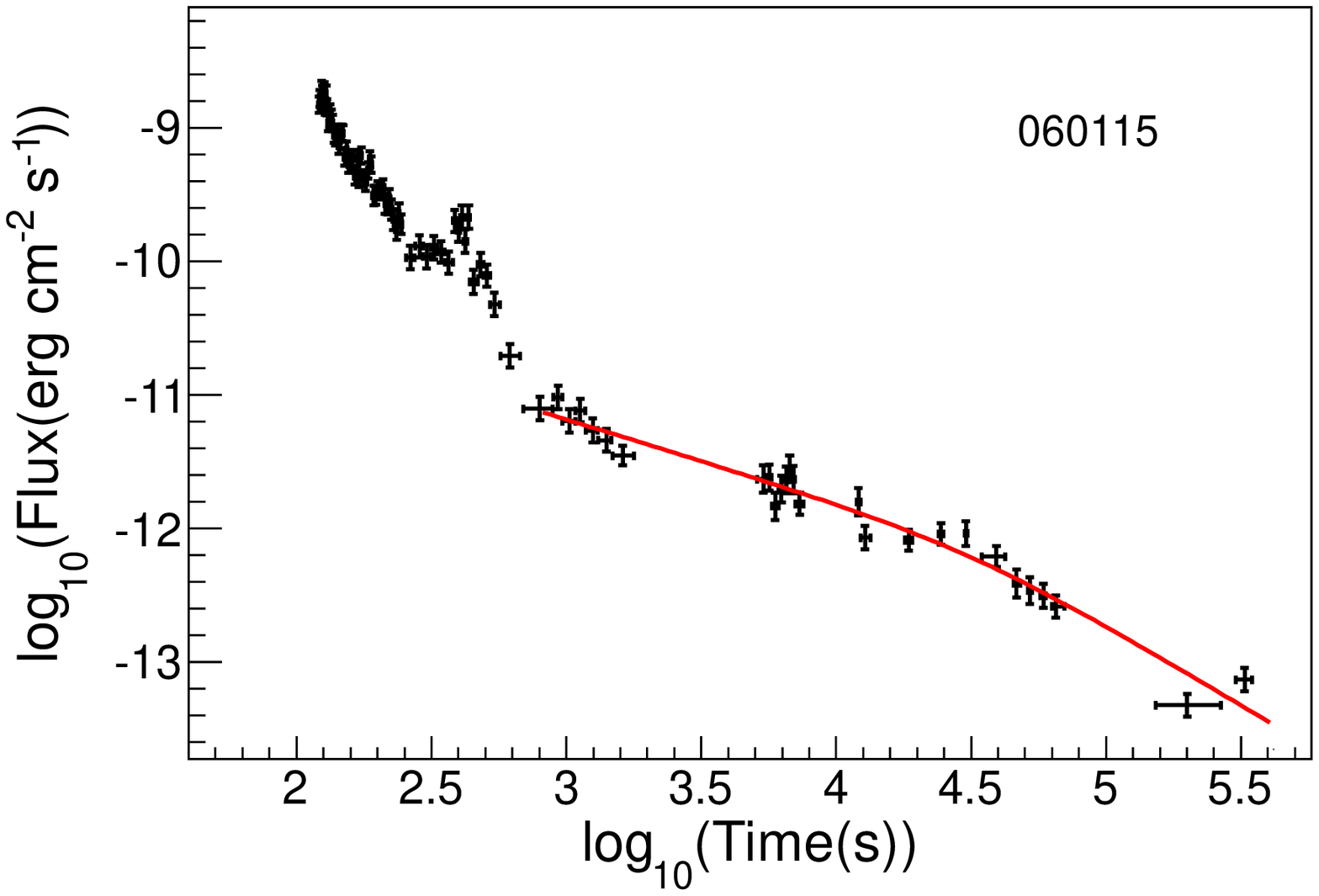}
\includegraphics[width=5.5cm,height=5cm]{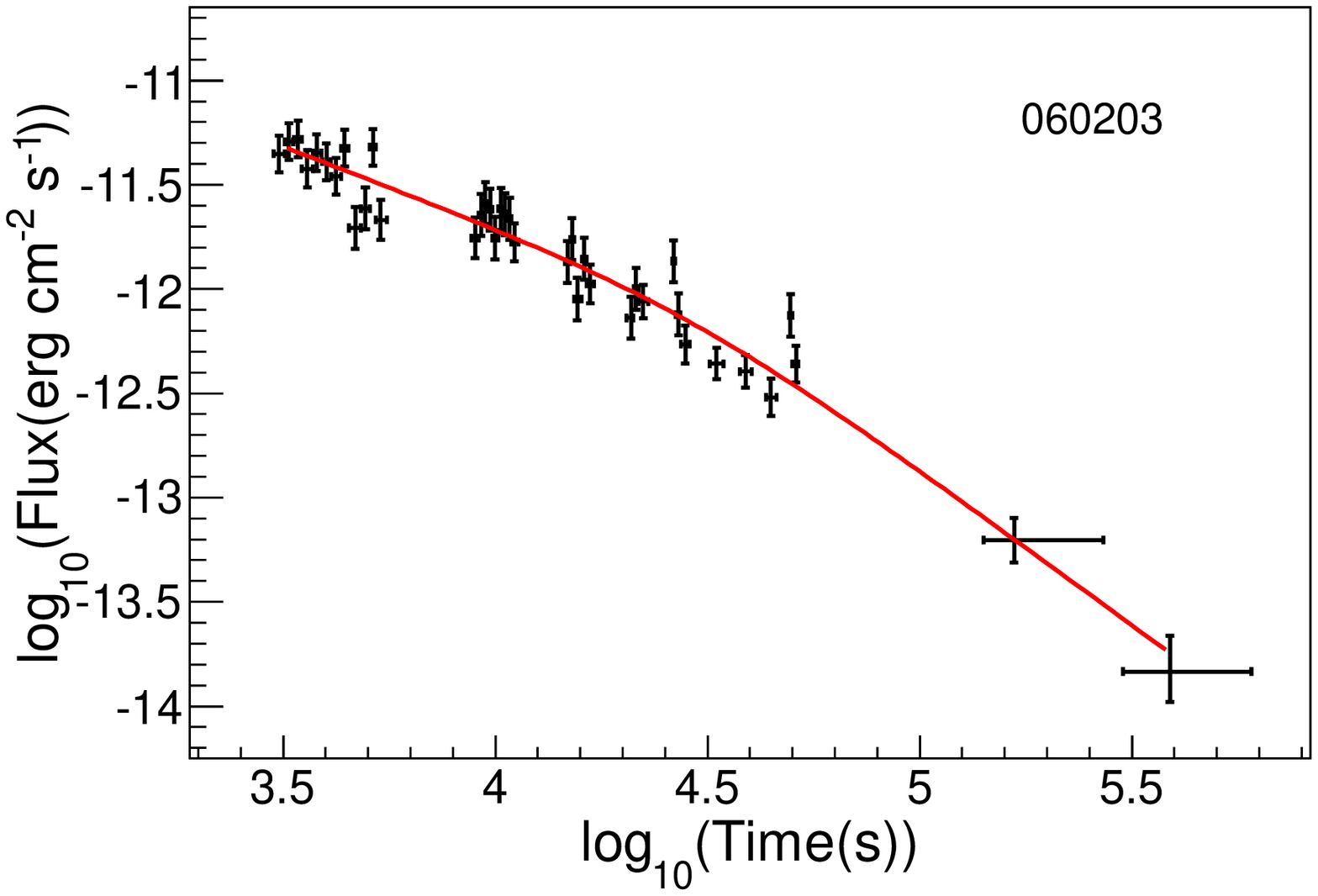}
\includegraphics[width=5.5cm,height=5cm]{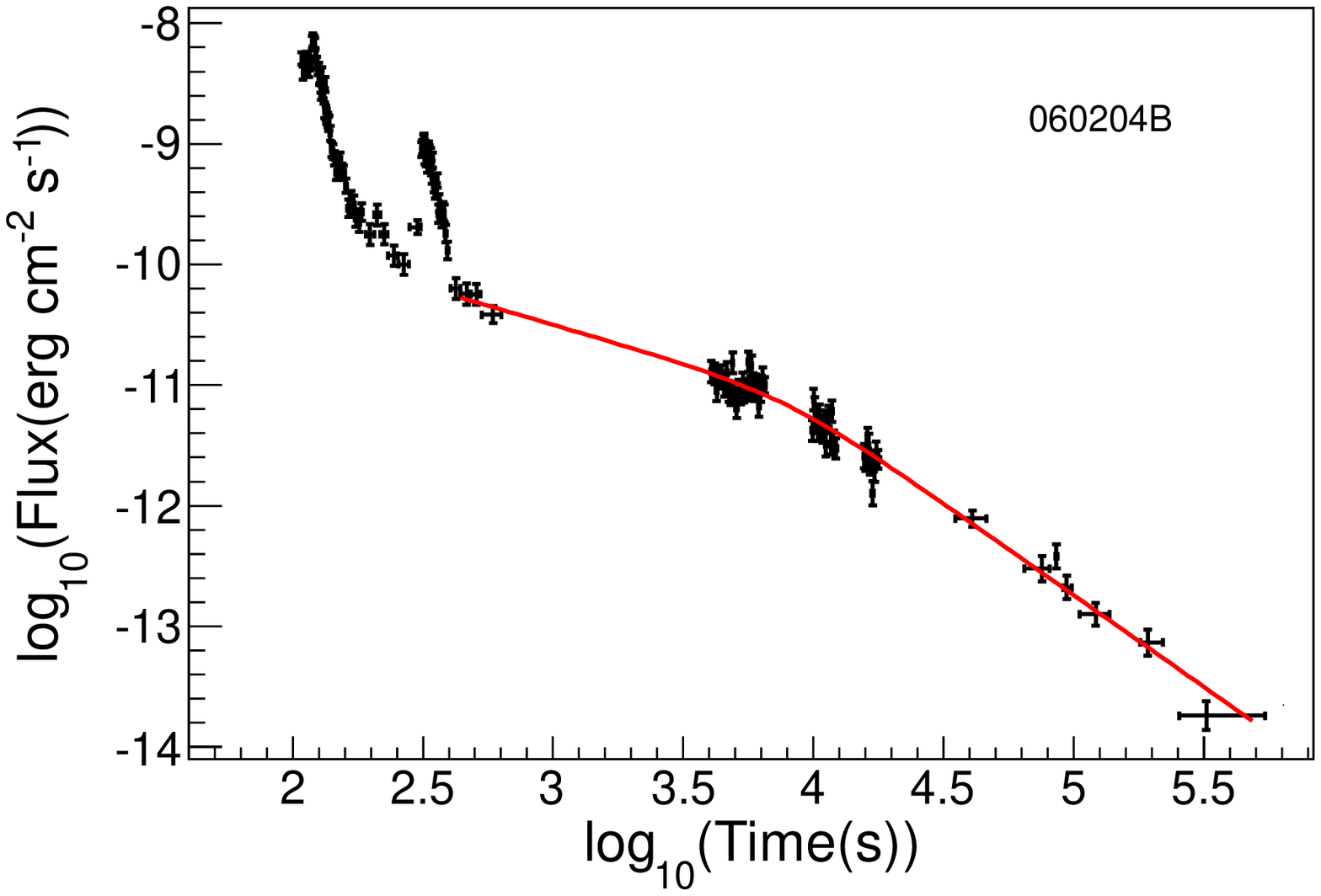}
\caption{ Continued.}
\label{fig-1-2}
\end{center}
\end{figure*}

\begin{figure*}
\begin{center}
\setlength{\abovecaptionskip}{0.cm}
\setlength{\belowcaptionskip}{-0.cm}
\figurenum{1}
\hspace{0cm}
\graphicspath{{lightcurve/}}
\includegraphics[width=5.5cm,height=5cm]{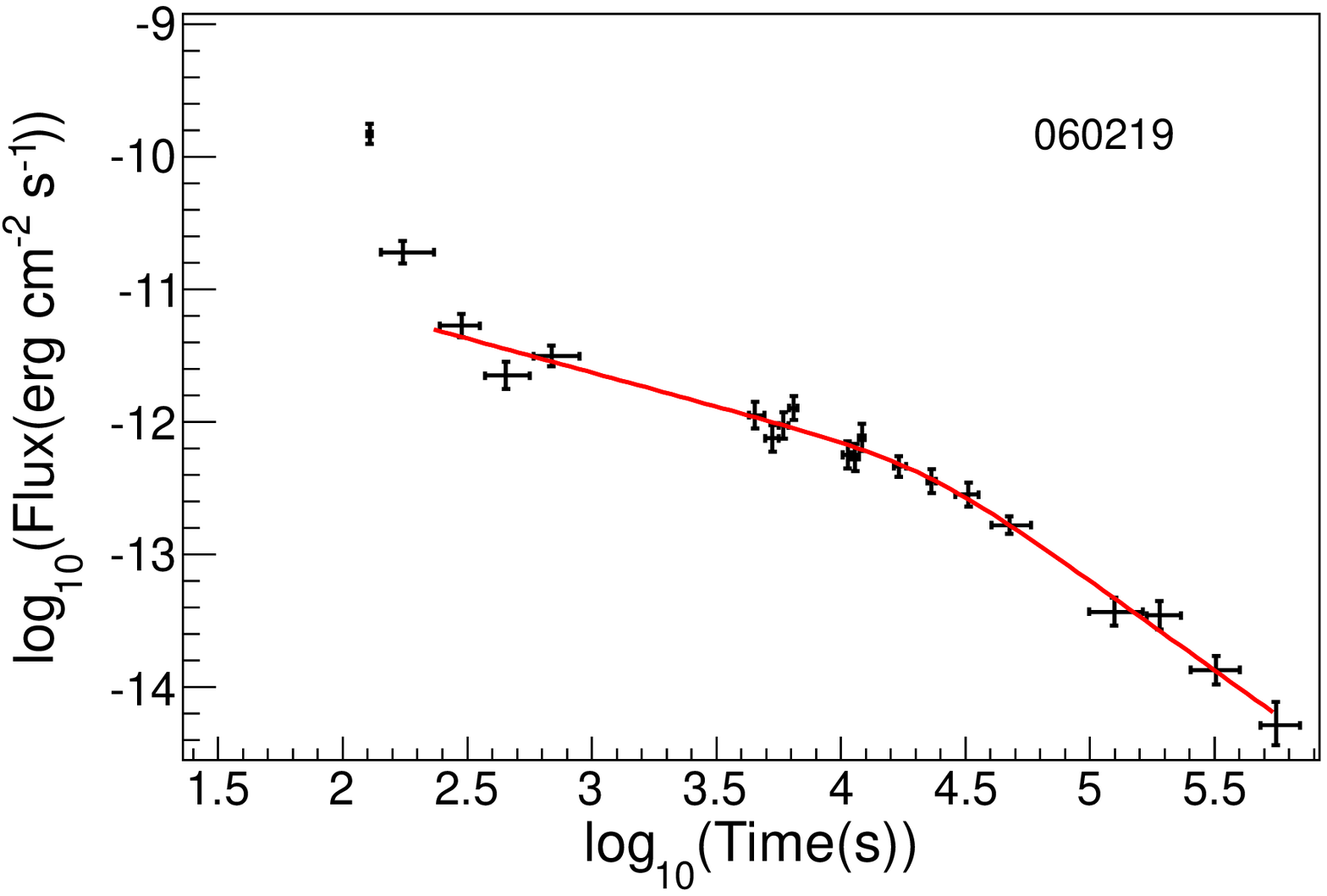}
\includegraphics[width=5.5cm,height=5cm]{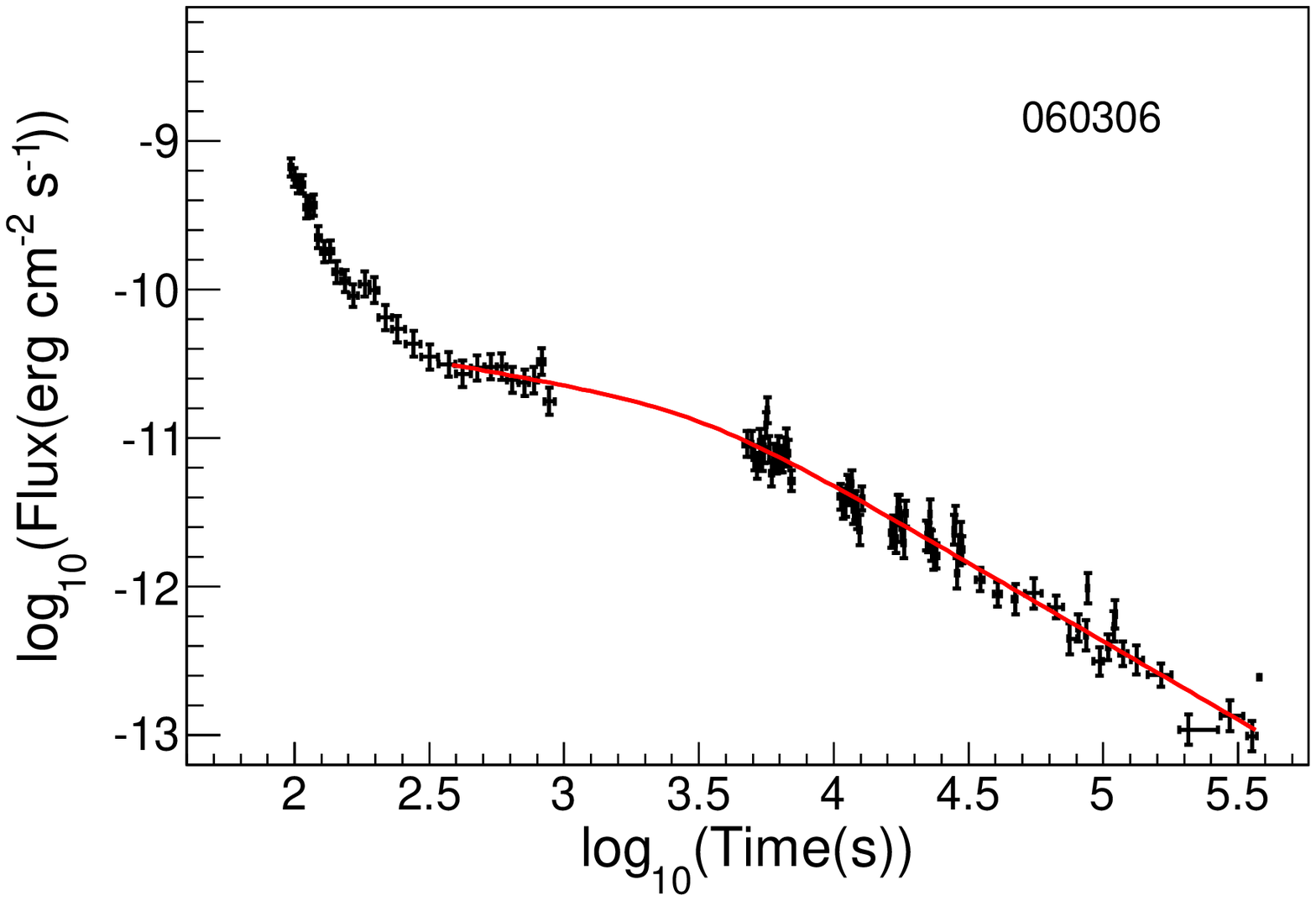}
\includegraphics[width=5.5cm,height=5cm]{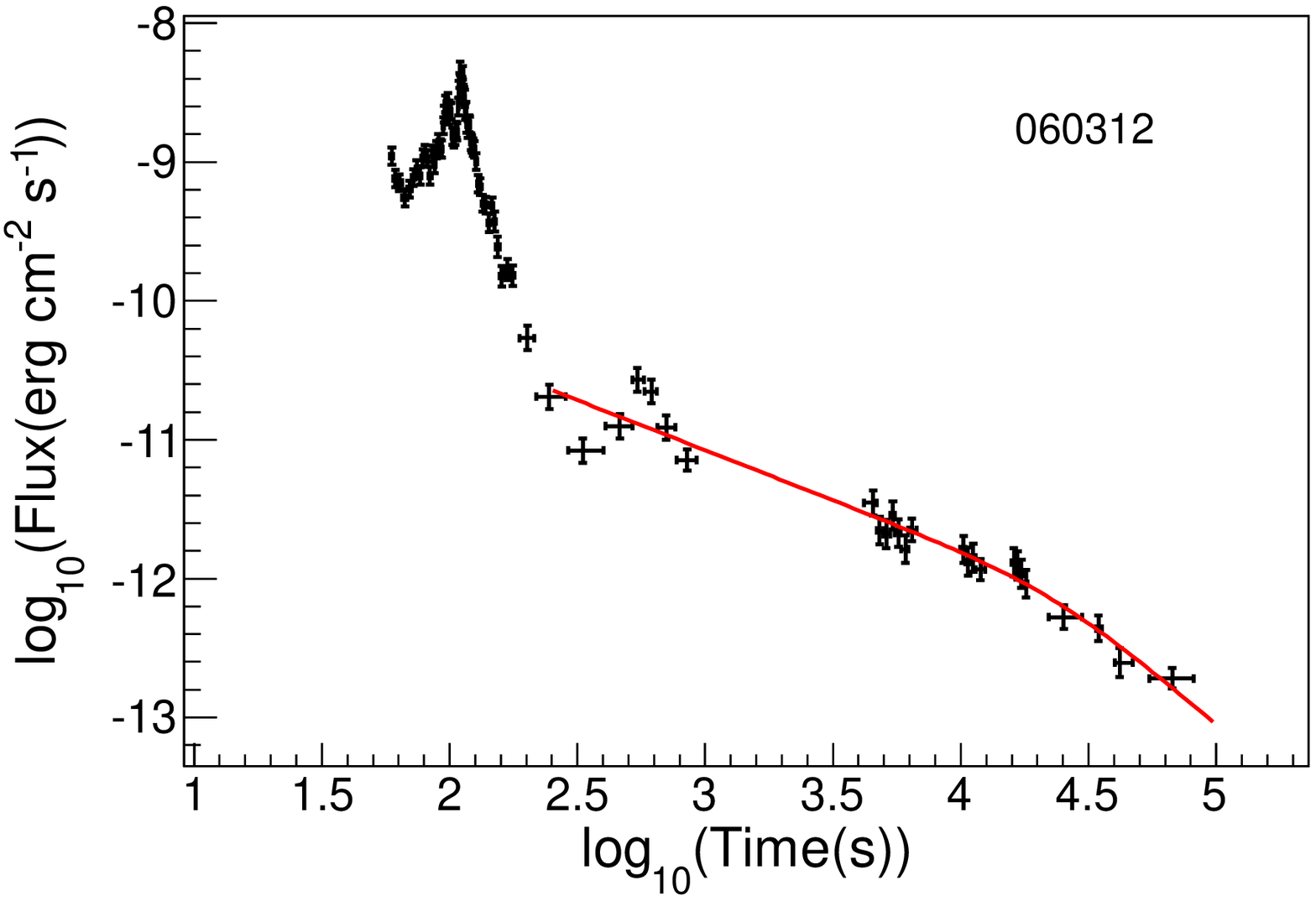}
\includegraphics[width=5.5cm,height=5cm]{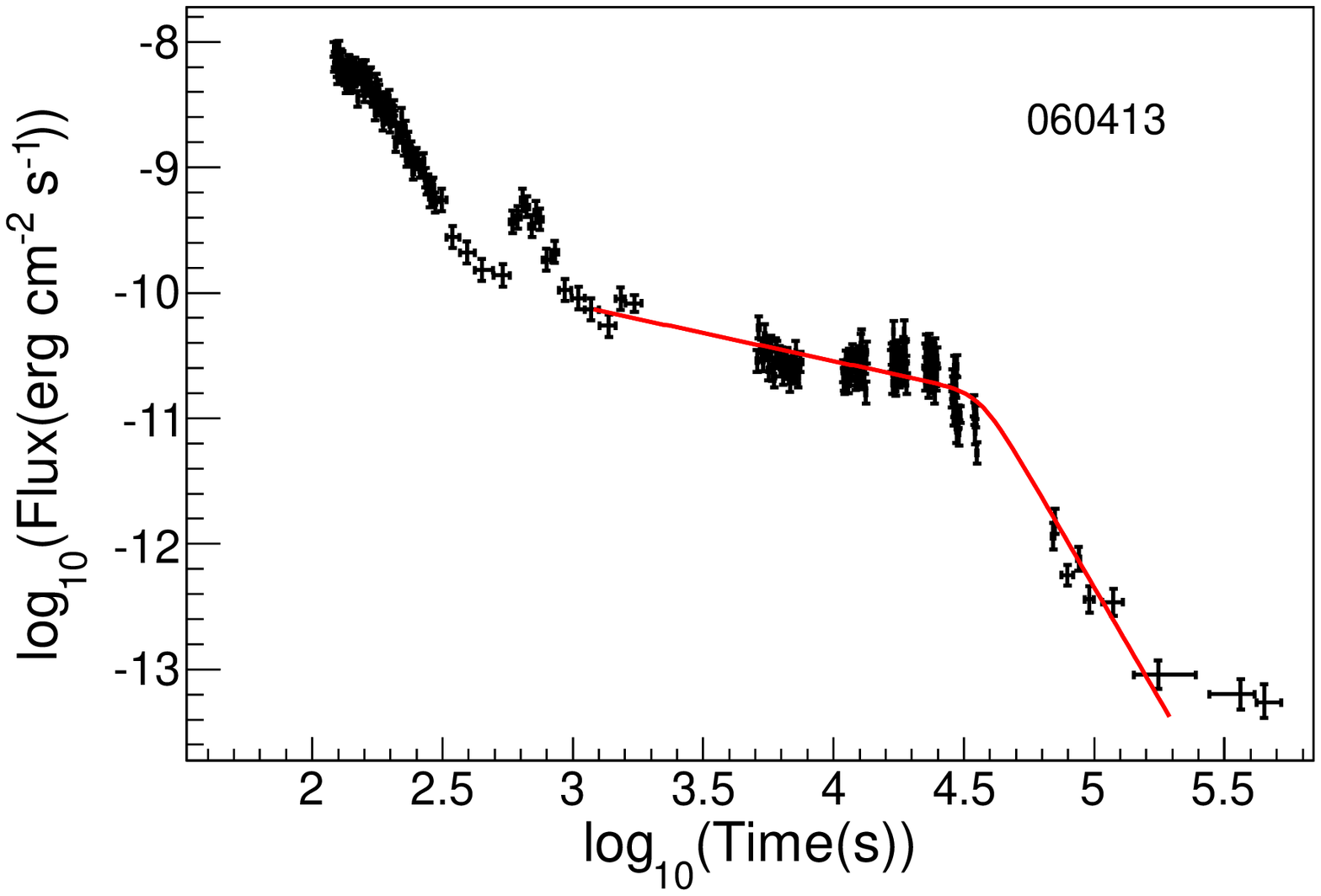}
\includegraphics[width=5.5cm,height=5cm]{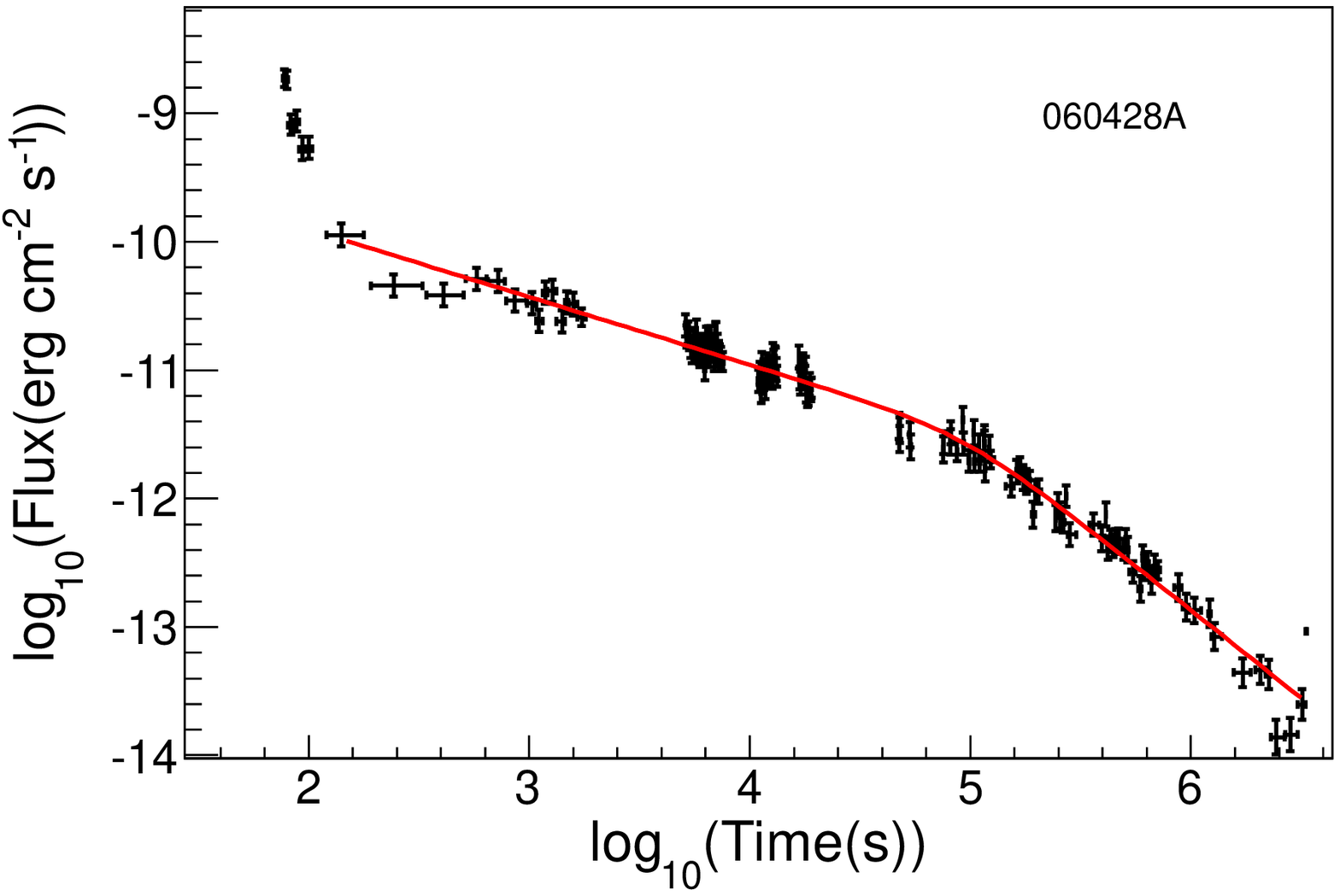}
\includegraphics[width=5.5cm,height=5cm]{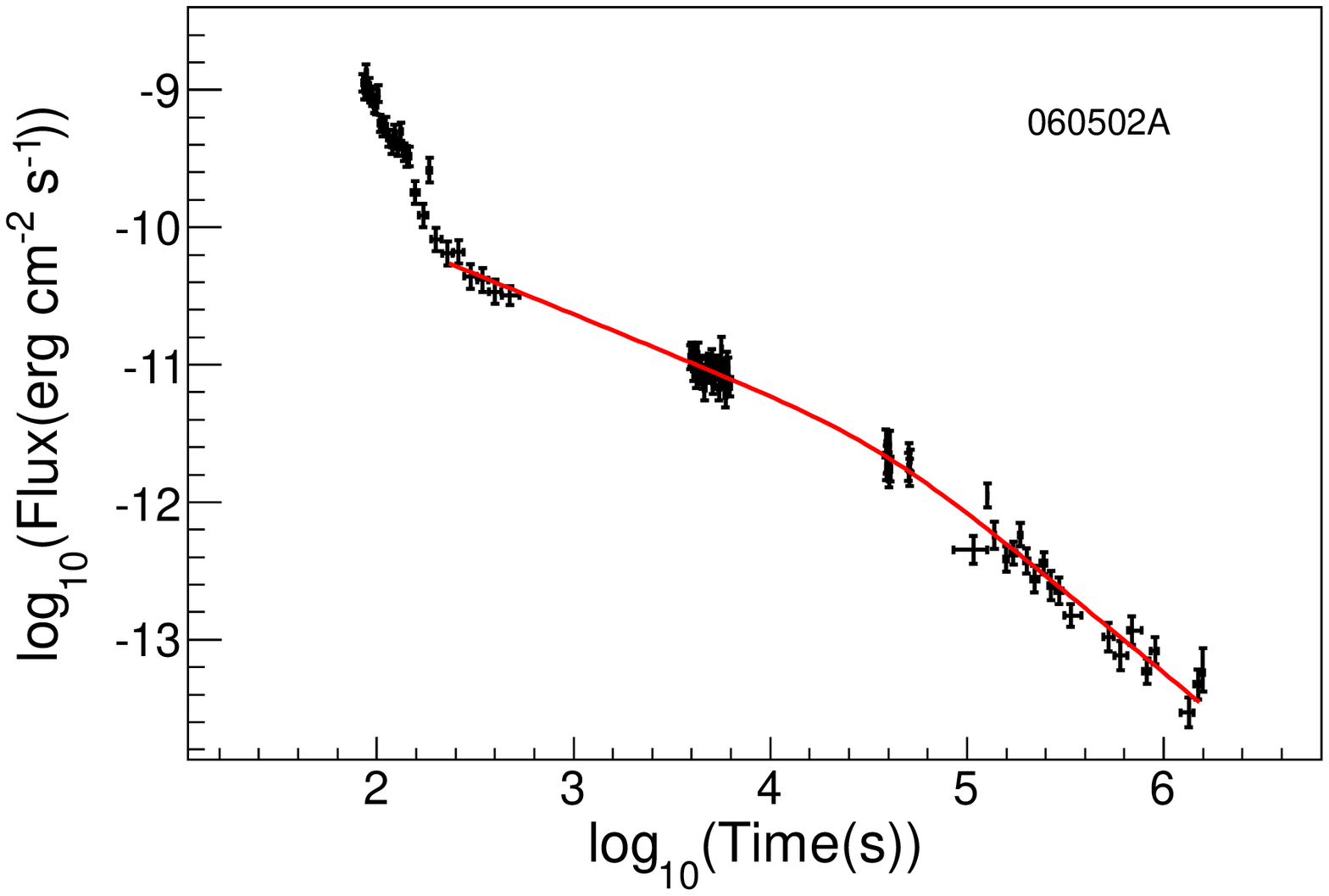}
\includegraphics[width=5.5cm,height=5cm]{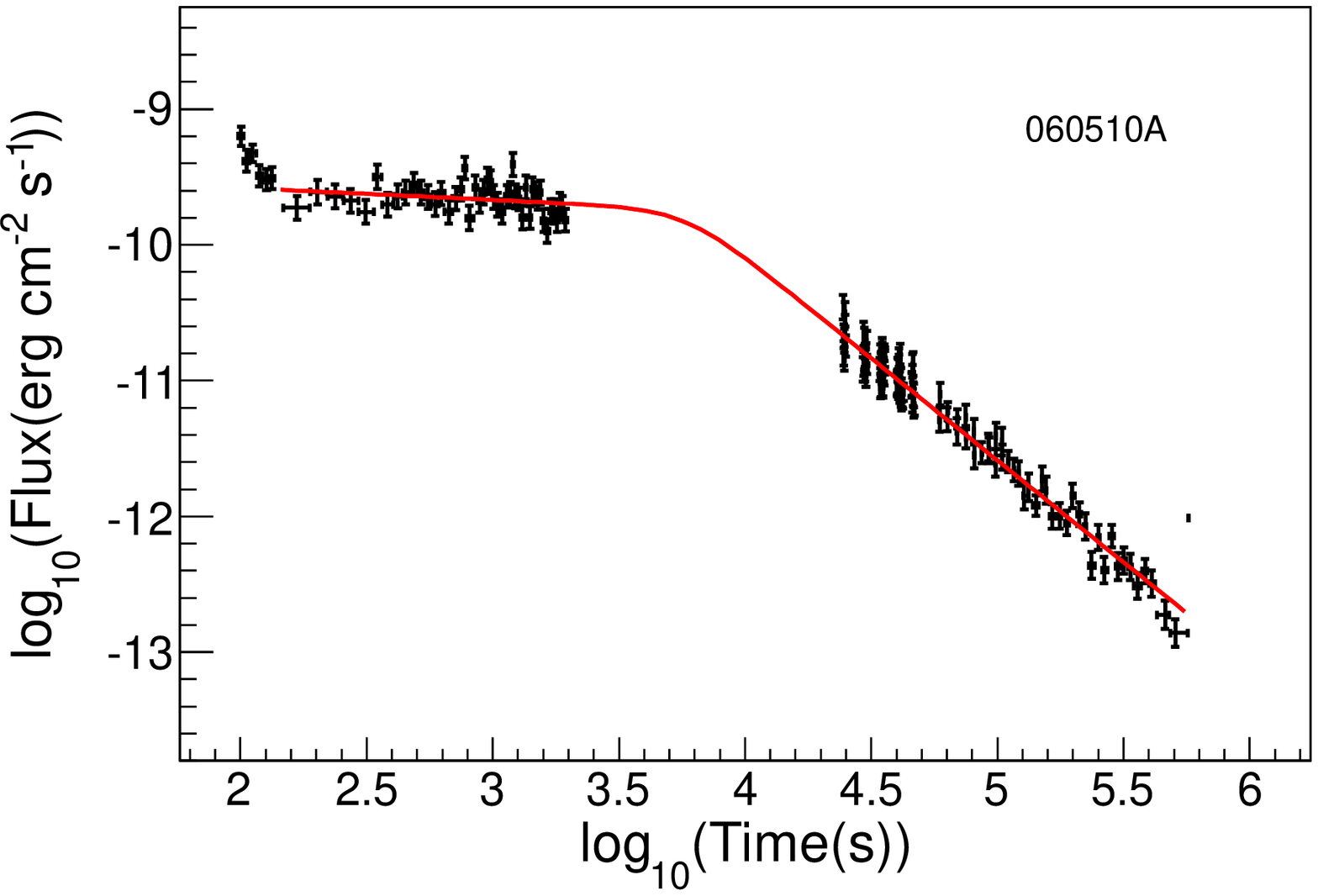}
\includegraphics[width=5.5cm,height=5cm]{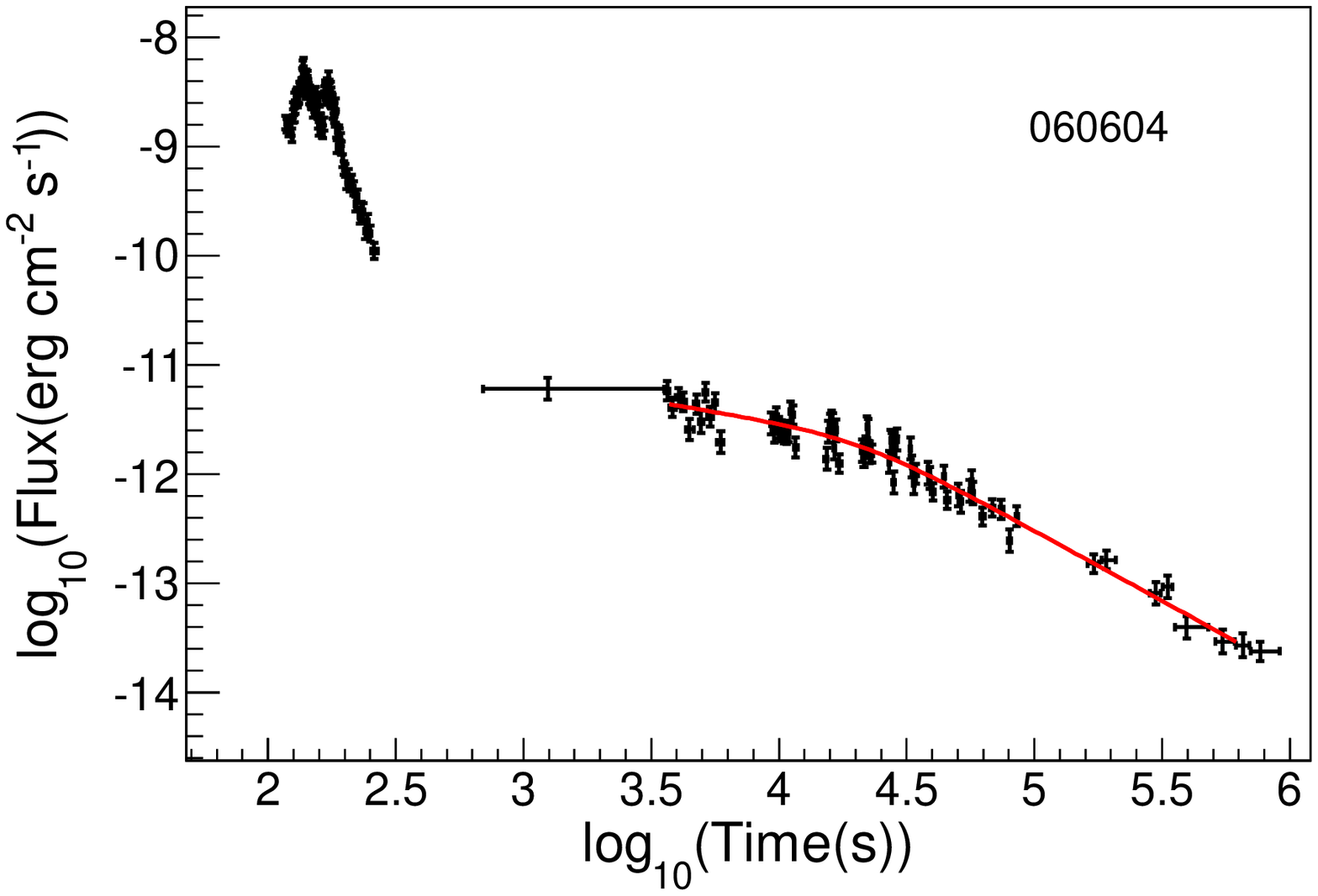}
\includegraphics[width=5.5cm,height=5cm]{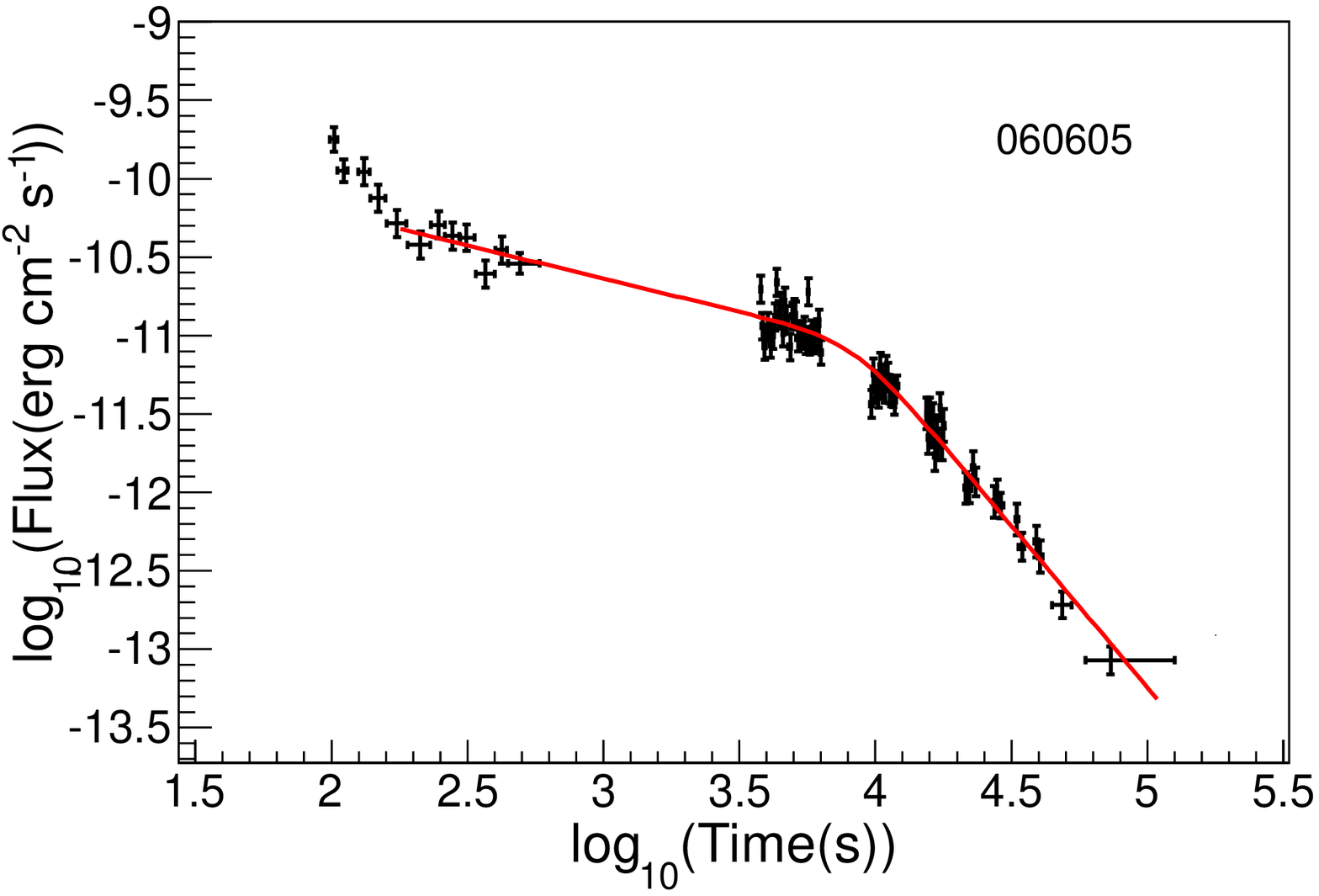}
\includegraphics[width=5.5cm,height=5cm]{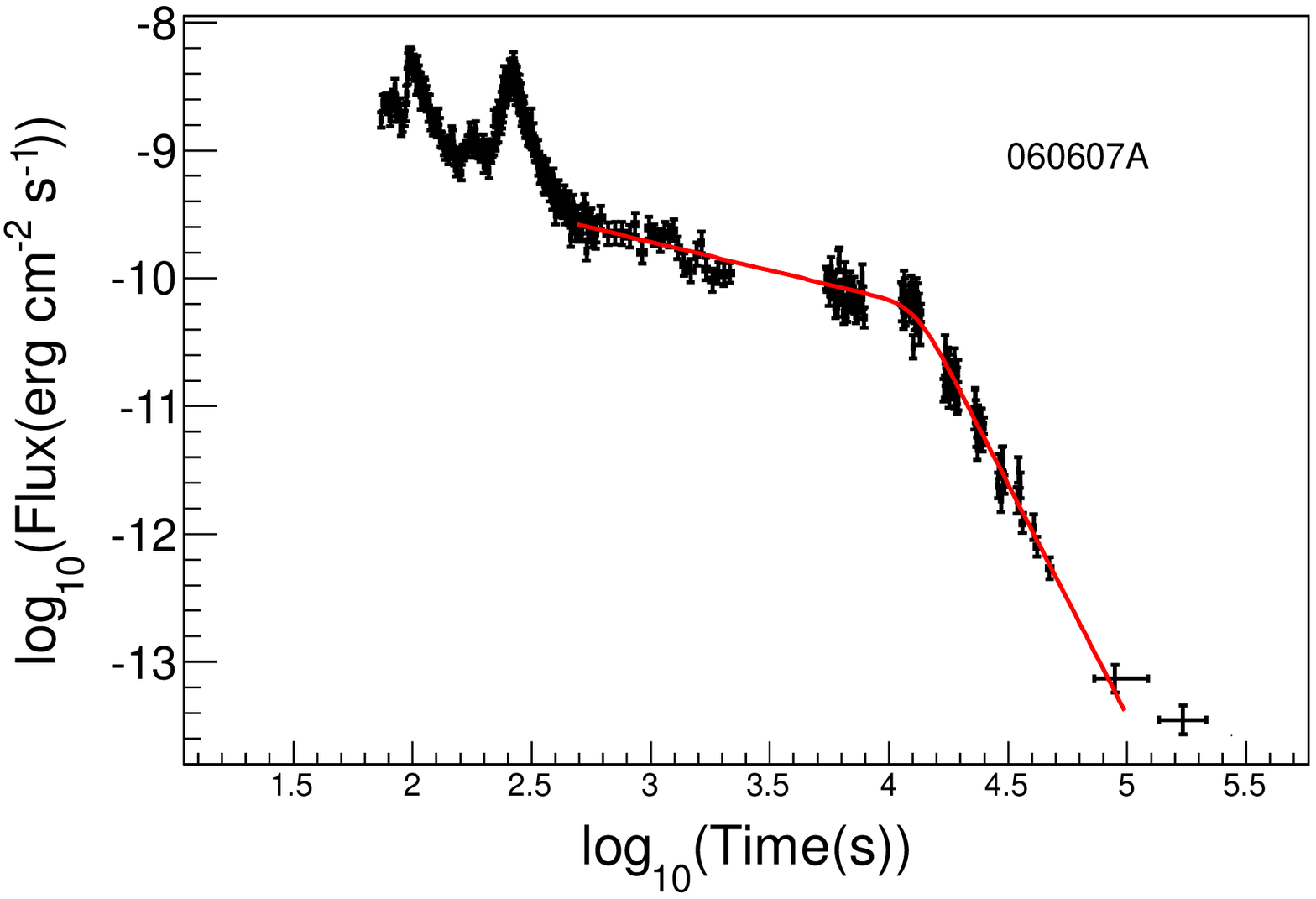}
\includegraphics[width=5.5cm,height=5cm]{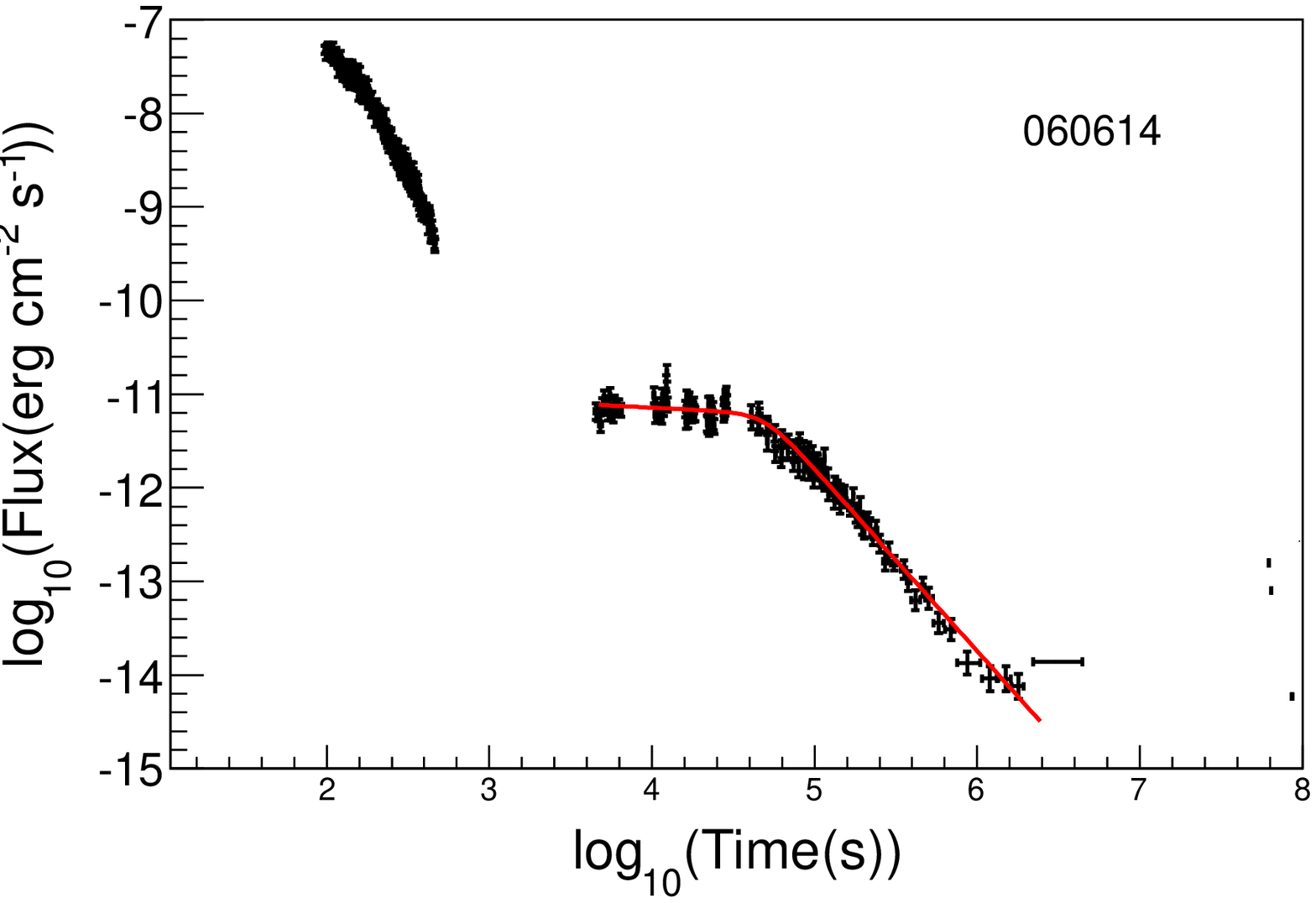}
\includegraphics[width=5.5cm,height=5cm]{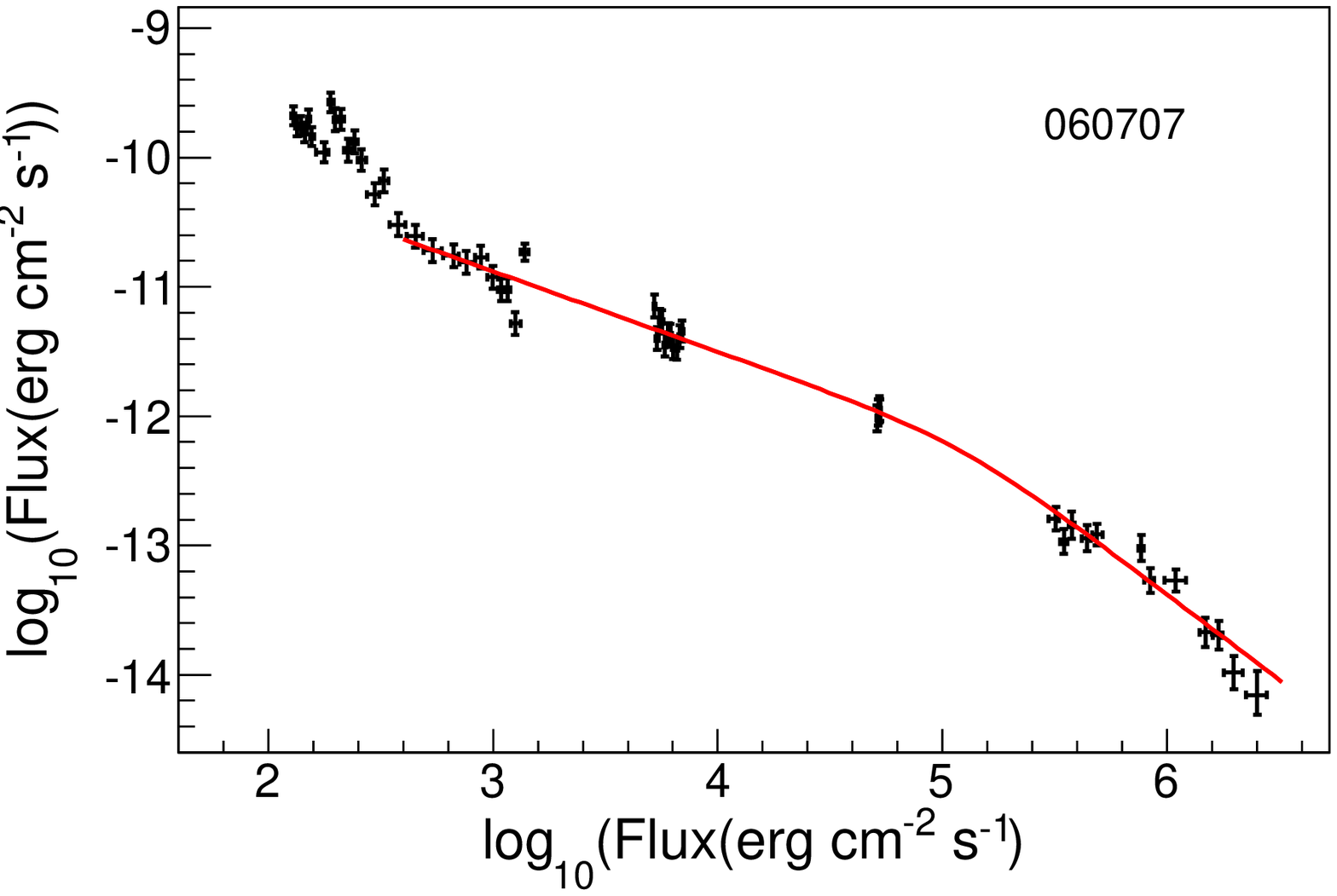}
\caption{ Continued.}
\label{fig-1-3}
\end{center}
\end{figure*}

\begin{figure*}
\begin{center}
\setlength{\abovecaptionskip}{0.cm}
\setlength{\belowcaptionskip}{-0.cm}
\figurenum{1}
\hspace{0cm}
\graphicspath{{lightcurve/}}
\includegraphics[width=5.5cm,height=5cm]{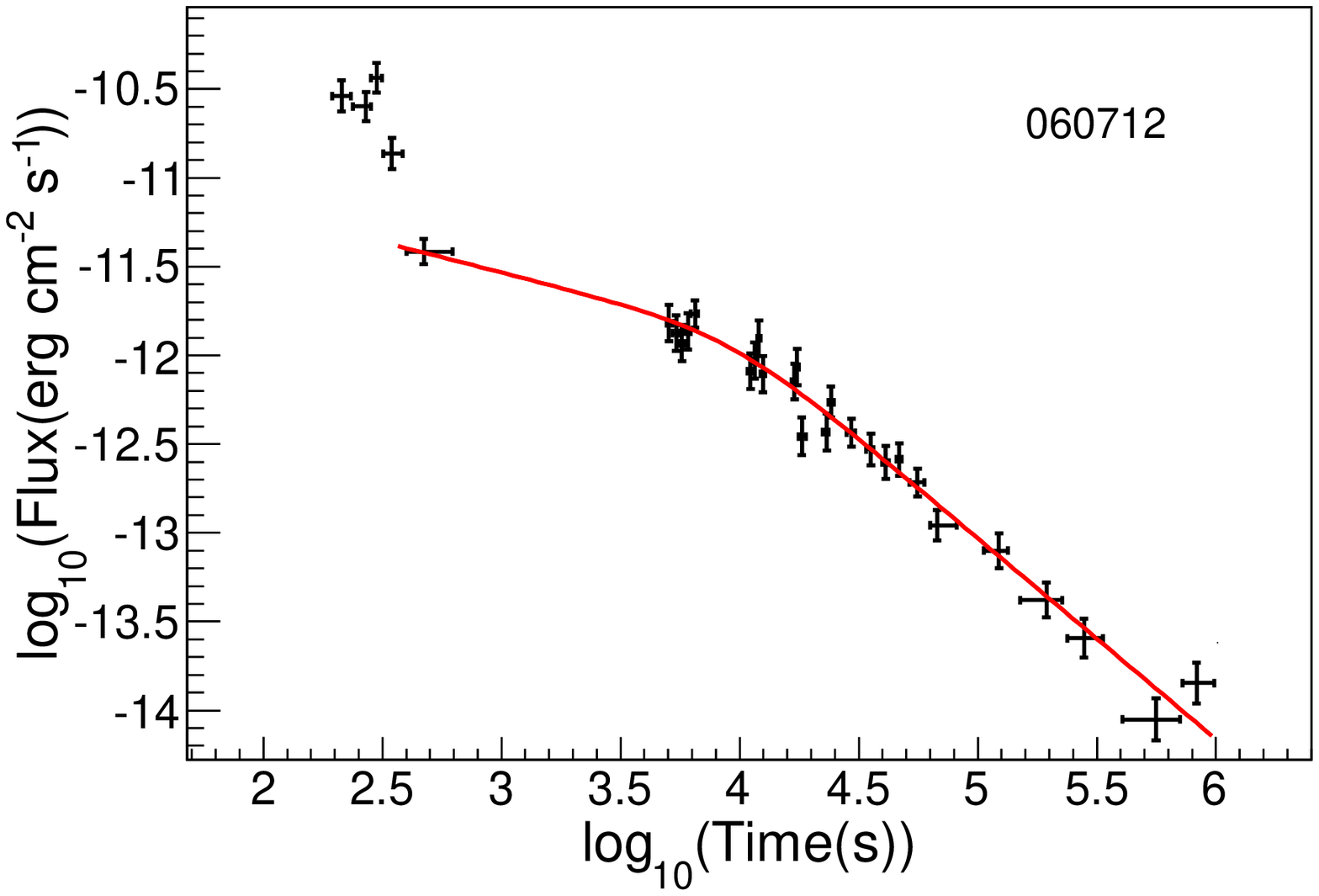}
\includegraphics[width=5.5cm,height=5cm]{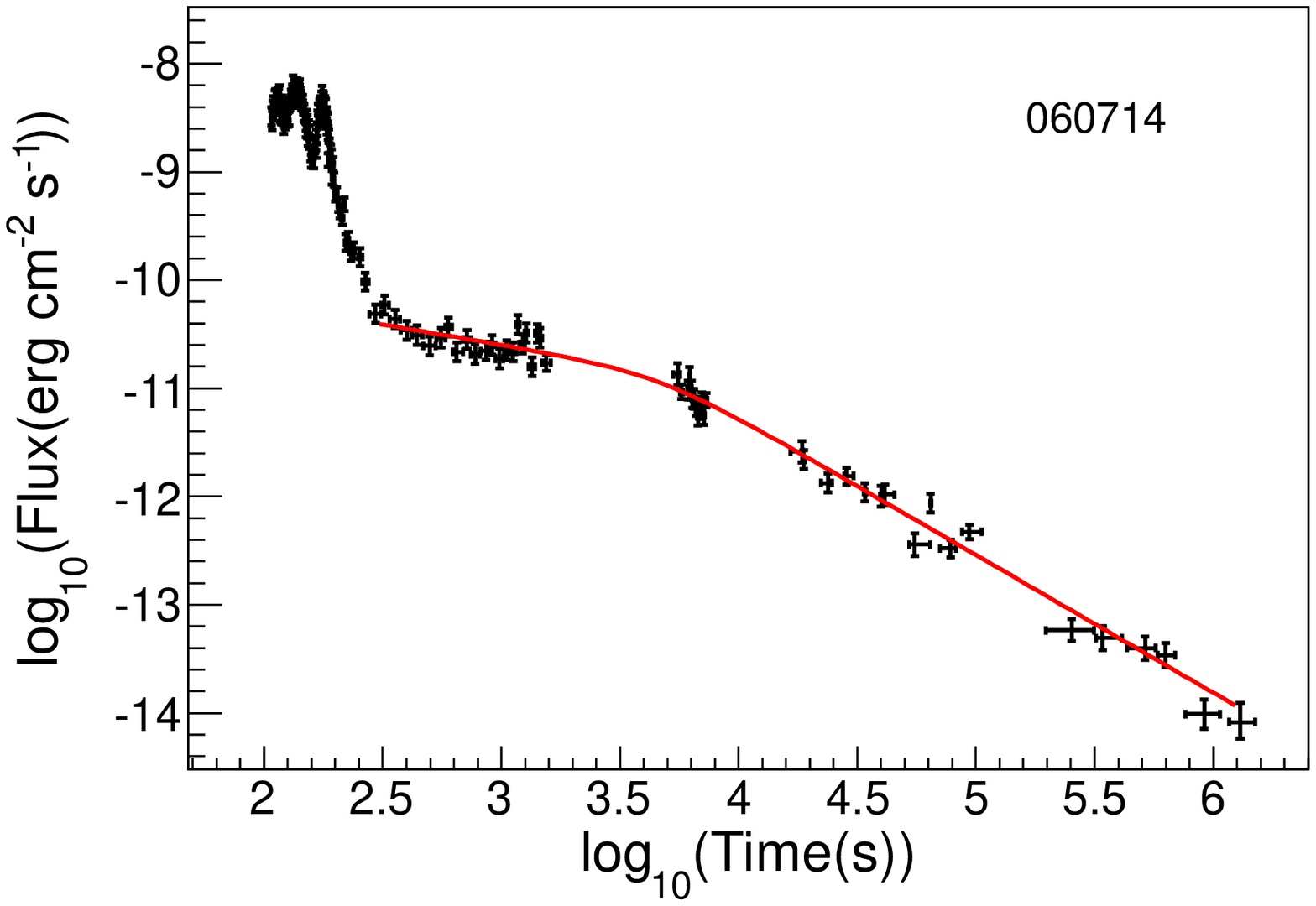}
\includegraphics[width=5.5cm,height=5cm]{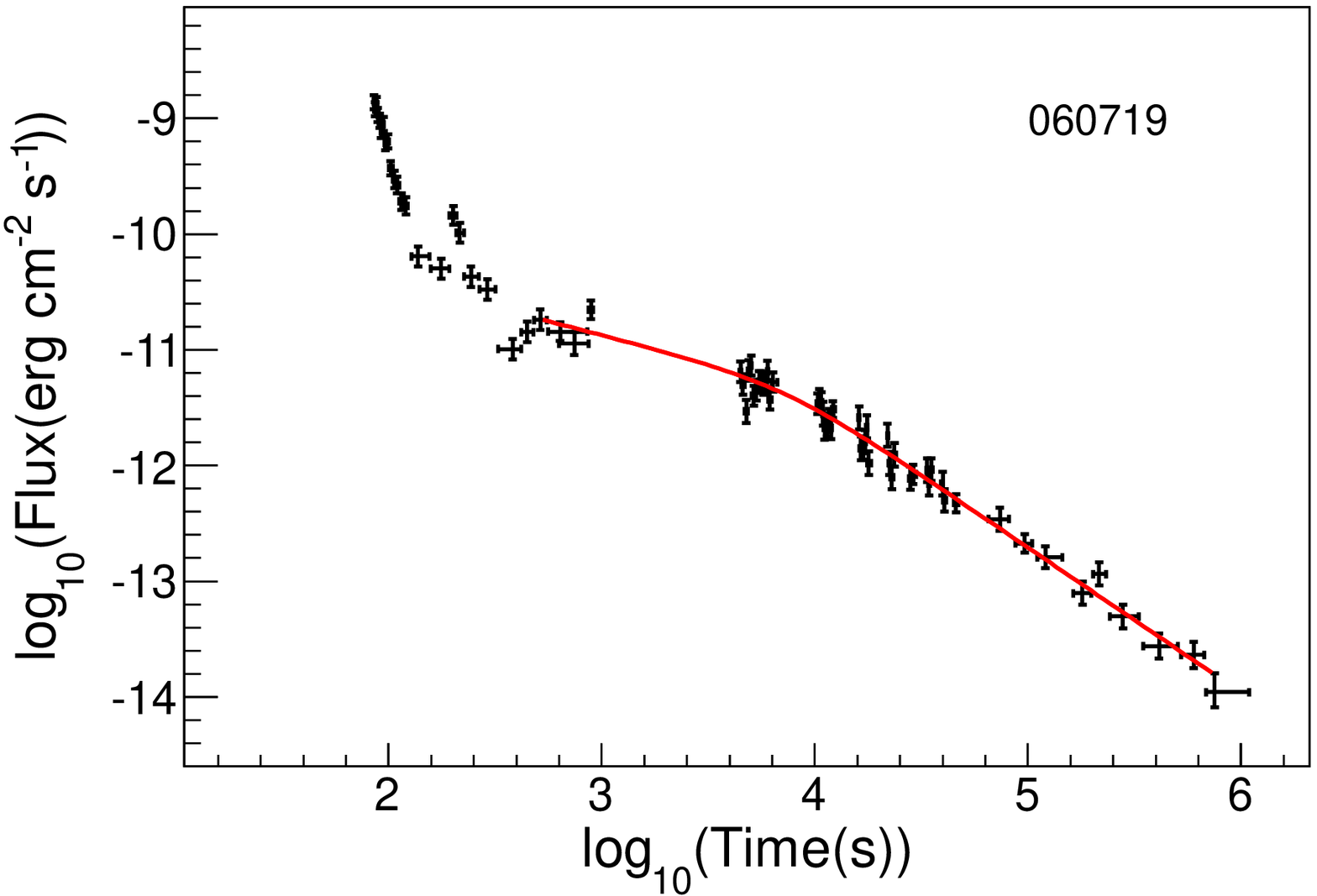}
\includegraphics[width=5.5cm,height=5cm]{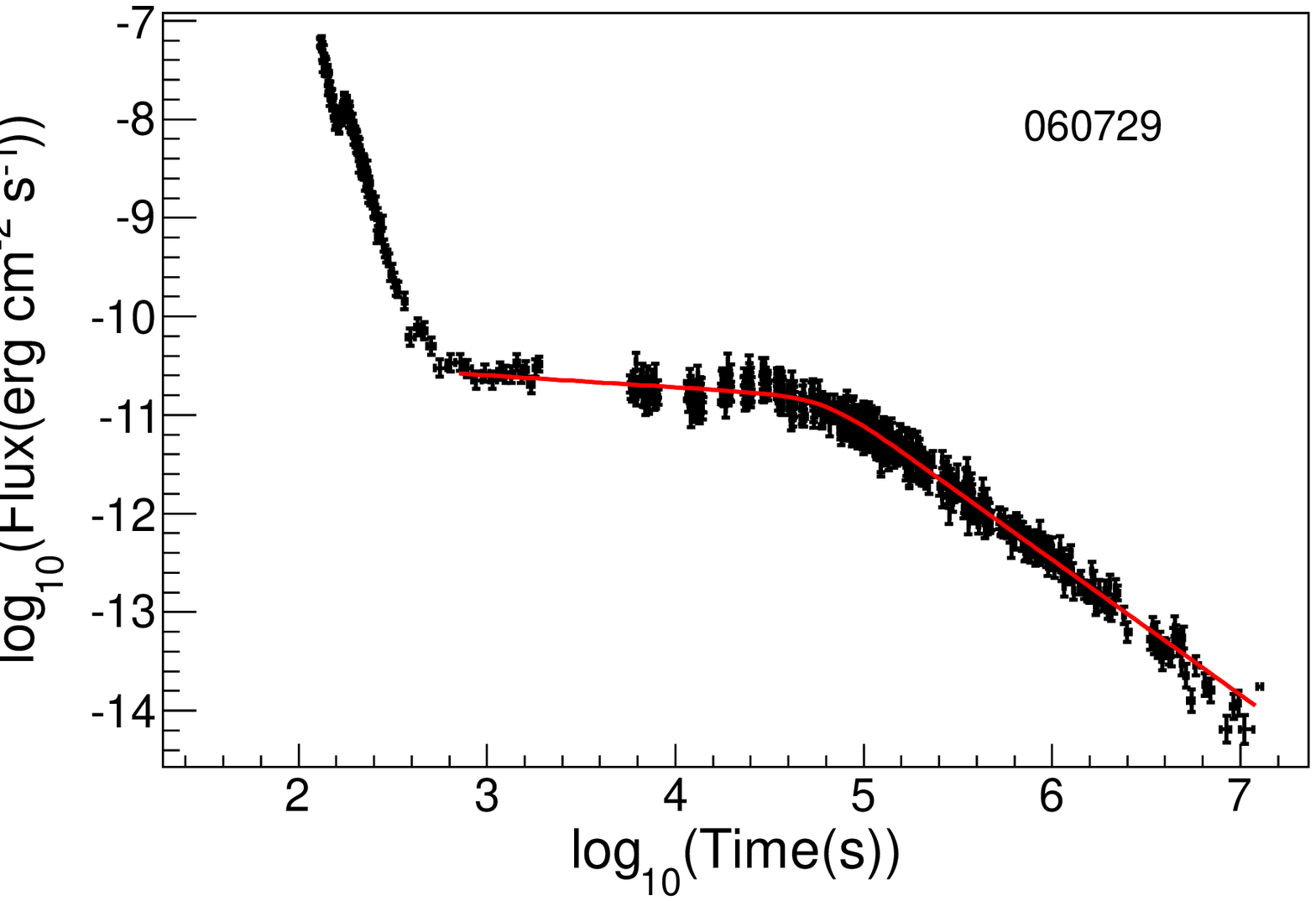}
\includegraphics[width=5.5cm,height=5cm]{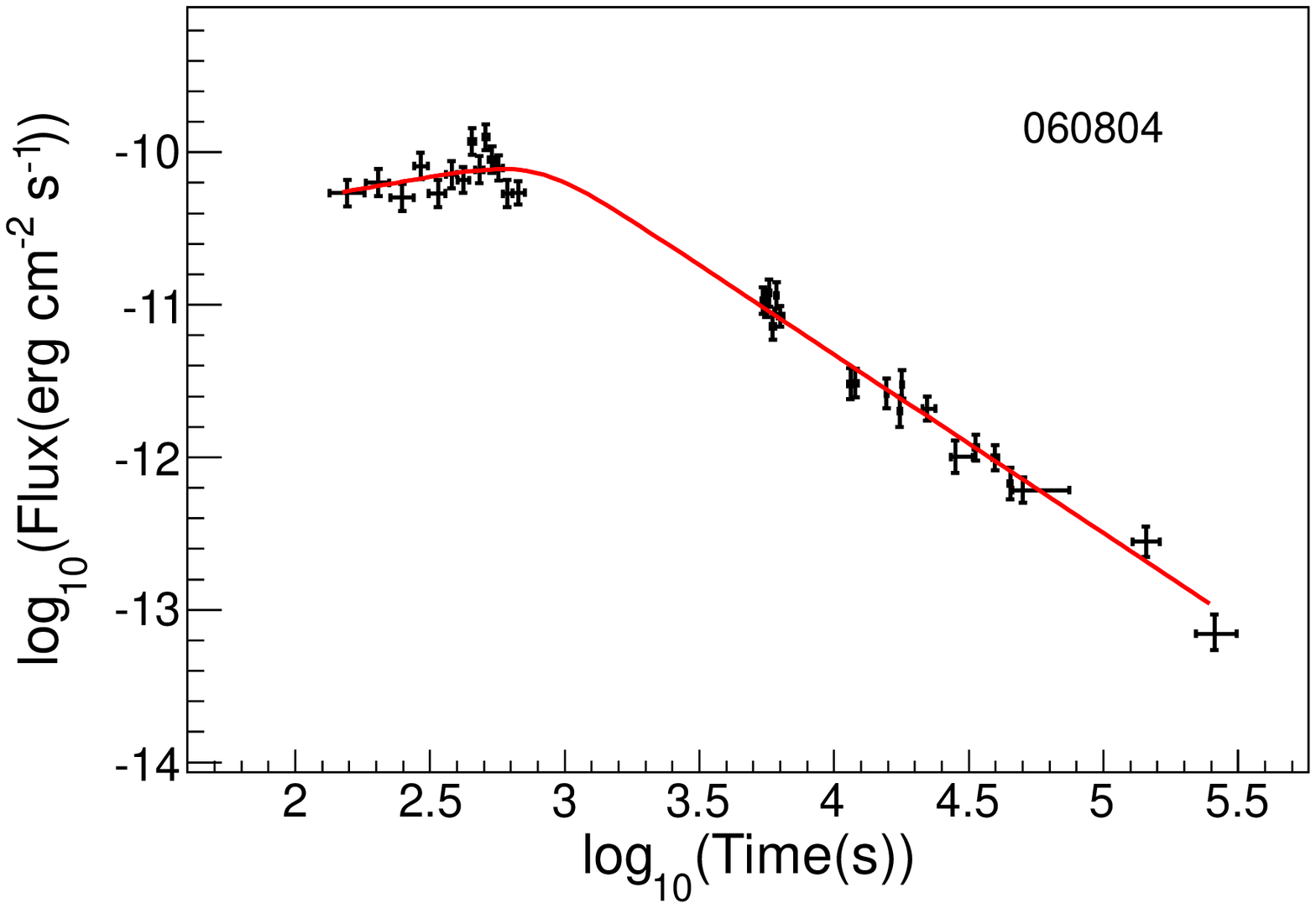}
\includegraphics[width=5.5cm,height=5cm]{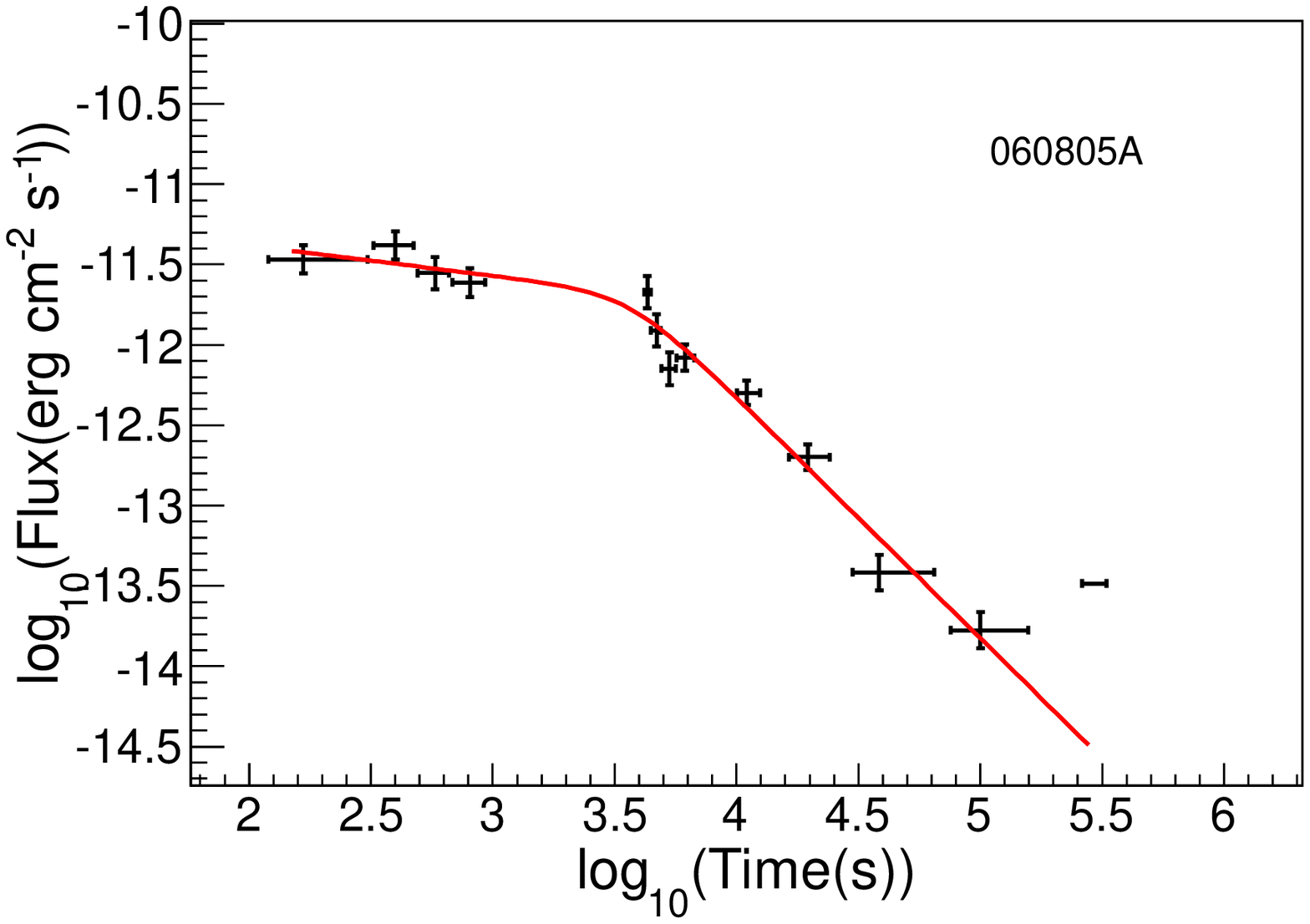}
\includegraphics[width=5.5cm,height=5cm]{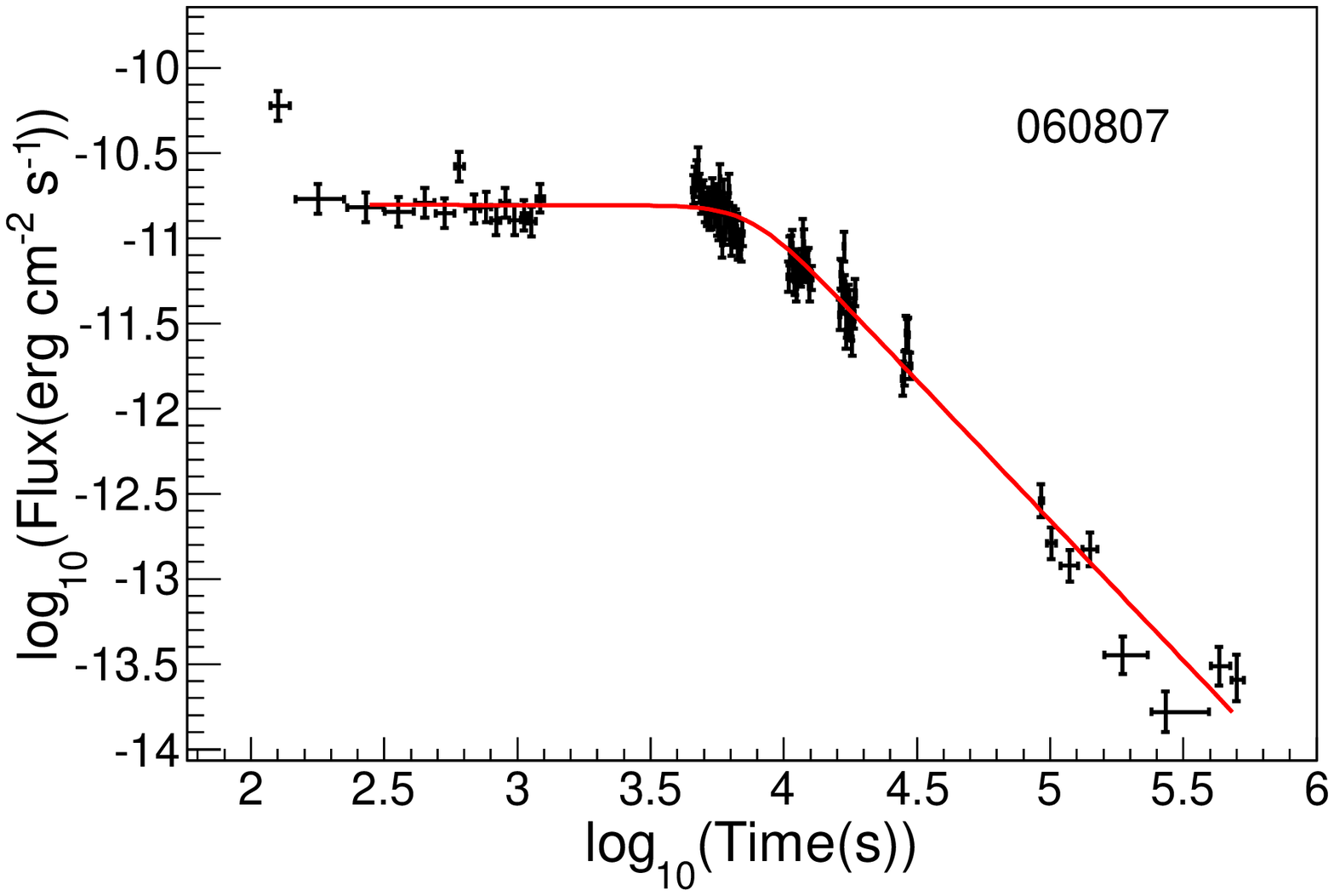}
\includegraphics[width=5.5cm,height=5cm]{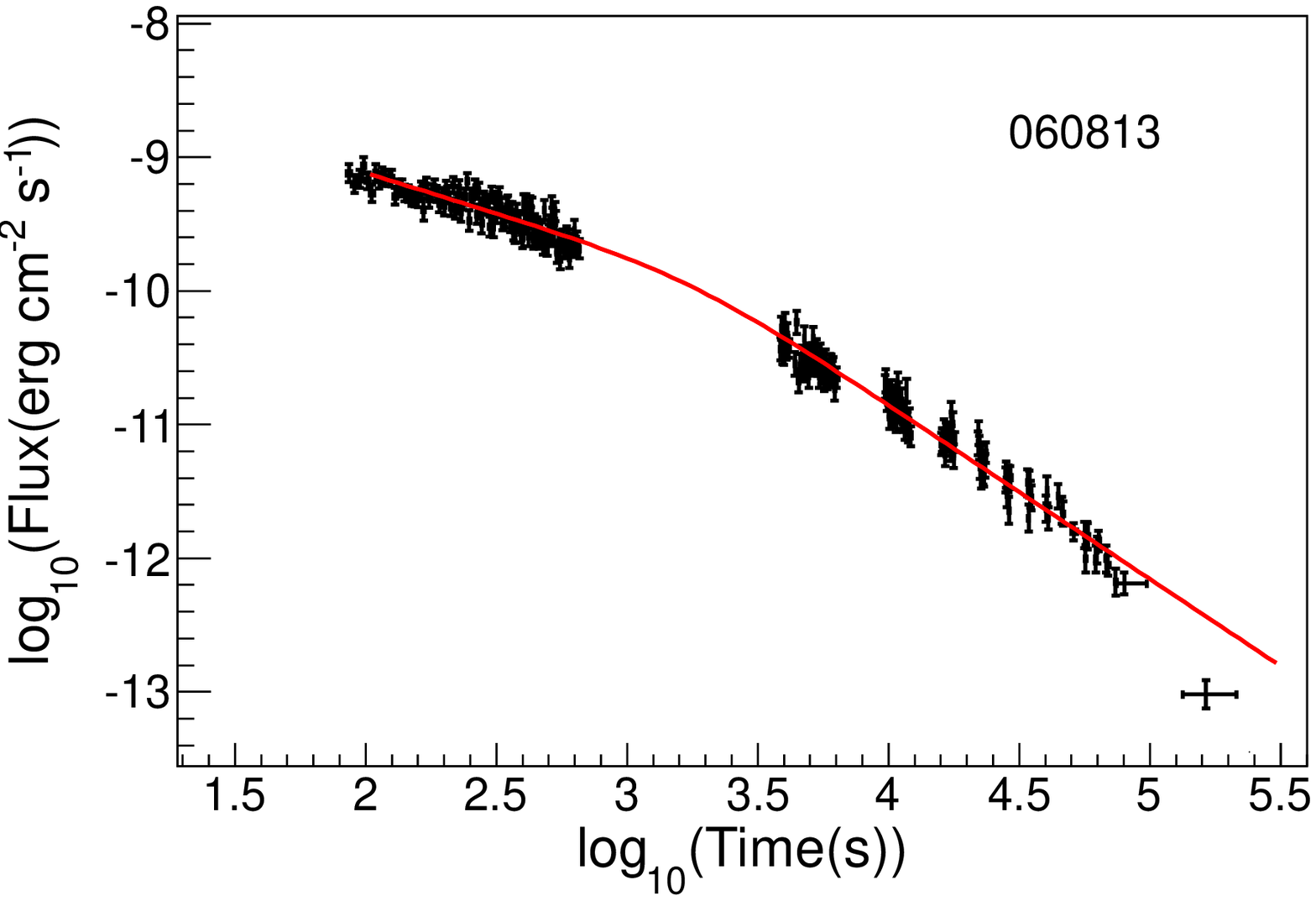}
\includegraphics[width=5.5cm,height=5cm]{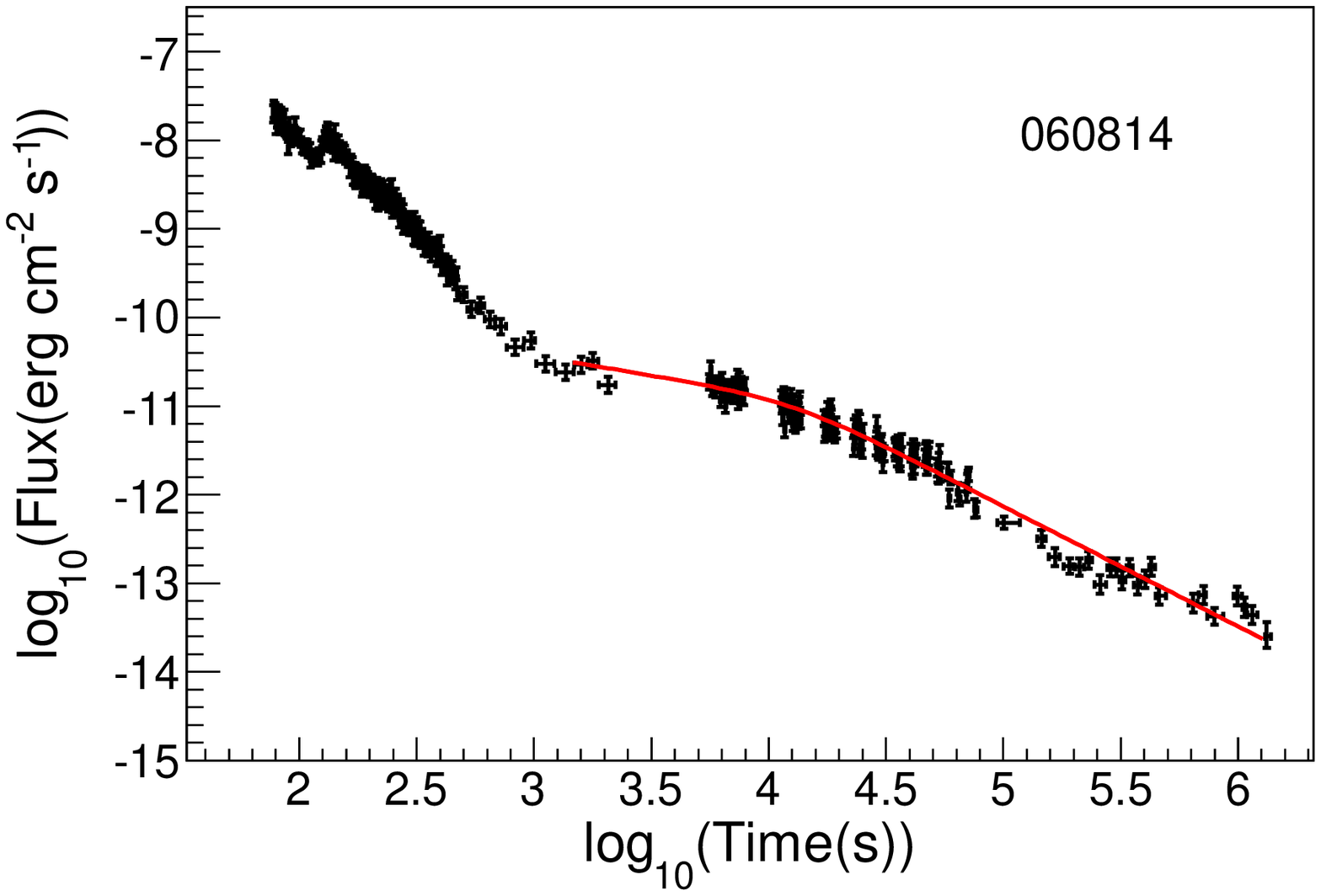}
\includegraphics[width=5.5cm,height=5cm]{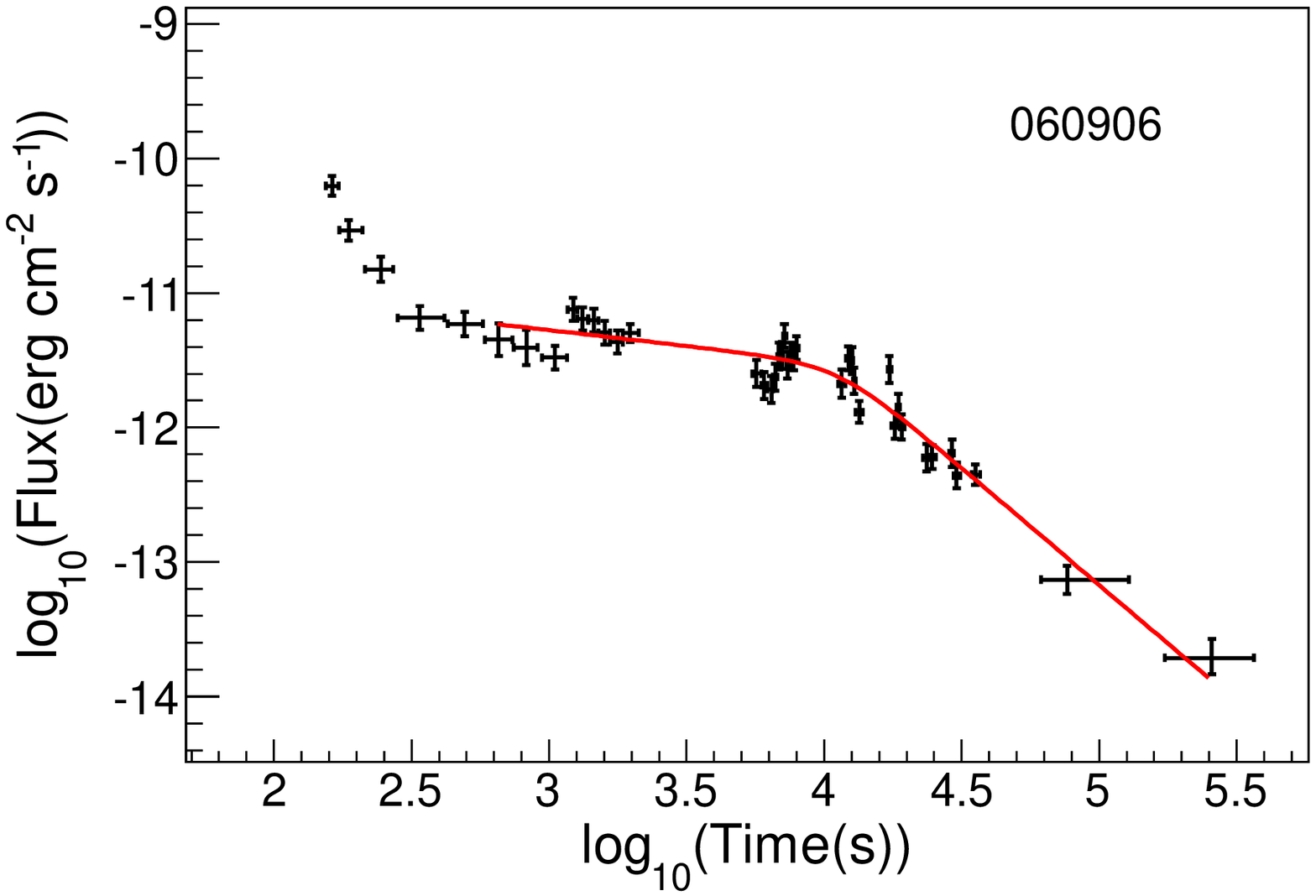}
\includegraphics[width=5.5cm,height=5cm]{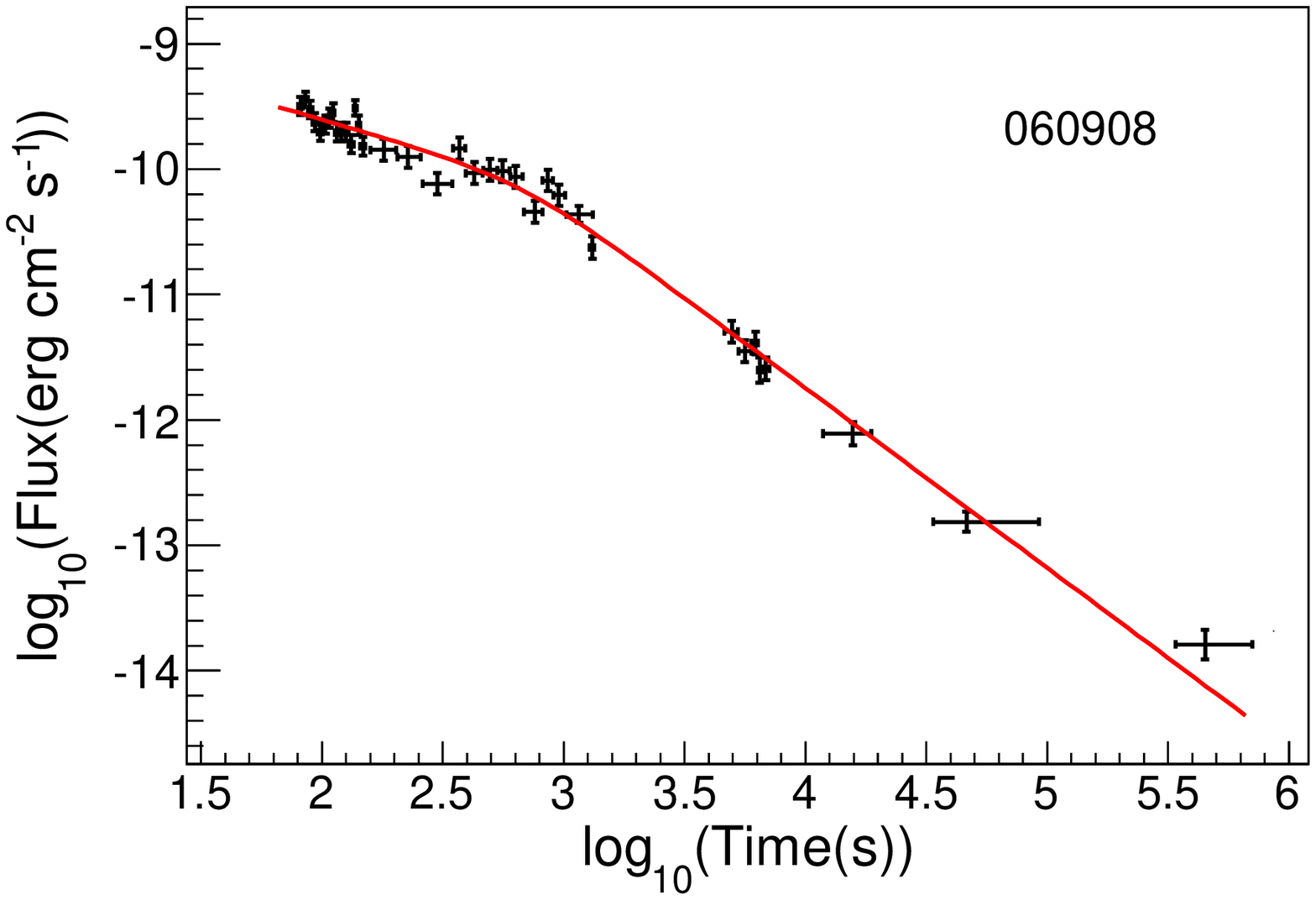}
\includegraphics[width=5.5cm,height=5cm]{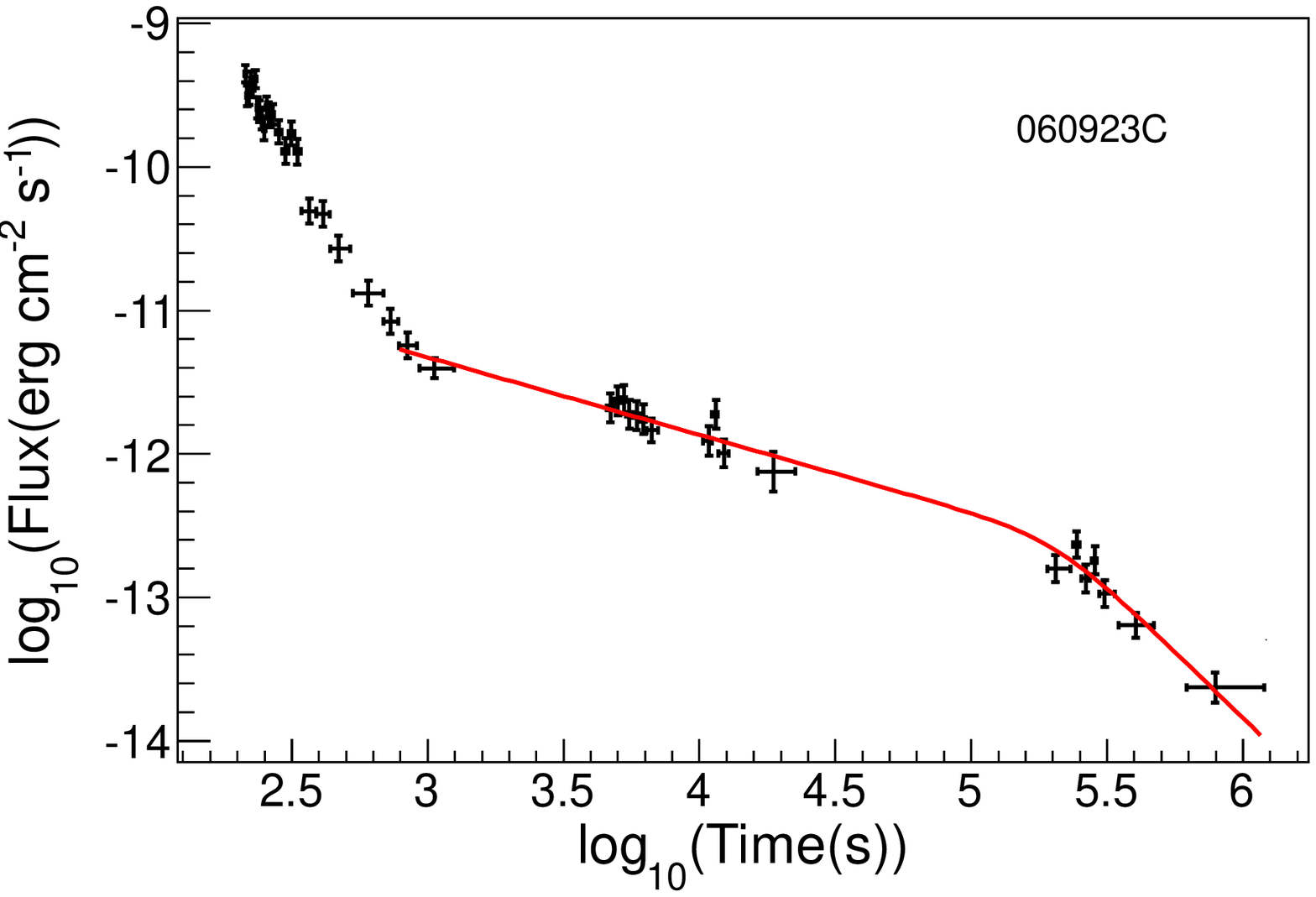}
\caption{ Continued.}
\label{fig-1-4}
\end{center}
\end{figure*}

\begin{figure*}
\begin{center}
\setlength{\abovecaptionskip}{0.cm}
\setlength{\belowcaptionskip}{-0.cm}
\figurenum{1}
\hspace{0cm}
\graphicspath{{lightcurve/}}
\includegraphics[width=5.5cm,height=5cm]{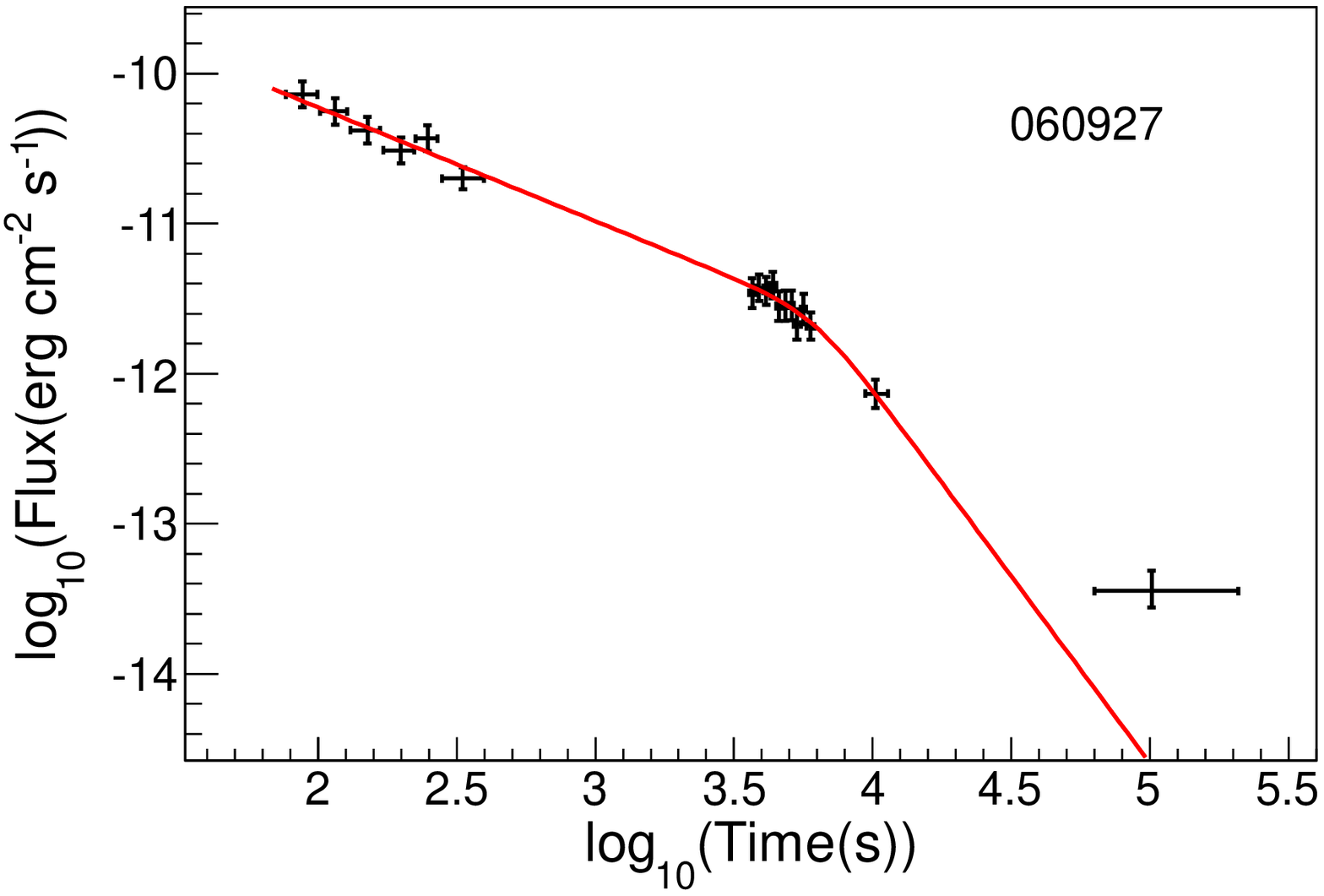}
\includegraphics[width=5.5cm,height=5cm]{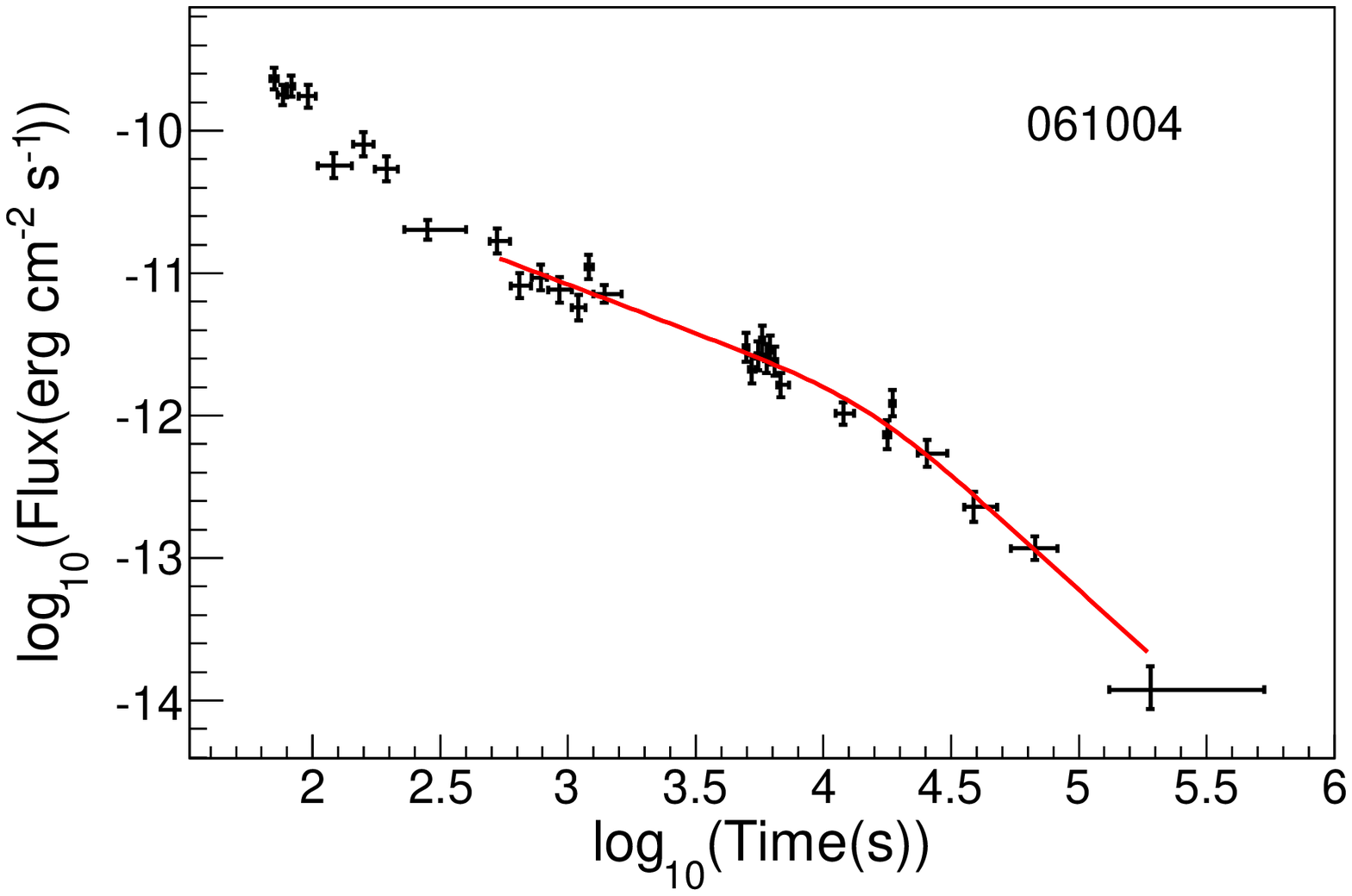}
\includegraphics[width=5.5cm,height=5cm]{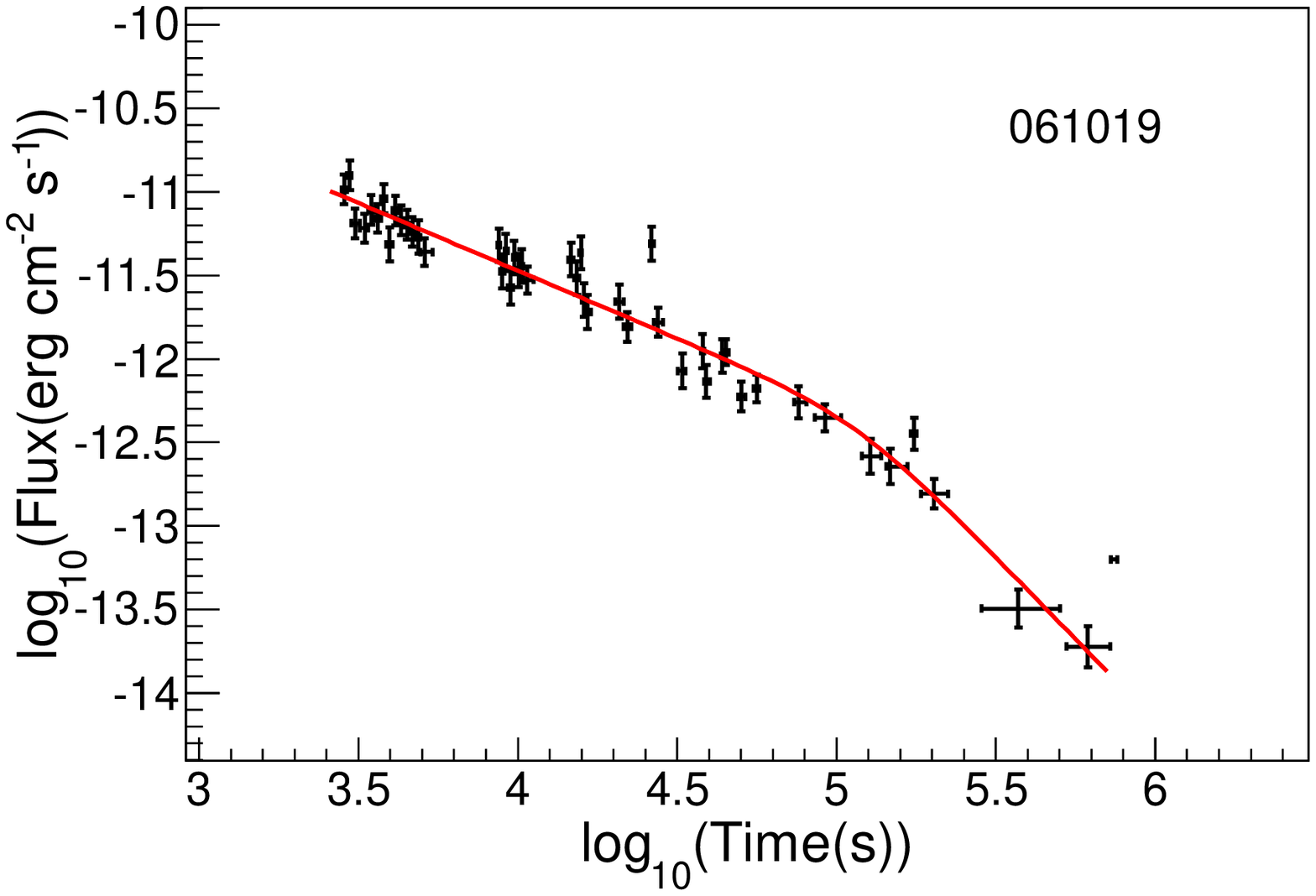}
\includegraphics[width=5.5cm,height=5cm]{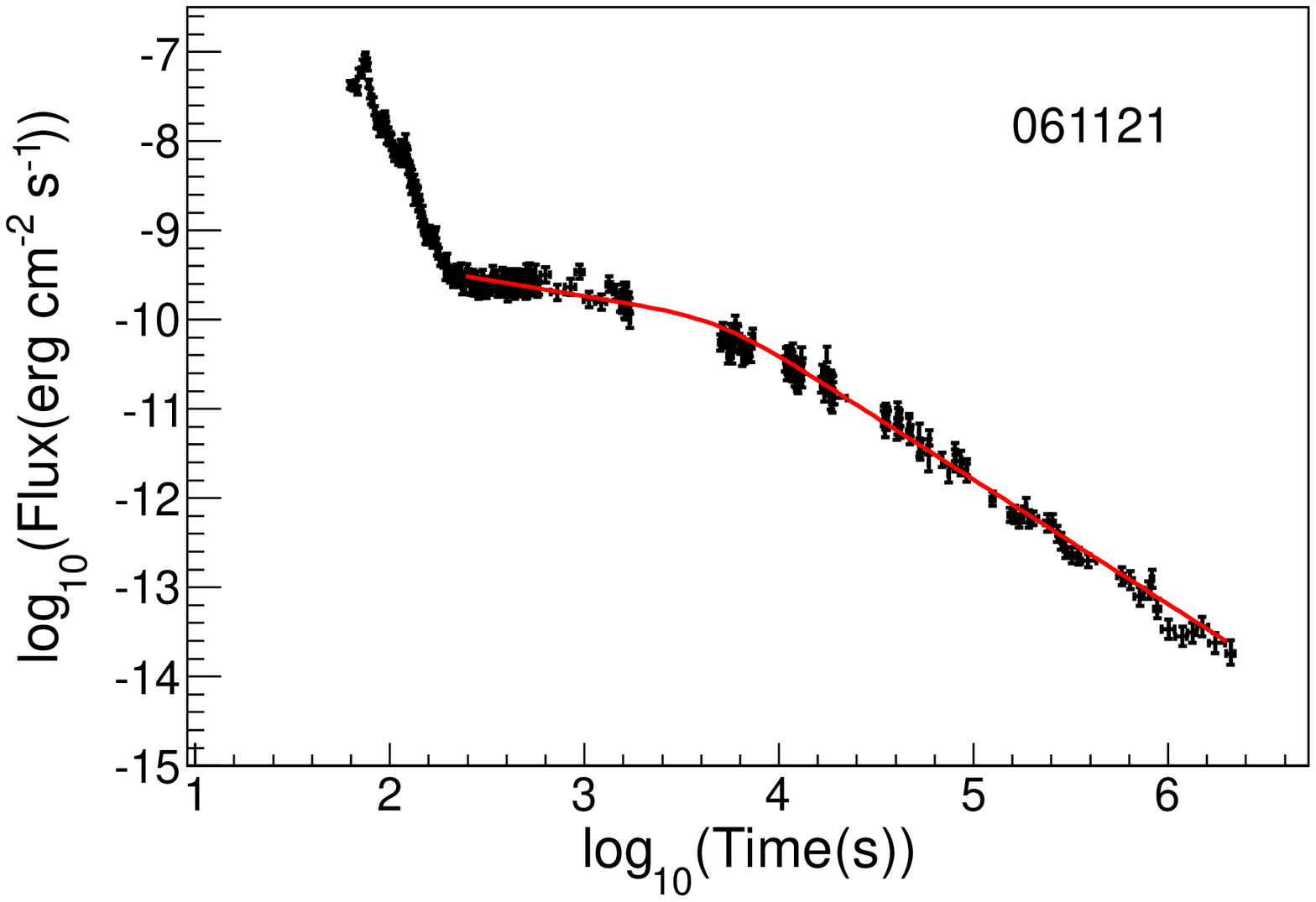}
\includegraphics[width=5.5cm,height=5cm]{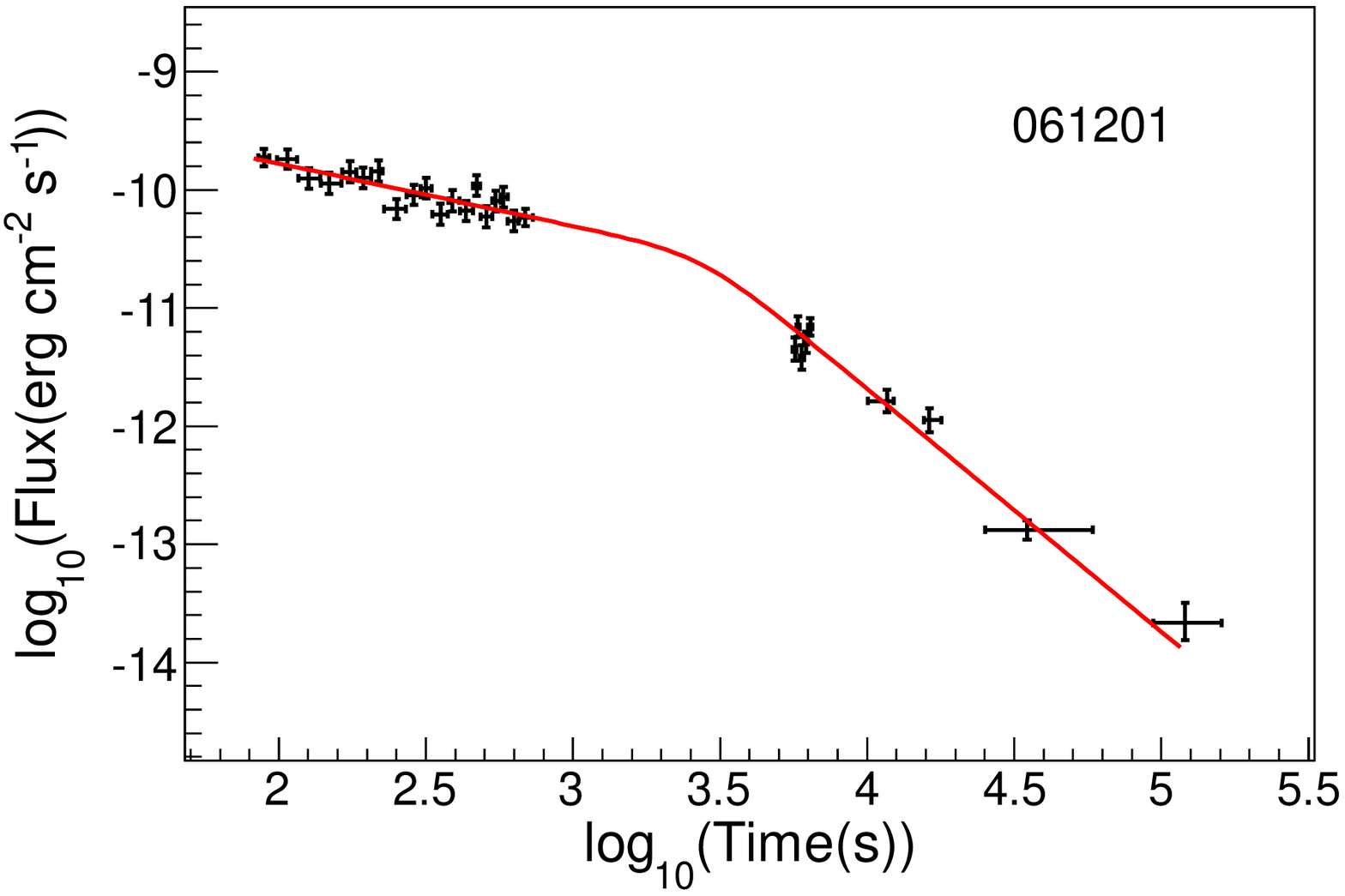}
\includegraphics[width=5.5cm,height=5cm]{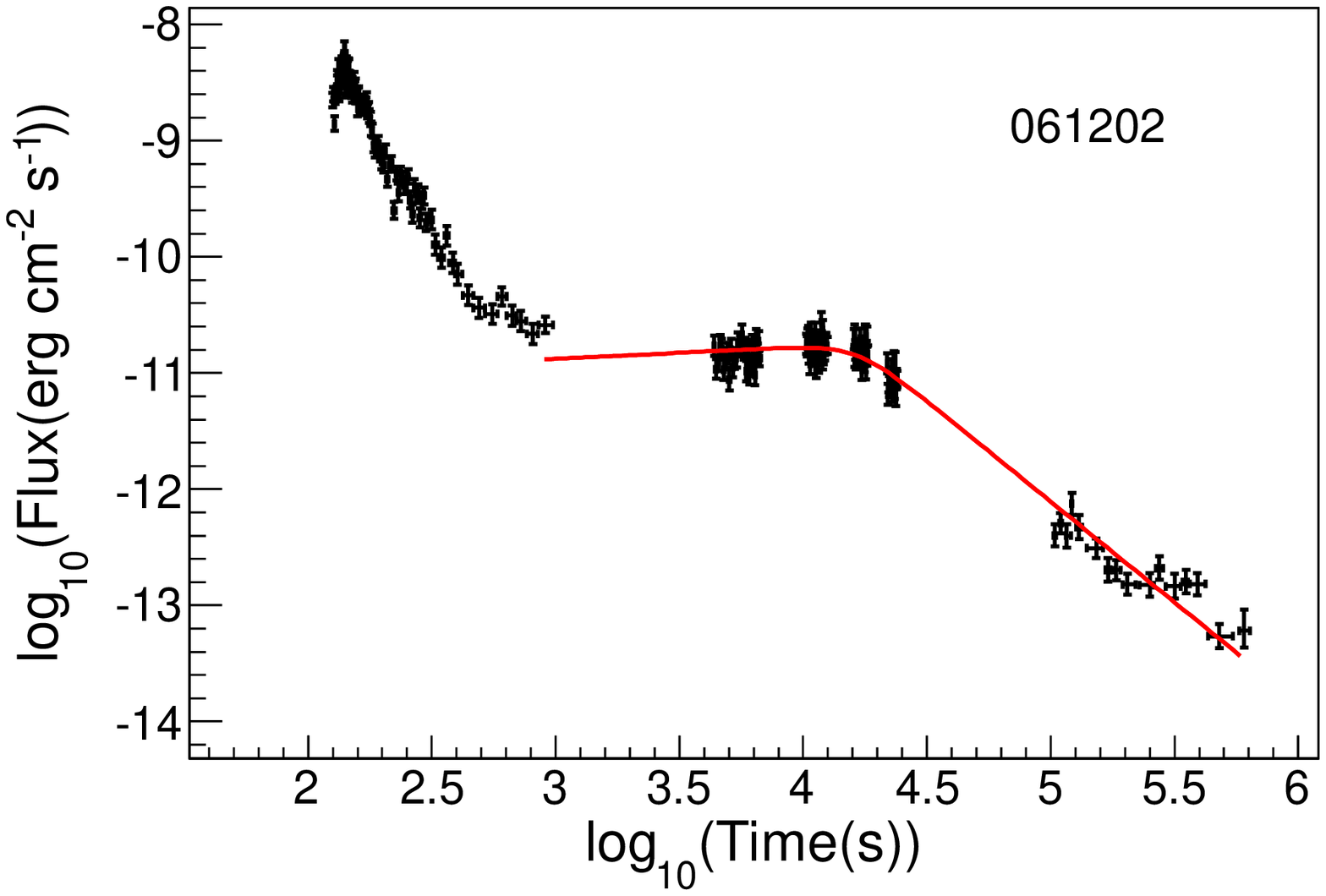}
\includegraphics[width=5.5cm,height=5cm]{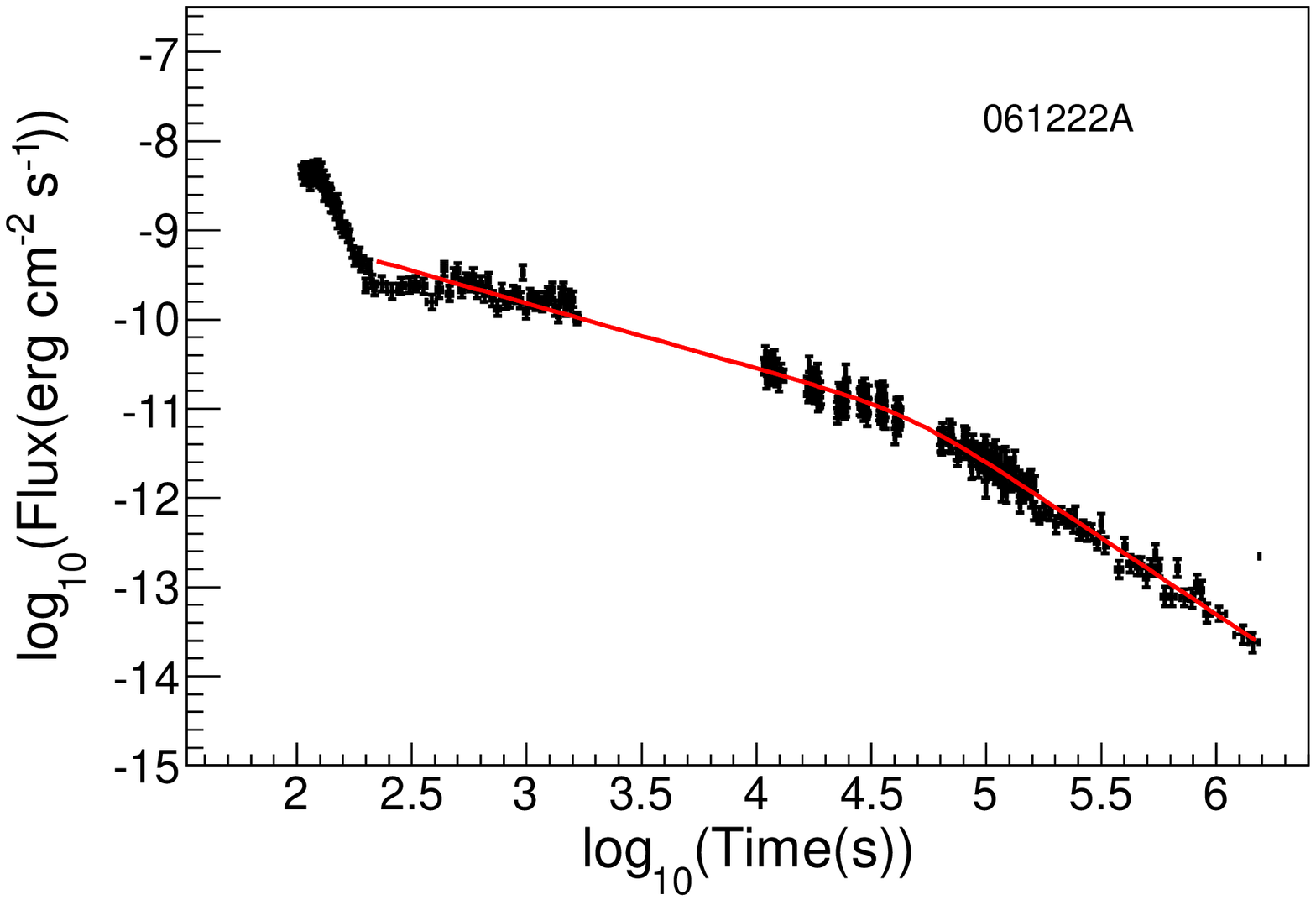}
\includegraphics[width=5.5cm,height=5cm]{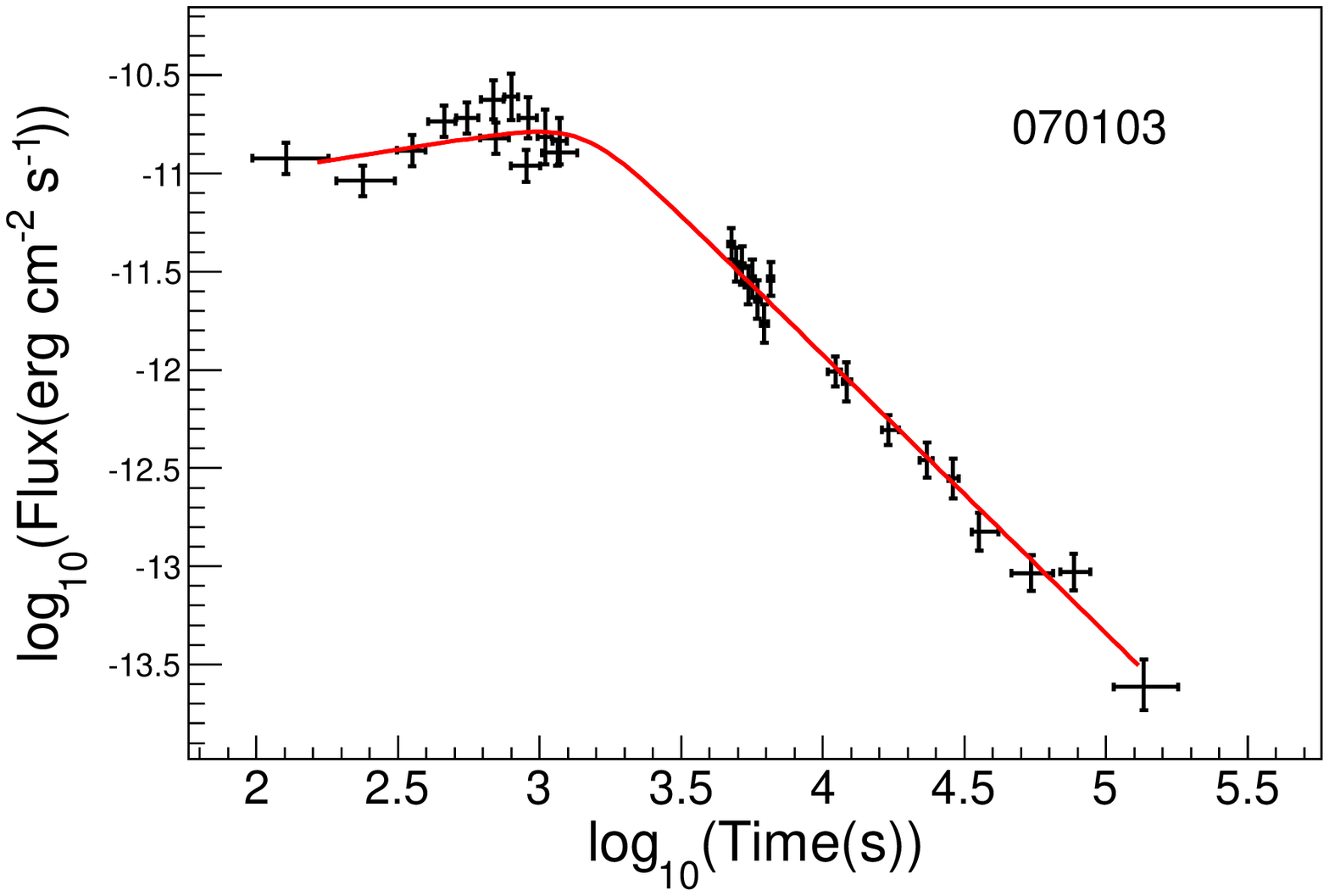}
\includegraphics[width=5.5cm,height=5cm]{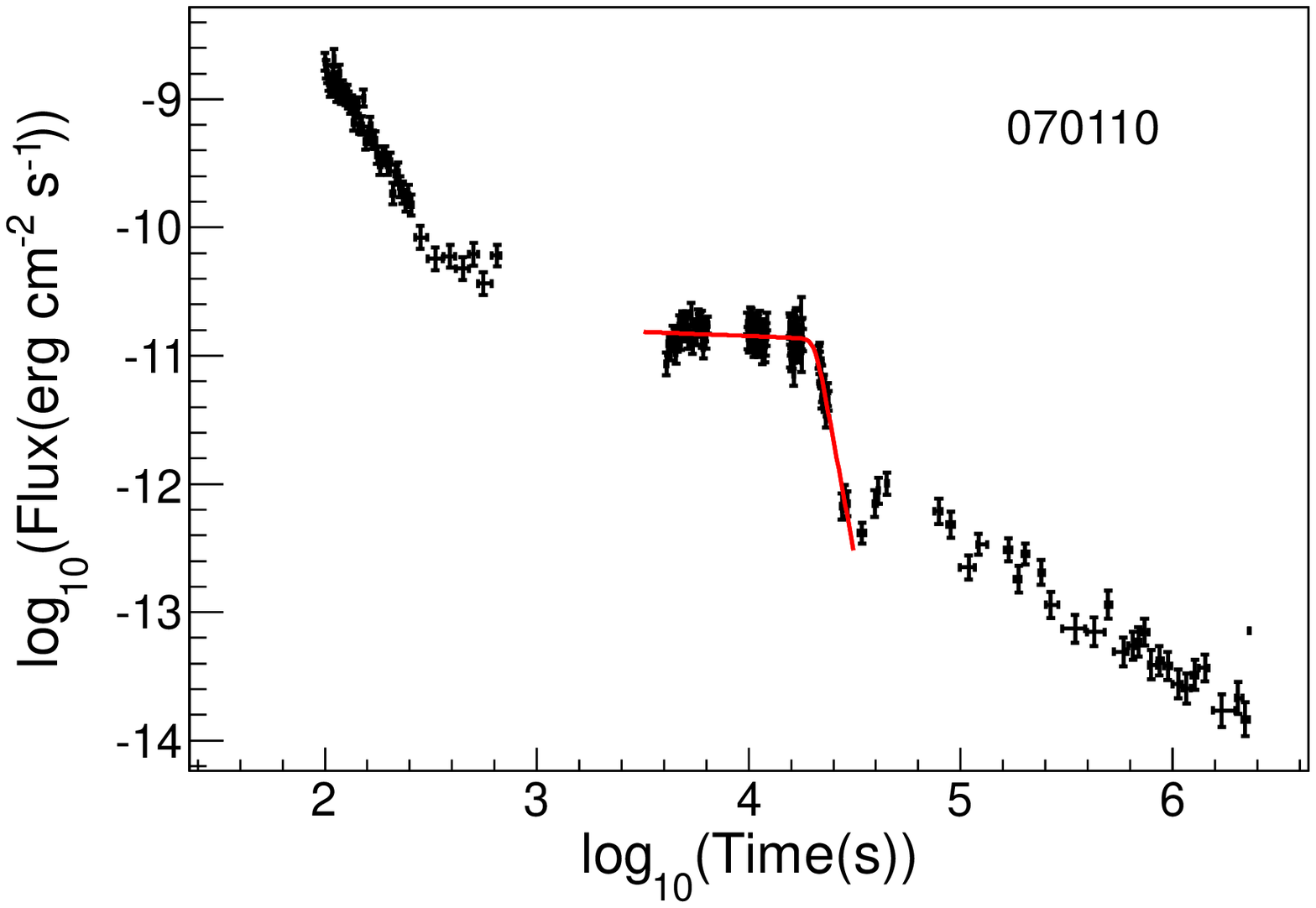}
\includegraphics[width=5.5cm,height=5cm]{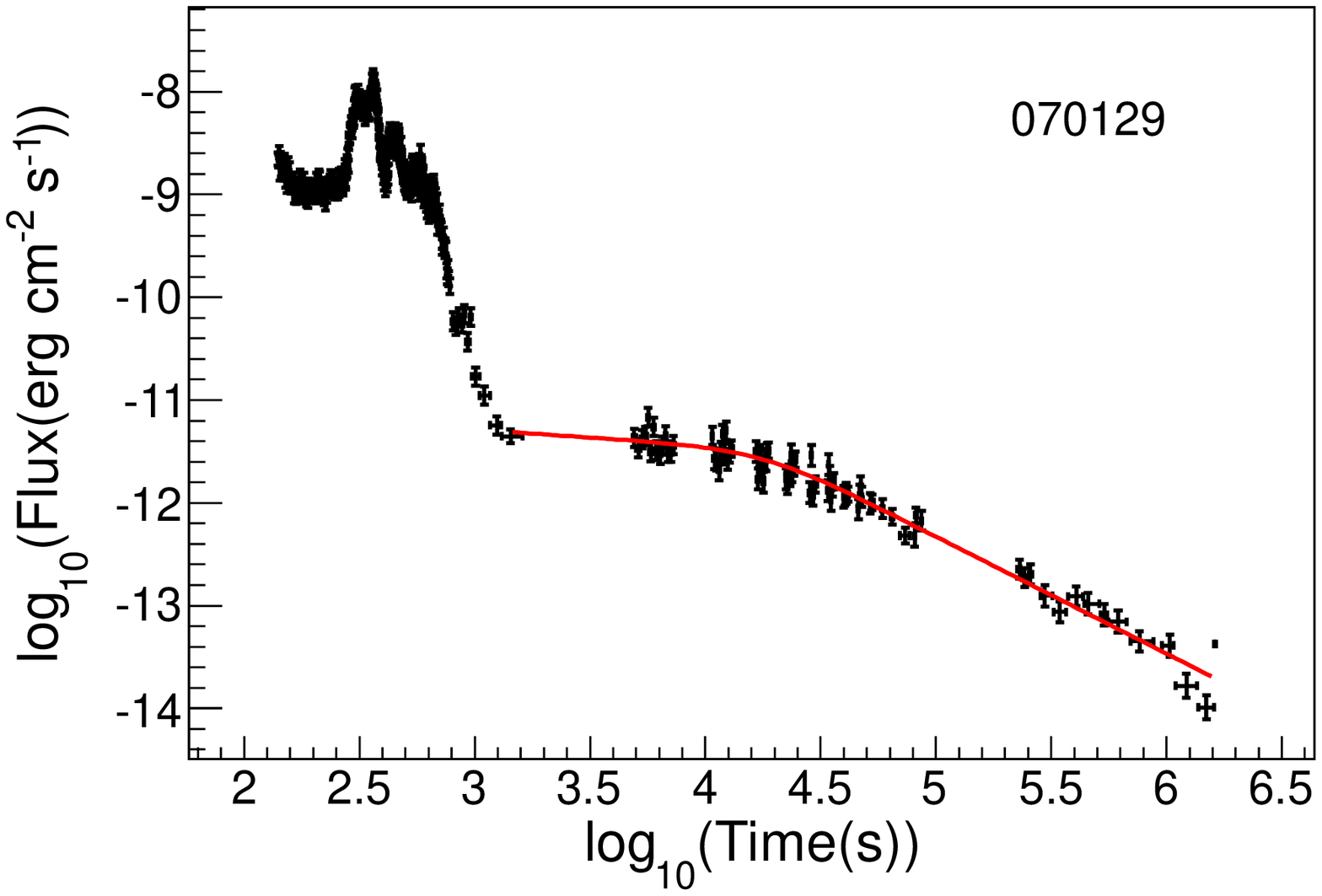}
\includegraphics[width=5.5cm,height=5cm]{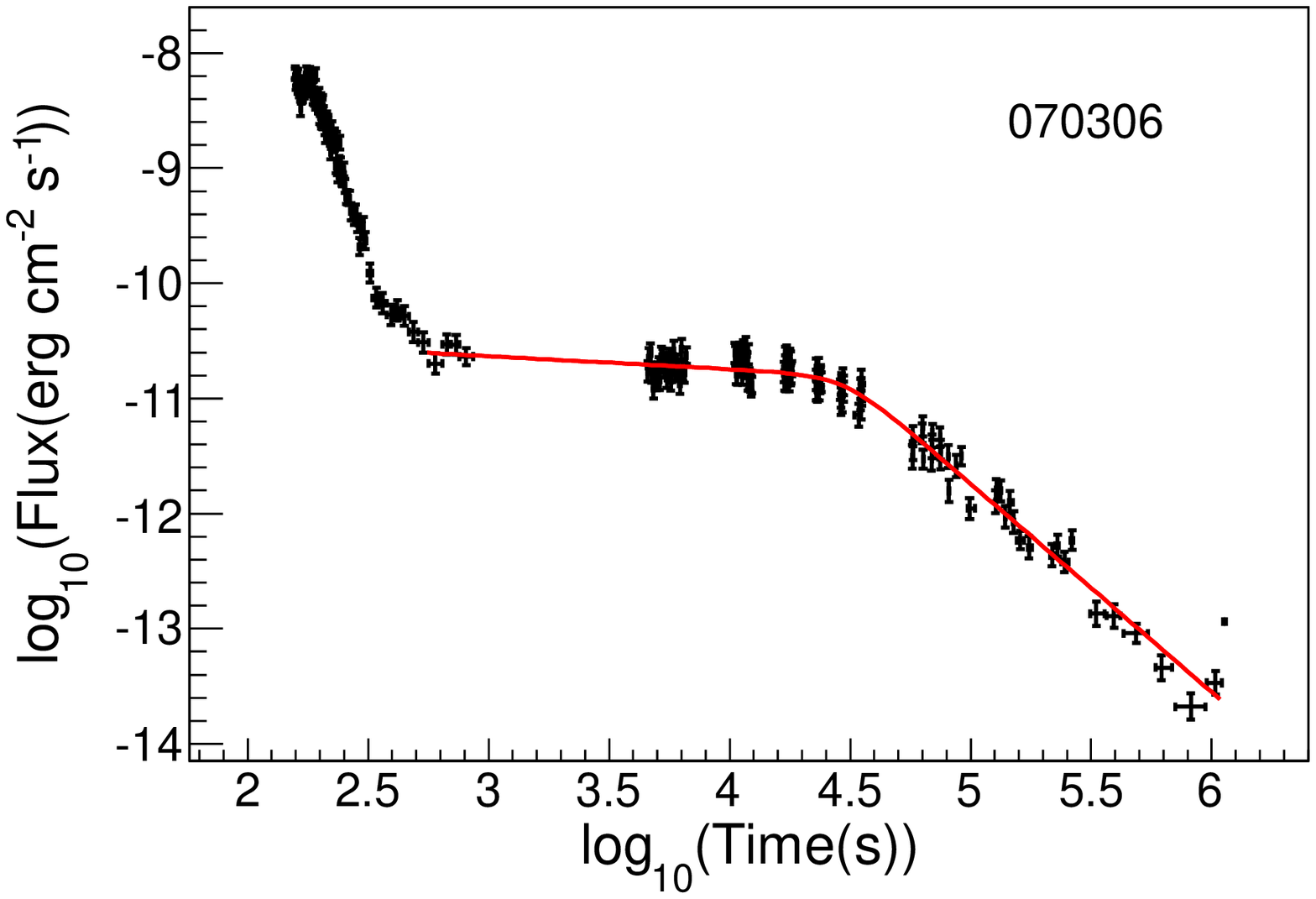}
\includegraphics[width=5.5cm,height=5cm]{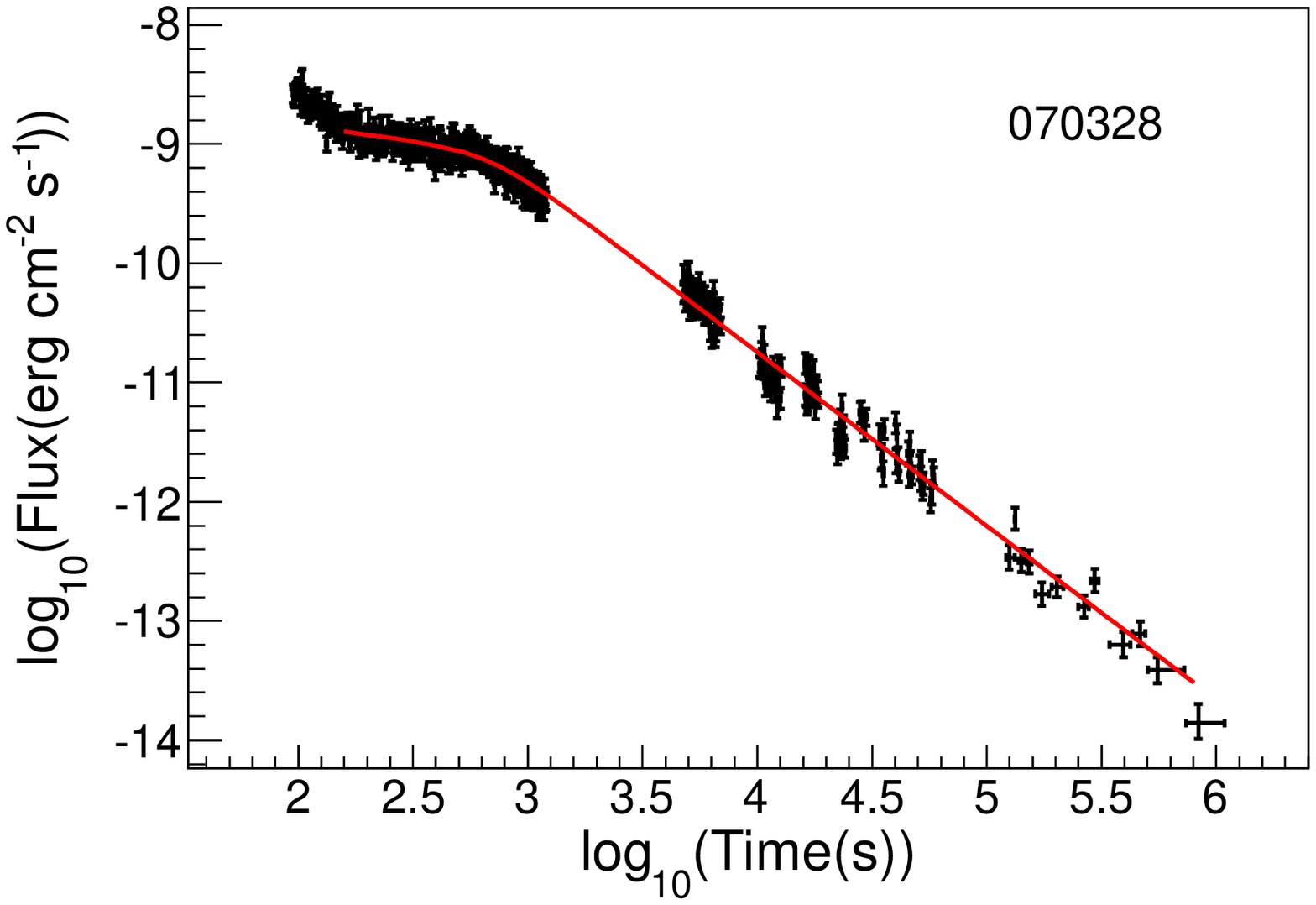}
\caption{ Continued.}
\label{fig-1-5}
\end{center}
\end{figure*}

\begin{figure*}
\begin{center}
\setlength{\abovecaptionskip}{0.cm}
\setlength{\belowcaptionskip}{-0.cm}
\figurenum{1}
\hspace{0cm}
\graphicspath{{lightcurve/}}
\includegraphics[width=5.5cm,height=5cm]{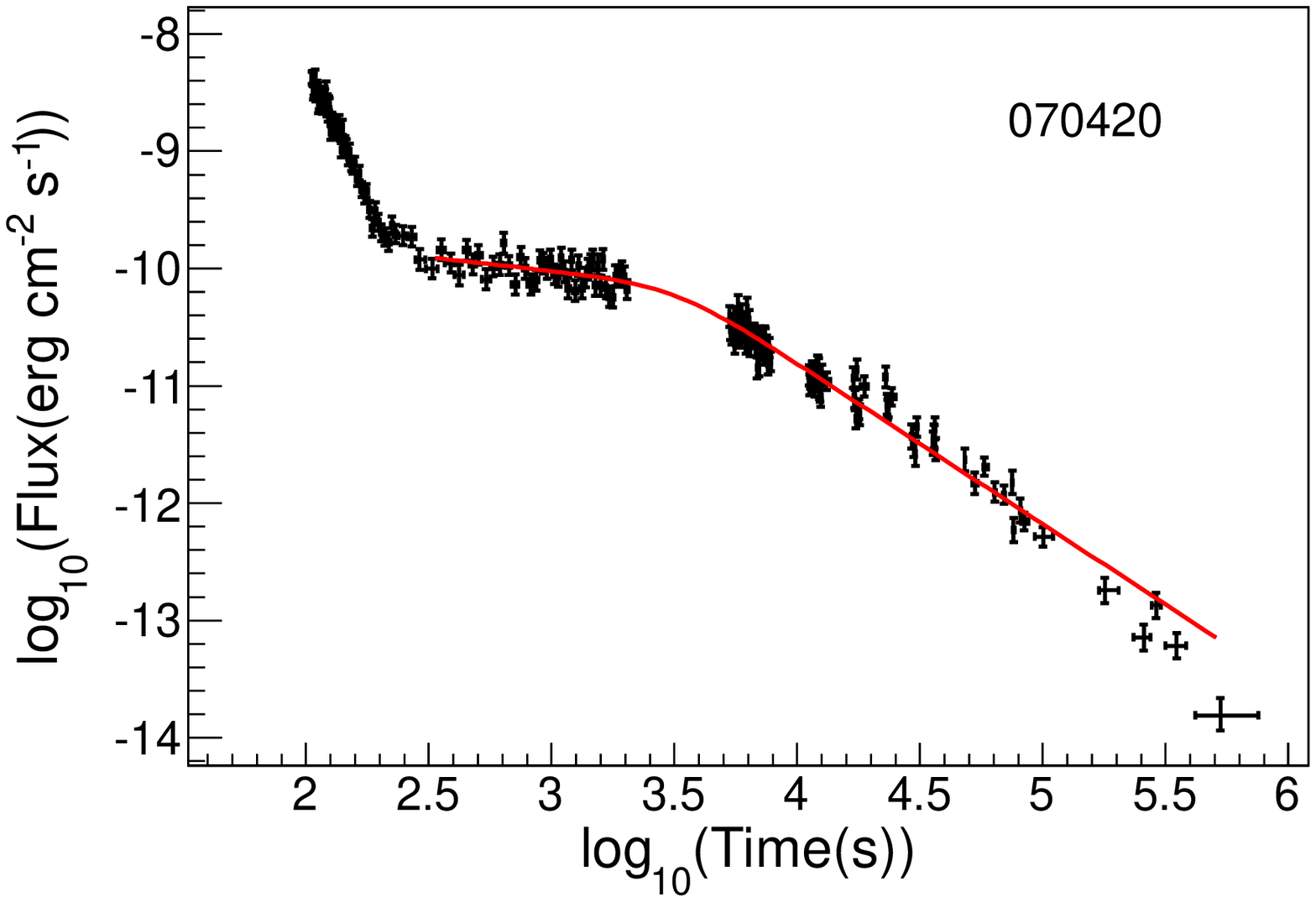}
\includegraphics[width=5.5cm,height=5cm]{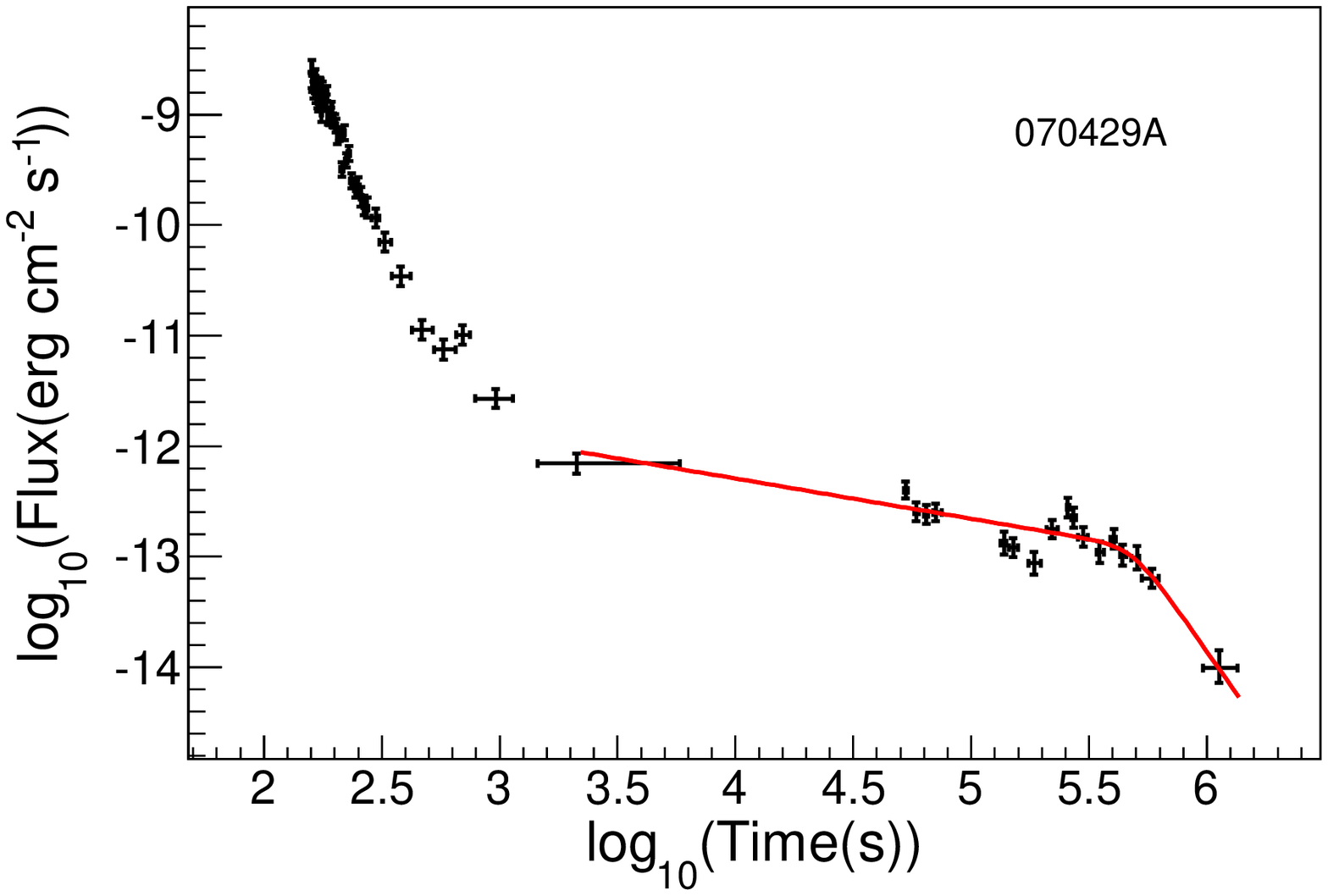}
\includegraphics[width=5.5cm,height=5cm]{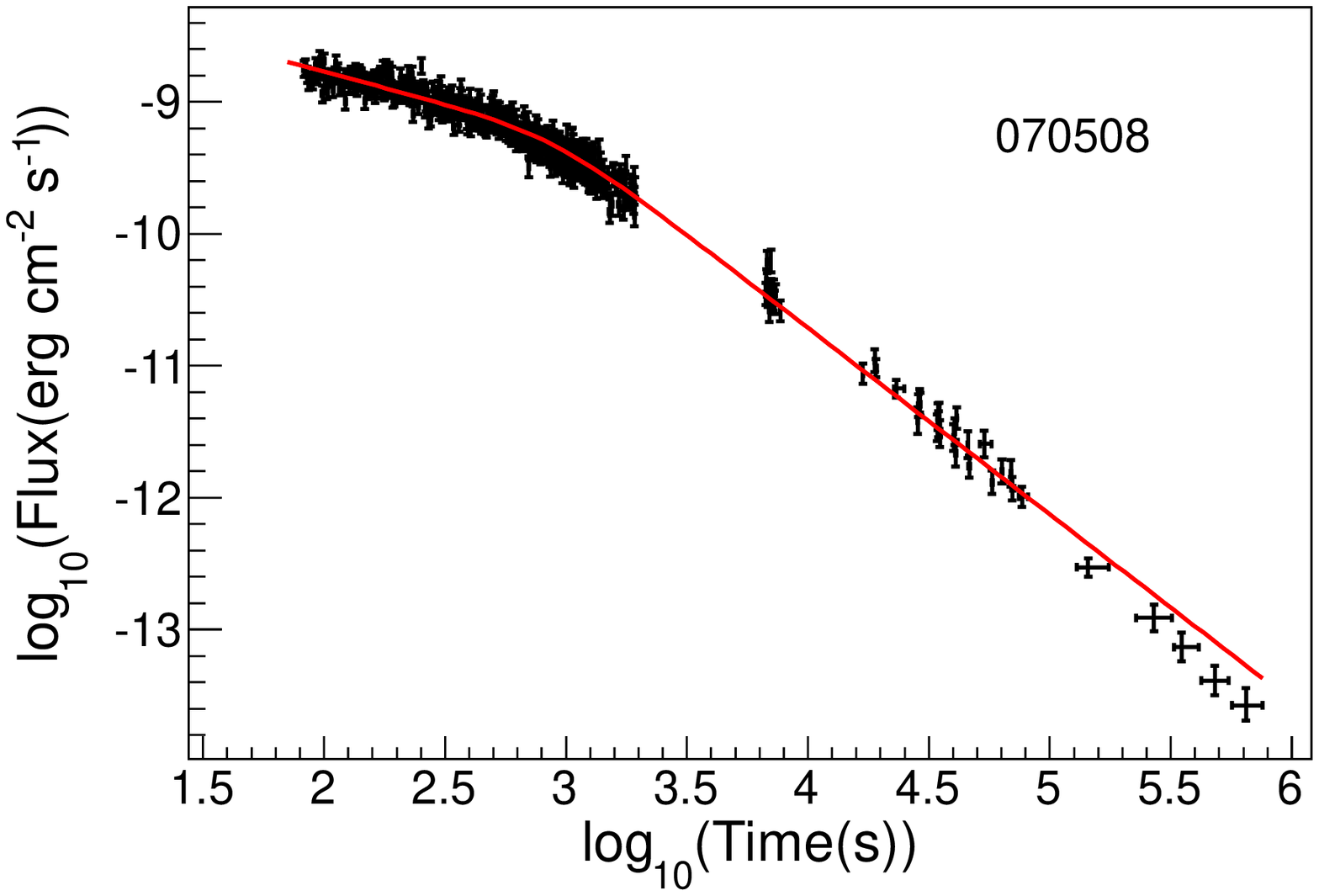}
\includegraphics[width=5.5cm,height=5cm]{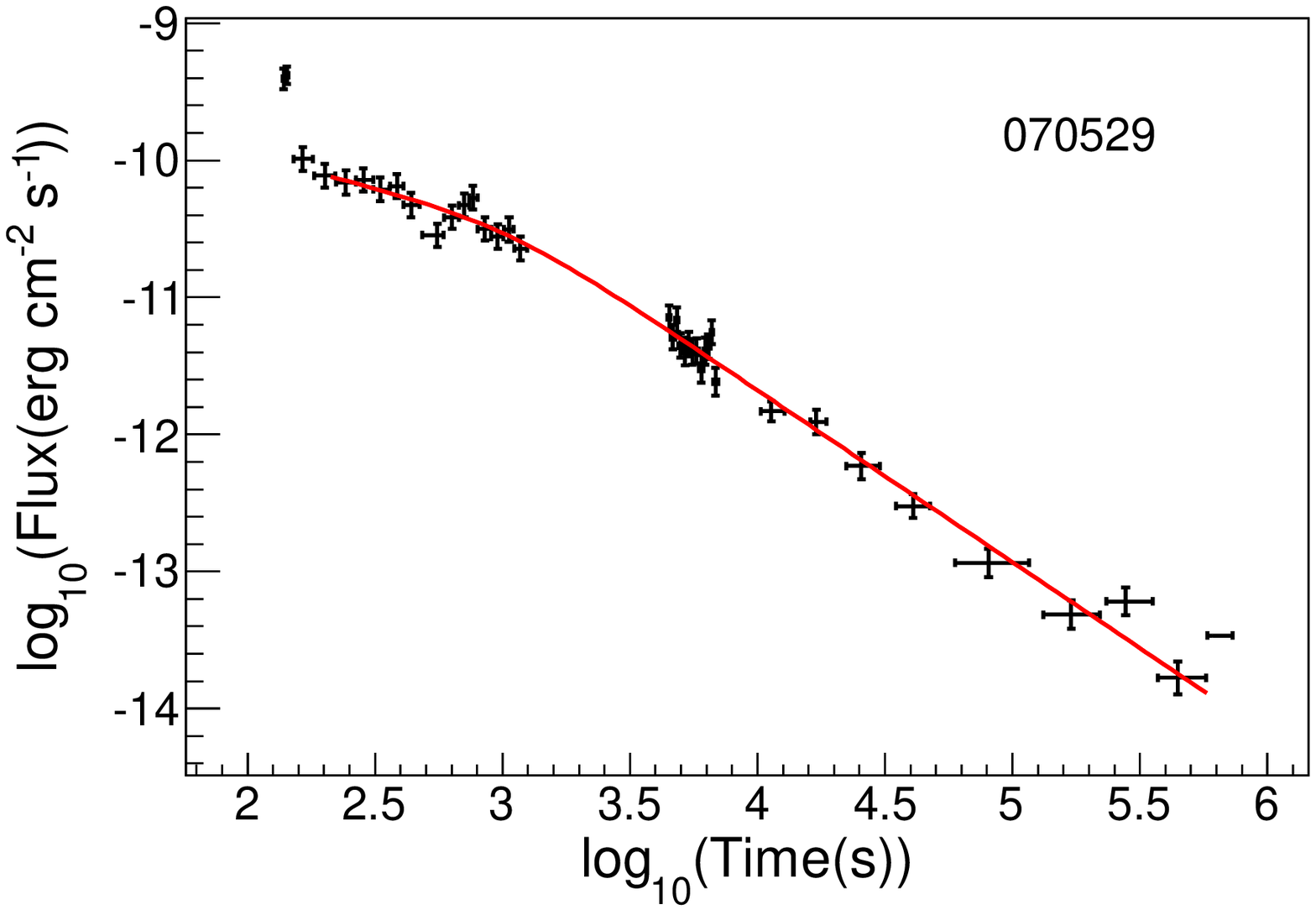}
\includegraphics[width=5.5cm,height=5cm]{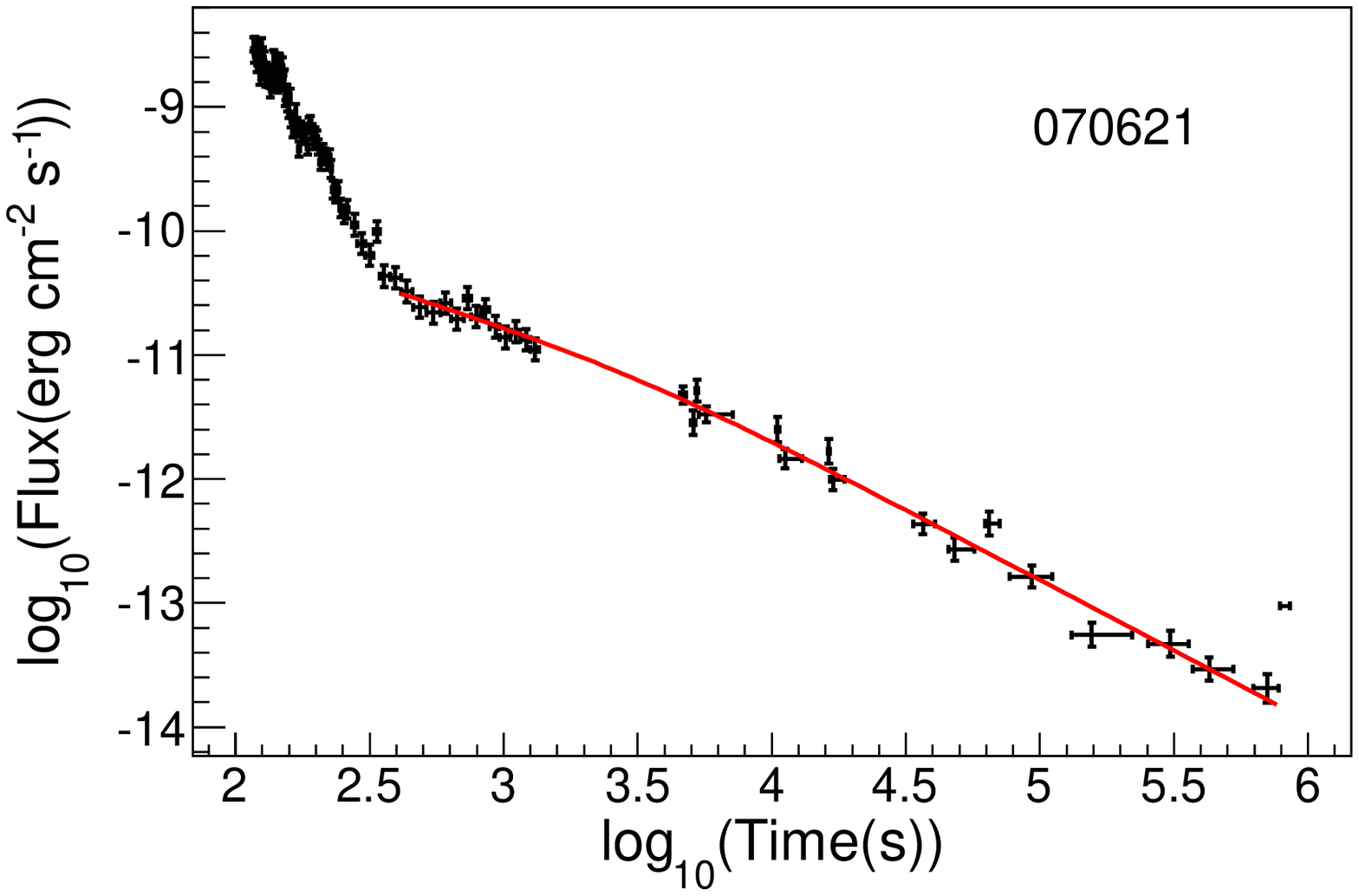}
\includegraphics[width=5.5cm,height=5cm]{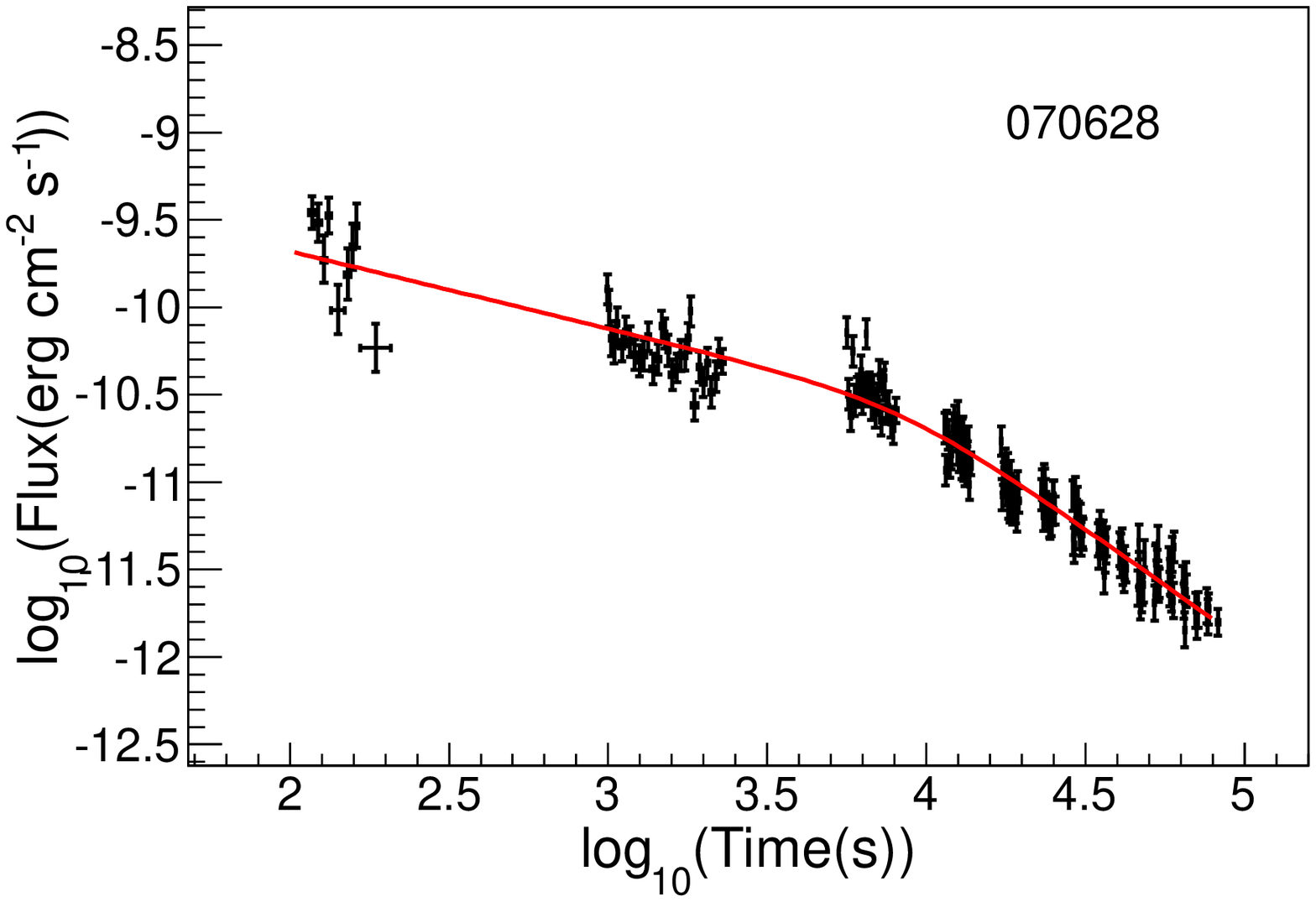}
\includegraphics[width=5.5cm,height=5cm]{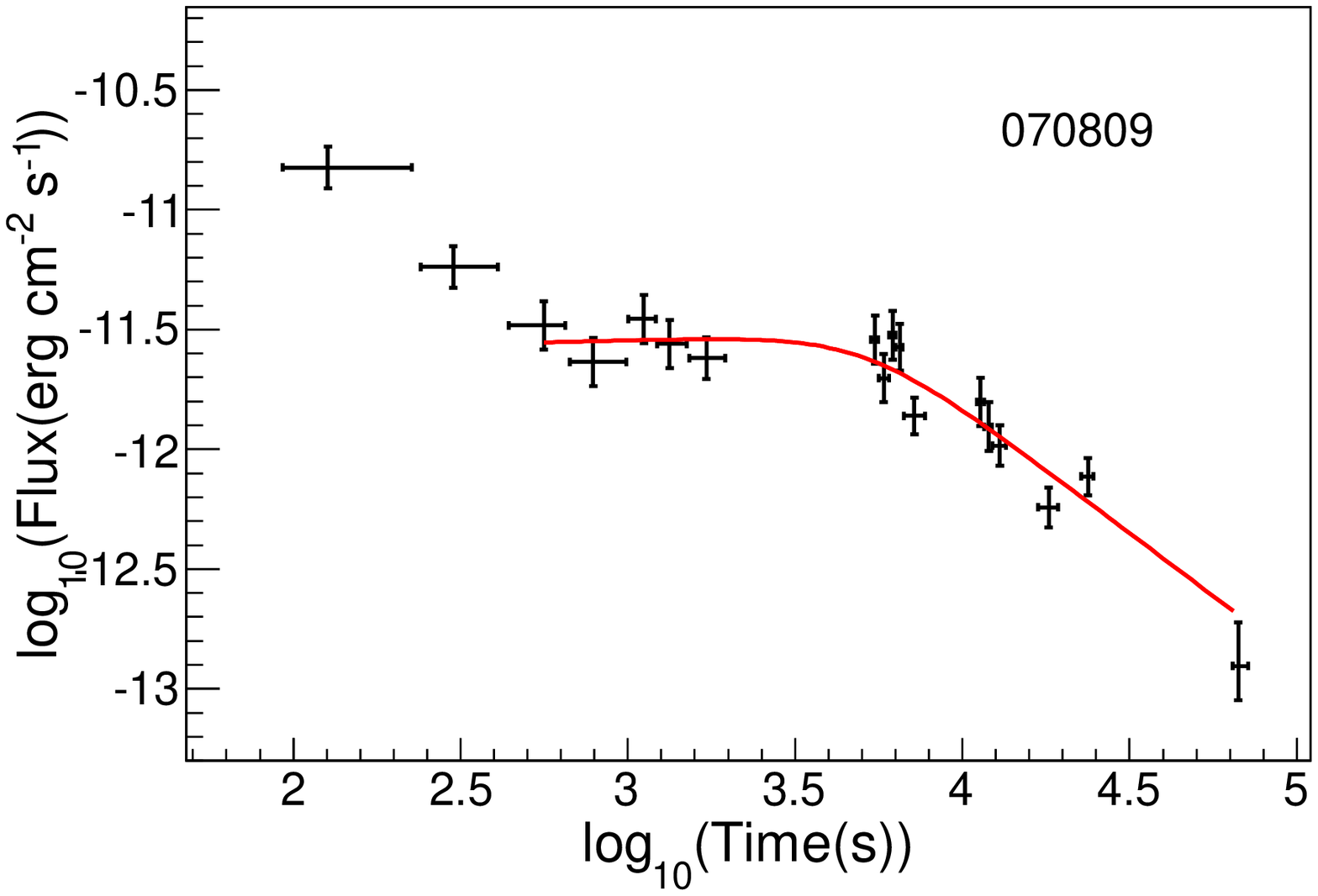}
\includegraphics[width=5.5cm,height=5cm]{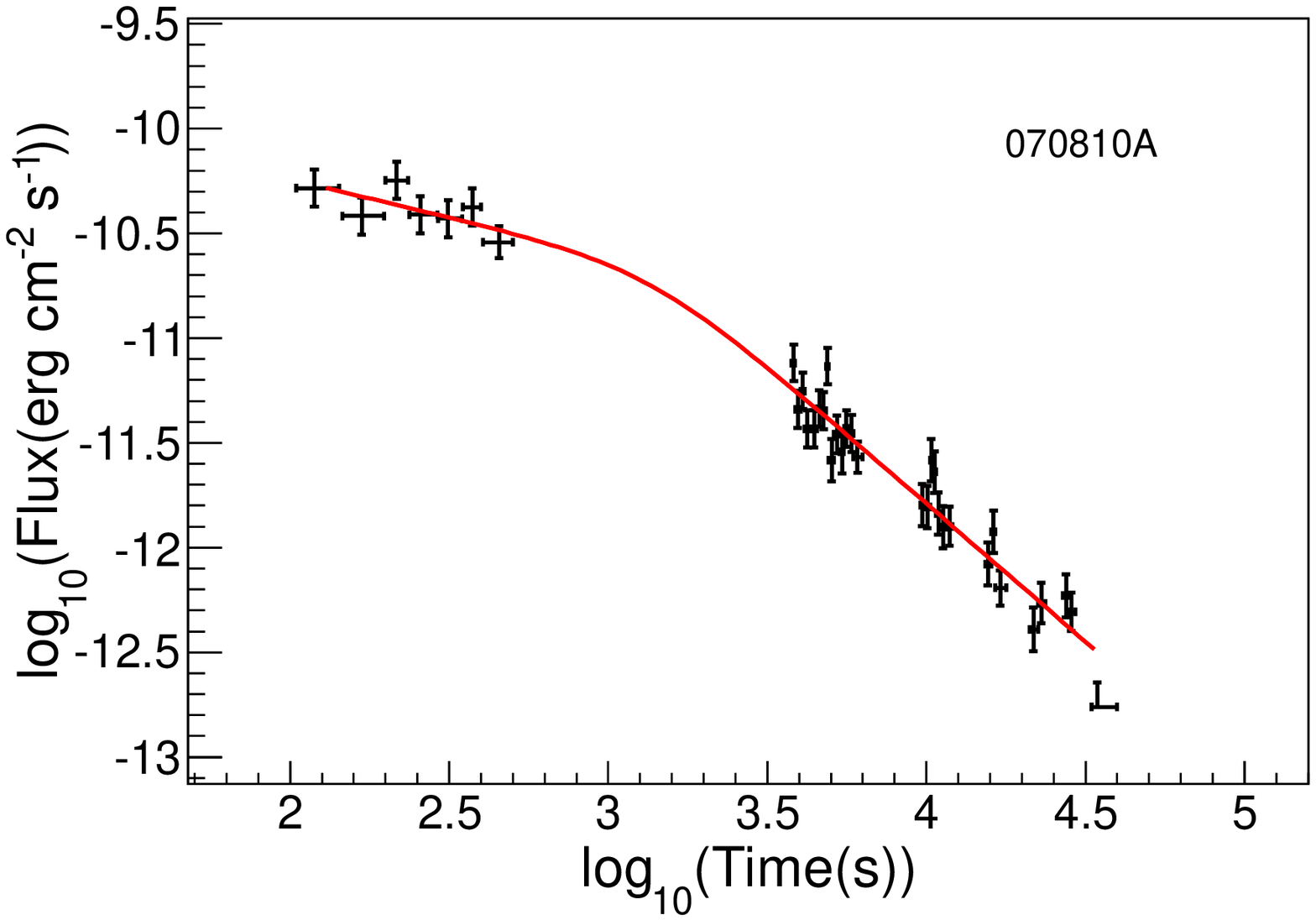}
\includegraphics[width=5.5cm,height=5cm]{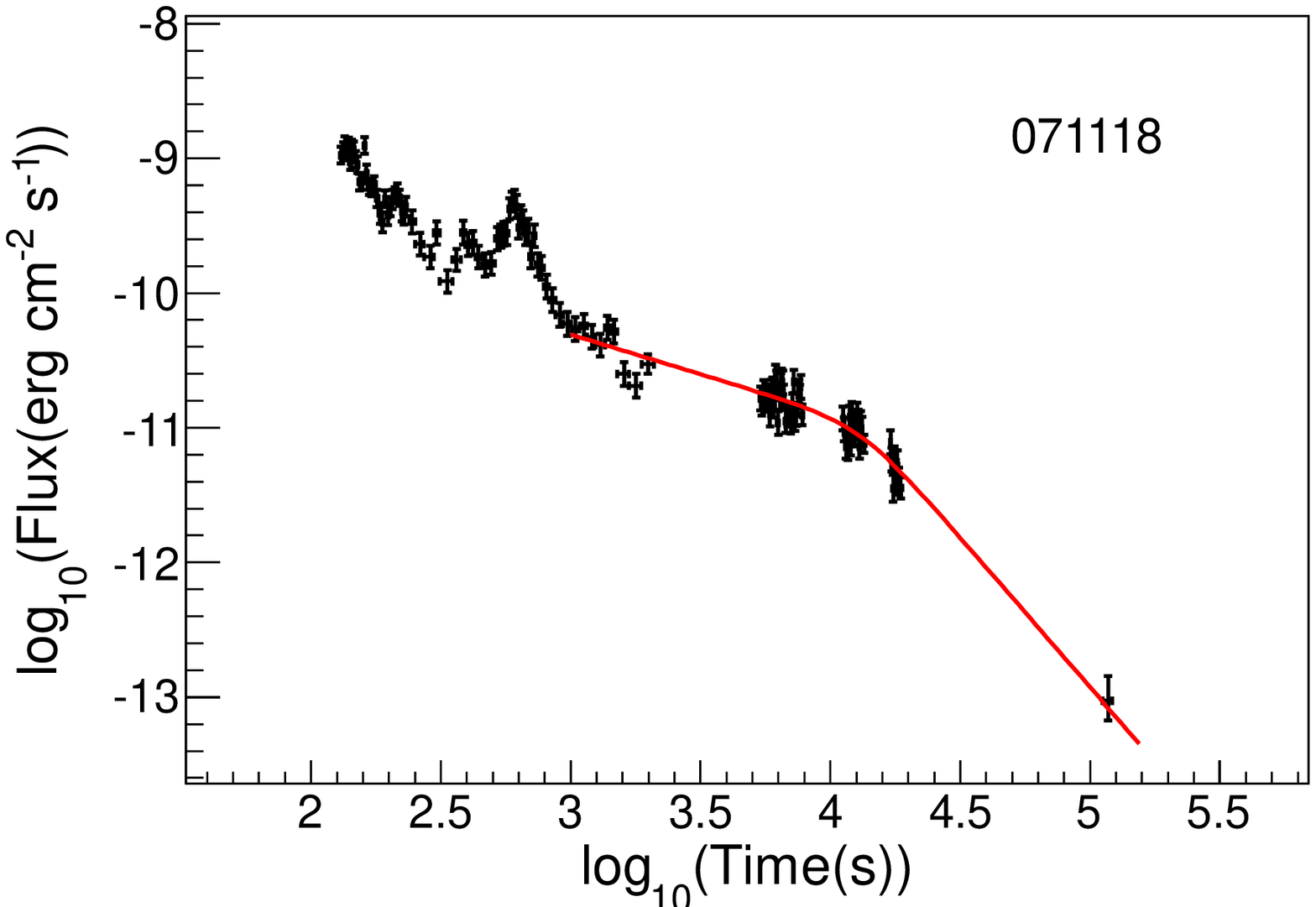}
\includegraphics[width=5.5cm,height=5cm]{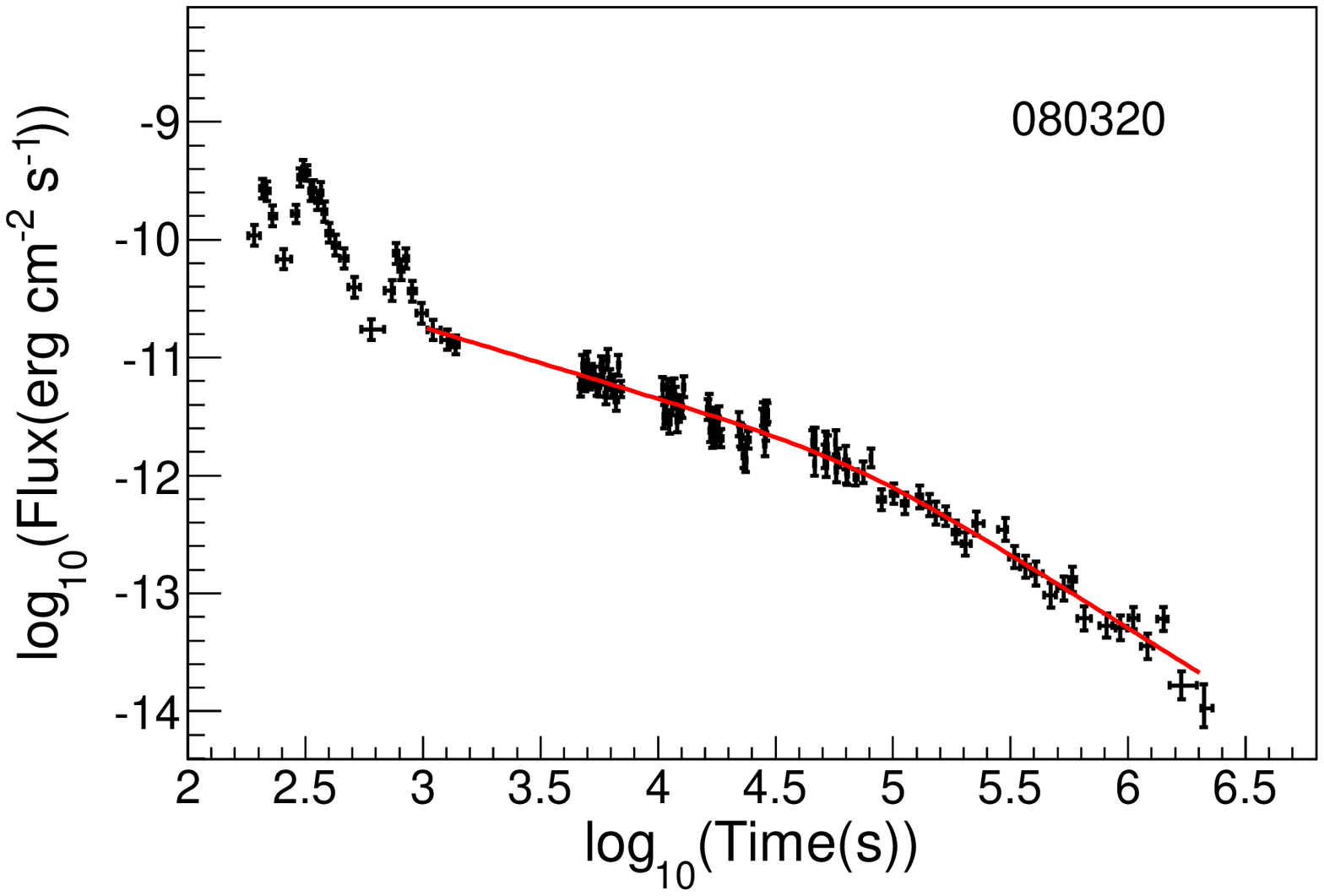}
\includegraphics[width=5.5cm,height=5cm]{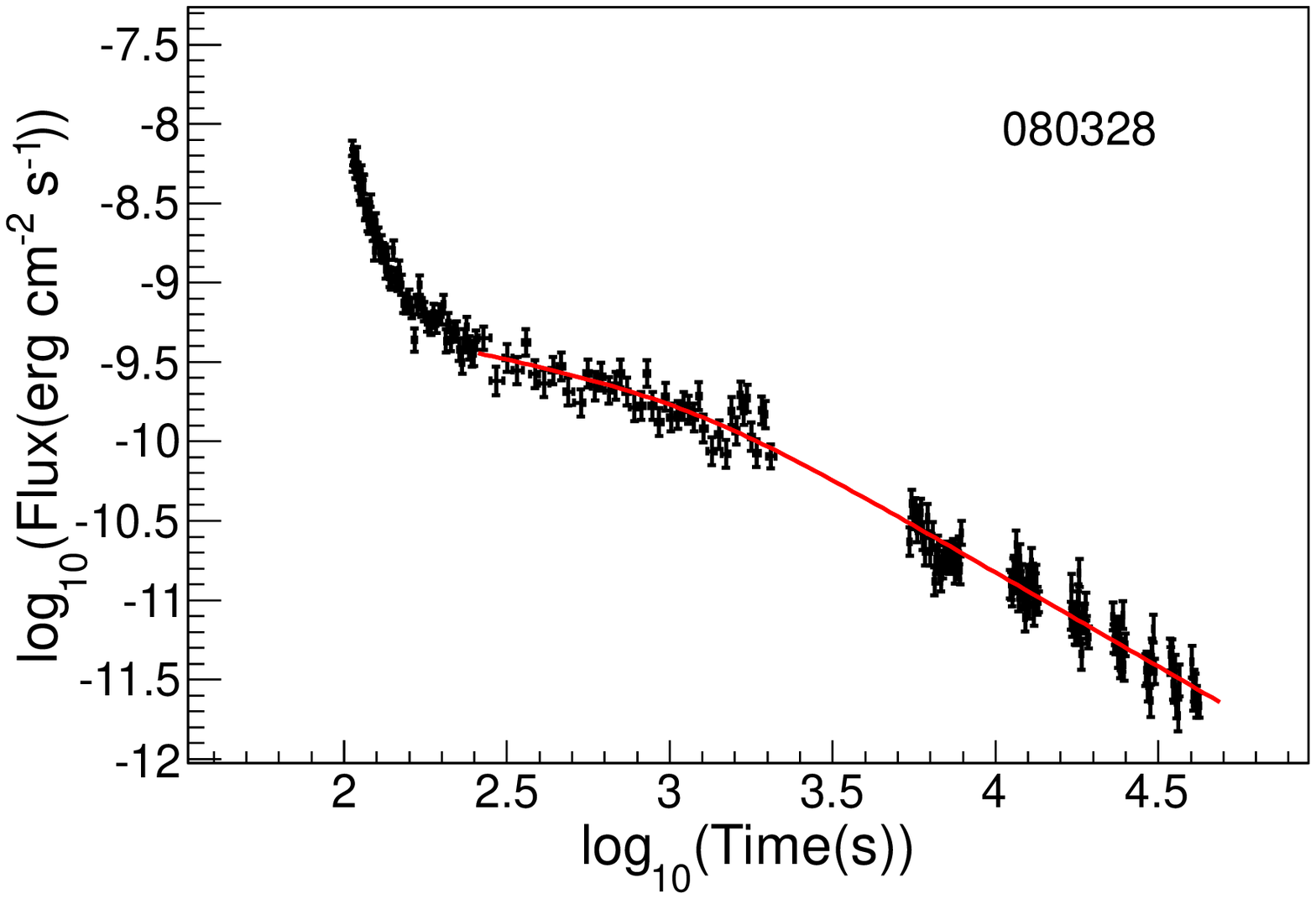}
\includegraphics[width=5.5cm,height=5cm]{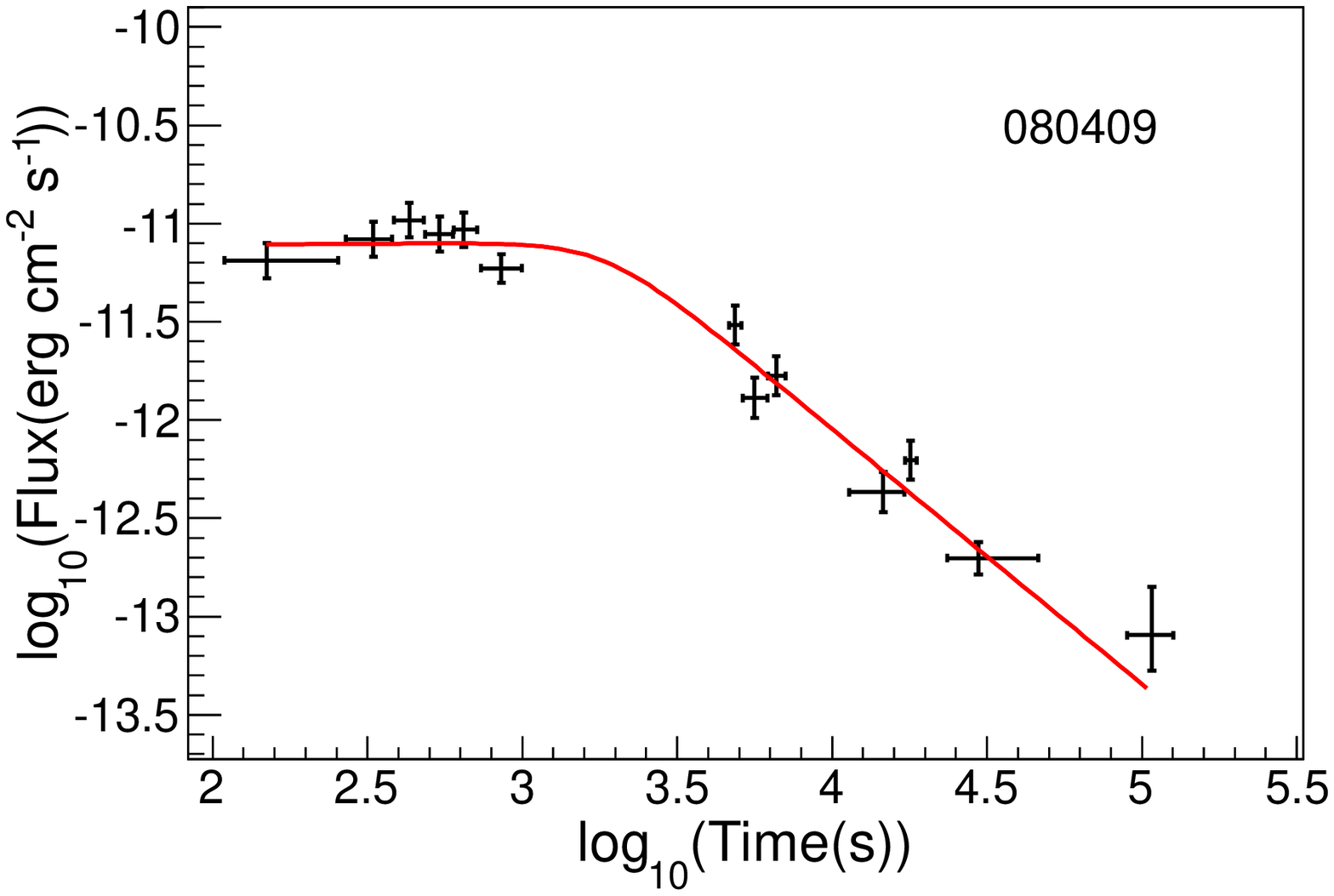}
\caption{ Continued.}
\label{fig-1-6}
\end{center}
\end{figure*}

\begin{figure*}
\begin{center}
\setlength{\abovecaptionskip}{0.cm}
\setlength{\belowcaptionskip}{-0.cm}
\figurenum{1}
\hspace{0cm}
\graphicspath{{lightcurve/}}
\includegraphics[width=5.5cm,height=5cm]{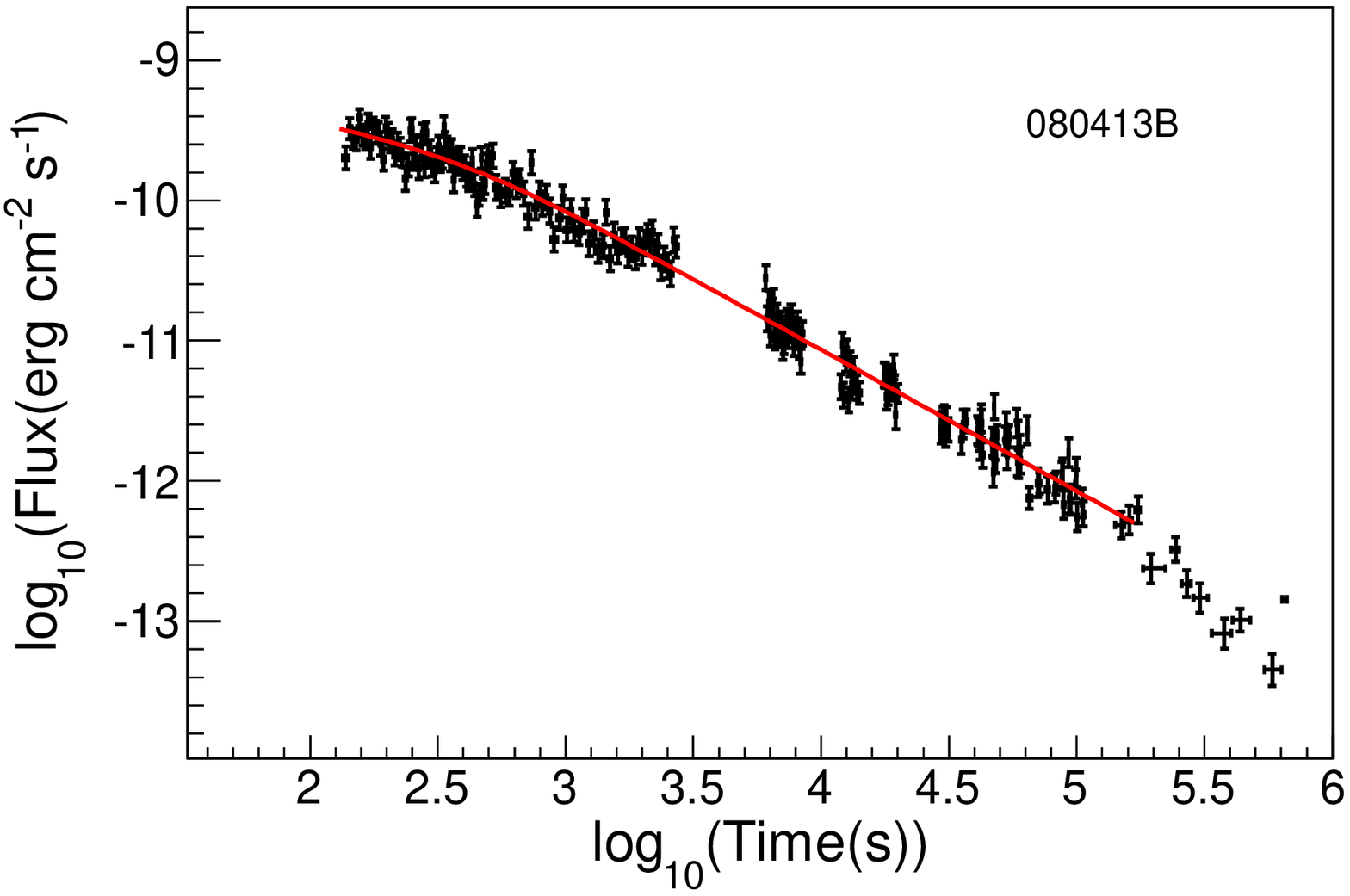}
\includegraphics[width=5.5cm,height=5cm]{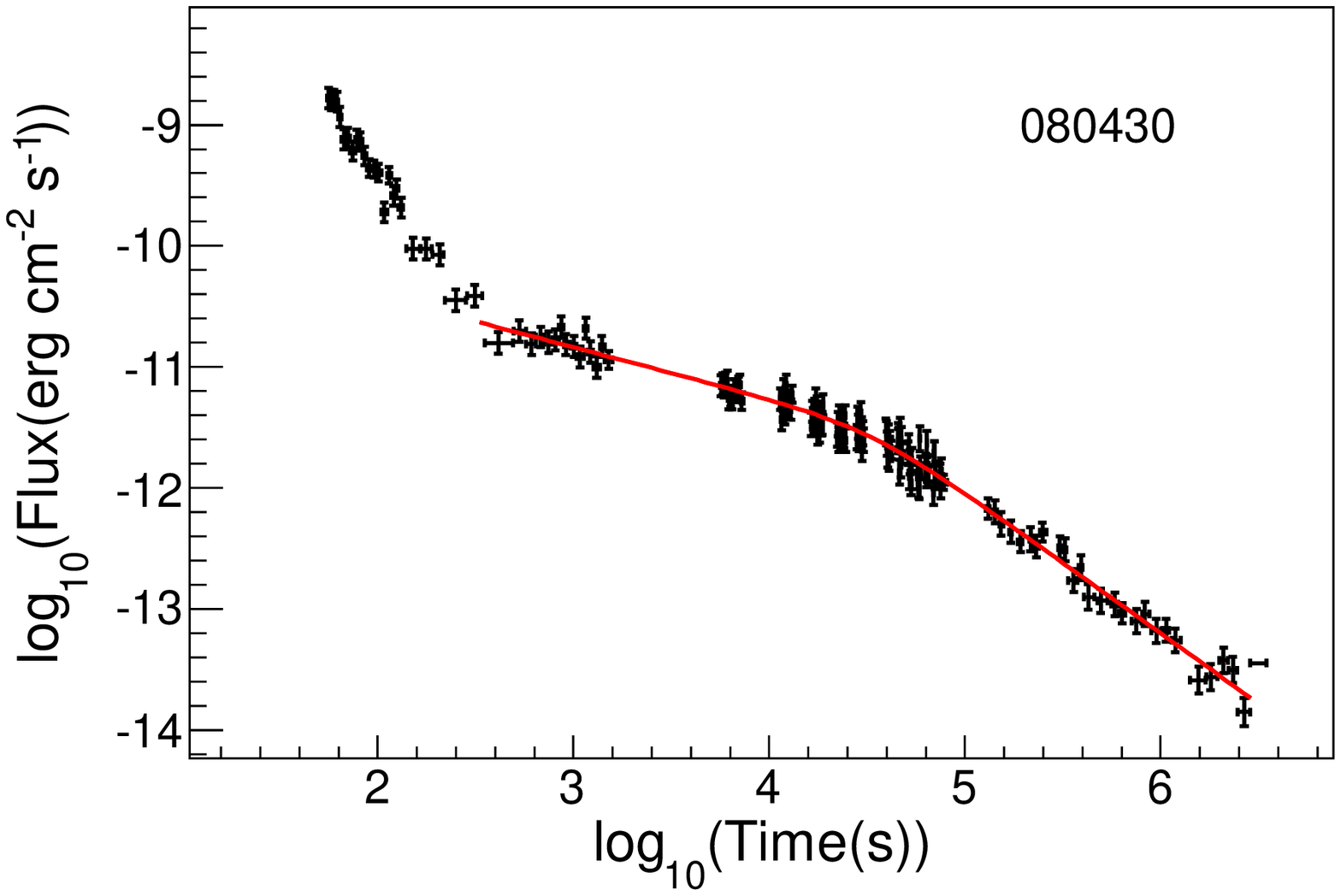}
\includegraphics[width=5.5cm,height=5cm]{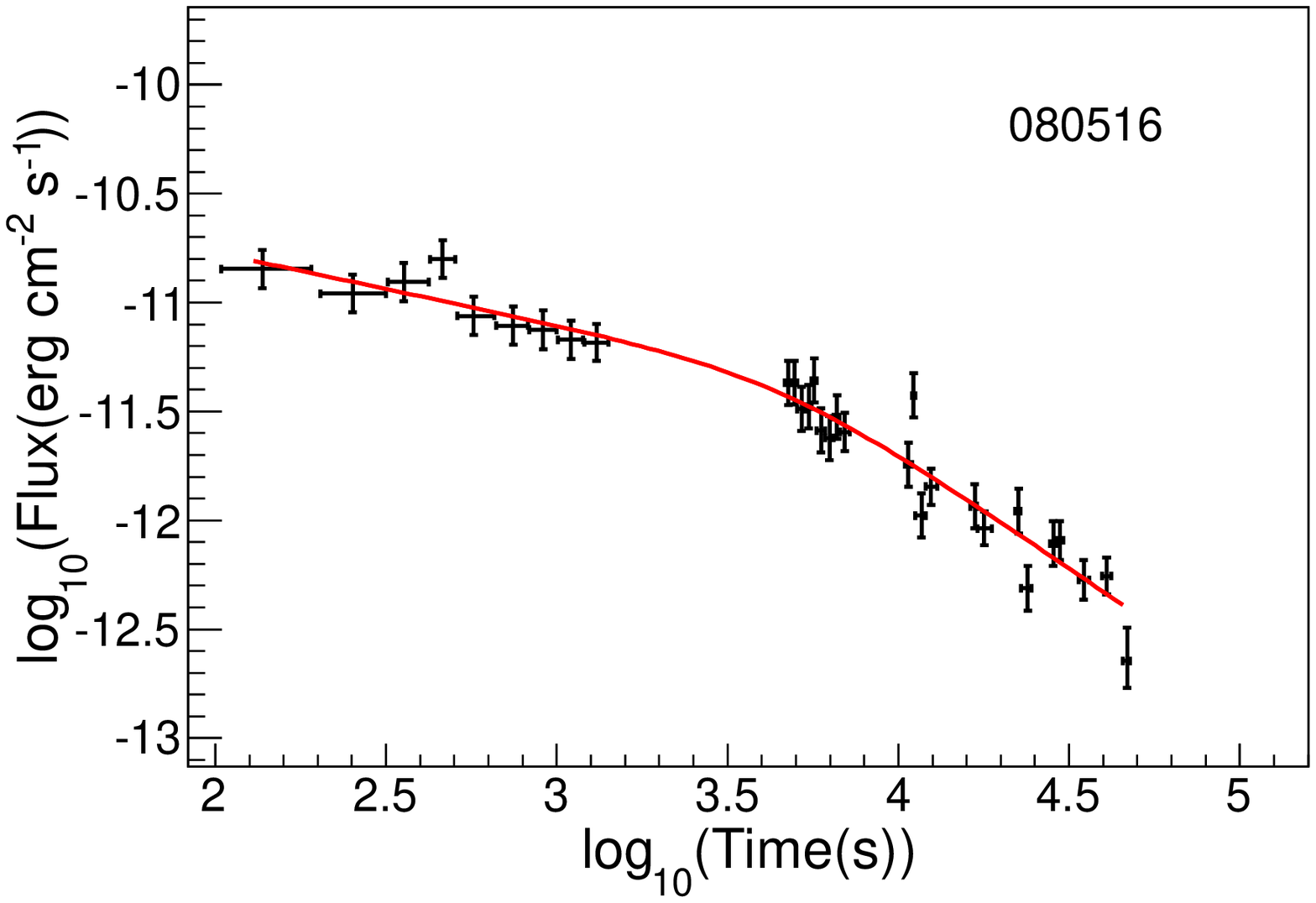}
\includegraphics[width=5.5cm,height=5cm]{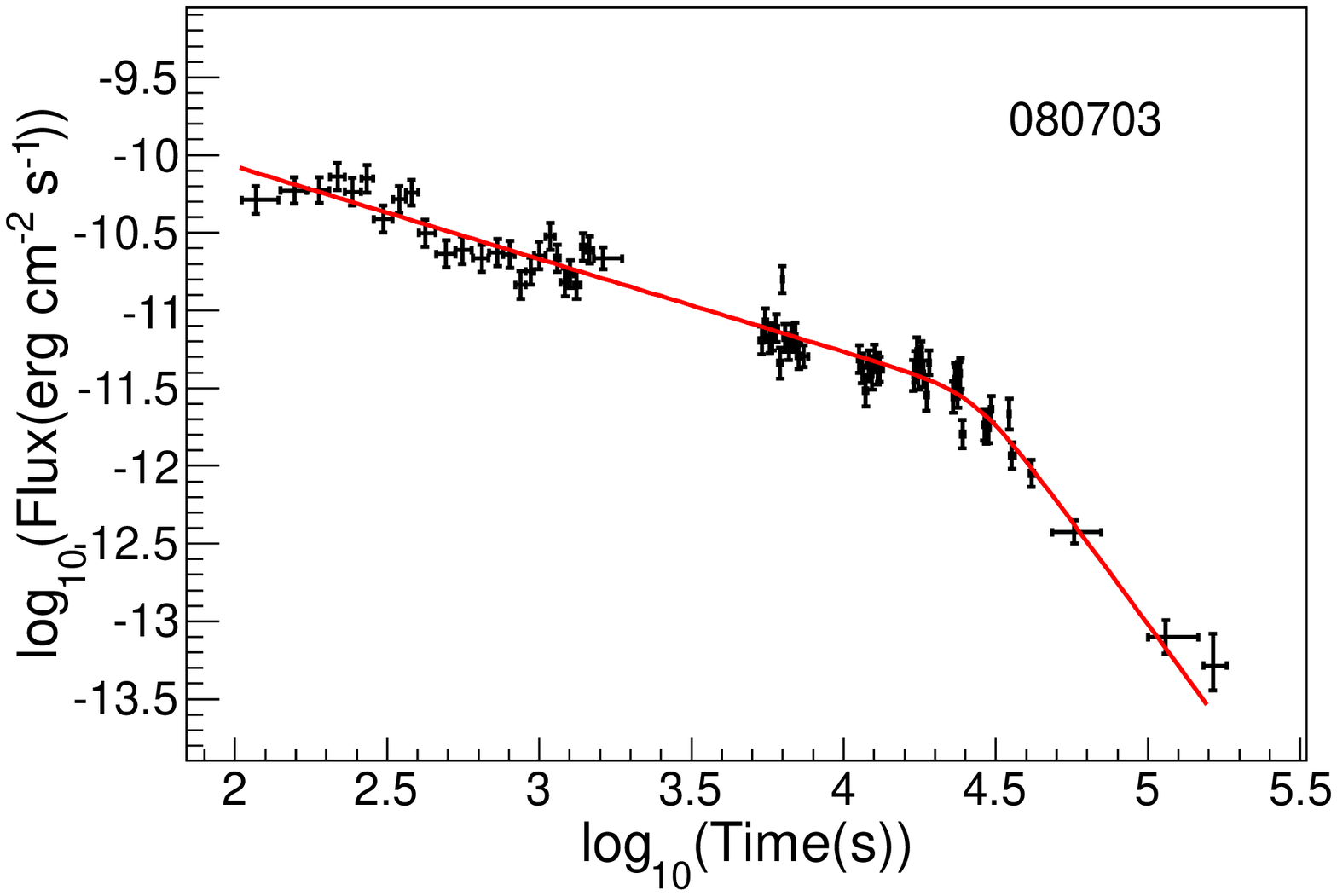}
\includegraphics[width=5.5cm,height=5cm]{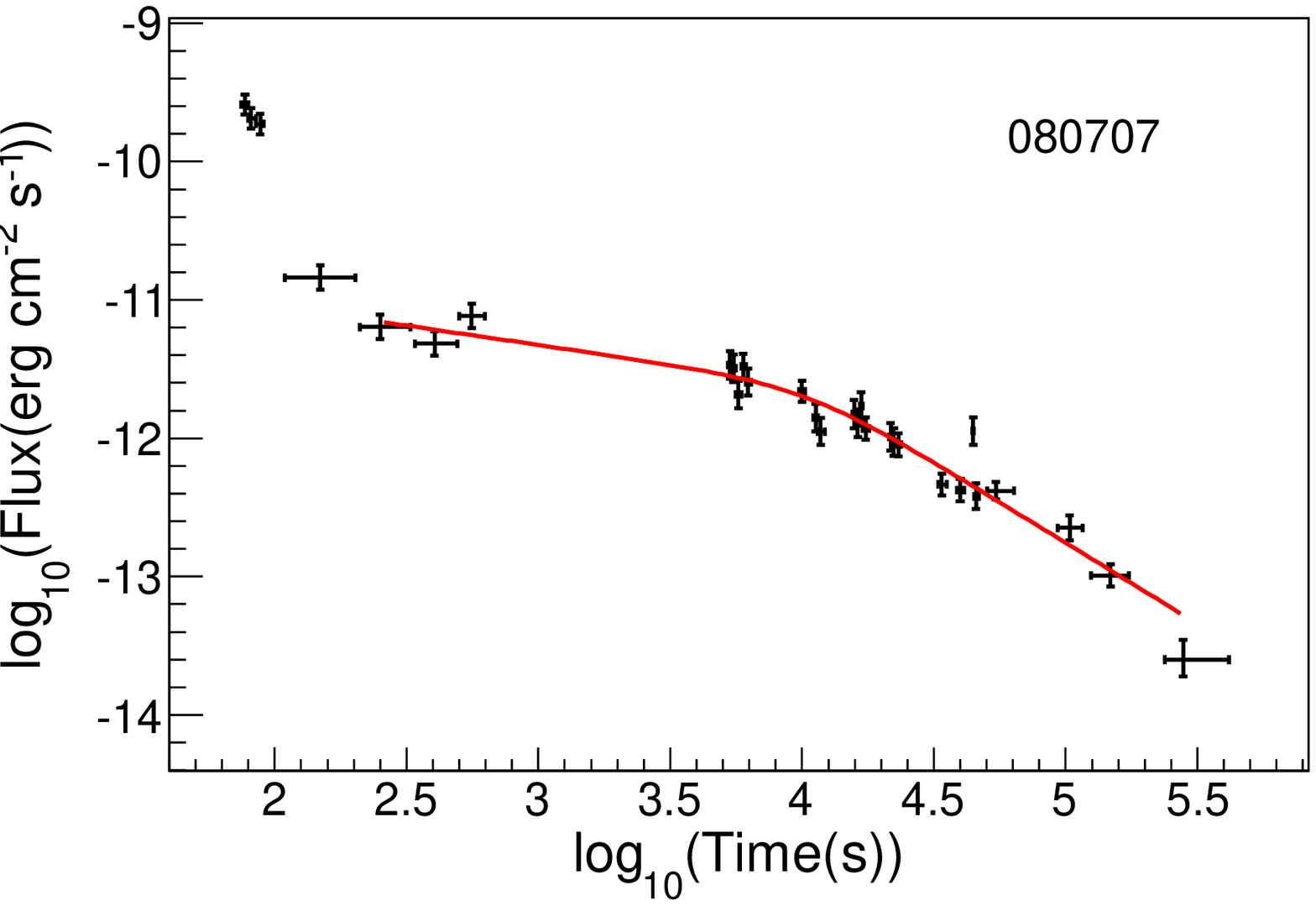}
\includegraphics[width=5.5cm,height=5cm]{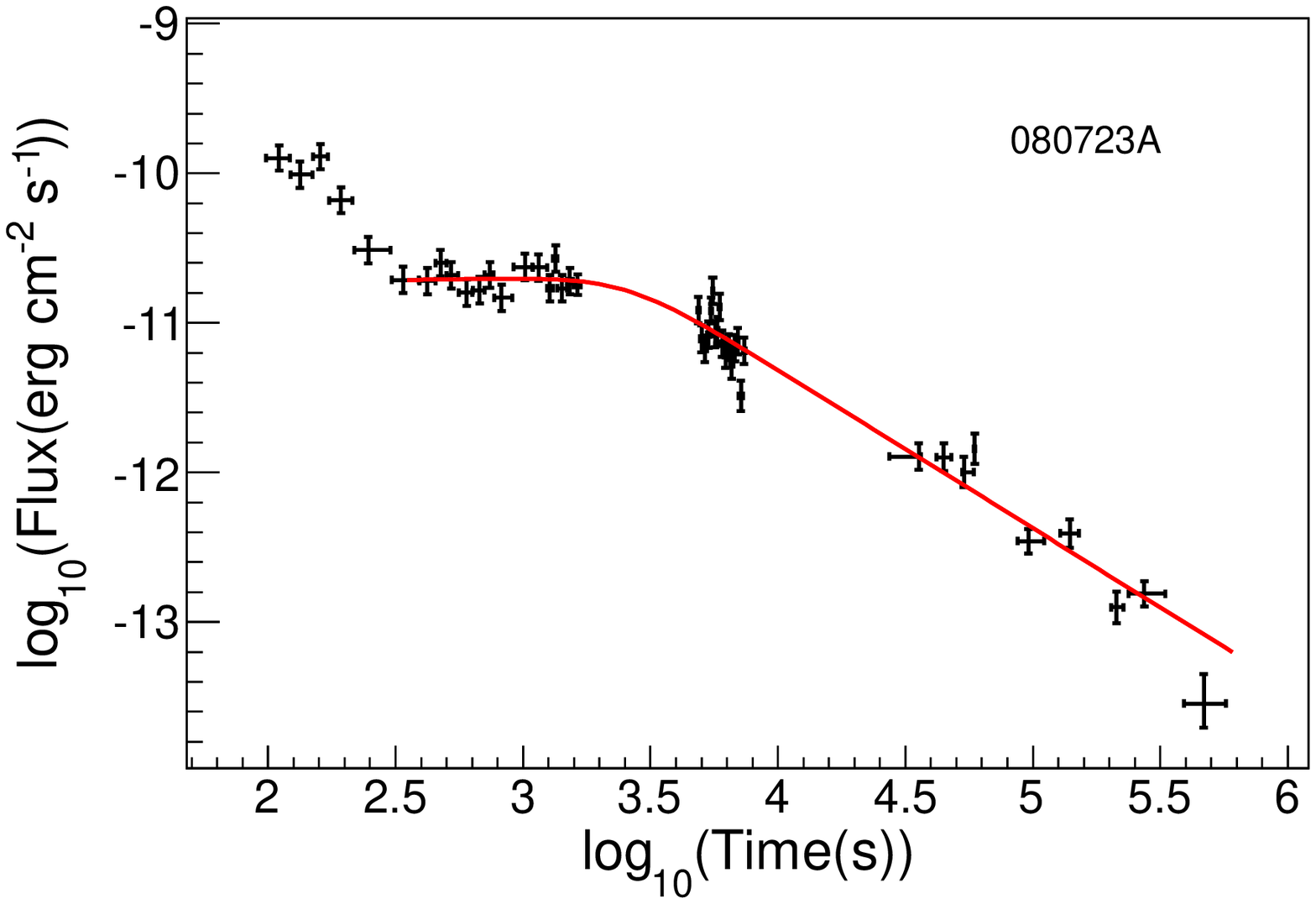}
\includegraphics[width=5.5cm,height=5cm]{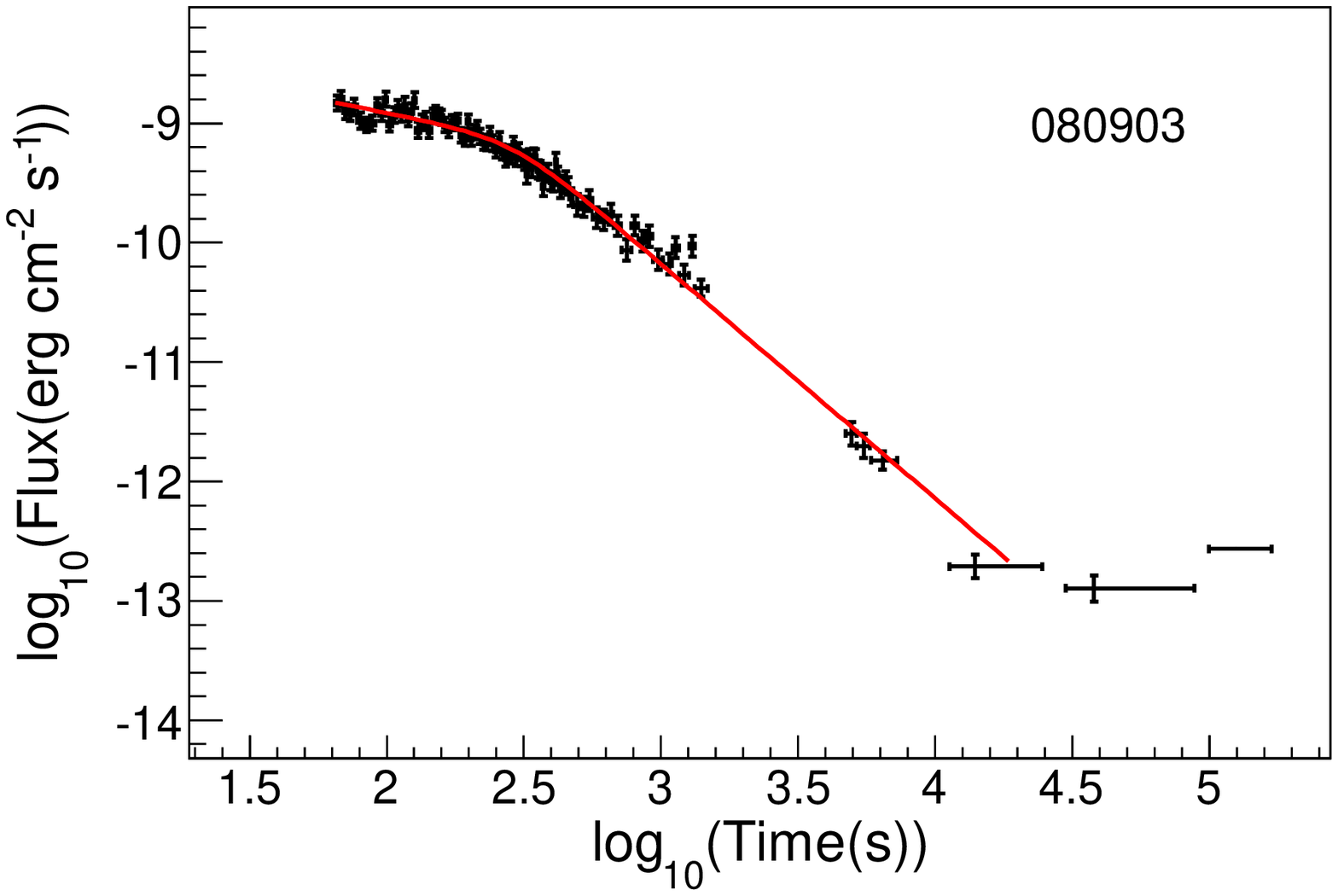}
\includegraphics[width=5.5cm,height=5cm]{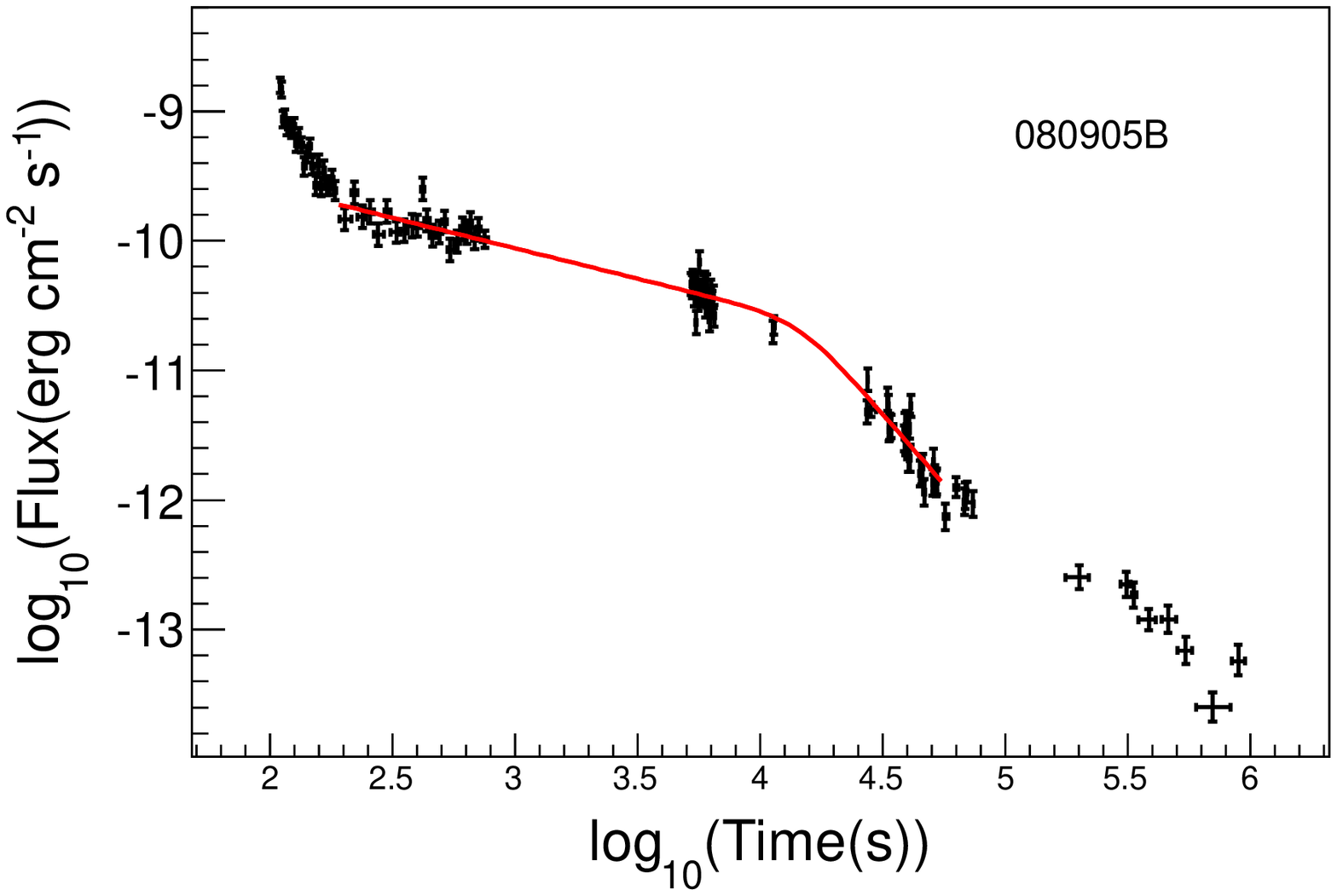}
\includegraphics[width=5.5cm,height=5cm]{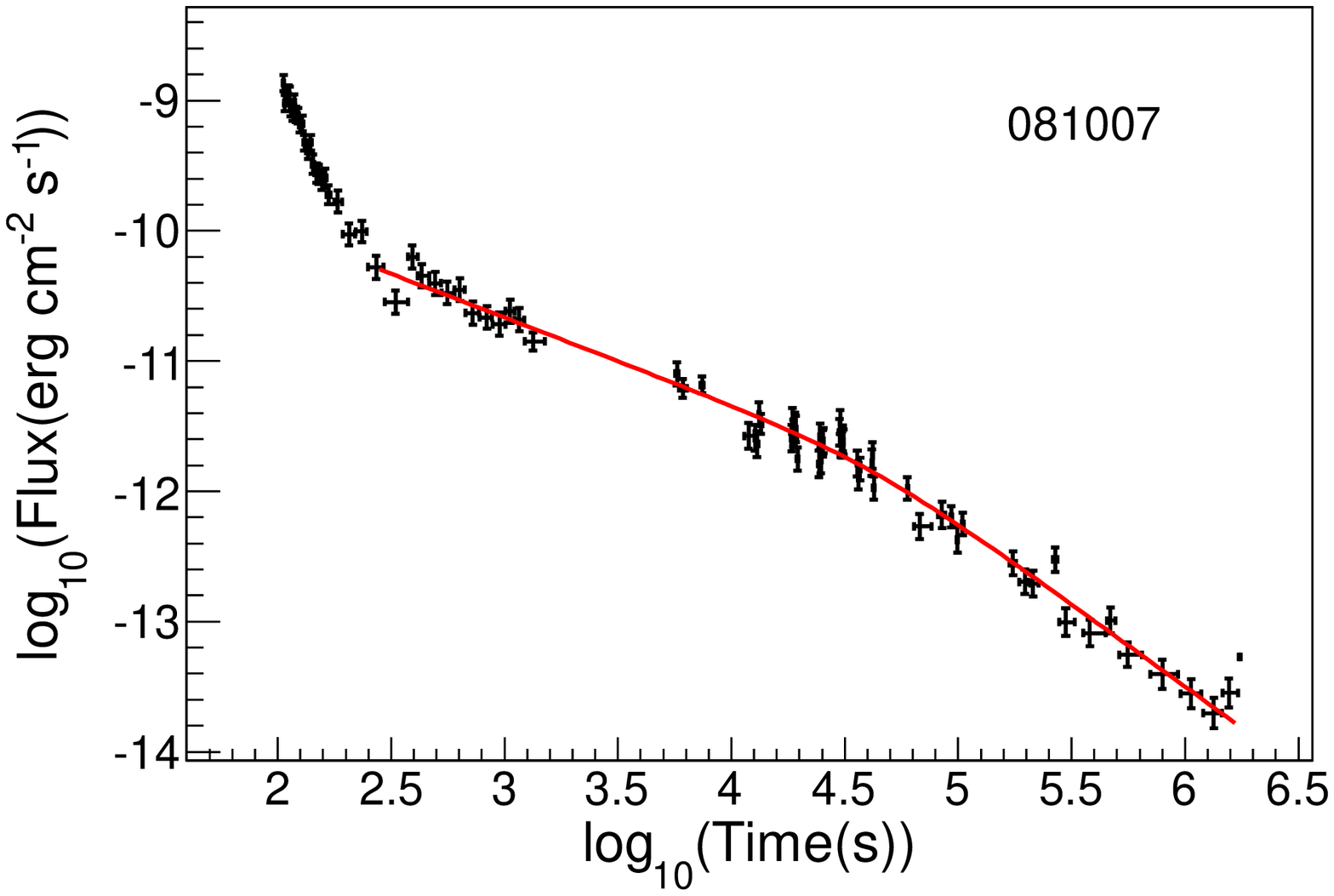}
\includegraphics[width=5.5cm,height=5cm]{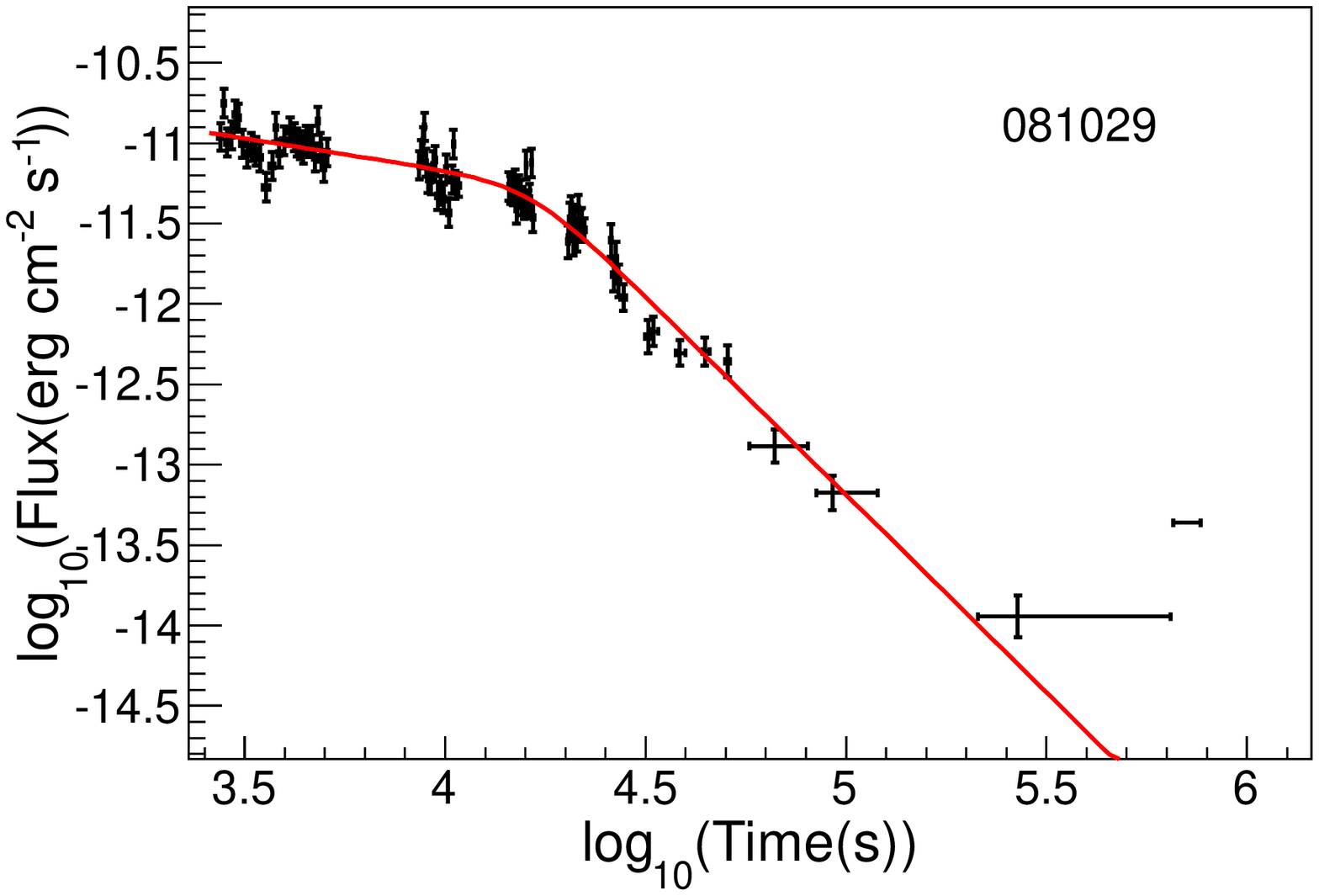}
\includegraphics[width=5.5cm,height=5cm]{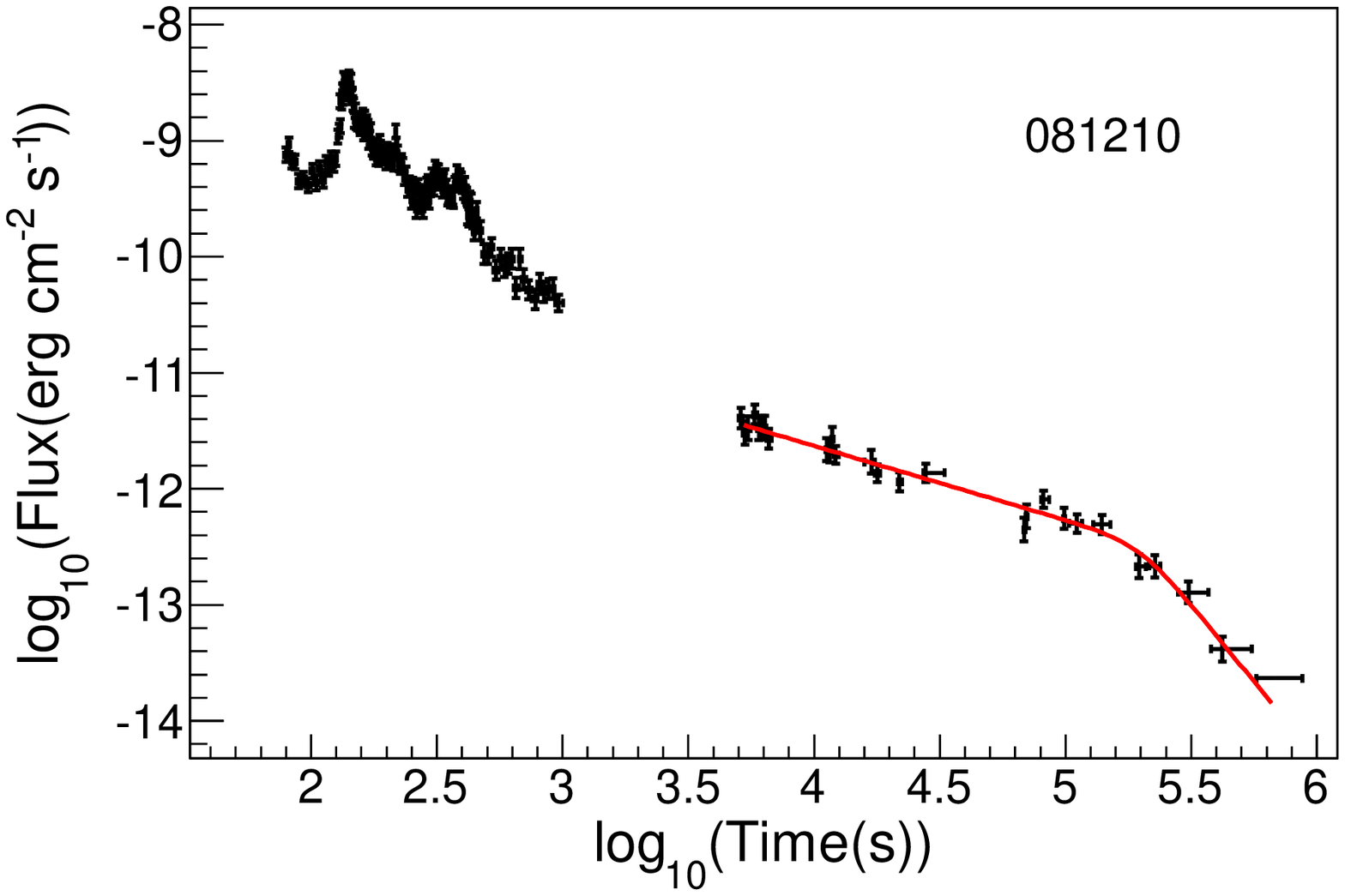}
\includegraphics[width=5.5cm,height=5cm]{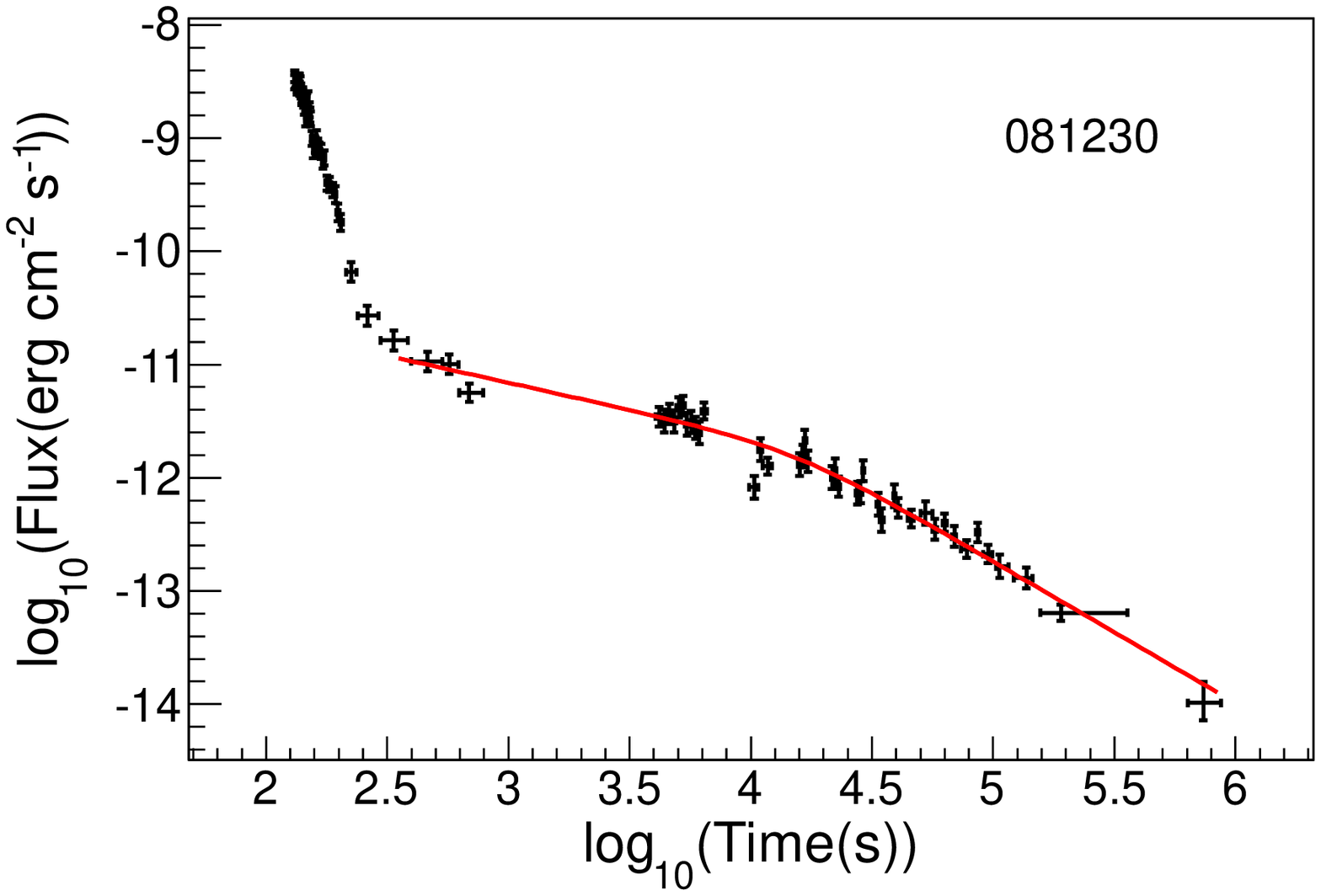}
\caption{ Continued.}
\label{fig-1-7}
\end{center}
\end{figure*}

\begin{figure*}
\begin{center}
\setlength{\abovecaptionskip}{0.cm}
\setlength{\belowcaptionskip}{-0.cm}
\figurenum{1}
\hspace{0cm}
\graphicspath{{lightcurve/}}
\includegraphics[width=5.5cm,height=5cm]{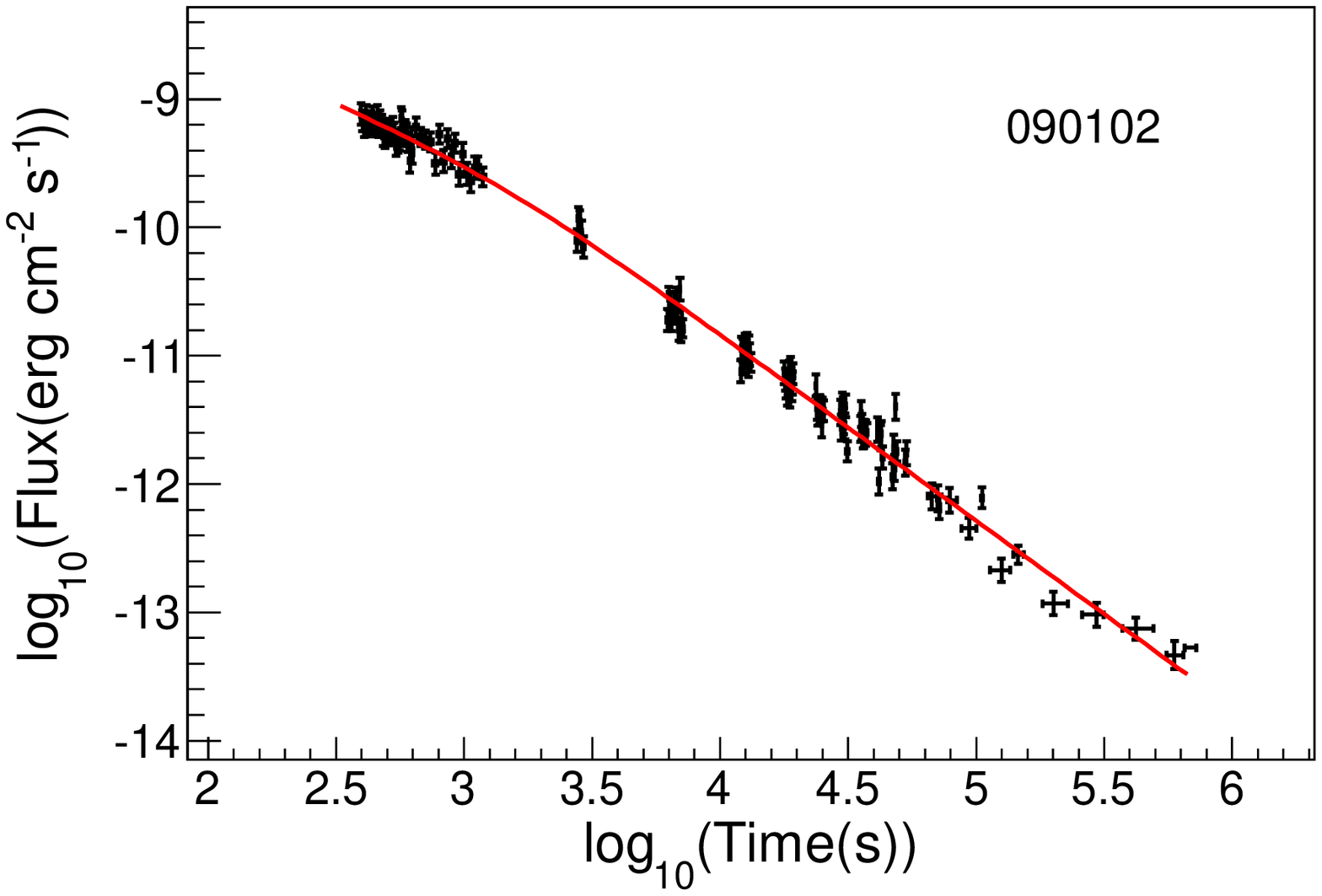}
\includegraphics[width=5.5cm,height=5cm]{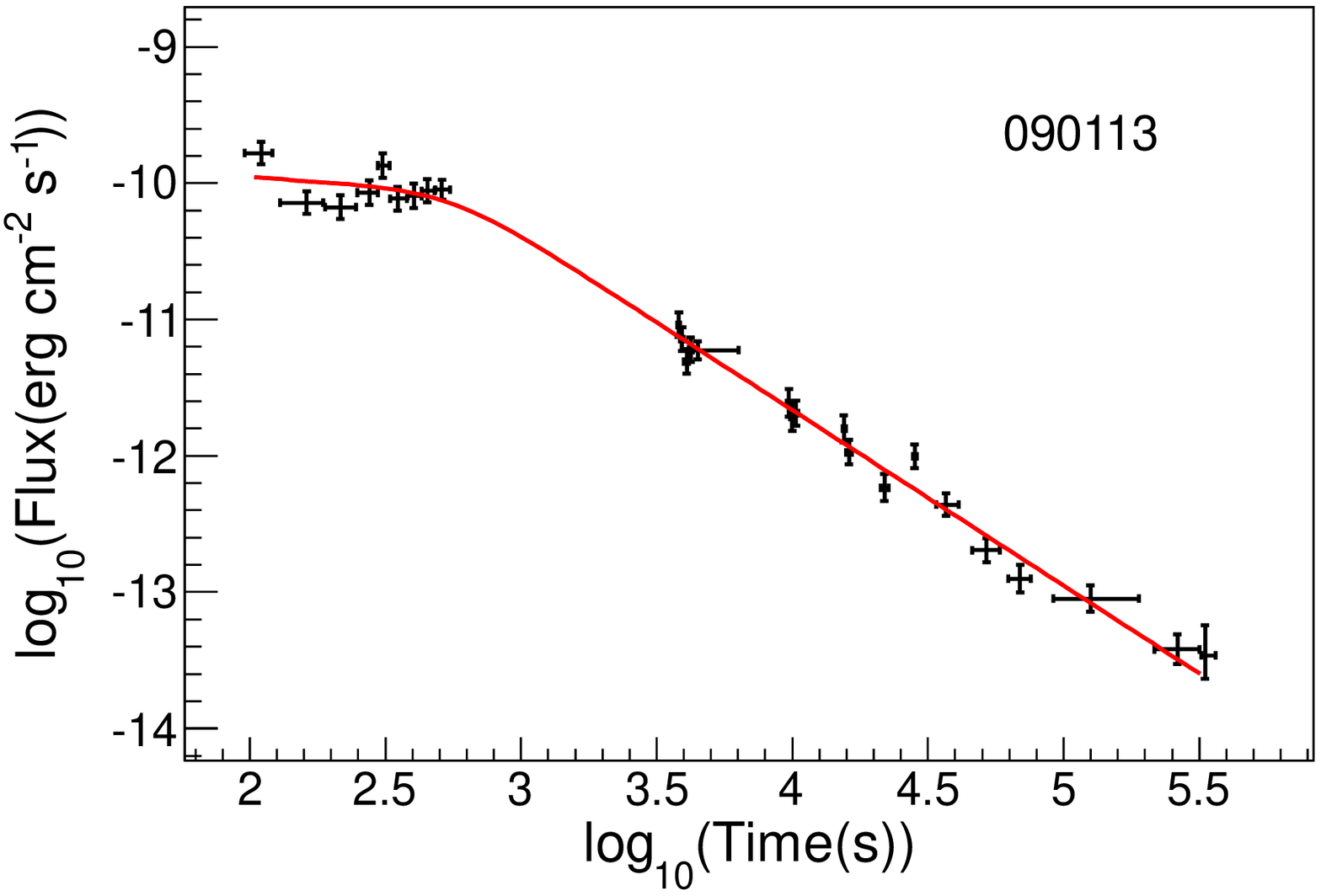}
\includegraphics[width=5.5cm,height=5cm]{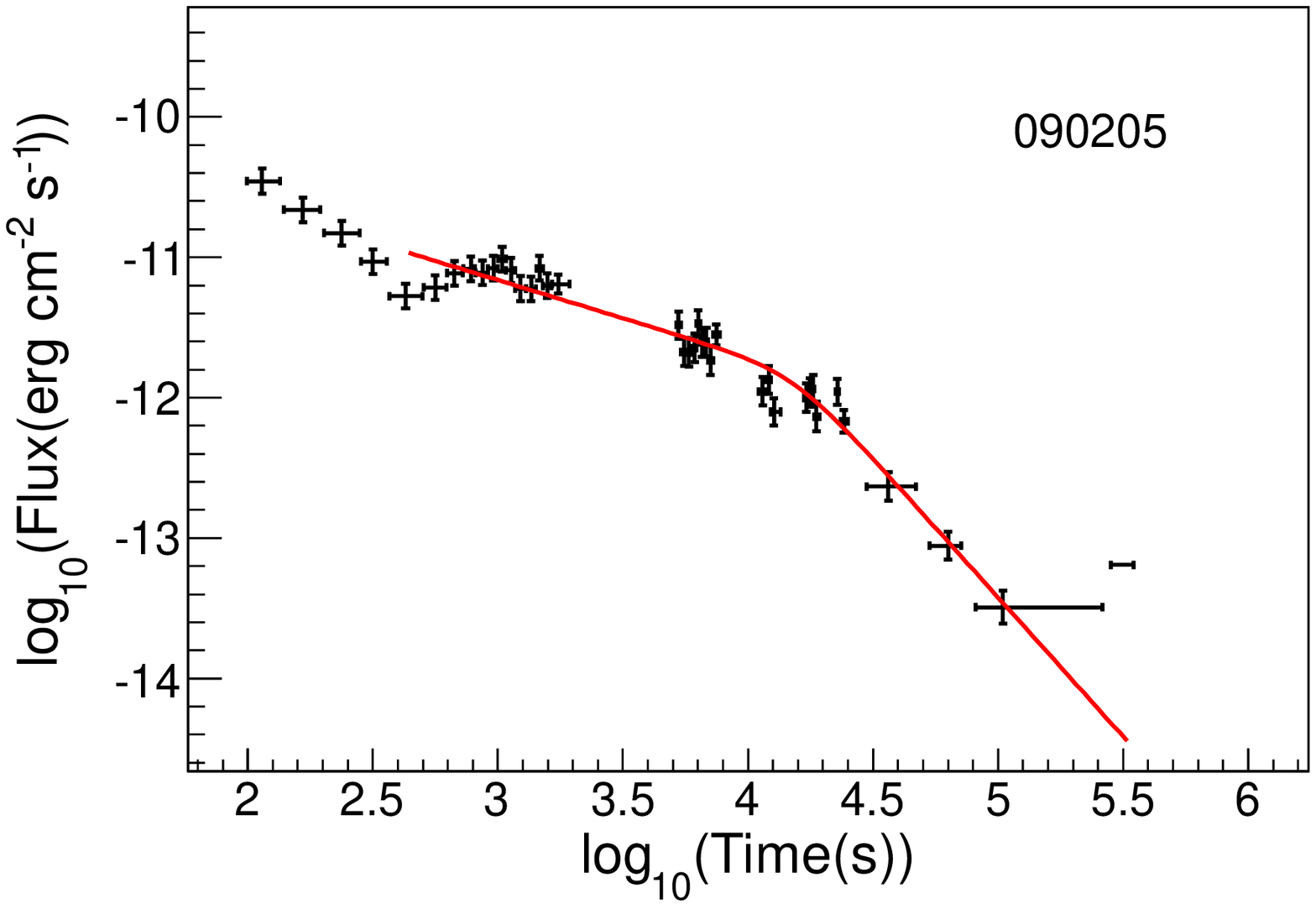}
\includegraphics[width=5.5cm,height=5cm]{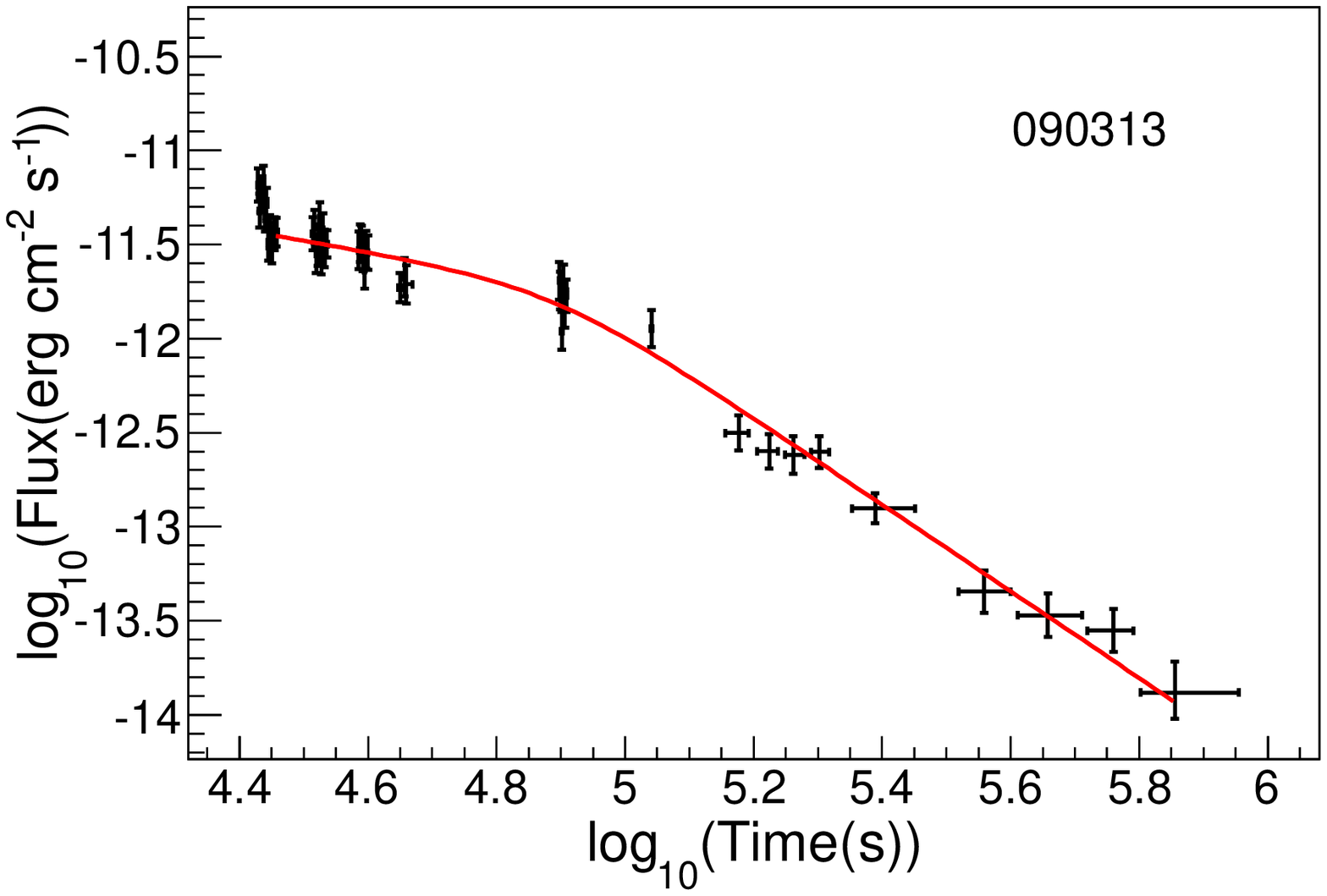}
\includegraphics[width=5.5cm,height=5cm]{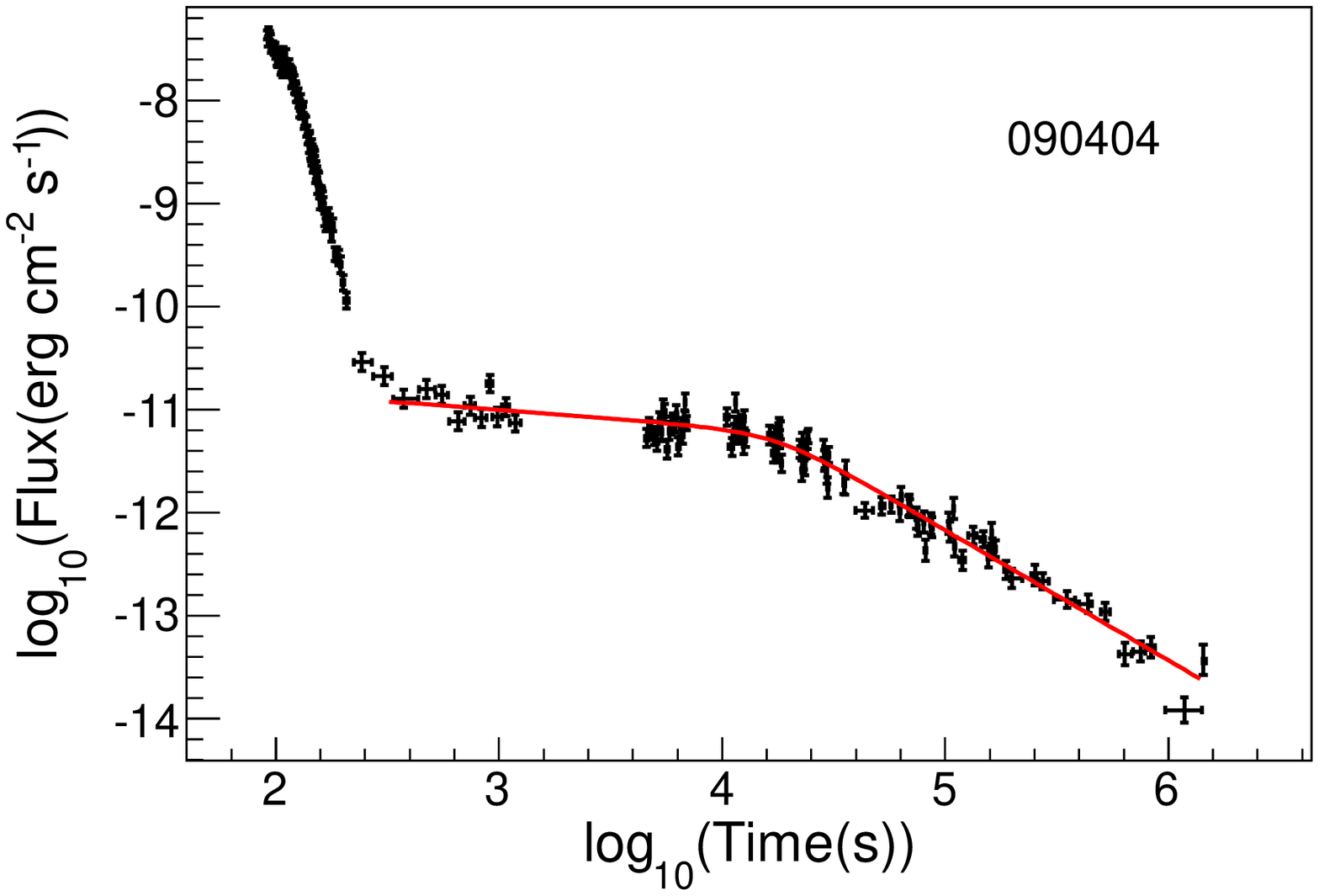}
\includegraphics[width=5.5cm,height=5cm]{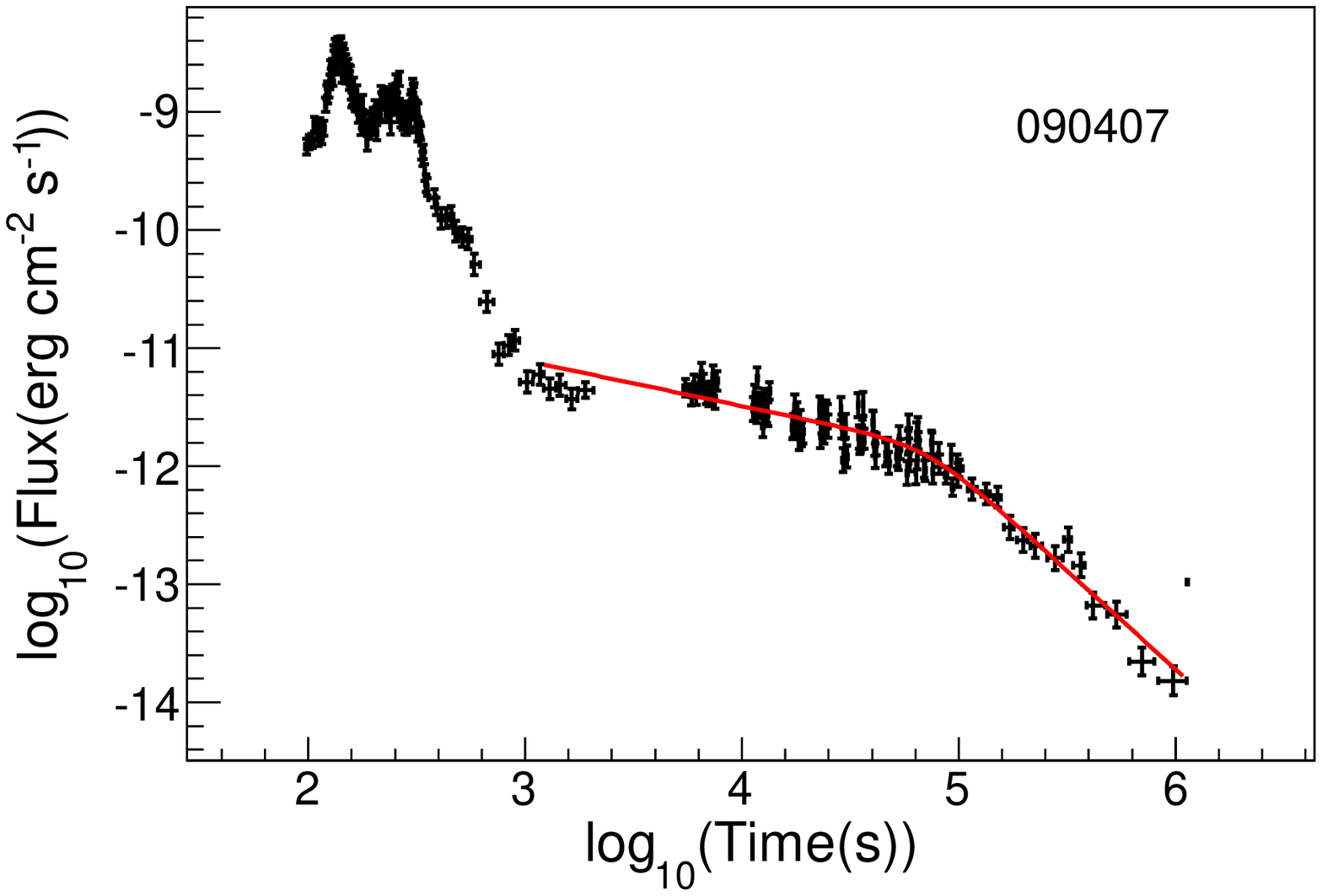}
\includegraphics[width=5.5cm,height=5cm]{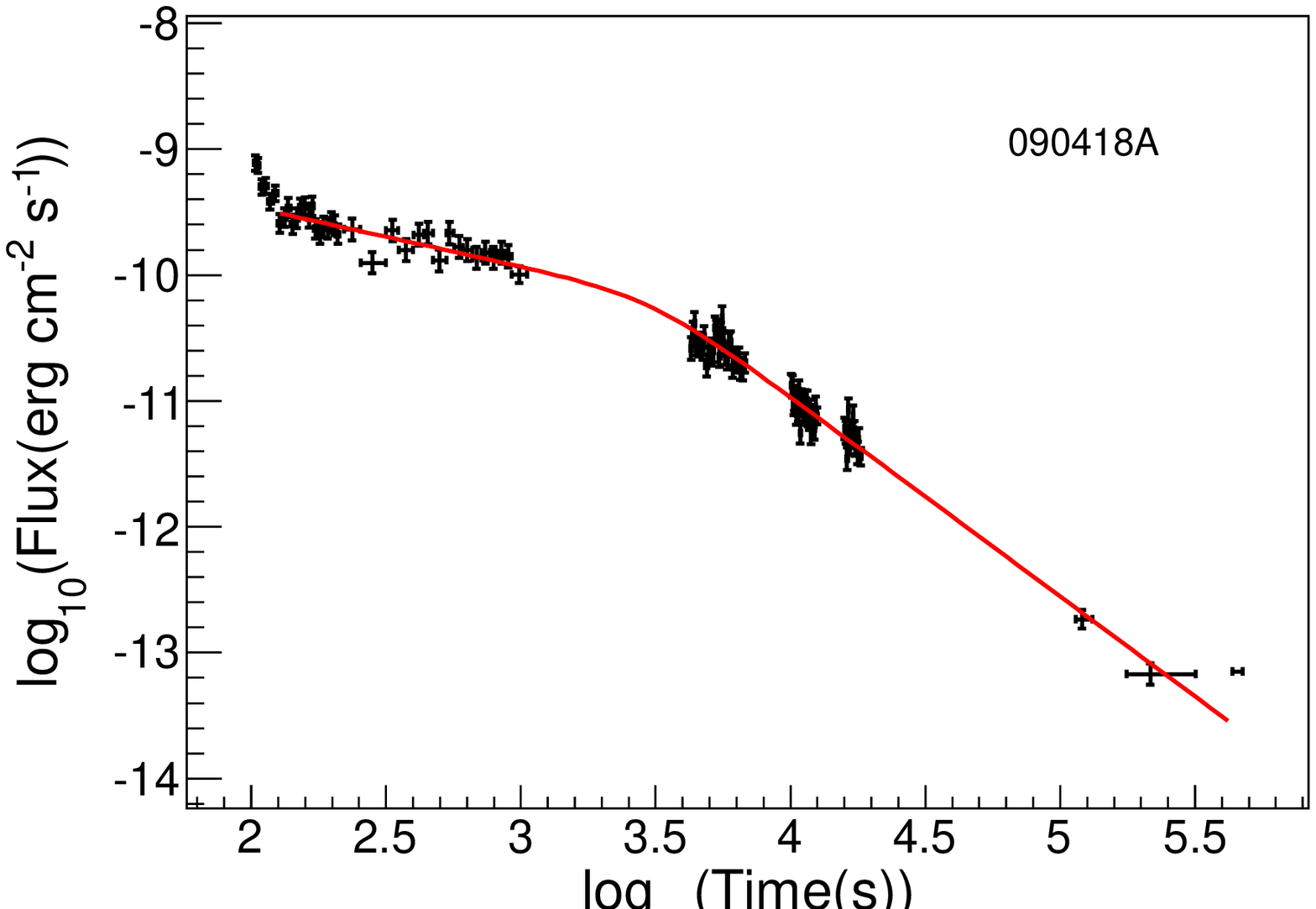}
\includegraphics[width=5.5cm,height=5cm]{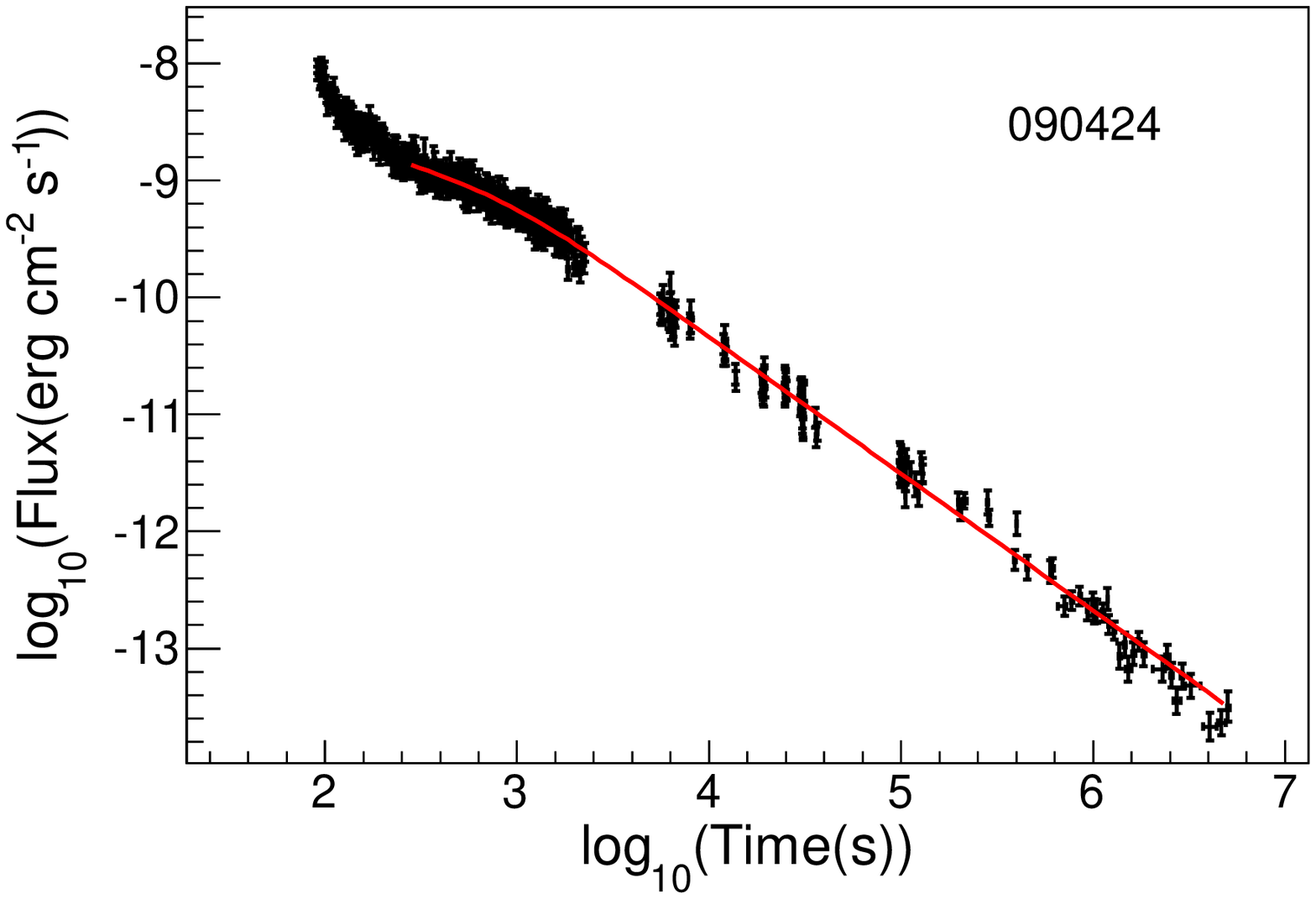}
\includegraphics[width=5.5cm,height=5cm]{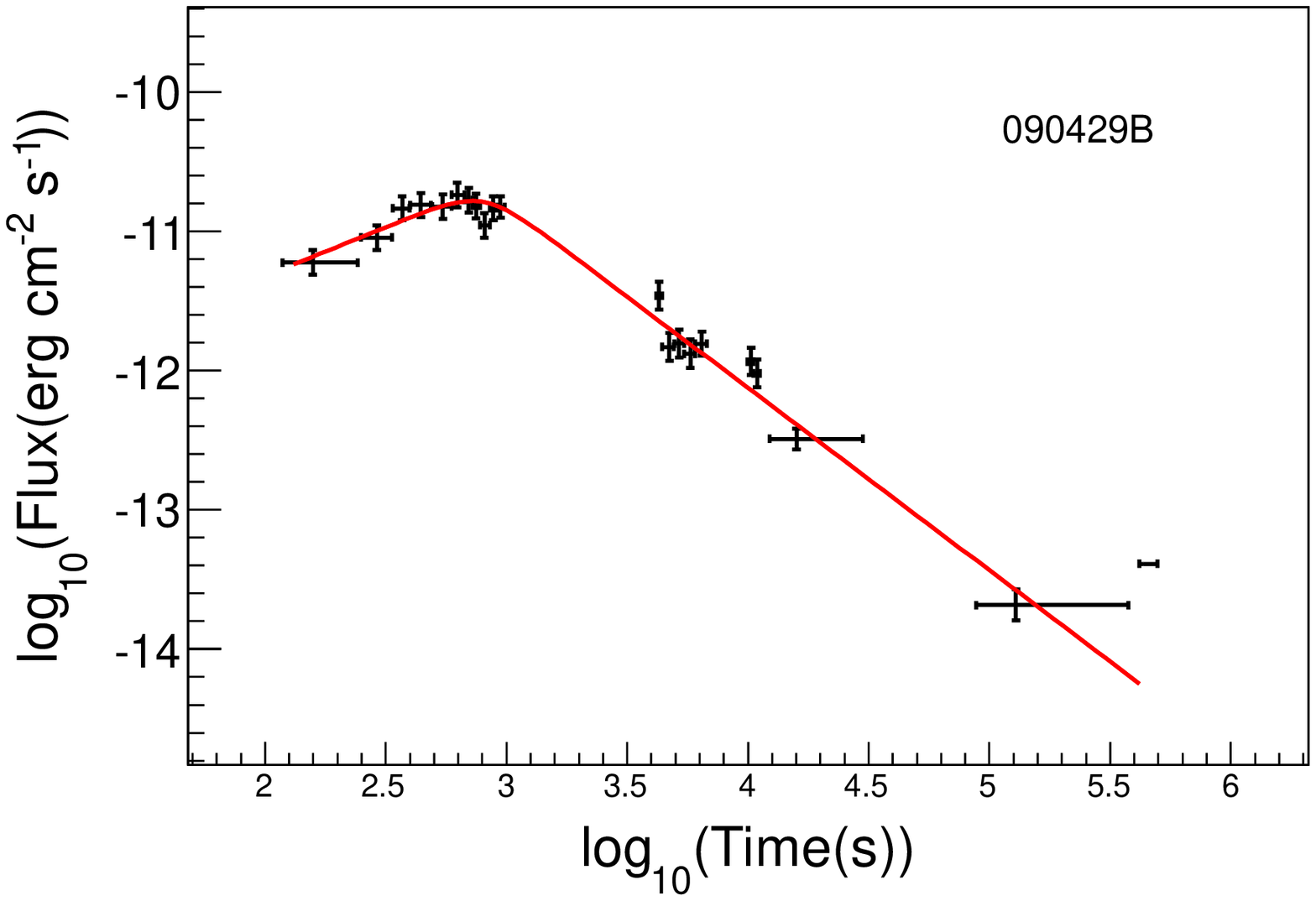}
\includegraphics[width=5.5cm,height=5cm]{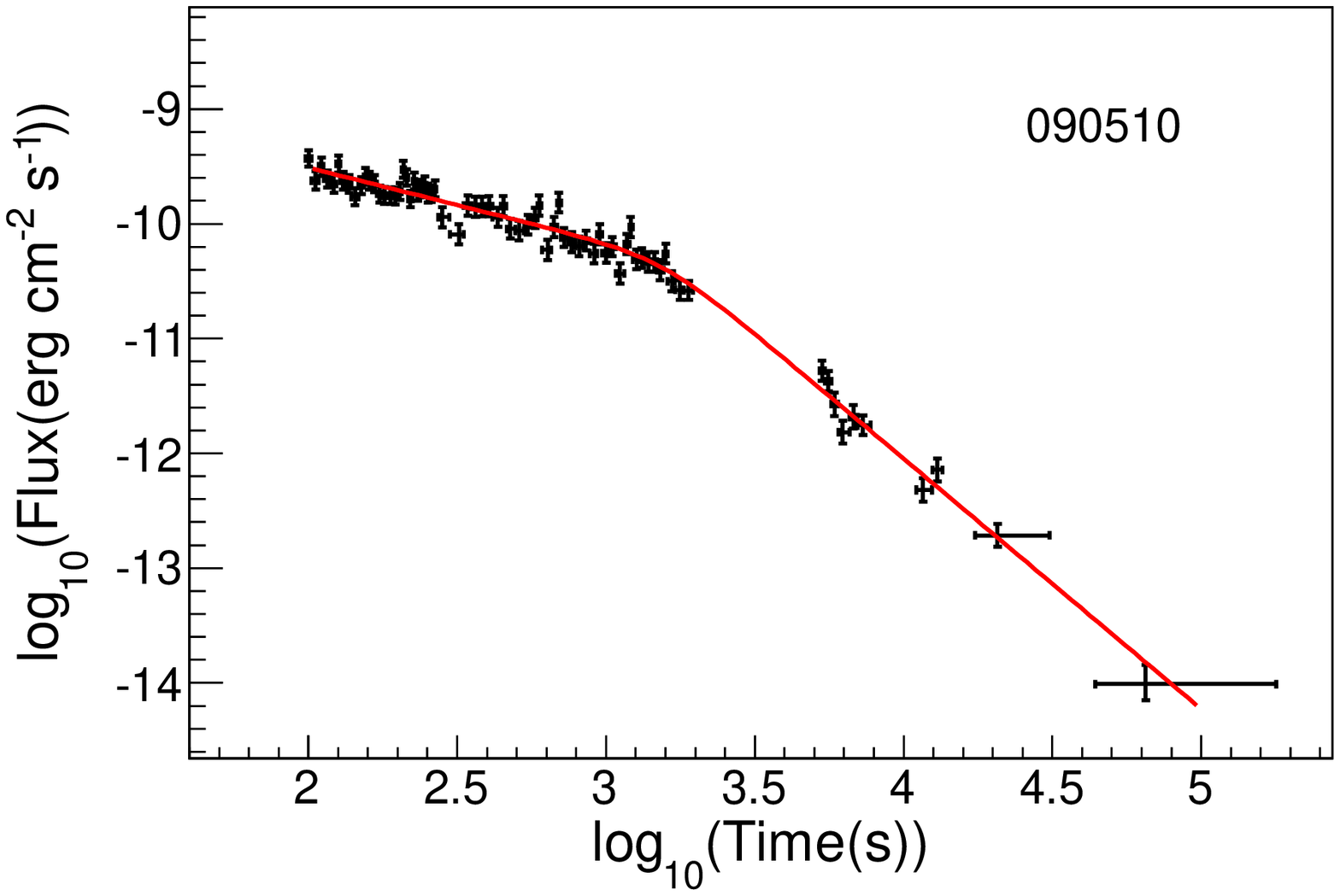}
\includegraphics[width=5.5cm,height=5cm]{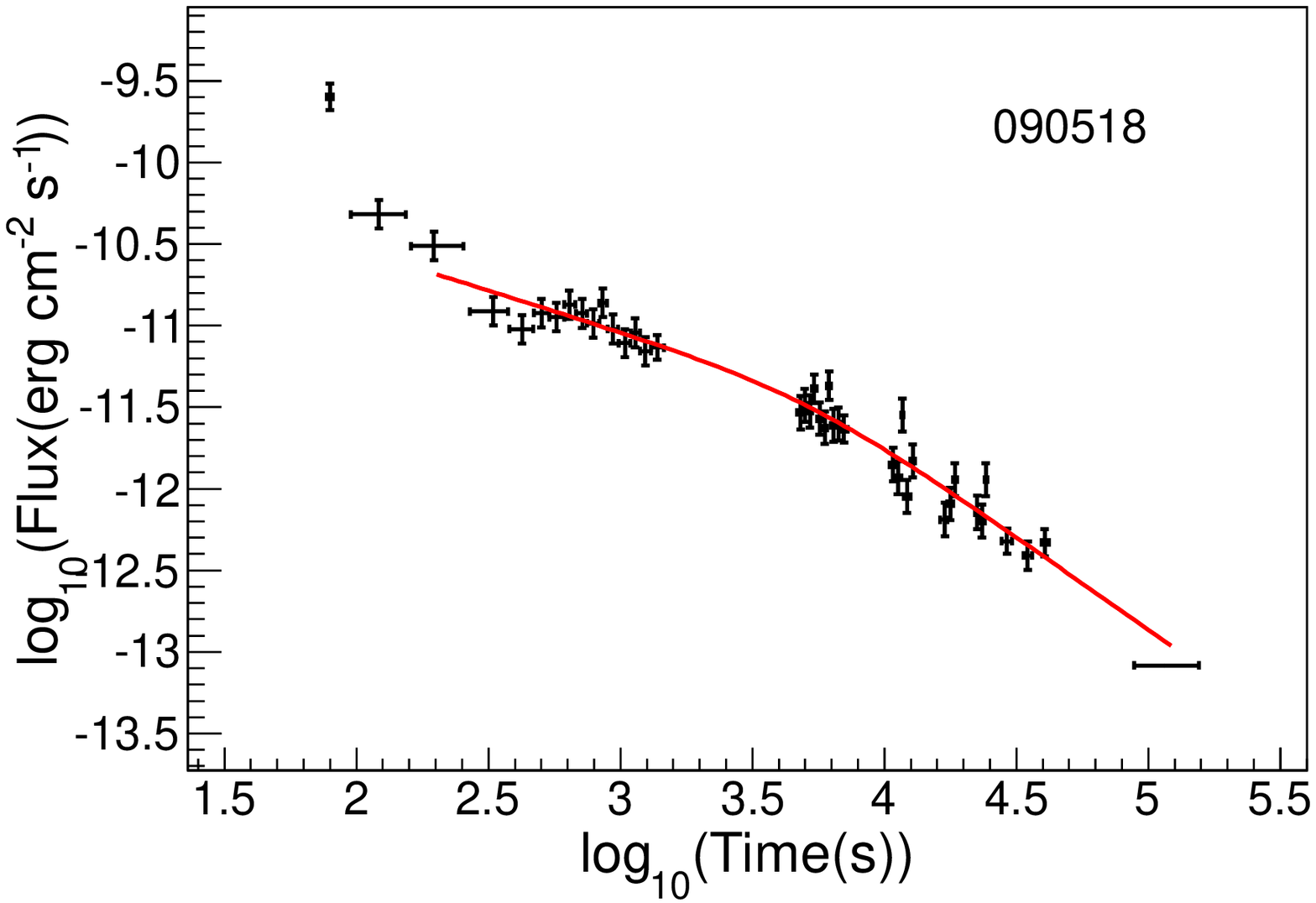}
\includegraphics[width=5.5cm,height=5cm]{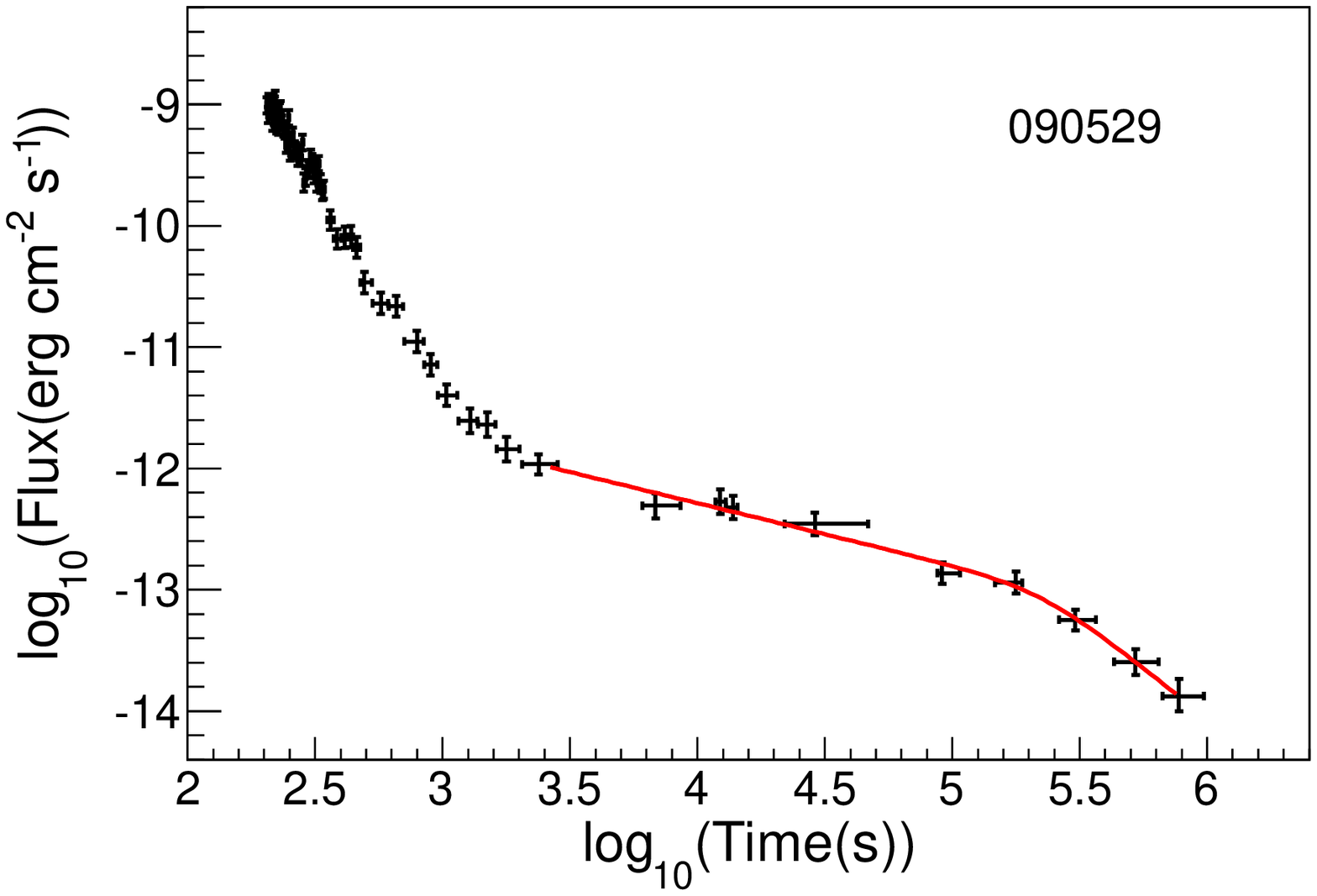}
\caption{ Continued.}
\label{fig-1-8}
\end{center}
\end{figure*}

\begin{figure*}
\begin{center}
\setlength{\abovecaptionskip}{0.cm}
\setlength{\belowcaptionskip}{-0.cm}
\figurenum{1}
\hspace{0cm}
\graphicspath{{lightcurve/}}
\includegraphics[width=5.5cm,height=5cm]{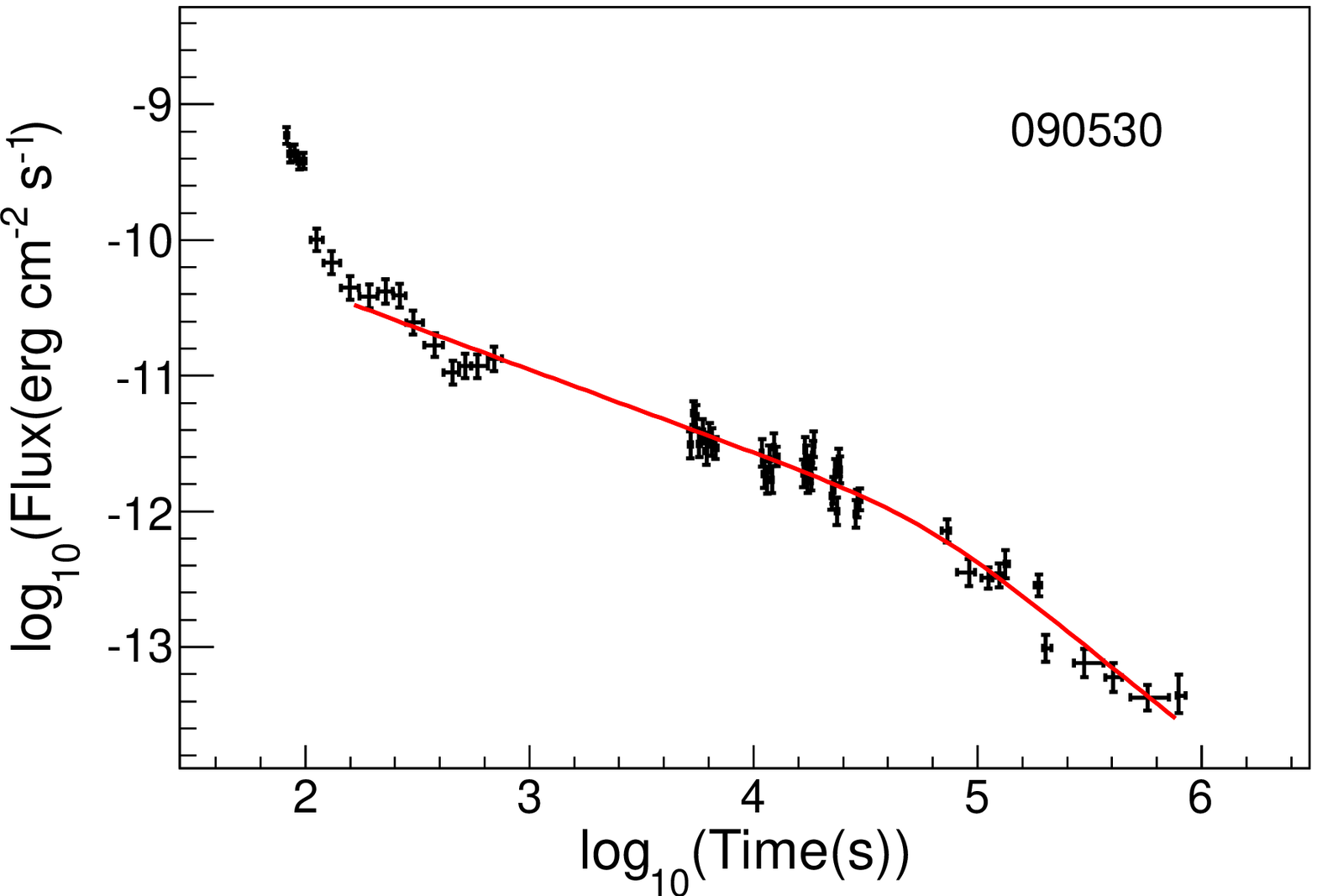}
\includegraphics[width=5.5cm,height=5cm]{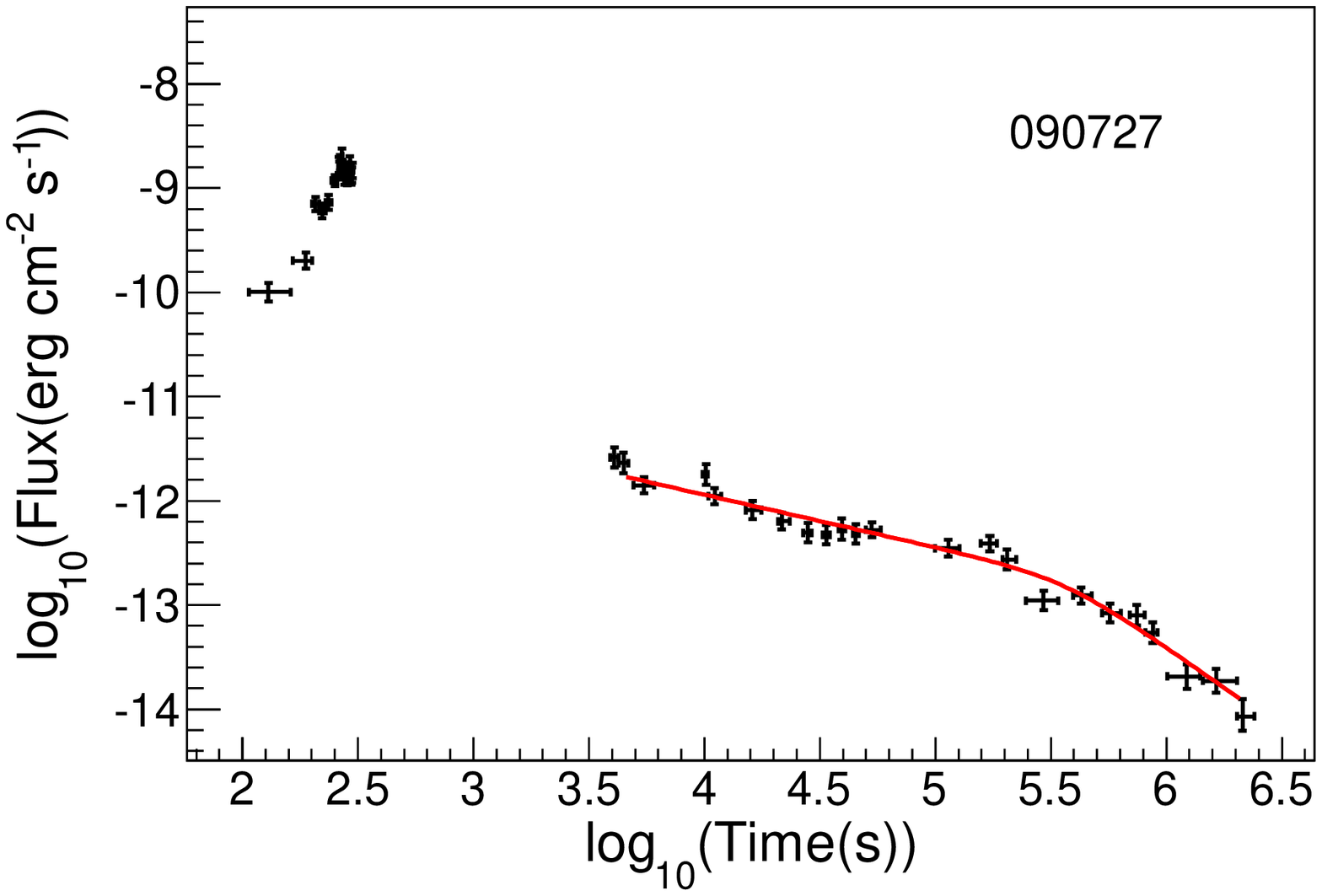}
\includegraphics[width=5.5cm,height=5cm]{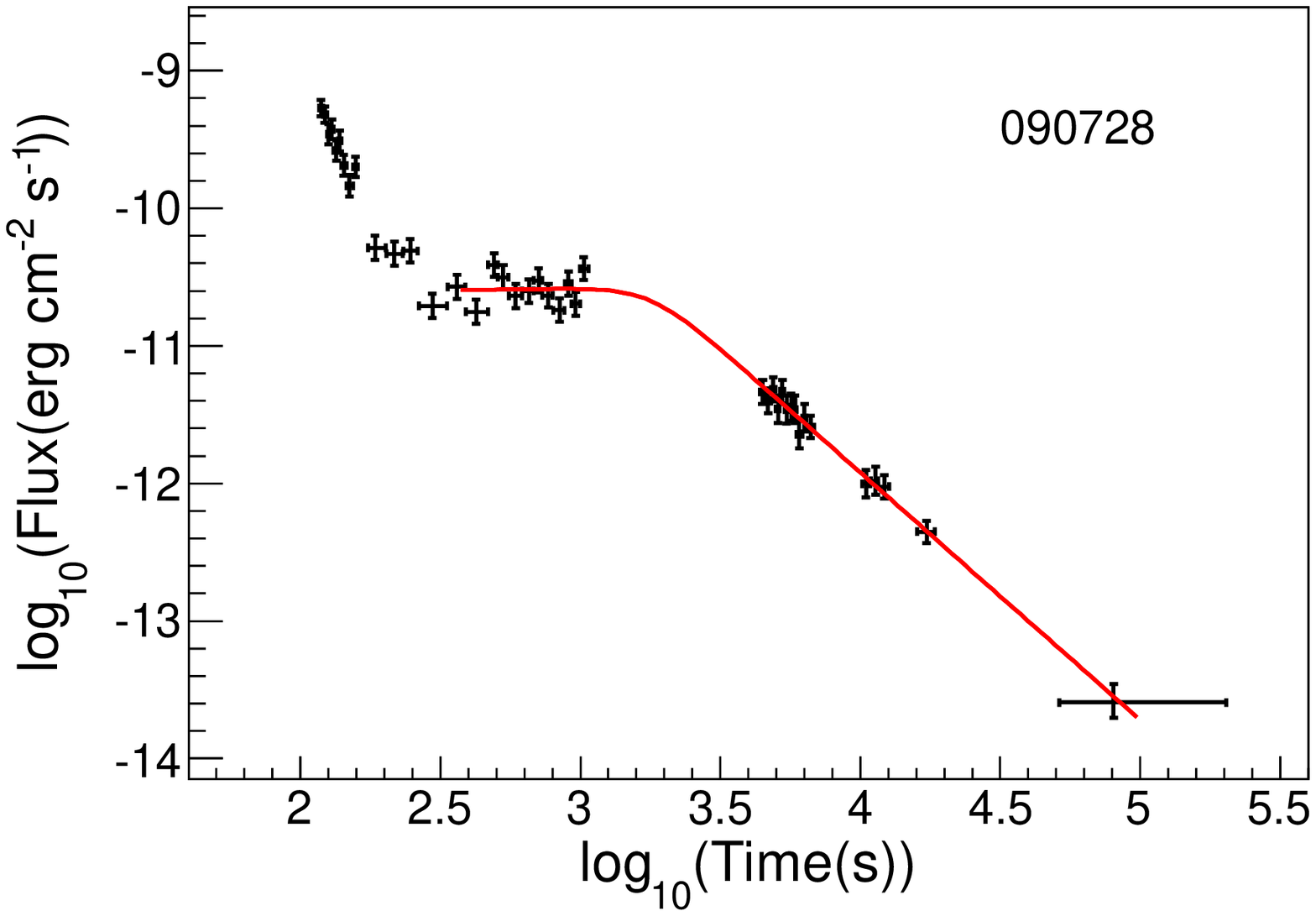}
\includegraphics[width=5.5cm,height=5cm]{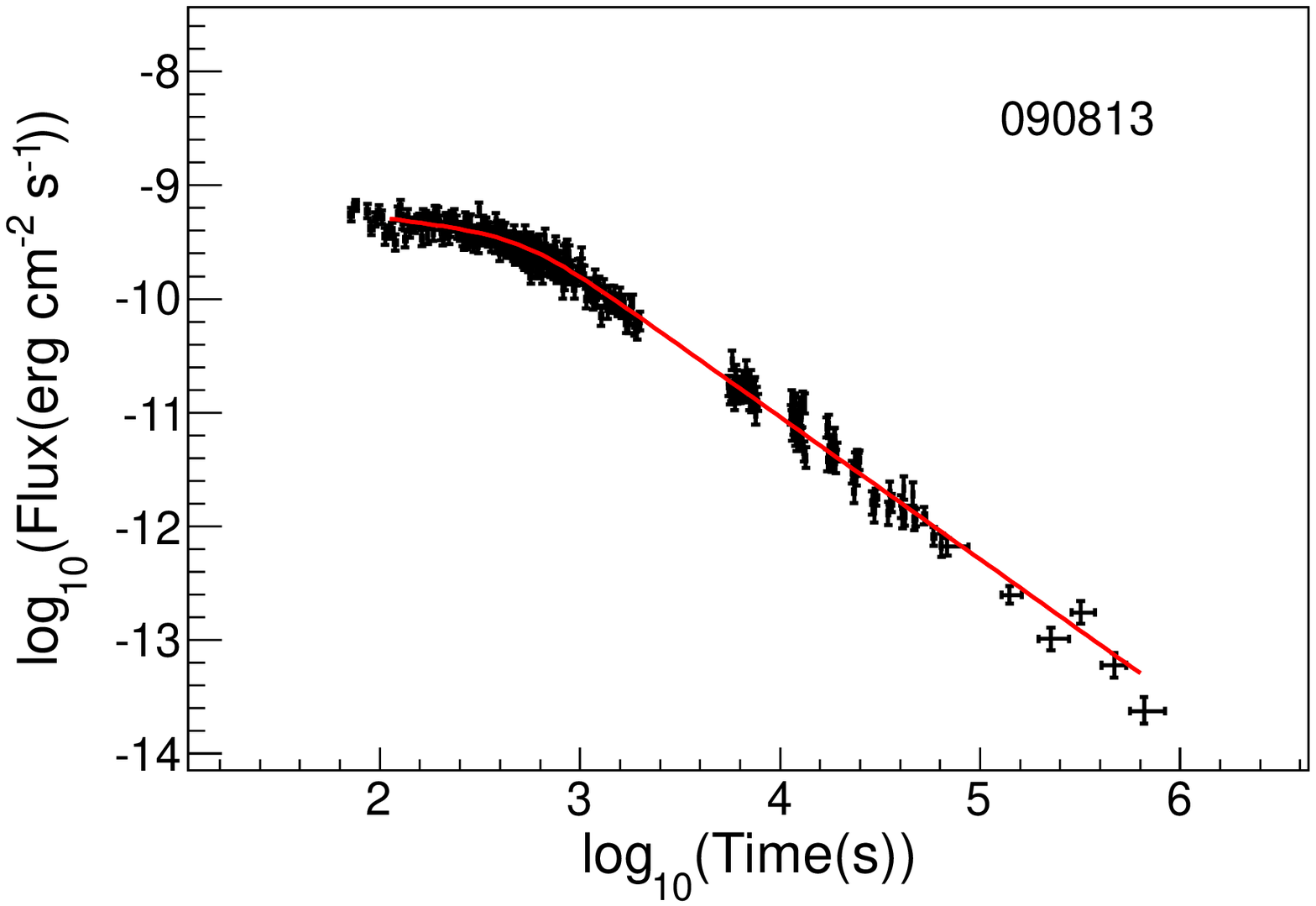}
\includegraphics[width=5.5cm,height=5cm]{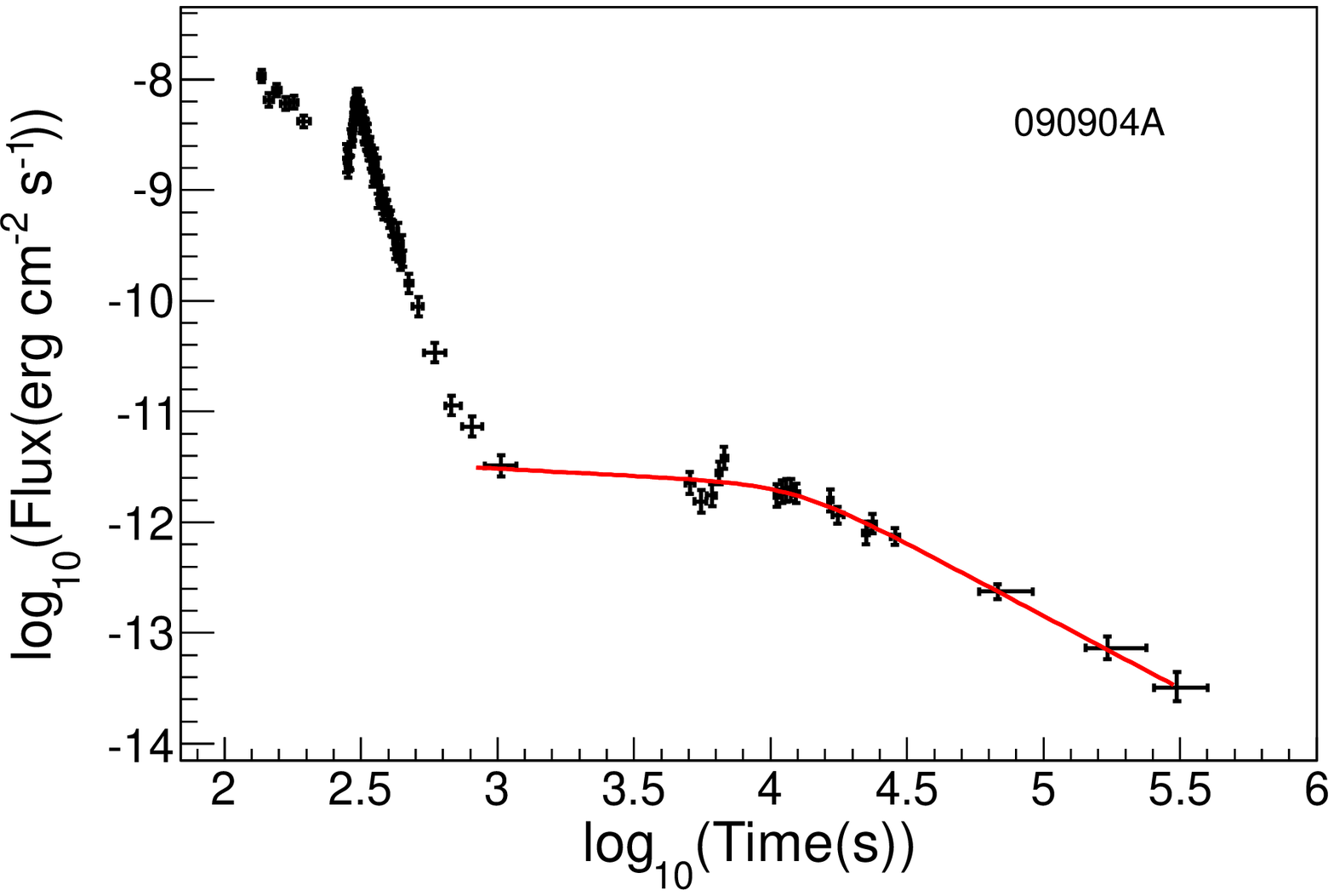}
\includegraphics[width=5.5cm,height=5cm]{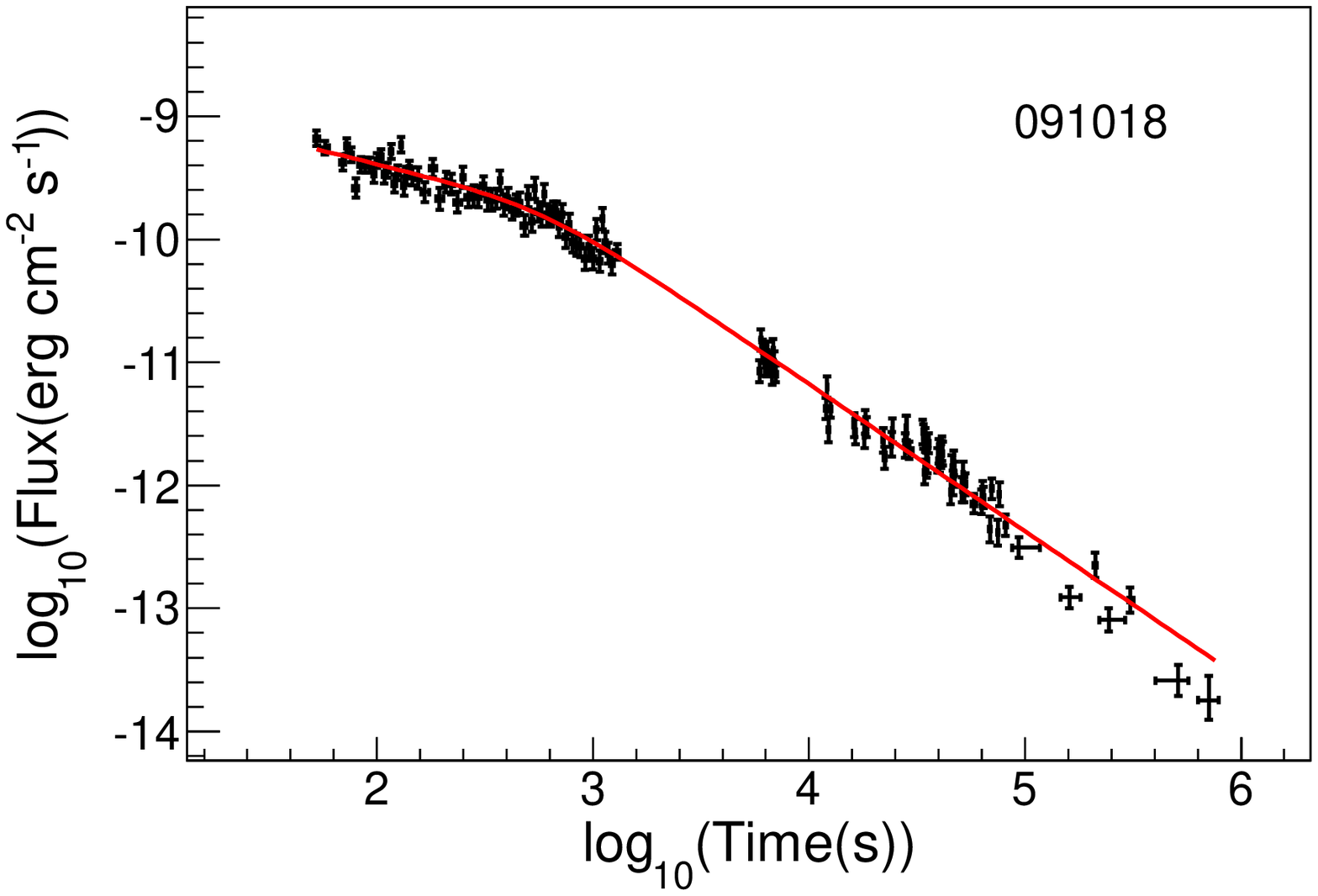}
\includegraphics[width=5.5cm,height=5cm]{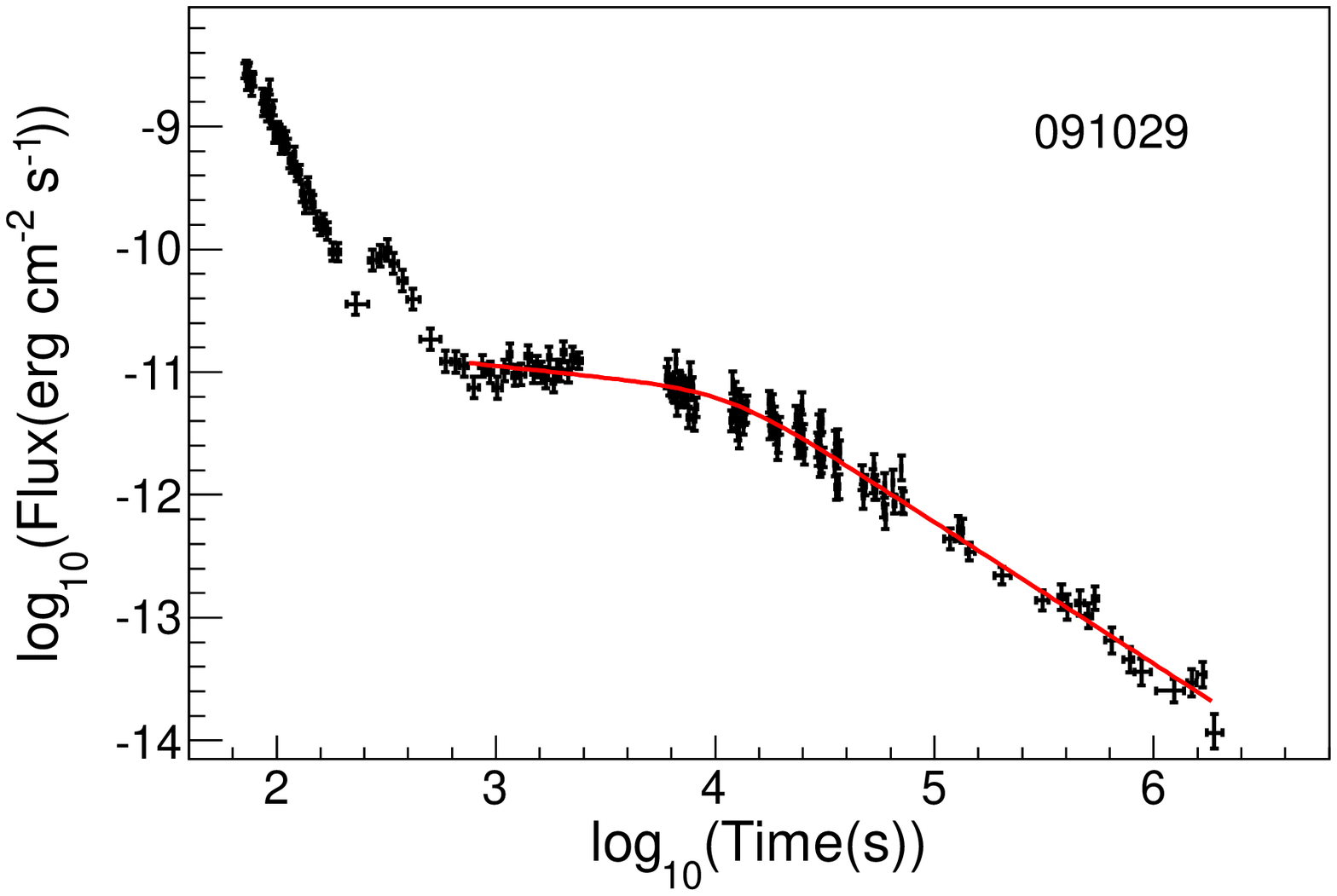}
\includegraphics[width=5.5cm,height=5cm]{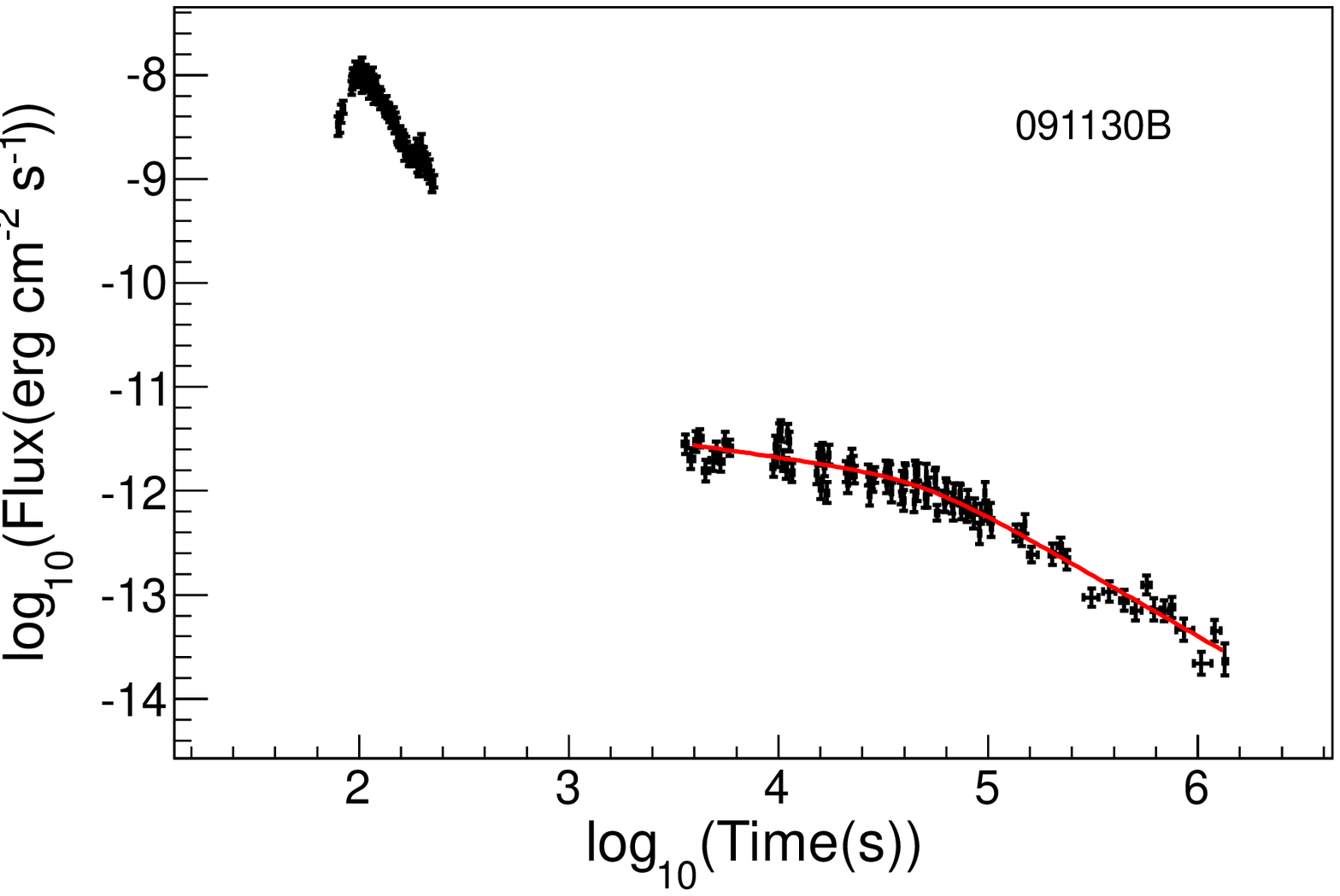}
\includegraphics[width=5.5cm,height=5cm]{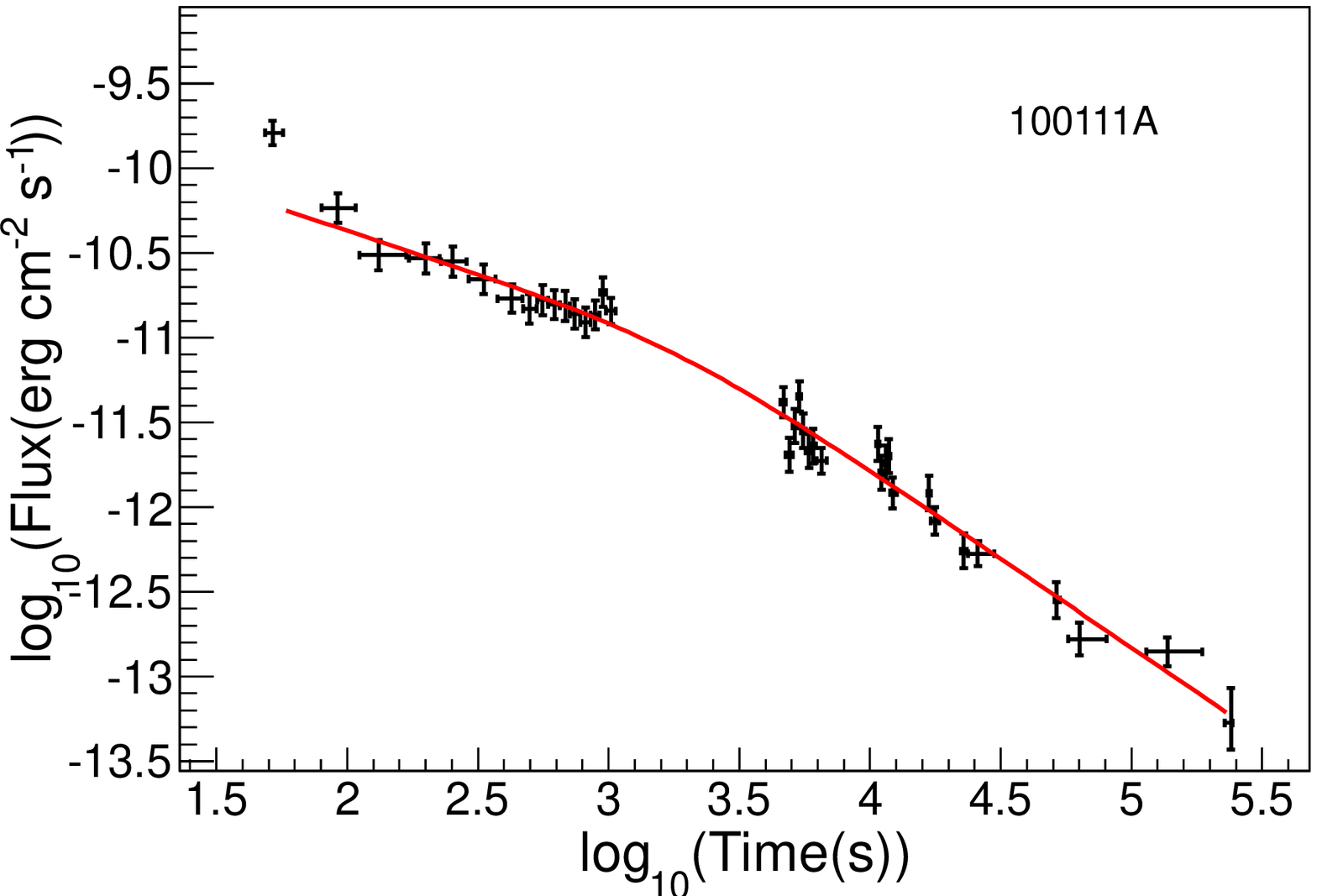}
\includegraphics[width=5.5cm,height=5cm]{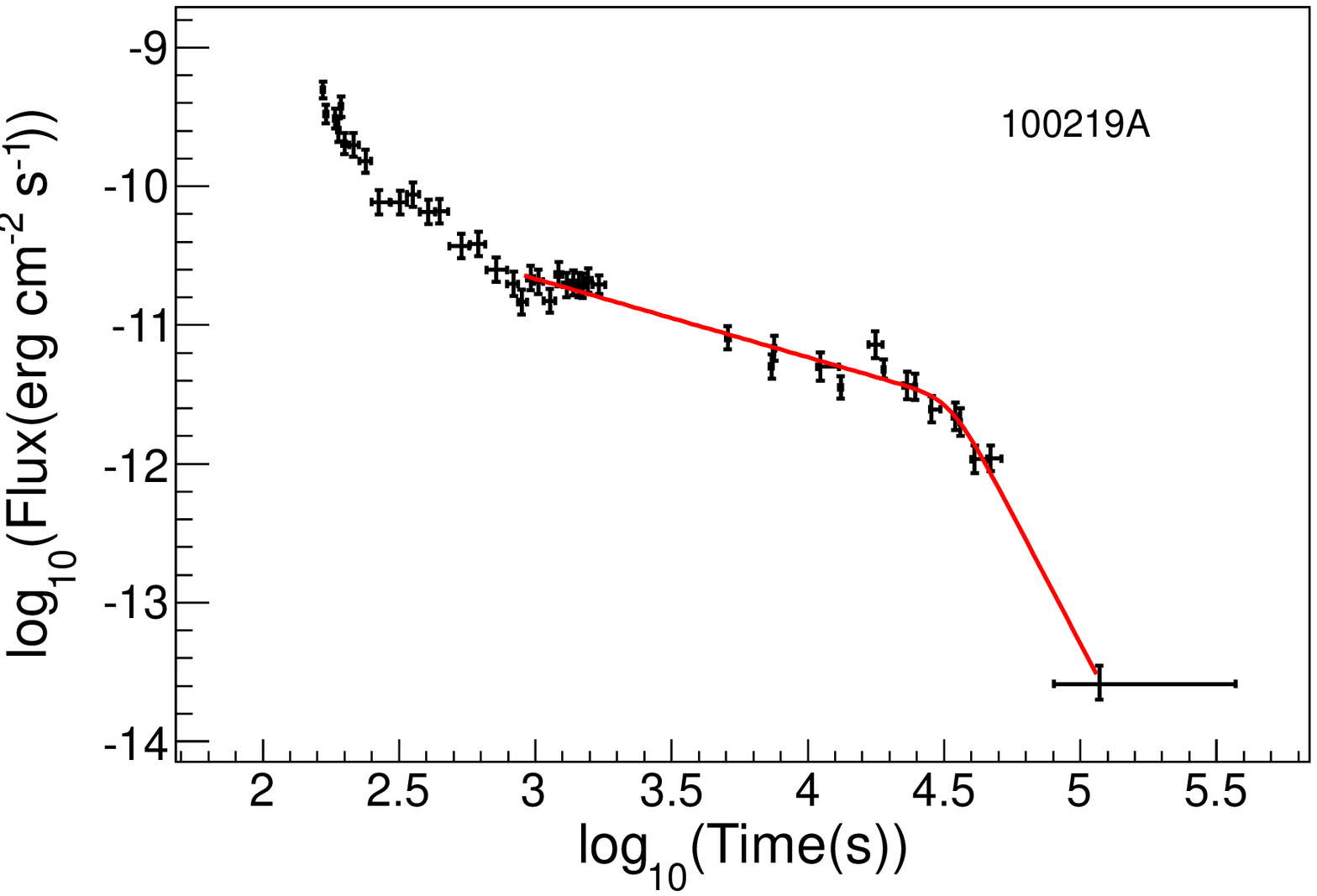}
\includegraphics[width=5.5cm,height=5cm]{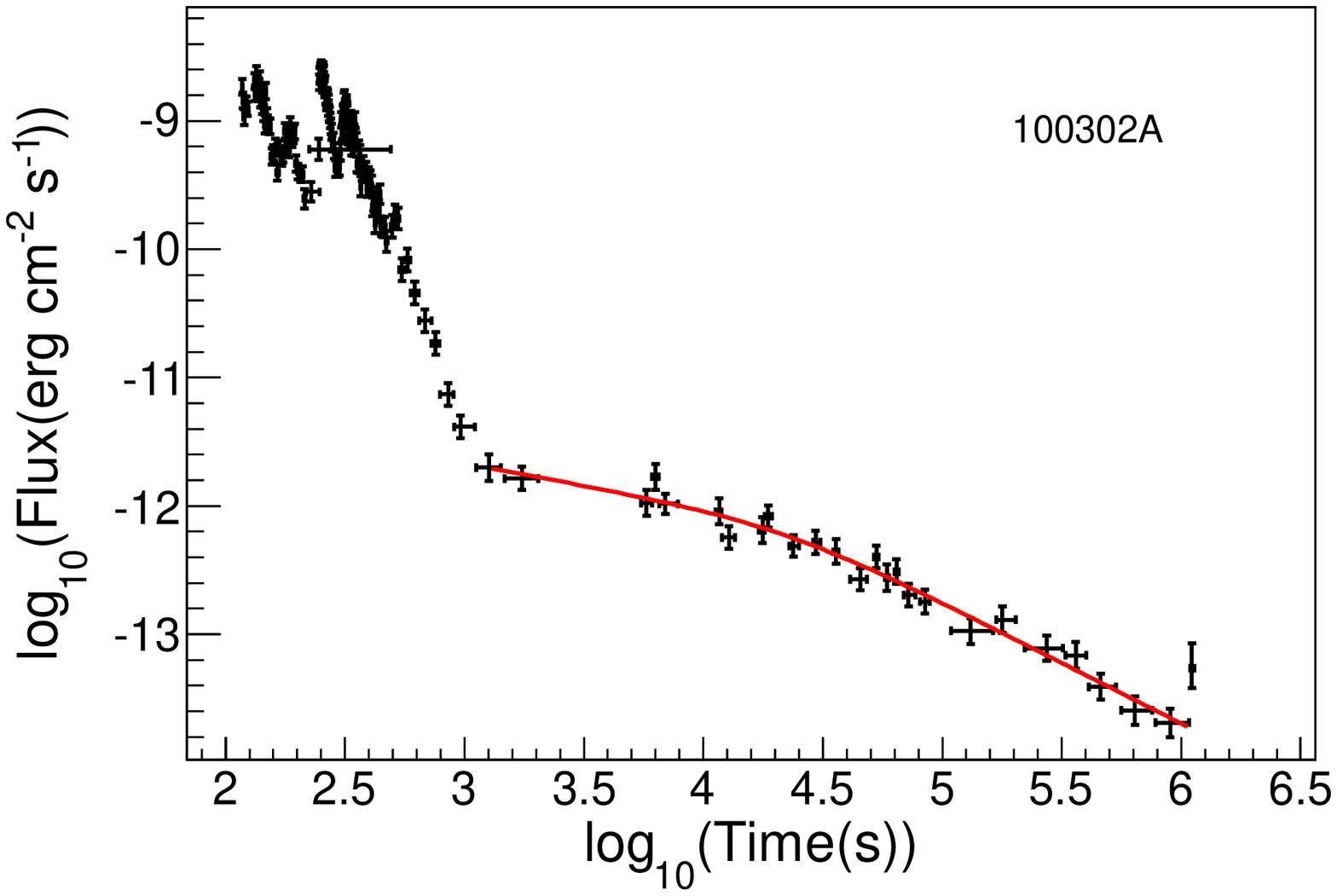}
\includegraphics[width=5.5cm,height=5cm]{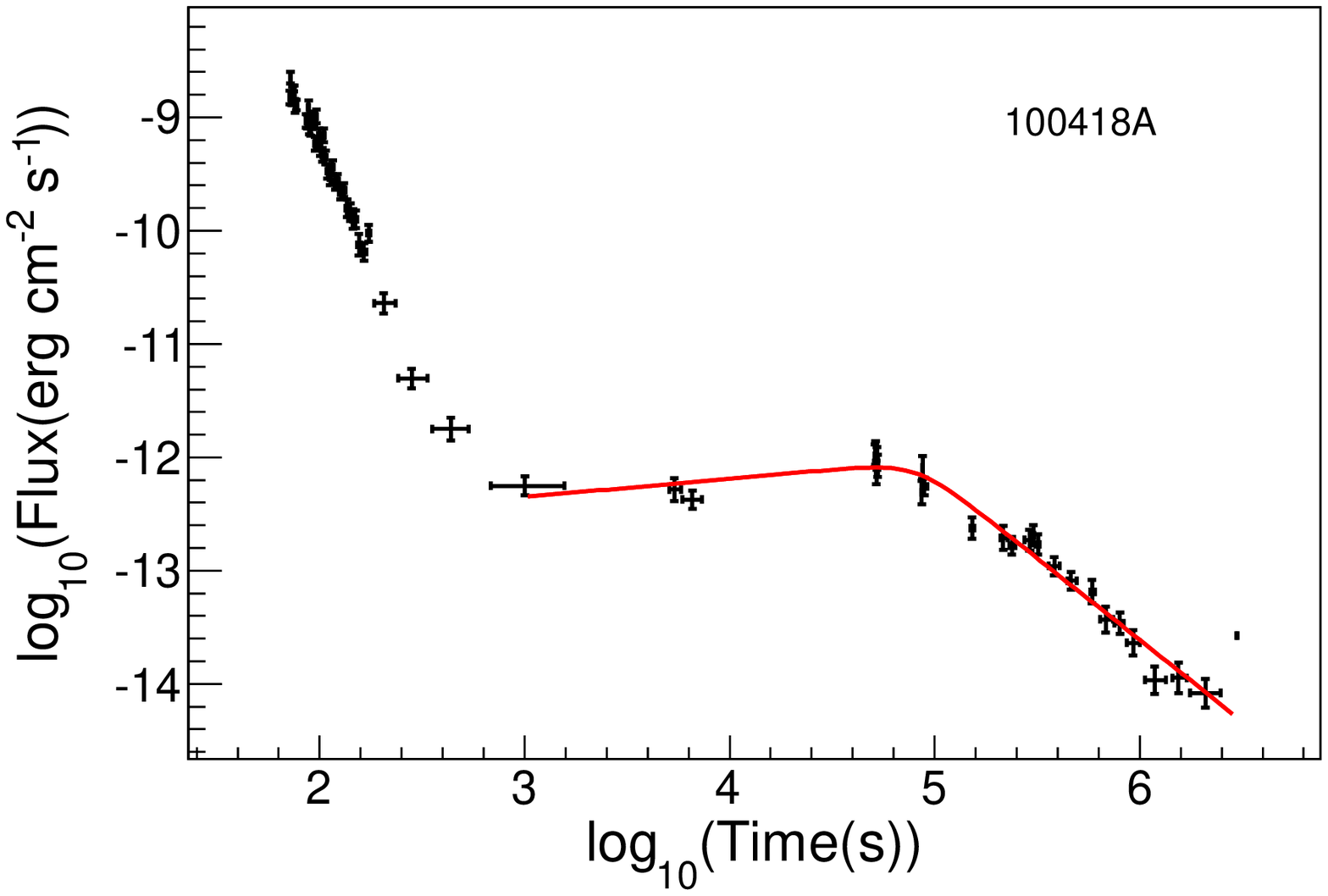}
\caption{ Continued.}
\label{fig-1-9}
\end{center}
\end{figure*}

\begin{figure*}
\begin{center}
\setlength{\abovecaptionskip}{0.cm}
\setlength{\belowcaptionskip}{-0.cm}
\figurenum{1}
\hspace{0cm}
\graphicspath{{lightcurve/}}
\includegraphics[width=5.5cm,height=5cm]{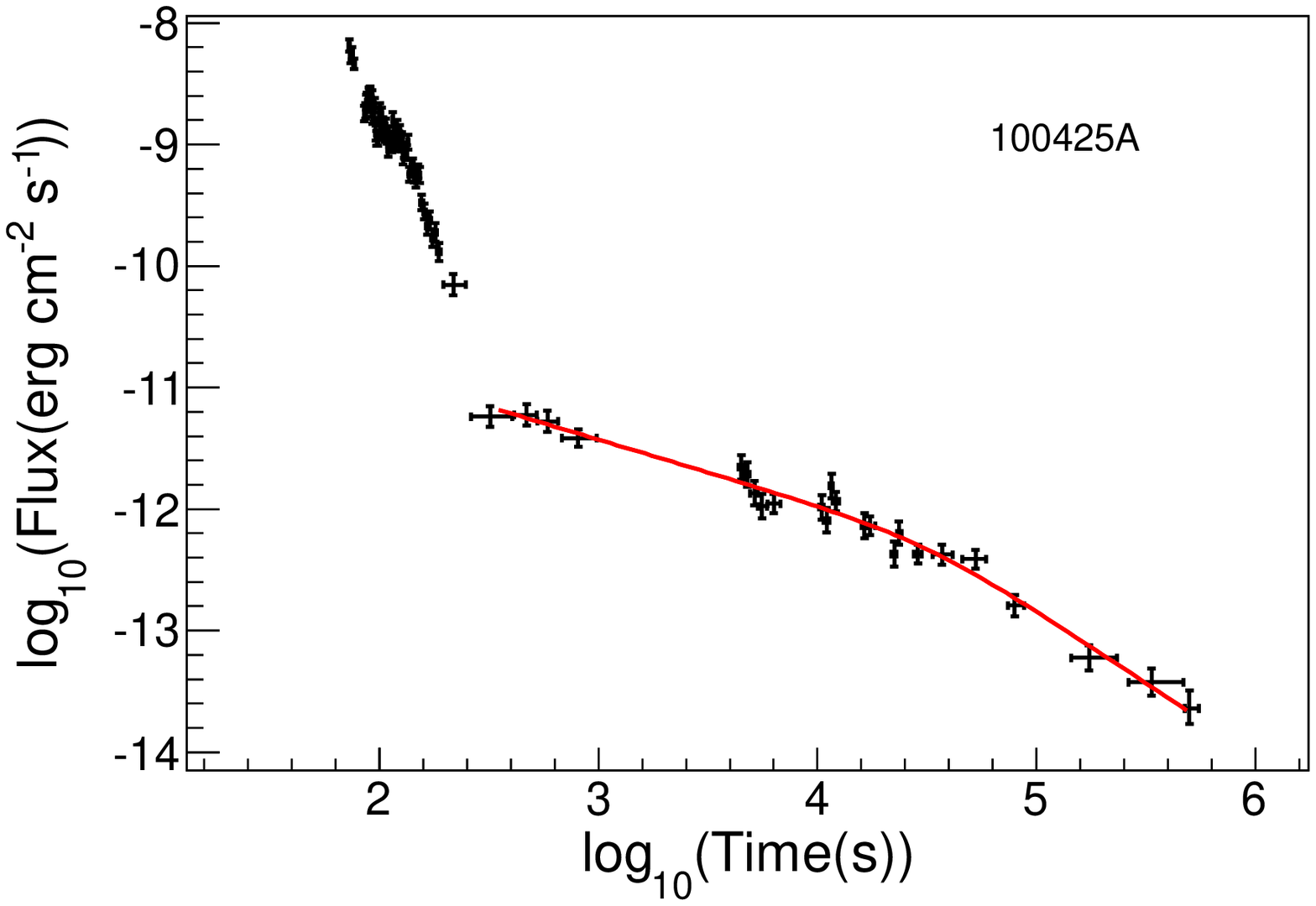}
\includegraphics[width=5.5cm,height=5cm]{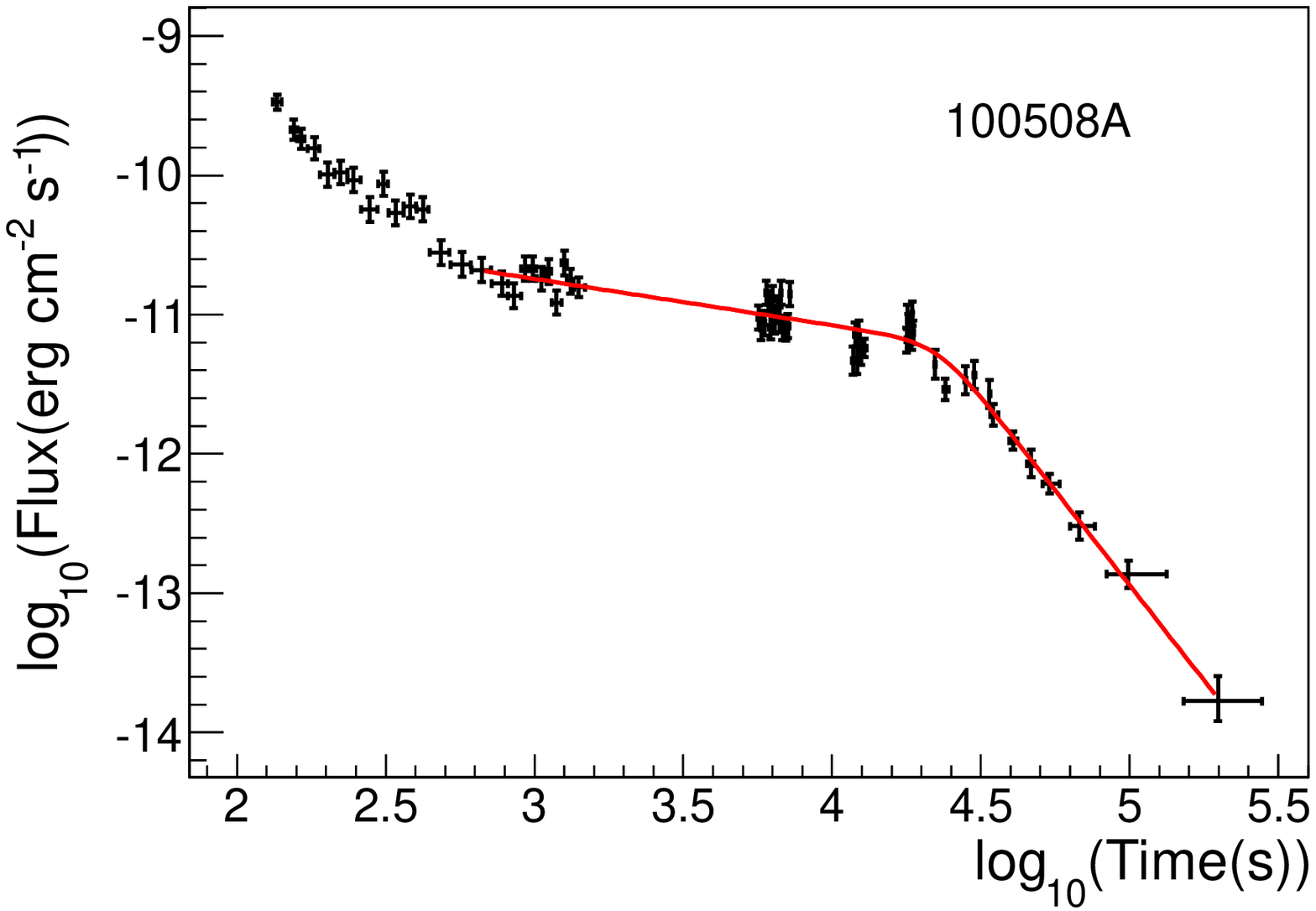}
\includegraphics[width=5.5cm,height=5cm]{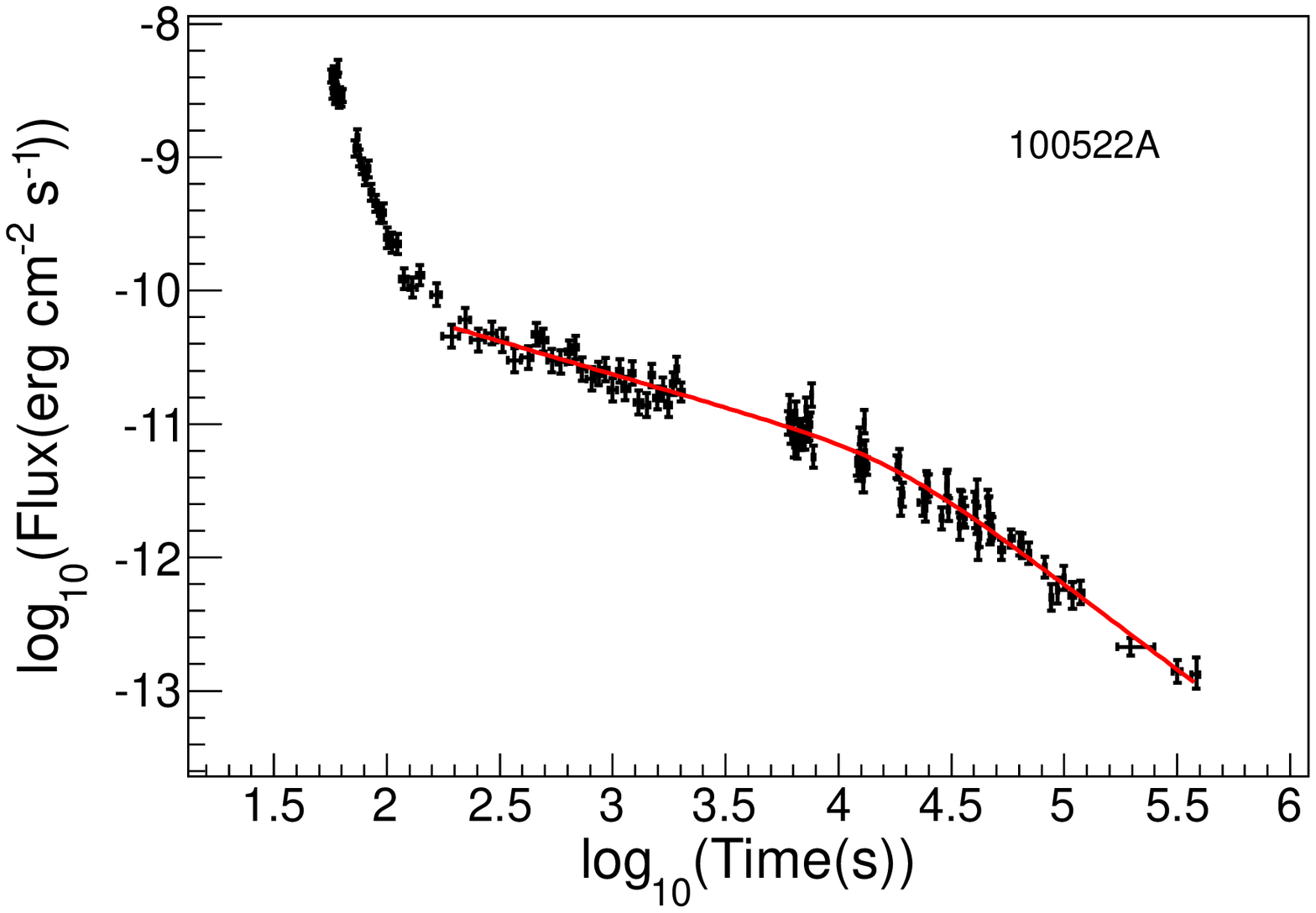}
\includegraphics[width=5.5cm,height=5cm]{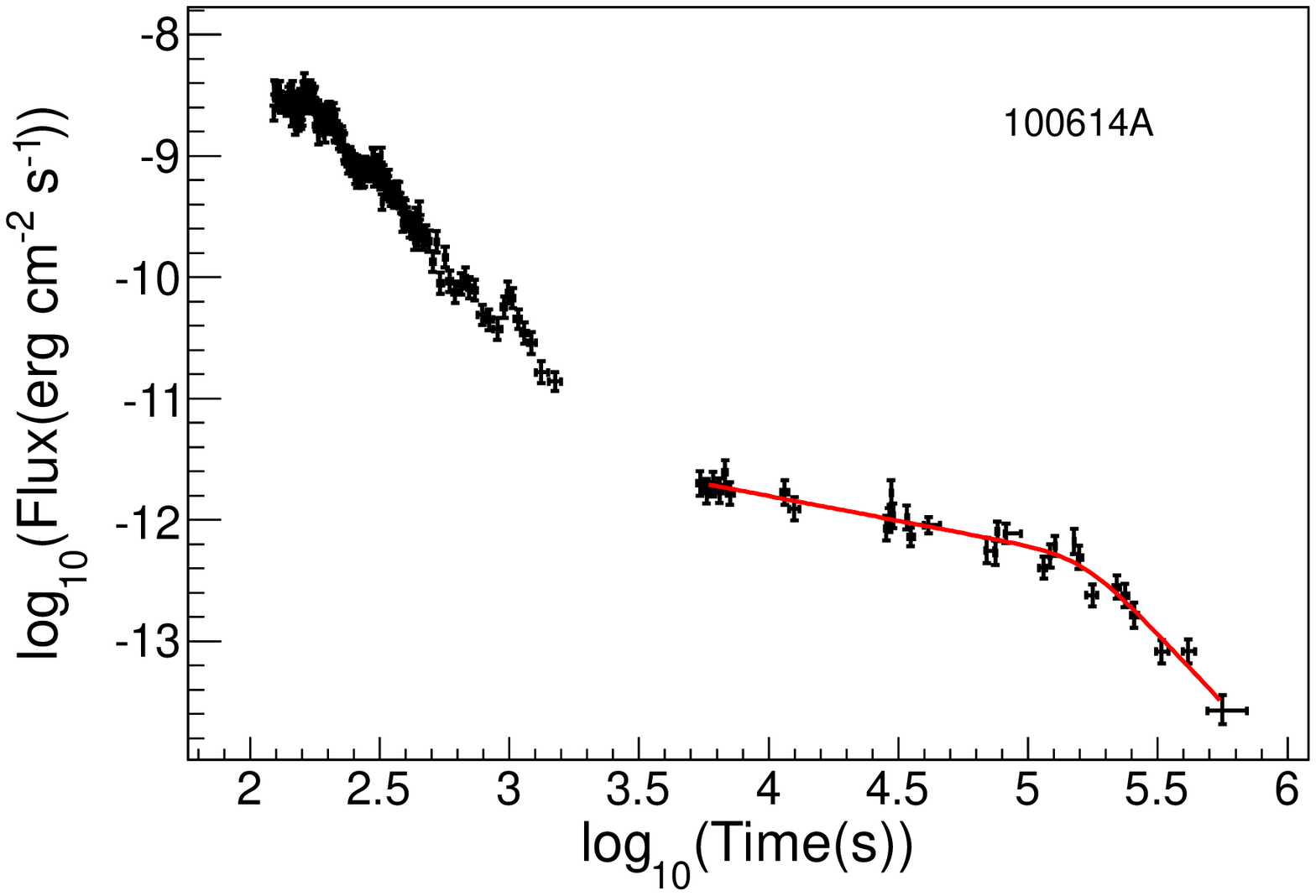}
\includegraphics[width=5.5cm,height=5cm]{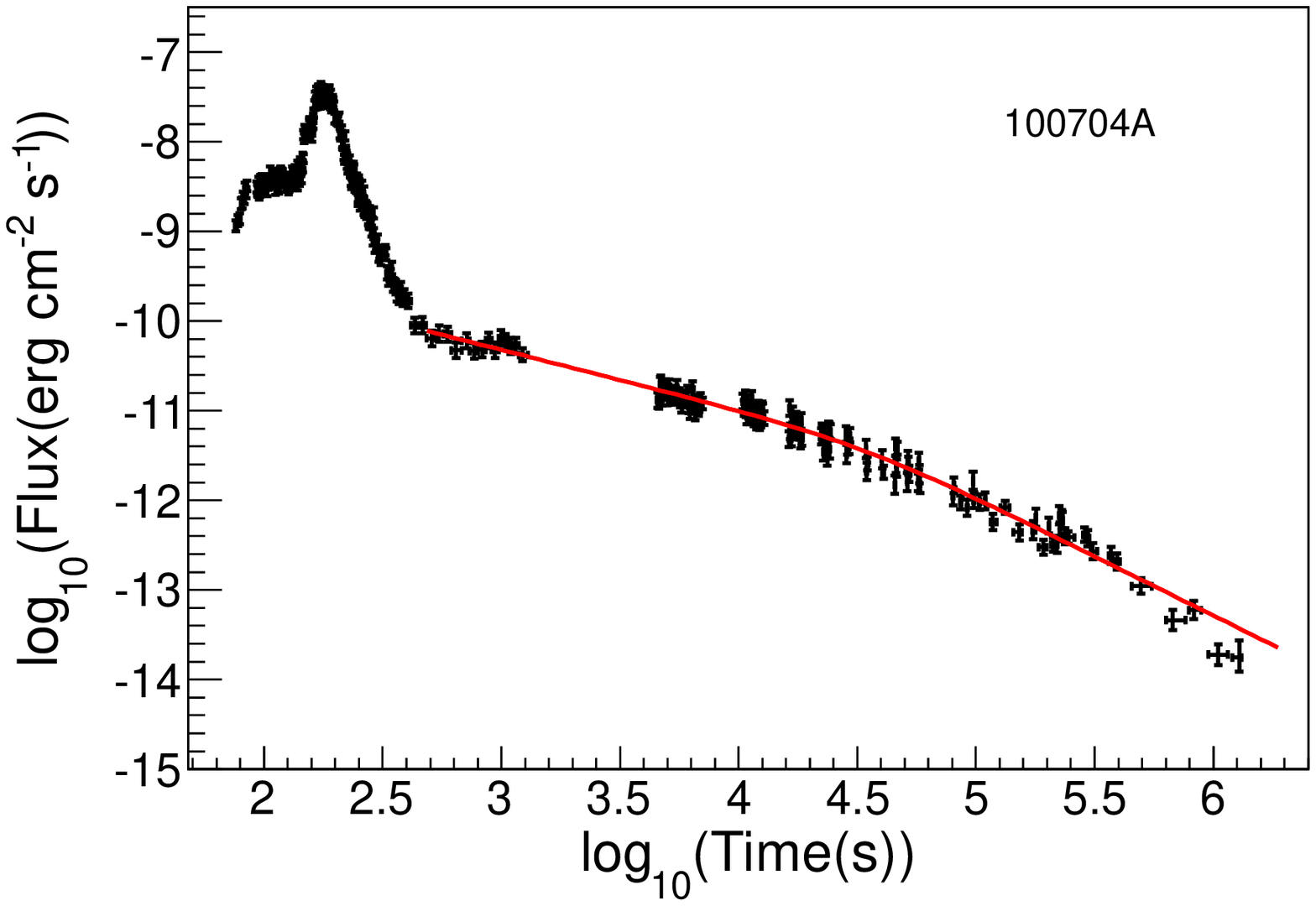}
\includegraphics[width=5.5cm,height=5cm]{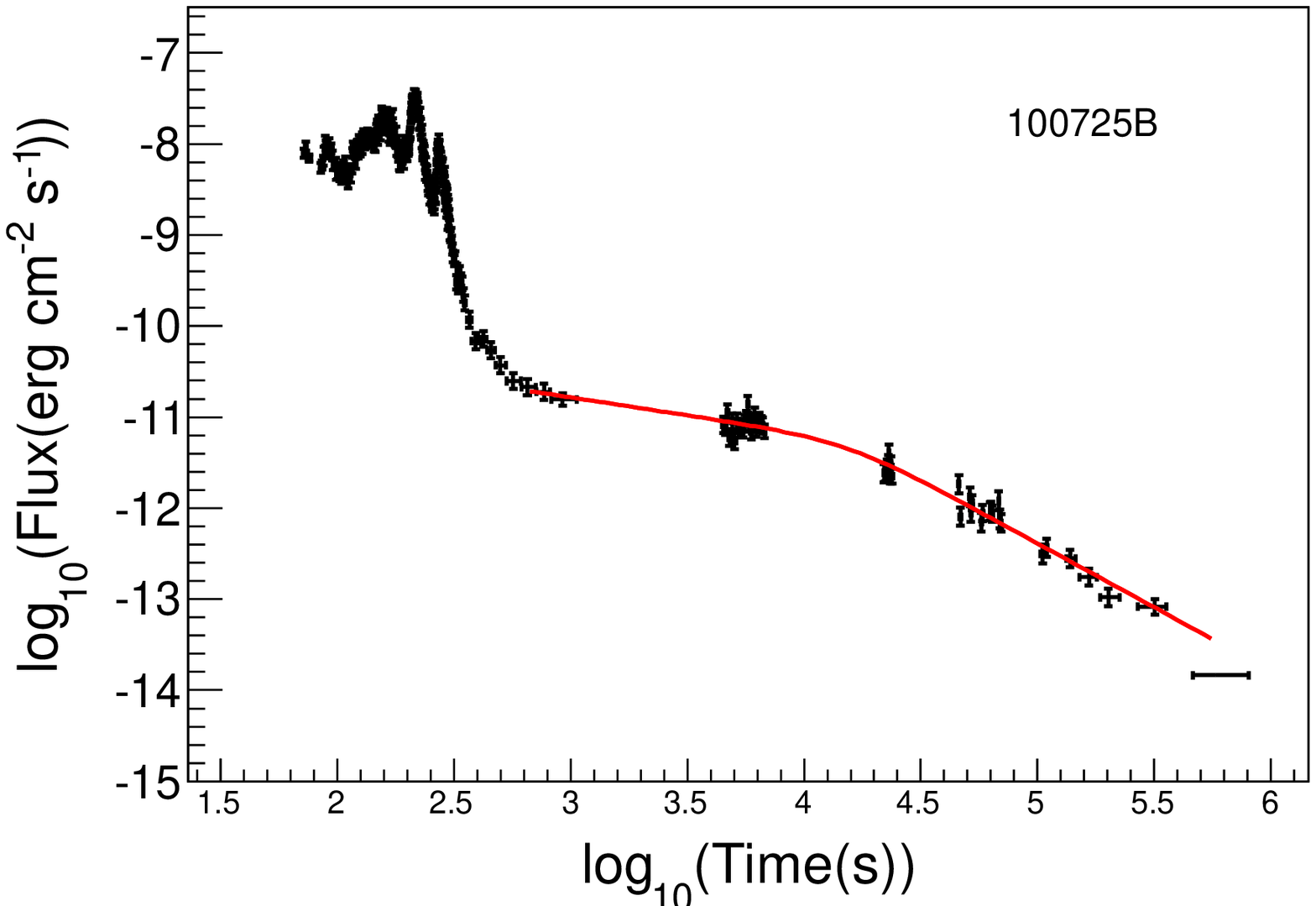}
\includegraphics[width=5.5cm,height=5cm]{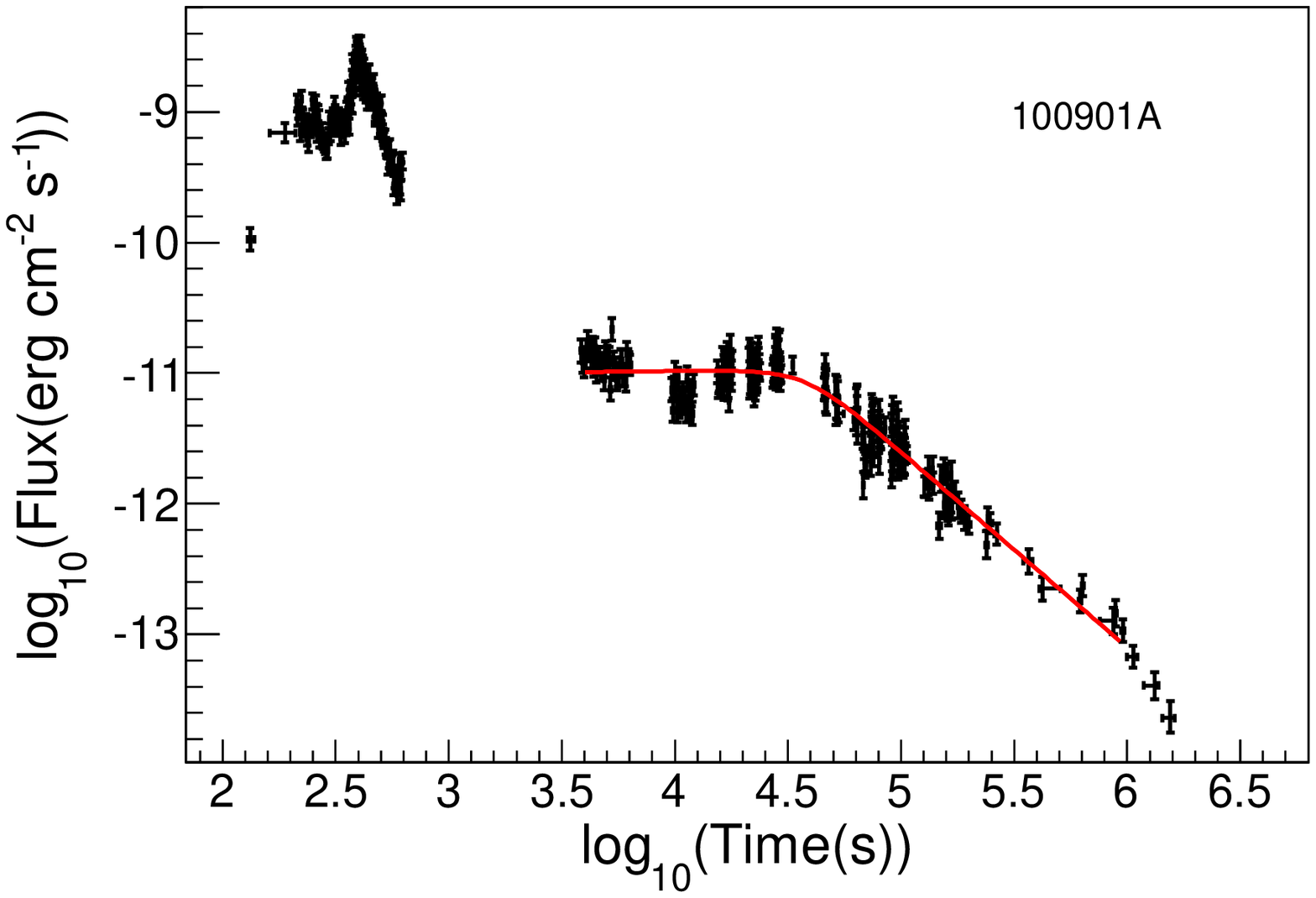}
\includegraphics[width=5.5cm,height=5cm]{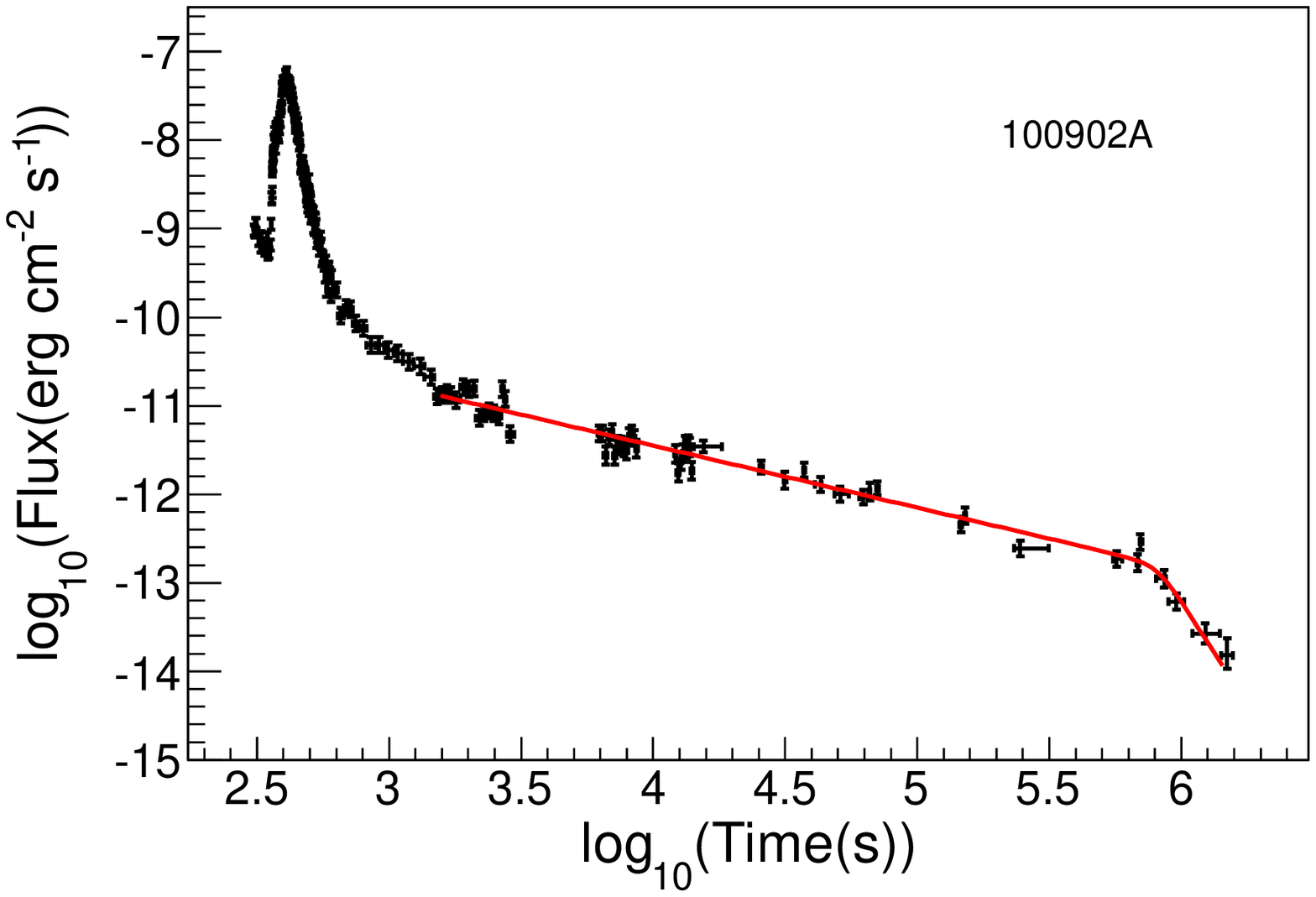}
\includegraphics[width=5.5cm,height=5cm]{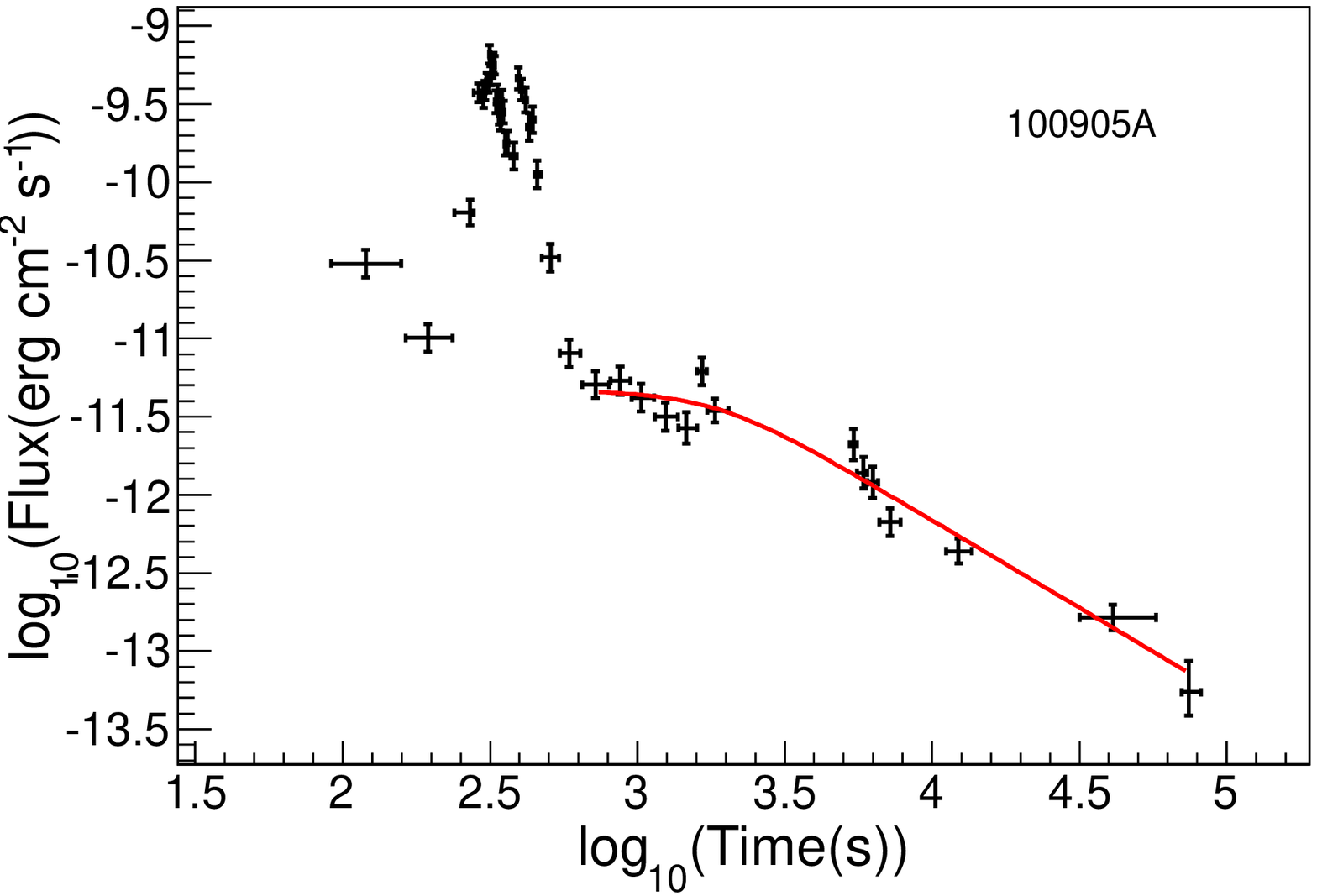}
\includegraphics[width=5.5cm,height=5cm]{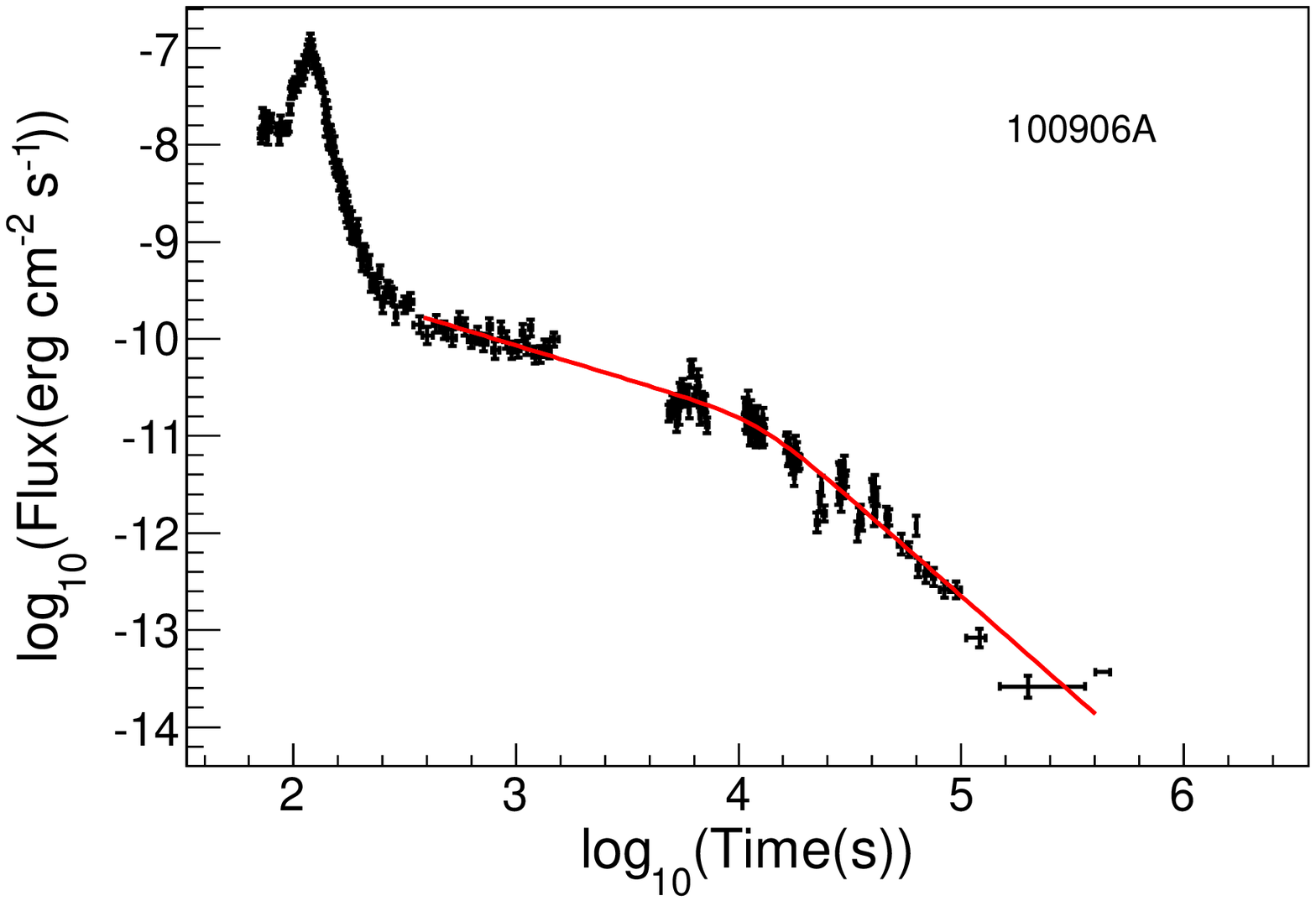}
\includegraphics[width=5.5cm,height=5cm]{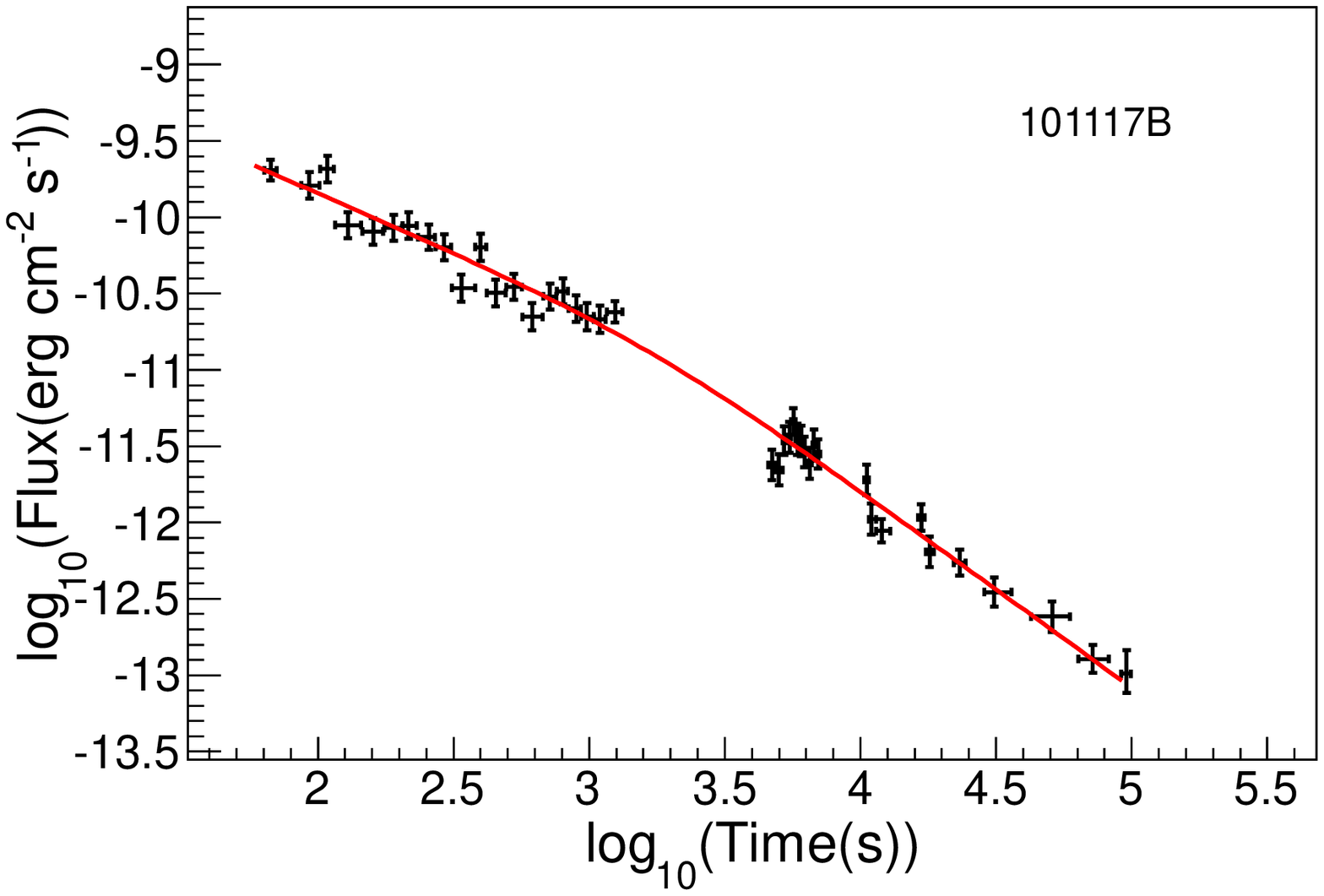}
\includegraphics[width=5.5cm,height=5cm]{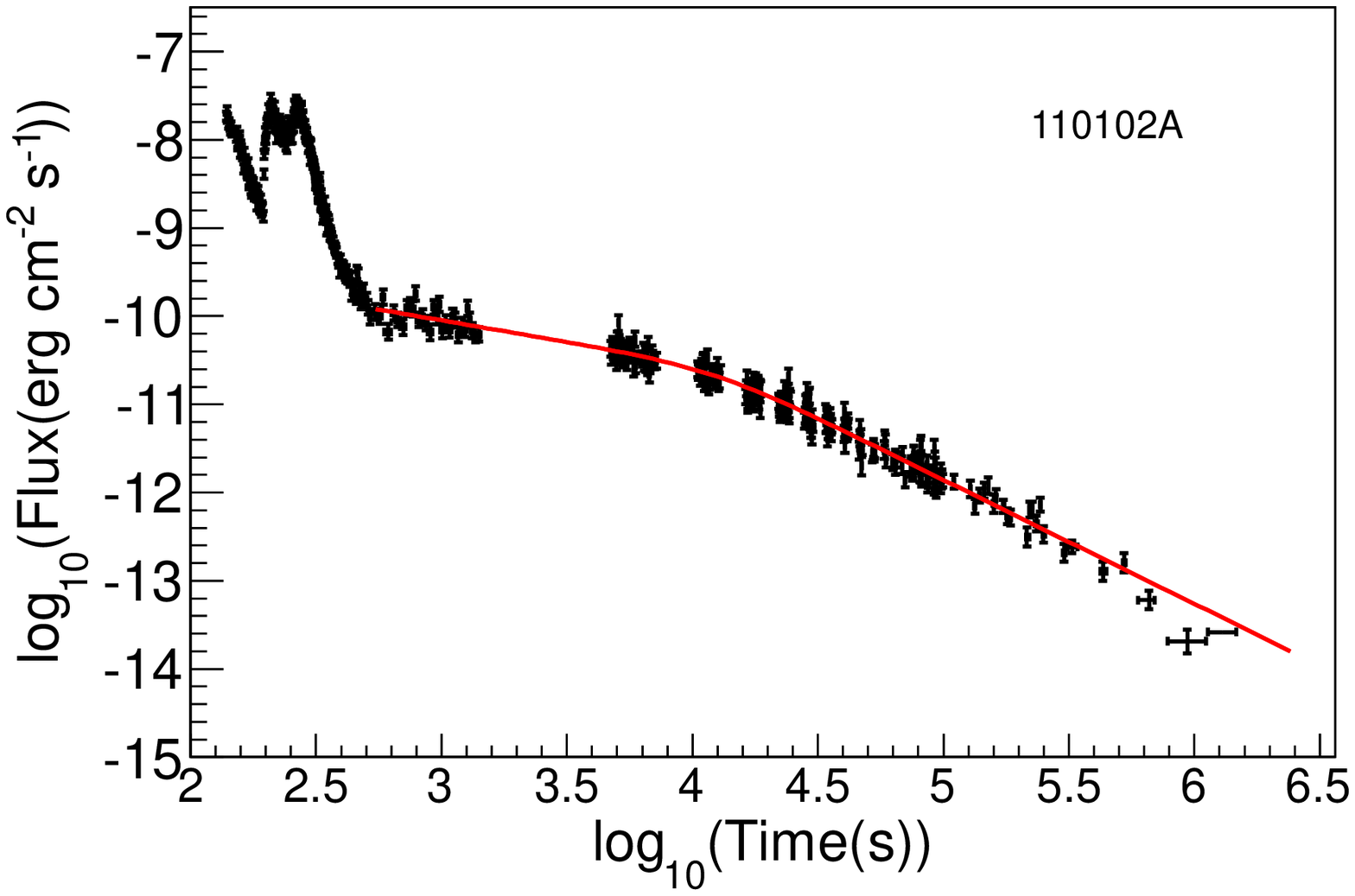}
\caption{ Continued.}
\label{fig-1-10}
\end{center}
\end{figure*}

\begin{figure*}
\begin{center}
\setlength{\abovecaptionskip}{0.cm}
\setlength{\belowcaptionskip}{-0.cm}
\figurenum{1}
\hspace{0cm}
\graphicspath{{lightcurve/}}
\includegraphics[width=5.5cm,height=5cm]{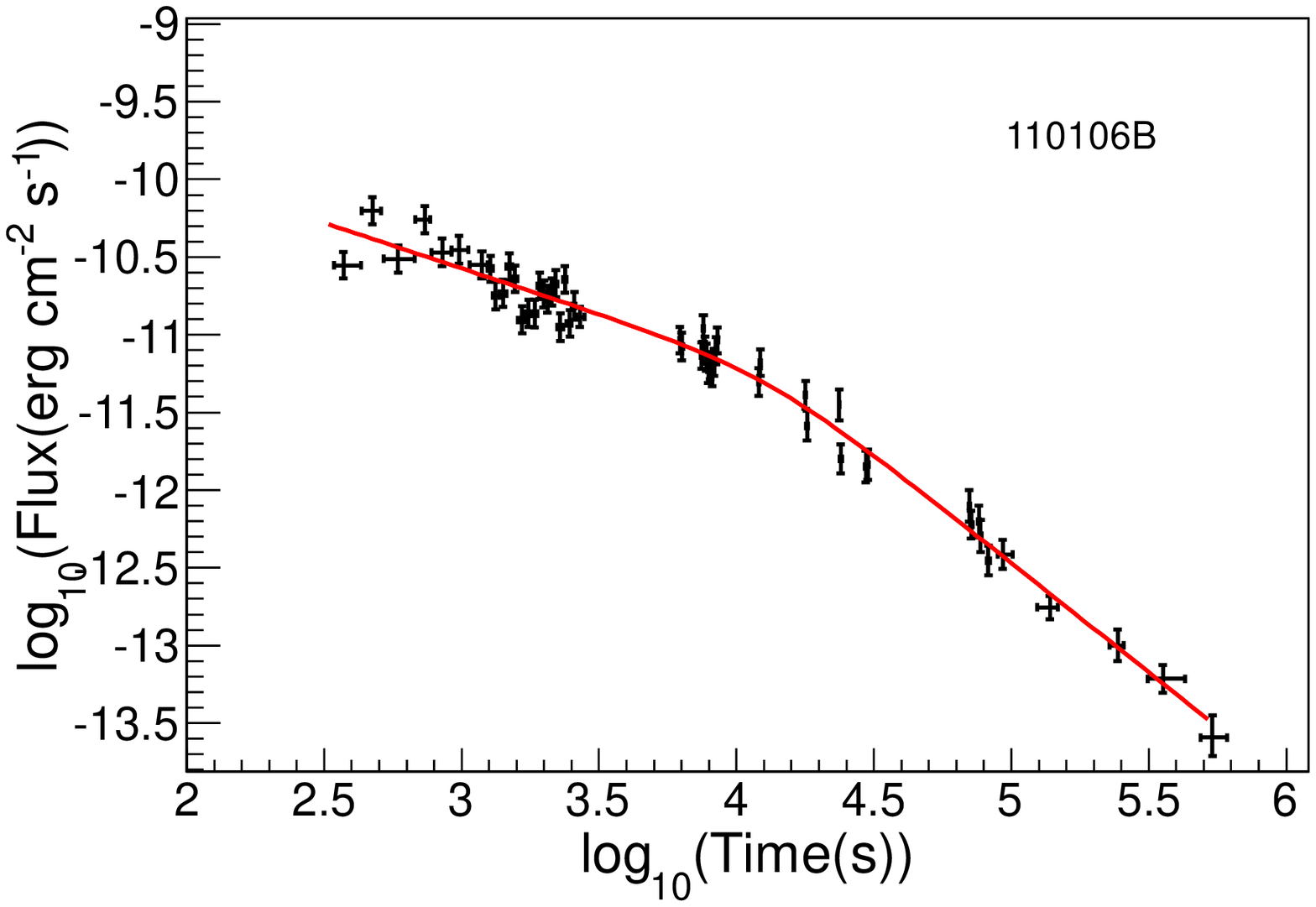}
\includegraphics[width=5.5cm,height=5cm]{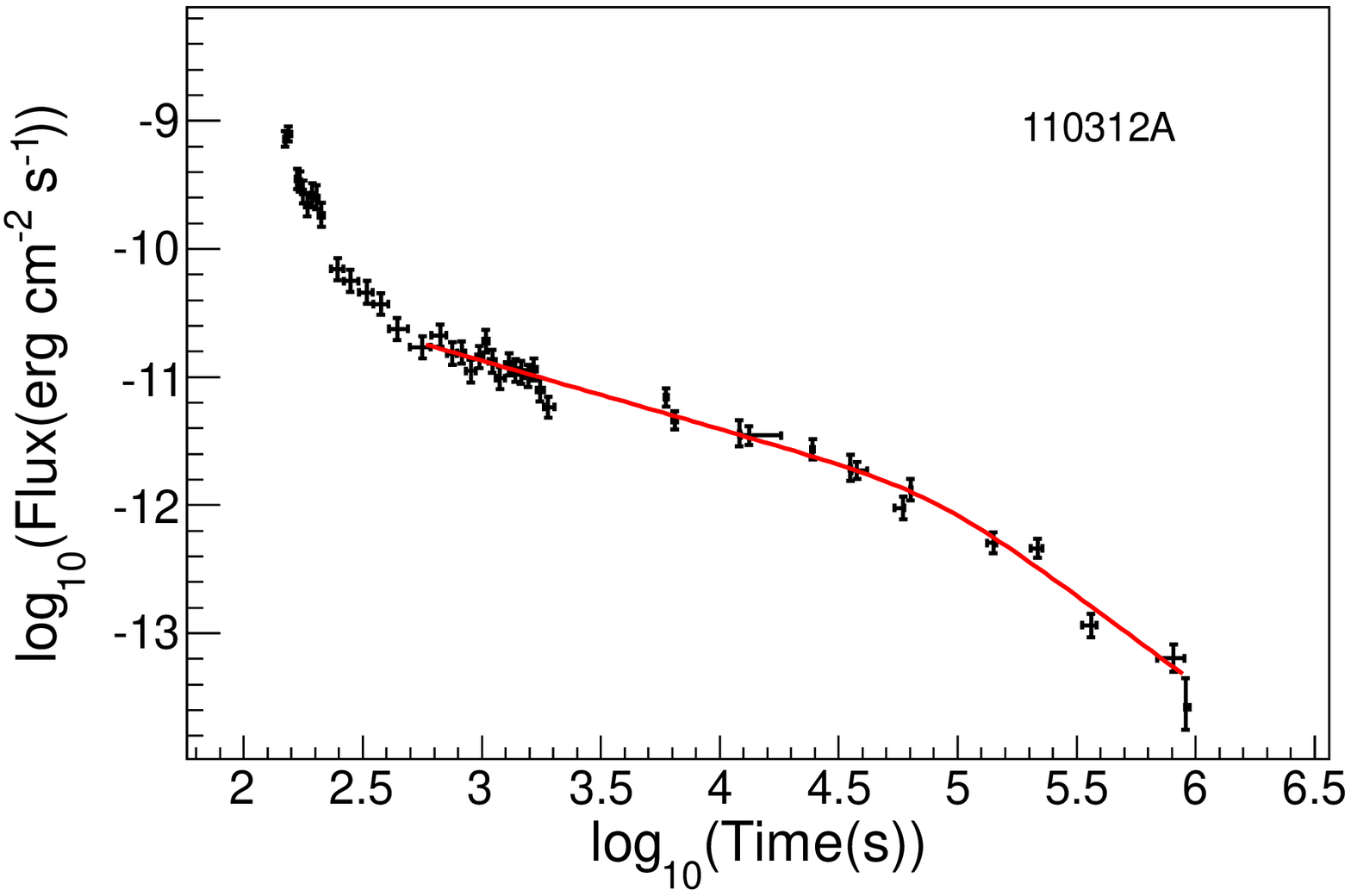}
\includegraphics[width=5.5cm,height=5cm]{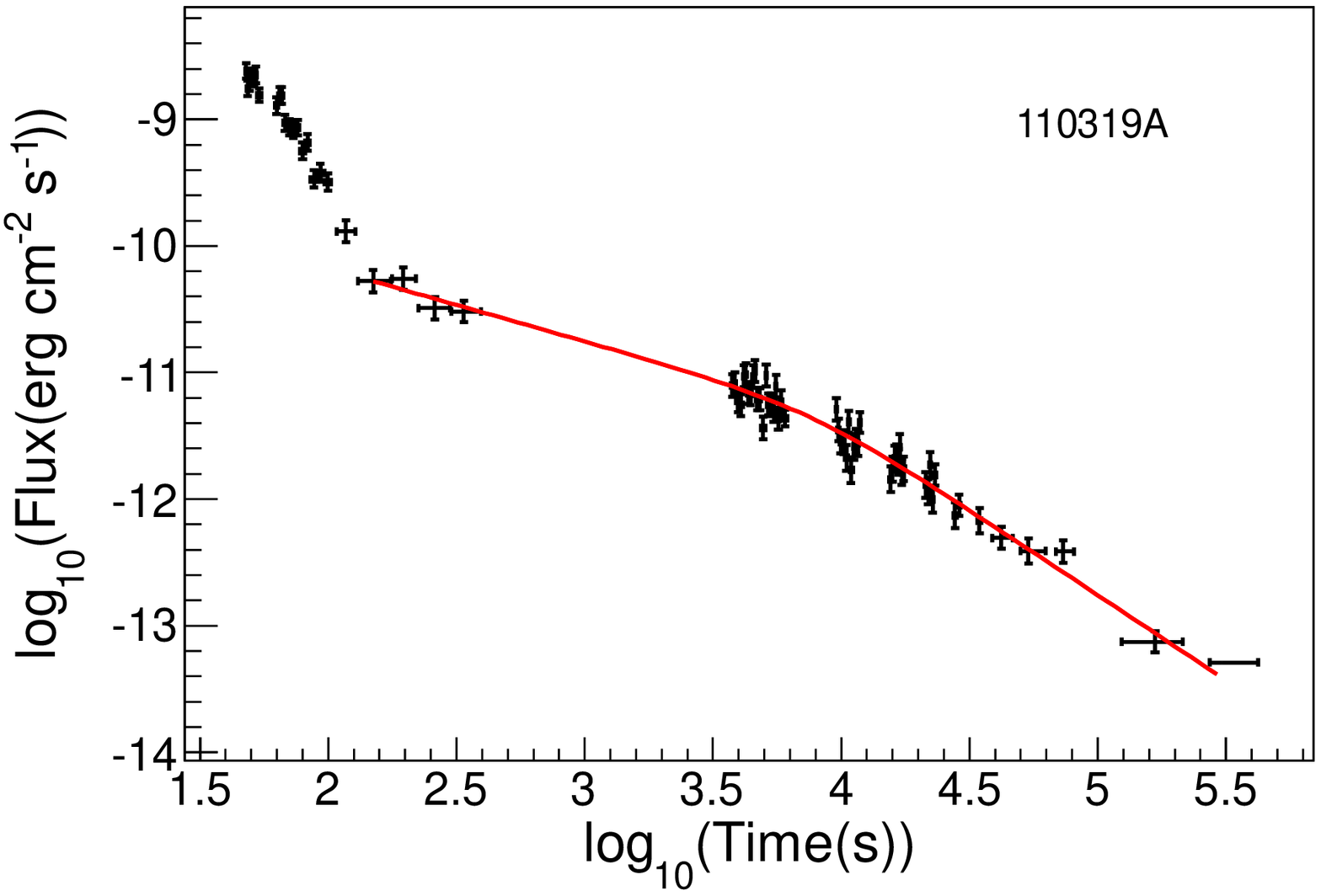}
\includegraphics[width=5.5cm,height=5cm]{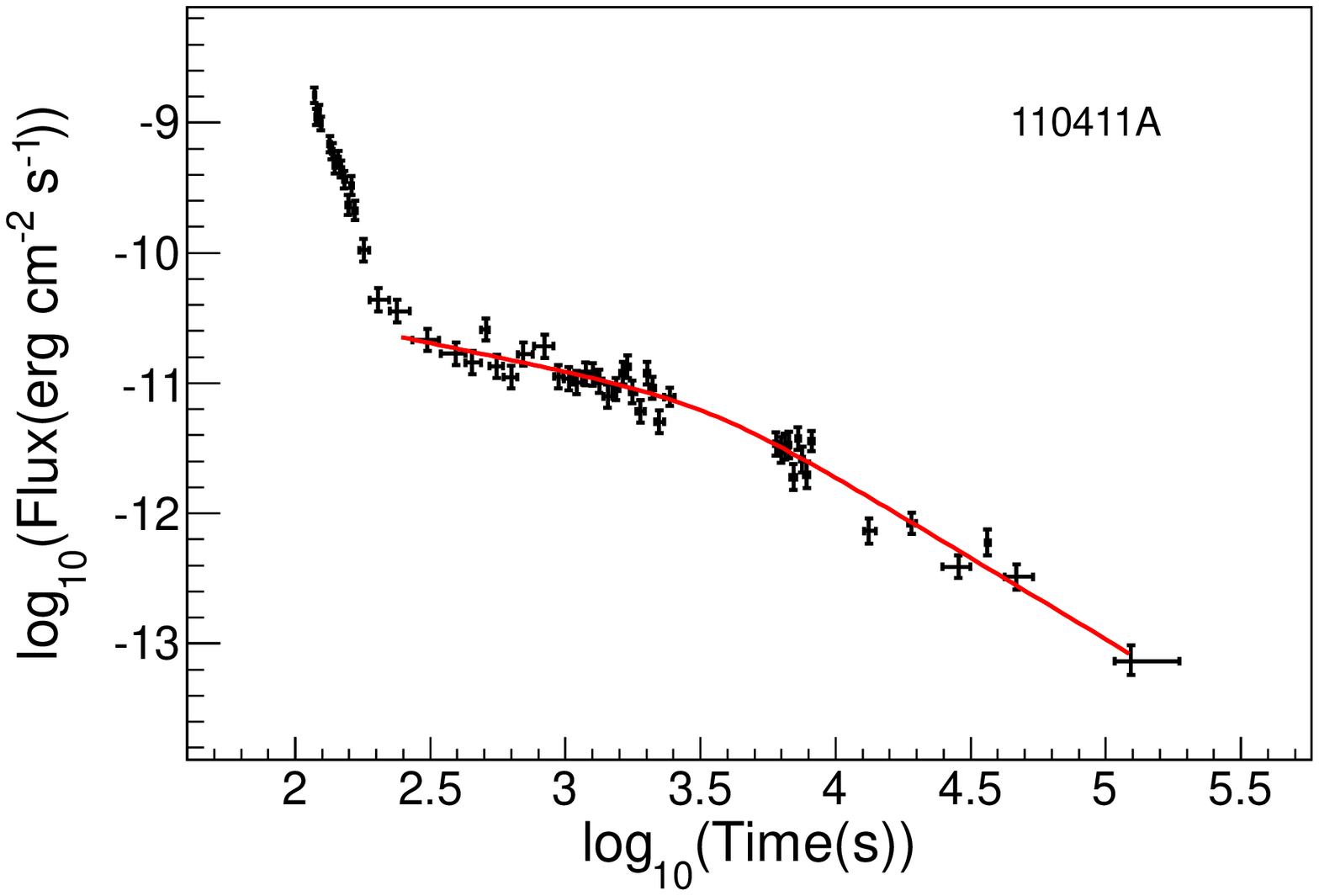}
\includegraphics[width=5.5cm,height=5cm]{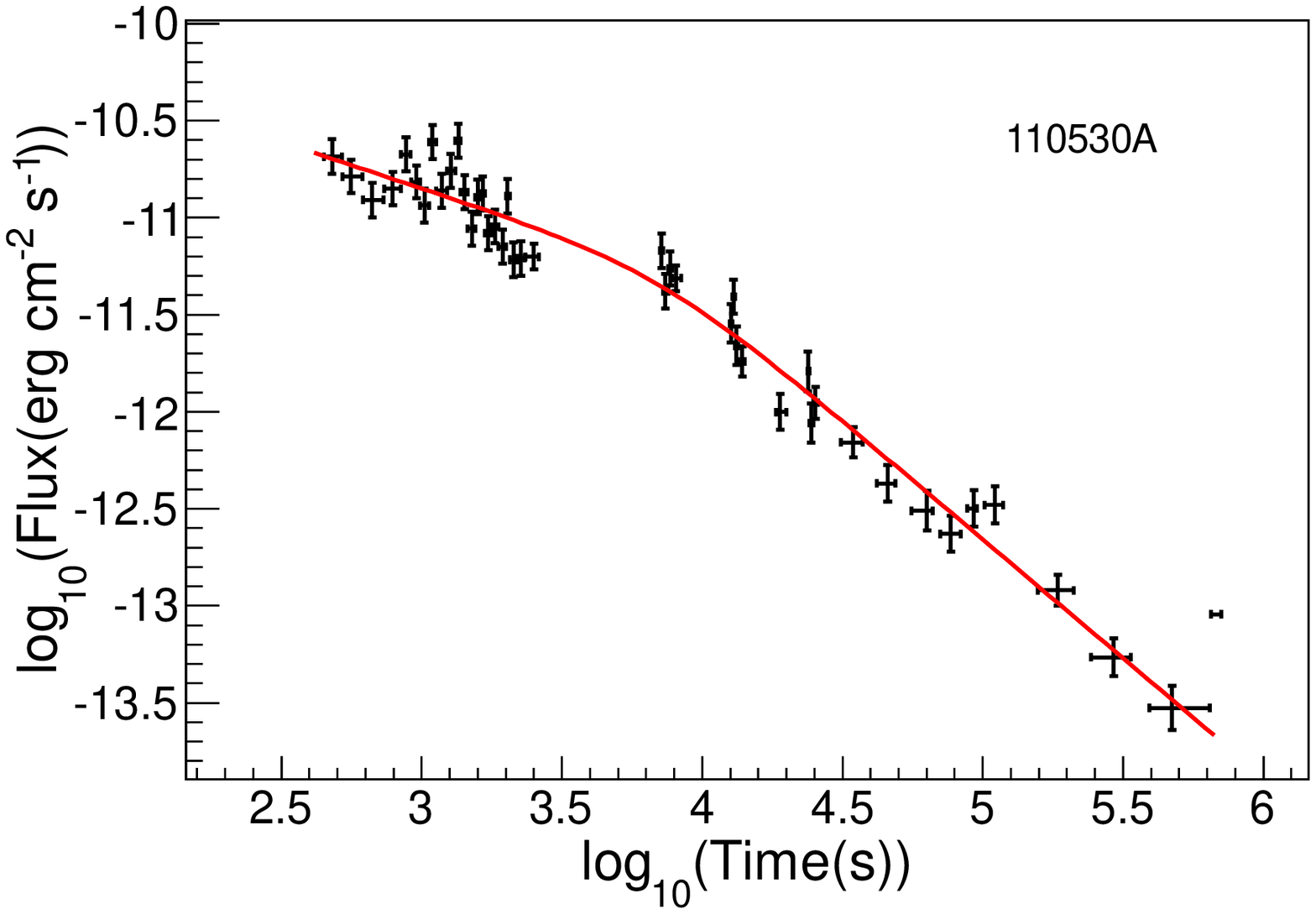}
\includegraphics[width=5.5cm,height=5cm]{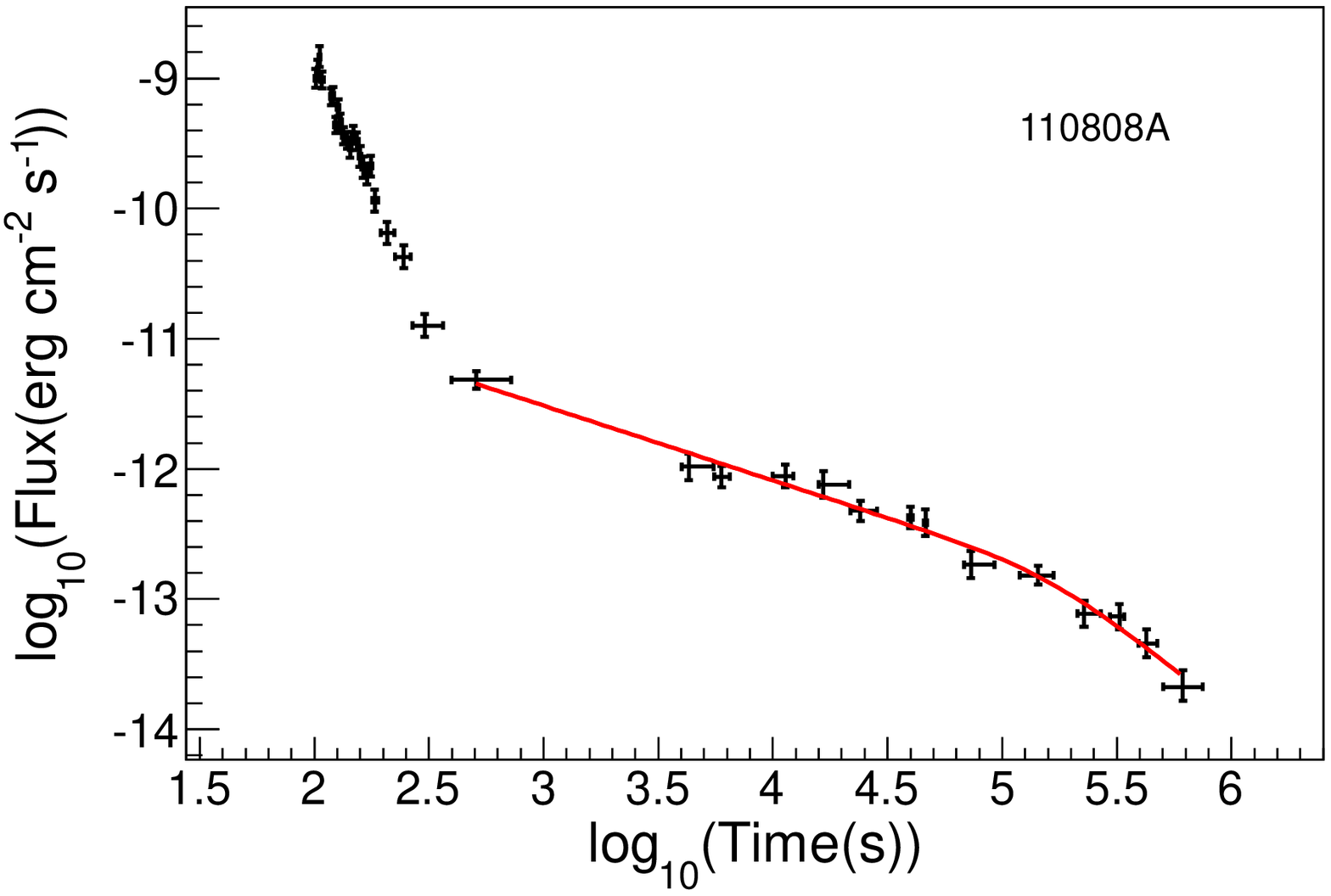}
\includegraphics[width=5.5cm,height=5cm]{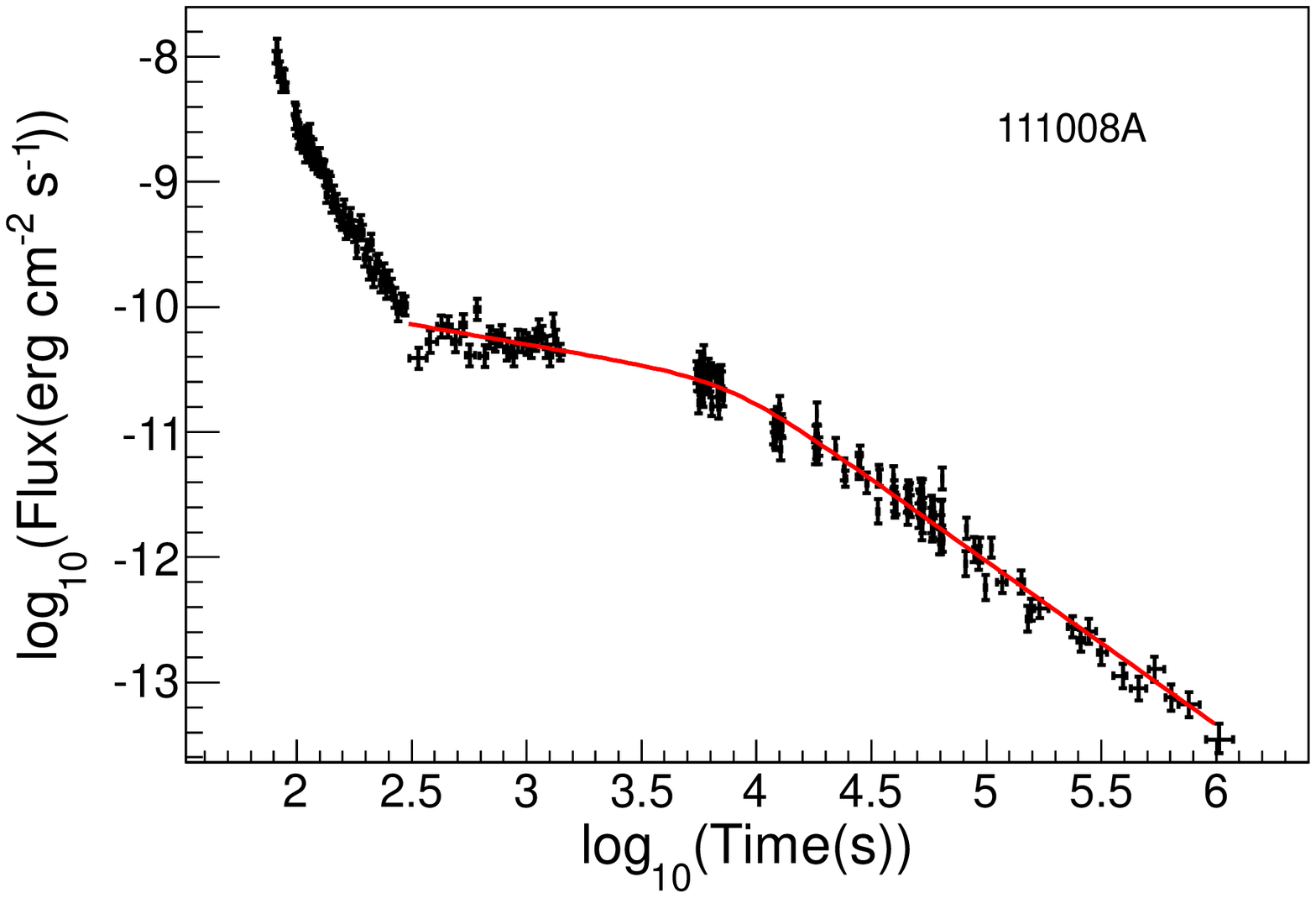}
\includegraphics[width=5.5cm,height=5cm]{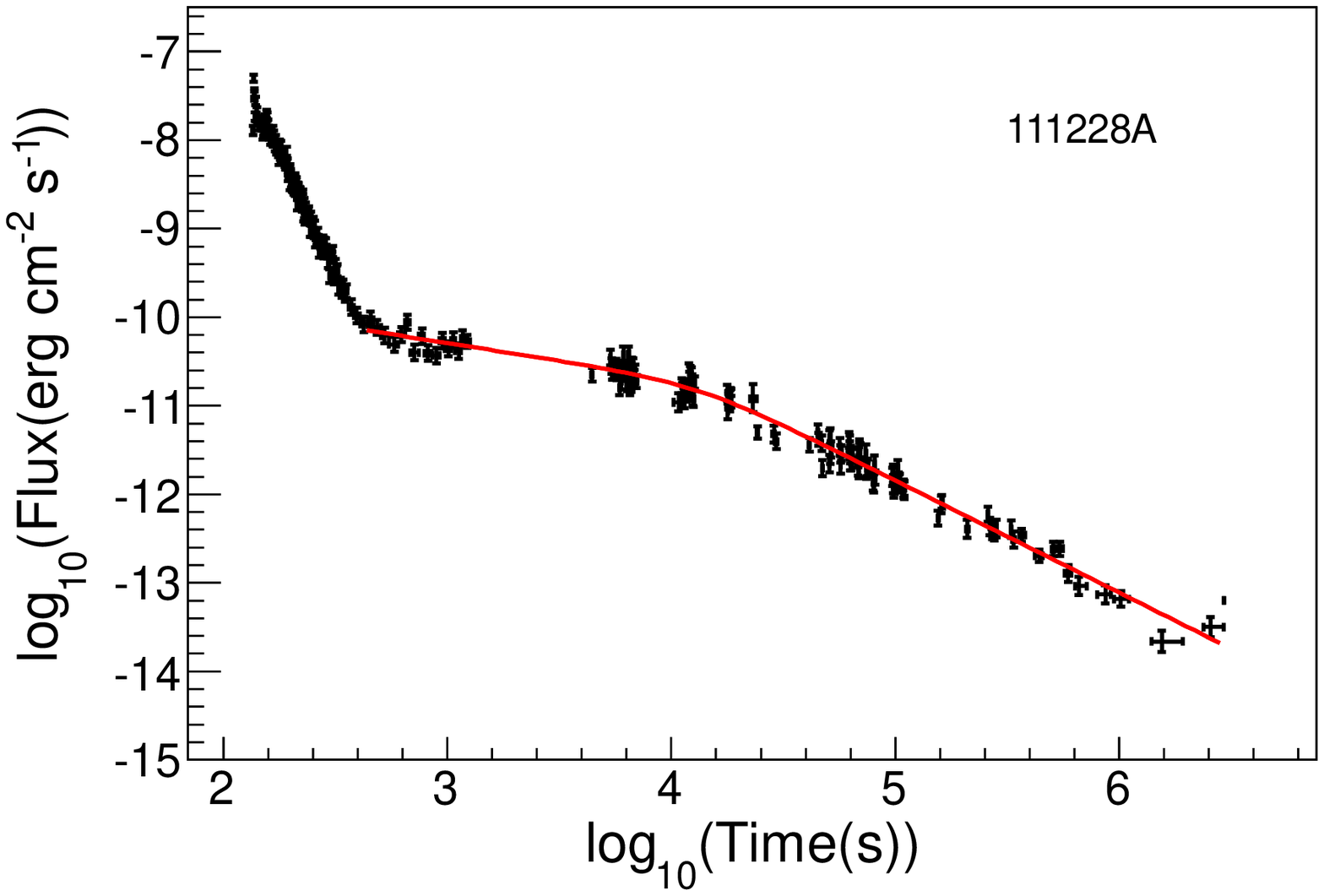}
\includegraphics[width=5.5cm,height=5cm]{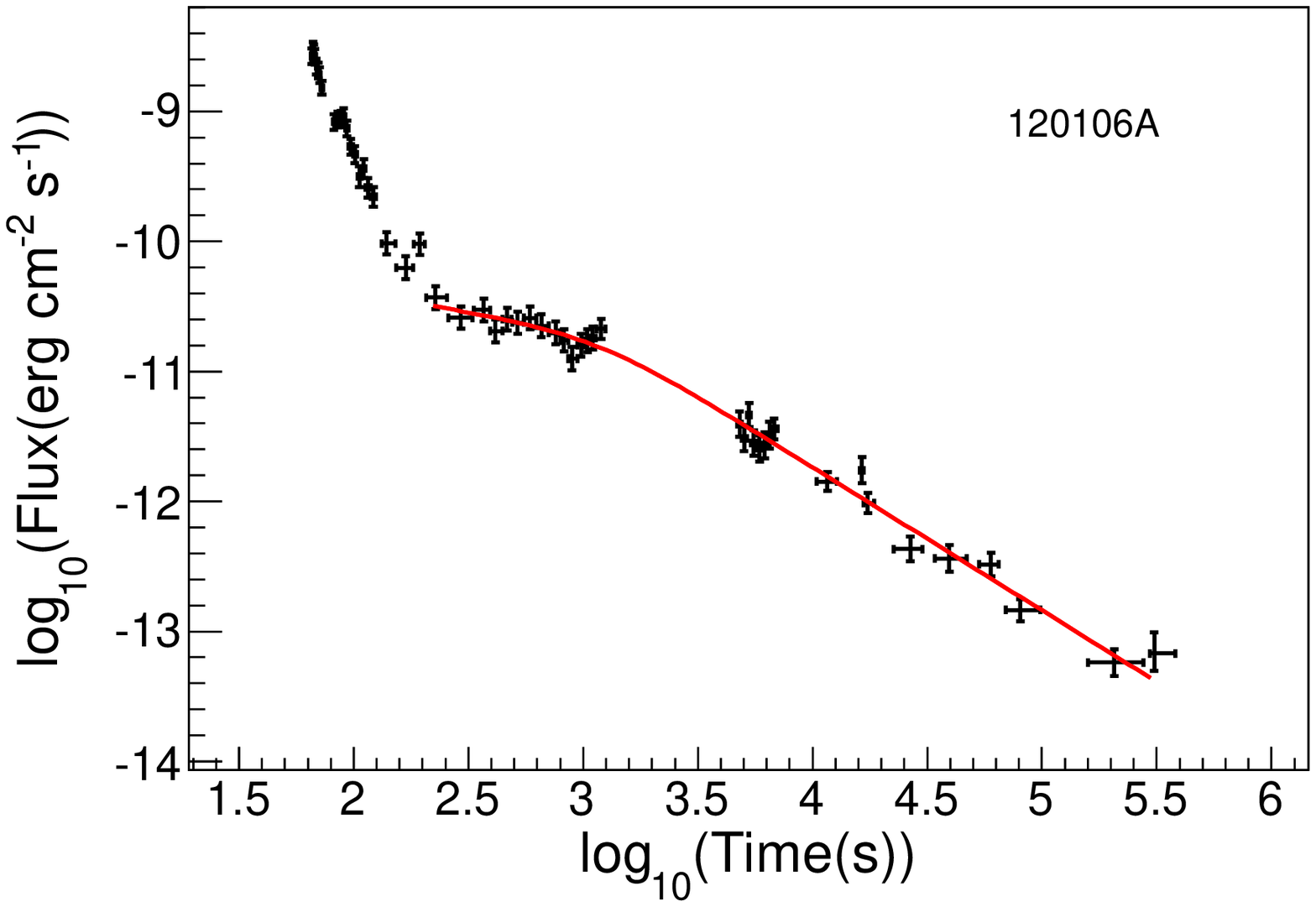}
\includegraphics[width=5.5cm,height=5cm]{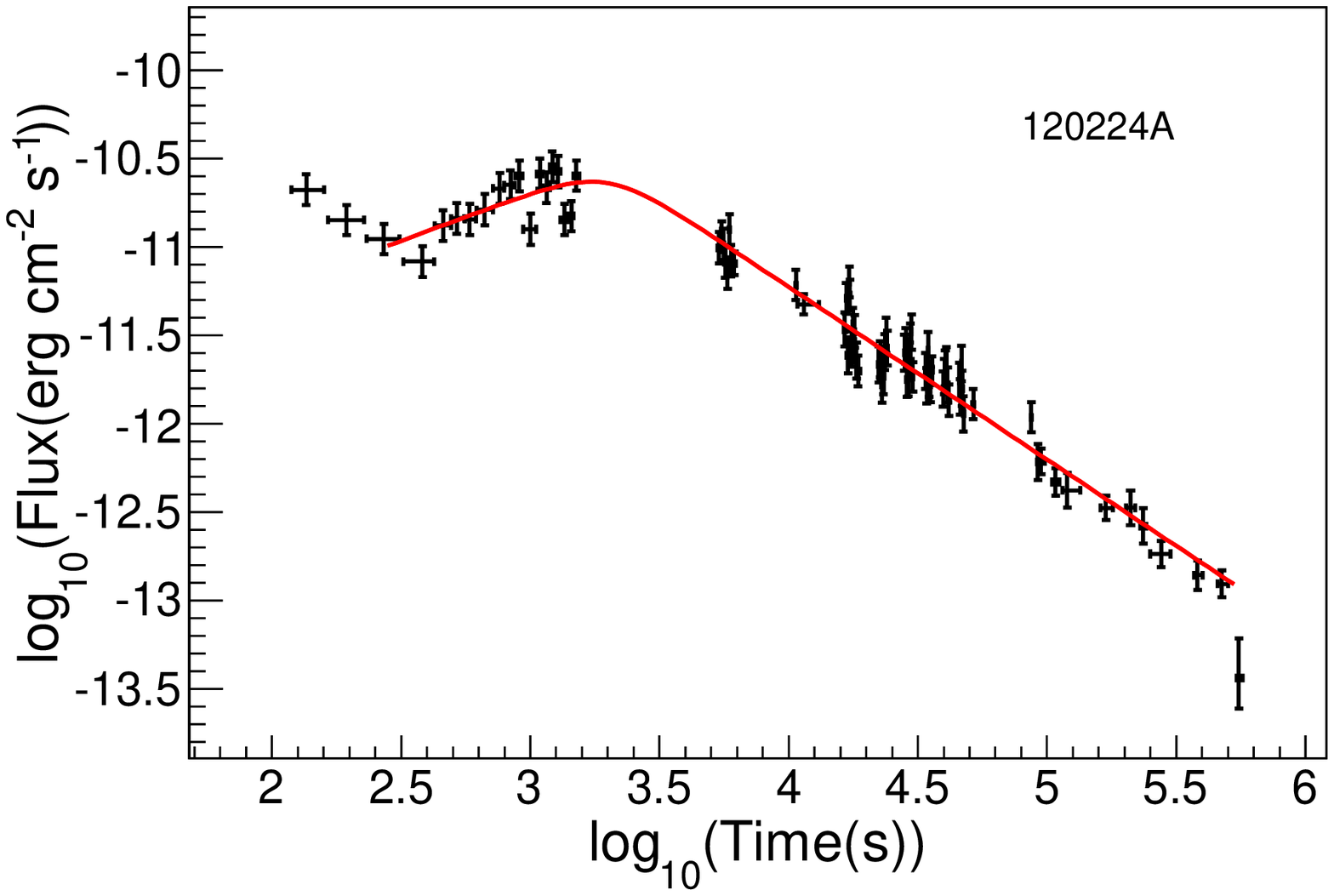}
\includegraphics[width=5.5cm,height=5cm]{101117B.eps}
\includegraphics[width=5.5cm,height=5cm]{110102A.eps}
\caption{ Continued.}
\label{fig-1-11}
\end{center}
\end{figure*}

\begin{figure*}
\begin{center}
\setlength{\abovecaptionskip}{0.cm}
\setlength{\belowcaptionskip}{-0.cm}
\figurenum{1}
\hspace{0cm}
\graphicspath{{lightcurve/}}
\includegraphics[width=5.5cm,height=5cm]{110106B.eps}
\includegraphics[width=5.5cm,height=5cm]{110312A.eps}
\includegraphics[width=5.5cm,height=5cm]{110319A.eps}
\includegraphics[width=5.5cm,height=5cm]{110411A.eps}
\includegraphics[width=5.5cm,height=5cm]{110530A.eps}
\includegraphics[width=5.5cm,height=5cm]{110808A.eps}
\includegraphics[width=5.5cm,height=5cm]{111008A.eps}
\includegraphics[width=5.5cm,height=5cm]{111228A.eps}
\includegraphics[width=5.5cm,height=5cm]{120106A.eps}
\includegraphics[width=5.5cm,height=5cm]{120224A.eps}
\includegraphics[width=5.5cm,height=5cm]{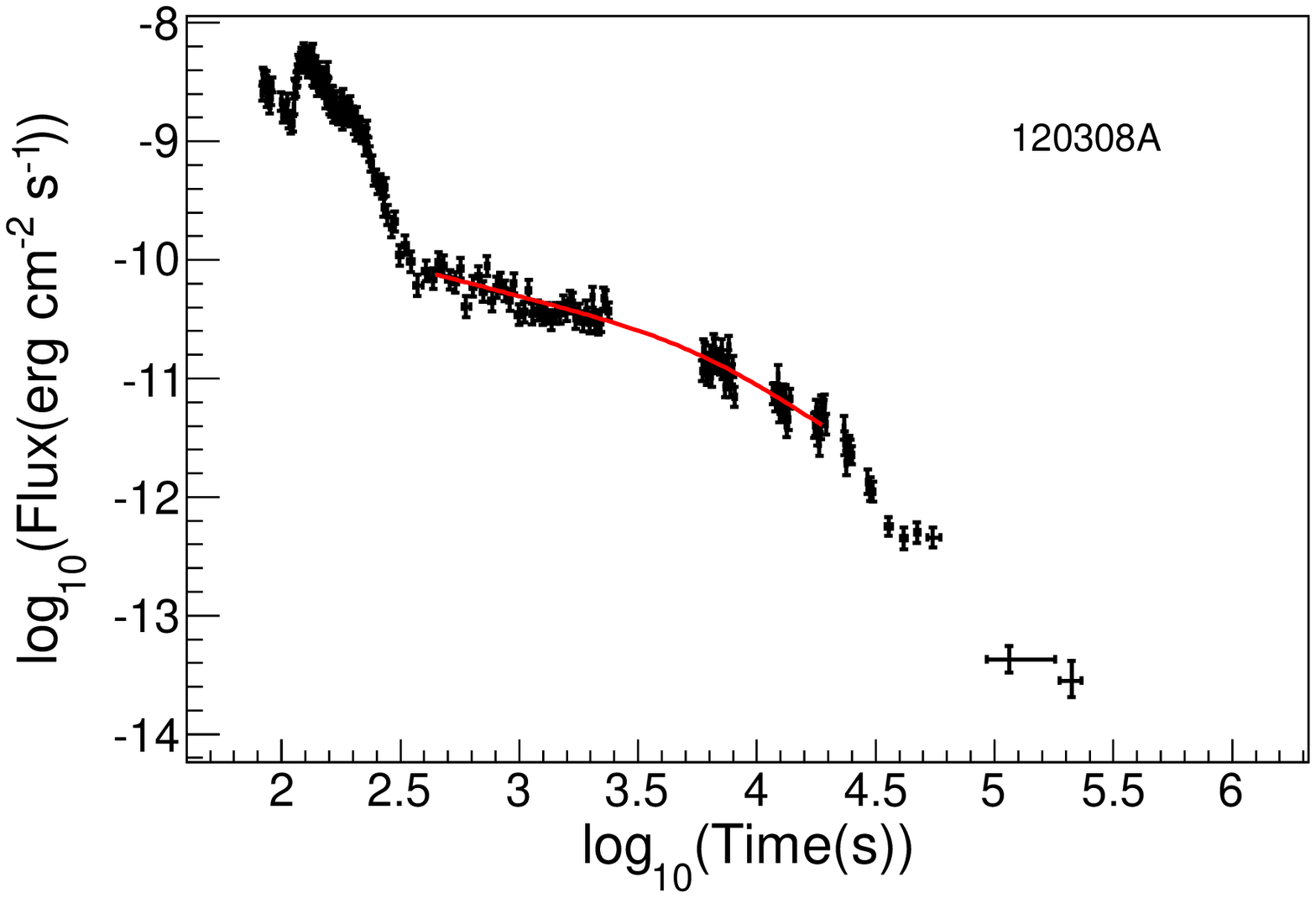}
\includegraphics[width=5.5cm,height=5cm]{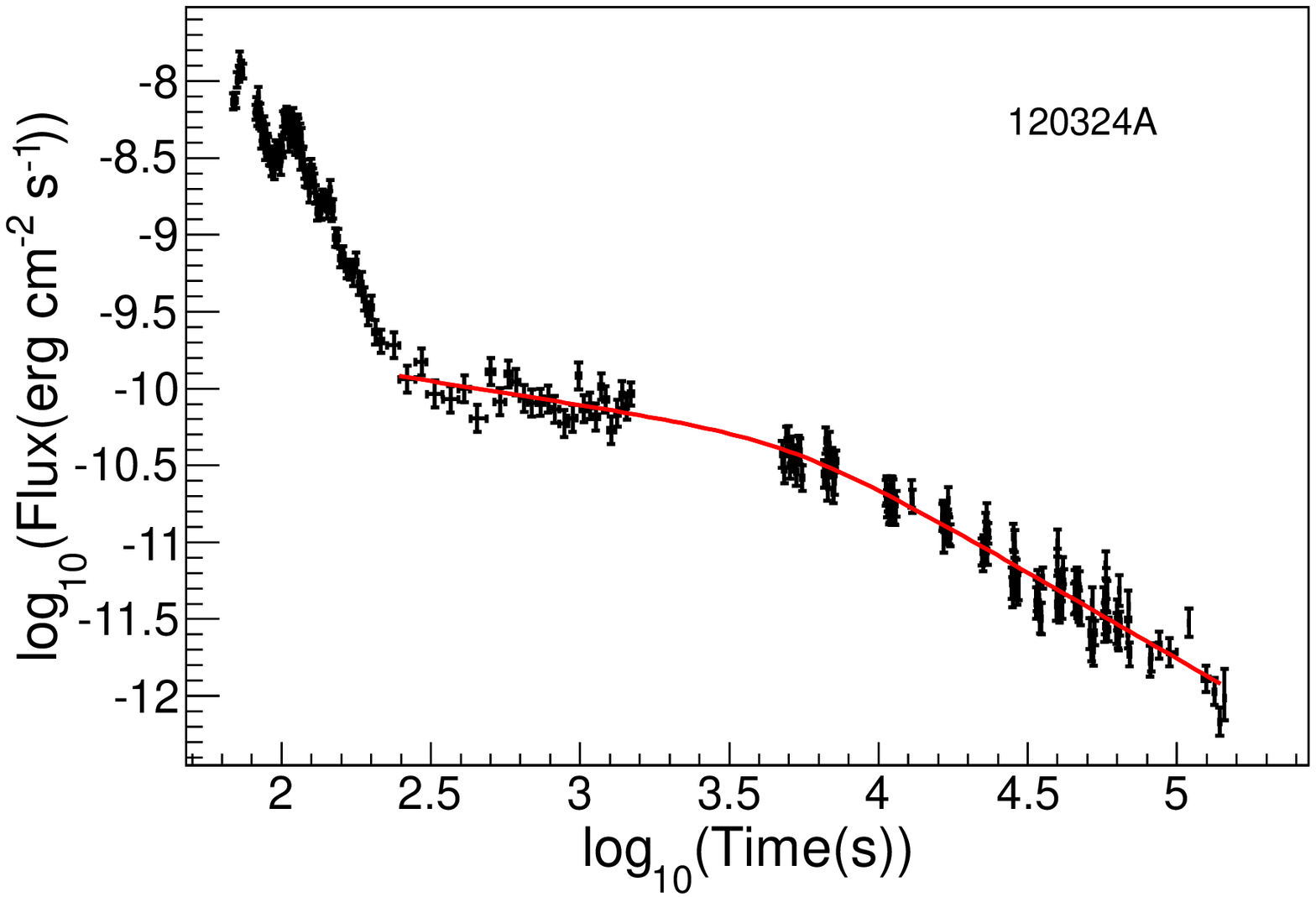}
\caption{ Continued.}
\label{fig-1-11}
\end{center}
\end{figure*}

\begin{figure*}
\begin{center}
\setlength{\abovecaptionskip}{0.cm}
\setlength{\belowcaptionskip}{-0.cm}
\figurenum{1}
\hspace{0cm}
\graphicspath{{lightcurve/}}
\includegraphics[width=5.5cm,height=5cm]{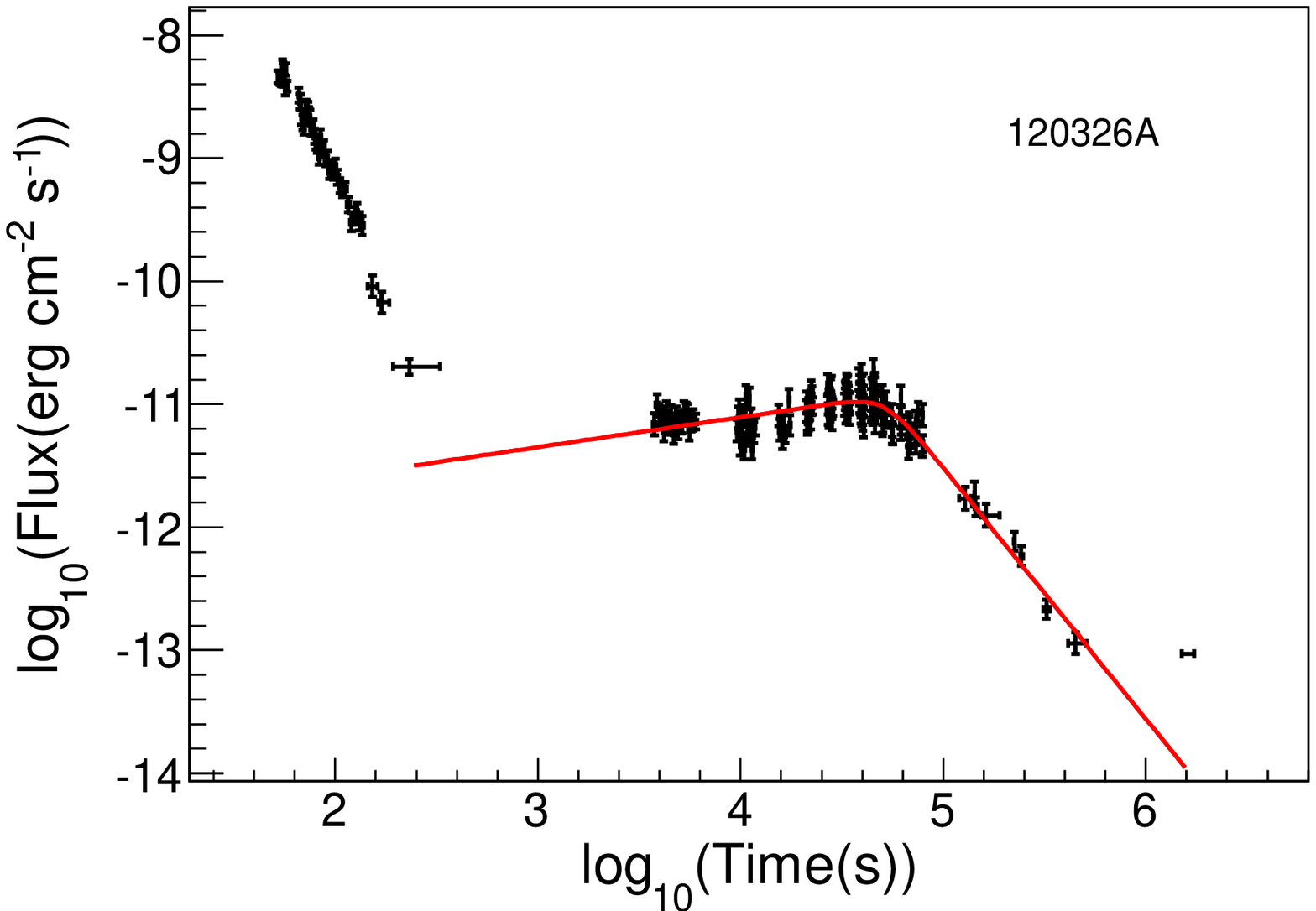}
\includegraphics[width=5.5cm,height=5cm]{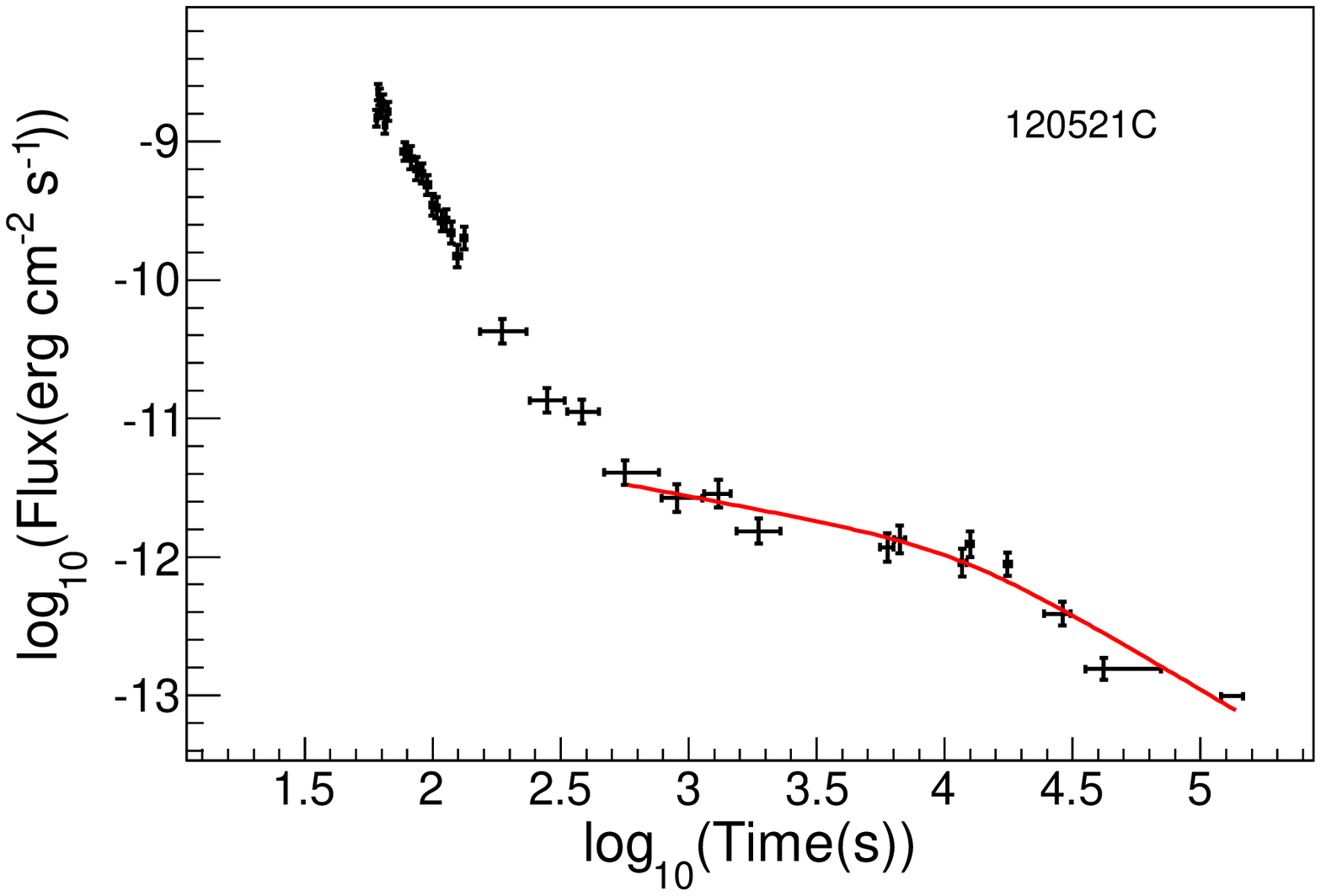}
\includegraphics[width=5.5cm,height=5cm]{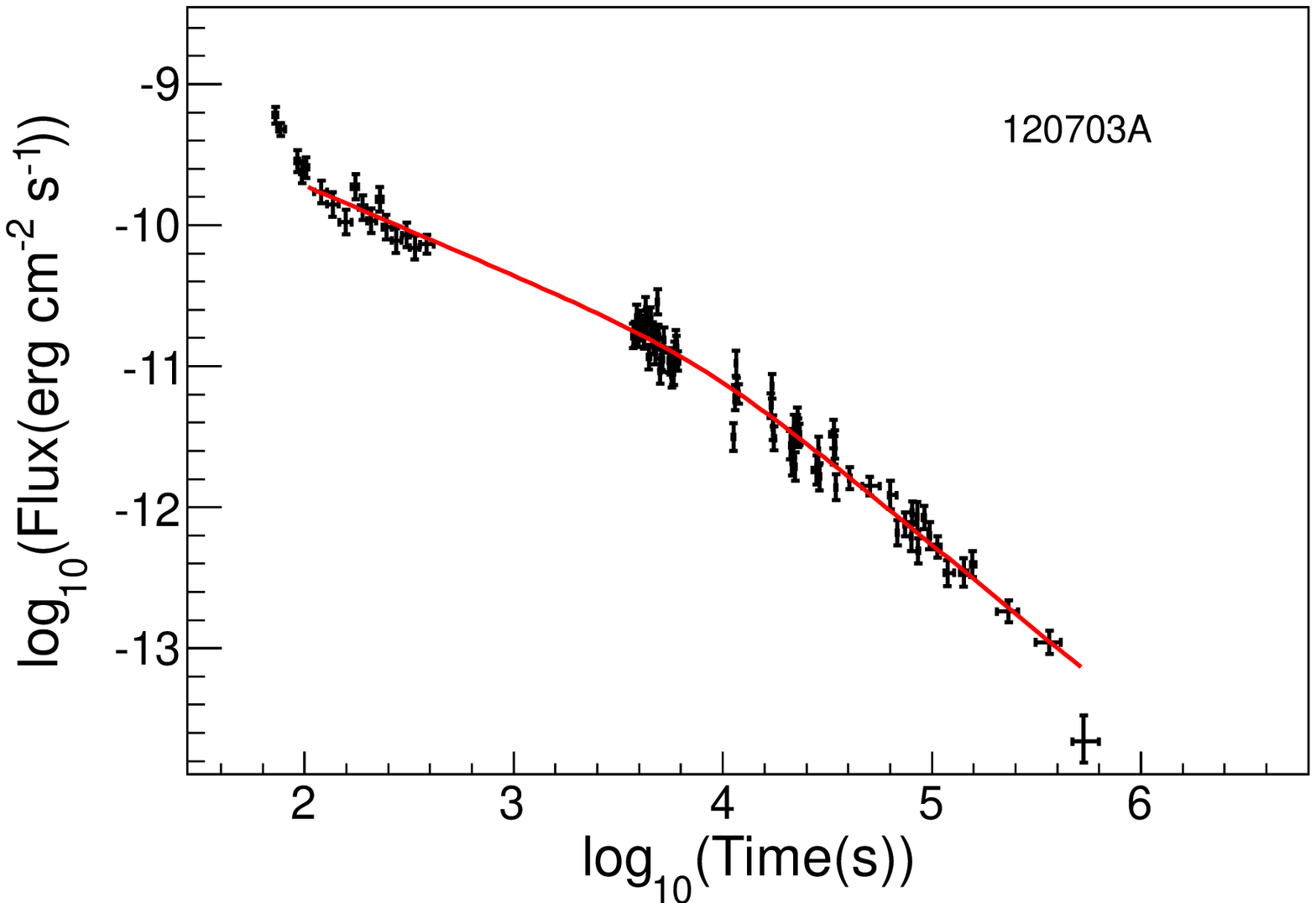}
\includegraphics[width=5.5cm,height=5cm]{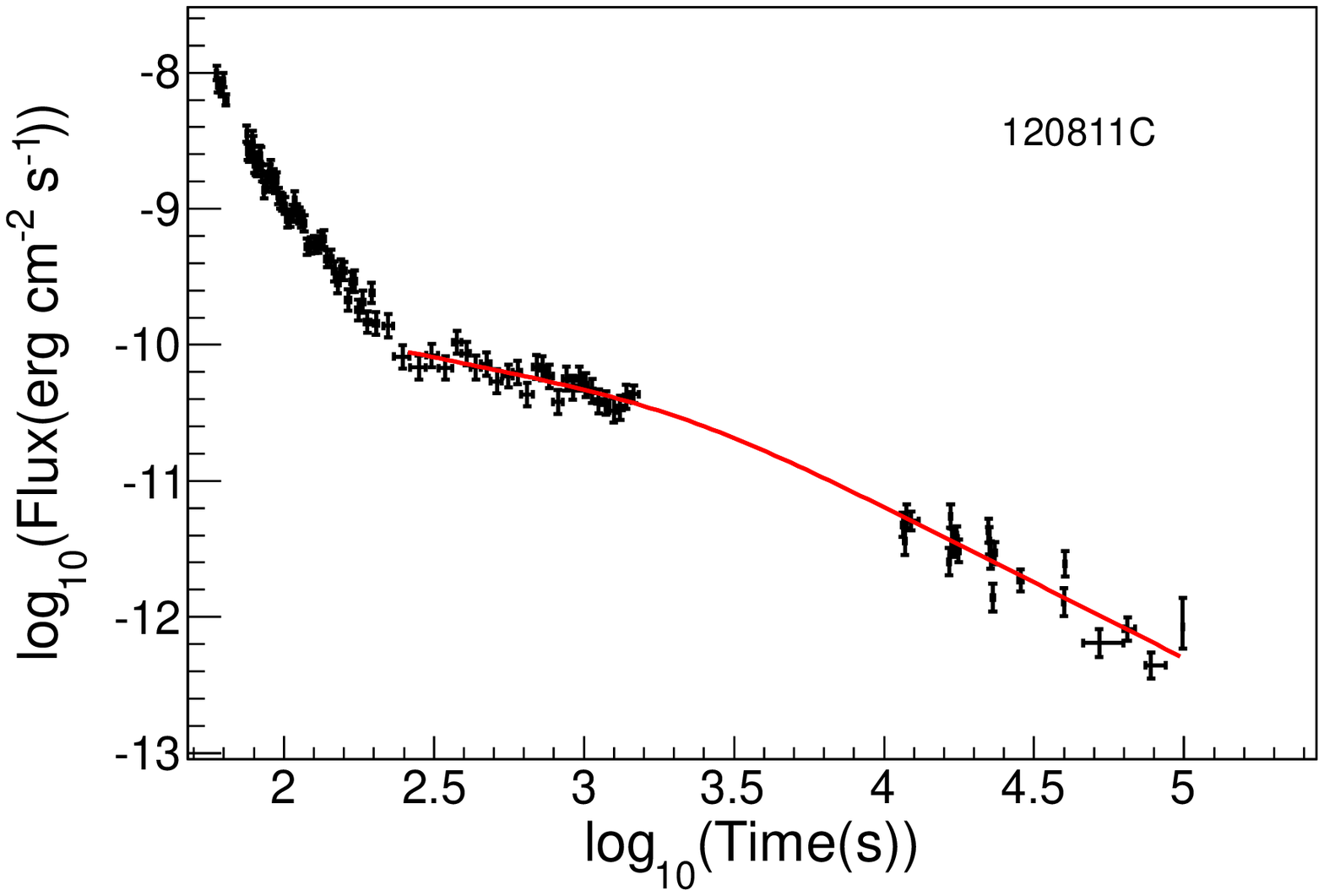}
\includegraphics[width=5.5cm,height=5cm]{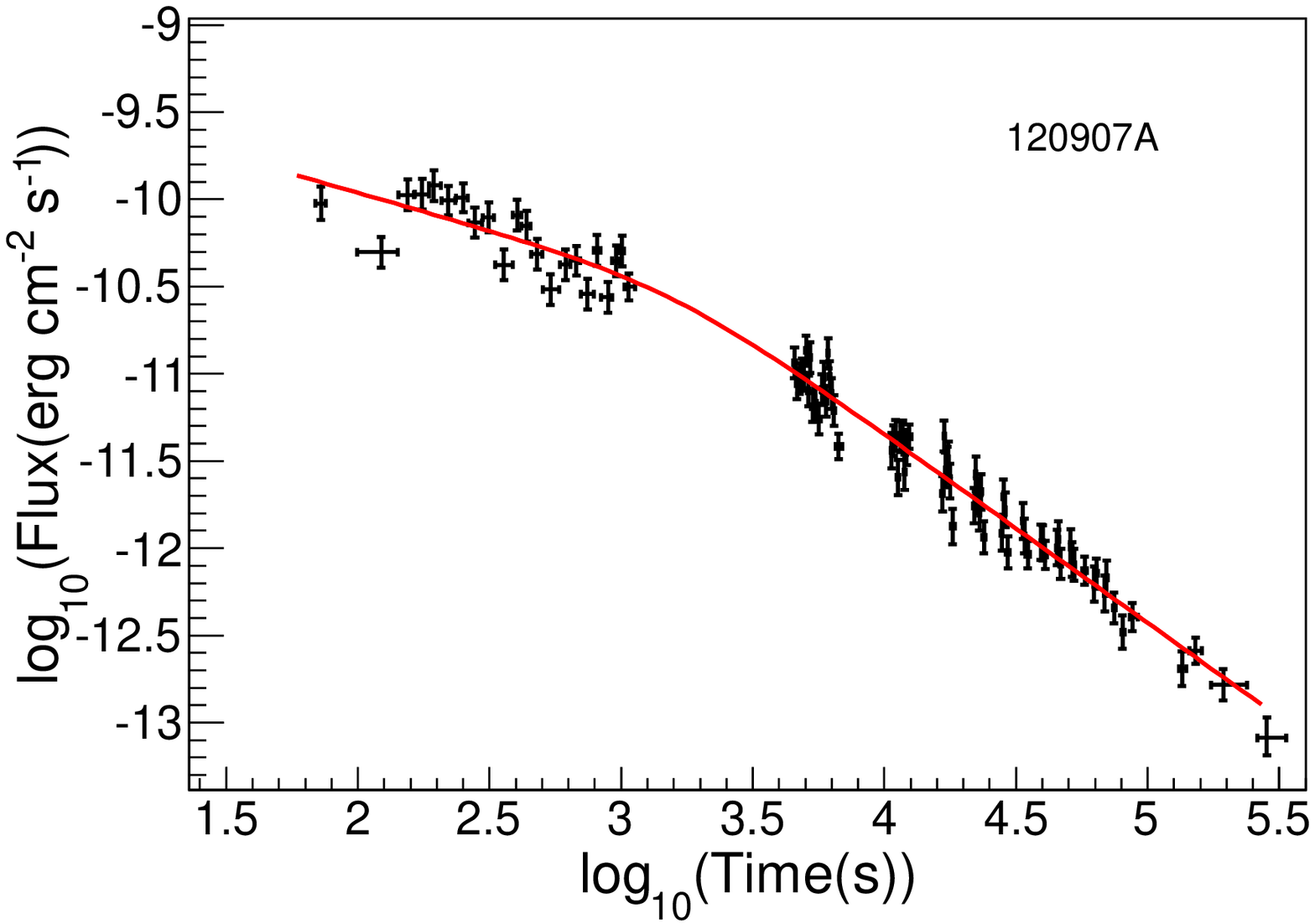}
\includegraphics[width=5.5cm,height=5cm]{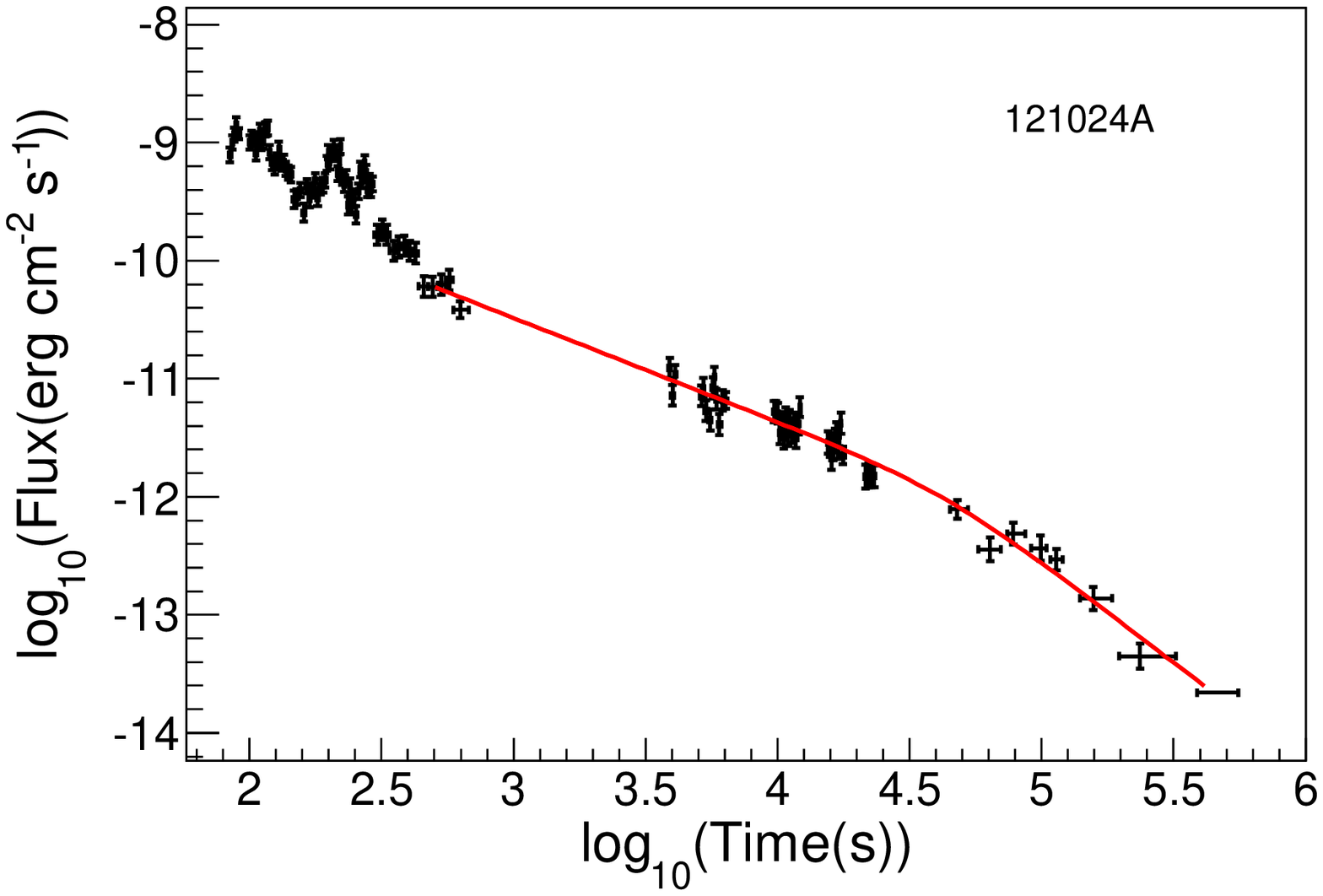}
\includegraphics[width=5.5cm,height=5cm]{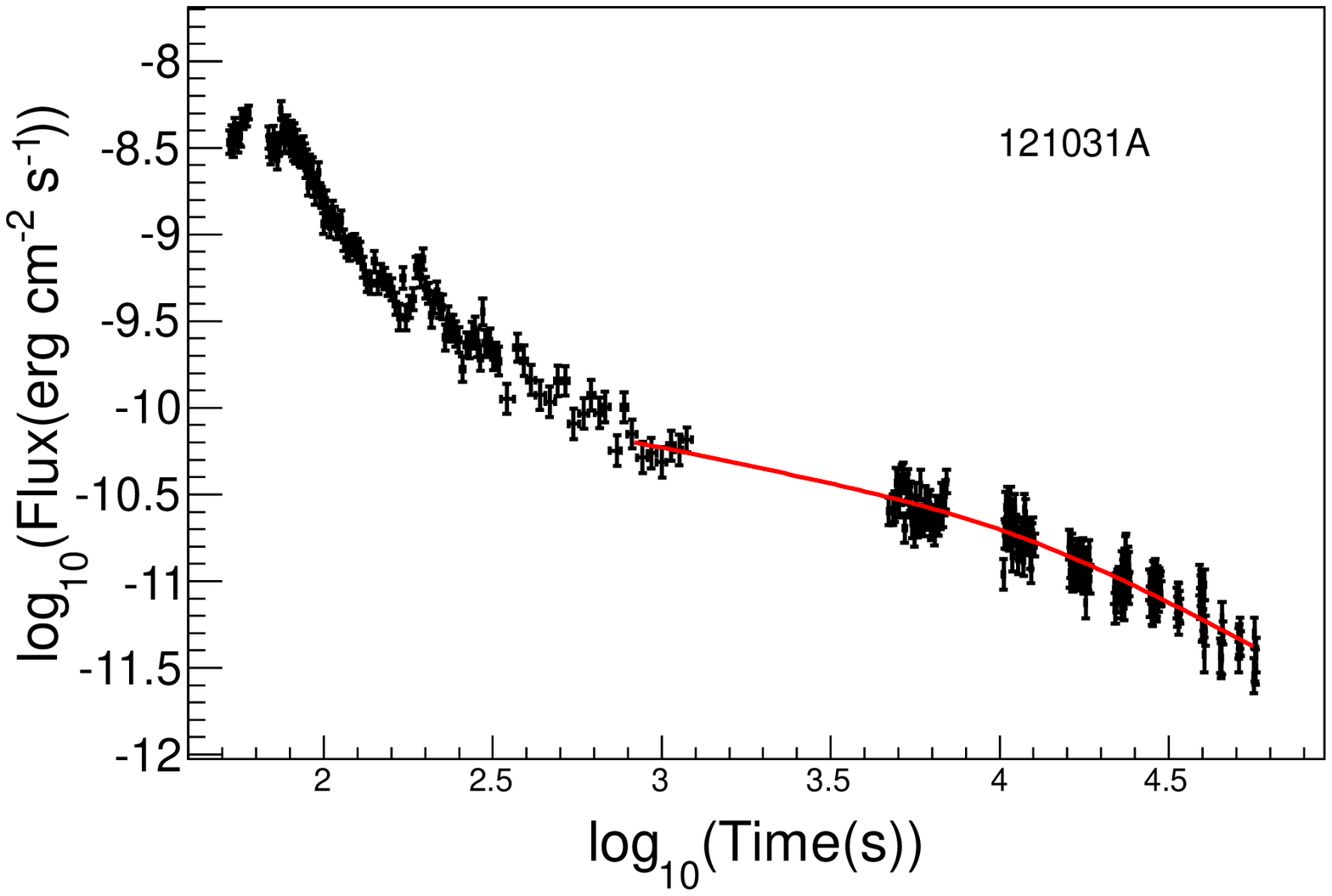}
\includegraphics[width=5.5cm,height=5cm]{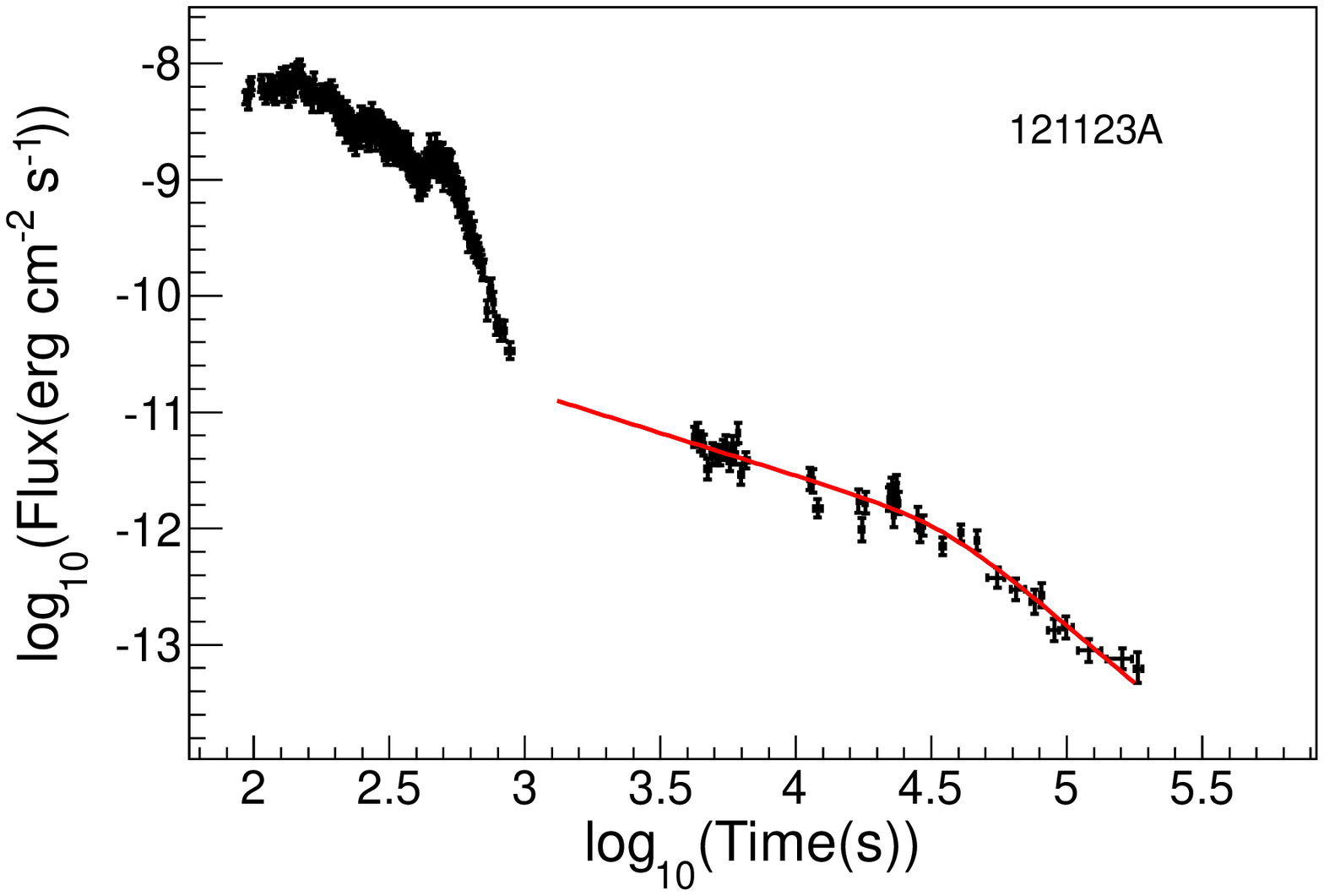}
\includegraphics[width=5.5cm,height=5cm]{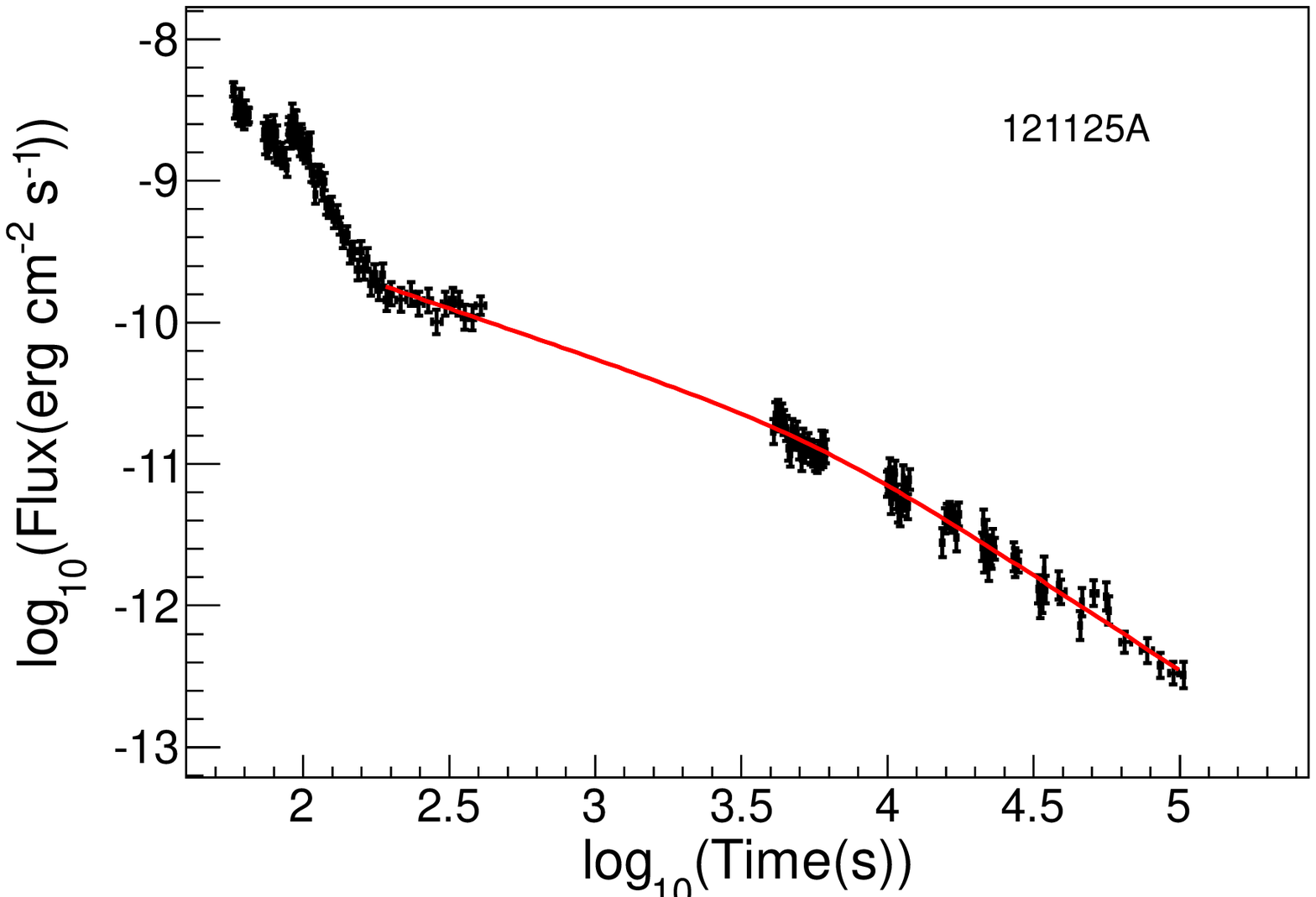}
\includegraphics[width=5.5cm,height=5cm]{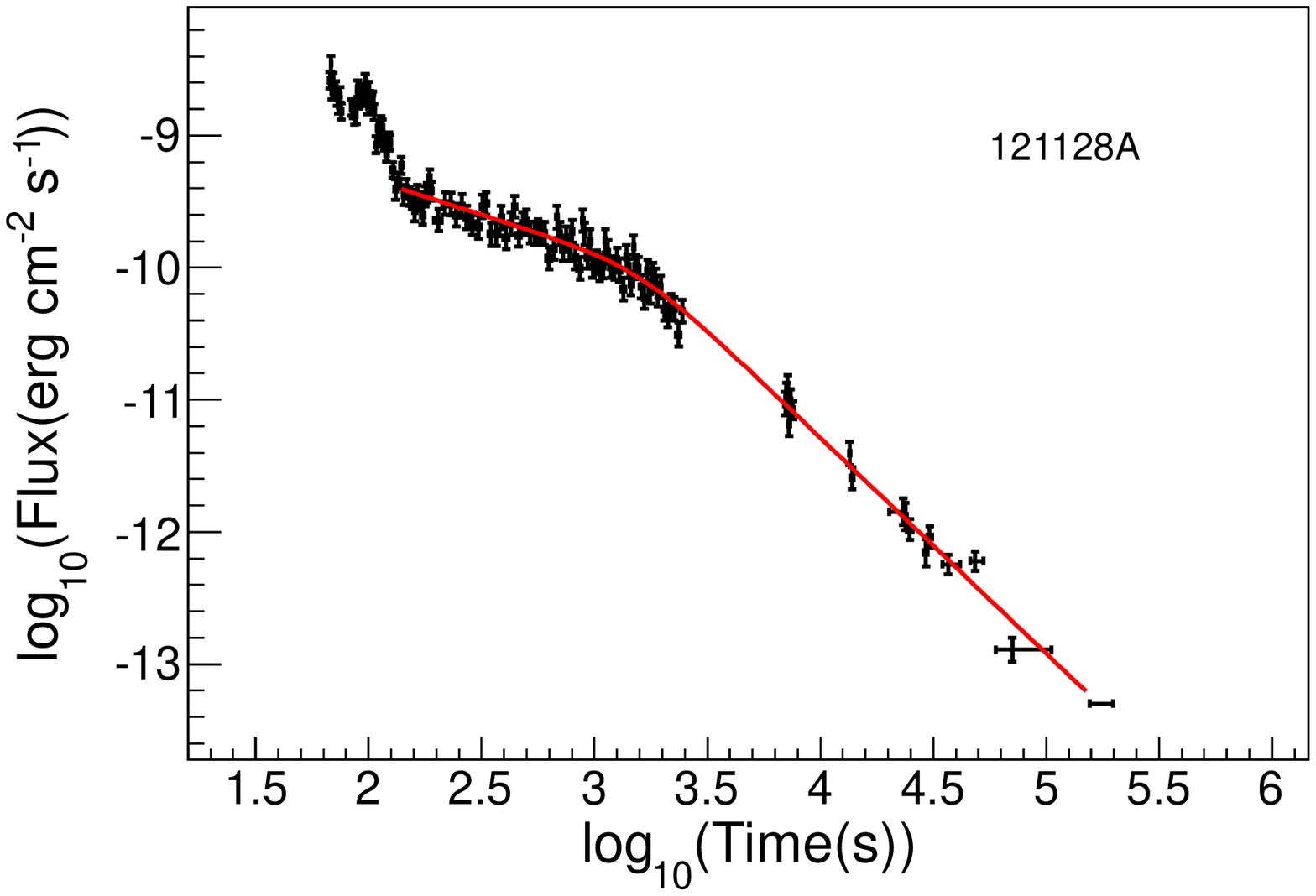}
\includegraphics[width=5.5cm,height=5cm]{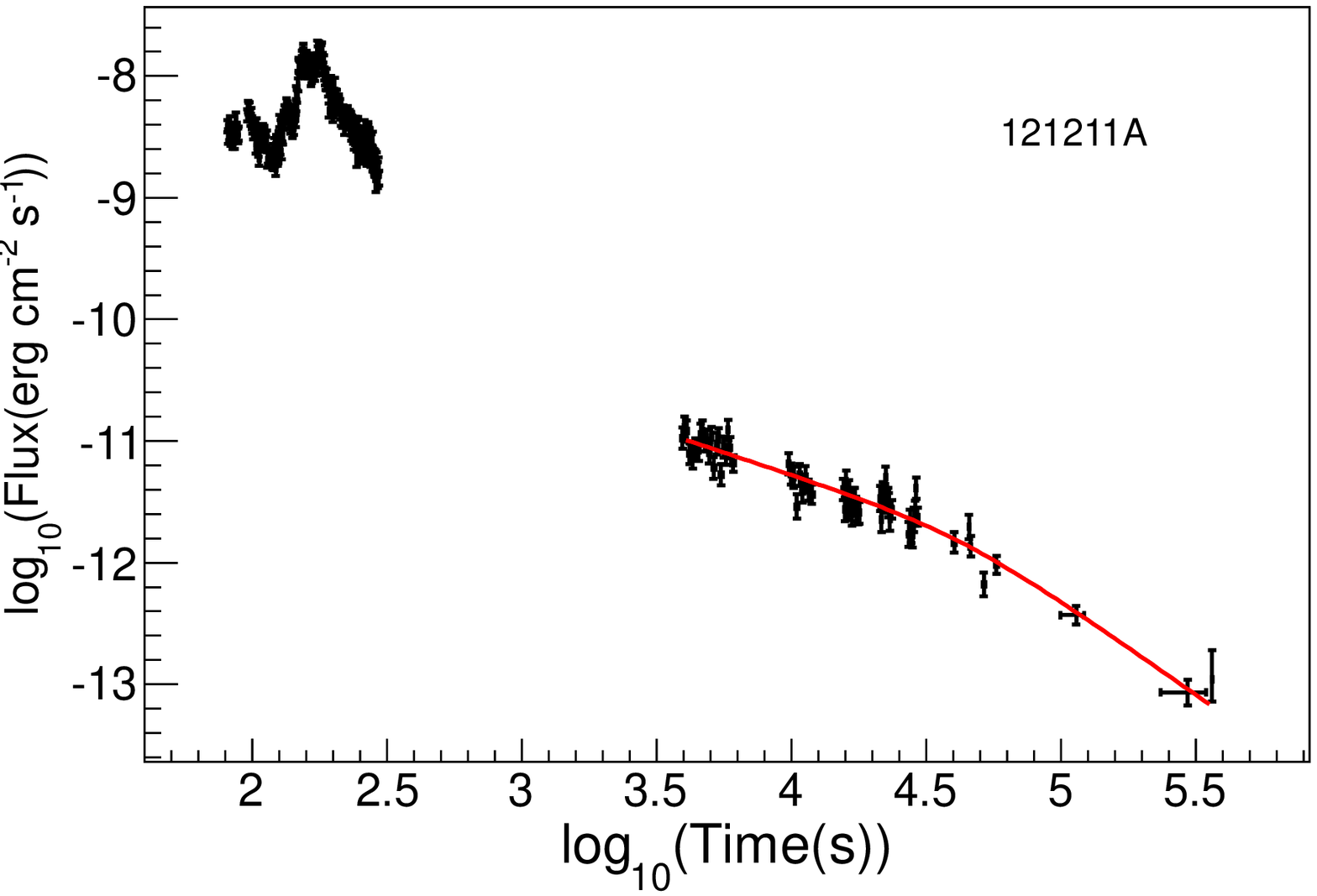}
\includegraphics[width=5.5cm,height=5cm]{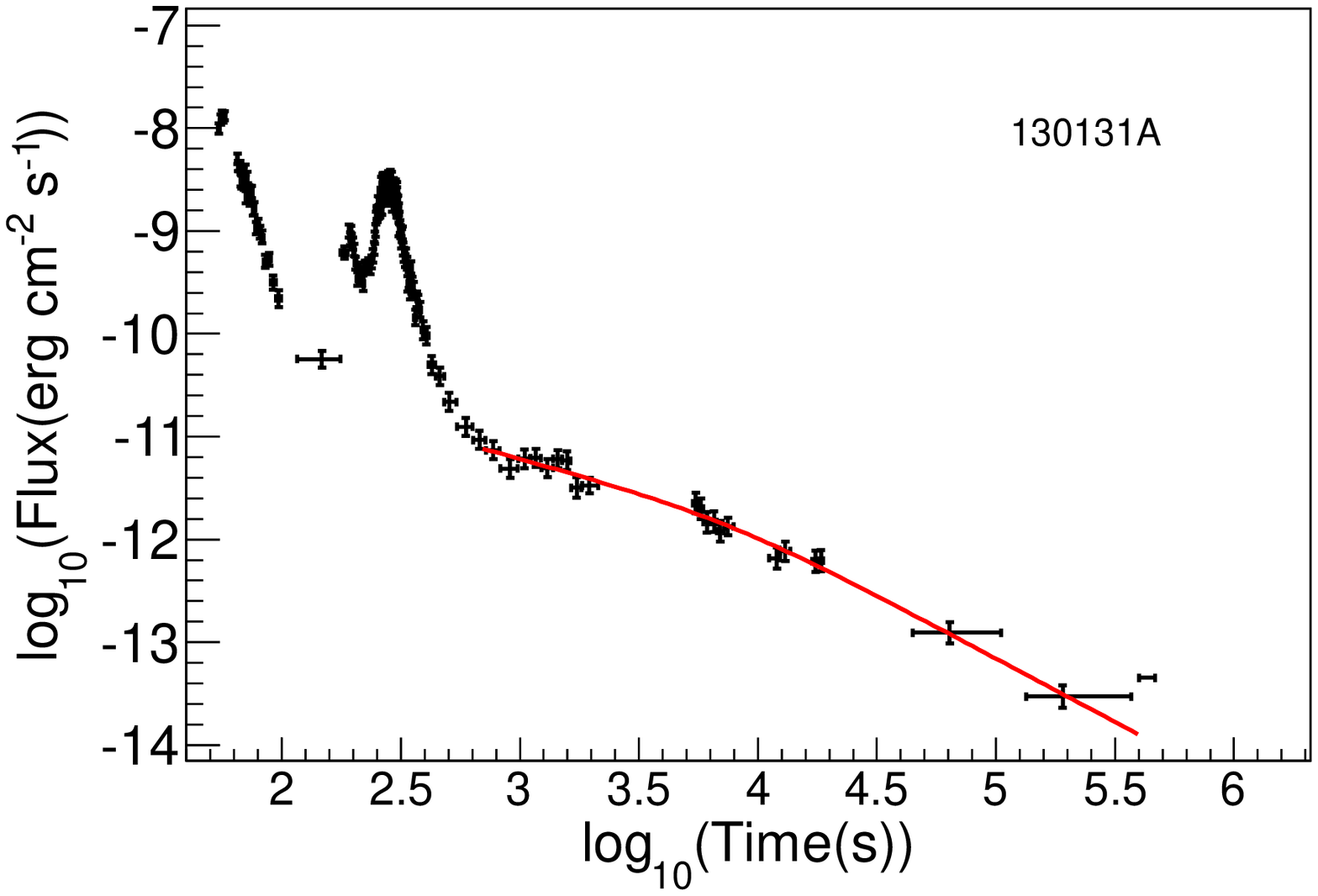}
\caption{ Continued.}
\label{fig-1-12}
\end{center}
\end{figure*}

\begin{figure*}
\begin{center}
\setlength{\abovecaptionskip}{0.cm}
\setlength{\belowcaptionskip}{-0.cm}
\figurenum{1}
\hspace{0cm}
\graphicspath{{lightcurve/}}
\includegraphics[width=5.5cm,height=5cm]{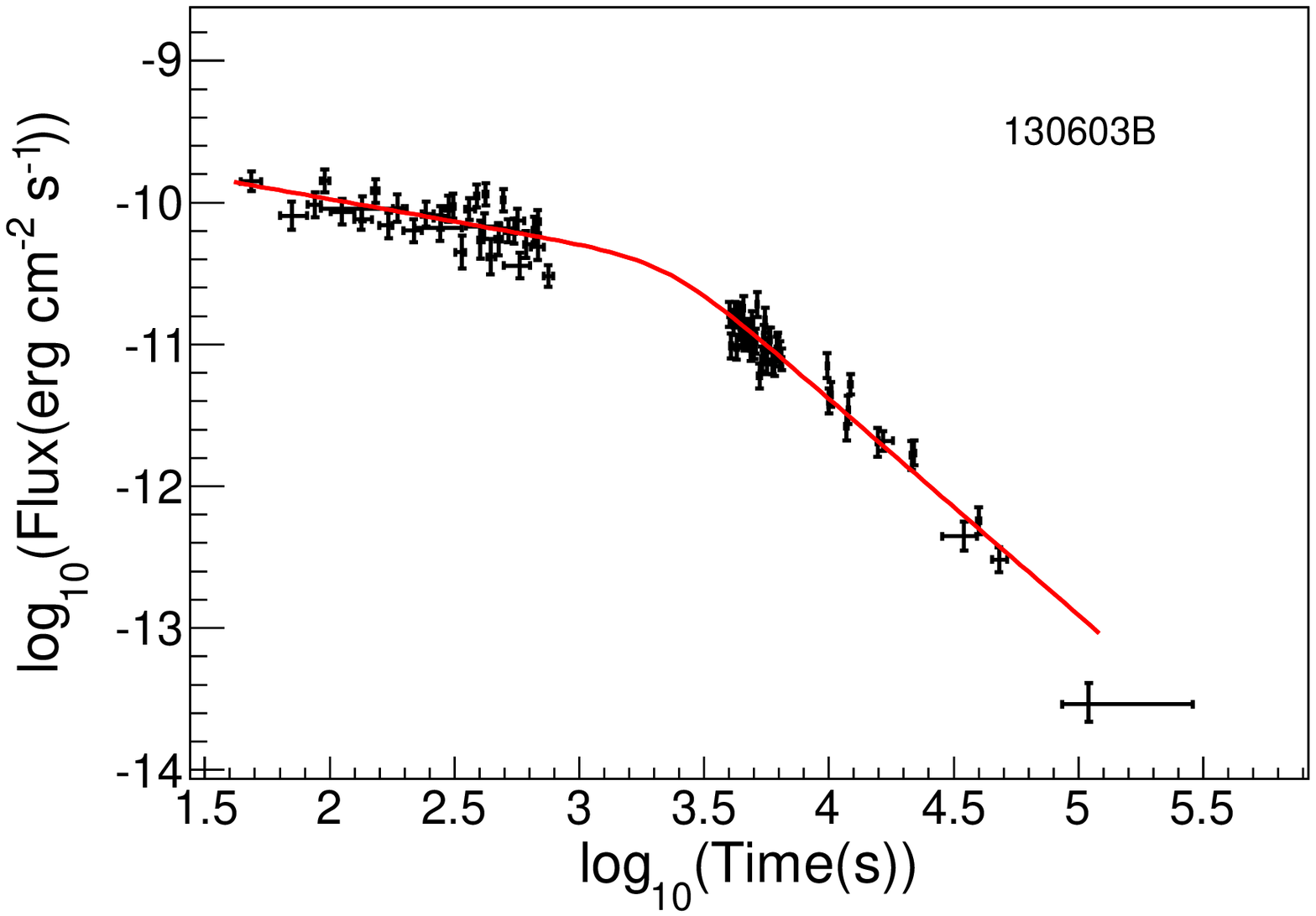}
\includegraphics[width=5.5cm,height=5cm]{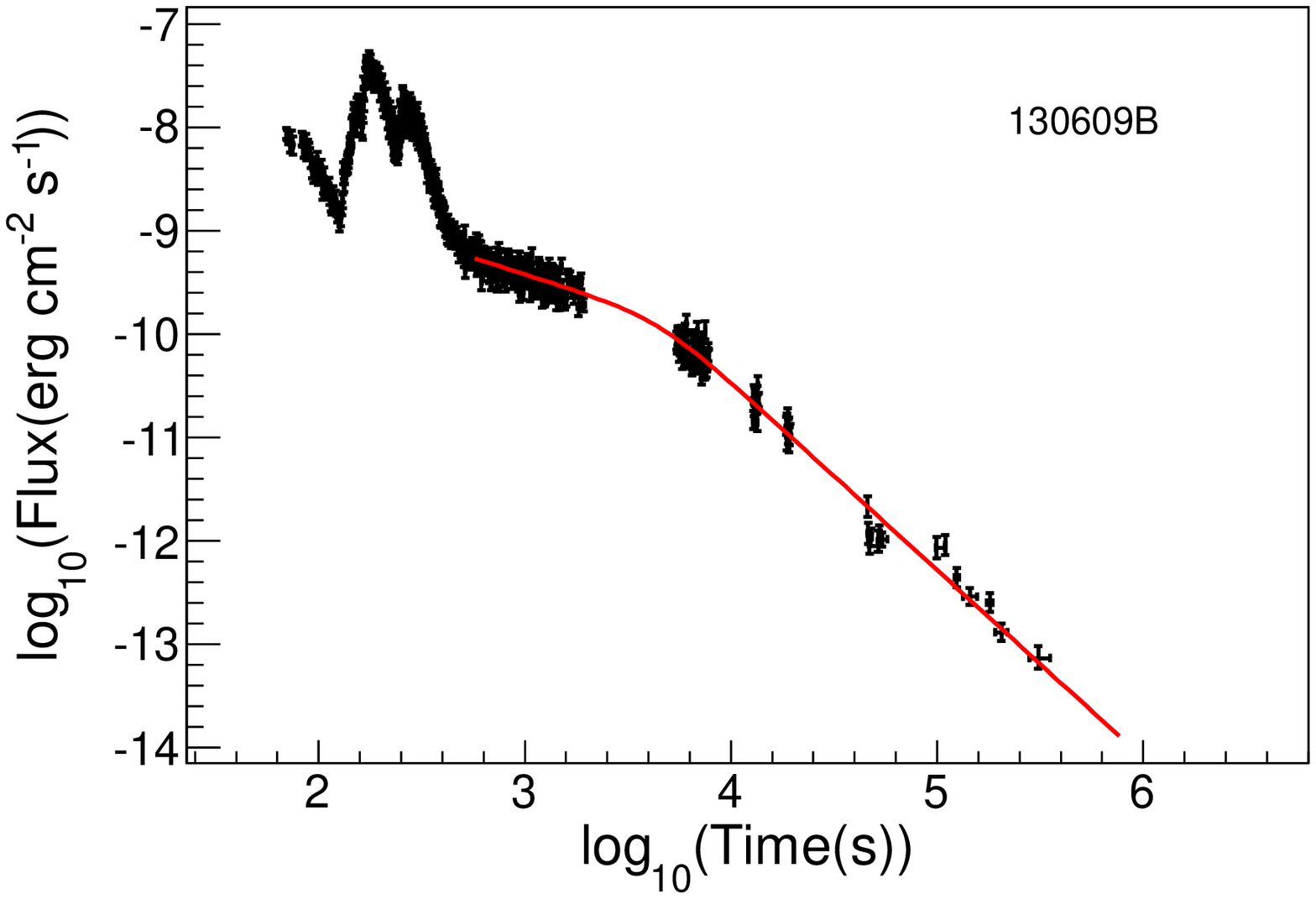}
\includegraphics[width=5.5cm,height=5cm]{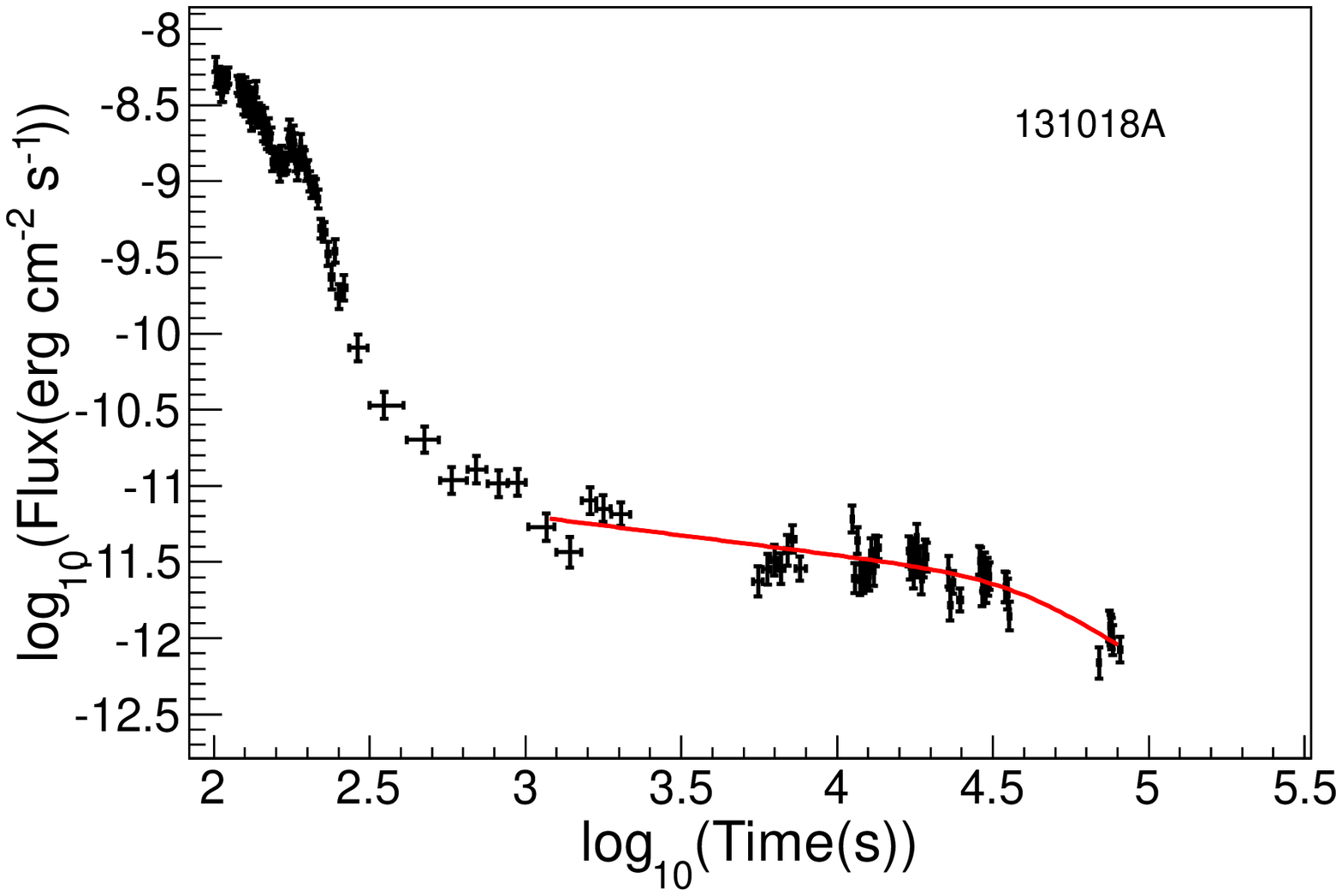}
\includegraphics[width=5.5cm,height=5cm]{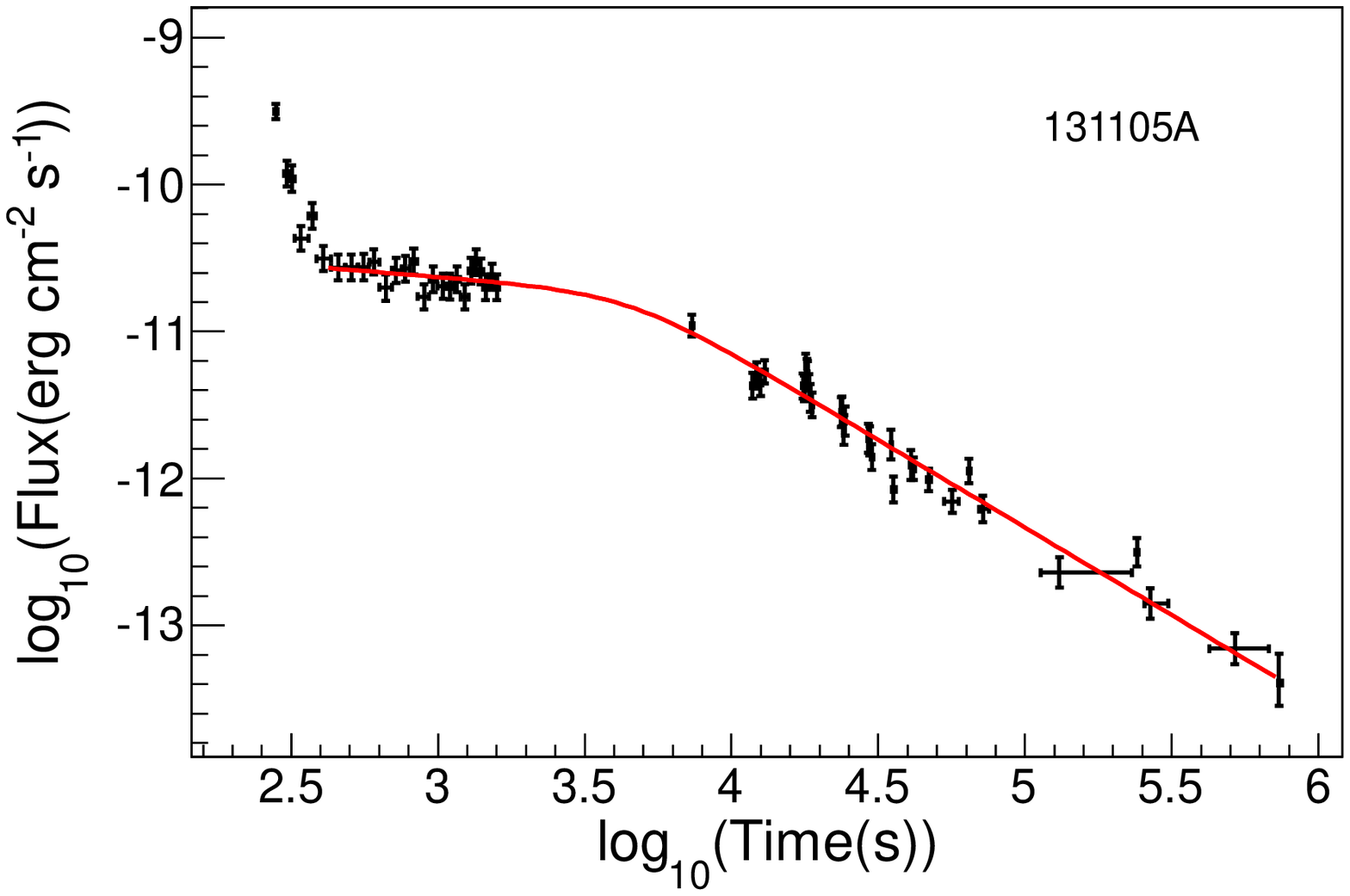}
\includegraphics[width=5.5cm,height=5cm]{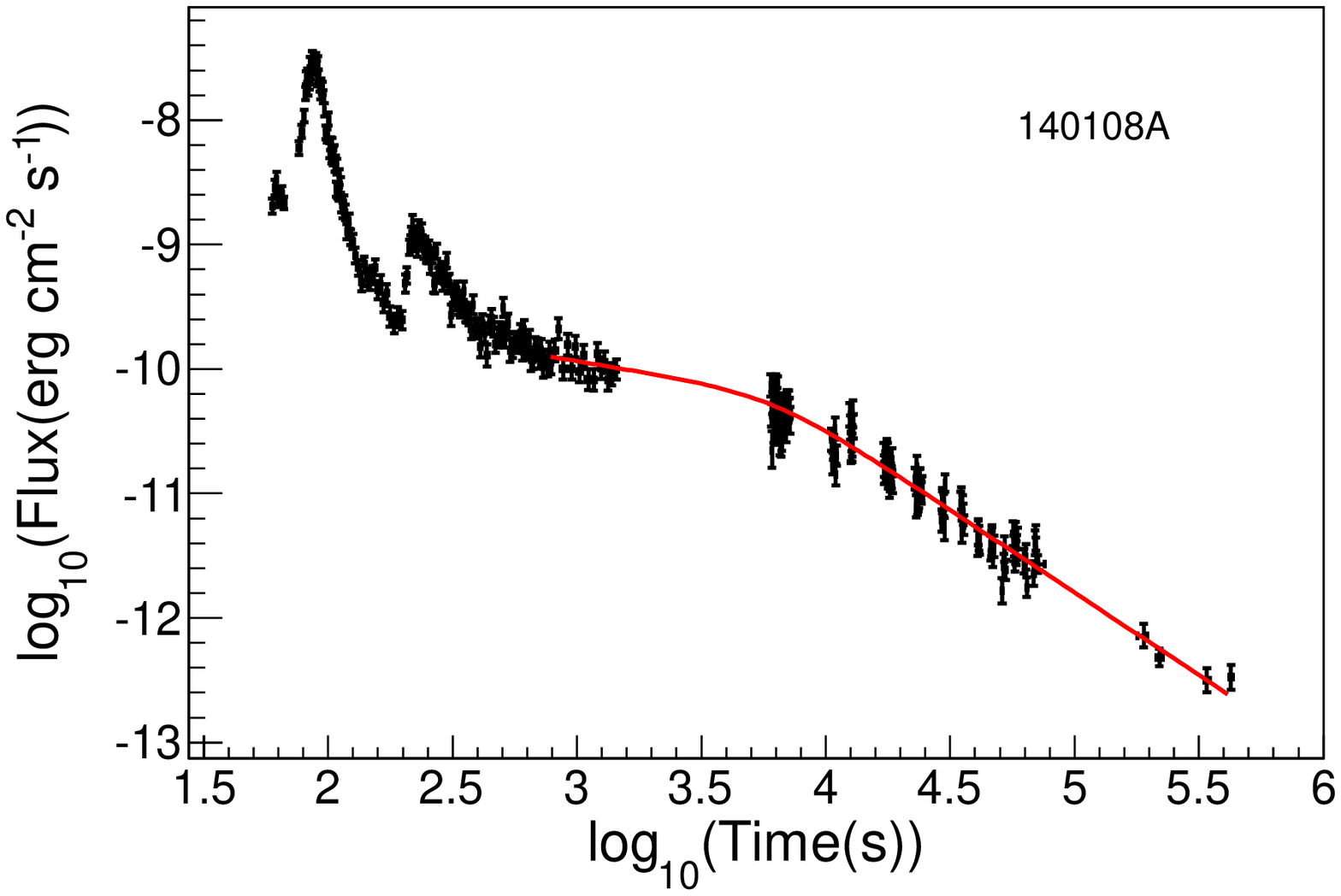}
\includegraphics[width=5.5cm,height=5cm]{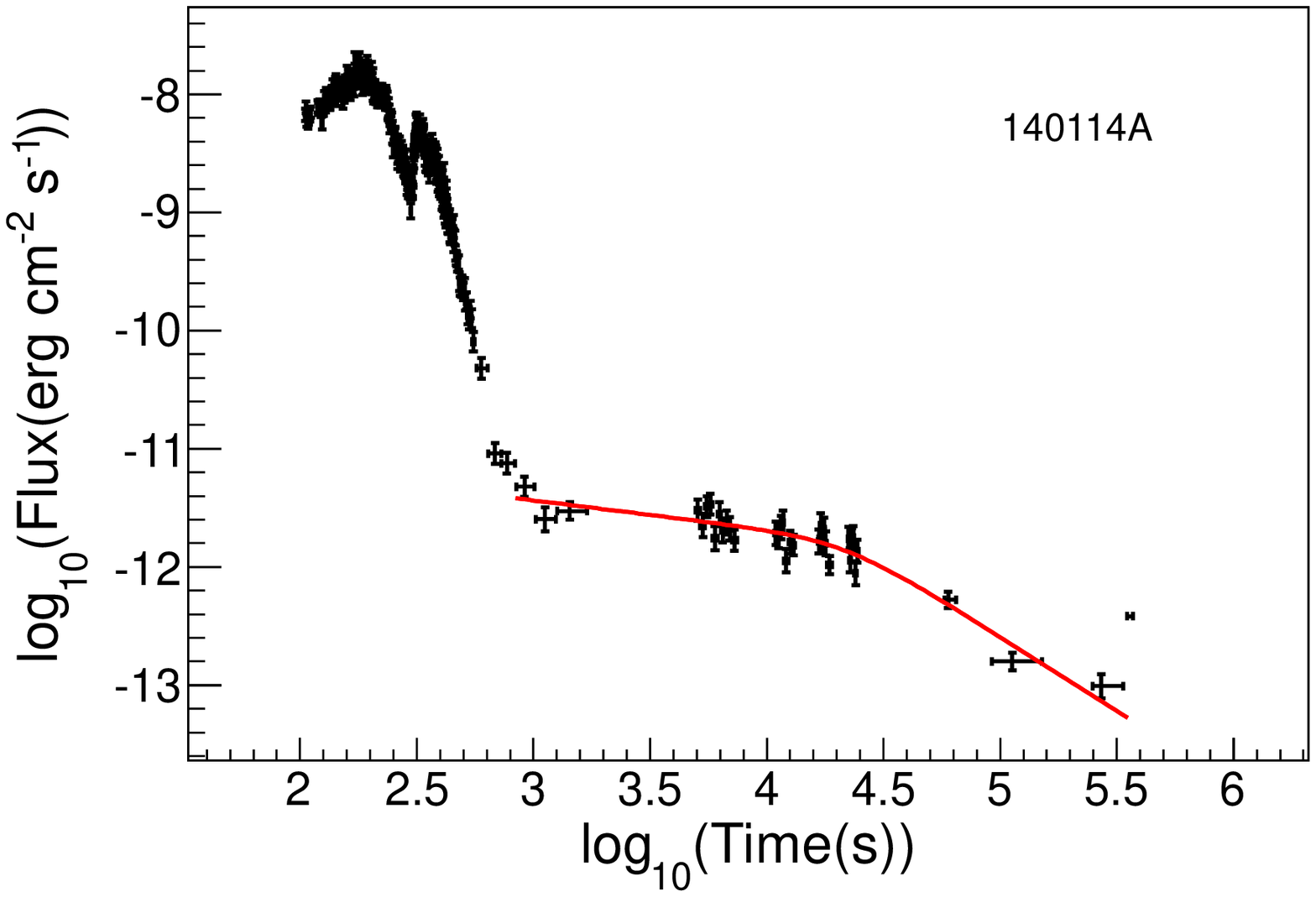}
\includegraphics[width=5.5cm,height=5cm]{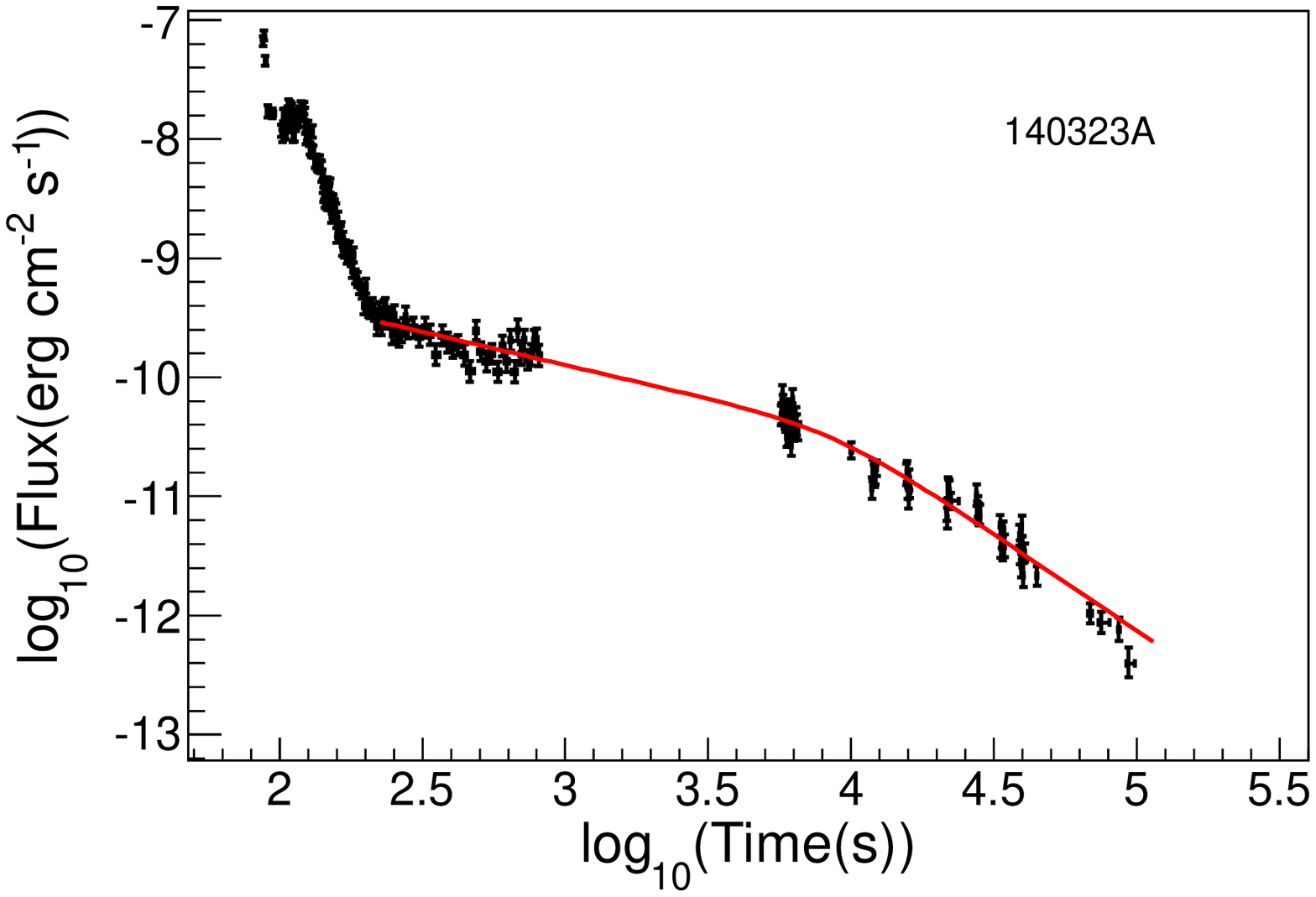}
\includegraphics[width=5.5cm,height=5cm]{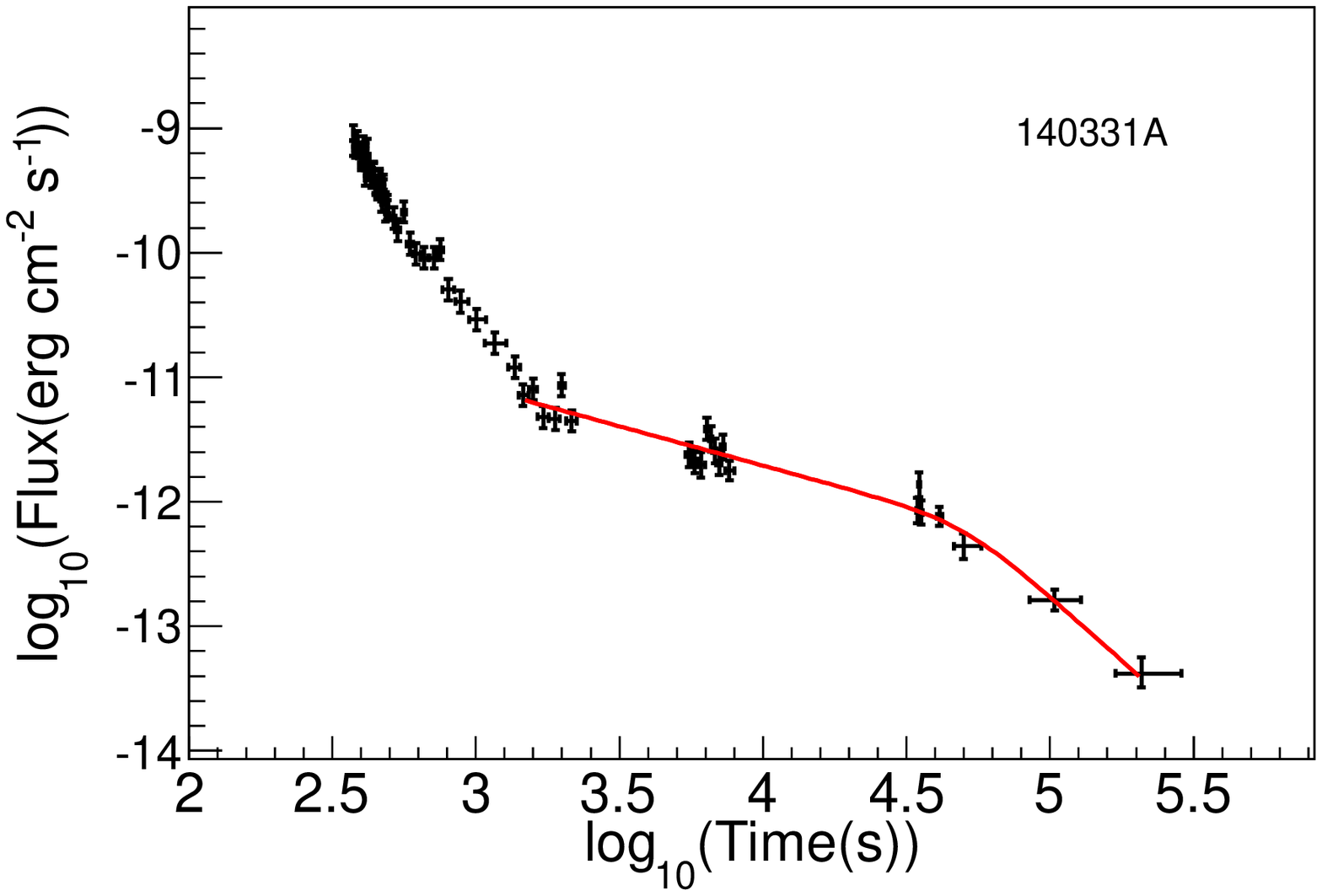}
\includegraphics[width=5.5cm,height=5cm]{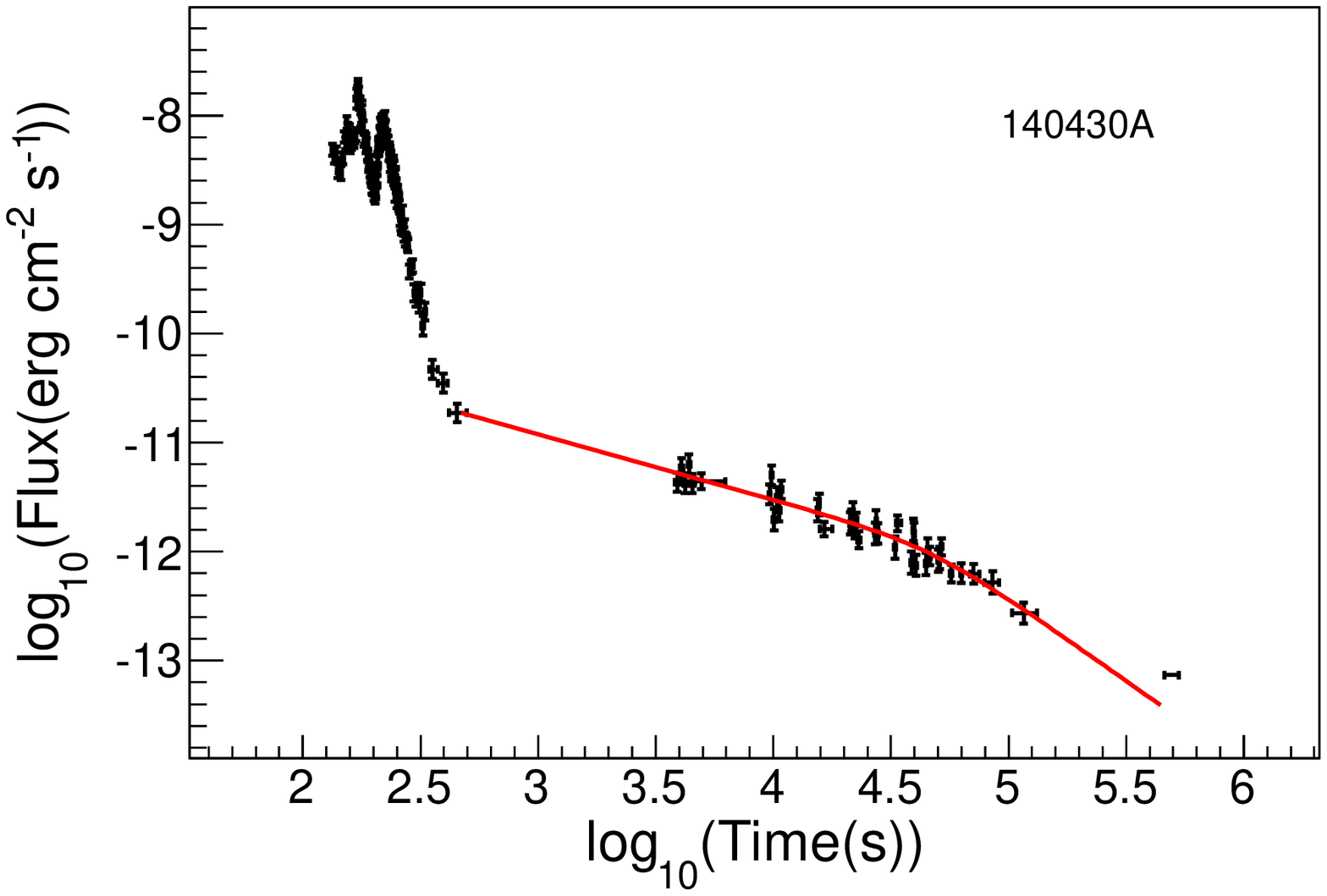}
\includegraphics[width=5.5cm,height=5cm]{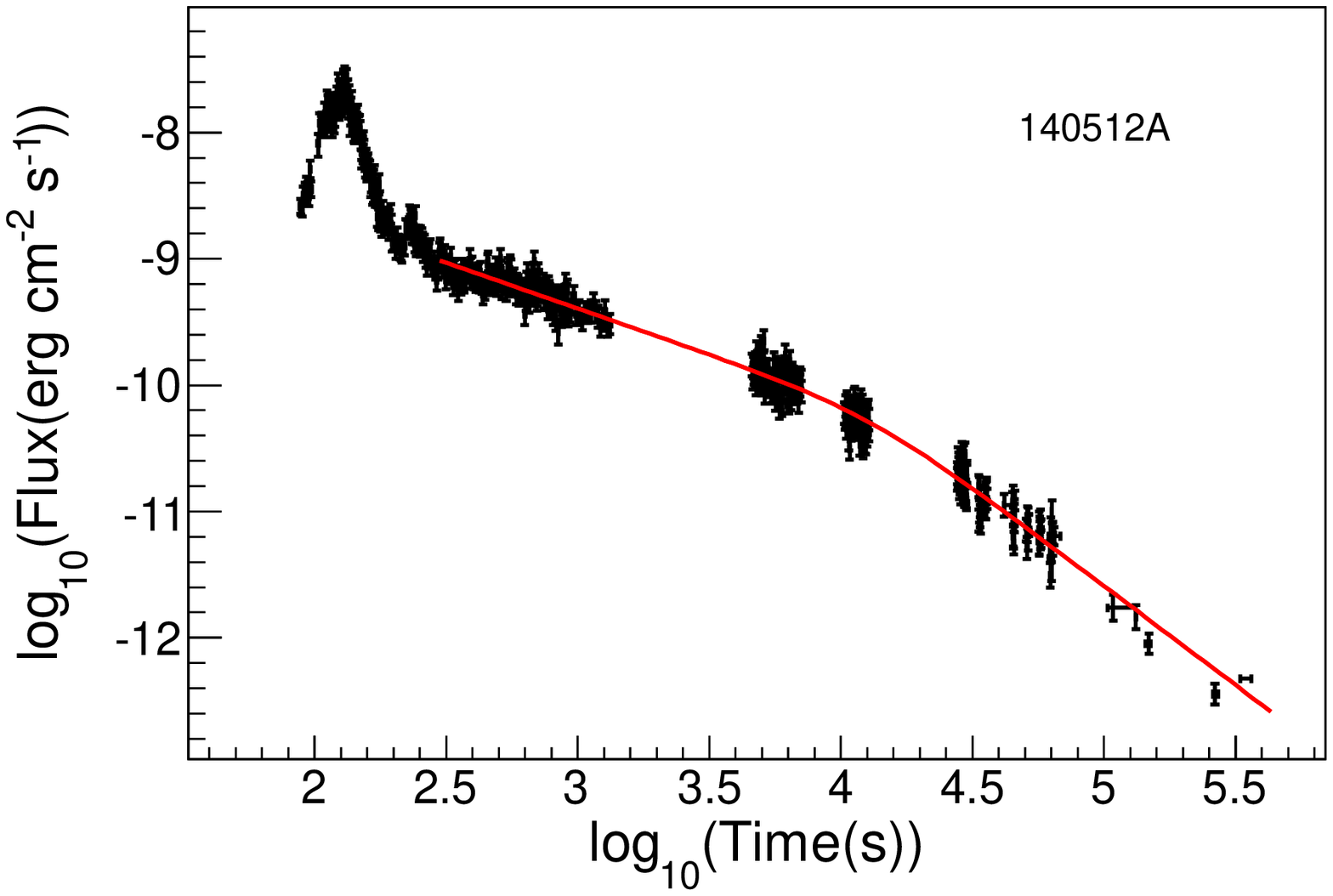}
\includegraphics[width=5.5cm,height=5cm]{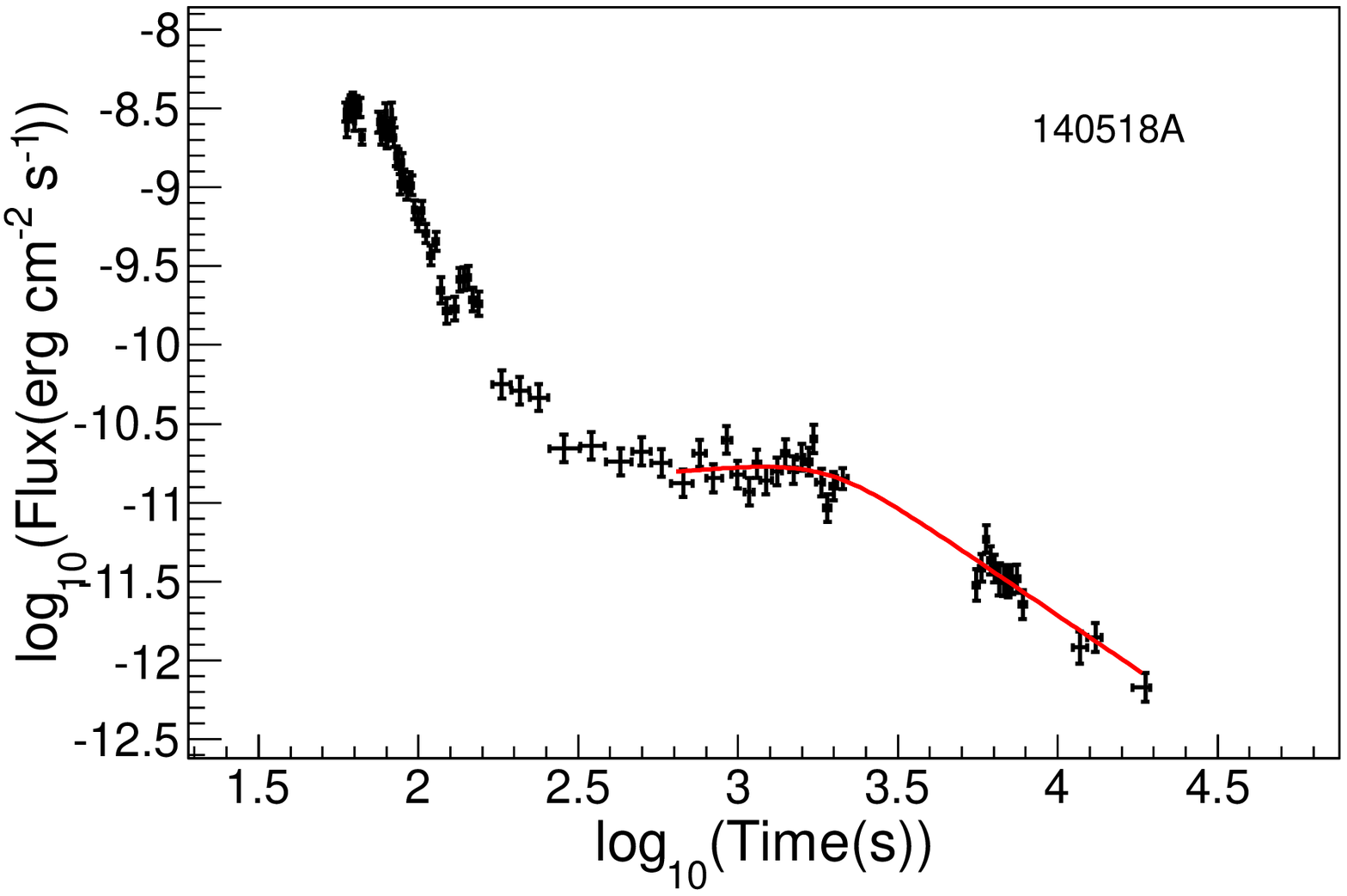}
\includegraphics[width=5.5cm,height=5cm]{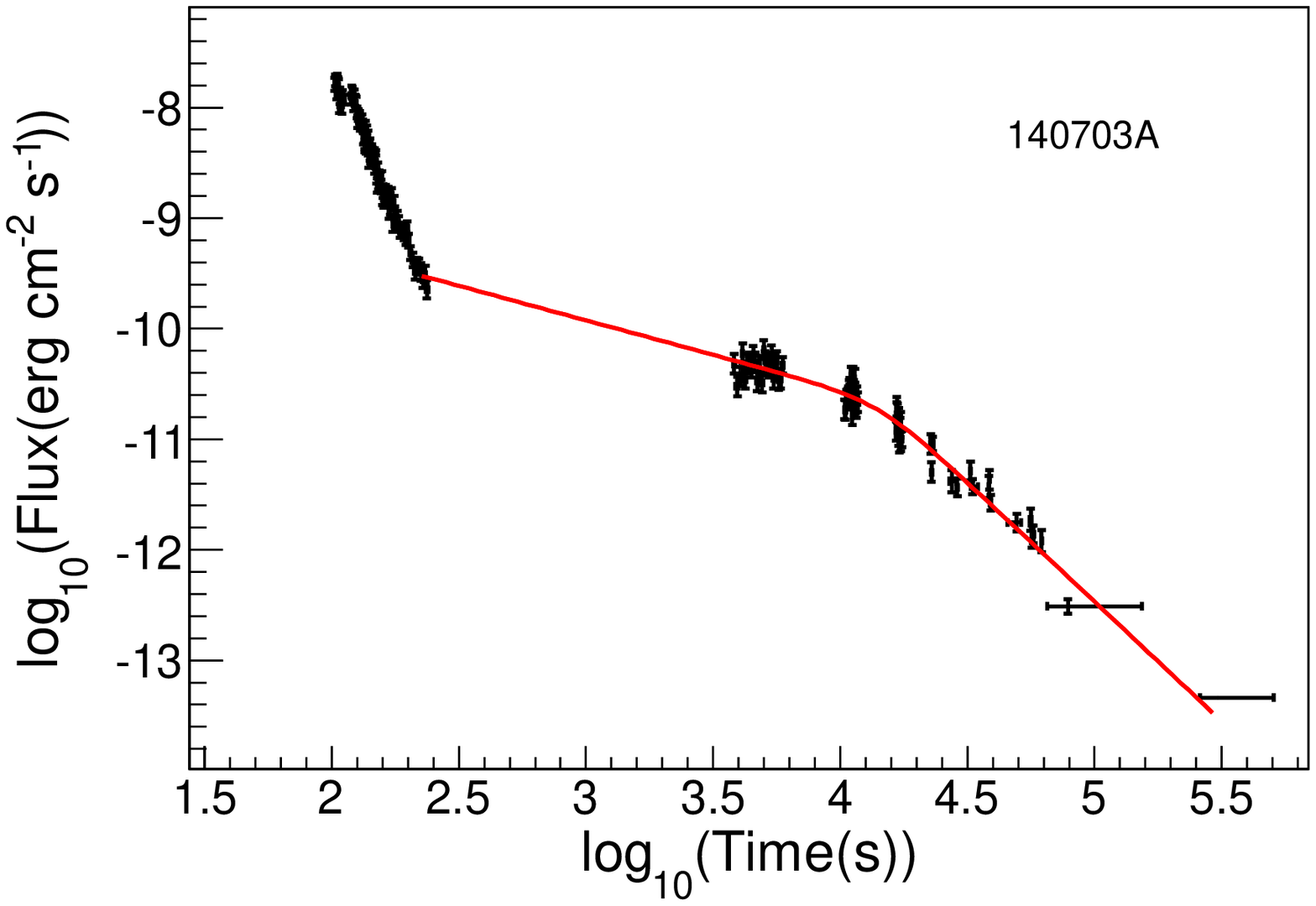}
\caption{ Continued.}
\label{fig-1-13}
\end{center}
\end{figure*}

\begin{figure*}
\begin{center}
\setlength{\abovecaptionskip}{0.cm}
\setlength{\belowcaptionskip}{-0.cm}
\figurenum{1}
\hspace{0cm}
\graphicspath{{lightcurve/}}
\includegraphics[width=5.5cm,height=5cm]{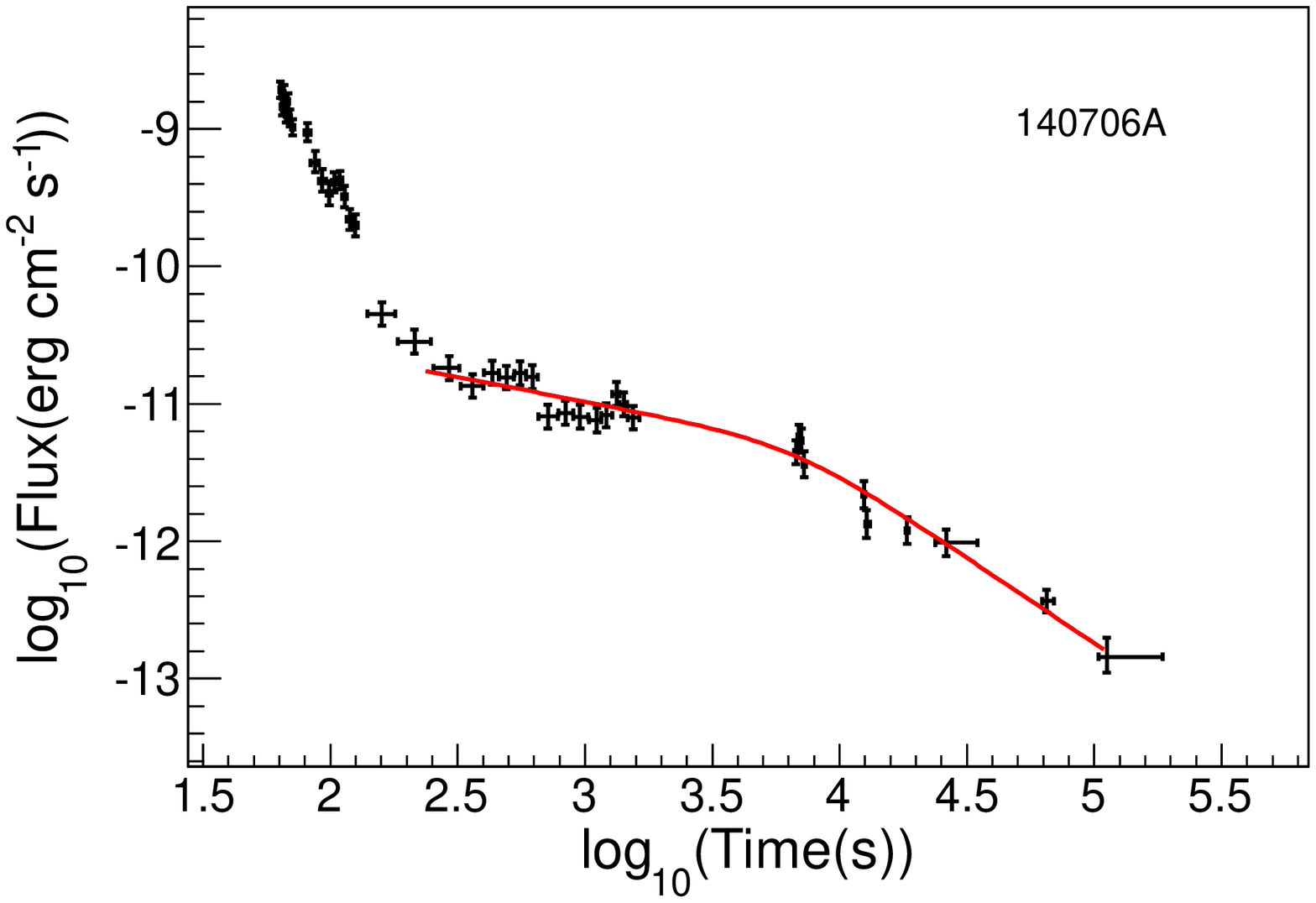}
\includegraphics[width=5.5cm,height=5cm]{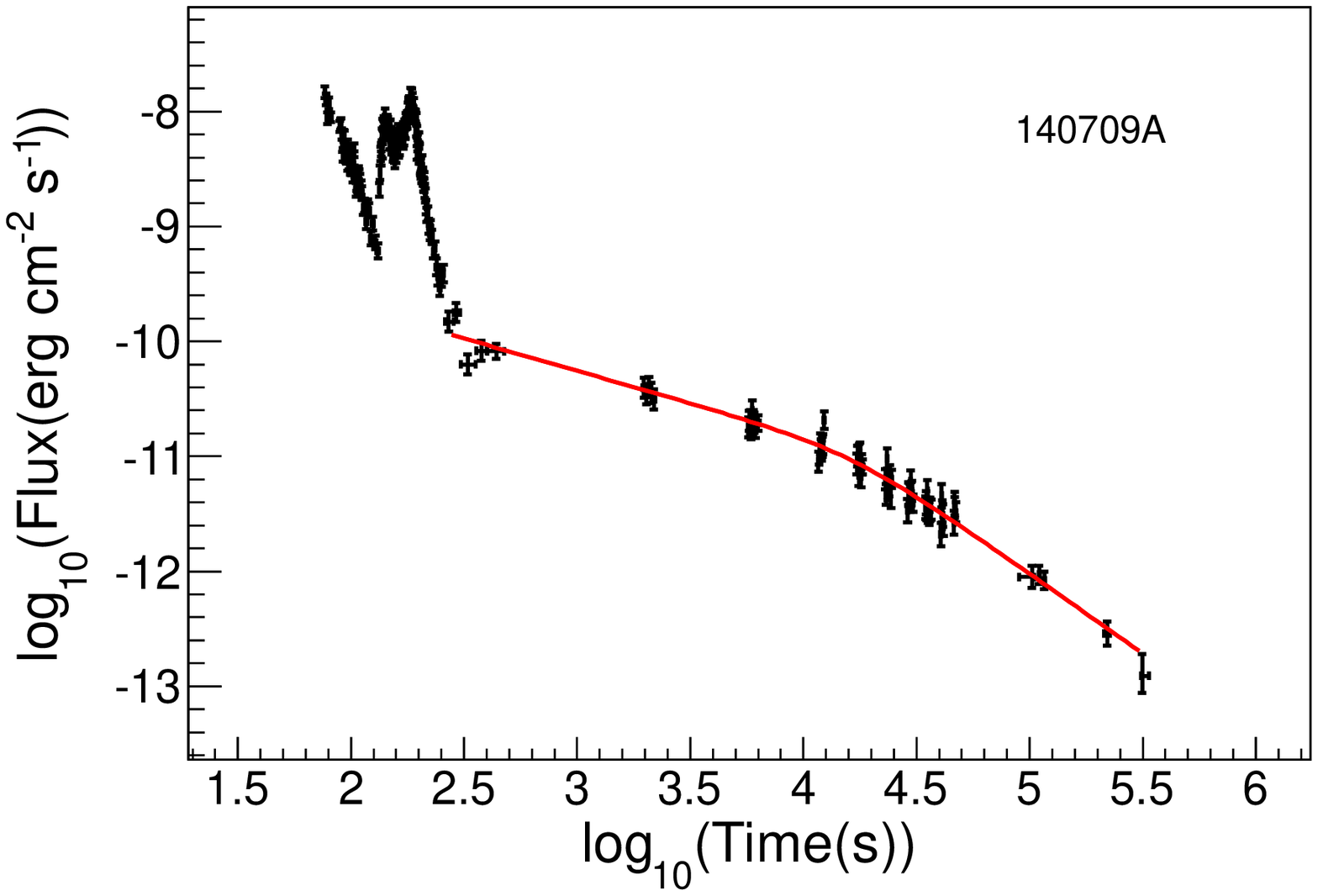}
\includegraphics[width=5.5cm,height=5cm]{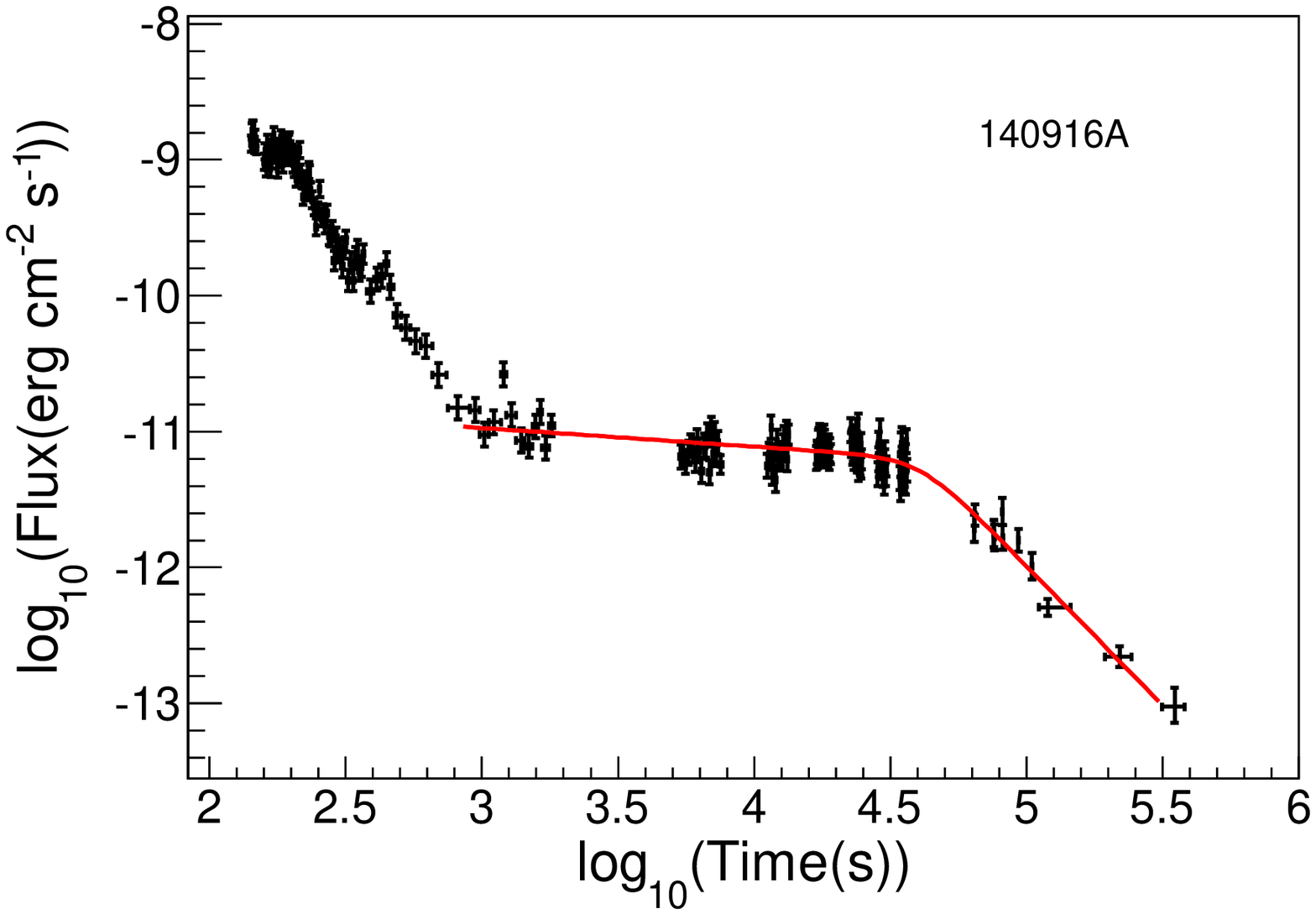}
\includegraphics[width=5.5cm,height=5cm]{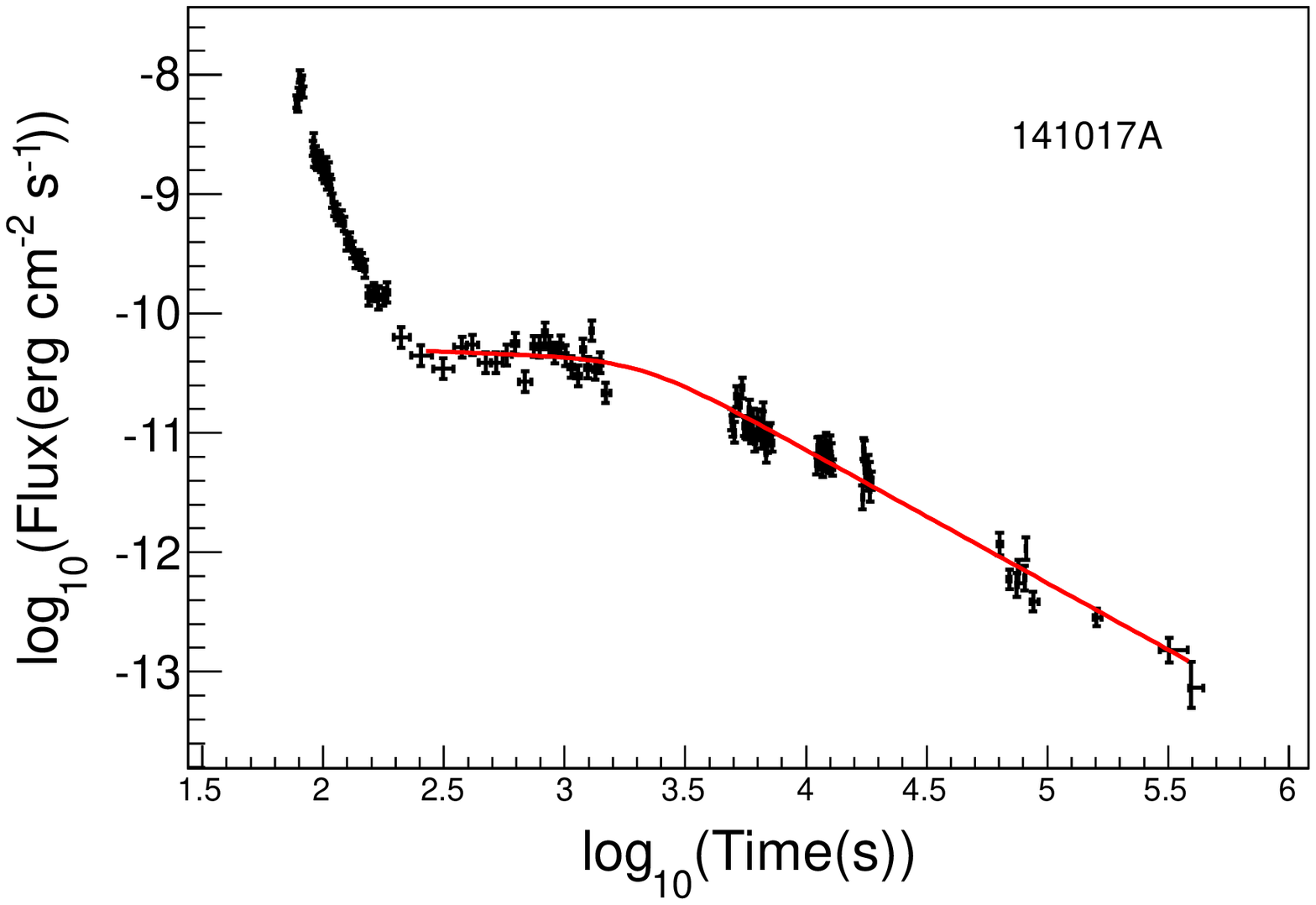}
\includegraphics[width=5.5cm,height=5cm]{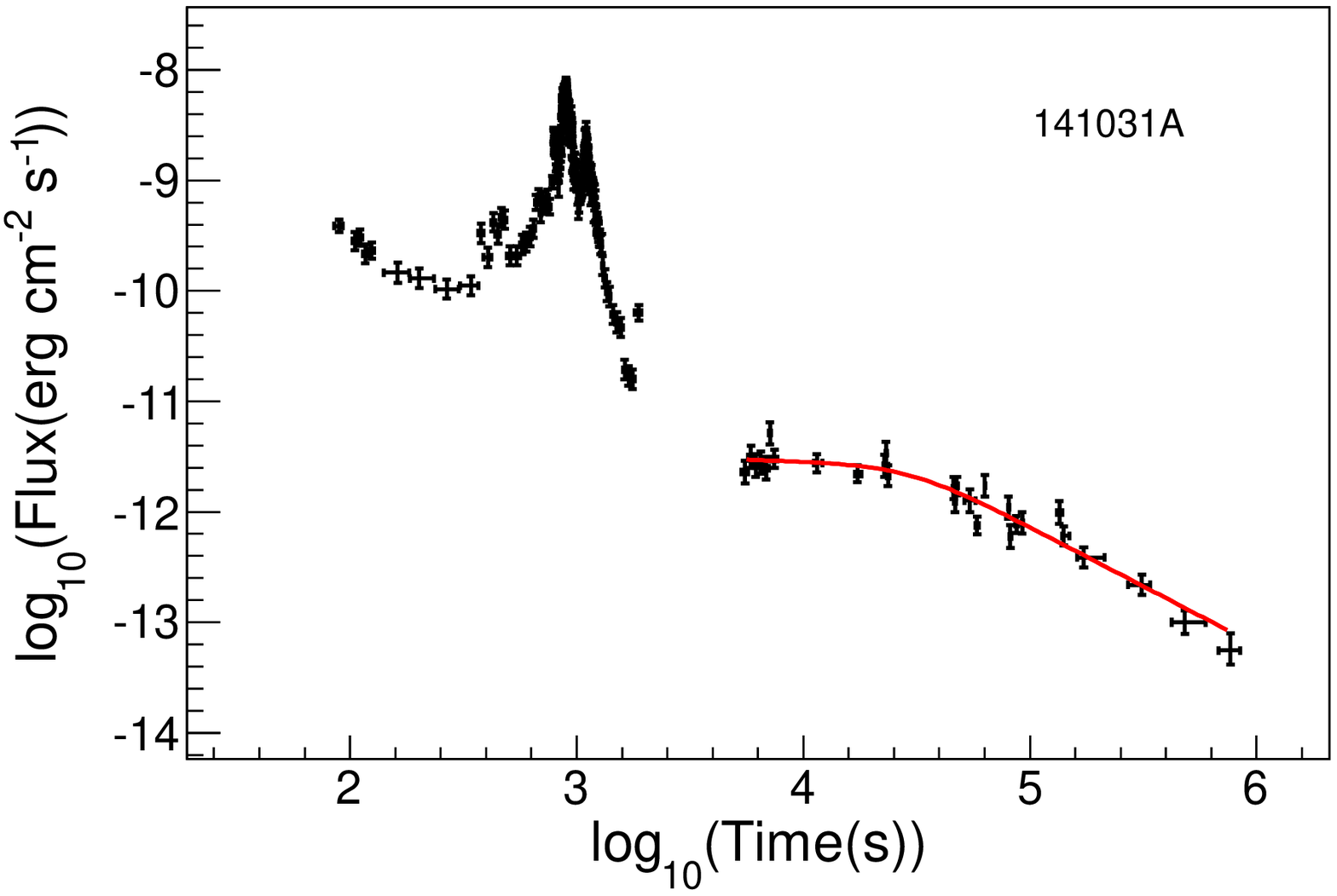}
\includegraphics[width=5.5cm,height=5cm]{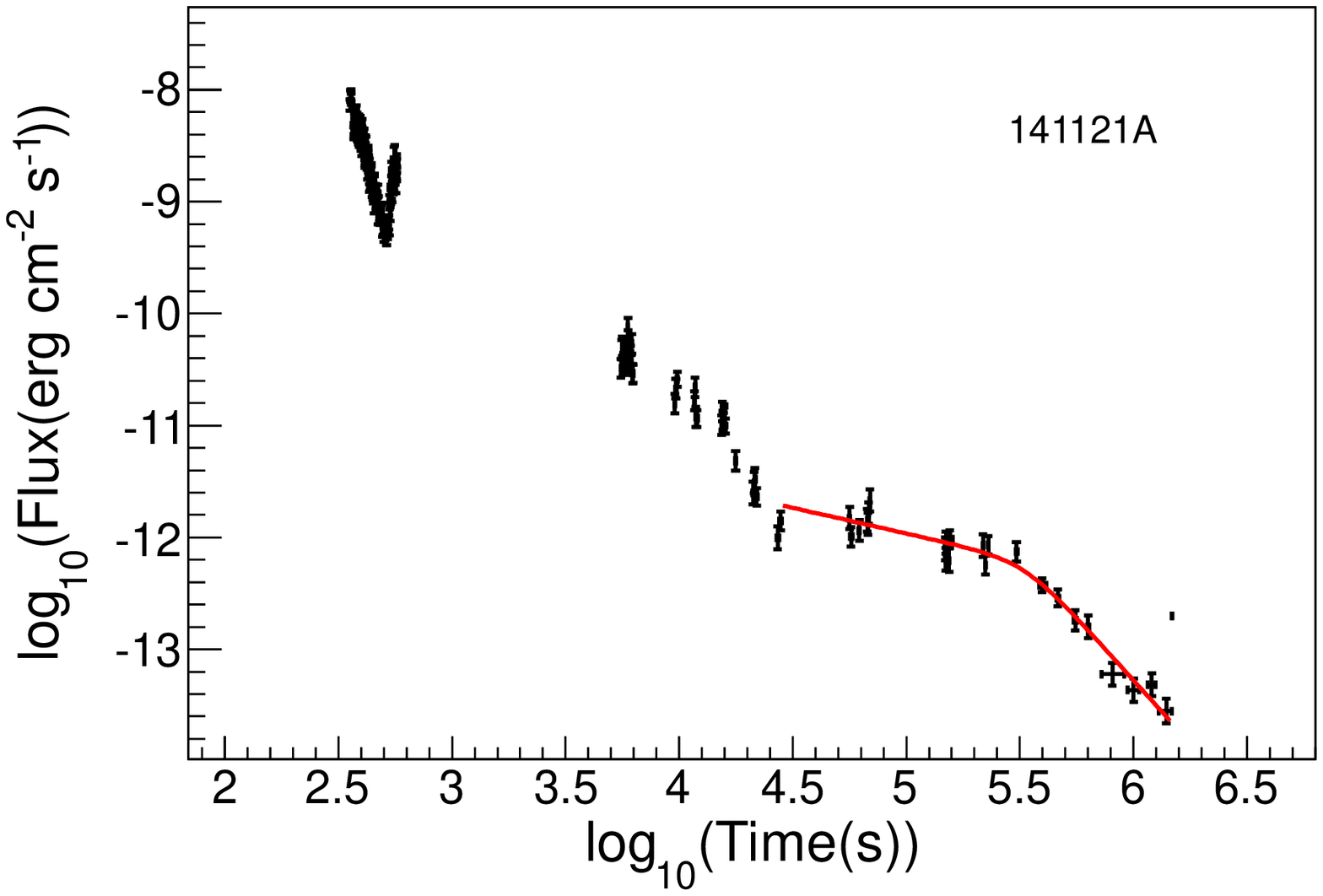}
\includegraphics[width=5.5cm,height=5cm]{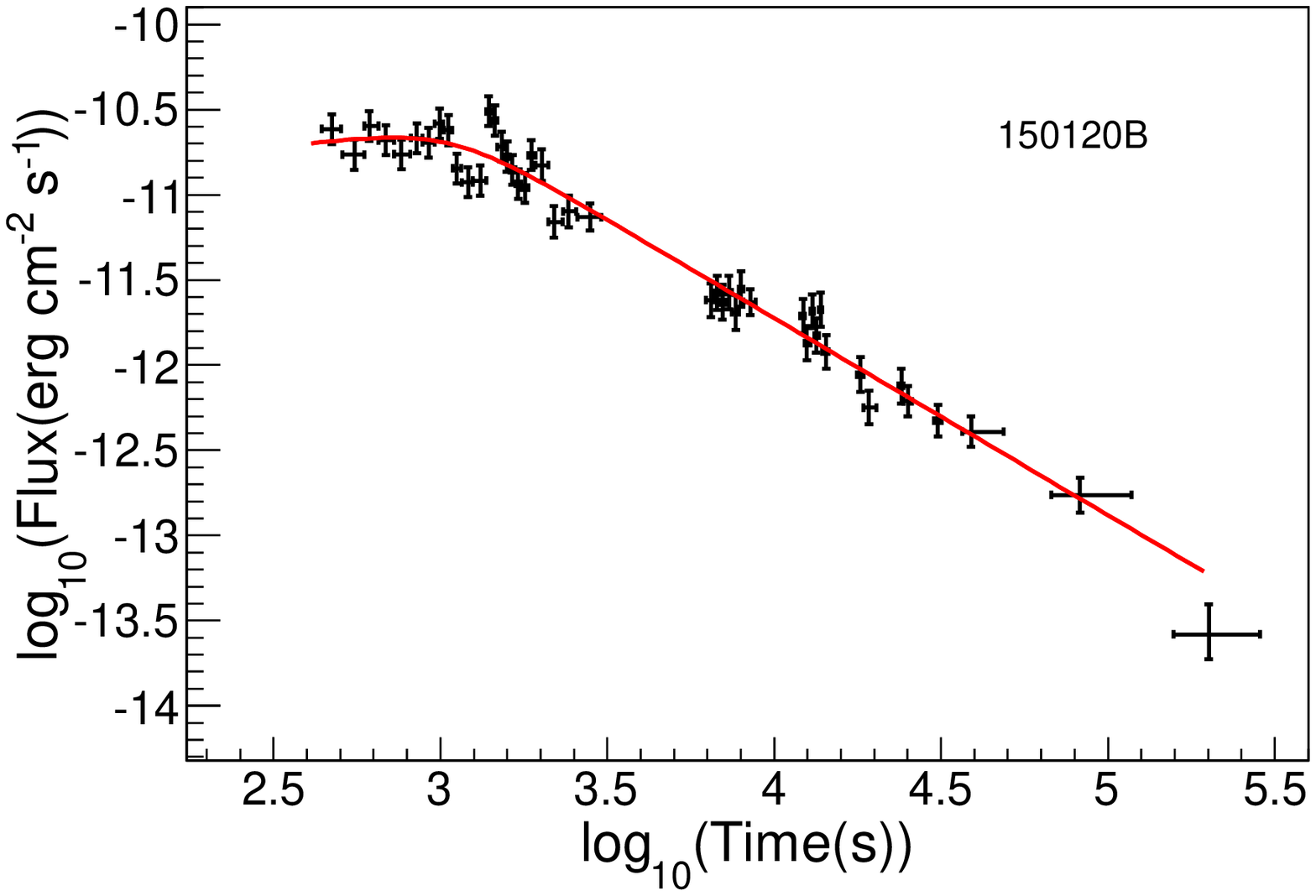}
\includegraphics[width=5.5cm,height=5cm]{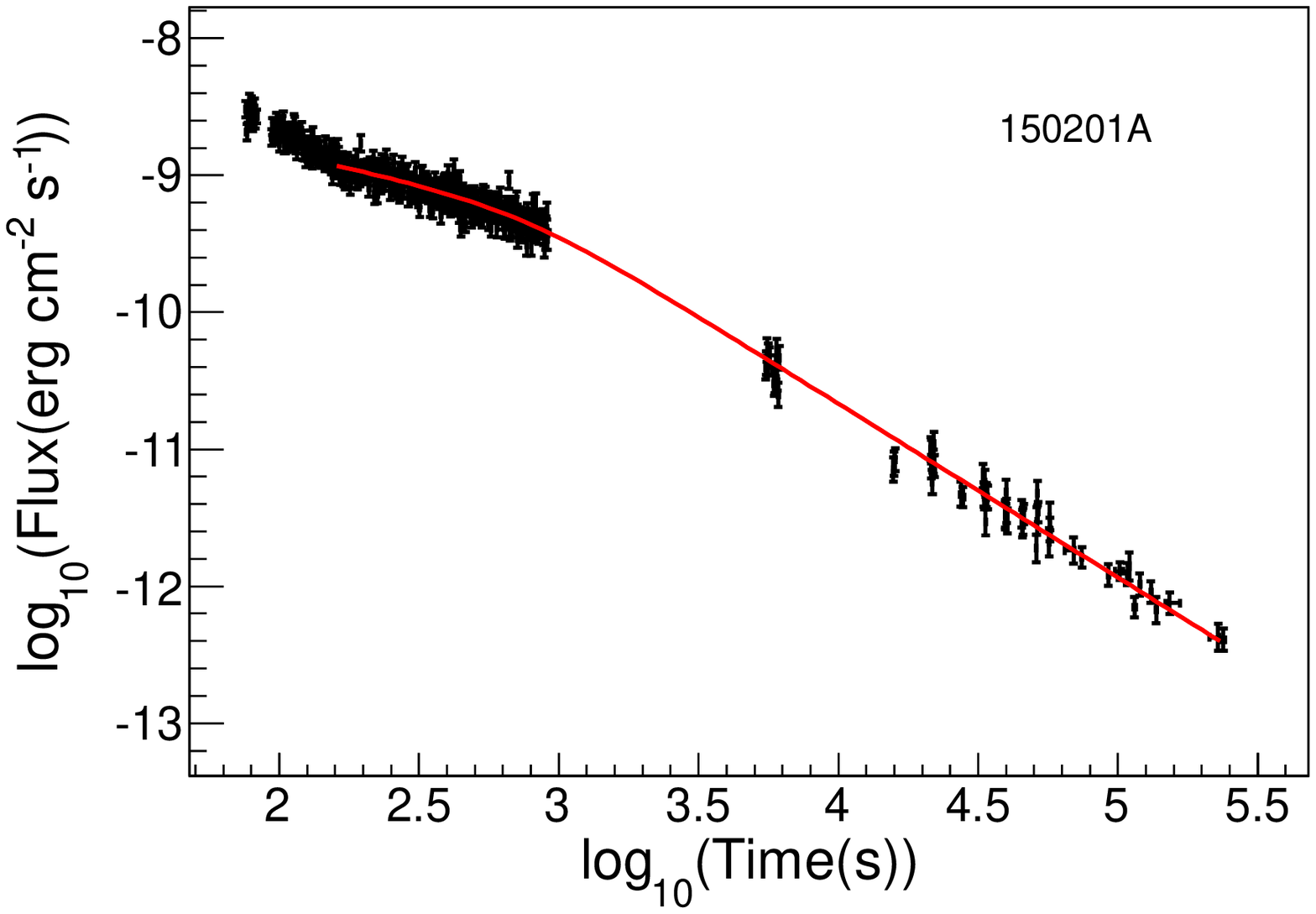}
\includegraphics[width=5.5cm,height=5cm]{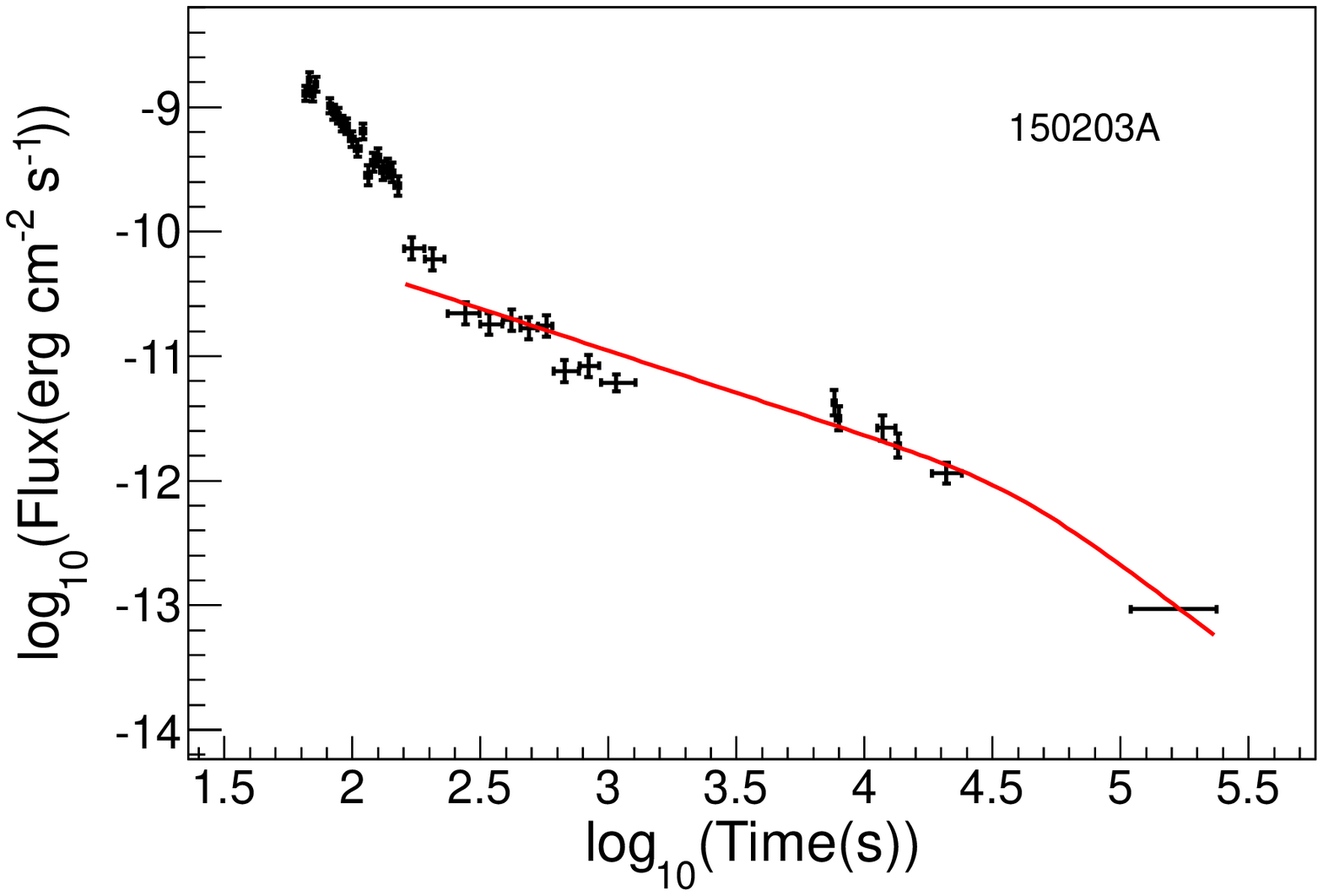}
\includegraphics[width=5.5cm,height=5cm]{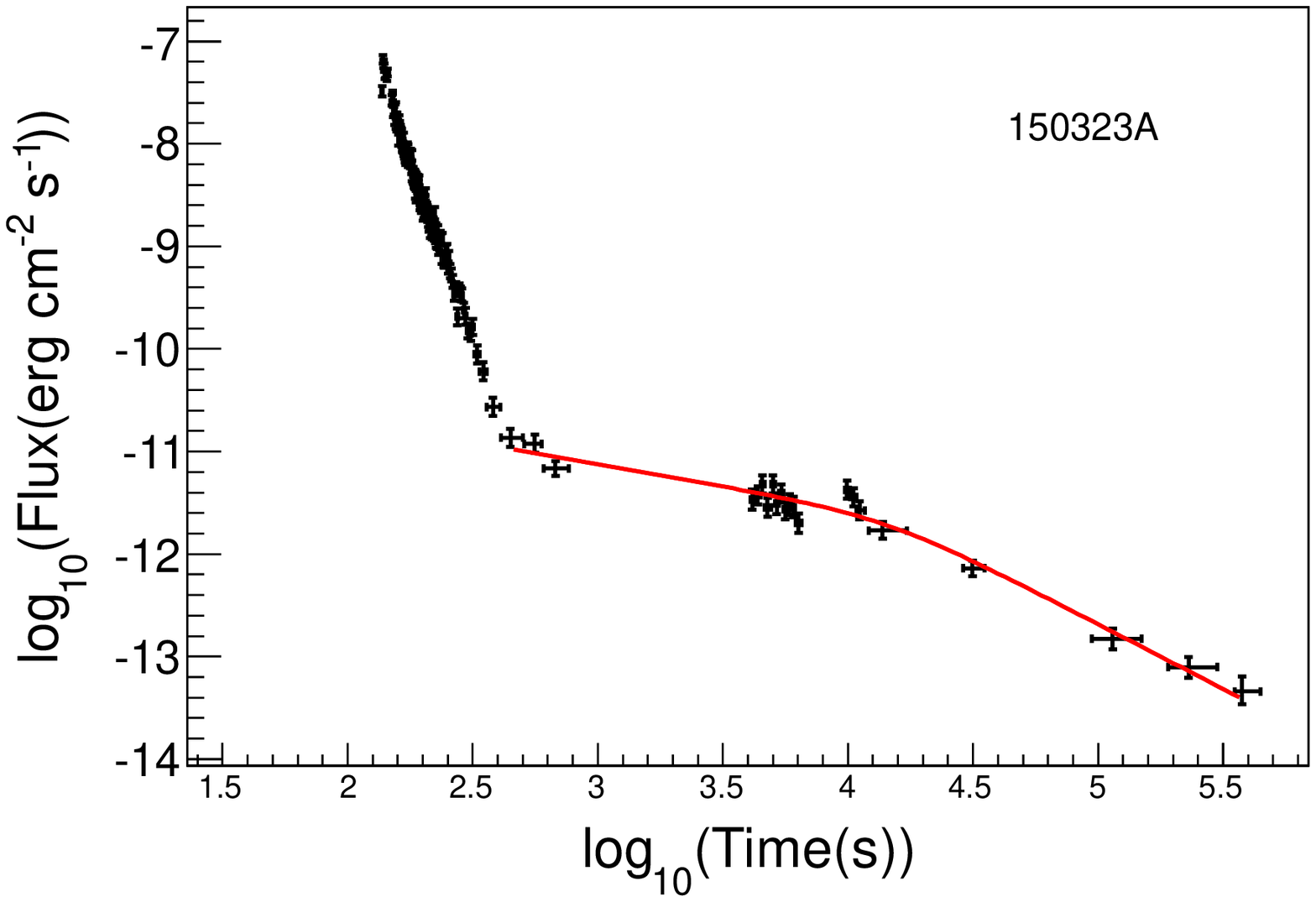}
\includegraphics[width=5.5cm,height=5cm]{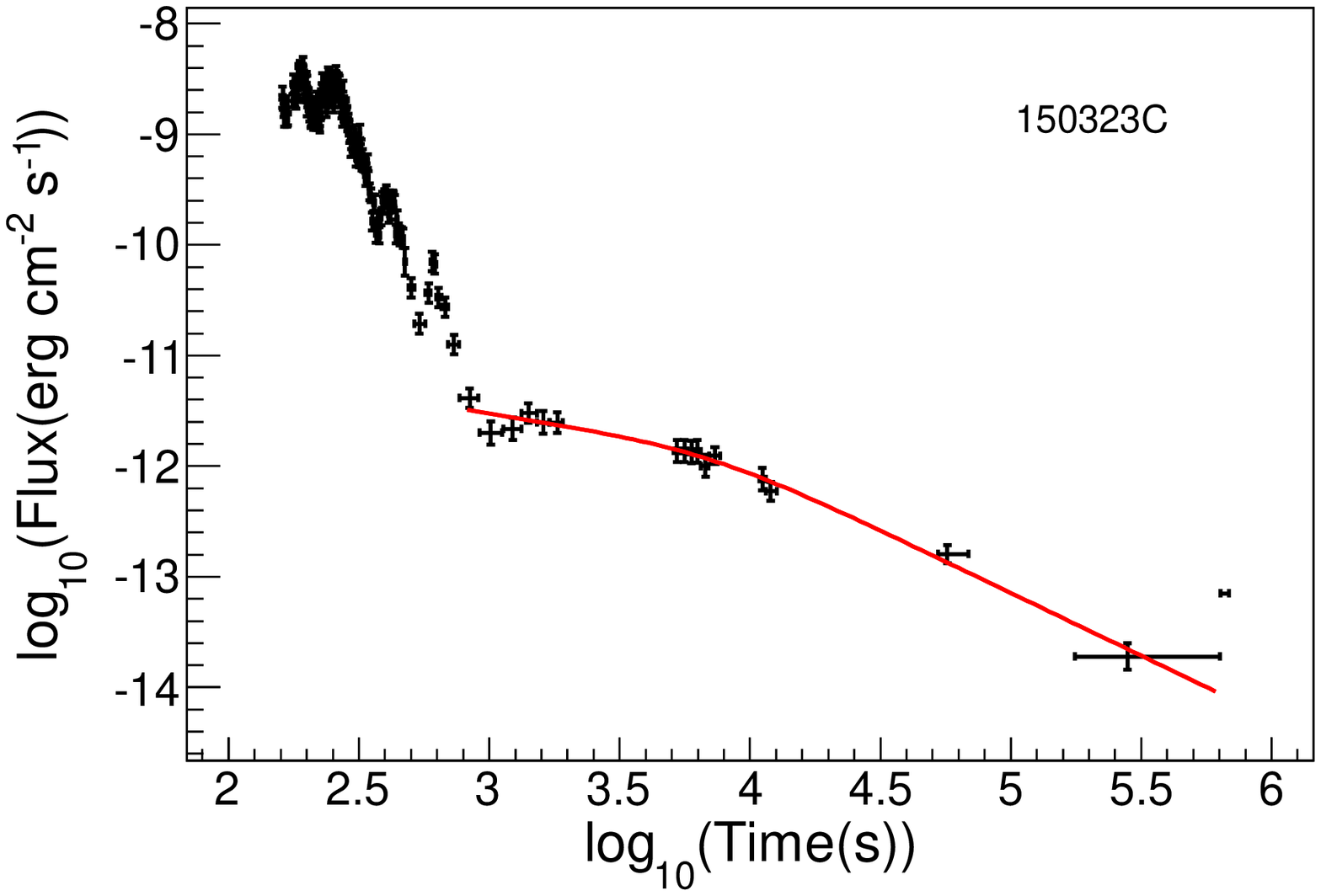}
\includegraphics[width=5.5cm,height=5cm]{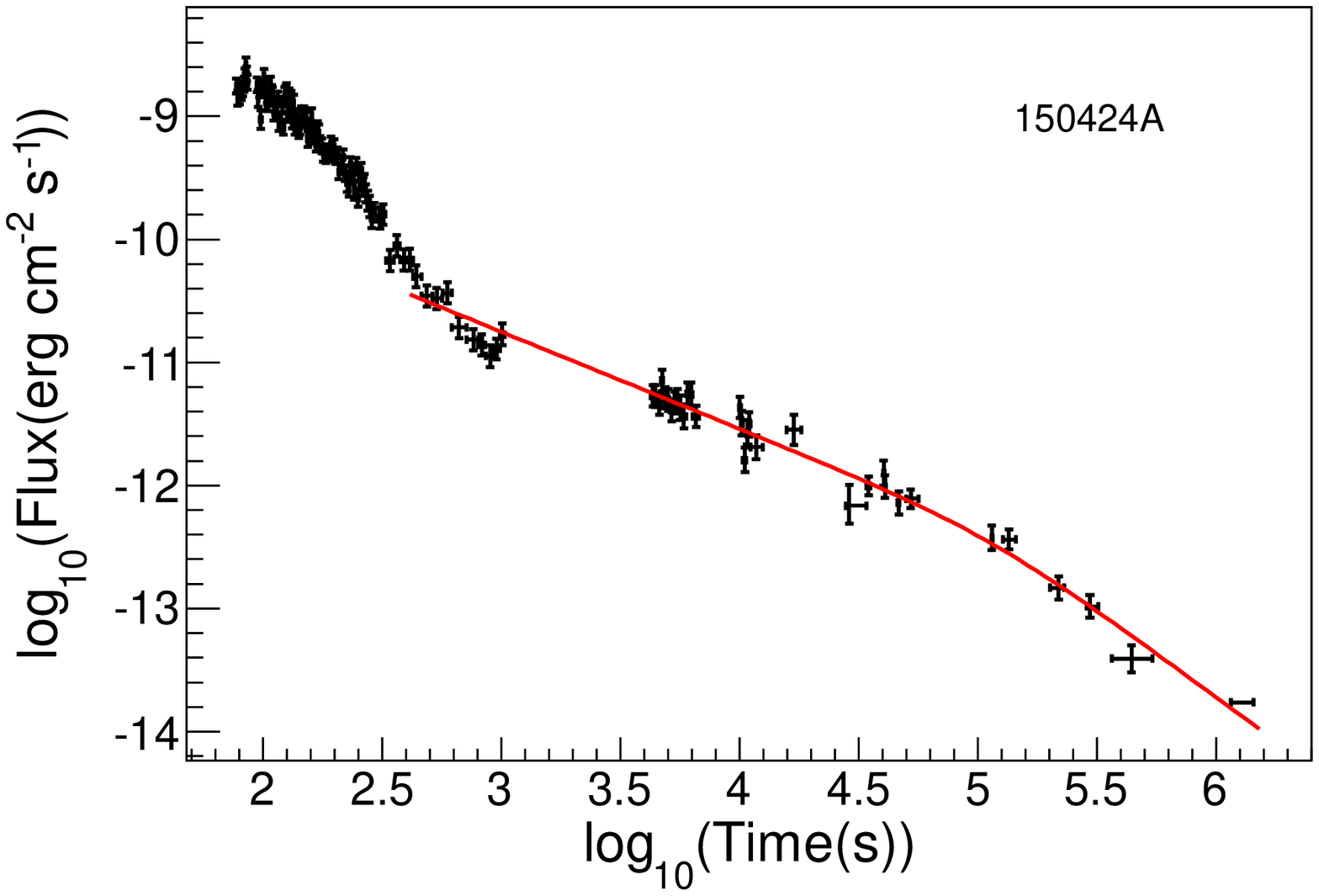}
\caption{ Continued.}
\label{fig-1-14}
\end{center}
\end{figure*}

\begin{figure*}
\begin{center}
\setlength{\abovecaptionskip}{0.cm}
\setlength{\belowcaptionskip}{-0.cm}
\figurenum{1}
\hspace{0cm}
\graphicspath{{lightcurve/}}
\includegraphics[width=5.5cm,height=5cm]{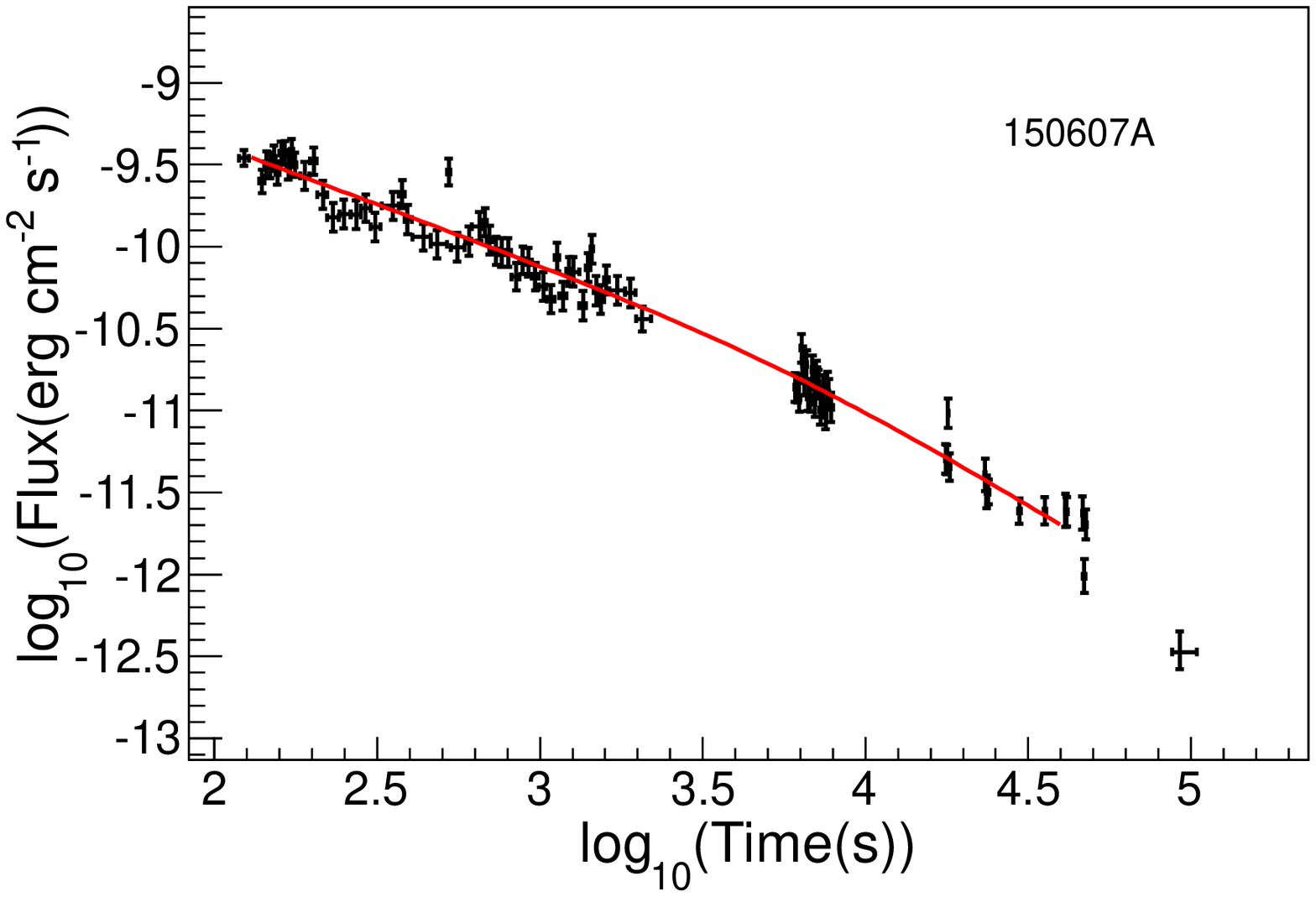}
\includegraphics[width=5.5cm,height=5cm]{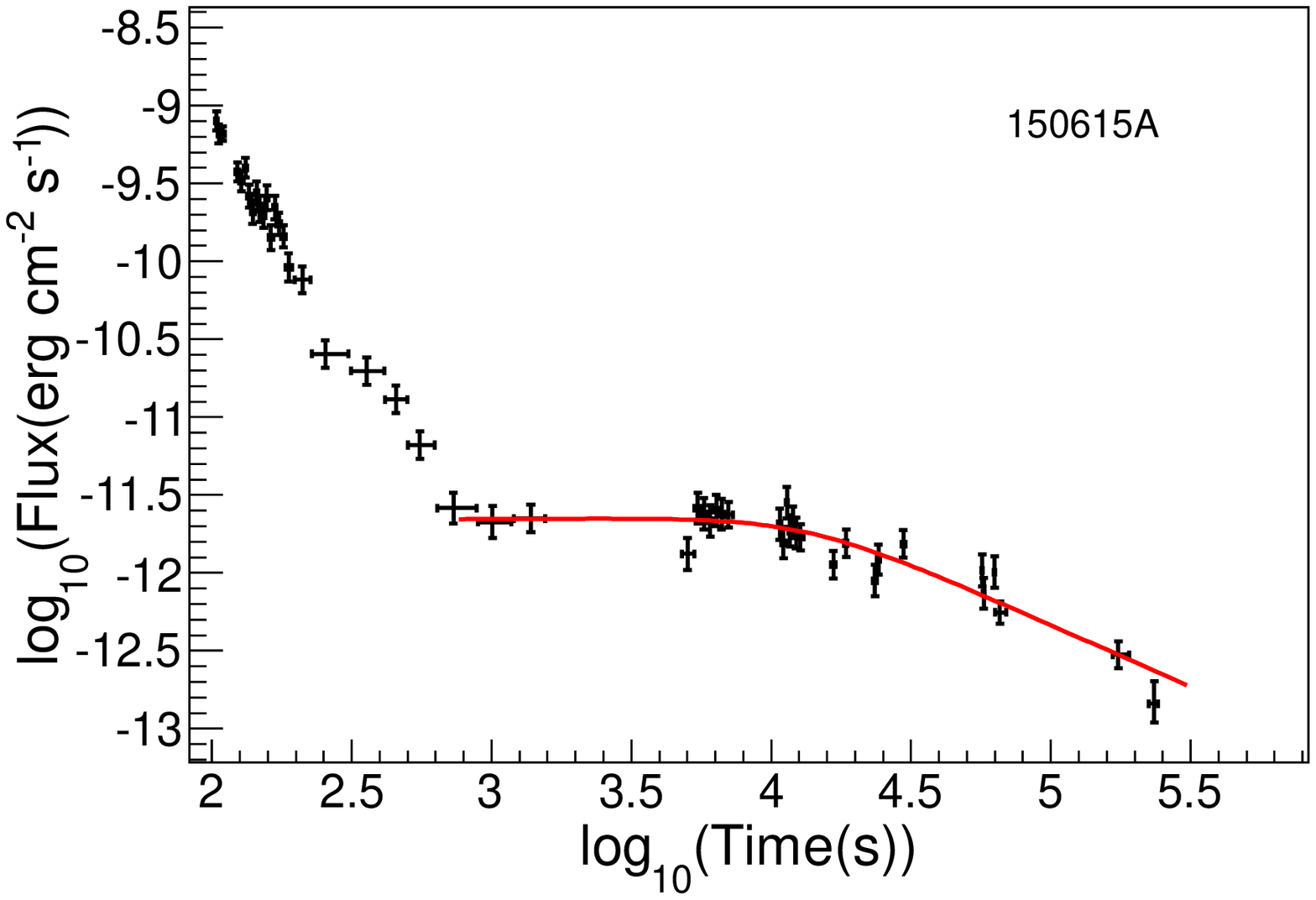}
\includegraphics[width=5.5cm,height=5cm]{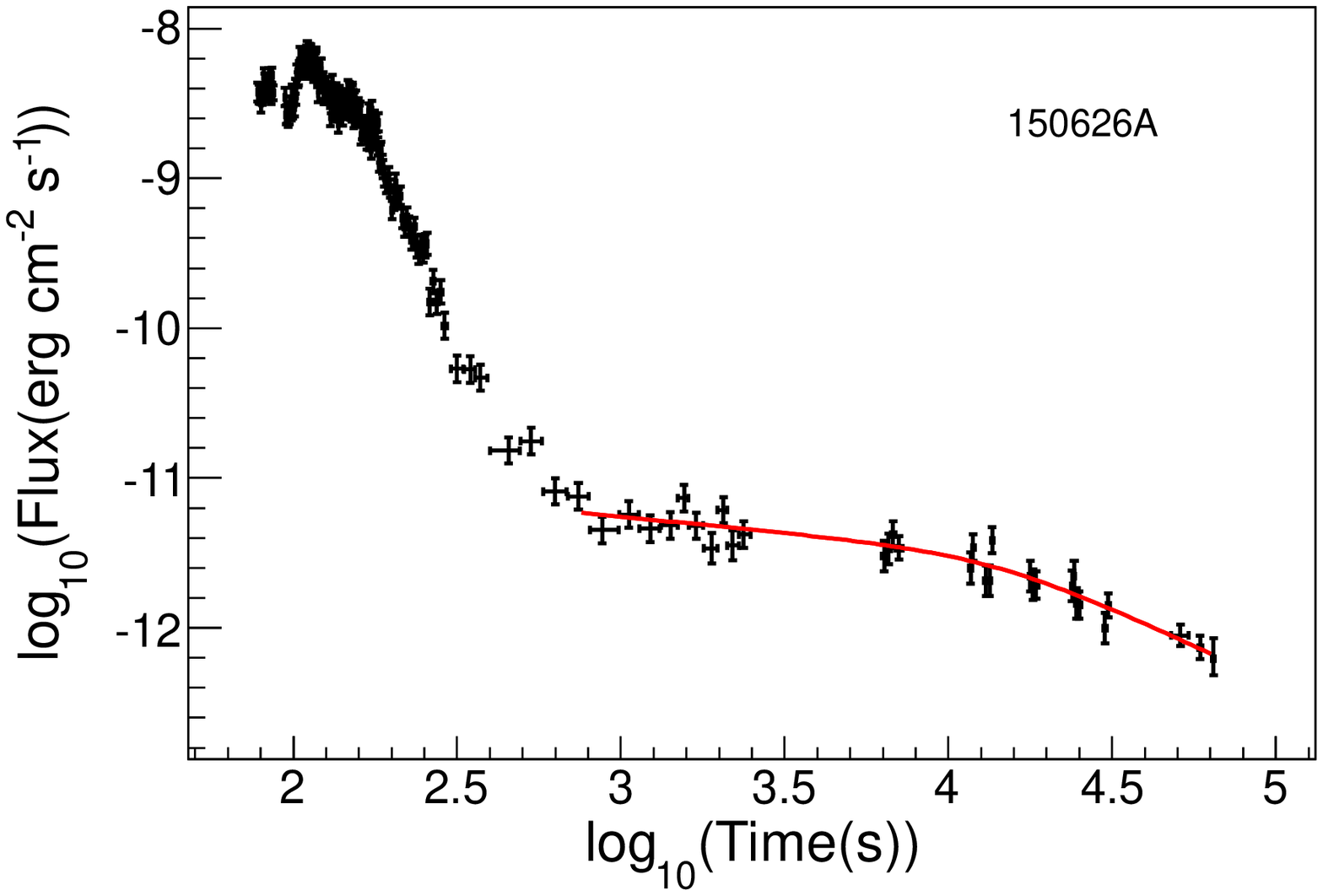}
\includegraphics[width=5.5cm,height=5cm]{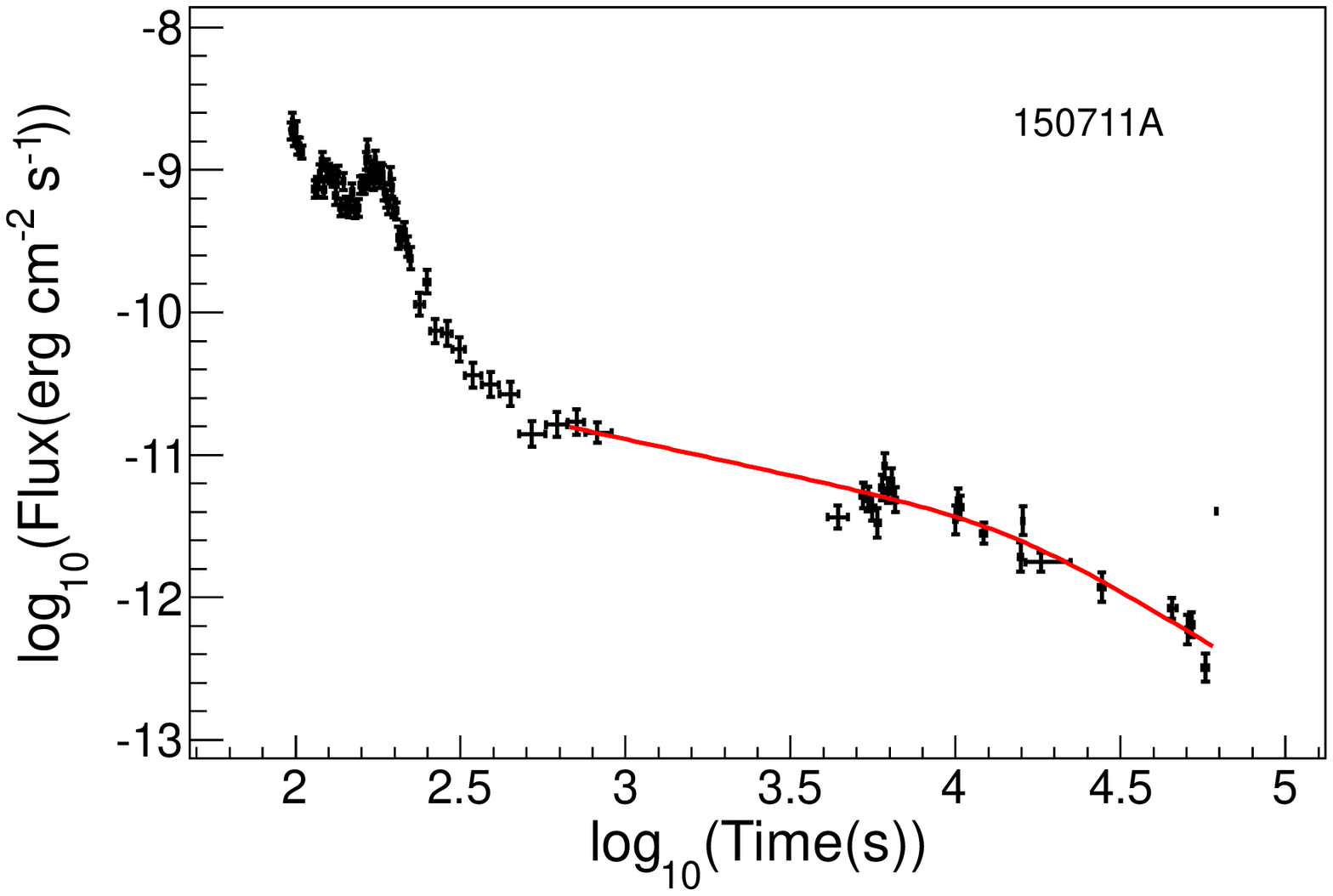}
\includegraphics[width=5.5cm,height=5cm]{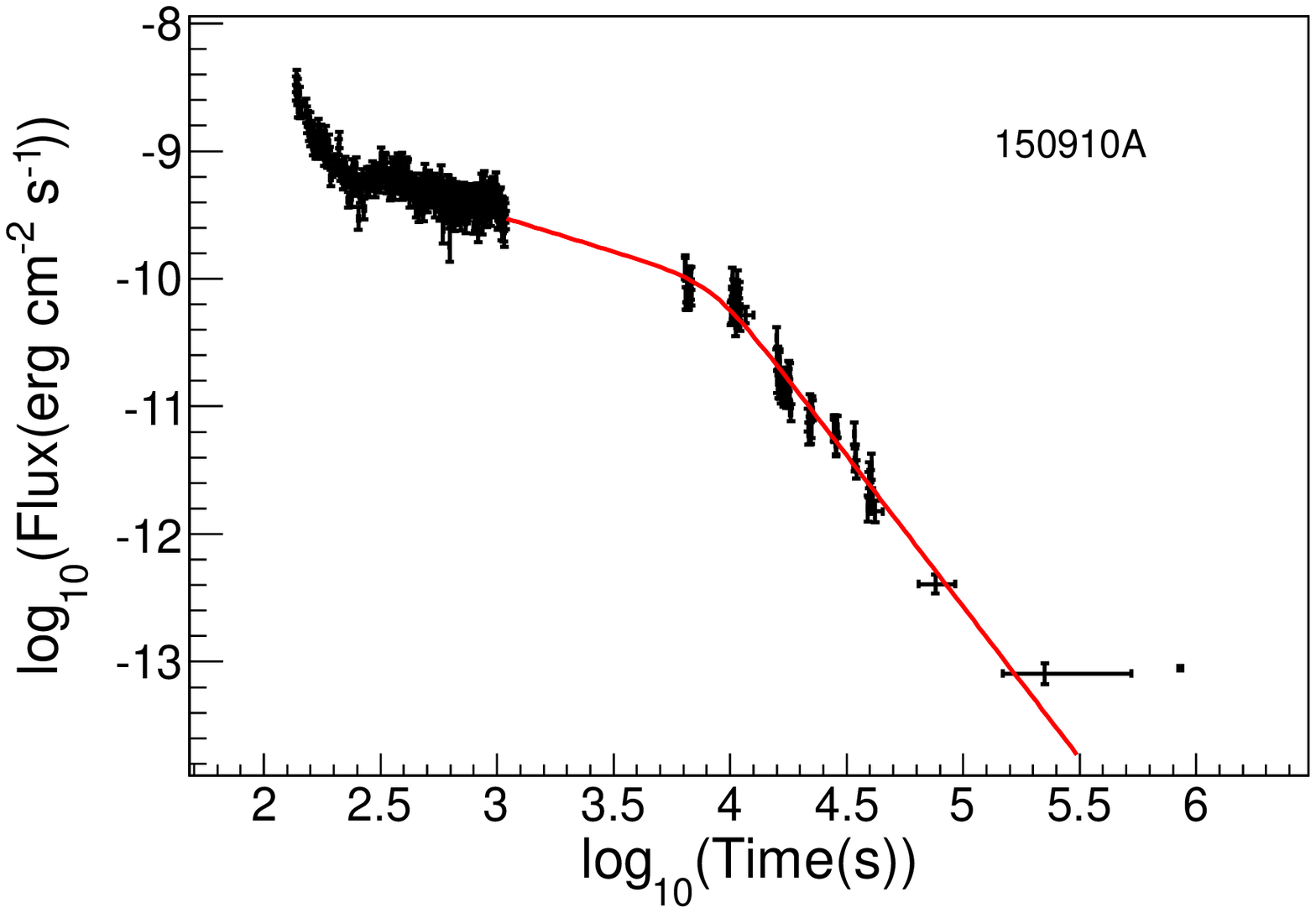}
\includegraphics[width=5.5cm,height=5cm]{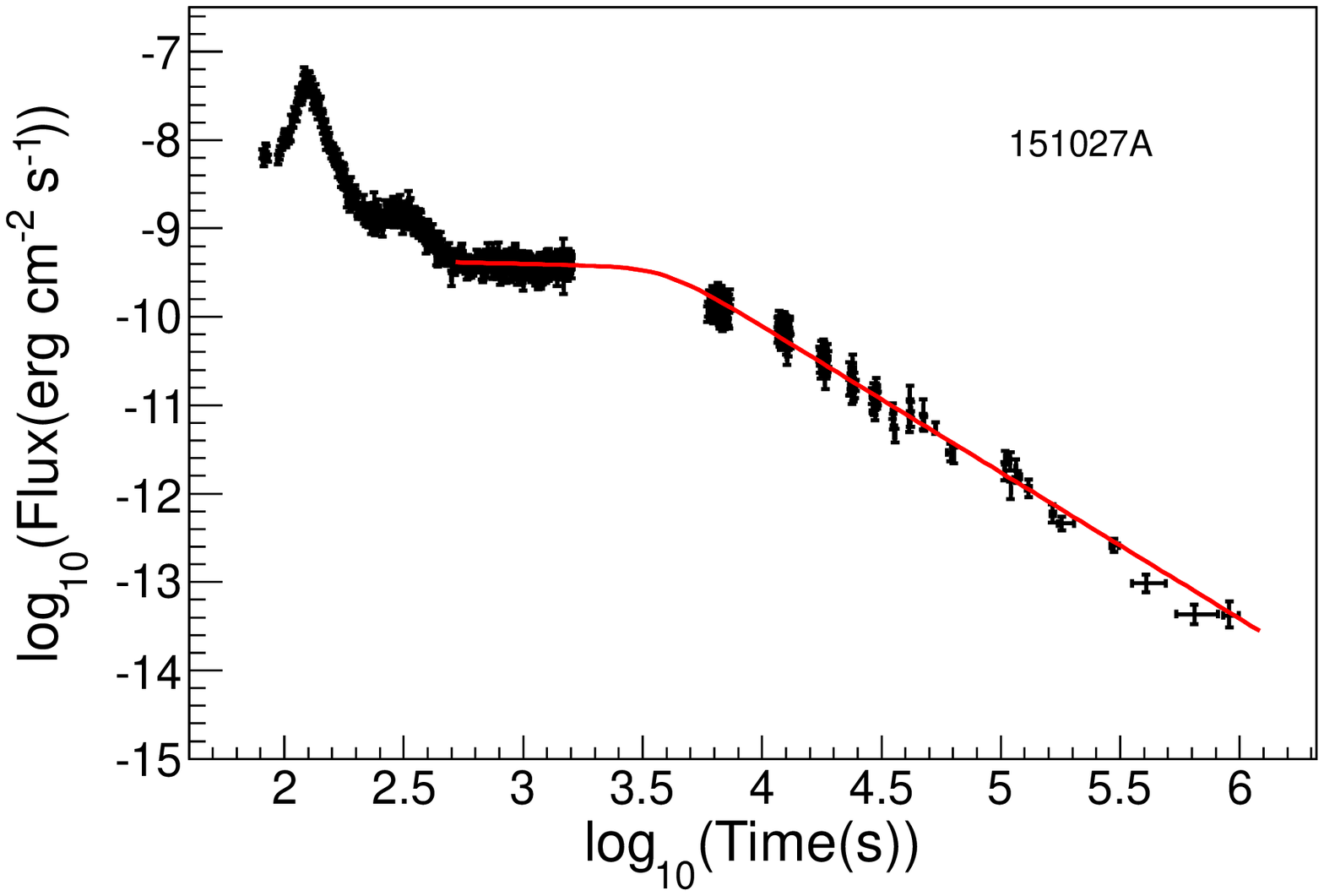}
\includegraphics[width=5.5cm,height=5cm]{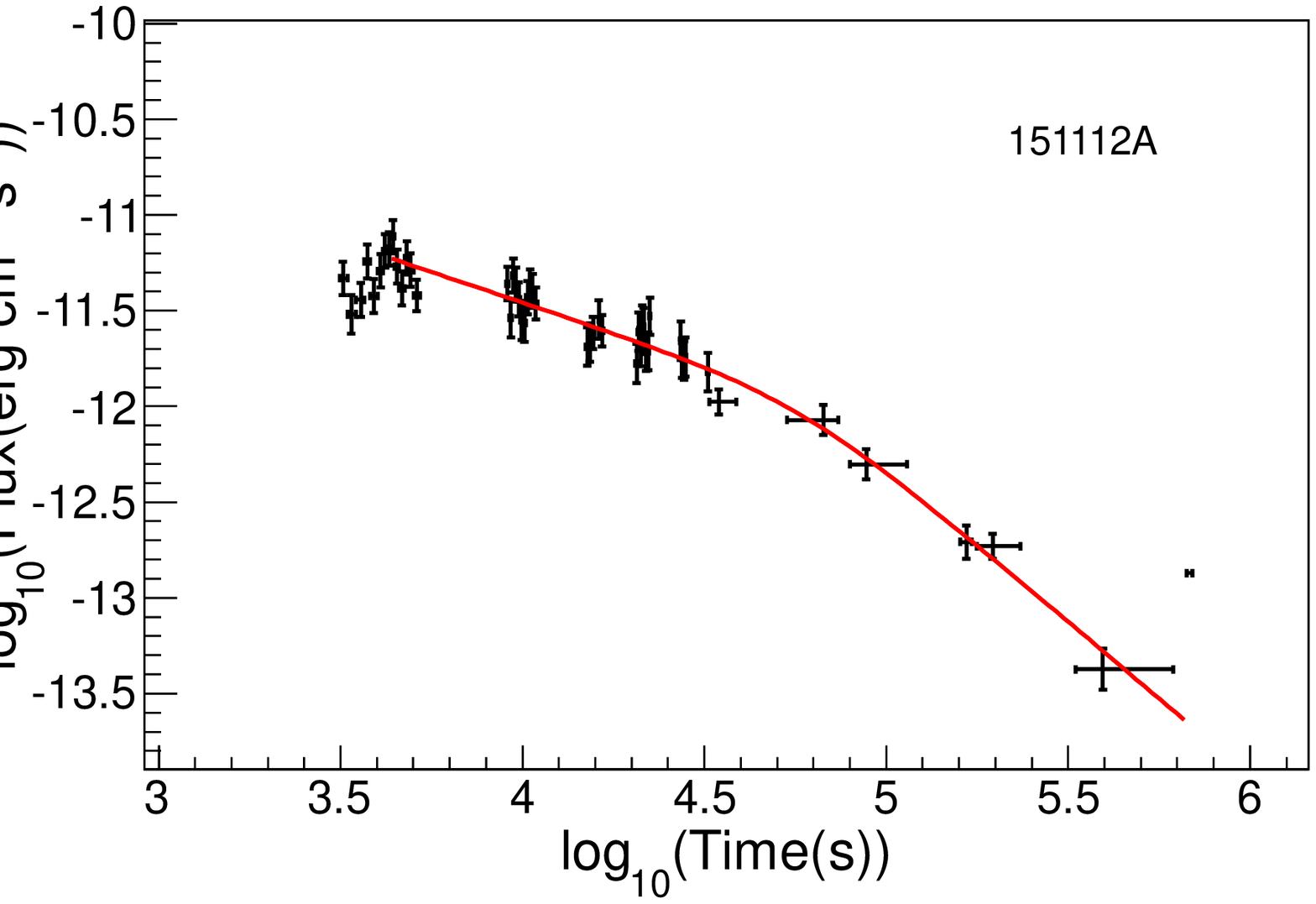}
\includegraphics[width=5.5cm,height=5cm]{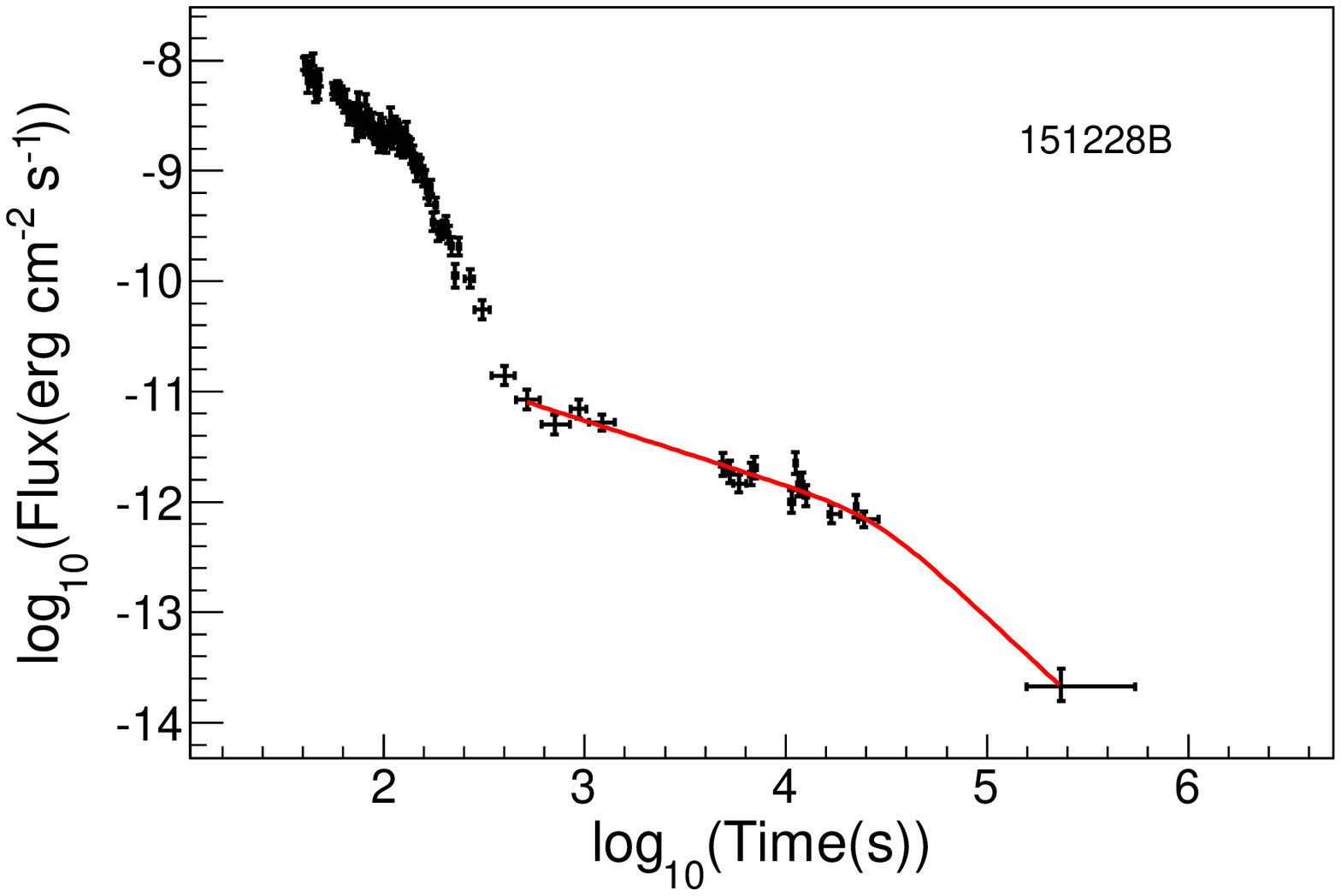}
\includegraphics[width=5.5cm,height=5cm]{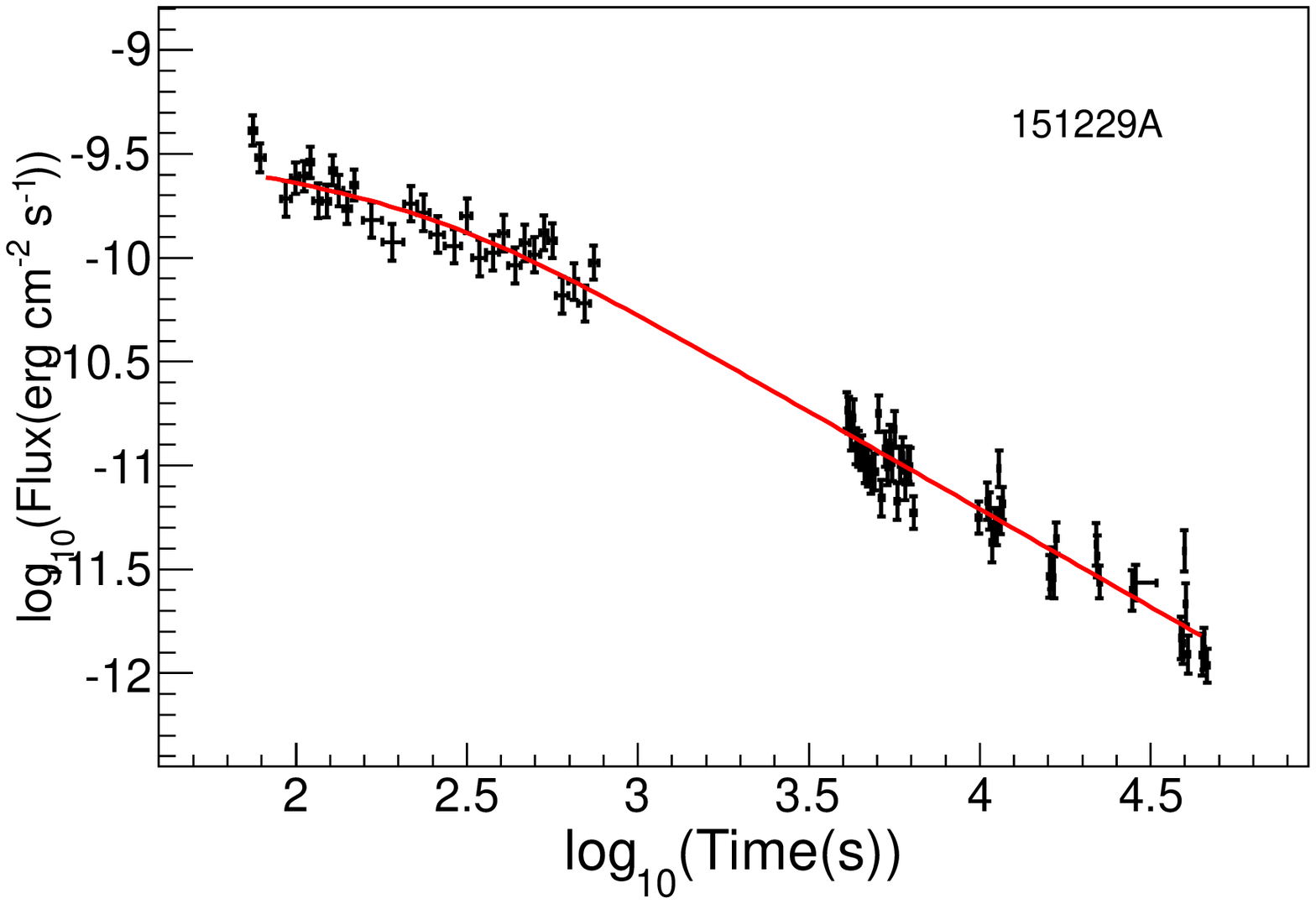}
\includegraphics[width=5.5cm,height=5cm]{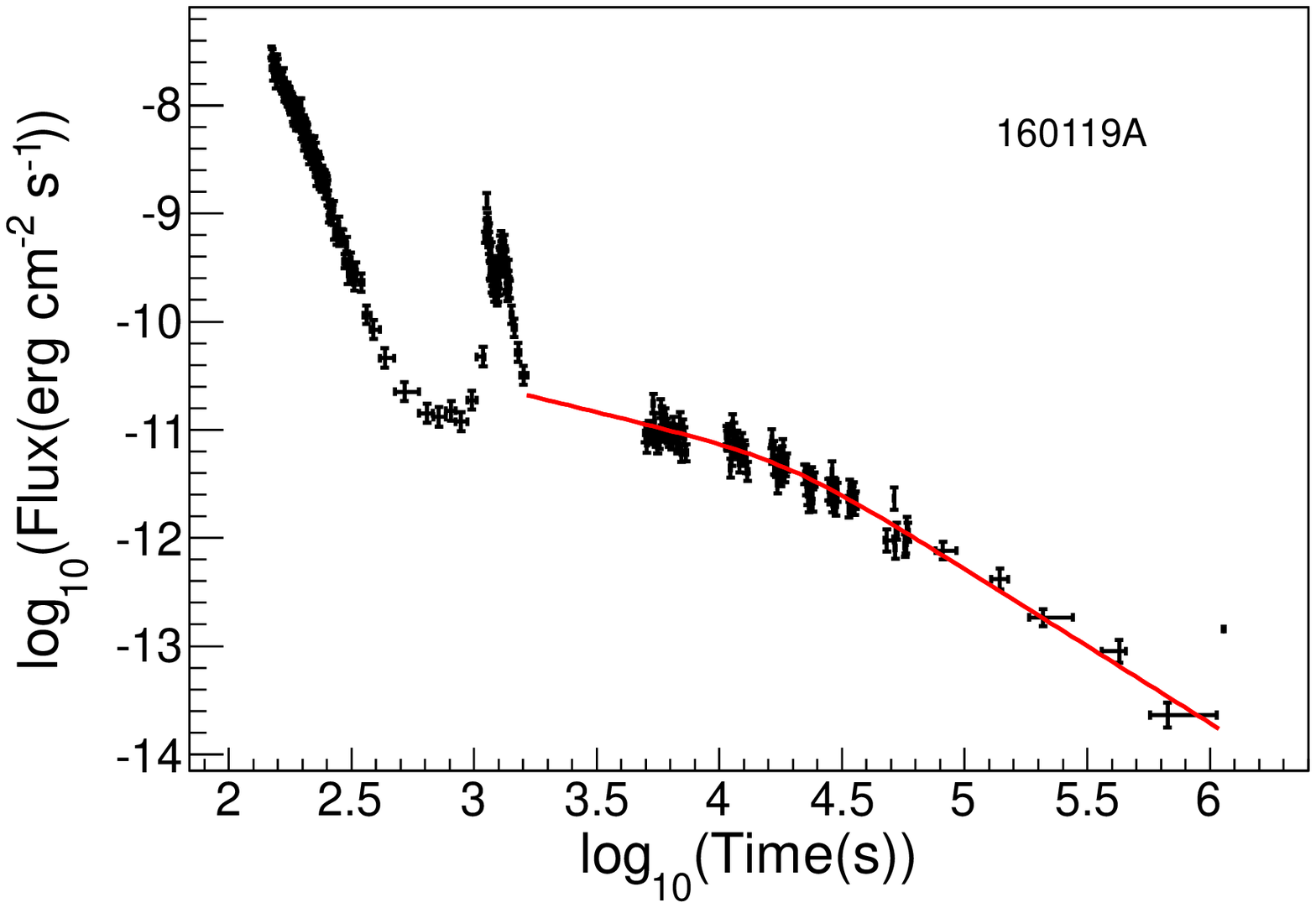}
\includegraphics[width=5.5cm,height=5cm]{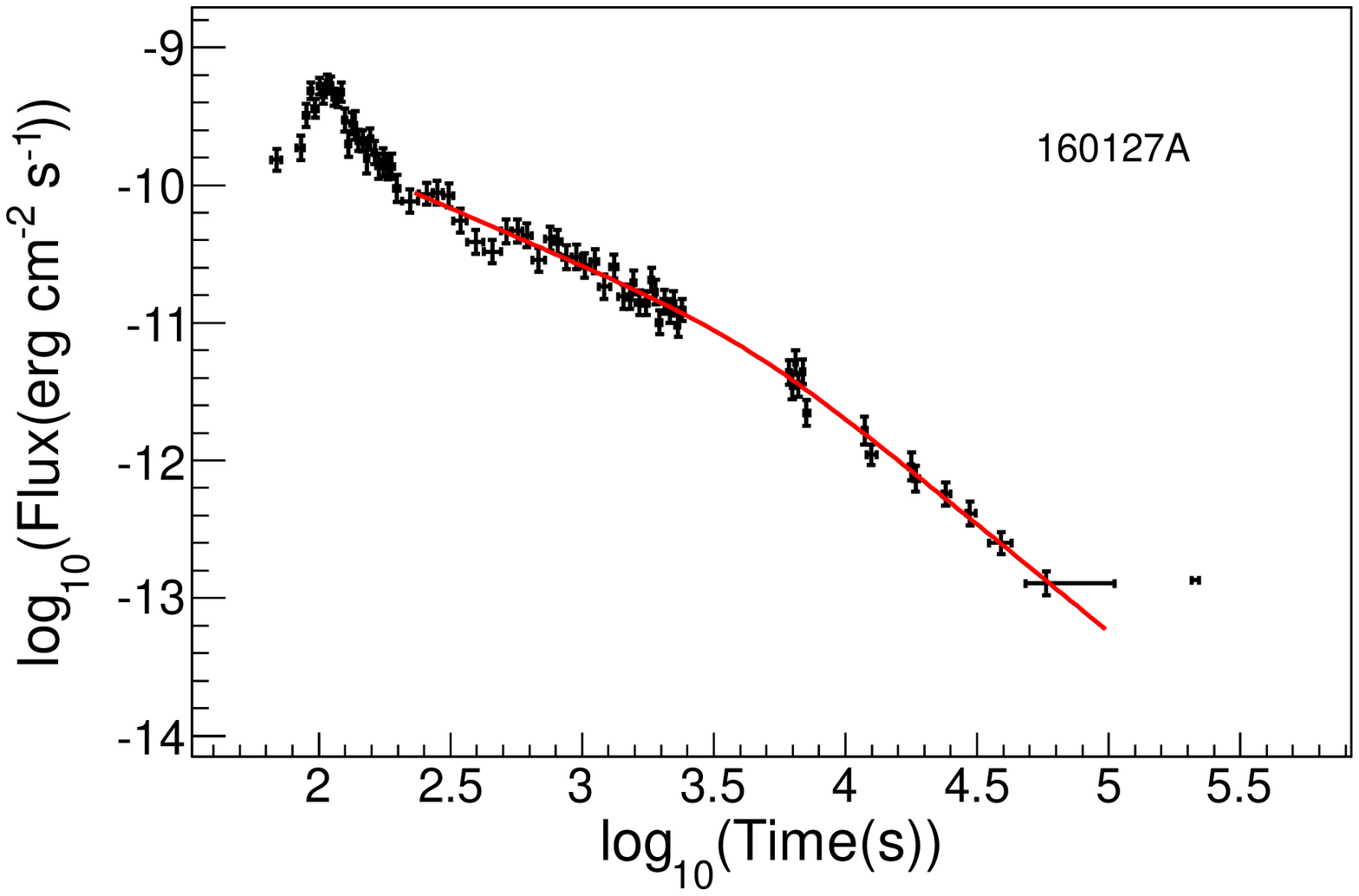}
\includegraphics[width=5.5cm,height=5cm]{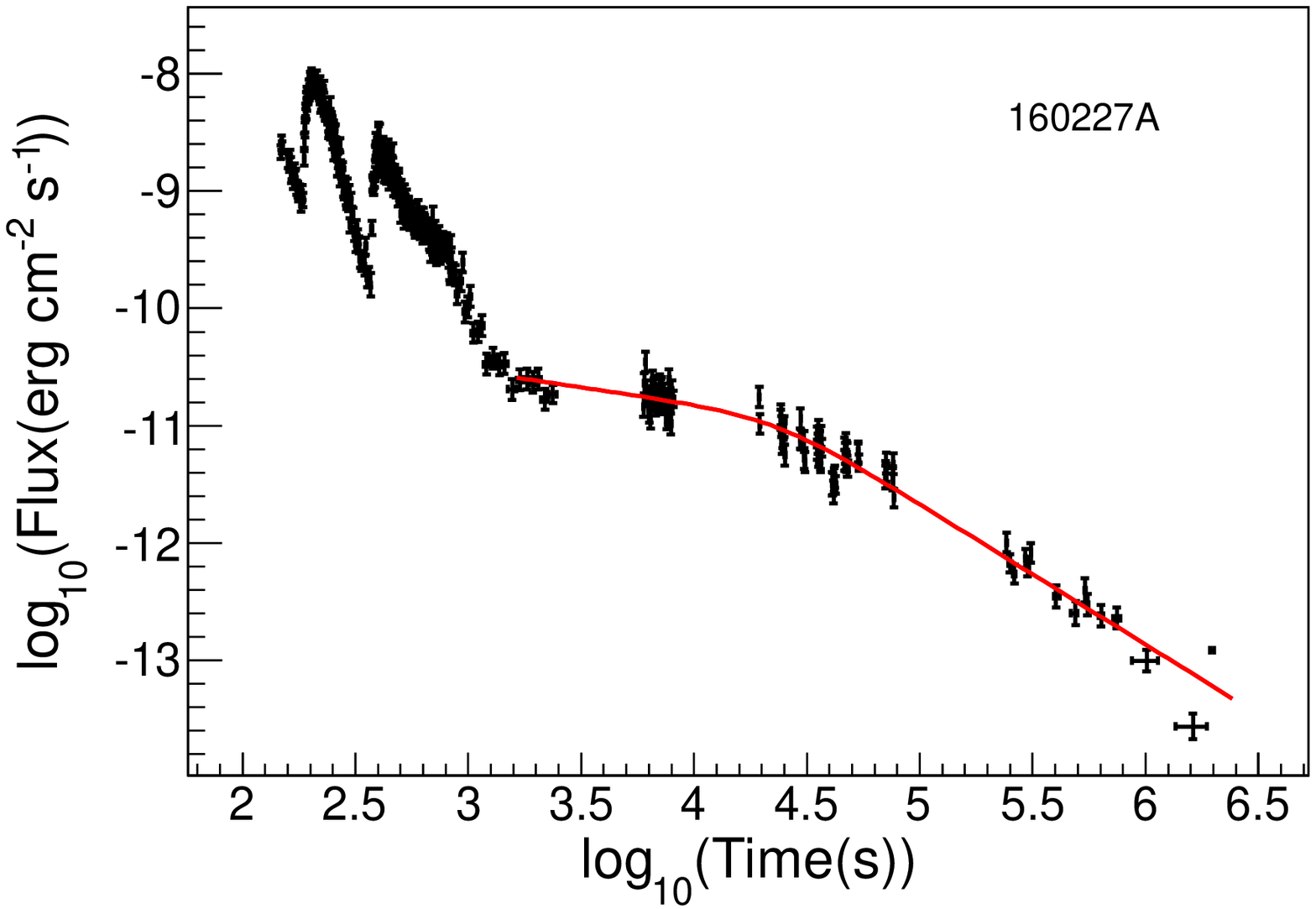}
\caption{ Continued.}
\label{fig-1-15}
\end{center}
\end{figure*}

\begin{figure*}
\begin{center}
\setlength{\abovecaptionskip}{0.cm}
\setlength{\belowcaptionskip}{-0.cm}
\figurenum{1}
\hspace{0cm}
\graphicspath{{lightcurve/}}
\includegraphics[width=5.5cm,height=5cm]{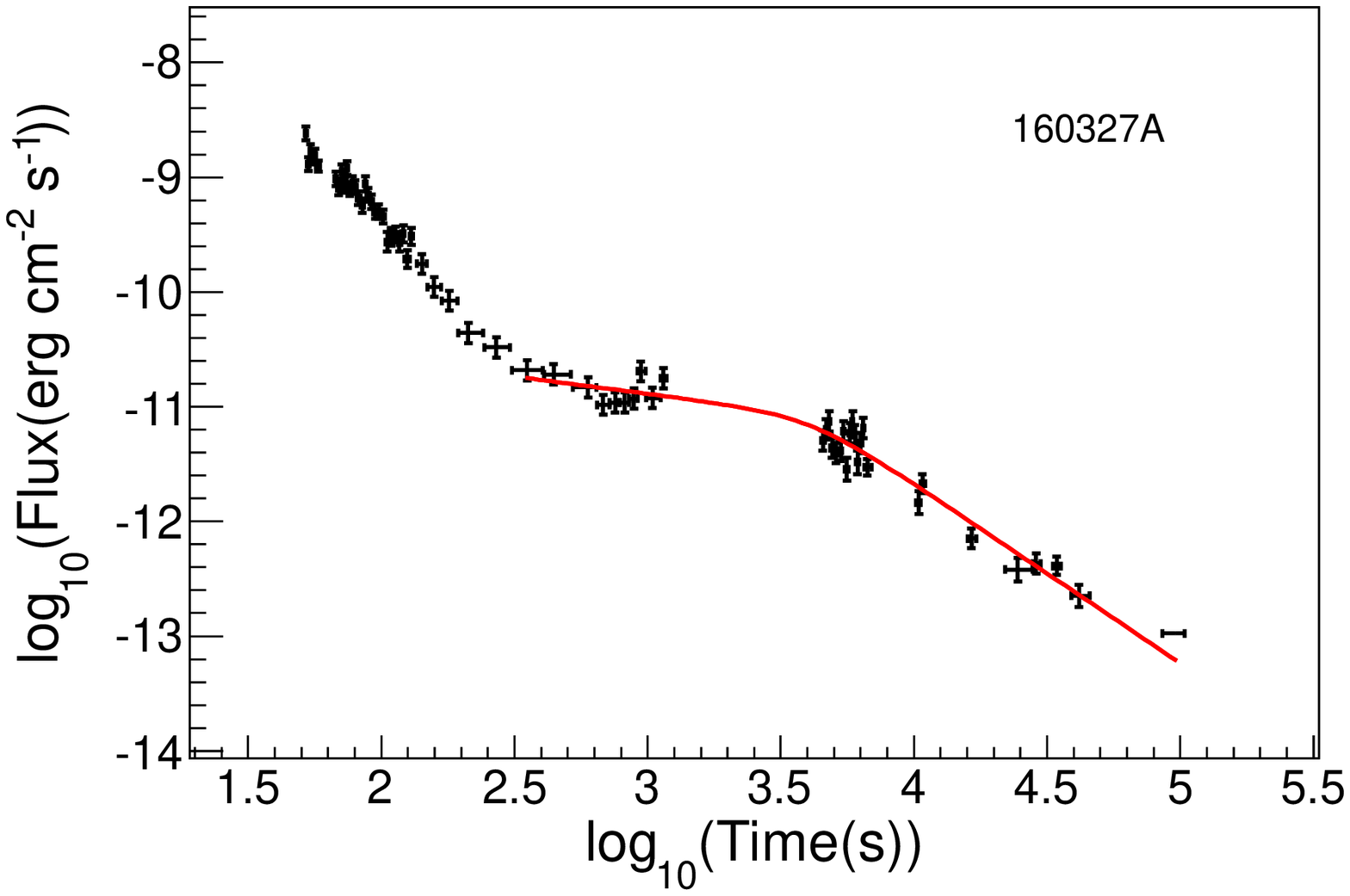}
\includegraphics[width=5.5cm,height=5cm]{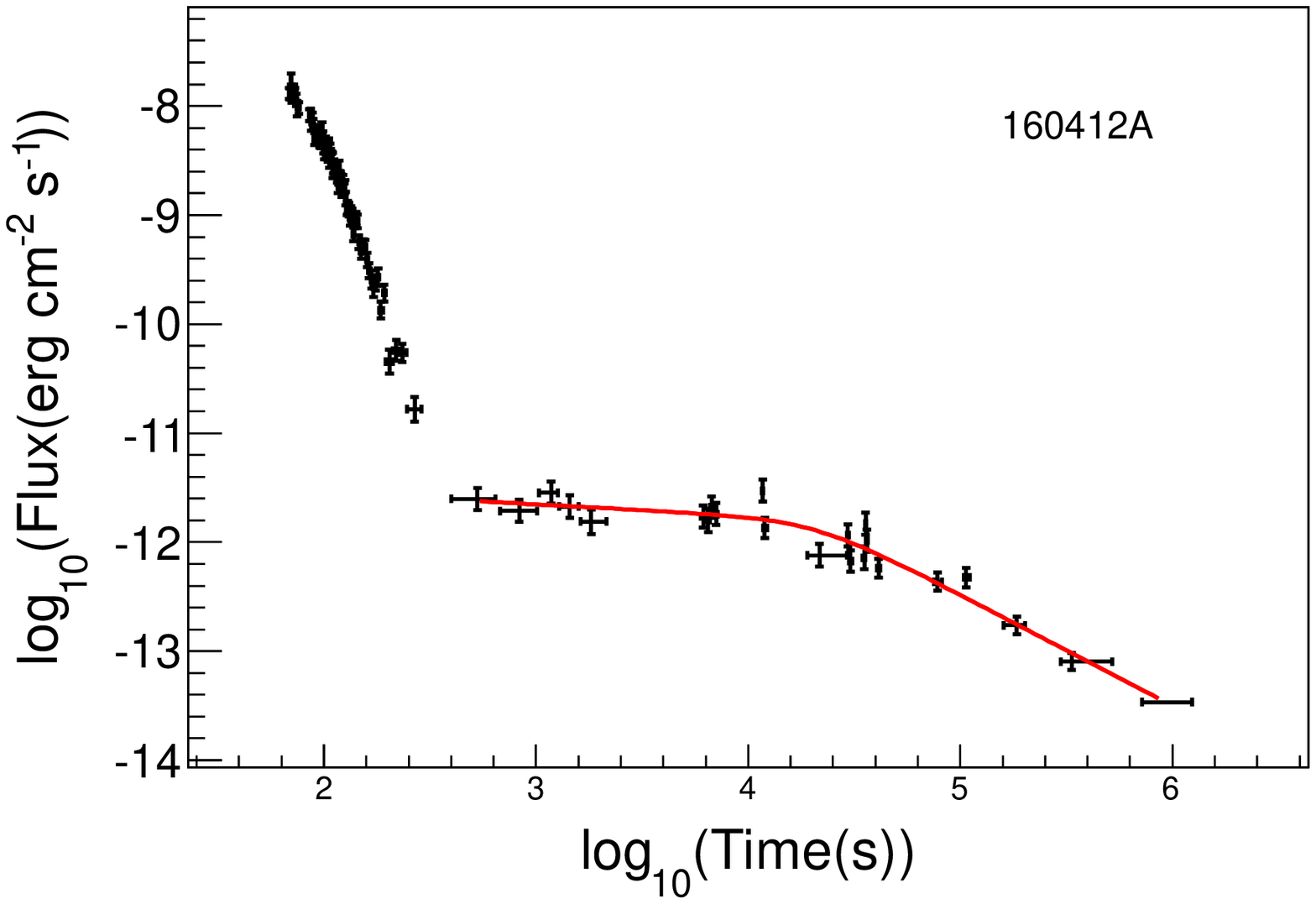}
\includegraphics[width=5.5cm,height=5cm]{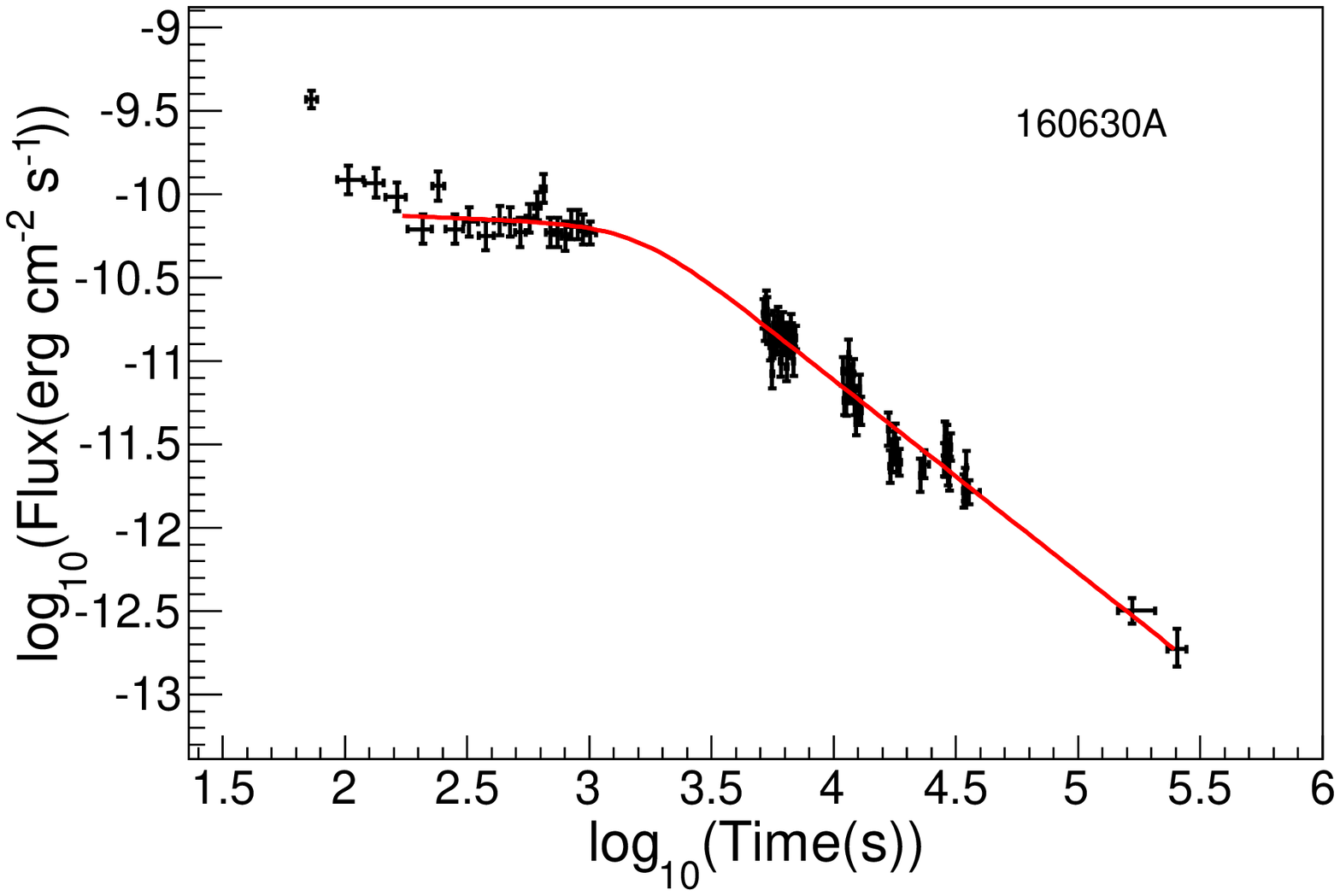}
\includegraphics[width=5.5cm,height=5cm]{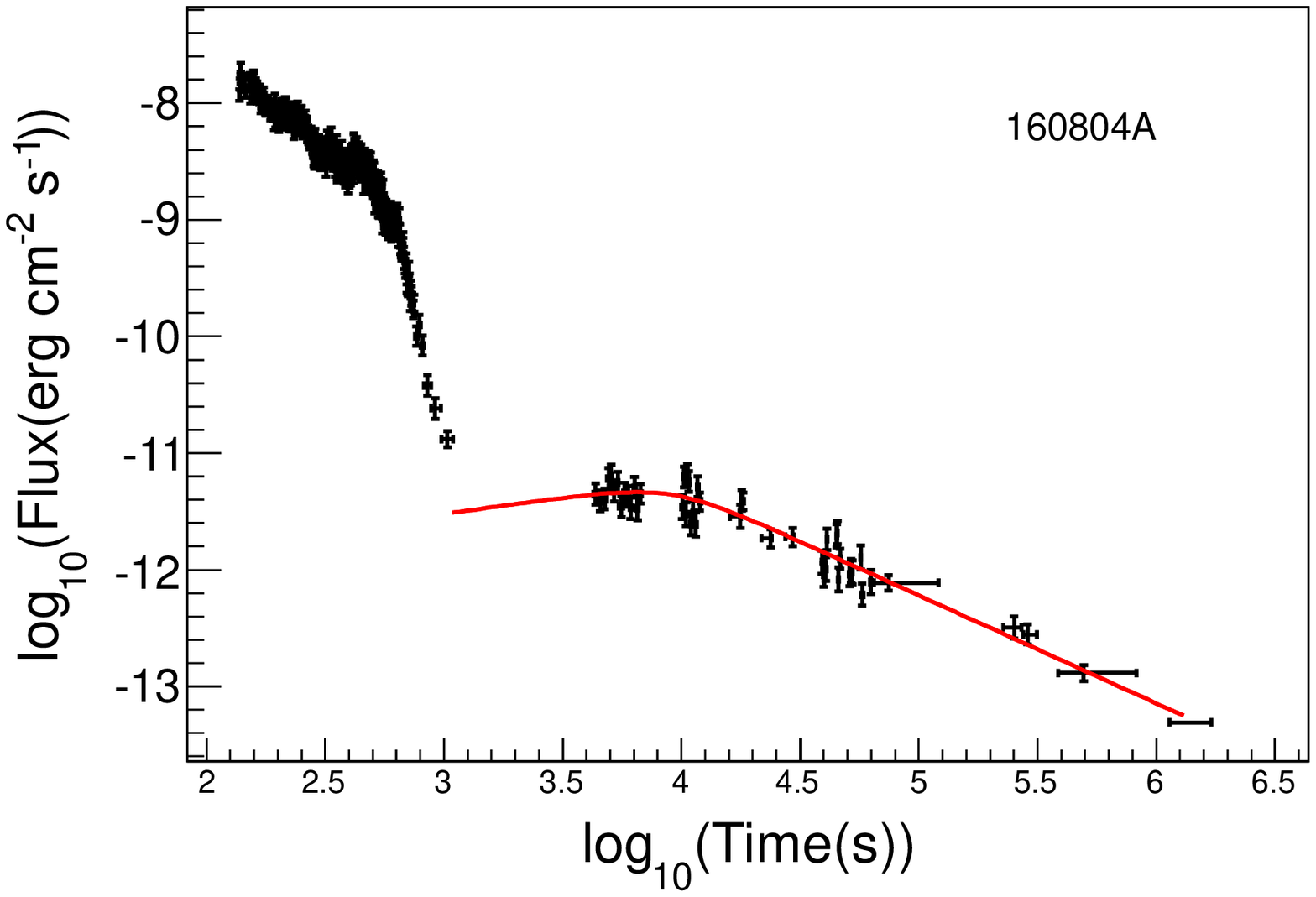}
\includegraphics[width=5.5cm,height=5cm]{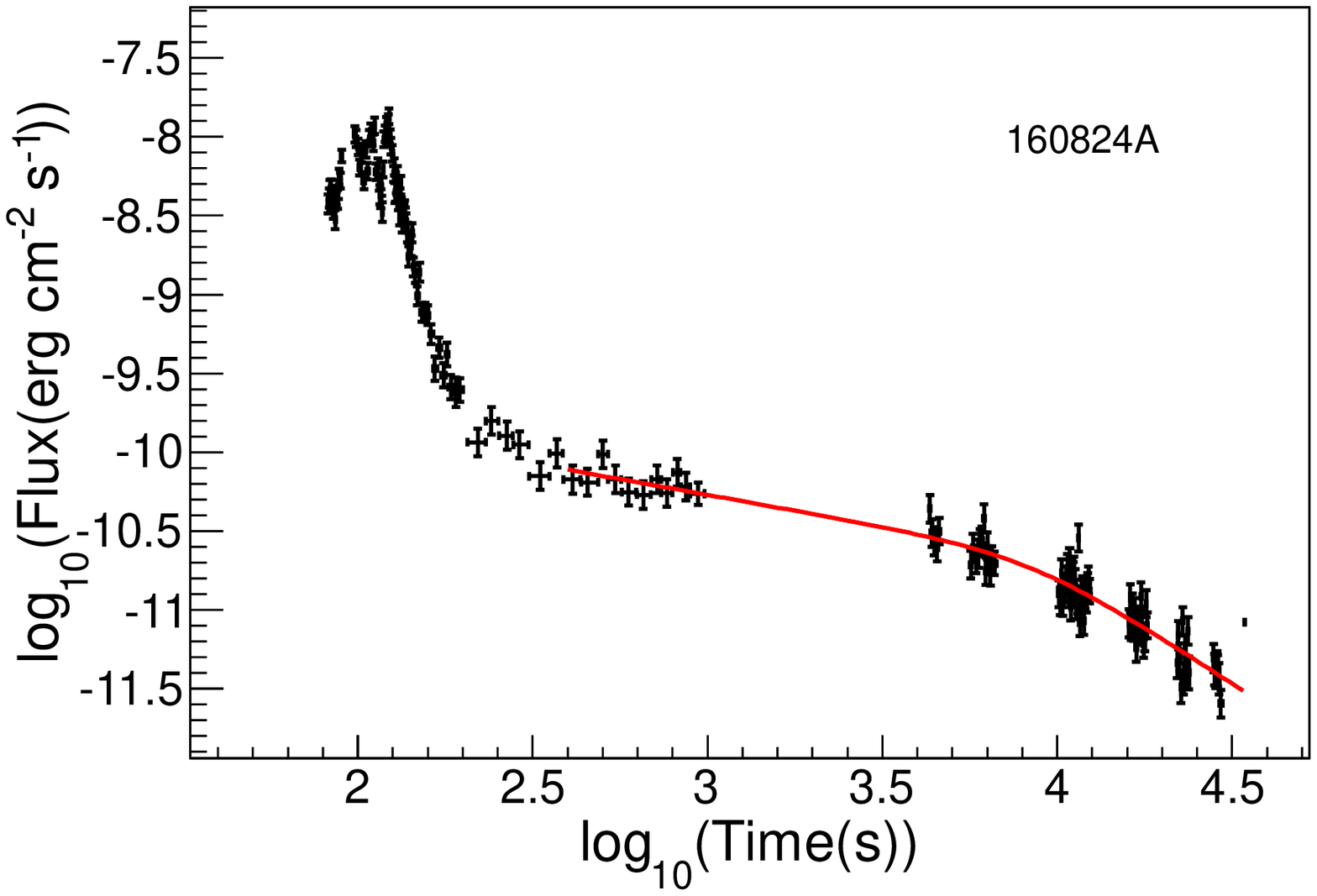}
\includegraphics[width=5.5cm,height=5cm]{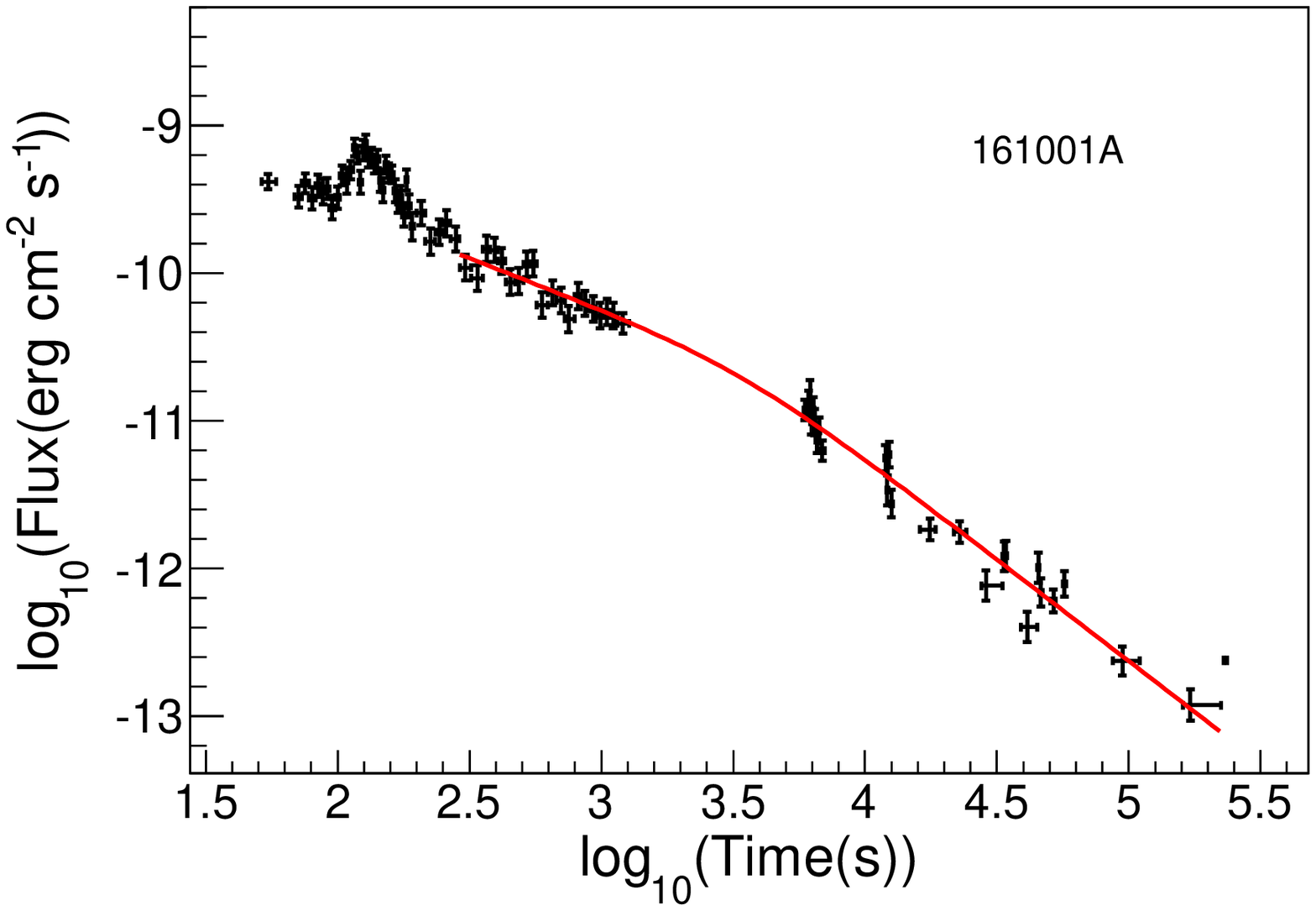}
\includegraphics[width=5.5cm,height=5cm]{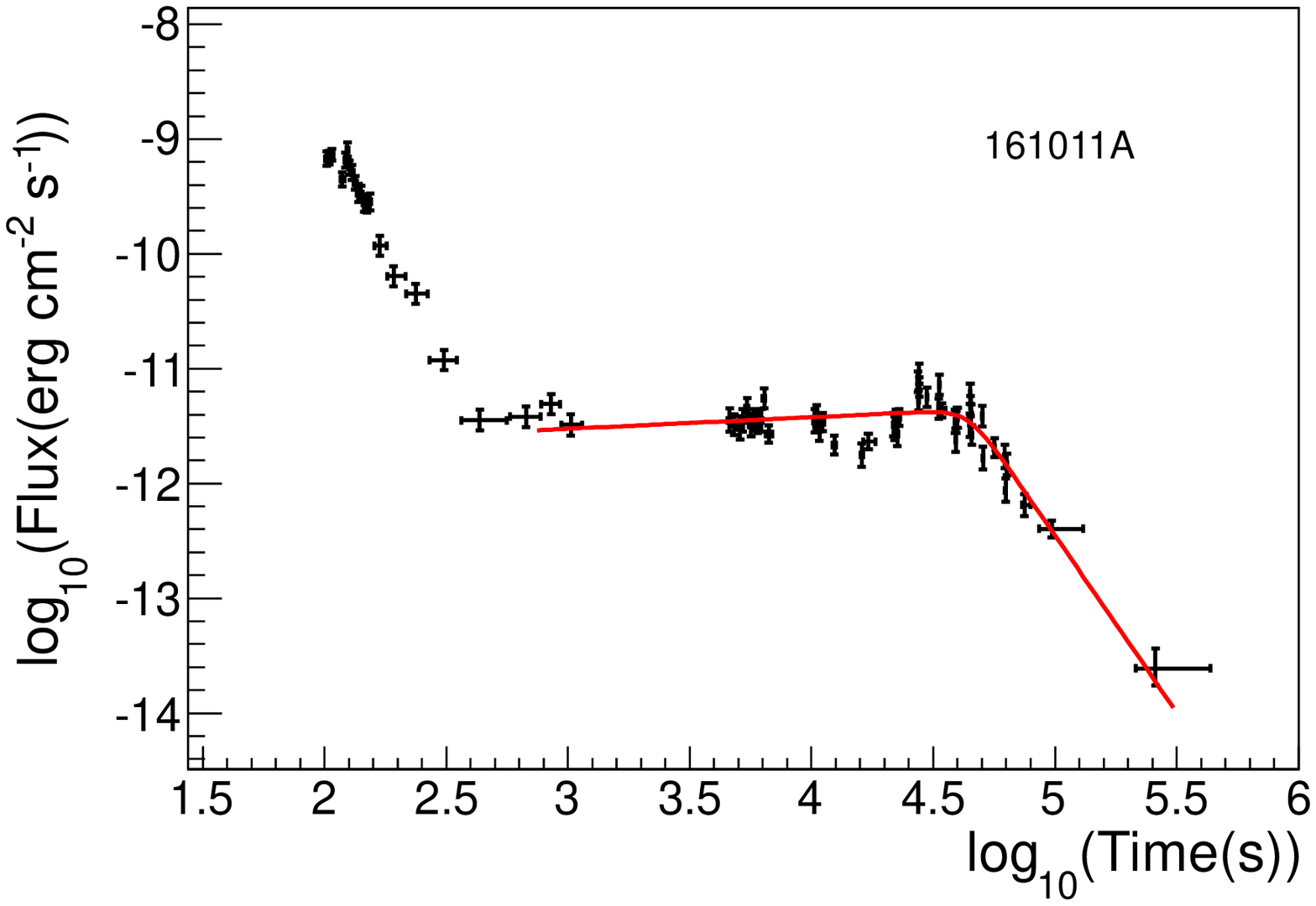}
\includegraphics[width=5.5cm,height=5cm]{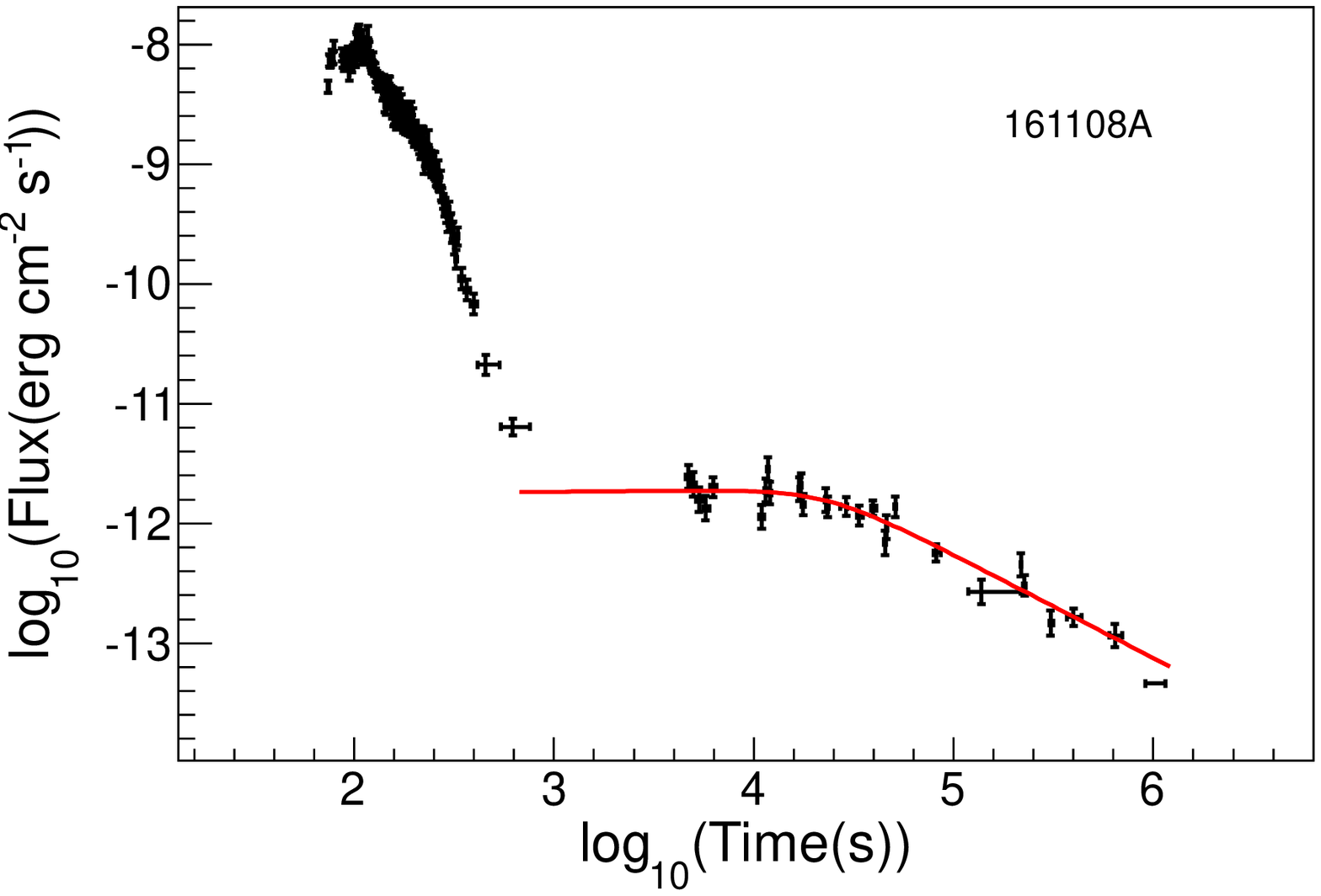}
\includegraphics[width=5.5cm,height=5cm]{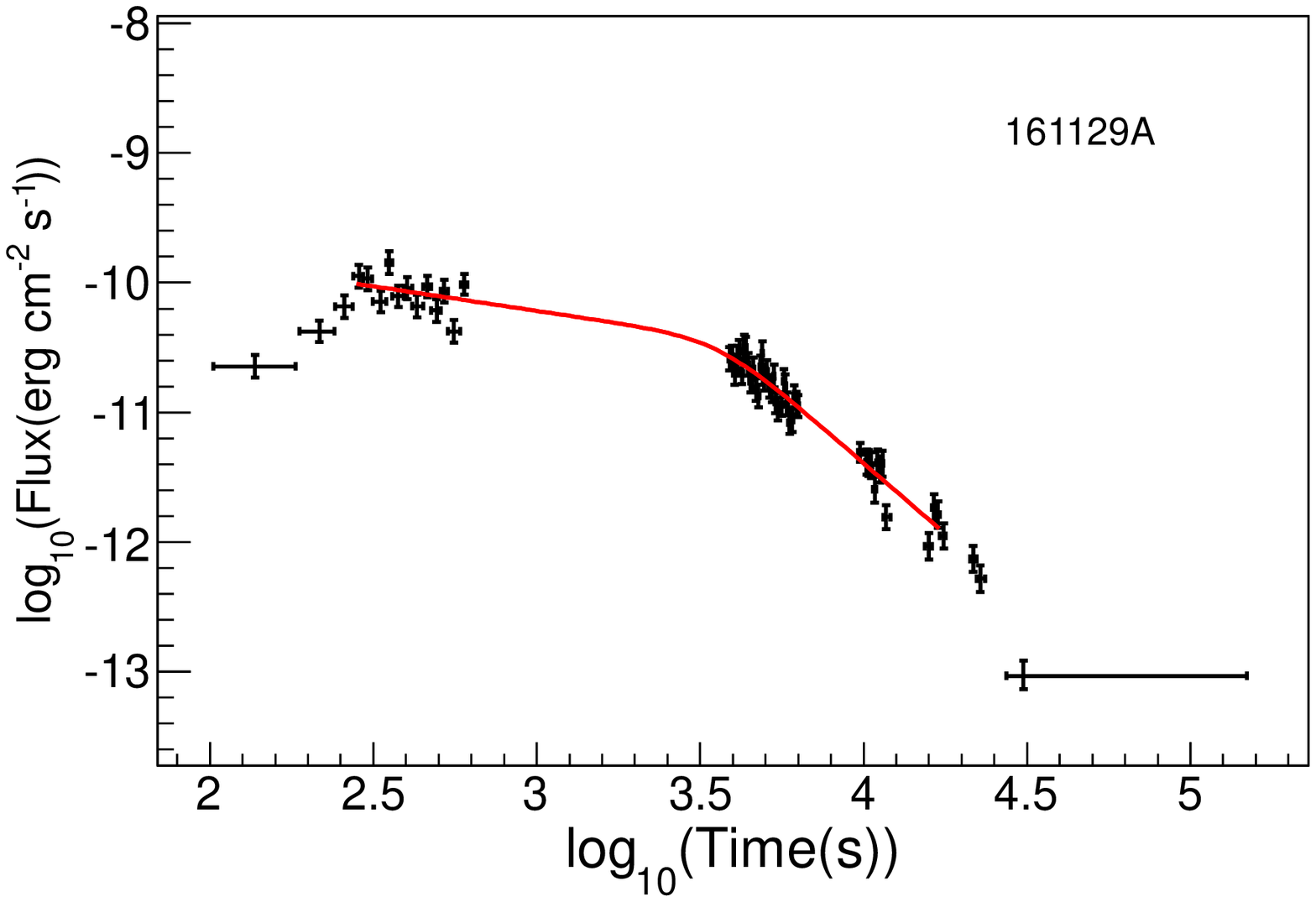}
\includegraphics[width=5.5cm,height=5cm]{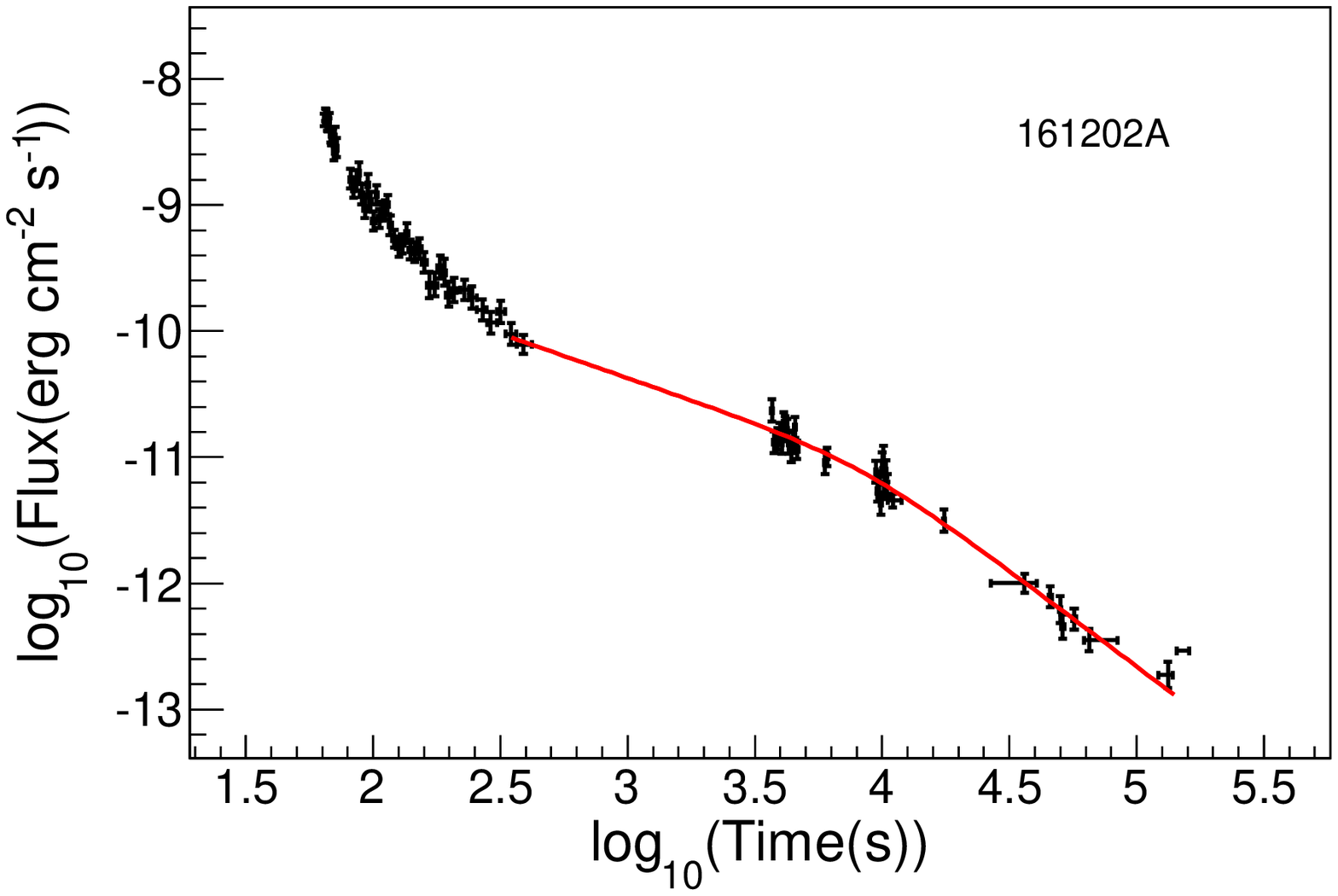}
\includegraphics[width=5.5cm,height=5cm]{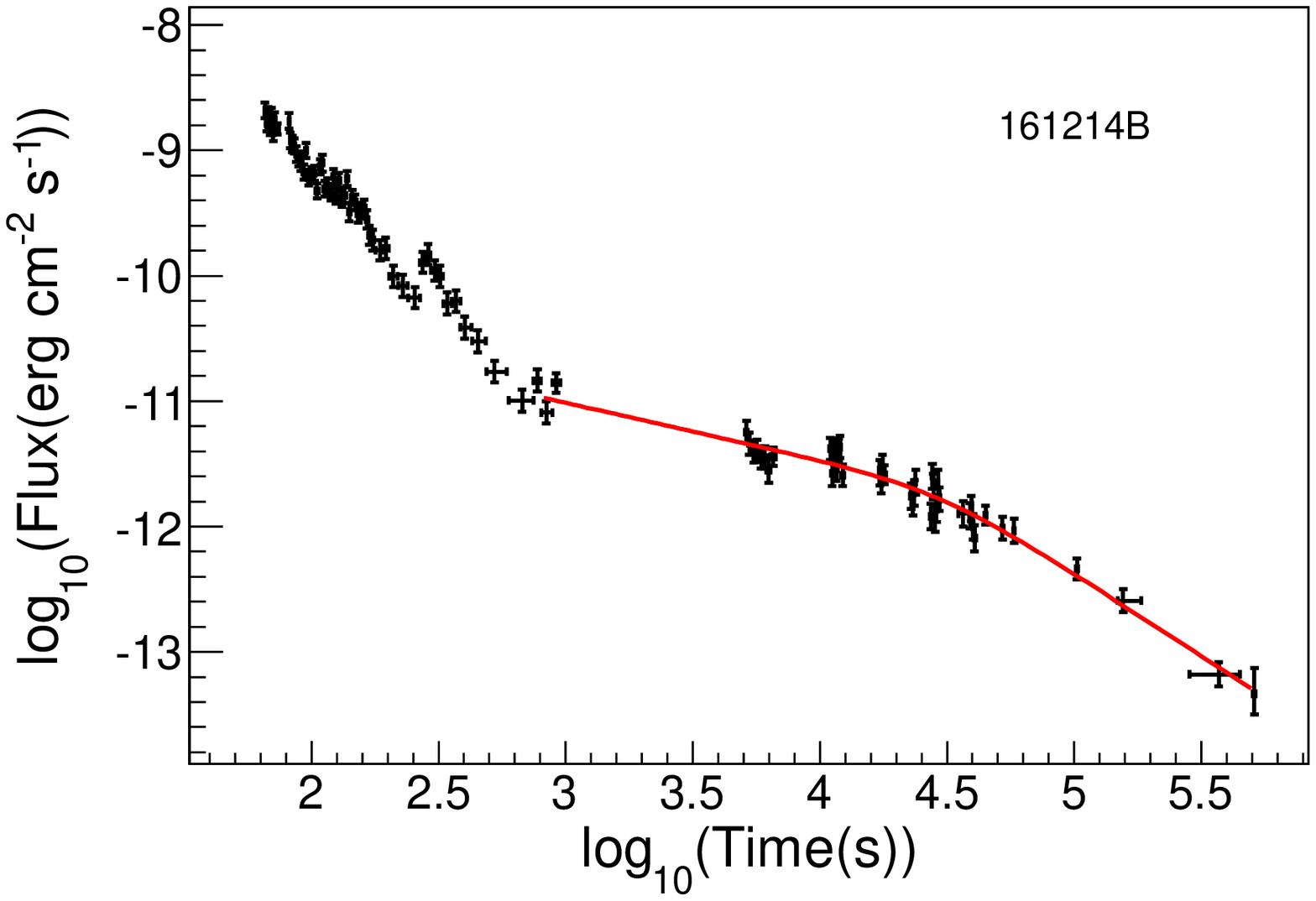}
\includegraphics[width=5.5cm,height=5cm]{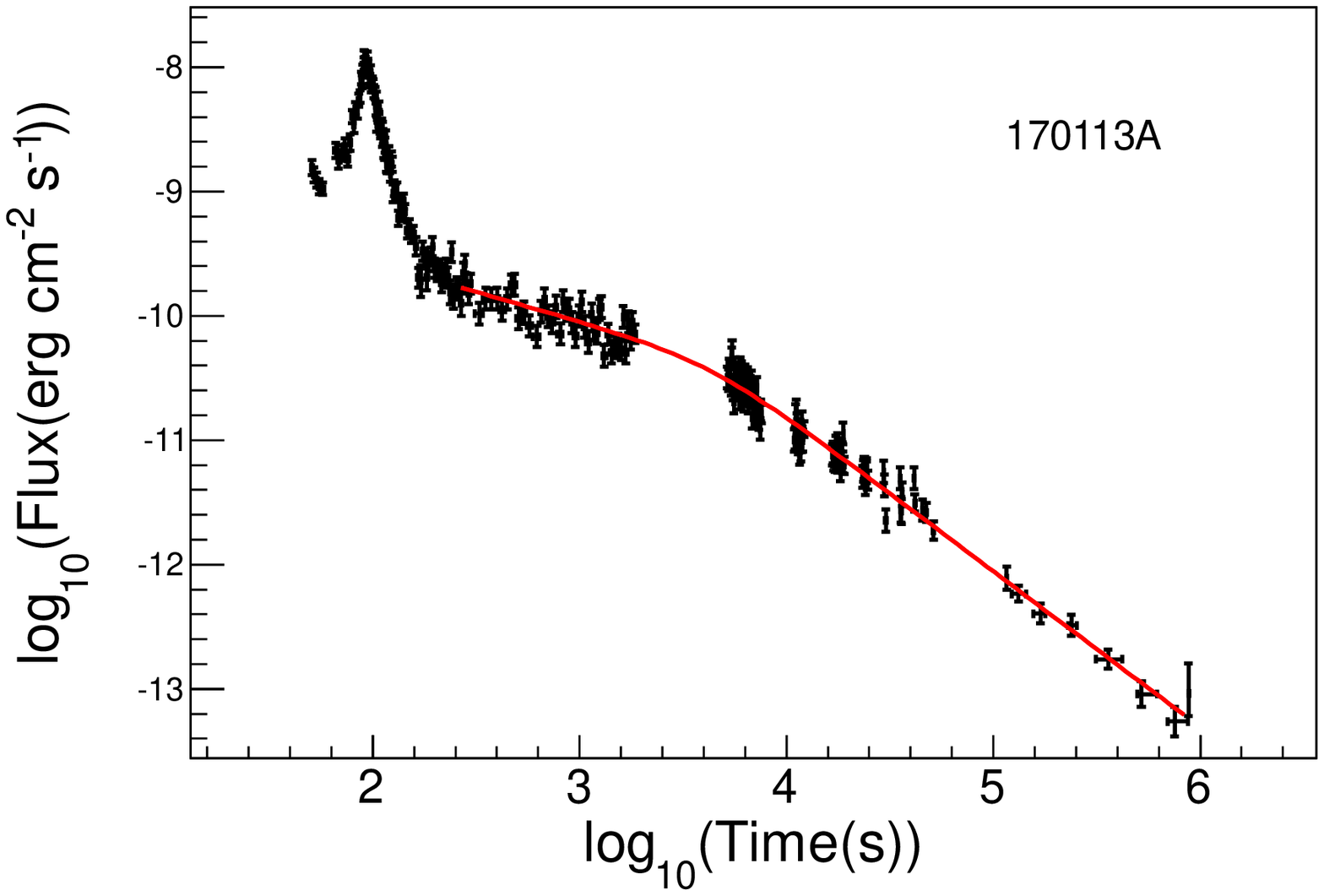}
\caption{ Continued.}
\label{fig-1-16}
\end{center}
\end{figure*}

\begin{figure*}
\begin{center}
\setlength{\abovecaptionskip}{0.cm}
\setlength{\belowcaptionskip}{-0.cm}
\figurenum{1}
\hspace{0cm}
\graphicspath{{lightcurve/}}
\includegraphics[width=5.5cm,height=5cm]{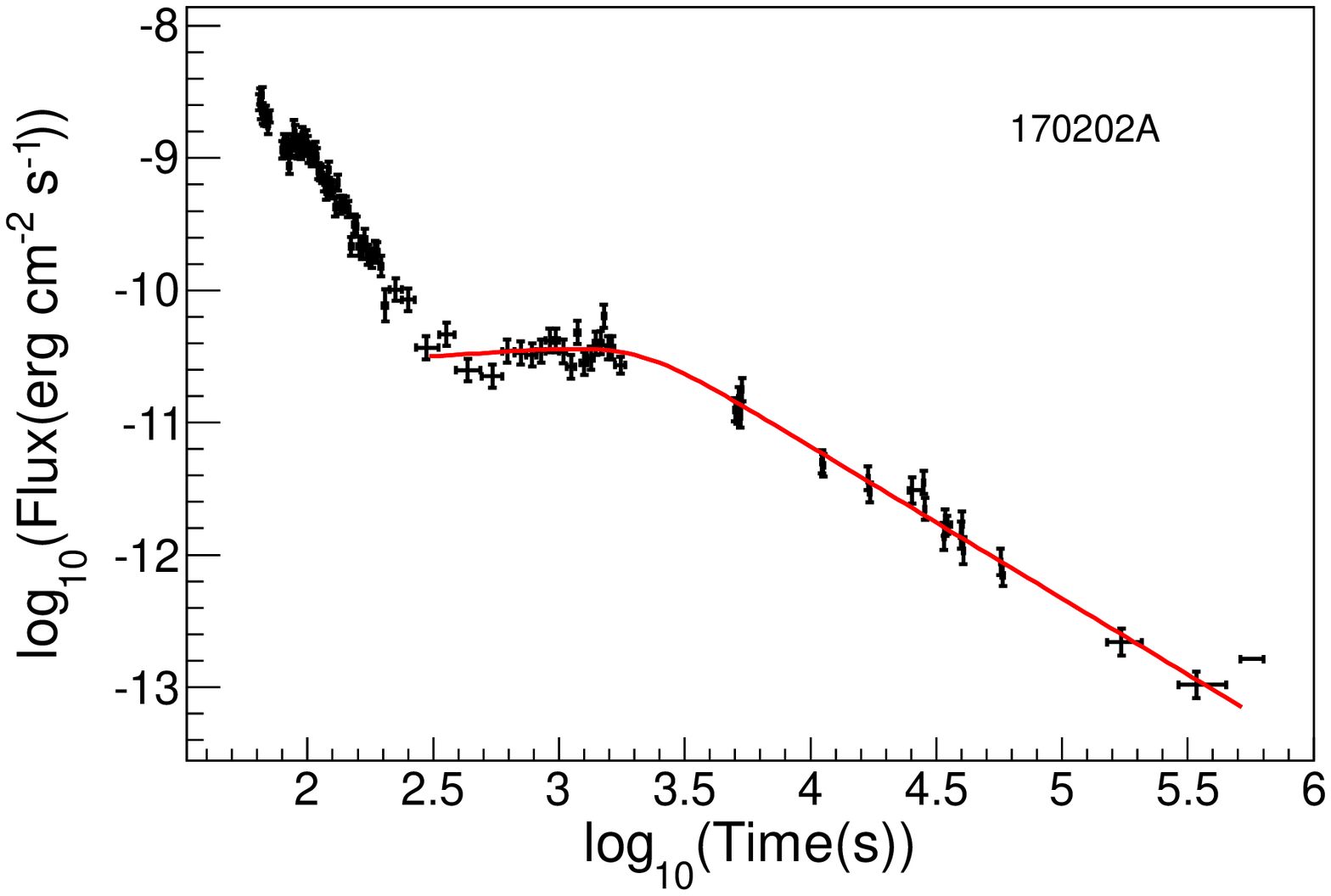}
\includegraphics[width=5.5cm,height=5cm]{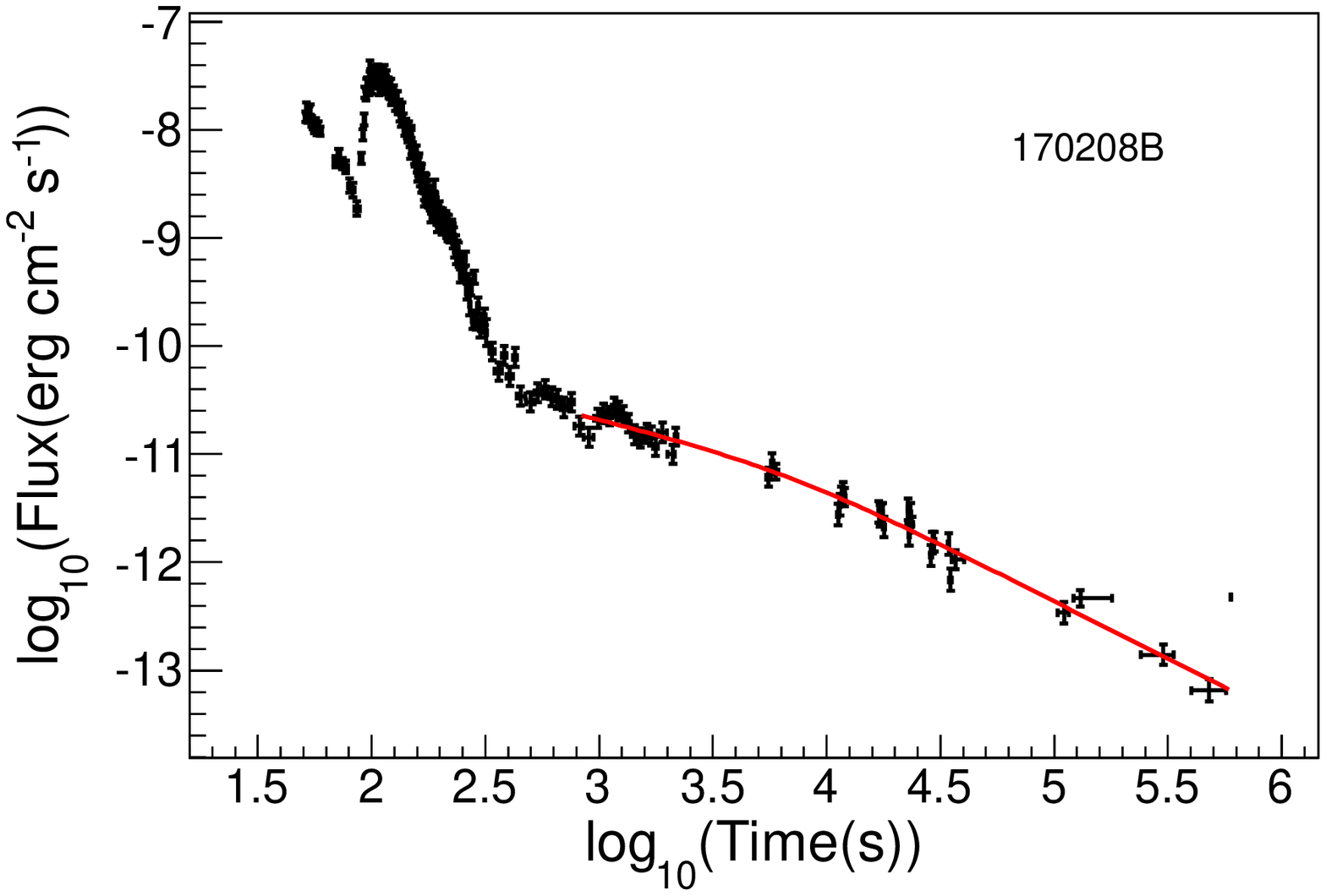}
\includegraphics[width=5.5cm,height=5cm]{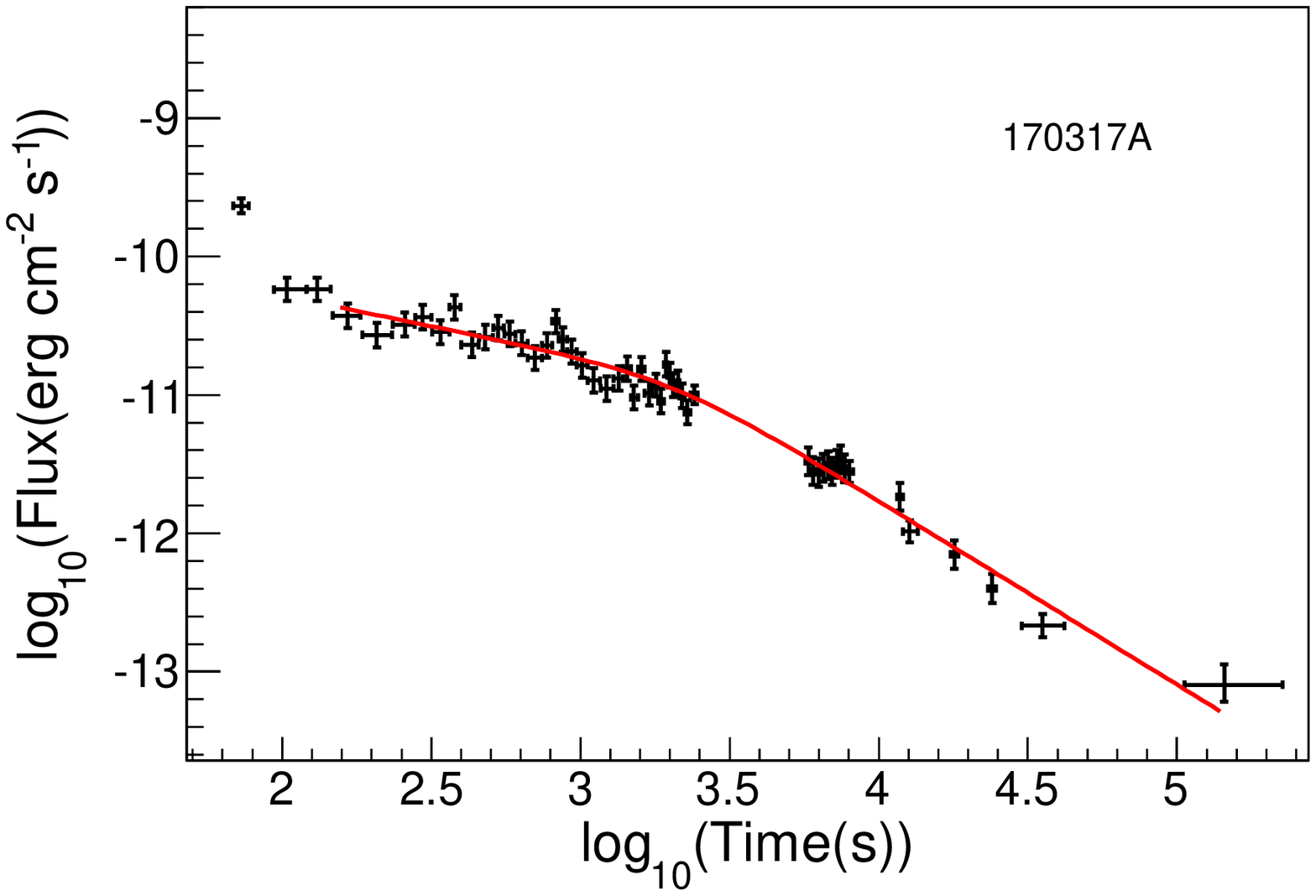}
\includegraphics[width=5.5cm,height=5cm]{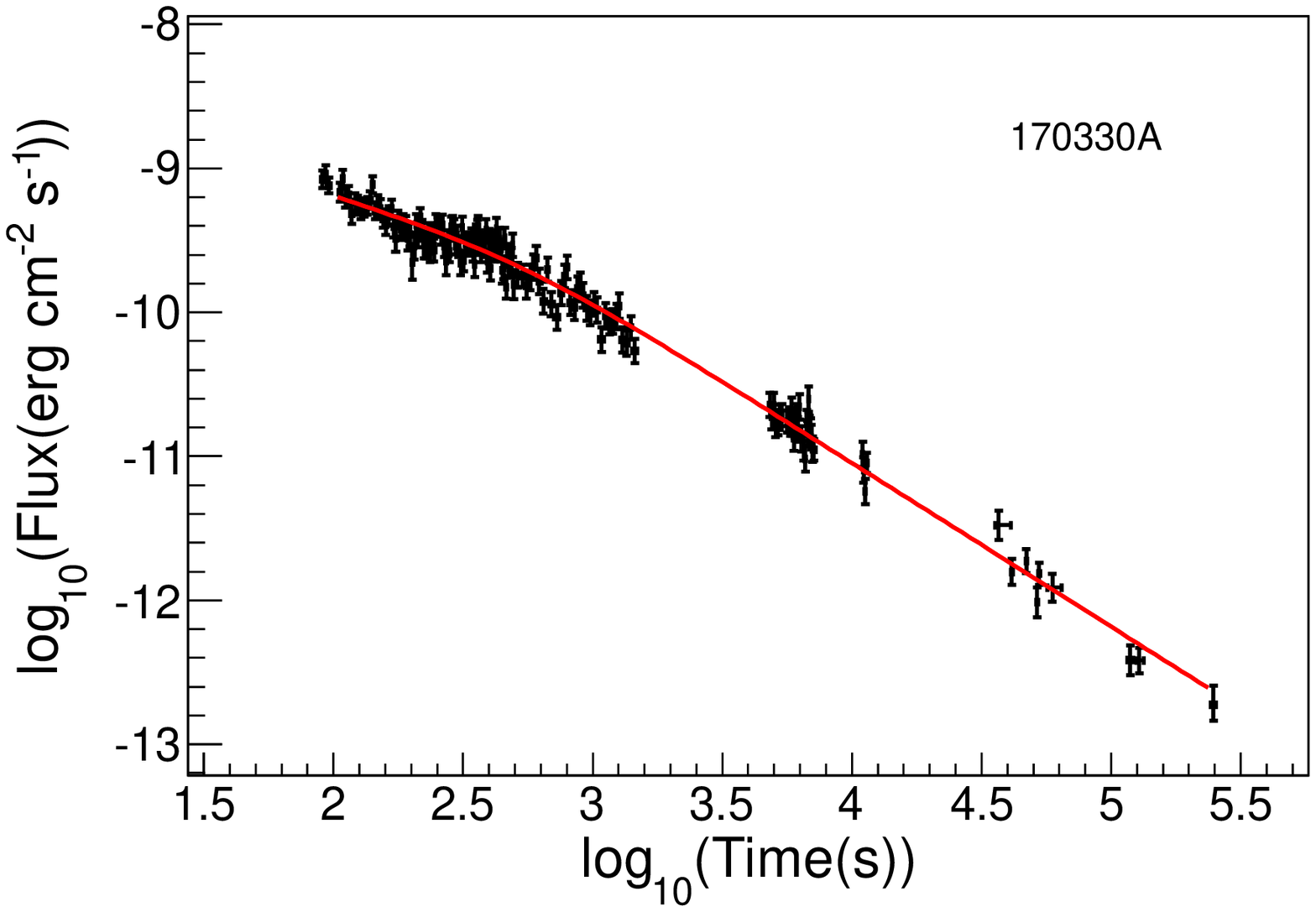}
\includegraphics[width=5.5cm,height=5cm]{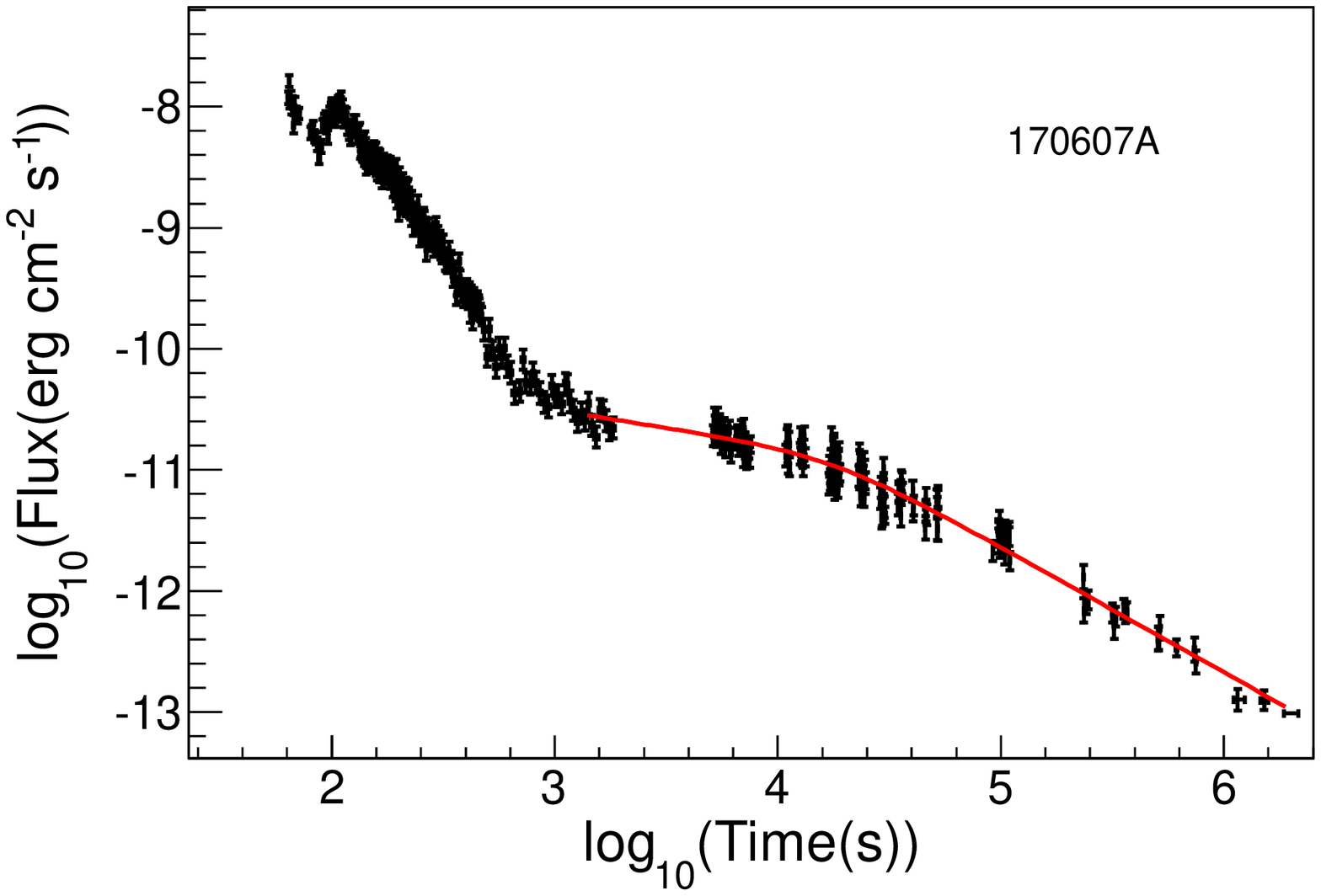}
\includegraphics[width=5.5cm,height=5cm]{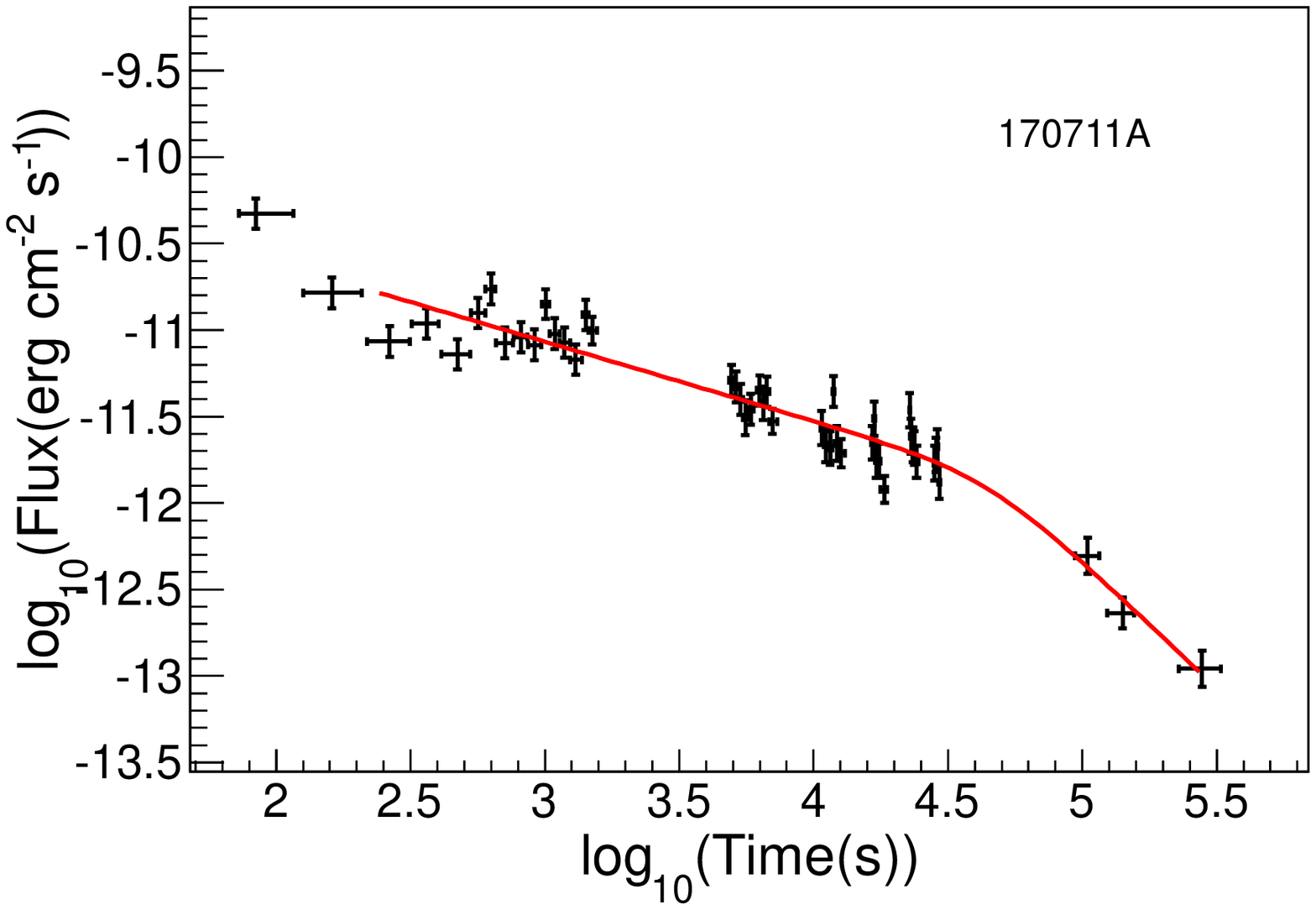}
\caption{ Continued.}
\label{fig-1-17}
\end{center}
\end{figure*}

\begin{deluxetable*}{ccCclccccc}[b!]
\tablecaption{Part I.  XRT observations and fitting results for our sample \label{table-1}}
\tablecolumns{6}
\tablenum{1}
\tablewidth{0pt}
\tablehead{
\colhead{GRBname} &
\colhead{$\rm{log}_{10}(t_{1})$\tablenotemark{a}} &
\colhead{$\rm{log}_{10}(t_{2})$\tablenotemark{a}} &
\colhead {$\rm{log}_{10}(t_{b})$\tablenotemark{b}} &
\colhead {$\alpha_{X,1}$\tablenotemark{b}} &
 \colhead{$\alpha_{X,2}$\tablenotemark{b}}&
\colhead{$\chi^{2}/\rm{dof}$\tablenotemark{b}} &
\colhead{$S_{X}$\tablenotemark{c}}&
\colhead{$\Gamma_{X,1}$\tablenotemark{d}}&
\colhead{$\Gamma_{X,2}$\tablenotemark{d}}\\
\colhead{} & \colhead{s} &\colhead{s} & \colhead{s} & \colhead{}&\colhead{}&\colhead{}& \colhead{$10^{-7}\rm{ergs} \ \rm{cm}^{-2}$}& \colhead{}
}
\startdata
050319& 2.61 & 6.32 & 4.52 &$0.5\pm0.03$&$1.54\pm0.11$&96.98/91&$ 3.1 \pm 0.46  $&$1.89^{+0.08}_{-0.07}$&$2.12^{+0.18}_{-0.17}$\\
050401& 2.1 & 5.91 & 3.67 &$0.55\pm0.02$&$1.43\pm0.04$&429.9/317&$ 6.51  \pm 0.8  $&$1.91\pm0.04$&$1.82^{+0.17}_{-0.16}$\\
050416A& 2.26 & 6.77 & 3.22 &$0.39\pm0.1$&$0.9\pm0.02$&111.4/91&$ 0.31  \pm 0.15  $&$1.98^{+0.21}_{-0.2}$&$2.08^{+0.15}_{-0.14}$\\
050505& 3.5 & 6.2 & 4.31 &$0.68\pm0.07$&$1.65\pm0.06$&233.1/188&$ 2.32  \pm 0.57  $&$1.97^{+0.09}_{-0.08}$&$2.01^{+0.09}_{-0.08}$\\
050607& 2.84 & 6.21 & 3.99 &$0.57\pm0.1$&$1.17\pm0.08$&19.9/20&$ 0.15\pm0.11  $&$1.71^{+0.21}_{-0.16}$&$2.17^{+0.52}_{-0.43}$\\
050712& 3.68 & 6.19 & 4.82 &$ 0.65\pm0.07$&$1.3\pm0.1$&75.22/41&$ 0.74\pm0.47  $&$2.04^{+0.19}_{-0.15}$&$1.85^{+0.35}_{-0.18}$\\
050713A& 2.41 & 6.23 & 3.99 &$ 0.58\pm0.03$&$1.29\pm0.03$&97.41/79&$ 2.43\pm0.64 $&$1.9^{+0.19}_{-0.18}$&$2.18^{+0.18}_{-0.17}$\\
050713B& 3.06 & 6.14 & 4.04 &$ -0.2\pm0.1$&$1.02\pm0.03$&151/107&$ 1.72\pm0.22 $&$2.03^{+0.15}_{-0.14}$&$1.83\pm0.13$\\
\enddata
\end{deluxetable*}

\begin{deluxetable*}{ccCclccccc}[b!]
\tablecaption{Part II.  XRT observations and fitting results for our sample \label{table-1}}
\tablecolumns{6}
\tablenum{1}
\tablewidth{0pt}
\tablehead{
\colhead{GRBname} &
\colhead{$\rm{log}_{10}(t_{1})$\tablenotemark{a}} &
\colhead{$\rm{log}_{10}(t_{2})$\tablenotemark{a}} &
\colhead {$\rm{log}_{10}(t_{b})$\tablenotemark{b}} &
\colhead {$\alpha_{X,1}$\tablenotemark{b}} &
 \colhead{$\alpha_{X,2}$\tablenotemark{b}}&
\colhead{$\chi^{2}/\rm{dof}$\tablenotemark{b}} &
\colhead{$S_{X}$\tablenotemark{c}}&
\colhead{$\Gamma_{X,1}$\tablenotemark{d}}&
\colhead{$\Gamma_{X,2}$\tablenotemark{d}}\\
\colhead{} & \colhead{s} &\colhead{s} & \colhead{s} & \colhead{}&\colhead{}&\colhead{}& \colhead{$10^{-7}\rm{ergs} \ \rm{cm}^{-2}$}& \colhead{}
}
\startdata
060927& 1.82 & 5  & 3.79  &$ 0.76\pm0.05$&$2.49\pm0.81$&5.479/13&$ 0.45\pm0.16$&$1.83^{+0.18}_{-0.11}$&$2.53^{+1.7}_{-0.59}$\\
061004& 2.72 & 5.28  & 4.19  &$ 0.68\pm0.07$&$1.64\pm0.29$&24.67/17&$ 0.4\pm0.24$&$1.95^{+0.23}_{-0.19}$&$2.32^{+0.68}_{-0.32}$\\
061019& 3.4 & 5.86  & 5.07  &$ 0.81\pm0.05$&$1.96\pm0.39$&80.89/42&$ 1.46\pm0.73$&$2.05^{+0.24}_{-0.23}$&$2.03^{+0.87}_{-0.69}$\\
061121& 2.37 & 6.32 & 3.71 &$ 0.36\pm0.03$&$1.4\pm0.02$&369.5/269&$ 7.04\pm0.84$&$1.94^{+0.13}_{-0.12}$&$1.79\pm0.06$\\
061201& 1.9 & 5.08 & 3.44&$ 0.54\pm0.07$&$2.05\pm0.11$&33.23/23&$ 1.39\pm0.27$&$1.3^{+0.18}_{-0.17}$&$2.02^{+0.4}_{-0.36}$\\
061202& 2.94 & 5.78 & 4.22 &$ -0.1\pm0.05$&$1.73\pm0.03$&202.2/116&$ 2.55\pm0.15$&$1.97^{+0.14}_{-0.13}$&$2.2^{+0.15}_{-0.14}$\\
061222A& 2.33 & 6.19 & 4.68 &$ 0.73\pm0.01$&$1.72\pm0.03$&516.3/327&$ 12.37\pm0.89$&$1.84\pm0.06$&$2.05\pm0.1$\\
070103& 2.2 & 5.13 & 3.16 &$ -0.23\pm0.14$&$1.42\pm0.06$&31.95/24&$ 0.21\pm0.04$&$1.91^{+0.29}_{-0.28}$&$2.15^{+0.30}_{-0.28}$\\
\enddata
\end{deluxetable*}

\begin{deluxetable*}{ccCclccccc}[b!]
\tablecaption{Part III.  XRT observations and fitting results for our sample\label{table-1}}
\tablecolumns{6}
\tablenum{1}
\tablewidth{0pt}
\tablehead{
\colhead{GRBname} &
\colhead{$\rm{log}_{10}(t_{1})$\tablenotemark{a}} &
\colhead{$\rm{log}_{10}(t_{2})$\tablenotemark{a}} &
\colhead {$\rm{log}_{10}(t_{b})$\tablenotemark{b}} &
\colhead {$\alpha_{X,1}$\tablenotemark{b}} &
 \colhead{$\alpha_{X,2}$\tablenotemark{b}}&
\colhead{$\chi^{2}/\rm{dof}$\tablenotemark{b}} &
\colhead{$S_{X}$\tablenotemark{c}}&
\colhead{$\Gamma_{X,1}$\tablenotemark{d}}&
\colhead{$\Gamma_{X,2}$\tablenotemark{d}}\\
\colhead{} & \colhead{s} &\colhead{s} & \colhead{s} & \colhead{}&\colhead{}&\colhead{}& \colhead{$10^{-7}\rm{ergs} \ \rm{cm}^{-2}$}& \colhead{}
}
\startdata
090530& 2.2& 5.90& 4.77&$0.61\pm0.02$&$1.35\pm0.1$&84.28/49&$1.27\pm0.35$&$1.97^{+0.14}_{-0.13}$&$1.87^{+0.3}_{-0.28}$\\
090727& 3.65& 6.33 & 5.57 &$0.5\pm0.05$&$1.56\pm0.25$&26.34/17&$1.22\pm0.45$&$1.9^{+0.27}_{-0.12}$&$2.1^{+0.63}_{-0.34}$\\
090728& 2.56& 4.90 & 3.25 &$-0.03\pm0.22$&$1.8\pm0.09$&22.97/22&$0.37\pm0.08$&$1.74^{+0.15}_{-0.12}$&$1.83^{+0.27}_{-0.25}$\\
090813& 2.03& 5.82& 2.73 &$0.23\pm0.06$&$1.25\pm0.01$&403.4/286&$1.78\pm0.21$&$2.01\pm0.07$&$1.79\pm0.09$\\
090904A& 2.91& 5.49 & 4.1&$0.13\pm0.12$&$1.31\pm0.09$&13.19/15&$0.29\pm0.07$&$2.38^{+0.44}_{-0.4}$&$2.04^{+0.35}_{-0.24}$\\
091018& 1.7 & 5.9 & 2.78 &$0.43\pm0.05$&$1.2\pm0.02$&232/137&$1.53\pm0.31$&$1.87^{+0.18}_{-0.17}$&$1.88\pm0.09$\\
091029& 2.86& 6.28 & 4.07  &$0.19\pm0.04$&$1.15\pm0.02$&183.4/125&$0.91\pm0.13$&$2.1\pm0.12$&$2.01\pm0.11$\\
091130B& 3.58& 6.13 & 4.70  &$0.29\pm0.06$&$1.17\pm0.05$&147.7/96&$0.78\pm0.17$&$2.3^{+0.17}_{-0.16}$&$2.27^{+0.19}_{-0.18}$\\
\enddata
\end{deluxetable*}

\begin{deluxetable*}{ccCclccccc}[b!]
\tablecaption{Part IV.  XRT observations and fitting results for our sample \label{table-1}}
\tablecolumns{6}
\tablenum{1}
\tablewidth{0pt}
\tablehead{
\colhead{GRBname} &
\colhead{$\rm{log}_{10}(t_{1})$\tablenotemark{a}} &
\colhead{$\rm{log}_{10}(t_{2})$\tablenotemark{a}} &
\colhead {$\rm{log}_{10}(t_{b})$\tablenotemark{b}} &
\colhead {$\alpha_{X,1}$\tablenotemark{b}} &
 \colhead{$\alpha_{X,2}$\tablenotemark{b}}&
\colhead{$\chi^{2}/\rm{dof}$\tablenotemark{b}} &
\colhead{$S_{X}$\tablenotemark{c}}&
\colhead{$\Gamma_{X,1}$\tablenotemark{d}}&
\colhead{$\Gamma_{X,2}$\tablenotemark{d}}\\
\colhead{} & \colhead{s} &\colhead{s} & \colhead{s} & \colhead{}&\colhead{}&\colhead{}& \colhead{$10^{-7}\rm{ergs} \ \rm{cm}^{-2}$}& \colhead{}
}
\startdata
130603B& 1.6 & 5.1 & 3.36 &$ 0.31\pm0.05$&$1.53\pm0.05$&154.8/74&$1.23\pm0.17$&$1.75^{+0.2}_{-0.19}$&$ 2\pm0.15$\\
130609B& 2.74 & 5.49 & 3.65 &$ 0.64\pm0.04$&$1.82\pm0.03$&373.4/278&$9.69\pm1.07$&$1.81\pm0.05$&$ 1.96^{+0.1}_{-0.09}$\\
131018A& 3.07 & 4.91 & 4.59 &$ 0.26\pm0.05$&$1.35\pm0.28$&76.53/53&$1.2\pm0.31$&$2.1^{+0.17}_{-0.16}$&$ 1.76^{+0.57}_{-0.49}$\\
131105A& 2.61 & 5.87 & 3 .66 &$ 0.16\pm0.11$&$1.19\pm0.04$&63.77/49&$0.88\pm0.22$&$1.92\pm0.17$&$ 1.99^{+0.18}_{-0.17}$\\
140108A& 2.88 & 5.63 & 3.8  &$ 0.33\pm0.06$&$1.33\pm0.03$&240.5/150&$4.5\pm0.85$&$1.76^{+0.32}_{-0.3}$&$ 2.08^{+0.17}_{-0.16}$\\
140114A& 2.91 & 5.56 & 4.34 &$ 0.24\pm0.06$&$1.25\pm0.12$&43.81/31&$0.45\pm0.13$&$2.17^{+0.2}_{-0.19}$&$ 1.7^{+0.56}_{-0.48}$\\
140323A& 2.34 & 5.07 & 3.95 &$ 0.56\pm0.03$&$1.62\pm0.08$&162.3/105&$6.07\pm1.18$&$2.1^{+0.17}_{-0.16}$&$ 2.2^{+0.68}_{-0.61}$\\
140331A& 3.16 & 5.32 & 4.71 &$ 0.3\pm0.06$&$2.07\pm0.47$&25.97/17&$0.71\pm0.24$&$1.73^{+0.22}_{-0.2}$&$ 2.19^{+1.22}_{-0.75}$\\
\enddata
\end{deluxetable*}

\begin{deluxetable*}{ccCclccccc}[b!]
\tablecaption{Part V.  XRT observations and fitting results for our sample\label{table-1}}
\tablecolumns{6}
\tablenum{1}
\tablewidth{0pt}
\tablehead{
\colhead{GRBname} &
\colhead{$\rm{log}_{10}(t_{1})$\tablenotemark{a}} &
\colhead{$\rm{log}_{10}(t_{2})$\tablenotemark{a}} &
\colhead {$\rm{log}_{10}(t_{b})$\tablenotemark{b}} &
\colhead {$\alpha_{X,1}$\tablenotemark{b}} &
 \colhead{$\alpha_{X,2}$\tablenotemark{b}}&
\colhead{$\chi^{2}/\rm{dof}$\tablenotemark{b}} &
\colhead{$S_{X}$\tablenotemark{c}}&
\colhead{$\Gamma_{X,1}$\tablenotemark{d}}&
\colhead{$\Gamma_{X,2}$\tablenotemark{d}}\\
\colhead{} & \colhead{s} &\colhead{s} & \colhead{s }& \colhead{}&\colhead{}&\colhead{}& \colhead{$10^{-7} \rm{ergs} \ \rm{cm}^{-2}$}& \colhead{}
}
\startdata
170202A& 2.47  & 5.73& 3.32 &$ -0.12\pm0.1$&$1.15\pm0.04$&50.64/39&$0.66\pm0.11$&$2.12^{+0.18}_{-0.17}$&$ 1.92^{+0.18}_{-0.17}$\\
170208B& 2.91  & 5.78& 3.83 &$ 0.53\pm0.14$&$1.07\pm0.06$&47/40&$0.56\pm0.75$&$2.19^{+0.23}_{-0.22}$&$ 2.54^{+0.28}_{-0.26}$\\
170317A& 2.18  & 5.16 & 3.32 &$ 0.44\pm0.08$&$1.33\pm0.08$&56.84/45&$0.39\pm0.14$&$1.97^{+0.17}_{-0.06}$&$ 2.15^{+0.31}_{-0.23}$\\
170330A& 2  & 5.39 & 2.71 &$ 0.56\pm0.08$&$1.33\pm0.08$&210.1/164&$1.59\pm0.14$&$2.01^{+0.39}_{-0.26}$&$ 1.92^{+0.13}_{-0.12}$\\
170607A& 3.13 & 6.29 & 4.25 &$ 0.29\pm0.05$&$1.03\pm0.03$&204.7/166&$2.86\pm0.63$&$2.08^{+0.1}_{-0.09}$&$ 1.93\pm0.1$\\
170711A& 2.37 & 5.45 & 4.66 &$ 0.46\pm0.03$&$1.48\pm0.3$&85.36/42&$1.17\pm0.51$&$2.07^{+0.38}_{-0.35}$&$ 2.11^{+0.22}_{-0.21}$\\
\enddata
\tablenotetext{a}{ $t_{1}$ and $t_{2}$ represent the beginning time ($t_{1}$) of the shallow decay segment and the end time ($t_{2}$) of its follow-up segment.}
\tablenotetext{b}{ Break time and decay slopes before and after the break, and the fit $\chi^{2}$/dof}
\tablenotetext{c}{ The X-ray fluence of shallow decay segment.}
\tablenotetext{d}{ Photon index of shallow decay segment and its follow-up segment.}
\end{deluxetable*}

\begin{deluxetable*}{cccclcccc}[b!]
\tablecaption{Part I  BAT  observations and redshift for our sample \label{table-2}}
\tablecolumns{6}
\tablenum{2}
\tablewidth{0pt}
\tablehead{
\colhead{GRBname} &
\colhead{$T_{90}(s)$}&
\colhead{$S_{\gamma}$\tablenotemark{a}} &
\colhead {$\Gamma_{\gamma ,1}$\tablenotemark{b}} &
\colhead{$E_{p}$\tablenotemark{b}}&
\colhead{BAT Ref\tablenotemark{c}} &
\colhead{z}&
\colhead{z Ref\tablenotemark{c}}&
  \colhead{ Group of GRBs\tablenotemark{d}} \\
\colhead{} & \colhead{s} &\colhead{$10^{-7}\rm{ergs} \ \rm{cm}^{-2}$} & \colhead{} & \colhead{\rm{keV}}& \colhead{}&\colhead{}&\colhead{}
}
\startdata
050319&10$\pm$2&8$\pm$0.8&2.2$\pm$0.2&33.41$\pm$12.22&3119&3.24&3136& $case^{1}$(regime 2)   \\
050401&33$\pm$2&140$\pm$14&1.5$\pm$0.06&133.14$\pm$29.35&3173&2.9&  3176&$case^{2}$\\\
050416A&2.4$\pm$0.2&69&2.9$\pm$0.2&12.32$\pm$4.45&3273&0.6528&3542&$case^{1}$(regime 1) \\
050505&60$\pm$2&41$\pm$4&1.5$\pm$0.1&133.14$\pm$38.15&3364&4.27&3368&$case^{1}$(regime 2)  \\
050607&26.5&8.9$\pm$1.2&1.93$\pm$0.14&53.6$\pm$16.4&3525&...&...&$case^{1}$(regime 1) \\
050712&48$\pm$2&18&1.56$\pm$0.18&115.56$\pm$42.64&3576&...&...&$case^{1}$(regime 2) \\
050713A&70$\pm$10&91$\pm6$&1.58$\pm$0.07&110.37$\pm$25.84&3597&...&...&$case^{1}$(regime 1) \\
050713B&54.2&82$\pm10$&1.56$\pm$0.13&115.56$\pm$34.64&3600&...&...&$case^{1}$(regime 2)  \\
\enddata
\end{deluxetable*}

\begin{deluxetable*}{cccclcccc}[b!]
\tablecaption{Part II  BAT  observations and redshift for our sample \label{}}
\tablecolumns{6}
\tablenum{2}
\tablewidth{0pt}
\tablehead{
\colhead{GRBname} &
\colhead{$T_{90}$}&
\colhead{$S_{\gamma}$\tablenotemark{a}} &
\colhead {$\Gamma_{\gamma ,1}$\tablenotemark{b}} &
\colhead{$E_{p}$\tablenotemark{b}}&
\colhead{BAT Ref\tablenotemark{c}} &
\colhead{z}&
\colhead{z Ref\tablenotemark{c}}&
  \colhead{ Group of GRBs\tablenotemark{d}} \\
\colhead{} & \colhead{s} &\colhead{$10^{-7}\rm{ergs} \ \rm{cm}^{-2}$} & \colhead{} & \colhead{\rm{keV}}& \colhead{}&\colhead{}&\colhead{}
}
\startdata
060927&$22.6\pm0.3$  &11$\pm$ 1 &0.93$\pm$0.38&$ 71.7\pm17.6$ &5639& 5.467 & 5651& $case^{1}$(regime 2) \\
061004&$6.2\pm0.3$  &5.7 $\pm$ 0.3 &1.81$\pm$0.10&$ 67.57\pm18.13$ &5964& ...& ...&$case^{1}$(regime 1)  \\
061019&$191\pm3$  &17 $\pm$ 2 &1.85$\pm$0.26&$ 62.45\pm31.63$ &5732& ...& ...&$case^{1}$(regime 3)  \\
061121&$81\pm5$  &137 $\pm$ 2 &1.32$\pm$0.05&$ 606 ^{+90}_{-72}$ &5831 5837& 1.3145& 5826&  $case^{1}$(regime 2) \\
061201&$0.8\pm0.1$  &3.3 $\pm$ 0.3&0.36$\pm$0.65&$ 873 ^{+458}_{-284}$ &5882 5890 & 0.111& 5952& $case^{1}$(regime 3) \\
061202&$91\pm5$  &35 $\pm$ 1&1.63$\pm$0.27&$ 98.63\pm23.83$ &5887 & ...& ...& $case^{1}$(regime 2) \\
061222A&$72\pm3$  &83 $\pm$ 2&1.02$\pm$0.11&$ 324 ^{+54}_{-41}$ &5984& 2.088&  P09 & $case^{1}$(regime 2) \\
070103&$19\pm1$  &3.4 $\pm$ 0.5&2$\pm$0.2&$ 47.13\pm17.93$ &5991 & 2.6208& K12  &$case^{1}$(regime 1)  \\
\enddata
\end{deluxetable*}

\newpage

\begin{deluxetable*}{cccclcccc}[b!]
\tablecaption{Part III  BAT  observations and redshift for our sample \label{}}
\tablecolumns{6}
\tablenum{2}
\tablewidth{0pt}
\tablehead{
\colhead{GRBname} &
\colhead{$T_{90}$}&
\colhead{$S_{\gamma}$\tablenotemark{a}} &
\colhead {$\Gamma_{\gamma ,1}$\tablenotemark{b}} &
\colhead{$E_{p}$\tablenotemark{b}}&
\colhead{BAT Ref\tablenotemark{c}} &
\colhead{z}&
\colhead{z Ref\tablenotemark{c}}&
  \colhead{ Group of GRBs\tablenotemark{d}} \\
\colhead{} & \colhead{s} &\colhead{$10^{-7}\rm{ergs} \ \rm{cm}^{-2}$} & \colhead{} & \colhead{\rm{keV}}& \colhead{}&\colhead{}&\colhead{}
}
\startdata
090530&$48\pm36$  &11$\pm$ 1&1.61$\pm$0.17&$ 103.12\pm34.85$& 9443&1.266&15571&$case^{1}$(regime 2) \\
090727&$302\pm23$  &14$\pm$ 2&1.24$\pm$0.24&$ 264.7\pm140.61$& 9724& ...&...&$case^{1}$(regime 2) \\
090728&$59\pm18$  &10$\pm$ 2&2.05$\pm$0.26&$ 43.11\pm19.87$& 9736& ...&...&$case^{1}$(regime 3) \\
090813&$9$  &13$\pm$ 1&1.43$\pm$0.08&$ 158.21\pm39.97$& 9792& ...&...& $case^{1}$(regime 2) \\
090904A&$122\pm10$  &30$\pm$ 2&2.01$\pm$0.1&$ 46.29\pm12.36$& 9888& ...&...&$case^{1}$(regime 2)  \\
091018&$4.4\pm0.6$  &14$\pm$ 1&1.77$\pm$0.24&$ 19.3^{+18}_{-11}$& 10040& 0.971&10038& $case^{1}$(regime 2) \\
091029&$39.2\pm5$  &24$\pm$ 1&1.46$\pm$0.27&$ 61.4\pm17.5$& 10103& 2.752&10100&$case^{1}$(regime 1)  \\
091130B&$112.5\pm17.1$  &13$\pm$ 1&2.15$\pm$0.15&$ 36.3\pm12.26$& 10221& ...&...&$case^{1}$(regime 1) \\
\enddata
\end{deluxetable*}

\newpage

\begin{deluxetable*}{cccclcccc}[b!]
\tablecaption{Part IV  BAT  observations and redshift for our sample \label{}}
\tablecolumns{6}
\tablenum{2}
\tablewidth{0pt}
\tablehead{
\colhead{GRBname} &
\colhead{$T_{90}$}&
\colhead{$S_{\gamma}$\tablenotemark{a}} &
\colhead {$\Gamma_{\gamma ,1}$\tablenotemark{b}} &
\colhead{$E_{p}$\tablenotemark{b}}&
\colhead{BAT Ref\tablenotemark{c}} &
\colhead{z}&
\colhead{z Ref\tablenotemark{c}}&
  \colhead{ Group of GRBs\tablenotemark{d}} \\
\colhead{} & \colhead{s} &\colhead{$10^{-7}\rm{ergs} \ \rm{cm}^{-2}$} & \colhead{} & \colhead{\rm{keV}}& \colhead{}&\colhead{}&\colhead{}
}
\startdata
130603B&$0.18\pm0.02$  &6.3$ \pm0.3$ &0.82$\pm$0.07&$ 1177.96\pm 322.01$& 14741& 0.356& 14744&$case^{1}$(regime 2)  \\
130609B&$210.6\pm15.1$  &160 &1.32$\pm$0.04&$ 211.22\pm 40.09$& 14867& ...& ...&$case^{1}$(regime 3)  \\
131018A&$73.22\pm18.97$  &11$ \pm1$ &2.24$\pm$0.14&$ 31.3\pm 9.47$& 15354& ...& ...& $case^{1}$(regime 2) \\
131105A&$112.3\pm4.1$  &71$ \pm5$ &1.45$\pm$0.11&$ 150.47\pm44.64 $& 15459& 1.686& 15450& $case^{1}$(regime 2) \\
140108A&$94\pm4$  &70$ \pm2$ &1.62$\pm$0.05&$ 100.84\pm22.11 $& 15717& ...& ...&  $case^{1}$(regime 1) \\
140114A&$139.7\pm16.4$ & 32$ \pm1$ &2.06$\pm$0.09&$ 42.36\pm11.23 $& 15738& ...& ...& $case^{1}$(regime 2)  \\
140323A&$104.9\pm2.6$ & $160 $ &1.35$\pm$0.17&$ 127.1\pm59.2 $& 16029& ...& ...&$case^{1}$(regime 2)  \\
140331A&$209\pm33$ & $6.7 \pm1.3$ &2.01$\pm$0.29&$ 46.29\pm26.12 $& 16063& ...& ...& $case^{1}$(regime 3) \\
\enddata
\end{deluxetable*}

\newpage

\begin{deluxetable*}{cccclcccc}[b!]
\tablecaption{Part V BAT  observations and redshift for our sample \label{}}
\tablecolumns{6}
\tablenum{2}
\tablewidth{0pt}
\tablehead{
\colhead{GRBname} &
\colhead{$T_{90}$}&
\colhead{$S_{\gamma}$\tablenotemark{a}} &
\colhead {$\Gamma_{\gamma ,1}$\tablenotemark{b}} &
\colhead{$E_{p}$\tablenotemark{b}}&
\colhead{BAT Ref\tablenotemark{c}} &
\colhead{z}&
\colhead{z Ref\tablenotemark{c}}&
  \colhead{ Group of GRBs\tablenotemark{d}} \\
\colhead{} & \colhead{s} &\colhead{$10^{-7}\rm{ergs} \ \rm{cm}^{-2}$} & \colhead{} & \colhead{\rm{keV}}& \colhead{}&\colhead{}&\colhead{}
}
\startdata
170202A&$46.2\pm11.5$ & $33\pm1$ &1.68$\pm$0.07&$ 88.44\pm20.87$& 20596& 3.645& 20584& $case^{1}$(regime 2) \\
170208B&$128.07\pm5.75$ & $61\pm2$ &1.18$\pm$0.21&$ 97.8\pm27.4$& 20654& ...& ...& $case^{1}$(regime 1) \\
170317A&$11.94\pm4.87$ & $13\pm1$ &0.8$\pm$0.49&$ 58.8\pm10.5$& 20896& ...& ...&$case^{1}$(regime 1)  \\
170330A&$144.73\pm3.51$ & $56\pm3$ &1.38$\pm$0.09&$ 179.9\pm45.88$& 20968& ...& ...& $case^{1}$(regime 2) \\
170607A&$23$ &$ 75\pm3$ &1.61$\pm$0.07&$ 103.12\pm23.51$& 21218& ...& ...& $case^{1}$(regime 1) \\
170711A&$31.3\pm7.6$ & $14\pm1$ &1.75$\pm$0.1&$ 76.32\pm20.03$& 21331& ...& ...&$case^{1}$(regime 2) \\
\enddata
\end{deluxetable*}

\clearpage

\tablecomments{References Continued:(10462)\cite{2010GCN.10462....1C}; (10466)\cite{2010GCN.10466....1C}; (10615)\cite{2010GCN.10615....1U}; (10620)\cite{2010GCN.10620....1A}; (10684)\cite{2010GCN.10684....1G}; (10685)\cite{2010GCN.10685....1M}; (10732)\cite{2010GCN.10732....1S}; (10788)\cite{2010GCN.10788....1B}; (10852)\cite{2010GCN.10852....1S}. (10932)\cite{2010GCN.10932....1C}; (10993)\cite{2010GCN.10993....1S}; (11164)\cite{2010GCN.11164....1C}; (11169)\cite{2010GCN.11169....1S}; ( 11202)\cite{2010GCN.11202....1S}; (11218)\cite{2010GCN.11218....1B}; (11230)\cite{2010GCN.11230....1T}; (11233)\cite{2010GCN.11233....1B}; (11414)\cite{2010GCN.11414....1B}; (11511)\cite{2011GCN.11511....1S}; (11533)\cite{2011GCN.11533....1U}; (11538)\cite{2011GCN.11538....1C};  (11691)\cite{2011GCN.11691....1S}; (11783)\cite{2011GCN.11783....1B}; (11811)\cite{2011GCN.11881....1F}; (11921)\cite{2011GCN.11921....1B}; (12049)\cite{2011GCN.12049....1B}; (12258)\cite{2011GCN.12258....1D}; (12262)\cite{2011GCN.12262....1S}; (12424)\cite{2011GCN.12424....1B}; (12431)\cite{2011GCN.12431....1W}; (12749)\cite{2011GCN.12749....1C}; (12761)\cite{2011GCN.12761....1C}; (12815)\cite{2012GCN.12815....1B}; (12983)\cite{2012GCN.12983....1B}; (13022)\cite{2012GCN.13022....1S}; (13096)\cite{2012GCN.13096....1C}; (13118 )\cite{2012GCN.13118....1T}; (13120)\cite{2012GCN.13120....1B}; (13333)\cite{2012GCN.13333....1M}; (13348)\cite{2012GCN.13348....1T}; (13414)\cite{2012GCN.13414....1S}; (13628)\cite{2012GCN.13628....1T}; (13634) \cite{2012GCN.13634....1K}; (13720) \cite{2012GCN.13720....1S}; (13723)\cite{2012GCN.13723....1S}; (13890)\cite{2012GCN.13890....1T}; (13899)\cite{2012GCN.13899....1B}; (13949)\cite{2012GCN.13949....1C}; (13990)\cite{2012GCN.13990....1B}; (13996)\cite{2012GCN.13996....1B}; (14011)\cite{2012GCN.14011....1P}; (14067)\cite{2012GCN.14067....1B}; (14163)\cite{2013GCN.14163....1P}; (14741)\cite{2013GCN.14741....1B}; (14867)\cite{2013GCN.14867....1K}; (15354)\cite{2013GCN.15354....1U}; (15459)\cite{2013GCN.15459....1B}; (15571)\cite{2013GCN.15571....1G}; (15717)\cite{2014GCN.15717....1C}; ( 15738)\cite{2014GCN.15738....1K}; (16029)\cite{2014GCN.16029....1S}; (16063)\cite{2014GCN.16063....1S}; (16194)\cite{2014GCN.16194....1K}; (16200)\cite{2014GCN.16200....1K}; (16258)\cite{2014GCN.16258....1S}; (16301)\cite{2014GCN.16301....1C}; (16306)\cite{2014GCN.16306....1U}; ( 16310)\cite{2014GCN.16310....1D}; (16505)\cite{2014GCN.16505....1C}; (16509)\cite{2014GCN.16509....1K}; (16539)\cite{2014GCN.16539....1M}; (16553)\cite{2014GCN.16553....1M}; (16827)\cite{2014GCN.16827....1S}; (16927)\cite{ 2014GCN.16927....1M}; (17010)\cite{2014GCN.17010....1U}; (17081)\cite{2014GCN.17081....1P}; (17083)\cite{2014GCN.17083....1K}; (17330)\cite{2015GCN.17330....1M}; (17374)\cite{2015GCN.17374....1P}; (17410)\cite{2015GCN.17410....1U}; (17616)\cite{2015GCN.17616....1P}; (17628)\cite{2015GCN.17628....1M}; (17637)\cite{2015GCN.17637....1P}; ( 17758)\cite{2015GCN.17758....1C}; (17761)\cite{2015GCN.17761....1B}; (17907)\cite{2015GCN.17907....1P}; (17930)\cite{2015GCN.17930....1S}; (17941)\cite{2015GCN.17941....1S}; (18020)\cite{2015GCN.18020....1K}; (18268)\cite{2015GCN.18268....1L}; (18273)\cite{2015GCN.18273....1Z}; (18487)\cite{2015GCN.18487....1P}; (18496)\cite{2015GCN.18496....1P}; (18593)\cite{2015GCN.18593....1K}; (18603)\cite{2015GCN.18603....1B}; (18751)\cite{2015GCN.18751....1L}; (18752)\cite{2015GCN.18752....1K}; (18899)\cite{2016GCN.18899....1S}; (18944)\cite{2016GCN.18944....1B}; ( 19106)\cite{2016GCN.19106....1S}; (19109)\cite{2016GCN.19109....1X}; (19240)\cite{2016GCN.19240....1M}; (19245)\cite{2016GCN.19245....1D}; (19301)\cite{2016GCN.19301....1U}; (19639)\cite{2016GCN.19639....1K}; (19765)\cite{2016GCN.19765....1K}; (19773)\cite{2016GCN.19773....1X}; (19861)\cite{2016GCN.19861....1S}; (19974)\cite{2016GCN.19974....1M}; (20032)\cite{2016GCN.20032....1U};  (20150)\cite{ 2016GCN.20150....1D}; (20151)\cite{2016GCN.20151....1P}; (20220)\cite{2016GCN.20200....1M}; (20238)\cite{2016GCN.20238....1K}; (20245)\cite{2016GCN.20245....1C}; (20270)\cite{2016GCN.20270....1K}; (20456)\cite{2017GCN.20456....1M}; (20458)\cite{2017GCN.20458....1X}; (20584)\cite{2017GCN.20584....1D}; (20596)\cite{2017GCN.20596....1B}; (20654)\cite{2017GCN.20596....1B}; (20896)\cite{2017GCN.20896....1B}; (20968)\cite{2017GCN.20968....1L}; (21218)\cite{2017GCN.21331....1M}.}

\newpage

\begin{deluxetable*}{cccclccc}[b!]
\tablecaption{Part I  Rest-frame properties for bursts with known redshifts \label{table-3}}
\tablecolumns{6}
\tablenum{3}
\tablewidth{0pt}
\tablehead{
\colhead{GRBname} &
\colhead{$\rm{log}_{10}(T'_{90})$}&
\colhead{$\rm{log}_{10}(E'_{p})$} &
\colhead {$\rm{log}_{10}(E_{\rm{iso},\gamma})$} &
\colhead{$\rm{log}_{10}(E_{\rm{iso},\gamma}^{b})$}&
\colhead{$\rm{log}_{10}(E_{\rm{iso},X})$} &
\colhead{$\rm{log}_{10}(t'_{b})$}&
\colhead{$\rm{log}_{10}((L_{X})$}\\
\colhead{} & \colhead{s} &\colhead{keV} & \colhead{ergs} & \colhead{\rm{ergs}}& \colhead{ergs}&\colhead{s}&\colhead{ergs/s}
}
\startdata
050319&$0.37\pm0.08$&$2.15\pm0.18$&$52.23\pm0.04$&$52.58\pm0.08$&$51.82\pm0.06$&$3.89\pm0.06$&$47.05\pm0.05$\\
050401&$0.93\pm0.03$&$2.72\pm0.1$&$53.39\pm0.04$&$53.77\pm0.04$&$52.06\pm0.05$&$3.08\pm0.04$&$48.13\pm0.03$\\
050416A&$0.16\pm0.03$&$1.31\pm0.17$&$50.47\pm0.03$&$50.91\pm0.06$&$49.52\pm0.17$&$3\pm0.14$&$46.21\pm0.11$\\
050505&$1.06\pm0.01$&$2.85\pm0.13$&$53.13\pm0.04$&$53.55\pm0.05$&$51.88\pm0.1$&$3.59\pm0.07$&$47.43\pm0.08$\\
050714B&$1.13$&$1.8\pm0.22$&$51.93\pm0.09$&$52.3\pm0.09$&$50.13\pm0.18$&$4.03\pm0.19$&$45.53\pm0.13$\\
050802&$0.68\pm0.06$&$2.46\pm0.13$&$52.29\pm0.04$&$52.65\pm0.06$&$51.18\pm0.18$&$3.29\pm0.04$&$47.38\pm0.06$\\
050814&$1.01\pm0.12$&$2.64\pm0.17$&$52.99\pm0.07$&$53.33\pm0.11$&$51.54\pm0.08$&$4.15\pm0.06$&$46.35\pm0.08$\\
050822&$1.62\pm0.01$&$1.71\pm0.14$&$52.24\pm0.04$&$52.59\pm0.04$&$50.7\pm0.09$&$3.89\pm0.08$&$46.34\pm0.07$\\
050824&$1.14\pm0.08$&$1.46\pm0.27$&$50.6\pm0.09$&$51\pm0.11$&$50.07\pm0.17$&$4.56\pm0.16$&$45.2\pm0.1$\\
050826&$1.43\pm0.09$&$2.59\pm0.38$&$49.95\pm0.07$&$50.55\pm0.5$&$48.39\pm0.18$&$4.51\pm0.15$&$43.83\pm0.15$\\
051109B&$1.14\pm0.03$&$1.74\pm0.23$&$48.58\pm0.06$&$48.92\pm0.09$&$47.25\pm0.12$&$3.46\pm0.1$&$43.72\pm0.09$\\
051221A&$-0.04$&$2.79\pm0.14$&$51.38\pm0.01$&$52.09\pm0.07$&$49.69\pm0.09$&$4.45\pm0.08$&$44.78\pm0.08$\\
060108&$0.68\pm0.03$&$2.15\pm0.17$&$51.55\pm0.05$&$51.89\pm0.05$&$50.53\pm0.1$&$3.63\pm0.1$&$46.34\pm0.08$\\
060115&$1.5\pm0.02$&$2.53\pm0.14$&$52.67\pm0.02$&$53\pm0.11$&$51.07\pm0.19$&$3.84\pm0.18$&$46.27\pm0.16$\\
060306&$1.38\pm0.01$&$2.2\pm0.13$&$52.11\pm0.02$&$52.46\pm0.08$&$50.54\pm0.17$&$3.11\pm0.16$&$46.97\pm0.14$\\
060502A&$1.12\pm0.06$&$2.6\pm0.12$&$52.09\pm0.02$&$52.53\pm0.05$&$51.08\pm0.11$&$4.17\pm0.11$&$46.2\pm0.1$\\
060604&$0.5\pm0.11$&$2.25\pm0.34$&$51.13\pm0.09$&$51.46\pm0.18$&$50.72\pm0.11$&$3.83\pm0.08$&$46.37\pm0.08$\\
060605&$0.5\pm0.05$&$2.98\pm0.19$&$52.1\pm0.04$&$52.57\pm0.1$&$51.52\pm0.03$&$3.25\pm0.03$&$47.41\pm0.03$\\
060607A&$1.39\pm0.02$&$2.79\pm0.11$&$52.71\pm0.02$&$53.16\pm0.05$&$52.37\pm0.03$&$3.5\pm0.01$&$48.08\pm0.02$\\
060614&$1.95             $&$2.53\pm0.26$&$51.16\pm0.02$&$51.71\pm0.13$&$49.03\pm0.03$&$4.64\pm0.02$&$44.34\pm0.03$\\
060707&$1.19\pm0.03 $&$2.47\pm0.62$&$52.6\pm0.05$&$52.95\pm0.09$&$51.64\pm0.16$&$4.45\pm0.16$&$46.18\pm0.15$\\
060714&$1.49\pm0.02 $&$2.25\pm0.13$&$52.68\pm0.03$&$52.99\pm0.01$&$51.13\pm0.11$&$3.1\pm0.1$&$47.35\pm0.11$\\
060719&$1.34\pm0.04 $&$2.08\pm0.13$&$51.97\pm0.03$&$52.28\pm0.01$&$50.49\pm0.12$&$3.47\pm0.1$&$46.47\pm0.1$\\
060729&$1.88\pm0.04 $&$1.98\pm0.15$&$51.3\pm0.03$&$51.63\pm0.07$&$50.9\pm0.02$&$4.61\pm0.01$&$46.04\pm0.01$\\
060805A&$0.2\pm0.04 $&$2.04\pm0.31$&$50.98\pm0.11$&$51.33\pm0.22$&$50.06\pm0.16$&$3.03\pm0.14$&$46.45\pm0.15$\\
060814&$1.66\pm0.01 $&$2.88\pm0.49$&$53.37\pm0.02$&$53.89\pm0.1$&$51.23\pm0.07$&$3.64\pm0.05$&$47.02\pm0.06$\\
060906&$0.97\pm0.01 $&$2.33\pm0.14$&$52.76\pm0.03$&$53.09\pm0.01$&$51.01\pm0.05$&$3.38\pm0.05$&$46.89\pm0.05$\\
060908&$0.83\pm0.01 $&$2.77\pm0.11$&$52.39\pm0.01$&$52.84\pm0.05$&$50.85\pm0.11$&$2.38\pm0.08$&$47.85\pm0.09$\\
060927&$0.54\pm0.01 $&$2.67\pm0.62$&$52.72\pm0.04$&$53.06\pm0.03$&$51.32\pm0.14$&$2.98\pm0.12$&$47.1\pm0.15$\\
061121&$1.54\pm0.03 $&$3.15\pm0.14$&$52.77\pm0.01$&$53.58\pm0.04$&$51.48\pm0.05$&$3.34\pm0.04$&$47.65\pm0.05$\\
061201&$-0.14\pm0.05 $&$2.99\pm0.2$&$48.96\pm0.04$&$49.86\pm0.12$&$48.58\pm0.08$&$3.4\pm0.06$&$44.9\pm0.09$\\
061222A&$1.37\pm0.02 $&$3\pm0.21$&$52.92\pm0.01$&$53.57\pm0.05$&$52.1\pm0.03$&$4.19\pm0.03$&$46.96\pm0.03$\\
070103&$0.72\pm0.02 $&$2.23\pm0.19$&$51.70\pm0.06$&$52.03\pm0.06$&$50.49\pm0.08$&$2.6\pm0.06$&$47.44\pm0.05$\\
070110&$1.4\pm0.02 $&$2.58\pm0.15$&$52.3\pm0.03$&$52.66\pm0.07$&$51.48\pm0.02$&$3.79\pm0.01$&$47.22\pm0.02$\\
070129&$2.14\pm0.02 $&$2.16\pm0.16$&$52.58\pm0.04$&$52.90\pm0.15$&$50.88\pm0.08$&$3.73\pm0.06$&$46.61\pm0.05$\\
070306&$0.48\pm0.09 $&$2.33\pm0.13$&$50.92\pm0.25$&$51.26\pm0.09$&$51.46\pm0.03$&$4.08\pm0.02$&$46.94\pm0.02$\\
070508&$1.06\pm0.02 $&$2.67\pm0.33$&$52.52\pm0.01$&$53.07\pm0.11$&$50.79\pm0.33$&$2.2\pm0.24$&$48.35\pm0.16$\\
070529&$1.49\pm0.01 $&$2.8\pm0.2$&$52.55\pm0.04$&$53\pm0.11$&$50.75\pm0.15$&$2.48\pm0.11$&$47.69\pm0.12$\\
070809&$0.03\pm0.03 $&$2.02\pm0.22$&$49.05\pm0.04$&$49.39\pm0.25$&$48.24\pm0.15$&$3.69\pm0.13$&$44.53\pm0.11$\\
070810A&$0.54\pm0.04 $&$2.14\pm0.15$&$51.87\pm0.04$&$52.21\pm0.04$&$50.61\pm0.21$&$2.64\pm0.17$&$47.38\pm0.19$\\
080413B&$0.58\pm0.05 $&$2.19\pm0.18$&$52\pm0.01$&$52.34\pm0.02$&$50.34\pm0.14$&$2.28\pm0.1$&$47.83\pm0.08$\\
080430&$0.96\pm0.06 $&$2.15\pm0.12$&$51.25\pm0.04$&$51.6\pm0.07$&$50.42\pm0.07$&$4.28\pm0.06$&$45.68\pm0.05$\\
080516&$0.14\pm0.02 $&$2.44\pm0.25$&$51.73\pm0.06$&$52.07\pm0.14$&$50.79\pm0.17$&$3.12\pm0.16$&$47\pm0.14$\\
080707&$1.08\pm0.02 $&$2.21\pm0.2$&$51.3\pm0.05$&$52.61\pm0.12$&$50.1\pm0.16$&$3.68\pm0.14$&$45.97\pm0.12$\\
080905B&$1.58\pm0.05 $&$2.38\pm0.16$&$52.35\pm0.05$&$52.7\pm0.12$&$51.88\pm0.07$&$3.64\pm0.07$&$47.5\pm0.05$\\
081007&$0.82\pm0.16 $&$1.5\pm0.17$&$50.7\pm0.05$&$51.06\pm0.07$&$50.1\pm0.3$&$4.44\pm0.12$&$45.09\pm0.12$\\
081029&$1.75\pm0.07 $&$2.88\pm0.21$&$52.77\pm0.04$&$53.22\pm0.15$&$51.46\pm0.03$&$3.53\pm0.02$&$47.2\pm0.03$\\
090102&$1.03\pm0.03$&$2.68\pm0.12$&$50.6\pm0.02$&$51.06\pm0.05$&$51.46\pm0.06$&$2.75\pm0.03$&$48.16\pm0.04$\\
\enddata
\end{deluxetable*}

\begin{deluxetable*}{cccclccc}[b!]
\tablecaption{Part II  Rest-frame properties for bursts with known redshifts \label{}}
\tablecolumns{6}
\tablenum{3}
\tablewidth{0pt}
\tablehead{
\colhead{GRBname} &
\colhead{$\rm{log}_{10}(T'_{90})$}&
\colhead{$\rm{log}_{10}(E'_{p})$} &
\colhead {$\rm{log}_{10}(E_{\rm{iso},\gamma})$} &
\colhead{$\rm{log}_{10}(E_{\rm{iso},\gamma}^{b})$}&
\colhead{$\rm{log}_{10}(E_{\rm{iso},X})$} &
\colhead{$\rm{log}_{10}(t'_{b})$}&
\colhead{$\rm{log}_{10}((L_{X})$}\\
\colhead{} & \colhead{s} &\colhead{keV} & \colhead{ergs} & \colhead{\rm{ergs}}& \colhead{ergs}&\colhead{s}&\colhead{ergs/s}
}

\startdata
090113&$0.52\pm0.04$&$2.46\pm0.13$&$51.75\pm0.02$&$52.12\pm0.05$&$50.51\pm0.1$&$2.31\pm0.08$&$47.81\pm0.07$\\
090205&$0.19\pm0.08$&$2.31\pm0.2$&$51.85\pm0.06$&$52.18\pm0.1$&$51.19\pm0.11$&$3.43\pm0.1$&$46.77\pm0.11$\\
090313&$1.25\pm0.09$&$2.39\pm0.26$&$52.5\pm0.06$&$52.80\pm0.1$&$51.46\pm0.1$&$4.25\pm0.06$&$46.64\pm0.09$\\
090404&$1.32\pm0.07$&$2.04\pm0.13$&$52.75\pm0.01$&$53.08\pm0.05$&$51.36\pm0.1$&$3.63\pm0.08$&$47.06\pm0.07$\\
090407&$2.1\pm0.09$&$2.29\pm0.27$&$51.75\pm0.07$&$52.12\pm0.27$&$50.93\pm0.05$&$4.47\pm0.05$&$45.89\pm0.04$\\
090418A&$1.33\pm0.04$&$2.56\pm0.11$&$52.46\pm0.02$&$52.87\pm0.04$&$51.31\pm0.08$&$3.07\pm0.06$&$47.64\pm0.08$\\
090424&$1.53               $&$2.44\pm0.01$&$52.35\pm0.25$&$52.8\pm0.01$&$50.78\pm0.18$&$2.79\pm0.17$&$47.73\pm0.16$\\
090429B&$-0.28\pm0.07$&$2.64\pm0.56$&$52.51\pm0.04$&$52.83\pm0.01$&$50.98\pm0.1$&$1.88\pm0.08$&$48.34\pm0.04$\\
090510&$-0.8\pm0.12$&$3.07\pm0.3$&$50.84\pm0.05$&$51.46\pm0.13$&$50.47\pm0.07$&$2.91\pm0.05$&$47.03\pm0.06$\\
090529&$1.44           $&$2.23\pm0.25$&$52\pm0.1$&$52.34\pm0.11$&$50.81\pm0.19$&$2.91\pm0.05$&$47.03\pm0.18$\\
090530&$1.33\pm0.24$&$2.37\pm0.19$&$51.64\pm0.04$&$52\pm0.26$&$50.7\pm0.11$&$4.42\pm0.1$&$45.47\pm0.09$\\
091018&$0.35\pm0.06$&$1.58\pm0.57$&$51.52\pm0.03$&$51.91\pm0.38$&$50.56\pm0.08$&$2.49\pm0.07$&$47.65\pm0.06$\\
091029&$1.02\pm0.05$&$2.36\pm0.41$&$52.59\pm0.02$&$52.93\pm0.05$&$51.17\pm0.06$&$3.5\pm0.05$&$47.06\pm0.04$\\
100219A&$0.52\pm0.1$&$3.05\pm0.29$&$52.14\pm0.07$&$52.61\pm0.31$&$51.83\pm0.07$&$3.77\pm0.03$&$47.05\pm0.06$\\
100302A&$0.49\pm0.04$&$2.67\pm0.2$&$52.09\pm0.05$&$52.4\pm0.15$&$50.94\pm0.22$&$3.58\pm0.22$&$46.47\pm0.16$\\
100418A&$0.63\pm0.06$&$1.76\pm0.21$&$50.52\pm0.6$&$50.83\pm0.13$&$49.79\pm0.22$&$4.7\pm0.06$&$44.95\pm0.05$\\
100425A&$1.13\pm0.03$&$1.81\pm0.24$&$51.54\pm0.08$&$51.91\pm0.21$&$50.41\pm0.18$&$4.07\pm0.18$&$45.63\pm0.14$\\
100901A&$2.26\pm0.03$&$2.41\pm0.22$&$52\pm0.06$&$52.35\pm0.32$&$51.24\pm0.02$&$4.2\pm0.02$&$46.72\pm0.01$\\
100906A&$1.62\pm0.01$&$2.29\pm0.1$&$53.12\pm0.25$&$53.46\pm0.04$&$51.46\pm0.04$&$3.66\pm0.03$&$47.02\pm0.05$\\
110106B&$1.19\pm0.08$&$2.08\pm0.14$&$51.28\pm0.02$&$51.62\pm0.1$&$50.14\pm0.11$&$3.91\pm0.1$&$45.76\pm0.1$\\
110808A&$1.31\pm0.17$&$1.81\pm0.3$&$51.17\pm0.09$&$51.52\pm0.24$&$50.43\pm0.21$&$4.85\pm0.21$&$44.86\pm0.16$\\
111008A&$1.03\pm0.01$&$2.56\pm0.12$&$53.35\pm0.02$&$53.68\pm0.01$&$52.05\pm0.06$&$3.12\pm0.06$&$48.04\pm0.05$\\
111228A&$1.77\pm0.02$&$1.71\pm0.12$&$52.04\pm0.01$&$52.4\pm0.07$&$50.66\pm0.05$&$3.89\pm0.05$&$46.37\pm0.04$\\
120326A&$1.4\pm0.05$&$2.06\pm0.19$&$52.3\pm0.05$&$52.64\pm0.01$&$51.57\pm0.05$&$4.27\pm0.01$&$46.96\pm0.01$\\
120521C&$0.58\pm0.07$&$2.75\pm0.13$&$52.78\pm0.04$&$53.12\pm0.12$&$51\pm0.24$&$3.22\pm0.23$&$46.81\pm0.17$\\
120811C&$0.86\pm0.06$&$2.2\pm0.2$&$52.67\pm0.04$&$52.96\pm0.01$&$51.17\pm0.2$&$2.77\pm0.17$&$47.74\pm0.13$\\
120907A&$0.93\pm0.18$&$2.2\pm0.2$&$51.2\pm0.07$&$51.53\pm0.2$&$50.27\pm0.2$&$4.35\pm0.17$&$45.45\pm0.1$\\
121024A&$1.54\pm0.17$&$2.52\pm0.25$&$51.42\pm0.04$&$51.87\pm0.25$&$50.63\pm0.2$&$4.35\pm0.17$&$45.45\pm0.18$\\
121128A&$0.86\pm0.03$&$2.31\pm0.14$&$52.88\pm0.03$&$53.22\pm0.01$&$51.43\pm0.07$&$2.7\pm0.05$&$48.06\pm0.06$\\
121211A&$1.95\pm0.08$&$1.72\pm0.21$&$51.53\pm0.08$&$51.88\pm0.19$&$50.56\pm0.14$&$4.32\pm0.11$&$45.69\pm0.11$\\
130603B&$-0.88\pm0.05$&$3.2\pm0.15$&$50.29\pm0.02$&$51.29\pm0.11$&$49.58\pm0.06$&$3.23\pm0.05$&$46.08\pm0.06$\\
131105A&$1.62\pm0.02$&$2.61\pm0.14$&$52.69\pm0.03$&$53.14\pm0.07$&$50.77\pm0.1$&$3.23\pm0.08$&$47.1\pm0.08$\\
140430A&$1.82\pm0.01$&$2.09\pm0.2$&$51.83\pm0.08$&$52.16\pm0.07$&$50.86\pm0.17$&$4.26\pm0.16$&$45.86\pm0.16$\\
140512A&$1.95\pm0.01$&$2.41\pm0.09$&$52.27\pm0.01$&$52.71\pm0.04$&$51.4\pm0.06$&$3.88\pm0.06$&$46.92\pm0.06$\\
140518A&$1.03\pm0.02$&$2.4\pm0.4$&$52.58\pm0.04$&$52.92\pm0.01$&$50.94\pm0.16$&$2.53\pm0.12$&$47.86\pm0.08$\\
140703A&$1.21\pm0.3$&$2.51\pm0.15$&$52.9\pm0.03$&$53.21\pm0.1$&$52.13\pm0.04$&$3.54\pm0.03$&$47.66\pm0.04$\\
141121A&$2.35\pm0.03$&$2.3\pm0.15$&$52.45\pm0.03$&$52.78\pm0.12$&$51.18\pm0.07$&$5.13\pm0.05$&$45.52\pm0.07$\\
150323A&$1.97\pm0.03$&$2\pm0.11$&$51.73\pm0.01$&$52.06\pm0.01$&$49.66\pm0.13$&$3.96\pm0.11$&$45.32\pm0.09$\\
150424A&$1.85\pm0.09$&$2.55\pm0.2$&$50.51\pm0.03$&$51.09\pm0.12$&$49.51\pm0.2$&$4.94\pm0.18$&$43.96\pm0.17$\\
150910A&$1.68\pm0.13$&$2.58\pm0.15$&$52.34\pm0.04$&$52.81\pm0.08$&$51.73\pm0.15$&$3.56\pm0.07$&$47.65\pm0.12$\\
151027A&$1.86\pm0.02$&$2.17\pm0.1$&$52.11\pm0.04$&$52.46\pm0.03$&$51.34\pm0.02$&$3.34\pm0.02$&$47.78\pm0.02$\\
151112A&$0.58\pm0.42$&$2.57\pm0.21$&$52.46\pm0.05$&$52.81\pm0.17$&$51.54\pm0.18$&$4.06\pm0.16$&$46.55\pm0.15$\\
160227A&$1.97\pm0.09$&$2.35\pm0.33$&$52.59\pm0.03$&$52.90\pm0.04$&$51.63\pm0.1$&$3.85\pm0.09$&$47.18\pm0.08$\\
160327A&$0.67\pm0.12$&$2.58\pm0.14$&$52.78\pm0.03$&$53.09\pm0.01$&$51.23\pm0.13$&$2.84\pm0.11$&$47.55\pm0.13$\\
160804A&$1.92\pm0.05$&$1.96\pm0.07$&$52.19\pm0.01$&$52.49\pm0.01$&$49.7\pm0.1$&$3.72\pm0.07$&$45.88\pm0.04$\\
161108A&$1.69\pm0.05$&$2.13\pm0.17$&$51.57\pm0.04$&$51.91\pm0.09$&$50.16\pm0.15$&$4.03\pm0.14$&$45.82\pm0.08$\\
161129A&$1.33\pm0.02$&$2.27\pm0.1$&$51.57\pm0.01$&$51.96\pm0.03$&$50.26\pm0.09$&$3.34\pm0.07$&$46.59\pm0.11$\\
170113A&$0.84\pm0.08$&$2.34\pm0.49$&$51.78\pm0.04$&$52.13\pm0.13$&$51.41\pm0.07$&$3.16\pm0.06$&$47.61\pm0.06$\\
170202A&$1\pm0.1$&$2.61\pm0.11$&$52.93\pm0.01$&$53.28\pm0.04$&$51.22\pm0.07$&$2.65\pm0.06$&$48.01\pm0.04$\\
\enddata
\end{deluxetable*}

\end{document}